\documentclass[aps,rmp,
twocolumn,  
longbibliography,superscriptaddress,
floatfix]{revtex4-2}

\usepackage[T1]{fontenc} 
\usepackage[english]{babel} 

\usepackage{times}

\usepackage[table,usenames,dvipsnames]{xcolor}
\usepackage[
	colorlinks=true,
	hypertexnames=false
	]{hyperref}
\PassOptionsToPackage{linktocpage}{hyperref} 

\hypersetup{
  colorlinks   = true, 
  urlcolor = magenta!90!black,
  linkcolor    = blue!60!black, 
  citecolor=black!60 
}

\usepackage{amsthm} 
\usepackage{amssymb}
\usepackage{amsmath,mathtools,bbm}

\usepackage{multirow}
\usepackage{booktabs}

\usepackage{enumitem}
\usepackage[capitalize,compress]{cleveref}

\usepackage{pgfplots}
\usetikzlibrary{pgfplots.groupplots}
\pgfplotsset{compat=1.11}
\usepgfplotslibrary{fillbetween}

\usepackage{tikz}
\usepackage{pgffor}
\usetikzlibrary{
  arrows.meta,
  backgrounds,
  chains,
  decorations.pathreplacing,
  decorations.pathmorphing,
  fit,
  fadings,
  intersections,
  through,
  positioning,
  shapes,
  shapes.geometric,
  shapes.arrows,
  calc,
  3d
  }

\usepackage{algorithm}
\usepackage{algpseudocode}

\makeatletter
\renewcommand{\ALG@name}{Protocol}

\newcommand{\ipic}[3][-0.5]{\raisebox{#1\height}{\scalebox{#3}{\includegraphics{figures/#2}}}}

\definecolor{jens}{rgb}{0,.8,.5}

\makeatother
\usepackage{complexity}

\newclass{\stoqma}{StoqMA}
\newclass{\classP}{P}
\newclass{\bqp}{BQP}
\newclass{\qcam}{QCAM}
\newclass{\postbqp}{postBQP}
\newclass{\posta}{postA}
\newclass{\postiqp}{postIQP}
\newclass{\classa}{A}
\newclass{\bpp}{BPP}
\newclass{\fbpp}{FBPP}
\newclass{\pp}{PP}
\newclass{\cocp}{coC_=P}
\newclass{\ph}{PH}
\newclass{\np}{NP}
\newclass{\conp}{coNP}
\newclass{\gapp}{GapP}
\newclass{\approxclass}{Apx}
\newclass{\gapclass}{Gap}
\newclass{\sharpP}{\#P}
\newclass{\ma}{MA}
\newclass{\am}{AM}
\newclass{\qma}{QMA}

\newclass{\hog}{HOG}
\newclass{\quath}{QUATH}
\newclass{\bog}{BOG}
\newclass{\xeb}{XEB}
\newclass{\xhog}{XHOG}
\newclass{\xquath}{XQUATH}
\newclass{\maxcut}{MAXCUT}
\newclass{\sat}{SAT}
\newclass{\maxtwosat}{MAX2SAT}
\newclass{\twosat}{2SAT}
\newclass{\threesat}{3SAT}
\newclass{\sharpsat}{\#SAT}
\newclass{\se}{Sign Easing}
\newclass{\classx}{X}

\newtheorem{theorem}{Theorem}

\newtheorem{conjecture}[theorem]{Conjecture}
\newtheorem{definition}[theorem]{Definition}
\newtheorem{lemma}[theorem]{Lemma}
\newtheorem{problem}[theorem]{Problem}

\newtheorem{task}[theorem]{Task}

\newtheorem{corollary}[theorem]{Corollary}

\DeclareMathOperator{\tr}{Tr}
\DeclareMathOperator{\gap}{gap}
\DeclareMathOperator{\ngap}{ngap}
\newcommand{\ket}[1]{\vert{#1}\rangle}
\newcommand{\bra}[1]{\langle{#1}\vert}
\newcommand{\braket}[2]{\langle{#1}\vert #2 \rangle}
\newcommand{\proj}[1]{\ket{#1}\bra{#1}}

\newcommand{\abs}[1]{\vert #1 \vert}

\newcommand{\norm}[1]{\Vert #1 \Vert}

\newcommand{\tvd}[1]{\norm{ #1 }_{\mathrm{TV}}}
\newcommand{\ce}{\mathrm{CE}}

\newcommand{\kl}{D_{\mathrm{KL}}}

\DeclareMathOperator{\median}{med}

\DeclareMathOperator{\diag}{diag}

\newcommand{\mc}{\mathcal}
\newcommand{\mb}{\mathbb}
\newcommand{\mr}{\mathrm}

\newcommand{\id}{\mathbbm{1}}
\newcommand{\ee}{\mathrm{e}}
\newcommand{\ii}{\mathrm{i}}

\newcommand{\e}{\mathrm{e}}

\newcommand{\pbos}{P_{\mathrm{bs},U}}
\newcommand{\Gpbos}{P_{\mathrm{Gbs},U}}

\DeclareMathOperator{\Perm}{Perm}
\DeclareMathOperator{\Haf}{Haf}
\DeclareMathOperator{\Sym}{Sym}

\DeclareMathOperator*{\Eb}{\mb E}
\newcommand{\Cb}{\mathbb{C}}
\newcommand{\Rb}{\mathbb{R}}

\usepackage{xcolor}

\usepackage{xprintlen}

\makeatletter 
 \hypersetup{
  pdftitle = {
    Computational advantage of quantum random sampling
    },
  pdfauthor = {
    Dominik Hangleiter, 
    Jens Eisert},
  pdfkeywords = {
    quantum supremacy,
    quantum advantage,
    quantum random sampling, 
    random circuit sampling, 
    IQP circuits, 
    boson sampling, 
    Sycamore,
    Jiuzhang, 
    Google Quantum, 
    USTC,
    Xanadu, 
    gapp,
    complexity theory}
  }
\makeatother

\newcommand{\QuICS}{
Joint Center for Quantum Information and Computer Science (QuICS), University of Maryland \& NIST, College Park, Maryland 20742, USA}

\newcommand{\jqi}{Joint Quantum Institute (JQI), University of Maryland \& NIST, College Park, Maryland 20742, USA}

\newcommand{\fu}{Dahlem Center for Complex Quantum Systems, Freie Universit\"{a}t Berlin, 
14195 Berlin, Germany}

\newcommand{\hzb}{Helmholtz-Zentrum Berlin f{\"u}r Materialien und Energie, 14109 Berlin, Germany}

\newcommand{\hhi}{Fraunhofer Heinrich Hertz Institute, 10587 Berlin, Germany}

\usepackage[normalem]{ulem}

\begin{document}

\title{
Computational advantage of quantum random sampling
}

\author{Dominik Hangleiter}
\email[Corresponding~author: ]{mail@dhangleiter.eu}

\affiliation{\QuICS}
\affiliation{\jqi}

\author{Jens Eisert}
\affiliation{\fu}
\affiliation{\hzb}
\affiliation{\hhi}


\begin{abstract}
Quantum random sampling is the leading proposal for demonstrating a computational advantage of quantum computers over classical computers. 
Recently, first large-scale implementations of quantum random sampling have arguably surpassed the boundary of what can be simulated on existing classical hardware. 
In this article, we comprehensively review the theoretical underpinning of quantum random sampling in terms of computational complexity and verifiability, as well as the practical aspects of its experimental implementation using
superconducting and photonic devices and its classical simulation.
We discuss in detail open questions in the field and provide perspectives for the road ahead, including potential applications of quantum random sampling.  
\end{abstract}


\maketitle
\tableofcontents


\section{Introduction}
\label{sec:intro}
Dating back as far as to the 1980s, researchers have been thinking about what the computational power would be of computers the constituents of which are not following the laws of classical physics but rather those of quantum physics \cite{benioff_computer_1980,
feynman_simulating_1982, 
feynman_quantum_1985,
deutsch_quantum_1985}. 
Given that quantum mechanical systems allow for superpositions and entanglement, this might give rise to quite a distinct model of computation compared to the paradigmatic Turing machine model that captures classical computations.

Within the model of quantum computation \cite{deutsch_quantum_1985,bernstein_quantum_1997}, certain computational tasks can indeed be achieved much more efficiently than is possible using classical computing devices.
While for some problems such as database search \cite{grover_fast_1996} quantum computation offers polynomial speedups over classical algorithms, 
for others such as factoring integer numbers \cite{shor_algorithms_1994,shor_polynomial-time_1997} and simulating quantum systems \cite{lloyd_universal_1996} it even offers presumably exponential speedups. 

Within the framework of computational complexity theory, quantum computation has also been exponentially separated from classical computation via so-called \emph{oracle separations} \cite{simon_power_1994,simon_power_1997,bernstein_quantum_1993,bernstein_quantum_1997,raz_oracle_2019,yamakawa_verifiable_2022}. 
The advent of quantum error correction \cite{shor_fault-tolerant_1996} and
the threshold theorem \cite{aharonov_fault-tolerant_1997} brought the notion of quantum computation closer to reality showing that---at least in principle---errors can be corrected faster than they are generated, provided their rate is low enough. 

Since these discoveries, the search for applications of quantum computation has flourished \cite{QuantumAlgorithms,martyn_grand_2021}. 
Quantum algorithms have been discovered for solving `classical problems' such as solving structured linear equations \cite{PhysRevLett.103.150502}, solving systems of non-linear differential equations \cite{liu_efficient_2021}, 
and performing optimization tasks \cite{farhi_quantum_2000,QAOA,QuantumSDP}.
More sophisticated methods for quantum simulation have been devised, such as higher-order Trotter formulae \cite{childs_theory_2021}, qubitization \cite{low_hamiltonian_2019}, or linear combination of unitaries approaches \cite{ChildsWiebe},  
and we have a much better understanding of computational primitives possible in quantum computing in terms of the quantum singular value transform \cite{gilyen_quantum_2019} as a general way to process quantum signals \cite{low_hamiltonian_2019,low_optimal_2017}.

Today, there already is strong evidence that the dream of a universal quantum computer may become a reality in the not-too-far future. 
Quantum devices have been developed in a plethora of experimental platforms, ranging from ultracold atoms trapped in an optical-lattice potential \cite{bloch_many-body_2008}, Rydberg atoms in optical tweezers \cite{bernien_probing_2017} and trapped ions \cite{blatt_quantum_2012} to superconducting qubits \cite{clarke_superconducting_2008}, photonic platforms \cite{kok_linear_2007,Fusion} and silicon quantum dots \cite{zwanenburg_silicon_2013}.
Already for more than ten years, special-purpose analogue quantum simulators have been able to qualitatively simulate variants of the Hubbard model \cite{jaksch_cold_1998}, the Heisenberg model \cite{friis_observation_2018}, and other 
classically
intractable Hamiltonians with high precision and tunability of parameters at scales of up to tens of thousands of atoms \cite{MonteCarloValidator}.
While much smaller still, universal quantum devices are advancing at
a rapid pace.
Moving beyond the proof-of-principle demonstrations of quantum algorithms on small scales \cite{vandersypen_experimental_2000,vandersypen_experimental_2001},
first steps towards error-corrected quantum devices are being made at the moment 
\cite{ofek_extending_2016,
ryan-anderson_implementing_2022,
krinner_realizing_2022,
acharya_suppressing_2022,
egan_fault-tolerant_2021}.
The quest to actually build a universal, fault-tolerant quantum computer has now also reached industry \cite{Rigetti2018,arute_quantum_2019,Fusion,jurcevic_demonstration_2021}.
Quantum computing has thus expanded from an area of primarily academic interest to the consistent subject of news headlines around the world. 

However, the devices at our availability right now remain far from the error-correctable regime in terms of both error rates and the sheer number of qubits and quantum operations required for quantum error correction \cite{haner_factoring_2017,ogorman_quantum_2017,gheorghiu_benchmarking_2019,gidney_how_2019}. 
Available today are noisy universal quantum devices with up to roughly $50$ to $100$ physical qubits \cite{zhu_quantum_2022,arute_quantum_2019}, as well as special-purpose quantum simulators which allow for larger system sizes but lack universal programmability. 
When engineering those devices one is faced with the challenge of controlling individual quantum systems with a high degree of accuracy over long times, making their improvement and scaling a monumental challenge.

Given this profound challenge associated with building a universal, fault-tolerant quantum computer, one may---and should---ask whether we should even believe that quantum computations that outperform classical computation are physically possible? 
This is the question at the heart of this review.
The so-called \emph{extended Church-Turing thesis} states that any physically implementable model of computation can be efficiently simulated by a classical computer \cite{vergis_complexity_1986,bernstein_quantum_1997}. 
In particular, this thesis implies that quantum computers which exponentially outperform classical computers should not be possible. 
And indeed, in the entire history of computation, and despite the significant evolution of computing devices, no counter example---other than quantum computing---has been found, lending significant credibility to the thesis. 
Vice versa, the physical possibility of quantum computers challenges the extended Church-Turing thesis.

We can think of the extended Church-Turing thesis as a computational analogue of the thesis that nature must have a description in terms of a local and realistic theory \cite{einstein_can_1935}. 
Bell's inequalities \cite{bell_einstein_1964} quantitatively capture how quantum theory violates this thesis and provide a concise experimental setting to test local realism. 
The experimental violation of a Bell inequality \cite{freedman_experimental_1972,aspect_experimental_1982,aspect_experimental_1982-1} has once and for all falsified this belief and fundamentally changed the way we think about the interactions between the (local) constituents of our world. 
Reasonable sceptics will have been convinced of this since the last closable loopholes have been closed \cite{hensen_loophole-free_2015,shalm_strong_2015,giustina_significant-loophole-free_2015}.

An experimental violation of the extended Church-Turing thesis, called \emph{quantum advantage} or \emph{``quantum supremacy''} \cite{preskill2012quantum}, would mark a similar milestone for the field of computing.
From the perspective of computer science, it would demonstrate the physical possibility of computations that are not efficiently simulable in a classical Turing machine model. 
From the perspective of physics, it would demonstrate that quantum theory is applicable even in regimes that are not accessible by the means of computation we currently have. 

This gives rise to the question what a computational analogue of a Bell inequality as a means to test local realism is. 
In other words, what is (i) a simple task that can be performed on noisy and intermediate-scale quantum devices which is at the same time computationally difficult to simulate for classical computers both (ii) asymptotically and (iii) in practice using available computing hardware? 
And what could be (iv) a simple test that this task has been successfully and unambiguously achieved so that a reasonable sceptic can be convinced? 

All of these requirements are extremely challenging at different levels. 
The central complexity-theoretic challenge is to prove an asymptotic speedup of quantum computers over classical computers, a challenge that has remained elusive for several decades now. 
Next, given the intrinsic complexity of the task by the first requirement, a direct verification using only classical computing resources seems impossible at first sight. 
The final challenge is to actually build an intermediate-scale quantum computer that is able to outperform the classical supercomputers available today.
At the same time, it is a conceptual challenge to identify ways to fairly compare near-term quantum and large-scale classical computations solving the same task since their limitations are very different in nature: 
Roughly speaking, near-term quantum devices are limited by noise, while large-scale classical devices are limited by the size of the available computers.

A conceptually simple way to achieve these theoretical requirements is to make use of the quantum algorithm for integer factoring. 
This is because factoring is believed to be a problem for which no efficient classical algorithm exists. 
In fact, a large part of the presently applied public-key cryptography is based on the hardness of factoring. 
Factoring is particularly suited to public-key cryptography because it is believed to define a so-called one-way function, that is, a function which can be computed easily (the product of two large prime numbers) but which is extremely difficult to invert (finding those numbers given their product). 
Vice versa, this means that verifying a successful implementation of Shor's algorithm is simple: 
One simply has to multiply the output and compare it to the input.  
While proof-of-principle demonstrations of Shor's algorithm have been achieved \cite{vandersypen_experimental_2001}, factoring a large $2048$ bit number as is used for public-key encryption via \texttt{RSA} is estimated to require a large-scale, error-corrected universal quantum computer using roughly $20$ million physical qubits \cite{haner_factoring_2017,ogorman_quantum_2017,gheorghiu_benchmarking_2019,gidney_how_2019},
placing this algorithm outside the realm of what can realistically be achieved in the near future.
Hence, while impressive progress is being made along these lines of thought
\cite{barends_superconducting_2014,ryan-anderson_implementing_2022,acharya_suppressing_2022}, factoring cannot serve as a simple and near-term test of the computational advantage offered by quantum devices.   

A particularly natural class of problems for quantum computers are \emph{sampling problems}. 
Indeed, any quantum mechanical experiment can be seen as just being a sampling experiment: 
given an experimental prescription, a repeated measurement will provide intrinsically random measurement outcomes according to a probability distribution determined by the Born rule.
Almost 20 years ago, it was first observed that the patterns of measurement outcomes resulting from certain quantum computations could in fact be so complicated that classical computers would not be able to reproduce them \cite{terhal_adaptive_2004}. 

A simple class of computations to consider as a test of quantum devices are \emph{random quantum computations}. 
Such computations are presumably not computations that solve a relevant computational problem, but they may be useful in themselves, serving at the same time as a benchmark of a given computing device and as a test of quantum computational advantage. 
The task of sampling from the output distribution of a random quantum computation is called \emph{quantum random sampling}. 

In the past 20 years, significant evidence has accumulated that for a large variety of computations, and in particular for non-universal computations, this task is computationally intractable for classical computers \cite{shepherd_temporally_2009,aaronson_computational_2010,bremner_classical_2010,bremner_averagecase_2016,boixo_characterizing_2016,bouland_complexity_2018,movassagh_quantum_2020,kondo_fine-grained_2021,bouland_noise_2021,krovi_average-case_2022,fujii_commuting_2017}.
At the same time, there is significant evidence that current-day supercomputers have a very hard time simulating this task even for small systems comprising roughly $50$ to $100$ subsystems \cite{markov_quantum_2018,neville_classical_2017,bulmer_boundary_2022,pan_solving_2022,huang_classical_2020}.
Very recently, quantum random sampling in a classically intractable regime has been claimed to be achieved experimentally on a universal quantum processor comprising 53 qubits \cite{arute_quantum_2019}, and up to 60 qubits \cite{zhu_quantum_2022,wu_strong_2021}, as well as using photonic systems \cite{zhong_phase-programmable_2021,zhong_quantum_2020,madsen_quantum_2022}.

In this review, we provide a detailed overview of quantum random sampling as a test of the presumed exponential computational advantage of quantum computers over classical ones. 
We show in what precise way quantum random sampling can be seen as a computation. 
We explain what that computation solves, in what way it outperforms classical computations, what methods of verification are available, and what challenges arise in this context.

In the first part, we focus on the theoretical aspects of quantum random sampling:
the question of how to prove an asymptotic quantum speedup, and the question of whether and how quantum random sampling can be verified. 
Here, we explain in detail how the key idea of \textcite{terhal_adaptive_2004} to relate the hardness of sampling to the hardness of computing probabilities has been developed further in recent years. 
Building on the idea to show a collapse of the so-called \emph{polynomial hierarchy} \cite{bremner_classical_2010,aaronson_computational_2010} based on the classical hardness of computing quantum probabilities \cite{valiant_complexity_1979,fujii_commuting_2017} and the assumed availability of an efficient classical sampler, 
this idea has been further developed to allow for certain errors in the implementation \cite{aaronson_computational_2010,bremner_averagecase_2016}, and brought closer to experimental implementation \cite{boixo_characterizing_2016,lund_boson_2014,hamilton_gaussian_2017,bermejo-vega_architectures_2018}. 
The question of how to verify quantum random sampling has first been addressed by \textcite{shepherd_temporally_2009}, and it has been pointed out that in its most restrictive forms, classical verification is unviable \cite{gogolin_bosonsampling_2013,aaronson_bosonsampling_2013,hangleiter_sample_2019}. 
This notwithstanding, weaker forms of classical verification turn out to be indeed possible \cite{aaronson_bosonsampling_2013,boixo_characterizing_2016,arute_quantum_2019}, albeit at a potentially prohibitive computational cost \cite{arute_quantum_2019}.

In the second part, we discuss the practical aspects of quantum random sampling, in particular, experimental implementations and concrete classical simulation algorithms for quantum random sampling. 
In the context of experimental implementation, it is key to fully understand and analyse the noise which remains present on the device in order to devise as-robust-as-possible schemes \cite{boixo_characterizing_2016,arute_quantum_2019}. 
Likewise, from the perspective of classical simulation, a central question is what features of a scheme obstruct classical algorithms \cite{markov_quantum_2018,aaronson_complexity-theoretic_2017}, and, vice versa, how to best exploit ``weaknesses'' of a scheme or a verification method in order to devise faster simulation algorithms \cite{pan_simulation_2022,gao_limitations_2021,clifford_faster_2020,bulmer_boundary_2022}.

It is important to stress that the topic at hand is highly conceptual in nature, so that a precise understanding of the underlying premises and an appreciation of the fine print that comes along are key. 
For this reason, we have made the deliberate choice of keeping the exposition precise and accurate in most places, sometimes using formal language, while at the same time pedagogically introducing all required concepts. 

What we do not discuss in this review are ways to demonstrate a quantum advantage by other means. 
Particularly prominent examples are the discovery of verifiable proofs of quantumness \cite{brakerski_cryptographic_2018,brakerski_simpler_2020,kahanamoku-meyer_classically-verifiable_2021} for which there are recent proof-of-principle demonstrations \cite{zhu_interactive_2021}. 
These schemes demonstrate access and control over a single qubit via a cryptographic encoding. 
Recent work by \textcite{yamakawa_verifiable_2022} makes great progress along these lines by devising a verifiable proof of computational quantum advantage based on certain random computations. 
In this sense, it is at the interface of quantum random sampling and cryptographic proofs of quantumness. 
Presumably, none of these methods can be implemented at a scale required for a quantum advantage in the intermediate term, however \cite{zhu_interactive_2021,liu_depth-efficient_2021,hirahara_test_2021}. 

Before we start, let us also point the reader to more concise and briefer reviews of quantum advantage \cite{HarrowSupremacy}, quantum random sampling \cite{lund_quantum_2017}, and implementations of boson sampling \cite{brod_photonic_2019} that may serve as starting points into the literature.
It is also worth mentioning the excellent textbook by \textcite{nielsen_quantum_2010}, which covers the basics of quantum computing that we do not explain here.

We begin this review by setting the stage and stating what a quantum random 
sampling scheme is in the first place in \cref{sec:sampling schemes}. 
Here, we define universal circuit sampling, instantaneous quantum polynomial time (IQP)
circuit sampling, boson sampling, and Gaussian boson sampling; but we also hint at other schemes.
\cref{sec:complexity theory} explains the basics of computational complexity to the extent they are needed in \cref{sec:hardness} to show the computational hardness of quantum random sampling on classical computers.
This detailed discussion constitutes the heart of this review: It is precisely this fine print that is needed to appreciate the significance of experimental implementations of quantum random sampling. 
\cref{sec:verification} is concerned with the question of how to verify the correctness of the implementation of a quantum random sampling scheme. 
In \cref{sec:experimental implementations}, we then detail the to-date experimental implementations of quantum random sampling. 
\cref{sec:classical simulation} then overviews methods of simulation 
run on classical supercomputers that aim to challenge quantum implementations in their computational power. 
Finally, in \cref{sec:perspectives}, we put the findings into perspective and discuss a wealth of open questions, as an invitation to taking further steps, in particular, to explore potential applications of quantum random sampling.


\section{Quantum random sampling schemes}
\label{sec:sampling schemes}

Every experiment in quantum physics can be viewed as a sampling experiment: 
Measurement outcomes are intrinsically random, sampled from a probability distribution determined by the Born rule. 
Sampling problems are therefore natural candidates exhibiting specifically quantum features. 
The most prominent example of a quantum-classical divide is for a specific quantum 
sampling problem that cannot be reproduced classically under locality constraints: 
the violation of a Bell inequality \cite{bell_einstein_1964}. 
Similarly, in terms of computational complexity, we expect it to be difficult to reproduce the experimental outcomes of generic quantum computations. 
And indeed, we can think of the corresponding experiments as violating a computational equivalent of the Bell inequality. 
The reasons for why we expect generic computations to be hard to simulate are manifold and not precisely understood---the exponentially growing Hilbert space dimension, quantum interference leading to non-positive amplitudes, and entanglement are only some examples of distinctly quantum features obstructing classical simulation algorithms.
Roughly speaking, generic quantum computations explore the entire state space available, providing no structure that can be exploited by a classical  simulation algorithm. 
Consequently, so the reasoning, the  runtime of such an algorithm must be determined by the exponential Hilbert space dimension. 

In order to make the intuition rigorous that generic quantum computations give rise to sampling problems that are classically intractable, the idea of quantum random sampling has been introduced. 
In quantum random sampling problems, a quantum computation is drawn at random according to some specification. 
The task is then to sample from the Born rule distribution generated by this random quantum computation. 
Crucially, there are now two notions of randomness at play: 
First, the randomness of the computation itself, which is classical randomness 
used to draw the computation at random.
Second, the intrinsically quantum randomness of individual outcomes sampled from the output distribution of that computation. 
Such quantum random sampling schemes are not only hard to simulate by the known classical simulation algorithms already at comparably small scales, but we can also give complexity-theoretic evidence for asymptotic intractability. 
Importantly, such evidence is \emph{independent} of specific algorithms and regards the intrinsic complexity of the problem by reducing it to a paradigmatic computational problem that can be independently studied and therefore much stronger than merely the failure of our known simulation algorithms.
Quantum random sampling schemes are particularly appealing for demonstrations of quantum advantage because, as we will see, the complexity-theoretic argument even applies to certain non-universal computations that may be comparably easy to experimentally implement. 

A quantum random sampling scheme is defined by the random choice of a quantum computation realized by a quantum circuit. 
A \emph{quantum circuit} describes an arrangement of quantum gates from a certain \emph{gate set} in some spatial and temporal order,
acting on a specific set of individual quantum systems, here often taken to be qubits.
In a random quantum circuit individual quantum logic gates are chosen at random from a given gate set and applied to input registers according to a certain rule.
For a fixed input size $n$, e.g., the number of qubits in a random quantum circuit, this gives rise to a family of computations, realized as a \emph{circuit family}, denoted by $\mc C_n$.
The classical sample space $\Omega$ comprises the possible measurement outcomes. 
\begin{task}[Quantum random sampling]
\label{task:quantum random sampling}
Given as input a problem size $n$ and a circuit $C$ chosen at random from a family 
$\mc C_n$, sample from the output distribution $p(C)$ of the circuit applied to a reference state $\ket 0$\footnote{Throughout this review, we use the term `state' both for density operators~$\rho$ and for state vectors~$\ket \psi$ in the underlying Hilbert space.}, with the probability
of an outcome $S \in \Omega$ given by
\begin{align}
  \label{eq:p(U)}
  p_S(C) = |\bra S C \ket 0 |^2 . 
\end{align} 
\end{task} 
Depending on whether the emphasis lies on the probability distribution over the circuits $C$ or the outcomes $S$ of a fixed circuit, we at times use $p_S(C)$ and at other times $p_C(S)$ for the outcome probabilities.

In the remainder of this section, we formally introduce the most important schemes---universal circuit sampling, IQP 
circuit sampling and boson sampling. 
Those schemes recurrently appear over the course of this review in which we discuss their and similar schemes' properties.
This includes, not only their complexity-theoretic analysis (\cref{sec:hardness}) and the question in how far classical samples from their output distributions can be verified (\cref{sec:verification}), but also their experimental implementations (\cref{sec:experimental implementations}) and specific classical simulation schemes (\cref{sec:classical simulation}).  

\subsection{Universal circuit sampling}
\label{subsec:universalcircuitsampling}

The most prominent example of a quantum random sampling scheme, or rather, family of 
random sampling schemes, is \emph{universal circuit sampling}. 
The rationale behind universal circuit sampling is to explore the entire Hilbert space available in 
small- or intermediate-scale experiments as quickly as possible. 
This is why it is also a universal circuit sampling scheme which has been implemented to experimentally demonstrate a computational quantum advantage for the first time~\cite{arute_quantum_2019}. 

In universal circuit sampling, quantum gates are drawn from a gate set which is universal for quantum computation: 
that is, any quantum computation could be implemented with gates drawn from this set. 
The gates are placed at certain positions in a quantum circuit architecture, which might be fixed or random. 
The circuit might also contain other non-random gates.

\begin{figure*}[t]
\includegraphics{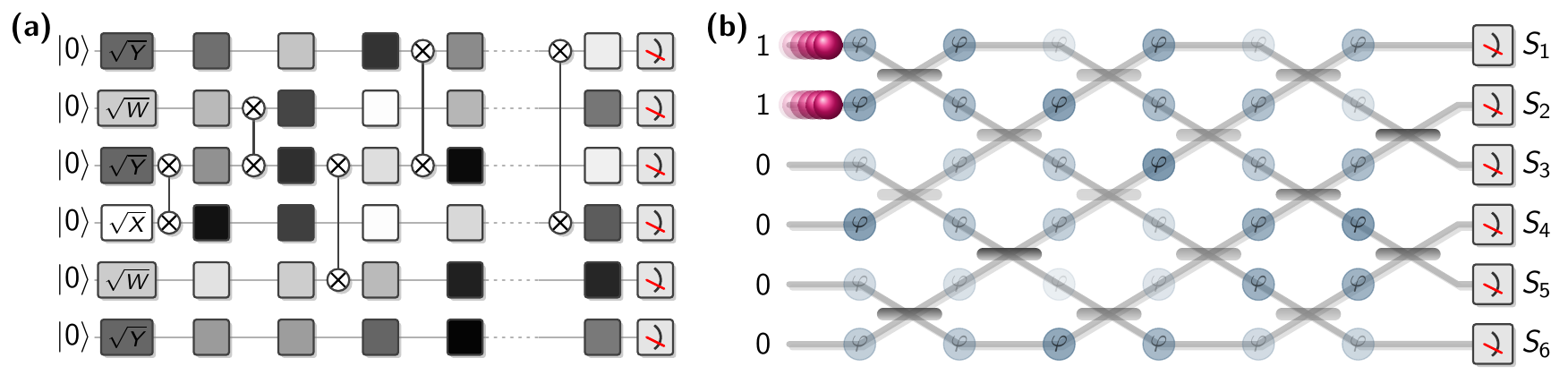}

\caption{
\label{fig:circuit diagrams google boson sampling} 
Circuit diagrams for \emph{(a)} random universal circuits as performed in the experiment by \textcite{arute_quantum_2019} with random single-qubit gates comprised from the gate set $\sqrt X, \sqrt Y, \sqrt W$ and fixed two-qubit entangling gates iSWAP$^*$ at fixed positions in the circuit, and \emph{(b)} boson sampling, where passive linear optics comprising beam splitters and phase shifters are applied to a Fock input state  $\ket{1^n}$ and then measured in the Fock basis with outcomes $S_{i}$. 
}
\end{figure*}

For example, in the experiment of \citet{arute_quantum_2019} a very specific type of random circuit is applied: in every layer of the circuit random single-qubit gates are applied to every qubit, and a specific two-qubit entangling gate is applied to each edge of a square lattice in a particular sequence, see \cref{fig:circuit diagrams google boson sampling}(a). 
The single-qubit gates are drawn from the set $\{ \sqrt X, \sqrt Y, \sqrt W\}$ in such a way that and the same single-qubit gate is not allowed to sequentially repeat. 
Here 
\begin{align}
   X = \begin{pmatrix} 0 & 1 \\ 1 & 0 \end{pmatrix}, \hspace{1em} Y = \begin{pmatrix} 0 & - \ii \\ \ii & 0 \end{pmatrix}, \hspace{1em} Z  =\begin{pmatrix} 1 & 0 \\ 0 & -1 \end{pmatrix}, 
\end{align} 
denote the \emph{Pauli matrices} and $W = (X + Y)/\sqrt 2$. 
The entangling gates are given by the \emph{iSWAP}-like gate 
\begin{align}
\label{eq:iswapstar}
  \text{iSWAP}^* = \begin{pmatrix}
    1 & 0 & 0 & 0 \\
    0 & 0 & -\ii & 0 \\
    0 & - \ii & 0 & 0 \\
    0 & 0 & 0 & \ee^{- \ii \pi/6}
  \end{pmatrix} .
\end{align}

As a toy model of random universal circuits which is theoretically very appealing, consider a continuous gate set $\mc G= U(4)$ comprising all two-qubit gates.
In this model, a depth-$N$ random circuit $C$ acting on $n$ qubits is constructed by choosing a uniformly random gate in $G \in \mc G$ according to the Haar measure and the pair of qubits it is applied to at random~\cite{brandao_local_2016}. 
Alternatively, we can apply the gates in a parallel architecture in which each layer of the circuit comprises random gates from $\mc G$ applied in parallel to all qubits.

\subsection{IQP circuit sampling}

A prominent family of random quantum sampling schemes 
that uses restricted gate sets is given by so-called 
\emph{instantaneous quantum polynomial time} (IQP) circuits~\cite{shepherd_temporally_2009}. 
An IQP circuit is a commuting quantum circuit which is diagonal in the Hadamard basis. 
Such a circuit can always be written as $C = H^{\otimes n } D H^{\otimes n}$, where $D$ is diagonal in the computational basis and 
\begin{align}
  H = \frac 1 {\sqrt 2} \begin{pmatrix}
    1 & 1 \\ 1 & -1
  \end{pmatrix}
\end{align} 
denotes the Hadamard gate.
IQP circuits appear naturally in the context of measurement-based quantum computation \cite{raussendorf_one-way_2001}. 
Instances of IQP circuit families are defined by diagonal circuits comprised of diagonal $2$-qubit gates with arbitrary phases on the diagonal \cite{nakata_generating_2014} and circuits of $Z$, controlled-$Z$ ($CZ)$, and controlled-controlled-$Z$ ($CCZ$) gates, which flip the phase of the target qubit iff the control qubit ($CZ$) or qubits ($CCZ$) are in the 
$\ket 1$ state \cite{bremner_averagecase_2016}.  
But one can also phrase IQP circuits in the language of Hamiltonian time evolution. 
In this language, an IQP circuit is given by the constant-time evolution under an Ising Hamiltonian with edge weights chosen in a specific way \cite{bremner_averagecase_2016}.
In this formulation, one can generalize IQP circuits to arbitrary multi-qubit interactions---so-called $X$-programs \cite{shepherd_temporally_2009}.
Another natural family of random computations in this model of computation is given by preparing a so-called \emph{cluster state}~\cite{raussendorf_one-way_2001,raussendorf_measurement-based_2003} on a square lattice
and performing random local rotations around the $Z$-axis \cite{,haferkamp_closing_2020}.
This model bridges a gap to quantum simulation as it can be implemented using translation-invariant Hamiltonians \cite{gao_quantum_2017,bermejo-vega_architectures_2018}.

Two specific examples of IQP circuit families, which are theoretically clean and help us 
illustrate important concepts in the subsequent sections, have been introduced by \textcite{bremner_averagecase_2016}. 
An instance $C_{f}$ of the first family is defined by a degree-$3$ Boolean polynomial $f: \{0,1\}^n \rightarrow \{0,1\}$ over the field $\mb F_2 = (\{0,1\}, \oplus, \cdot)$ as
\begin{align}
\label{eq:iqp boolean polynomial}
  f(x) = \sum_{i,j,k}\alpha_{i,j,k} x_i x_j x_k + \sum_{i,j} \beta_{i,j} 
  x_i x_j + \sum_i \gamma_i x_i , 
\end{align}
with Boolean coefficients $\alpha_{i,j,k}, \beta_{i,j}, \gamma_i \in \{0,1\}$ denoting whether or not a $CCZ$, $CZ$ and $Z $ gate is applied to qubits $(i,j,k)$, $(i,j)$ and $i$, respectively. 

An instance of the second family is defined by an adjacency matrix $w$
with entries chosen from a set of angles $A = \{0, \pi/4, \ldots, 7\pi/4\}$ as 
\begin{equation}
\label{eq:iqp circuit weights}
C_{w} 
= \exp \left[\mr i \left (\sum_{i < j} w_{i,j} X_i X_j + \sum_i w_{i,i} X_i \right) \right] \ ,
\end{equation}
where $X_i$ is the Pauli-$X$ matrix acting on site $i$.
In other words, on every edge $(i,j)$ of the complete graph on $n$ qubits, a gate $\exp(\mr i w_{i,j} X_i X_j) $ with edge weight $w_{i,j}$ and on every vertex $i$ a gate $\exp(\mr i w_{i,i} X_i) $ with vertex weight $w_{i,i}$ is performed.

\subsection{Boson sampling}

The boson sampling scheme, due to \citet{aaronson_computational_2010}, is one of the most prominent and historically earliest quantum random sampling schemes. 
The conception of this scheme has its origins in the computational difficulty of computing the permanent of a matrix. 
The permanent turns out to describe the output distributions of interfering free bosons, such as single photons interfering on a beam splitter. 
The complexity of computing the permanent has its correspondence in a surprising physical effect---photon bunching.
The experimental observation of photon bunching in the famous Hong-Ou-Mandel experiment \cite{hong_measurement_1987} is one of the landmark experiments of quantum optics, being among the first to  experimentally confirm quantum entanglement.
In this experiment, two photons interfere on a beam splitter and are measured in the photon-number basis. 
Surprisingly, for indistinguishable photons one only ever observes zero or two photons in one of the modes but never one photon in each mode. 

The boson sampling problem generalizes this experiment. Now, we increase the number of photons and let them interfere in a complex network of beam splitters: 
$n$ photons are injected into the first $n$ of $m \in \poly(n)$ modes.
Those photons interfere in a linear-optical network comprising beam splitters and phase shifters which is chosen in such a way that it gives rise to a Haar-random unitary transformation of the input modes, given by $U \in U(m)$. 
Finally, the $m$ output modes of the network are measured in the photon-number basis; see \cref{fig:circuit diagrams google boson sampling}(b).
As unitary mode transformations conserve the total photon number, the sample space of boson sampling is given by
\begin{equation} \label{eq:bosonsamplingsamplespace}
  \Phi_{m,n} = \Big\{ (s_1,\dots,s_m) : \sum_{j=1}^m s_j = n \Big\} ,
\end{equation}
i.e., the set of all sequences of non-negative integers of length $m$ which 
sum to $n$.
Its output distribution  is
\begin{equation} \label{eq:bosonsamplingdistribution}
  p_U(S) \equiv \pbos(S) = |\bra S \varphi(U) \ket{1_n} |^2 .
\end{equation}
Here, the state $\ket S$ is the Fock state corresponding to a measurement outcome $S \in \Phi_{m,n}$, $\ket{1_n}$ is the initial state with $1_n = (1,\dots, 1,0,\dots, 0)$, and $\varphi(U)$ the Fock space representation of the mode transformation $U$.

In order to clearly distinguish the boson sampling protocol of \textcite{aaronson_computational_2010} with output probabilities given by \cref{eq:bosonsamplingdistribution} from its variants---discussed below---we will henceforth refer to it as \emph{Fock boson sampling}.

\subsection{Gaussian boson sampling}

Variants of the boson sampling protocol play with the input state and measurement basis. 
Most importantly, so-called \emph{Gaussian boson sampling} protocols start from a Gaussian 
quantum state, 
where the input modes are prepared in single-mode or two-mode squeezed states~\cite{lund_boson_2014,rahimi-keshari_what_2015,hamilton_gaussian_2017,kruse_detailed_2019,grier_complexity_2022}, or displaced squeezed states \cite{huh_boson_2015,quesada_franck-condon_2019}. 
The distribution of outcomes $S \in \Phi_{m}$ is given analogously to \cref{eq:bosonsamplingdistribution} by 
\begin{equation} 
\label{eq:Gaussianbosonsamplingdistribution}
\Gpbos(S) = |\bra S \varphi(U) \ket{g} |^2 ,
\end{equation}
where $\ket{g}$ is the initial Gaussian quantum state. 
Here, the sample space
\begin{equation}\label{eq:Gaussianbosonsamplingsamplespace}
	\Phi_{m} =\Bigl\{
	(s_1,\dots, s_m) \in \mb N_0^{m}
	\Bigr\}
\end{equation}
reflects an unbounded photon number, as Gaussian states do not feature a fixed photon number.
Similarly, we can also think of the reverse, where a photon-number state is prepared in the input and Gaussian measurements are performed \cite{lund_exact_2017,chabaud_continuous-variable_2017,chakhmakhchyan_boson_2017}. 

Gaussian boson sampling protocols are appealing in comparison to the original proposal as Gaussian states and measurements are experimentally much easier to implement than photon-number states and measurements. And, indeed, it is those protocols for which large-scale experiments have been performed recently~\cite{zhong_quantum_2020,zhong_phase-programmable_2021,madsen_quantum_2022}.

\subsection{Further schemes}
\label{ssec:further schemes}

Since the first quantum random sampling schemes---IQP sampling \cite{bremner_classical_2010} and boson sampling \cite{aaronson_computational_2010}---have been conceived, many more proposals 
for quantum random sampling schemes have been put forward. 
A theoretically particularly clear proposal is so-called ``Fourier sampling'' \cite{fefferman_power_2015}, 
which is a qubit analogue of boson sampling. 
Another analogue of boson sampling is \emph{fermion sampling} \cite{oszmaniec_fermion_2020}, for which so-called ``magic states'' are required in the input, and the closely related matchgates with magic state inputs \cite{hebenstreit_all_2019}. 
The fermionic schemes that make use of resource states as an input find their qubit analogue in Clifford circuits with magic-state inputs \cite{yoganathan_quantum_2018,hangleiter_anticoncentration_2018}. 
The so-called 
\emph{one clean qubit}
(DQC1) model is a model in which all but one qubit are initialized in the maximally mixed state \cite{fujii_impossibility_2018,morimae_hardness_2014,morimae_hardness_2017}. 
This model is motivated by mixed-state quantum computations, which is a
suitable framework to capture, for instance, nuclear magnetic resonance 
quantum processors \cite{negrevergne_liquid-state_2005}. 
Other proposals include Clifford circuits which are conjugated by arbitrary product 
unitaries \cite{bouland_complexity_2018}, and permutations of distinguishable particles in specific conditions \cite{aaronson_computational_2016}. 
Finally, certain models have also been 
proposed with the goal to close loopholes such as the 
necessity to certify the correct 
implementation of a quantum supremacy experiment 
\cite{hangleiter_direct_2017,miller_quantum_2017}, or to make such an 
experiment more error-tolerant \cite{fujii_noise_2016,kapourniotis_nonadaptive_2019}. 

In what follows, we discuss the properties of those schemes with respect to the possibility of using them to demonstrate a computational advantage over classical computations. 
Before we dive into the main focus of this review, the complexity-theoretic argument for the classical 
intractability of \cref{task:quantum random sampling}, let us review some basics of computational complexity theory in the next section.


\section{Computational complexity of simulating quantum devices}
\label{sec:complexity theory}

The quantum random sampling schemes introduced above have been devised to show computational quantum advantages of quantum devices over classical supercomputers.
There are two ways in which we can understand this goal: 
First, we can understand it in terms of the actual time required to simulate an actual experiment performing quantum random sampling. 
This is the realm of concrete algorithm development and a quantum advantage in this sense is reached as soon as available supercomputers running state-of-the-art algorithms are no longer capable of providing samples from the desired distribution. 
Second, we can understand it in terms of the asymptotic scaling of the best possible classical simulation algorithm. 
This is the realm of computational complexity theory. 
Computational complexity theory studies classes of problems in terms of their intrinsic complexity in an algorithm-agnostic way.
We can therefore supplement evidence towards the first type of quantum advantage using computational complexity theory. 
This can help us to hedge against a ``lack of imagination'' in classical algorithm development.

Think of the related context of cryptography: in order for us to be confident in the security of a certain cryptographic scheme, it is key that this scheme is not just based on some problem on which known algorithms do not perform well. 
Rather, we want to collect additional evidence and---ideally---underlying reasons that in fact no algorithm can efficiently solve the problem on which the scheme is based.
It is such additional, independent evidence that computational complexity theory can contribute to quantum random sampling. 

Here, we will precisely explicate the available evidence for the classical intractability of quantum random sampling, making the intuition that quantum devices
are more powerful than classical ones more rigorous. 
We will see which ingredients come together in a strategy to provide  complexity-theoretic evidence for the hardness of sampling from, or \emph{weakly simulating}, the sampling schemes defined above.
These results will constitute the complexity-theoretic underpinning of experimental prescriptions designed to demonstrate quantum computational supremacy, that is, to experimentally violate the extended Church-Turing thesis.

The argument is rather intricate, however, and builds on some basic results about the computational complexity of approximately computing the output probabilities of, or \emph{strongly simulating}, quantum circuits, and algorithms for this task.  
In this section, we review those results, before we leverage them to weak simulation in the next section, \cref{sec:hardness}. 

\subsection{Basics of computational complexity theory}

In order to provide theoretical evidence for quantum advantage,
we have to enter the realm of theoretical computer science. 
There, classes of problems, so-called \emph{complexity classes}, are studied with respect to their \emph{computational complexity}, that is, the resources that an 
algorithm designed to solve problem instances from such a class would require in the worst case. 
In computational complexity theory, we can discern distinct problem classes defined by certain resource restrictions, most importantly the runtime and the memory requirement of algorithms. 
Understanding the relations between different complexity classes, that is, separations and inclusions between them is the main subject of study in the theory of computational complexity. 
For convenience, most often \emph{decision problems} are considered, where the task is to decide whether a given string\footnote{We write the set of all finite-length bit strings as $\{0,1\}^* = \bigcup_{n\in \mb N} \{0,1\}^n$. } $x \in\{0,1\}^*$ is in a so-called \emph{language} $L \subset \{0,1\}^*$, which is a set of bit strings. 
A machine that computes the Boolean  function $f_L:\{0,1\}^* \rightarrow \{0,1\}$, which satisfies $f_L(x) = 1 \Leftrightarrow x \in L$, decides $L$. 
For example, a language $L$ could be given by the set of all graphs for which there exists a path that visits each vertex once, in binary encoding, and a string $x \in L$ is the binary encoding of a particular graph instance.

The central concept of computational complexity theory is that of an \emph{algorithm}. 
In a simplified picture, we can think of an algorithm as computing a Boolean function $f:\{0,1\}^* \rightarrow \{0,1\}$ for arbitrary-length inputs. 
Abstractly speaking, an algorithm is a set of rules according to which a machine acts on any given input. 
In the case of classical algorithms, formalized as a Turing machine, those rules may involve \emph{reading} bits of the input or a scratch pad and \emph{writing} bits to that scratch pad, \emph{choosing} a new rule according to which to continue, or \emph{stopping} and outputting either $0$ or $1$ \cite{arora_computational_2009}. We say that an algorithm is efficient if its runtime scales polynomially in the input size, given by the length~$|x|$ of $x$.  

On an actual silicon-chip computer, those rules can be implemented using certain elementary logic operations that are applied sequentially (or in parallel) to some of the input registers (bits) at a time. 
The elementary logical operations might act on a single register or bit such as the $\texttt{NOT}$ operation, on two such as $\texttt{OR}$ and $\texttt{AND}$
or even more registers. 
A set of such operations is said to be universal if an arbitrary Boolean function $f: \{0,1\}^n \rightarrow \{0,1\}$ can be expressed as a classical circuit using $\poly(n)$ many input registers. 
A \emph{classical circuit} is a mathematical model of an arrangement of classical gates implementing a logical operation that is chosen from a certain set in some spatial and temporal order
computing a Boolean function.
Examples of such universal sets of logical operations are $\{\texttt{AND},\texttt{NOT}\}$ and the singleton $\{\texttt{NAND}\}$. 
Using a sequence of universal logical operations, one can therefore express any other elementary logical operation. 
A classical circuit $C_n$ effectively computes a function of the values of its $n$ input registers, potentially using additional auxiliary registers. 
On input $x \in \{0,1\}^n$, its outcome $C_n(x) \in \{0,1\}$ is given by its value on a single---say, the first---output register. 
The size of a circuit $|C_n|$ is given by the number of gates in it. 
We call the model of computation in which we can execute classical circuits the circuit model.

Notice that any given circuit takes inputs of a fixed size $n$, while of an algorithm we demand that it works for any input size. 
We can turn a family of circuits $\{C_n\}_{n \in \mb N}$ into a meaningful algorithm\footnote{Indeed, if we ask merely for the existence of a circuit family as opposed to an efficient algorithm then this allows us to solve undecidable problems using polynomial-size circuits.} by supplementing it with an efficient instance-generating procedure that given the input size $n$ efficiently produces a description of $C_n$, which is then run on the input $x\in \{0,1\}^n$.
We call circuit families for which such a procedure is possible \emph{uniform circuit families}. Uniform circuit families are therefore a realisation of an algorithm in the circuit model. 

The fundamental class of problems in computational complexity theory  is the class \classP, the class of problems which can be solved efficiently on a deterministic classical computer. 

\begin{definition}[\classP]
	A language $L\subset \{0,1\}^*$ is in the class \classP\ if there exists a classical algorithm $\mc A$ that, given $x \in\{0,1\}^*$ as an input, decides whether $ x \in L $ in polynomial runtime in $|x|$: 
  \begin{align}
    x \in L \quad \Leftrightarrow \quad \mc A(x)=1.
  \end{align}
\end{definition}
Relations between complexity classes are typically studied with respect to polynomial reductions---so-called \emph{Cook reductions}---where access to a machine in \classP\ is granted. 
A key problem in the theory of computational complexity is that the relation between different complexity 
classes defined with very different resource restrictions in mind is inherently hard to pin down. 
For this reason, basic relations between complexity classes are therefore often merely conjectured based on the available evidence. 
The most basic and at the same time most fundamental separation in complexity theory is the belief that $\classP \neq \np$. 
While \classP\ is the class of problems which can be efficiently \emph{computed} on a classical computer, \np\ is the class of problems which can be efficiently \emph{verified}.  
\begin{definition}[\np]
\label{def:np}
A language $L \subset \{0,1\}^*$ is in the class \np\ if there exists a polynomial $p : \mb N \rightarrow \mb N$ and a polynomial-time classical algorithm $\mc V$ (called the verifier for $L$) such that for every $x \in \{0,1\}^*$, 
\begin{align}
x \in L \quad \Leftrightarrow \quad \exists y \in \{0, 1\}^{p(|x|)} :  \mc V(x, y) = 1 . 
\end{align}
We call $y$ the \emph{proof} of $x$. 
\end{definition}

When gathering evidence for a separation between quantum and classical computation, quantum and classical sampling in particular, we want to try and keep as close to problems that have been well-studied such as the conjecture $\classP \neq \np$. 
The main challenge is that, at the same time, the computational task must be such that it can realistically be realized on near-term quantum devices in as easy and error resilient a way as possible. 

\subsection{Where to look for a quantum-classical separation?}
\label{sec:quantum-classical separation}

In order to better understand the complexity theory of quantum computing we compare to its closest cousin,  randomized classical computation.\footnote{In this section, we 
follow a line of thought which to the best of our knowledge is due to 
Scott Aaronson @ \url{https://www.scottaaronson.com/blog/?p=3427}.} 
We formalize randomized classical and quantum computations in terms of decision problems as complexity classes \bpp\ and \bqp. 
\begin{definition}[Classical and quantum computation]
\label{def:bpp,bqp}
	\bpp\ (\bqp) is the class of all languages $L \subset \{ 0,1\}^* $ for which there exists a polynomial-time randomized classical (quantum) algorithm with uniform circuit family $\{C_n\}_{n \in \mb N }$ such that for all $n \in \mb N $ and all inputs $x \in \{ 0,1 \}^n$
	\begin{align}
		x \in L \quad  & \Rightarrow \quad \Pr [C_n(x) = 1] \geq 2/3, \\
		x \notin L \quad  & \Rightarrow \quad \Pr [C_n(x) = 1] \leq 1/3,
	\end{align}
	where the probability is taken over the internal randomness of the algorithm. 
\end{definition}

Classical computations are modelled as intrinsically deterministic; only by artificially introducing randomness into the circuit do we construct a randomized classical algorithm using elementary logic gates. 
A randomized algorithm for a Boolean function $f: \{0,1\}^n \times \{0,1\}^{\ell}
\rightarrow \{ 0,1\}$ acts on both the problem input $x \in \{0,1\}^n$ and a uniformly random bit string $r \in \{0,1\}^{\ell}$ with $\ell \in \poly(n)$.
Clearly, randomized algorithms are at least as powerful as deterministic one, as such a function can simply disregard the random inputs, giving rise to a deterministic algorithm. 
In many practical situations, randomized algorithms turn out to be much more efficient than deterministic algorithms, however.

While classical logical gates are not generally \emph{reversible} in that the mapping from input to output is injective, it turns out that one can implement any classical computation in a circuit that uses only reversible operations~\cite{toffoli_reversible_1980,fredkin_conservative_1982}. 
In other words, there are sets of reversible operations such as the three-bit Toffoli, or controlled-controlled-\texttt{NOT}, gate \texttt{TOF}~\cite{toffoli_reversible_1980} such that an arbitrary Boolean function can be expressed using the outcome of a single register in a computation involving only those operations. 

By taking the leap to reversible classical computation we have already made it halfway to quantum computation. 
Indeed, the question about the possibility of reversible classical computation has originally been motivated by the observation that the laws of physics are reversible~\cite{fredkin_conservative_1982}.
Hence, so the thought, a physical model of computation should be, too. 

Quantum circuits are a generalization of reversible classical circuits. 
A quantum circuit acts on \emph{qubits} the state space of which is given by $\mb C^2$.
The elementary operations or quantum gates are unitary matrices acting on a $k$-qubit input space $(\mb C^{2})^{\otimes k}$, where $k$ is a small number; typically $k=2$. 
A quantum circuit acting on $m \in \poly(n)$ qubit registers produces not a single bit string as an output but a quantum state in $(\mb C^{2})^{\otimes m}$, which only upon a quantum measurement in some basis---typically the standard basis---produces a bit string as an output. 
Indeed, we notice that classical computation is a special case of quantum computation: 
If we restrict to state preparations and measurements in the standard basis and permutation matrices in that basis (which are in particular unitary), then we recover classical computation.

A quantum gate set $\mc G$ is said to be \emph{computationally universal} if an arbitrary quantum circuit acting on $n$-qubits and using $t$ gates can be simulated by a circuit composed of gates from $\mc G$ up to error $\epsilon$ with overhead $\polylog(n,t,  1/\epsilon)$ in terms of both the number of registers and gates~\cite{aharonov_simple_2003}. 
With polynomial overhead in $n$ and $t$, computational universality therefore tolerates errors of the order $2^{-\poly(n,t)}$. 
A computationally universal gate set that will serve us well in due course is the set $\{H,\texttt{TOF}\}$ consisting of the Hadamard and the Toffoli gate. 
This gate set is universal for $n$-qubit computations when acting on $n+1$ many qubits~\cite{aharonov_simple_2003}. 

In contrast to classical computations, quantum computations are intrinsically randomized---the probability that an $n$-qubit quantum circuit $C_n$ applied to an input state $\ket{x} \in \mb C^n$ results in a particular outcome $y$ after a measurement is given by the \emph{Born rule} as $| \bra y  C_n \ket x |^2 $. 
We also call these probabilities the \emph{output probabilities} of $C_n$. 
Indeed, it is (presumably) not possible to separate out the randomness from the computation as it is for classical computations. 

A key but very subtle difference between quantum and randomized classical computations presents itself in the guise of the probability that such computations accept.
This difference is a lever that allows us to separate the two types of algorithms in terms of their computational power.

\subsection{Computing acceptance probabilities of randomized algorithms}
\subsubsection{Classical acceptance probabilities}

We start by discussing acceptance probabilities of classical randomized algorithms before turning to quantum algorithms.
The acceptance probability 
\begin{align}
\label{eq:sharppsum}
	\Pr[C_n(x) = 1] = \frac{1}{2^{p(|x|)}} \sum_{r \in \{0,1\}^{p(|x|)}} f_x(r), 
\end{align}
of a classical randomized circuit $C_n(x)$ computing a Boolean function $f_x$ is given by the fraction of accepting random inputs $r \in \{0,1\}^{p(|x|)}$, where $p: \mb N \rightarrow \mb N$ is some polynomial.
Computing the (unnormalized) acceptance probability of classical circuits is therefore clearly a \sharpP-complete problem.\footnote{
Given a complexity class \classx, we say that a problem is \classx-hard if it is at least as hard as any problem in \classx\ in the sense that all problems in the class are polynomial-time reducible to it. 
We say that it is \classx-\emph{complete} if it is in \classx\ and \classx-hard. 
}
\begin{definition}[\sharpP~\cite{arora_computational_2009}]
	\label{def:sharpp}
	The \emph{function class} \sharpP\ is the class of all functions $f: \{0,1\}^* \rightarrow \mb N$ for which there exists a polynomial-time classical algorithm $C$ and a polynomial $p: \mb N \rightarrow \mb N$ such that
	\begin{align}
		f(x)  = \left | \left \{ y \in \{ 0,1 \}^{p(|x|)}: C(x,y) = 1 \right \} \right | . 
	\end{align}
\end{definition}
In other words, \sharpP\ functions by definition count the number of accepting inputs to a polynomial-time computation $C$.
In contrast to \bpp\ and \bqp, which are classes of \emph{decision problems}, \sharpP\ is therefore a class of \emph{counting problems}.
In turn, we can view the decision class \np\ (Def.~\ref{def:np}) as asking to decide whether there exists \emph{any input} such that a computation $C$ accepts.

\subsubsection{Quantum acceptance probabilities}

We say that a quantum computation with circuit $C_n$ accepts an input $x$, if a measurement on $C_n \ket x$ results in one of a set of accepting outcomes $\Gamma_{\rm acc}$. 
The acceptance probability of the computation is then given by  
\begin{align}
  \Pr[C_n(x)=1] = \sum_{y \in \Gamma_{\rm acc}} | \bra y C_n \ket x|^2. 
\end{align}
For the following argument, it will be sufficient to consider the set of accepting outcomes to be $\Gamma_{\rm acc} = \{0\}$, where $0 \equiv 0^n$ denotes the all-zero outcome string, cf.\ \cite{fenner_determining_1998}.
The acceptance probability of $C_n$ is then just given by a single output probability $ \Pr[C_n(x)=1] = | \bra 0 C_n \ket x|^2$. 

We can express the acceptance probabilities of a polynomial-size quantum circuit $C_n$ on input $x \in \{0,1\}^\ell$ via a function $g_x: \{ 0,1\}^{p(\ell)} \rightarrow \{ +1,-1\}$ for some polynomial $p$ as \cite{dawson_quantum_2005,montanaro_quantum_2017}\footnote{We highly recommend the introduction to Boolean functions and their relation to quantum output probabilities by \textcite{montanaro_quantum_2017}.
} 
\begin{align}
	\label{eq:gappsum}
	\Pr[C_n(x) = 1] = \frac 1 {2^{p(\ell)}} \sum_{y\in \{ 0,1\}^{p(\ell)}} g_x(y). 
\end{align}
This is easily seen using the fact that the gate set comprising the Hadamard and the Toffoli gate is universal for quantum computing\footnote{As discussed above, since Hadamard and Toffoli are computationally universal~\cite{shi_both_2002,aharonov_simple_2003}, the acceptance probability of an arbitrary polynomial-size computation can be expressed as the acceptance probability of such a circuit up to an error $\epsilon$ with an overhead of $\polylog (1/\epsilon)$. 
This means that we can obtain an $O(2 ^{-\poly(n)})$ approximation of this acceptance probability. 
We will shortly come to a more detailed discussion of such approximations and the question how hard it is to compute them.}. 
In this gate set, we can express the all-zero amplitude of an $n$-qubit computation $C_n = C^{(t)} \cdots C^{(1)}$ using $t$ quantum gates $C^{(1)}, \ldots, C^{(t)}$ \cite{dawson_quantum_2005}
\begin{align}
\label{eq:path integral unitary time evolution}
	\bra 0 C_n \ket x & = \sum_{\lambda_1, \ldots, \lambda_t} \bra 0 C^{(t)} \ket{ \lambda_1} \cdots \bra {\lambda_t} C^{(1)} \ket x \\
	& = \frac 1 {\sqrt {2^h}} \sum_{y} s_x(y), 
\end{align}
in terms of the number $h$ of Hadamard gates and a signed function $s_x$, with input space size given by the number of paths leading from $x$ to $0$, which is bounded by $4^{t}$ for a circuit consisting of two-qubit gates, and hence for polynomial-size circuits as $2^{\poly(n)}$.
This is because the matrix elements of the Toffoli gate are binary and those of the Hadamard gate are $\pm 1/\sqrt{2}$ so that each entry of the matrix product $C^{(t)} \cdots C^{(1)}$ is a sum of numbers $(\pm 1) \cdot 2^{- h/2} $. 
We thus obtain
\begin{align}
\label{eq:acceptance probability quantum circuit}
	| \bra 0 C_n \ket x |^2 = \frac 1 {2^h} \left| \sum_{y} s_x(y) \right|^2 = \frac 1 {2^h} \sum_{y,z} g_x(y,z),    
\end{align}
where $g_x(y,z) = s_x(y) s_x(z)$.

Notice the subtle difference in the range of the function $g_x$ versus the range of the function $f_x$ arising in classical computation: 
while $f_x$ is Boolean, $g_x$ takes values in $\{ +1,-1\}$. 
We can view this difference between Boolean and signed functions as a signature of \emph{quantum interference} as it allows for the possibility of cancelling paths famously demonstrated in the Hong-Ou-Mandel experiment which we discussed in the introduction. 

But we can easily translate back and forth between signed and Boolean functions via the map 
$g'_x(y) =   (g_x(y) +1)/2$ and reexpress 
\begin{align}
\label{eq:gapp}
	\sum_{y \in \{0,1\}^{p(|x|)}} g_x(y)  = \left |\{y:  g'_x(y)  = 1 \} \right| - \left |\{y:  g'_x(y)  = 0 \} \right|.  
\end{align}
Notice that $g'_x$ is again a Boolean \sharpP\ function.
The sum \eqref{eq:gappsum} can be viewed as the difference between the accepting paths of the function $g'_x$ and its rejecting paths or, in other words, the \emph{gap} of that function. 
For a Boolean function $f: \{0,1\}^n \rightarrow \{0,1\}$ the gap is defined as 
\begin{align}
	\gap(f) = \left |\{y:  f(y)  = 1 \} \right| - \left |\{y:  f(y)  = 0 \} \right|, 
\end{align}
which we normalize to 
\begin{align}
	\ngap(f) = \frac{1}{2^{n}} \gap(f). 
\end{align}
This is why computing functions whose values can be written as the gaps of \sharpP\ functions is complete for a class called \gapp. 
\begin{definition}[\gapp~\cite{fenner_gap-definable_1994}]
\label{def:gapp}
	Define the \emph{function class} \gapp\ as the class of all functions $f:\{0,1\}^* \rightarrow \mb Z$ for which there exist $g,h \in \sharpP$ such that $f = g - h$. 
\end{definition}
Conversely, given a \gapp\ function $g: \{0,1\}^\ell \rightarrow \{ -2^{p(\ell)}, \ldots ,2^{p(\ell)}\}$ for a polynomial $p$, we can find an $n$-qubit quantum circuit $Q_g(x)$ with $n \in \poly(\ell)$ which has acceptance amplitude $ \bra {0^n} Q_g(x) \ket {0^n} =g(x)/2^n$ \cite{fenner_determining_1998,kondo_fine-grained_2021}.
To see this, we observe that for every \gapp\ function $g$ there is a polynomial-time computable function $G(x,y)$ such that $g(x) = |\{y \in \{0,1\}^{p(|x|)}: G(x,y) = 1\}| - |\{y \in \{0,1\}^{p(|x|)}: G(x,y) = 0\}|$. 
With the diagonal poly-size circuit $D_x = \sum_{y \in \{0,1\}^{n}} (-1)^{G(x,y)} \proj y$, we then find that $Q_g(x) =  H^{\otimes n} D_x H^{\otimes n}$ has acceptance amplitude $g(x)/2^{n}$. 

Altogether, we have found that acceptance probabilities of a classical circuit are given by the fraction of accepting paths of \sharpP\ functions, while the acceptance probabilities of a quantum circuit $C_n$ can be expressed as the absolute value of the normalized gap of a \sharpP\ function $f_0$ as
\begin{align}
	|\bra {0^n} C_n \ket {0^n} |^2 = |\ngap(f_0)|^2.
\end{align}

How are \gapp\ and \sharpP\ related in terms of their computational complexity? 
We have already seen a simple mapping between the two, which implies that computing \gapp\ and \sharpP\ functions is equivalent under Cook reductions\footnote{We write a complexity class \textsf{X} in the exponent of another class \textsf{Y} to mean that a machine in \textsf{Y} can call an \emph{oracle} with access to a machine solving arbitrary problems in the class \textsf{X} at unit time cost.} which we write as 
\begin{align}
\label{eq:gapp sharpp equivalence}
	\classP^\gapp = \classP^\sharpP . 
\end{align}
So in this sense the two classes are very similar. 
But they actually turn out to be very distinct once we turn to the hardness of \emph{approximating} the respective sums \eqref{eq:sharppsum} and \eqref{eq:gappsum} up to a multiplicative error $c$. 

\subsection{Approximating \gapp}
\label{sec:approximating gapp}

Hereafter, we distinguish the following notions of approximation: 
We say that for $c \in (0,1]$ an estimator $s$ provides a \emph{$c$-multiplicative approximation} of the value $S$ if 
\begin{align}
 	c S \leq s \leq S/c  . 
 \end{align} 
We say that for $r > 0 $ it is a \emph{$r$-relative approximation} if 
\begin{align}
	(1 - r) S \leq s \leq (1 + r) S, 
\end{align}
and an \emph{$\epsilon$-additive approximation} for $\epsilon > 0 $ if 
\begin{align}
	| S - s | \leq \epsilon . 
\end{align}
To intuitively see why there might be a difference in approximability, notice
that a \sharpP\ sum over $m$-bit strings takes on values between $0$ and $2^m$. 
Typically, the values will therefore be on the order of $2^m$ so that a constant relative error is also of that order. 
Conversely, \gapp\ sums take on values between $-2^m$ and $+2^m$, but as the corresponding \sharpP\ function takes on an exponentially large value, the value of the \gapp\ function is the difference between two such exponentially large numbers. 
This difference will in general be much smaller than each individual value so a relative error is, too. 

Importantly, a relative-error approximation of a quantity is guaranteed to have the correct sign. 
In contrast to relative-error approximations of \sharpP\ functions which always have a positive sign, relative-error approximations of \gapp\ function therefore teach us nontrivial sign information. 
In fact, this information is already sufficient to learn the \emph{exact value} of any \gapp\ function up to arbitrary relative error.
\begin{lemma}[Approximating \gapp]
\label{lem:gapp approximation}
	Let $f$ be a \sharpP\ function. 
	Then approximating $\gap(f)$ up to \emph{any} constant multiplicative error is \gapp-hard.
\end{lemma}

A detailed proof of \cref{lem:gapp approximation} is provided, for example, by \textcite[Chapter 2.2]{hangleiter_sampling_2021}. 
The basic idea is to use a \gapp\ oracle to iteratively compute the gap of a function $f_s$ that is shifted compared to the gap of $f$ by $s2^n$. 
We can then compare the signs of the two gaps and vary the value of $s$ to perform binary search.

For a function class \classx\ we define $\approxclass_{\cdot c} \classx$ as the class of problems which can be solved by approximating $\sum_xf(x)$ up to a multiplicative error $c$ for $f \in \classx$. 
We have now found that for any $c \in (0,1)$
\begin{align}
	\classP^{\approxclass_{\cdot c} \gapp} = \classP^\gapp. 
\end{align}

The attentive reader will have noticed that in our discussion of the hardness of approximating \gapp\ using the sign information we have glossed over the fact that, of course, acceptance probabilities of quantum circuits are \emph{non-negative}. 
And indeed, it seems unlikely that those acceptance probabilities are hard to approximate up to any constant multiplicative error. 

Nevertheless, using a similar proof strategy one can prove \gapp-hardness of approximations for the square of the output amplitudes of quantum circuits~\cite{terhal_adaptive_2004,aaronson_computational_2010,goldberg_complexity_2014,fujii_commuting_2017}. 
This strategy notices that not only do multiplicative-error approximations get the sign correct, but certainly also the instances in which the true value is exactly zero.
What is more, there is a trivial additive-error robustness given by the spacing of the values of a (normalized) \sharpP\ function. 

\begin{lemma}[Approximating the absolute value of $\ngap$]
\label{lem:normalized gap approximation}
Let $f:\{0,1\}^\ell$ be a \sharpP\ function. Then approximating $\ngap(f)^2$ up to 
  \begin{itemize}
    \item[(a)] any relative error $\epsilon < 1/2 $, or 
    \item[(b)] additive error $1/2^{2n} $ with $n \in \poly(\ell)$,
  \end{itemize}
is \gapp-hard. 
\end{lemma}
\begin{proof}[Proof sketch]
For part (b) we note that additive-error robustness $1/2^{2n}$ is trivial since the spacing of the function $\ngap(f)$ is given by $2/2^n$, i.e., twice the normalization of $\gap(f)$ in the definition of $\ngap(f)$. 

For part (a) of the proof we proceed similarly to the proof of Lemma~\ref{lem:gapp approximation} above, following \textcite[Proposition~8]{bremner_averagecase_2016}.
The idea of the proof is to estimate $\ngap(f)$ by using the fact that given a  guess $c$, an algorithm that outputs relative-error approximation to $|\ngap(f) - c |$ can certify the correctness of $c$.

In the first step, we show that there is a polynomial-size classical circuit $C$ acting on $p(\ell)+\ell+1$ registers for some polynomial $p: \mb N \rightarrow \mb N$ that computes a shifted function $f_c$ such that $\ngap(f_c)= (\ngap(f) - c)/2$ for some $c \in [-1,1]$ such that $c = 2 k/2^{p(\ell)} $ with $k \in \mb N$.
To this end we make use of the following: 
for any polynomial $p:\mb N \rightarrow \mb N$ there is a polynomial-size circuit $D_c$ acting on $p(\ell)$ registers computing a function $g$ such that $\ngap(g) = -c$. 
Now consider the polynomial-size circuit $Q_c$ acting on $p(\ell) + \ell + 1$ registers which executes either $C$ or $D_c$ depending on the control register. 
This circuit computes a function $f_c$ as desired.

Assume we have an efficient algorithm $\mc A$ that given a circuit $C$ approximates $\ngap(f_c)$
up to relative error $\epsilon < 1$. 
On input $Q_c$ this machine can certify whether $\ngap(f) = c$. 
We now use $\mc A$ to estimate $\ngap(f)$ using a sequence of guesses $c_0, c_1,\dots\,$ for its value, until we have found its exact value.
At each step, we have a guess $c_i$ for $c$, starting with $c_0= 0$. 
We use $\mc A$ to output an estimate $d_i$ to $|\ngap(f) - c_i|$ and then apply it again to output an estimate $d_i^{\pm}$ of $|\ngap(f) - (c_i \pm d_i)|$.
Define $c_{i+1}  = c_i + d_i$ if $d_i^+ \leq d_i^-$ and as $c_i - d_i$ otherwise. 

The algorithm acts contractively: 
Assuming $c < \ngap(f)$ we find that an estimate $d = ( 1 + \gamma) | c - \ngap(f)|$ for some $|\gamma| < \epsilon$ satisfies 
\begin{align}
	| c + d - \ngap(f)| = | \gamma(\ngap(f)-c)| \leq \epsilon | c- \ngap(f)| ,
\end{align}
and a similar inequality holds for $c -d $ if $c > \ngap(f)$. 
Consequently, since $\ngap(f) $ is an integer multiple of $2/2^n$, if the correct choice of $c \pm d$ is made in each step, the algorithm halts after $O(n)$ many steps. 

It remains to be shown that the algorithm indeed halts after $O(n)$ steps.
This can be seen from the equivalence
\begin{multline}
(1 + \epsilon)| c + d - \ngap(f) | < (1-\epsilon) | c - d - \ngap(f)|
\\
\Leftrightarrow ( 1 + \epsilon) |\gamma| < ( 1- \epsilon) | 2 + \gamma| , 
\end{multline}
which holds for $|\gamma| \leq \epsilon < 1/2 $. 
The same argument immediately holds for $|\ngap(f)|$ as we have not used the sign of $\ngap(f)$. 
\end{proof}

Given the mapping of gaps to output amplitudes of quantum circuits described above, it therefore follows directly from \cref{lem:normalized gap approximation} that approximating the output probabilities of quantum circuits is \gapp. 
\begin{corollary}[Approximating output probabilities of quantum circuits]
\label{lem:gapp hardness quantum probabilities}
Approximating the output probabilities $| \bra{0^n} C \ket{0^n}|^2$ of an $n$-qubit quantum circuit comprising $m$ gates is \gapp-complete up to 
\begin{enumerate}[label = (\alph*)]
  \item any relative error $\epsilon <1/2$, or 
  \item exponentially small additive error $1/2^{2n}$.
\end{enumerate}
\end{corollary}

\subsection{Approximating \sharpP: Stockmeyer's algorithm}
\label{sec:stockmeyer}

For many \sharpP-complete problems such as computing the value of the permanent 
of a matrix taking values in $\{0,1\}$, there are efficient randomized approximation schemes, 
including the so-called 
\emph{fully polynomial randomized approximation scheme}
FPRAS~\cite{jerrum_polynomial-time_2004}.
Many such algorithms for approximate counting are based on Markov-chain Monte Carlo methods~\cite{jerrum_random_1986,jerrum_polynomial-time_1990}.
The property that those algorithms exploit is the fact that each element of the sum~\eqref{eq:sharppsum} is non-negative. 
Thus, the sum can be estimated by \emph{importance sampling}, 
that is, sampling its elements according to their (normalized) weight in the sum. 
Insofar, the intricate sign structure of \gapp\ functions is what makes their relative-error approximation via such sampling algorithms hard. 

Going beyond specific algorithms, in this section, we will get to know a powerful general result on the approximability of such functions by a computationally restricted algorithm with access to an \np\ oracle due to \textcite{stockmeyer_complexity_1983}. 
Stockmeyer's algorithm is able to approximately count the number of accepting paths of \sharpP\ functions up to small multiplicative errors even though it is not able to exactly compute this number. 
It thus provides a rigorous foundation for the distinction between the approximability of \gapp\ and \sharpP. 
In the next section, \cref{sec:hardness}, we leverage the power of this algorithm to derive rigorous separations between classical and quantum \emph{sampling algorithms}. 

Before we are able to make those statements precise, however, we need to dive a little further into the depths of computational complexity theory and define what is called the \emph{polynomial hierarchy}. 
Stockmeyer's algorithm lies in the third level of the polynomial hierarchy. 
This class is much more powerful than \np, but much less powerful than \sharpP.

\subsubsection{The polynomial hierarchy}

We have already seen the most important classes in the theory of computational complexity, namely, \classP\ and \np. 
It is no exaggeration to say that the conjecture that $\classP \subsetneq \np$ is indeed one of the if not \emph{the} most tested and studied unproven statement that scientists across a range of disciplines are confident in. 
Among other things, this intuition rests on the presumed existence of problems whose solutions are hard to find but easy to verify.
In particular, the possibility of public-key cryptography is based on the existence of such problems. 
It is a generalization of this statement that forms the complexity-theoretic grounding of claims to quantum supremacy. 
This generalization posits that the levels of an infinite hierarchy of complexity classes---the so-called \emph{polynomial hierarchy}---are strict subsets of one another.
Considering hypothetical algorithms within and outside of this hierarchy also allows us to understand the computational complexity of approximating \sharpP\ functions.  

\begin{definition}[The polynomial hierarchy~\cite{arora_computational_2009}]
\label{def:ph}
	For $i \in \mb N_0$ a language $L \subset \{ 0,1\}^*$ is in $\Sigma_i$ if there exists a polynomial $q$ and a uniform polynomial-time circuit family $\{ C_n\}_{n\geq 1}$ such that $x \in L$ if and only if
	\begin{multline}
		\exists u_1 \in \{ 0,1\}^k \, \forall u_2 \in \{ 0,1\}^k \cdots Q_i u_i \in \{ 0,1\}^k : \\ 
		C_{|x|}(x,u_1, \ldots, u_i) = 1, 
	\end{multline}
	where $k = q(|x|)$ and $Q_i$ denotes a  $\forall $ or $\exists$ quantifier depending on whether $i$ is even or odd, respectively. 
	The \emph{polynomial hierarchy} $\ph $ is the set $\cup_i \Sigma_i$. 
\end{definition}

\begin{figure}
  \includegraphics{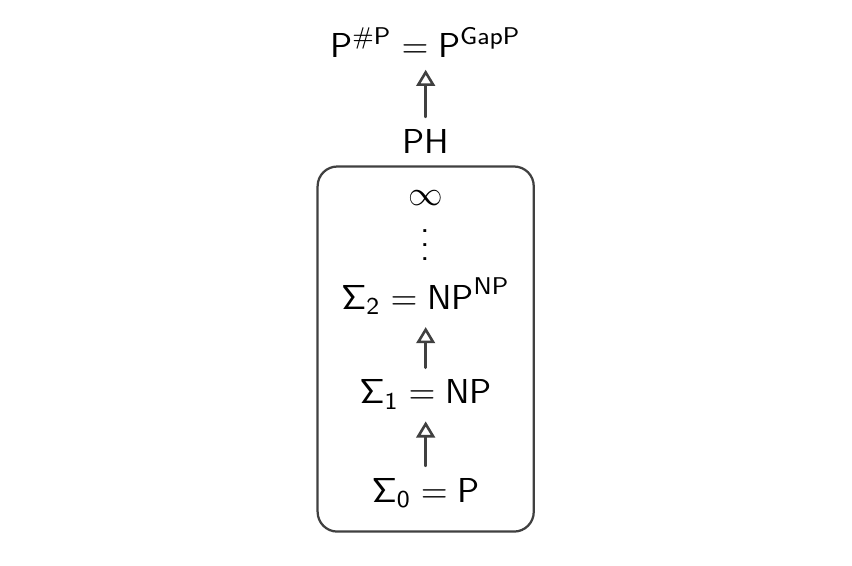}
  \caption{
  \label{fig:ph} The polynomial hierarchy is a hierarchy of complexity classes defined by adding consecutive \np\ oracles, where any layer is presumed to strictly contain all lower-lying layers. Toda's theorem (\cref{thm:toda}) states that the polynomial hierarchy is contained in $\classP^\sharpP$.  
  }
\end{figure}

Clearly $\Sigma_i \subset \Sigma_{i+1}$. Notice that \np $= \Sigma_1$ since in its definition (\ref{def:np}) there is only a single $\exists$ quantifier.
We can then equally characterize $\Sigma_i$ as $\Sigma_{i-1}^\np$, so in each level an additional \np-oracle is added, see Figure~\ref{fig:ph}.  
Intuitively, as we add alternating $\exists$ and $\forall$ quantifiers, the complexity of the problems solved by the circuit family $\{C_n\}$ strictly increases. 
Conversely, if two levels of the hierarchy coincide then so will all other levels above those.
Indeed, it is a central conjecture that the polynomial hierarchy is infinite, i.e., that every level strictly contains the previous levels. 
Stated in other words, the conjecture is that ``the polynomial hierarchy does not collapse''. 

\subsubsection{Stockmeyer's approximate counting algorithm}

Indeed, it is no surprise that, given access to \np\ oracles one can solve an enormously rich class of computational problems.
Nevertheless, it is quite surprising that one can efficiently approximate exponentially large sums up to any inverse polynomial \emph{multiplicative error}.
Stockmeyer's approximate counting algorithm \cite{stockmeyer_complexity_1983} achieves this task in a low level of the polynomial hierarchy---the third level. 
We are now ready to state this result. 

\begin{theorem}[\cite{stockmeyer_complexity_1983,aaronson_computational_2010}]
\label{thm:stockmeyer} 
Given a Boolean function $f : \{ 0, 1\}^n \rightarrow \{0,1\}$, let 
\begin{equation}
    p = \Pr_{x\in \{0,1\}^n} [f(x) = 1] = \frac{1}{2^n}
    \sum_{x\in \{0,1\}^n} f(x)\ . 
\end{equation} 
Then for all $c \geq 1 + 1/\mathrm{poly}(n)$, there exists an $\fbpp^\np$ machine\footnote{\fbpp\ is the function-class equivalent of the decision class \bpp, that is, the class of functions computable in probabilistic polynomial time with bounded failure probability.}
that approximates $p$ to within a multiplicative factor of~$c$. 
\end{theorem} 
See \cite{trevisan_lecture_2008} and \cite[Chapter 2.3]{hangleiter_sampling_2021} for a sketch of the proof. 
\cref{thm:stockmeyer} characterizes the complexity of approximately counting up to an inverse polynomially small multiplicative error: 
Since $\bpp \subset \Sigma_2$ \cite{lautemann_bpp_1983} and therefore $\bpp^\np \subset \Sigma_3$, this task lies within the third level of the polynomial hierarchy. 
But where does this complexity class lie in relation to exactly computing a \sharpP\ sum? 
For the answer, we refer to a final fact in complexity theory, namely that exactly computing \sharpP\ functions lets one solve \emph{any task} in \ph. 
\begin{theorem}[Toda's theorem~\cite{toda_pp_1991}]
\label{thm:toda}
	\begin{align}
	\ph \subset \classP^\sharpP . 
	\end{align}
\end{theorem}
The complexity of counting \sharpP\ sums is therefore significantly easier when considering multiplicative approximations as opposed to exact computation. 
Conversely, we have already seen above in Eq.~\eqref{eq:gapp sharpp equivalence} and Lemma~\ref{lem:gapp approximation} that \gapp\ does not change its complexity under multiplicative approximations so that the following inclusions hold 
\begin{align}
	P^{\approxclass_{\cdot c} \sharpP} \subset \Sigma_3 \subsetneq \ph \subset \classP^{\gapp} =  \classP^{\approxclass_{\cdot c}\gapp} , 
\end{align}
for any constant $c > 0$, since 
$\classP^\gapp = \classP^\sharpP \supset \ph $.
The separation $\Sigma_3 \subsetneq \ph$ marks the conjectured non-collapse of the polynomial hierarchy to any finite level.
The same inclusions hold true when restricted to \gapp-functions with non-negative gap for values of $c < 1/2$.

We have now carved out a substantial difference in complexity between quantum and randomized classical algorithms in terms of the computational complexity of approximating the respective acceptance probability to high precision. 
To describe quantum acceptance probabilities, negative signs are required and hence they are \gapp-hard to approximate up to relative error. 
Conversely, classical acceptance probabilities can be expressed as sums over nonnegative numbers and hence approximating them up to relative error is in the class $\Sigma_3$. 
Let us stress again that neither the quantum nor the classical algorithm should be able to multiplicatively approximate the respective acceptance probabilities because the classes $\Sigma_3$ and $\gapp$ are not expected to be contained in \bpp\ and \bqp, respectively. 
Nevertheless, this difference in complexity serves as an important tool using which we can amplify harder-to-pin-down differences in the runtime of actual classical and quantum algorithms. 
Following this route, we will arrive at a (conditional) exponential separation for sampling tasks.

\section{Computational complexity of quantum random sampling}
\label{sec:hardness}

\subsection{Sampling versus approximating outcome probabilities}
\label{ssec:sampling vs approximating}

Our goal in this section is to prove not only that there is an exponential quantum/classical divide in \emph{approximating output probabilities} of computations, but also that this divide reappears when it actually comes to \emph{performing such computations}, that is, perform the corresponding sampling. 
Randomized algorithms indeed seem to be the perfect playground, where we might see a quantum advantage since any quantum computation naturally produces random samples from the distribution determined by the Born rule, while classical randomized algorithms require external randomness. 

In order to make a rigorous statement about randomized computations, we consider the task of sampling from a given distribution, not caring about a specific outcome of the computation. 
To be able to apply the machinery of complexity theory and Stockmeyer's algorithm in particular, it in addition proves useful to consider the task of sampling from \emph{randomly chosen quantum computations}. 

The key idea that we use to make a rigorous statement about the complexity of classical and quantum sampling is to relate the task of sampling from a distribution to computing its output probabilities.
In doing so, we leverage the complexity-theoretic difference between computing classical and quantum output probabilities to classical and quantum sampling.
The key technical ingredient when doing so is Stockmeyer's algorithm. 
We observe that Stockmeyer's counting theorem (\cref{thm:stockmeyer}) can be directly applied to estimating the acceptance probability, and in fact all output probabilities, of so-called \emph{derandomizable sampling algorithms}, which are  deterministic algorithms with random inputs as discussed above~\cite[cf.][Def.~3.11 and the proof of Thm.~1.1]{aaronson_computational_2010}. 
\begin{definition}[Derandomizable sampling]
\label{def:derandomizable}
A \emph{derandomizable sampling algorithm} is an algorithm $\mathcal{A}$ that takes as an input a particular instance $y \in \{0,1\}^n$ of a problem, 
as well as a uniformly random string $r \in \{0,1 \}^{\poly(|y|)}$ and outputs a random bit string 
$x = \mathcal{A}(y,r)$ distributed according to 
\begin{align} 
p_{y}(x) = \Pr_r [ \mathcal{A}(y,r) = x ] \ .
\label{eq:derandomizable distribution}
\end{align}
\end{definition}

If $\mathcal{A}$ is such a derandomizable algorithm we can use Stockmeyer's algorithm to estimate its output probabilities \eqref{eq:derandomizable distribution}. 
To do so, we define its input function as
\begin{equation}
\label{eq:derandomizable function}
\begin{split}
f_{y}: & \ \{0,1\}^{\poly(|y|)} \rightarrow \{0,1\}\\
r &\mapsto  \begin{cases} 1 & \text{ if } \mathcal{A}(y,r) = x \\
0 & \text{ else} \end{cases} \ . 
\end{split}
\end{equation}
The output of Stockmeyer's approximation algorithm will then be a $ 1 - 1/\poly(|y|)$-multiplicative approximation to the probability $ p_{y}(x) $.
This provides the sought for connection between sampling and approximation of probabilities that forms the basis of the proofs of sampling hardness below. 

\subsection{Strongly simulating quantum computations}

For the specific schemes presented in \cref{sec:sampling schemes},  approximating the output probabilities is in fact a \gapp-hard task and thus just as hard as for arbitrary quantum computations. 
Generally, and this is in particular true for universal random circuits, the output probabilities of a circuit family $\mc C$ are \gapp-hard to approximate if the circuit family generates the whole of \bqp\ after so-called postselection~\cite{fujii_commuting_2017}. 
In a postselection argument we compare two probabilistic complexity classes by granting ourselves the ability to restrict attention to a certain subset of desired outcomes even if that subset has exponentially small probability. 
A postselected class \posta\ is defined as a class of decision problems which we can solve by using a computation within \classa\ and postselecting on certain outcomes with a bounded error~\cite{fujii_commuting_2017}. 
\begin{definition}[Postselected class~\cite{fujii_commuting_2017}]
A language $L$ is in the class \posta\ if there exists a uniform family of circuits $\{C_x \}$ associated with \classa\ for which there are a single output register $O_x$ and a $\poly(|x|)$-size postselection register $P_x$ such that 
\begin{enumerate}[label =\roman*.]
  \item if $x \in L$ then $\Pr(O_x = 1 | P_x = 00\ldots 0) \geq 2/3$, and 
  \item if $x \notin L$ then $\Pr(O_x = 1 | P_x = 00\ldots 0) \leq 1/3$ .  
\end{enumerate}
\end{definition}

\textcite{Aaronson-ProcRS-2005} showed that $\postbqp = \pp$, where \pp\ is the decision-problem equivalent of \sharpP\ which asks whether at least half of the inputs are accepted. This implies $\classP^\postbqp = \classP^\pp = \classP^\sharpP \supset \ph$. 
Building on this result, \textcite{fujii_commuting_2017} have demonstrated that if $\posta\ = \postbqp$ then a machine that approximates the output probabilities of circuits associated with \classa\ up to a multiplicative error $1/\sqrt 2 < c < 1$ can be used to decide any problem in \pp\ and hence any problem in \gapp. 
This is because the $\posta =  \postbqp$ condition ensures that \classa\ is rich enough to encode the output probabilities of arbitrary quantum computations and hence gaps of \sharpP\ functions. 

Taking a different perspective, one can show that the output probability of a universal quantum circuit can encode hard instances of the 
\emph{Jones polynomial}~\cite{kuperberg_how_2015,goldberg_complexity_2014,mann_complexity_2017} as well as \emph{Tutte polynomials}~\cite{kuperberg_how_2015,goldberg_complexity_2014} and certain Ising model partition functions~\cite{bremner_averagecase_2016,boixo_characterizing_2016}. 
In particular, estimating those quantities up to a relative error $1/4 + o(1)$ is \sharpP-hard.\footnote{Notice that achieving a relative error $1/4 + o(1)$ is slightly more demanding than a multiplicative error $1/\sqrt{2}$. }
Expressing the output probabilities in terms of such quantities, which have been studied in detail in the literature, will also prove to be extremely useful once we get to approximate sampling hardness. 

Similarly, the output probabilities of several restricted quantum computational models including the ones discussed above, can be expressed in terms of universal quantities which are \gapp-hard to approximate~\cite{fefferman_power_2015,morimae_hardness_2014,morimae_hardness_2017,fujii_impossibility_2018,bouland_complexity_2018,miller_quantum_2017,
gao_quantum_2017,bermejo-vega_architectures_2018,yoganathan_quantum_2018}.
In the following, we illustrate how this is achieved using the paradigmatic schemes introduced in \cref{sec:sampling schemes}. 

\subsubsection{IQP circuits}

As a particularly neat example of such reasoning, for IQP circuits, one finds that $\postiqp = \postbqp$.\footnote{This can be shown using a gadget to implement the Hadamard gate via teleportation, the idea being that what IQP circuits are lacking for universality is the possibility to switch between $X$ and $Z$ bases. 
By measuring a single output line one can teleport a Hadamard gate to an arbitrary position in the circuit using gate teleportation~\cite{bremner_averagecase_2016,montanaro_quantum_2017}.} 
What is more, for IQP circuits defined by a weighted adjacency matrix $W$ (cf.\ \cref{eq:iqp circuit weights}) the output amplitude 
\begin{align}
\label{eq:iqp output amplitude partition function}
  \bra 0 C_W \ket 0  = \frac{1}{2^{n}}Z_W, 
\end{align}
can be expressed as an  imaginary-temperature \emph{partition function} of an Ising 
model~\cite{bremner_averagecase_2016,fujii_commuting_2017}
\begin{align}
\label{eq:iqp ising partition function}
  Z_W   =
  \sum_{z \in \{\pm1\}^n} \exp\left[ \ii \bigg( \sum_{i<j}w_{i,j} z_i z_j + \sum_i w_{i,i} z_i \bigg) \right] . 
\end{align}
An analogous reduction can be made for  the universal circuits of \textcite[SM III.B]{boixo_characterizing_2016} with $CZ$ gates.
The modulus square $|Z_W|^2$ of such partition functions has been shown to be \gapp-hard to approximate up to a relative error $1/4 + o(1)$~\cite{goldberg_complexity_2014,fujii_commuting_2017}.

For an IQP circuit $C_f$ defined by a Boolean degree-$3$ polynomial $f$ with coefficient vectors  $\alpha,\beta, \gamma$ (cf.\ Eq.~\eqref{eq:iqp boolean polynomial}), one finds that the all-zero amplitude is given by the gap of $f$\footnote{To see this,  notice that 
\begin{align}
  Z_i \ket x & = (-1)^{x_i},\\
  CZ_{i,j} \ket x & = (-1)^{x_ix_j},\\
  CCZ_{i,j,k} \ket x & = (-1)^{x_ix_jx_k} .
\end{align}
}
\begin{multline}
\label{eq:iqp probabilities gap}
   \bra 0  H^{\otimes n} C_f H^{\otimes }\ket 0 
   = \frac{1} {2^n} \sum_{x,y} \bra y C_f \ket x \\ 
   = \frac{1}{2^n} \sum_x (-1)^{f(x)} = \ngap(f). 
\end{multline}
We have already seen above that approximating the gaps of arbitrary \sharpP\ functions $f$ up to multiplicative errors $1/\sqrt 2$ is \gapp-complete. 
This remains true when restricting the function $f$ to a degree-$3$ Boolean polynomial over the field $\mb F_2$, since IQP-circuits are universal with postselection \cite{bremner_averagecase_2016}.

\subsubsection{Fock boson sampling}
\label{ssec:boson sampling permanents}

\begin{figure}
  \includegraphics{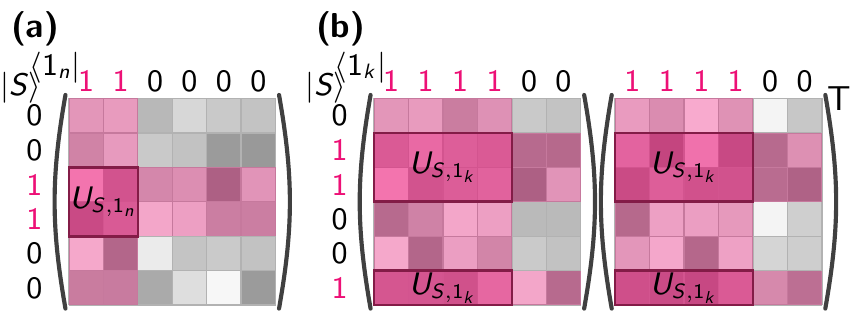}
  \caption{\label{fig:submatrices bs gbs}
    \emph{(a)} The output probabilities of Fock boson sampling \eqref{eq:bosonsamplingdistribution permanent} can be expressed as the modulus squared of the permanent of a submatrix $U_{S,1_n}$ of the Haar-random unitary $U$ constructed by discarding rows and columns according to the outcome and input registers $\ket S$ and $\ket 1_k$. 
    \emph{(b)} Analogously, the output probabilities of Gaussian boson sampling \eqref{eq:gbs squeezed state hafnian} with squeezed state inputs on $k$ modes are proportional to the modulus squared of the Hafnian of $U_{S,1_k}U_{S,1_k}^T$. 
    }
\end{figure}

The output distribution $\pbos$ of a Fock boson sampling experiment (cf.\ \cref{eq:bosonsamplingdistribution}) can be 
expressed as~\cite{scheel_permanents_2004}
\begin{equation} \label{eq:bosonsamplingdistribution permanent}
  \pbos(S) = \frac{|\Perm(U_{S,1_n})|^2}{\prod_{j=1}^m (s_j!)},
\end{equation}
in terms of the \emph{permanent} of the matrix $U_{S,1_n} \in \mb C^{n\times n}$ which can be obtained from $U$ according to the following prescription. 
Define the submatrix $U_{S,S'}$ with $S,S' \in \mb N^m$ as follows: for all $j,k \in [m]
=
\{ 1, 2, \ldots, m\}$, keep a matrix comprising $S_j$ copies of the $j^{\text{th}}$ row of $U$, and now write $S_j'$ copies of the $k^{\text{th}}$ column of that matrix into $U_{S,S'}$, see \cref{fig:submatrices bs gbs}(a). 
For so-called \emph{collision-free} outcomes $S \in \Phi_{m,n}$, that is, outcomes with only $0$ and $1$ entries, $U_{S,1_n}$ is therefore a certain submatrix of $U$. 
The permanent of a matrix $X = (x_{j,k}) \in \mb C^{n \times n }$ is defined analogously to the determinant but without the negative signs as
\begin{equation} \label{eq:definitionpermanent}
  \Perm(X) = \sum_{\tau \in \Sym_n} \prod_{j=1}^n x_{j,\tau(j)},
\end{equation}
where $\Sym_n$ labels all permutations of the set $[n] 
= \{1, 2, \ldots, n\}$.

It is a well-known fact that computing the permanent of a matrix is a problem that is \sharpP-hard even when restricting to binary matrices~\cite{valiant_complexity_1979}. 
At the same time, its close cousin, the determinant, can be exactly computed in polynomial time.
\textcite[Thm.~4.3]{aaronson_computational_2010} extend this famous result of \textcite{valiant_complexity_1979} to approximations of the \emph{modulus squared} of the permanent up to  multiplicative errors.
More precisely, they show that for any $c \in [1/\poly(n), 1 ]$, approximating $\Perm(X)^2$ up to multiplicative error $c$ for $X \in \mb R^{n \times n}$ remains \gapp-hard by a reduction similar to the one used to prove Lemma~\ref{lem:gapp approximation} on multiplicative-error \gapp-hardness of computing the modulus of the gap of a \sharpP\ function.

\subsubsection{Gaussian boson sampling}

Similarly, the output distribution of Gaussian
 boson sampling 
(cf.\ \cref{eq:Gaussianbosonsamplingdistribution})
can be expressed as \cite{hamilton_gaussian_2017,kruse_detailed_2019}
\begin{equation}\label{GaussianBosonSamplingDistribution}
\Gpbos(S) = \det(\sigma_Q)^{-1/2} \frac{ \Haf(M_S)}{\prod_{j=1}^m (s_j!)},
\end{equation}
in terms of the so-called \emph{Hafnian} of a matrix $M_S$ constructed as follows. 
Let $\sigma \in \mb C^{2m\times 2m}$ be the covariance matrix\footnote{See the textbook by \textcite{kok_introduction_2010} for an introduction to continuous-variable quantum information processing. } of the Gaussian state $\varphi(U) \ket g$ prior to the measurement and $\sigma_Q = \sigma + \id_{2m}/2$. 
Set  
\begin{equation}\label{eq:M GBS}
M =  
\begin{pmatrix}
0_m & \id_m\\
\id_m & 0_m
\end{pmatrix}
\left(
\id_{2m}
- \sigma_Q^{-1}
\right),
\end{equation}
where $\id_m$ denotes the $m\times m$ identity matrix. 
Analogously to how we construct a submatrix $U_{S,S'}$ from $U$, we obtain the submatrix $M_S$ of $M$ as follows: 
for every $j \in [m]$, $M_{S}$ comprises $S_j$ copies of the $j$-th and $m+j$-th row and column of $M$, respectively, see \cref{fig:submatrices bs gbs}(b). 
Hence, if $n = \sum_j S_j$-many photons are detected, then $M_S$ is a symmetric $2n \times 2n$ complex matrix.
Like the permanent, the Hafnian of a matrix is a certain polynomial in its matrix entries and defined for a matrix $A \in \mb C^{2n \times 2n} $ as  
\begin{align}
\label{eq:definition hafnian}
\Haf(A) = \sum_{\sigma \in \text{PMP}(2n)} \prod_{j = 1}^n A _{\sigma(2j -1), \sigma(2j)},
\end{align}
where $\text{PMP}(2n)$ is the set of all perfect matching permutations of $2n$ elements, that is, permutations $\sigma: [2n] \rightarrow [2n]$ that for every $i$ satisfy $\sigma(2i-1) < \sigma(2i)$ and $\sigma(2i-1) < \sigma(2i+1)$ \cite{barvinok_combinatorics_2016}. 
In particular, the permanent of $A$ can be written as a special case of the Hafnian as 
\begin{align}
\label{eq:permanent hafnian}
  \Perm(A)  = \Haf \left[ \begin{pmatrix}0 & A \\ A^T & 0 \end{pmatrix}\right] , 
\end{align}
and hence approximating the Hafnian is at least as hard as approximating the permanent, namely \gapp-hard, in the worst case. 

The output probabilities of Gaussian boson sampling take a particularly simple form  if the input state  $\ket g$ is a product of single-mode squeezed states with squeezing parameters $r_i$, which is the setting that has been studied in experiments~\cite{zhong_quantum_2020,zhong_phase-programmable_2021}.
In this case, the covariance matrix $\sigma$ of the Gaussian state before detection can easily be derived
to be given by
\begin{equation}
\sigma = \frac12
\begin{pmatrix}
U & 0\\
0 & U^\ast
\end{pmatrix}
\Sigma \Sigma^\dagger
\begin{pmatrix}
U^\dagger & 0\\
0 & U^T
\end{pmatrix},
\end{equation}
with $U\in U(m)$ the Haar-random unitary transformation of the input modes,
and 
\begin{equation}
	\Sigma  = 
\begin{pmatrix}
\oplus_{i=1}^m \cosh (r_i) & \oplus_{i=1}^m \sinh (r_i)\\
\oplus_{i=1}^m \sinh (r_i) & \oplus_{i=1}^m \cosh (r_i)
\end{pmatrix}.
\end{equation}
The output probabilities can then be written in terms of the matrix $A = U (\oplus_{i=1}^m \tanh(r_i)) U^T$ as 
\begin{align}
\label{eq:SMSS GBS probabilities}
  \Gpbos(S) = \frac{1}{\prod_{j=1}^m\cosh (r_j)} \left|\Haf(A_{S,S})\right|^2, 
\end{align}
recalling the definition of $A_{S,S'}$ from \cref{ssec:boson sampling permanents}; see also \cref{fig:submatrices bs gbs}(b). 
These probabilities take a particularly simple form whenever $k$ out of the $m$ modes are prepared in single-mode squeezed states with uniform squeezing parameter $r$ and the other $m-k$ modes are prepared in the vacuum state. In this case 
\begin{align}
\label{eq:gbs squeezed state hafnian}
  \Gpbos(S) = \frac{\tanh^k(r)}{\cosh^k (r)} \left|\Haf(U_{S,1_k}U_{S,1_k}^T)\right|^2.  
\end{align}

\subsection{Hardness argument}
\label{ssec:approximate sampling hardness}

We are now in the position to prove that under certain conditions on the quantum circuit family $\mc C$, sampling from the output distribution of a random instance $C \in \mc C$ cannot be done in classical polynomial time in the size of $C$, i.e., polynomial in the number of qubits. 
The idea of the proof is to exploit the fact that approximating output probabilities of unitaries in $\mc C$ is \gapp-hard. 
In contrast, if there was an efficient (derandomizable) sampling algorithm for a random $C \in \mc C$ then we could approximate its output probability using Stockmeyer's algorithm.
But because Stockmeyer's algorithm lies in the third level of the polynomial hierarchy, the existence of such an algorithm implies that $\Sigma_3 \supset \classP^\gapp \supset \ph$---the polynomial hierarchy collapses to its third level. 
Assuming the generalized $\classP \neq \np$ conjecture that the polynomial hierarchy is infinite, this rules out the existence of an efficient sampling algorithm for circuits from $\mc C$. 
In the following we present this argument, which is due to \textcite{bremner_classical_2010,bremner_averagecase_2016} and \textcite{aaronson_computational_2010}, in detail. 

\subsubsection{Exact sampling and worst-case hardness}

We formalize the idea sketched above in the following theorem. 
\begin{theorem}[Exact sampling hardness]
\label{thm:exact sampling hardness}
	Let $\mc C$ be a family of quantum circuits such that there exists a constant $c \in (0,1]$ for which approximating the output probabilities up to multiplicative error $c$ is \gapp-hard. 
	If there was an \emph{exact derandomizable sampling algorithm} for circuits in $\mc C$ then the polynomial hierarchy would collapse to its third level $\Sigma_3$. 
\end{theorem}
\begin{proof}
	Suppose there is a derandomizable sampling algorithm $\mc A$ that, given as an input a description of a circuit $C \in \mc C$ could efficiently sample from its output distribution $p(C)$ as defined in \cref{eq:p(U)}. 
	Then we can apply Stockmeyer's algorithm (\cref{thm:stockmeyer}) to the function $f_{C}$ defined in \cref{eq:derandomizable function}. 
	In time $\poly(1/c)$ and within the third level $\Sigma_3$ of the polynomial hierarchy, the output of this procedure will produce a multiplicative-error estimate $q_0(C)$ of the output probability $p_0(C)$ that satisfies 
	\begin{align}
	\label{eq:mult error stockmeyer proof}
	 	p_0(C)c \leq q_0(C) \leq p_0(C)/c . 
	\end{align} 
	But since approximating $p_0(C)$ is a \gapp-hard task by assumption, this implies that the polynomial hierarchy collapses to $\Sigma_3$.
\end{proof}

\begin{figure}[t]
\includegraphics{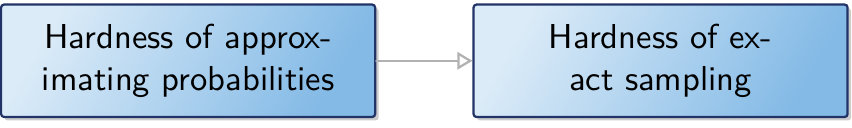}
  \caption{
  \label{fig:proof structure exact sampling}
  In the proof of Theorem~\ref{thm:exact sampling hardness} the idea is to relate the hardness of approximating the probabilities in a distribution to the hardness of 
  exactly
  sampling from that distribution. 
  }
\end{figure}

Notice two important subtleties of the argument: 
In order to prove exact sampling hardness, it is crucial that the output probabilities are not only \gapp-hard to compute exactly but even to approximate up to some constant relative error; see Fig.~\ref{fig:proof structure exact sampling}. 
Meanwhile, it is sufficient for exact sampling hardness that there is no algorithm which efficiently computes all instances of the output probabilities. 
In other words, the argument relies on \emph{worst-case hardness} of approximating the output probabilities since a single ``hard instance'' is sufficient for it.

What happens, though, once the sampling algorithm is allowed to make some error as compared to the ideal target distribution? 
Indeed, while an ideal quantum device samples from the ideal distribution no such device can exist. 
Every physical realization of the ideal model, be it in terms of a classical simulation algorithm, or a quantum implementation, will inevitably lead to errors so that it is only able to \emph{approximately sample} from the target distribution. 
Such errors may be due to finite precision issues intrinsic to computation, or noise in the physical implementation of quantum random sampling using near-term quantum devices. 

Does hardness of sampling still hold in the presence of errors on the sampled distribution? 
And if so, what types and magnitudes of errors are tolerated?

\subsubsection{Multiplicative-error sampling hardness}

As a first step, and quite naturally, the proof of sampling hardness can be extended from exact sampling, to sampling from a probability distribution $p$ that is \emph{multiplicatively close} to the target distribution $p(C)$ in the sense that for some constant $d \in (0, 1]$ each probability $p_x$ satisfies 
\begin{align}
\label{eq:d-multiplicative}
 d p_x(C) \leq p_x \leq p_x(C)/d. 
\end{align}
We can then easily amend the proof of Theorem~\ref{thm:exact sampling hardness} for this case to prove multiplicative-error robustness.

\begin{proof}[Multiplicative-error robustness of Thm~\ref{thm:exact sampling hardness}]
Assume there is an efficient classical sampling algorithm $\mc A$ that achieves the following task: 
Given as an input a description of a circuit $C \in \mc C$ produce a sample from a probability distribution $p$ that approximates the distribution $p(C)$ defined in \cref{eq:p(U)} up to a multiplicative error $d$ as in \cref{eq:d-multiplicative}. 
Then we can use Stockmeyer's algorithm to generate an approximation $q_0$ of the output probability $p_0$ that is correct up to any constant multiplicative error $c$
\begin{align}
	 cp_0  \leq q_0 \leq p_0/c. 
\end{align}
But the probability $p$ was multiplicatively close to the ideal output probability $p_0(C)$ to begin with so that we obtain
\begin{align}
	cd p_0(C) \leq c p_0 \leq q_0 \leq  p_0/ c \leq p_0(C)/(cd), 
\end{align}
that is, an overall multiplicative-error approximation to the probability $p_0(C)$ with constant multiplicative error $cd$.
If $c$ and $d$ are chosen such that  the probability $p_0(C)$ is \gapp-hard to approximate up to multiplicative error $cd$ the existence of an efficient sampling algorithm with multiplicative error guarantee $cd$ implies the collapse of the polynomial hierarchy.  
\end{proof}

\subsubsection{From multiplicative to additive errors}
\label{ssec:realistic errors}

We saw in our discussion about the approximability of \gapp\ how extraordinarily demanding multiplicative errors are in the guise of Lemma~\ref{lem:gapp approximation}. 
There, we used that such approximations always preserve the sign of a quantity and, moreover, attain 100\% accuracy if the quantity is $0$.
Similarly, for the sampling task, there is no difference in complexity when allowing for constant multiplicative errors compared to the exact case. 
And indeed, to satisfy such a notion of approximation, an algorithm would need to account for the size of all of the exponentially many probabilities, some of which may be computer-precision close to zero to begin with. 
While this notion of approximation may be achievable using a fault-tolerant quantum device, and a computation using ultra-high precision that scales with the system size, this state of affairs seems implausible in practice. 

What is a more plausible notion of approximation then? 
In the following, we consider approximations $q$ to a target distribution $p$ in terms of the \emph{total-variation distance} (TVD) 
\begin{align}
\norm{p - q}_{\rm TV} = \frac 1 2 \sum_x |p(x) - q(x)|
\end{align}
between $p$ and $q$.
The TVD measures the maximal distinguishability of two probability distributions in terms of the optimal distinguishing strategy \cite{watrous_theory_2018} and is therefore a natural measure of statistical distance.
But why is the TVD a sensible measure to consider when consider quantum advantage via quantum random sampling? 
While the answer to this question is not entirely clear, there are several arguments that one might make.

\paragraph{Why the total-variation distance?}
The first argument argues from the perspective of classical simulation algorithms. 
Indeed, a fundamental notion of imprecision of a randomized algorithm, such as sampling algorithms, is given by additive errors. 
To see why, observe that a classical computer makes use of a constant precision representation of numbers. 
This gives rise to an additive error on all computations that is exponentially small in the number of digits of the representation.
Going a step further, imperfections in an algorithm often give rise to additive errors on the result. 
One may therefore argue that the precision that is achievable by classical algorithms is fundamentally---and often in practice---just an additive error, and the TVD is a natural way of capturing this error.  
At the same time, this line of reasoning implies that the precision of computing individual probabilities in the process of sampling (see \cref{ssec:sampling vs approximating} for details) needs to scale with the size of the system. 

The second argument argues from the perspective of the noisy quantum device. 
This argument observes that any device error is reflected in an additive error on the distribution. 
To see this, we observe that for both coherent and incoherent errors, we can write an erroneously prepared quantum state $\rho_\epsilon$ as a convex mixture of the target state $\rho = U \proj 0 U^\dagger $ and some other quantum state $\sigma$ orthogonal to it as 
\begin{align}
  \rho_\epsilon = (1-\epsilon) U \proj 0 U^\dagger  + \epsilon \, \sigma. 
\end{align}
Consequently, the output distributions of the noisy and ideal states when measured in the standard basis, $p(\rho)$ and $p(\rho_\epsilon)$, satisfy 
\begin{align}
  \label{eq:tv trace distance}
  \norm{p(\rho) -p(\rho_\epsilon)}_{\text{TV}} 
  & = \frac 12 \sum_x | p_x(\rho) - p_x(\rho_\epsilon) |  \\
  & \leq \frac 12 \max_{\{M_x\}_x} |\tr[(\rho - \rho_\epsilon)M_x]  \\
  & = \norm{\rho - \rho_\epsilon}_{\tr}  = \epsilon, 
\end{align}
where the maximization runs over arbitrary positive operator-valued measures (POVMs) $\{M_x\}_x$. 
Here, we have defined the trace distance $\norm{\cdot}_{\tr}$ which is identical
to the TVD for diagonal quantum states.
The trace distance, analogously to the TVD, measures the maximal distinguishability of two quantum states in terms of the optimal quantum distinguishing strategy \cite{watrous_theory_2018}.
Since the trace distance maximizes over all possible measurement strategies, it upper-bounds the TVD between the outcome distributions, which is given by fixing a measurement basis. 

However, it is important to note that trace or total-variation distance are not good models of physically realistic errors occuring on a noisy quantum device. 
Such errors---like the imprecision of classical computers---are independent of the size of the system. 
Hence, as constant local gate errors occur in a quantum circuit, the trace distance of its output state scales linearly in the number of gates which quickly increases to a trivial value. 
In order to make TVD meaningful from the perspective of a  noisy quantum device we thus need to scale down the local errors as we scale up the circuit size. 

Finally, as we will see, it turns out that the TVD arises naturally when considering exact sampling algorithms that work only in the average case. 
Since average-case algorithms are natural for random quantum circuits, this provides further justification for the TVD. 
Compared to other statistical distances such as the Kullback-Leibler (KL) divergence, the TVD also turns out to be the measure that is amenable to the proof technique which we present in detail in the following.

To summarize this discussion, the TVD is a notion of robustness for both classical and near-term quantum algorithms solving the sampling task. 
The smallest meaningful, and nontrivial notion of approximation may be to consider the task of sampling up to \emph{constant} TVD. 
This only requires relatively mild error or precision scalings of the individual components of the respective algorithms on the order of $1/m$. 
Of course, such scaling of local algorithmic errors is already extremely demanding, however.

\paragraph{Showing TVD robustness.} 
In the following, we consider the task of sampling from a distribution $q$ that is $\epsilon$-close to a target output distribution $p(C)$ of a quantum circuit $C$ in the sense that
\begin{align}
	\norm{p(C) - q}_{\rm TV} \leq \epsilon . 
\end{align}
Our goal is to show that this task is hard for classical computers. 
Compared to exact and multiplicative-error sampling hardness, this endeavour is faced with the challenge that as $\epsilon $ increases, so does the legroom for classical simulation: 
to show hardness we have to prove that sampling from any distribution within an $\epsilon$ TVD neighborhood of the target distribution is classically intractable. 
We are faced with a dramatically increased burden in the proof as hardness needs to be shown for an entire \emph{volume} of probability distributions rather than a single point. 
Importantly, as $\epsilon \rightarrow 1 $ the output state of the computation becomes classically simulable as, in particular, the uniform distribution is always within this error bound of the target distribution. 
But the uniform distribution is easy to sample from even on an exponentially large sample space by repeated unbiased coin tosses.

Given what we have seen so far, there is a fundamental discrepancy between how the proof of exact sampling hardness can naturally be made
robust to noise and the errors that naturally occur in realistic settings. 
The discrepancy is one between the utterly unrealistic notion of multiplicative errors on all probabilities and the more realistic notion of additive errors on the global outcome distribution. 
The question we will focus on now is whether we can overcome this hurdle?

In technical terms, what we would like to prove is that no efficient classical algorithm $\mc A$ taking as an input an efficient description of $C$ exists that samples from any distribution $q$ such that $\norm{q - p(C)}_{\rm TV} \leq \epsilon$ for a constant $\epsilon > 0 $. 
Again, we will make use of Stockmeyer's approximate counting algorithm with a derandomizable sampling algorithm as an input in order to take the step from hardness of approximating probabilities. 
So how can we take the leap from proving robust hardness-of-sampling results for multiplicative to ones for additive errors?

To approach an answer to this question let us conceive of the sampling algorithm $\mc A$ as an adversarial party that, given $U$ as an input, tries to adversarially obstruct the approximate counting algorithm in its goal to approximate specific probabilities. 
The adversarially acting sampling algorithm is, however, constrained to sample from a distribution satisfying the respective error bounds. 
A few observations regarding the nature of additive errors in contrast to multiplicative ones are instructive.
\begin{enumerate}
	\item
	\label{it:error budget}
	When the sampling algorithm is constrained to multiplicative errors on individual probabilities, the total \emph{additive error} it can make depends strongly on the shape of the distribution. 
	In particular, every individual probability will be correct up to an error that depends on its size. 
	In contrast, the additive-error constraint allows the adversarial party much more flexibility. 
	An additive error can be viewed as a total \emph{error budget} that may be distributed across the individual probabilities at will. 
	In particular, a few probabilities  can come with large (relative) errors supposing
	that the other ones are correct up to a very small additive error.

	\item
	\label{it:volume} 
	When proving multiplicative-error robustness, the shape and volume of the region in the space of probability distributions on a sample space $\Omega$ of which hardness is proven depend heavily on the specific shape of the distribution. 
	In contrast, for additive-error robustness volume and shape of this region are only sensitive to boundaries of $\Omega$. 

	\item 
	\label{it:bqp-hardness of additive}
	Approximating output probabilities of quantum computations up to an inverse polynomial
	additive error does not remain hard for \gapp\ but only for \bqp~\cite[Thm.~3]{de_las_cuevas_quantum_2011}.
	Only for inverse exponentially small additive errors $\pm 1/2^n$ those approximations become again \gapp-hard. 
	This is easily seen using the fact that normalized gaps of Boolean functions acting on $\{0,1\}^n$ only take on values that are integer multiples of $2/2^n$. 
	Approximating those gaps up to an additive error $< 1/2^n$ is therefore just as hard as exactly computing them.\footnote{See also the Supplementary Material of \textcite{bremner_averagecase_2016}. }
\end{enumerate}

What can we take away from those observations? 
Point~\ref{it:bqp-hardness of additive} implies that to prove a polynomial-hierarchy collapse via Stockmeyer's algorithm, we must still rely on the hardness to approximate output probabilities of circuit families up to relative errors \emph{or} exponentially small additive errors. 

Points~\ref{it:error budget} and~\ref{it:volume} shine light on two sides of the same coin. 
In contrast to the case of multiplicative robustness, we cannot rely on the hardness of estimating individual probabilities that might be very small. 
In particular, it cannot be the case that only one of the circuits within $\mc C$ has a single output probability on which all classical algorithms fail. 
Instead, we must rely on circuit families for which not only single outcome probabilities of some members of the family are hard to compute, but rather a large---constant---fraction of all output probabilities of the circuit family must be hard to compute. 
This idea is formalized within the notion of \emph{average-case complexity}: 
Approximating the outcome probabilities of quantum circuits must be hard for a large fraction of the instances, where an instance is defined by a specific quantum circuit. 

In particular, average-case complexity therefore requires that not all but very few of those hard probabilities can be tiny, i.e., smaller than, say, doubly exponentially small while very few large ones are easy to approximate. 
Indeed, if this were the case, since tiny quantities have tiny relative errors, the adversarially acting sampling algorithm could easily distribute the better part of its constant error budget on the few large probabilities but at the same time still pass the relative-error threshold on the tiny probabilities. 
In this way they would meet the constraint imposed by the global additive error, but not achieve a provably hard task as the error on the computationally intractable probabilities would be too large. 
Rather, there must be a large fraction of hard instances that are reasonably large, say, at least as large as uniform probabilities $\sim 1/|\Omega|$ on the sample space $\Omega$. 
This idea is formalized within the notion of \emph{anticoncentration}, which is a condition on the probability that a randomly drawn problem instance---again, specified by a circuit and an outcome string---is reasonably large.
Anticoncentration constrains how the adversarial player can distribute their error budget: 
they can choose between getting many probabilities right with tiny errors, but making larger errors on a few outcomes, say, inverse polynomial errors on polynomially many probabilities, or getting all probabilities right with reasonably small inverse exponential errors. 
These observations have been made by \textcite{aaronson_computational_2010} who observed that the natural problem in boson sampling, namely, computing a permanent, is an average-case hard problem.

In the discussion above, we have been touching upon on a point that we had glossed over in our discussion of exact sampling hardness: 
it is key to random circuit sampling schemes that there are two notions of probability at play. 
First, there is the random choice of a circuit from the family $\mc C$, and second, there is the random choice of an outcome string $S$ that is distributed according to $p(U)$. 
Equally, there are two probability distributions---the distribution according to which random circuits are drawn, and the outcomes distribution of each such random circuit. 
These notions are crucially distinct. 

As we will see, the choice of random circuit instances is essential to providing evidence for the additive-error robust hardness of simulating quantum circuits. 
The second notion of probability is intrinsic to our choice of problem. 
In the end, we aim to prove hardness of a sampling task. 
This is a task requiring randomness: we want to obtain a random sample from a distribution that we, in turn, chose at random from another ensemble. 

\subsubsection{Additive-error sampling hardness}
\label{ssec:additive error sampling hardness}

\begin{figure}
\centering
\includegraphics{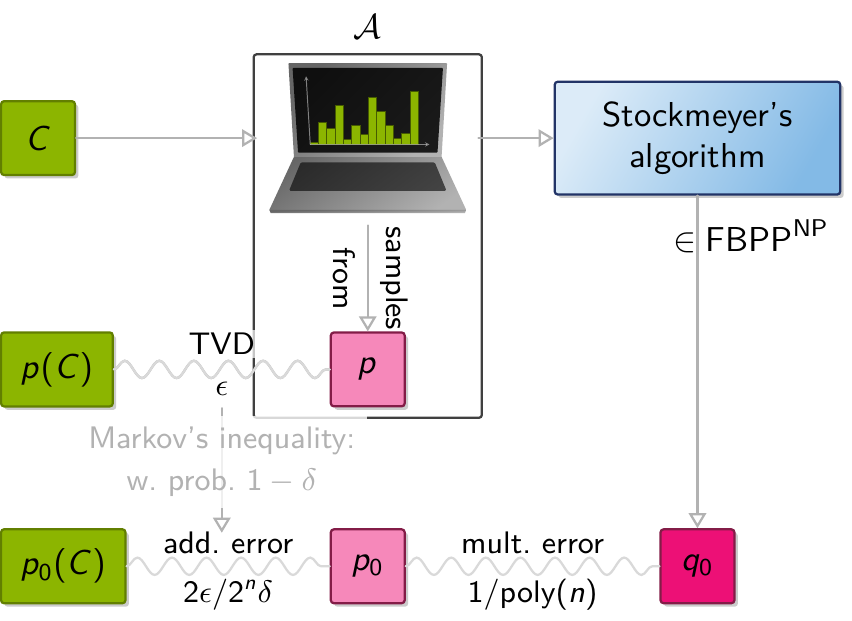}
\caption{
  \label{fig:additive error stockmayer proof}
  Outline of the proof strategy for additive-error sampling hardness: 
  a derandomizable sampling algorithm $\mc A$, given $C$ as an input, samples from a distribution $p$ that is $\epsilon$-close in total-variation distance (TVD) to the target distribution $p(C)$. 
  Using Markov's inequality and the hiding property this implies that the output probability $p_0$ of $p$ is within additive error $2\epsilon/(2^n\delta)$ of the ideal output probability $p_0(C)$ with probability at least $1-\delta$.
  Given $\mc A$ as an input, Stockmeyer's algorithm can infer a $1/\poly(n)$-multiplicative approximation $q_0$ of the approximate output probability $p_0$ in the third level of the polynomial hierarchy. 
  }
\end{figure}

Given average-case hardness of approximating the output probabilities, we can prove a hardness-of-sampling result that is robust to constant additive errors. 
We proceed analogously to the proof of multiplicative robustness, following the sketch in Fig.~\ref{fig:additive error stockmayer proof}. 

\begin{proof}[Additive-error robustness of Thm~\ref{thm:exact sampling hardness}]
Assume there is an efficient, derandomizable classical algorithm that takes as an input a description of a circuit instance $C$ from a family $\mc C $ and outputs samples distributed according to a probability distribution $p$ that satisfies 
\begin{align}
\label{eq:tv bound additive proof}
	\norm{p - p(C)}_{\text{TV}} \leq \epsilon. 
\end{align}
Here, $p(C)$ is the ideal target distribution defined in Eq.~\eqref{eq:p(U)}.
We want to use this sampling algorithm in order to approximate a random problem instance as given by the output probability $p_0(C) = | \bra 0 C \ket 0 |^2$ of $C$. 

According to \cref{task:quantum random sampling}, we generate an instance by drawing $C\in \mc C$ at random.  
To estimate the value of this instance, we use Stockmeyer's approximate counting algorithm with input given by the algorithm $\mc A$, the circuit instance $C$, and the outcome string $0^n$.
Using access to its \np\ oracle, Stockmeyer's algorithm will output a multiplicative-error approximation $q_0$ of the noisy output probability $p_0$ satisfying 
\begin{align}
	\label{eq:rel error additive proof}
	| q_0 - p_0 | \leq c p_0  , 
\end{align}
in time $\poly(n, 1/c)$ within the third level $\Sigma_3$ of the polynomial hierarchy. 

Our goal is to bound the error 
\begin{align}
\label{eq:triangle inequality probabilities}
	| q_0 - p_0(C) | \leq | q_0 - p_0 |+ |p_0 - p_0(C) | .  
\end{align}
\cref{eq:rel error additive proof} already provides the first half of this bound. 
For the second bound we need to leverage the total-variation-distance bound~\eqref{eq:tv bound additive proof} on the global distributions $p$ and $p(C)$ to an error bound on the individual probabilities $p_0$ and $p_0(C)$. 

To obtain such a bound, consider again the sampling algorithm $\mc A$. 
Remember that \emph{qua} being a derandomizable algorithm, on input $U,r$ with uniformly random $r \in \{0,1\}^{\poly(n)}$ it will output a random sample from $p$ so that 
\begin{align}
	p_x(C) & =  \Pr[ C \text{ outputs } x],\\
	p_x  & = \Pr_r[\mc A \text{ outputs } x \text{ on input } C]. 
\end{align}
Acting adversarially, the algorithm $\mc A$ wants to maximize the error $|p_0 - p_0(C)|$. 
To do so, it needs to have some prior information about which of the outcome strings are more likely to be queried in Stockmeyer's algorithm given a certain input $C$ so that it can distribute more of its constant error budget on those outcomes.  
Such information would manifest itself in a distribution of outcomes $x$ that is non-uniform---and in fact concentrated on the single all-zero outcome---from the perspective of $\mc A$ given $C$~\cite[p.~51]{aaronson_computational_2010}.
This is because the all-zero outcome is the one we are \emph{always} interested in. 
But if it was able to distribute all of its \emph{constant} error budget on this single outcome, then it would not be able to achieve a hard task, which is what we are trying to show. 

\paragraph{Hiding problem instances.}
To see how we can achieve that this distribution over outcomes is not biased towards a few outcomes but uniform over all outcomes, consider the distribution over circuits $C_y$ obtained by drawing $C \in \mc C$ at random and then appending $X$ gates $X_1^{y_1} \cdot X_2^{y_2} \cdots X_n^{y_n}$ for uniformly random $y \in \{0,1\}^n$ to the end of the circuit~\cite{bremner_averagecase_2016}. 
We can then re-express the outcome probabilities of $C_y$ as 
\begin{align}
	p_x(C_y) 
  =  |\bra {x } C_y \ket 0|^2 &  
  =  | \bra {0} C_{x \oplus y} \ket 0 |^2
  =  p_0(C_{x \oplus y}).
\end{align} 
Consequently, the very same problem instance $C$ can be equivalently obtained when providing the adversary $\mc A$ with an instance $C_{y}$ for uniformly random $y$ and then querying Stockmeyer's algorithm on the outcome $y$. 
When aiming to estimate the problem instance $p_0(C)$ we can therefore \emph{hide the instance} $C$ in the circuit $C_{y}$ by randomly appending $X$ gates according to a uniformly random $y$, and then querying Stockmeyer's algorithm on outcome $y$. 
But since $y$ is hidden from $\mc A$, the distribution over outcomes on which we are going to query Stockmeyer's algorithm to obtain the output probability is uniformly random, and it cannot bias its error towards any given outcome. 

For this to work, it is of course crucial that $\mc A$ cannot distinguish whether we have directly generated a random problem instance $C$ for which we are directly interested in the all-zero outcome, or whether we have first drawn a random $C \in \mc C$ and then \emph{hidden this instance} by constructing the unitary $C_y$ with uniformly random $y$ and query on the outcome $y$~\cite[p.~51]{aaronson_computational_2010}).
Hence, the probability of directly drawing $C_y$ must be the same as that of drawing $C$ and then appending uniformly random $X$ gates according to $y$. 

Generally, we therefore say that a circuit family $\mc C $ has the \emph{hiding property} if 
\begin{enumerate}[label=\alph*)]
  \item  there is an efficient instance-generating procedure that converts a given problem instance $C \in \mc C$ and a uniformly random outcome $y$ into another problem instance $C_{y}$, and 

   \item the distribution distribution over circuits is invariant under this procedure, i.e., 
\begin{align}
  \Pr_{C_y \sim \mc C} [C_{y}] = \Pr_{C \sim \mc C, y \sim \{0,1\}^n} [C_{y}]. 
\end{align}

\end{enumerate}
The hiding property holds very naturally for most random circuit families, and in particular so for universal random circuits where each gate is drawn from the Haar measure. 
This is because the Haar measure is left- and right-invariant under arbitrary unitaries and the Pauli-$X$ gate is one particular such unitary.

If the hiding property holds, without loss of generality, we can therefore always query Stockmeyer's algorithm on the all-zero outcome of $C$, making use of the fact that this outcome is indistinguishable from a uniformly random one from the perspective of $\mc A$. 
Conversely, we can conceive of the outcomes of the circuits we are going to query Stockmeyer's algorithm on as being uniformly distributed from the perspective of $\mc A$. 
In this case, we can apply Markov's inequality to obtain a bound on the error for individual probabilities. 
For uniformly random $x$ we obtain that 
\begin{multline}
\label{eq: markov error bound stockmeyer}
	\Pr_{x \in \{0,1\}^n}\left[ |p_x - p_x(C)| \geq \frac 1 \delta \Eb_{x \in \{0,1\}^n} [|p_x - p_x(C)|]\right] \\
	= \Pr_{x \in \{0,1\}^n}\left[ |p_x - p_x(C)| \leq \frac {2\epsilon} {\delta 2^n}\right] \leq \delta   , 
\end{multline}
since 
\begin{multline}
	\Eb_{x \in \{0,1\}^n} \left[|p_x - p_x(C)|\right] = \frac 1 {2^n} \sum_{x \in \{0,1\}^n} |p_x - p_x(C)| \\
	 = \frac 2 {2^n} \norm{ p - p(C)}_{\text{TV}} = \frac {2 \epsilon}{2^n} . 
\end{multline}
Putting together Eqs.~\eqref{eq:triangle inequality probabilities} and \eqref{eq: markov error bound stockmeyer},
we have now found that with probability at least $ 1- \delta$ over the inputs, the error of the estimate $q_0$ output by Stockmeyer's approximate counting algorithm satisfies
\begin{align}
	|q_0 - p_0(C)| & \leq \frac{1}{\poly(n)}p_0 + \frac{2 \epsilon } {2^n \delta}
	\\
	&\leq \frac{1}{\poly(n)}p_0(C) + \frac{2 \epsilon } {2^{n} \delta}\left( 1 + \frac 1 {\poly(n)} \right) .  
	\label{eq:stockmeyer error mixture}
\end{align}
This bound is a mixture of an exponentially small additive and inverse polynomially small multiplicative error. 
However, the error bound does not hold for all possible inputs to Stockmeyer's algorithm; it only holds for a $(1-\delta)$ fraction of the inputs. 
By the hiding property this corresponds to a $(1-\delta)$ fraction of the problem instances.

\paragraph{Approximate average-case hardness.}
To show hardness of the sampling task, we need to show that achieving this error on an arbitrary $(1-\delta)$ fraction of the outputs is sufficient for a collapse of the polynomial hierarchy.\footnote{This is because Markov's inequality does not control the \emph{instances} on which the bound fails.} 
Indeed, our procedure involving Stockmeyer's algorithm is precisely such an algorithm (in the third level of the polynomial hierarchy). 
A sufficient condition to show such a polynomial-hierarchy collapse is then the following: 
The problem of estimating the probabilities remains \gapp-hard even when using a polynomial-time algorithm that only succeeds on a constant fraction of the instances. 
Phrasing this in other words, an algorithm solving the estimation problem for $p_0(C)$ with error \eqref{eq:stockmeyer error mixture} and success probability given by the respective fraction of the instances (i.e. $1-\delta$) is as powerful as an arbitrary \gapp\ algorithm. 
This contrasts with the proof of exact sampling, where it was merely required that the estimation problem is \gapp-hard in the worst case, that is, for a machine that is required to succeed on \emph{all} instances.

Making this intuition rigorous is the idea of average-case hardness. 
\begin{definition}[Approximate average-case hardness]
\label{def:average-case hardness}
Let $\Gamma \in (0,1)$, $\varepsilon > 0$.  
A function class \textsf{F} is average-case hard with constant $\Gamma$ and error $\varepsilon$, if approximating any $\Gamma$ fraction of the instances in \textsf{F} up to error $\varepsilon$ is \gapp-hard. 
\end{definition}
If approximate average-case hardness holds with respect to the error \eqref{eq:stockmeyer error mixture}, the existence of an efficient sampling algorithm~$\mc A $ for the output distribution of a random instance~$C \in \mc C$ implies that we can approximate \gapp-hard probabilities in the third level of the polynomial hierarchy using Stockmeyer's algorithm. 
The polynomial hierarchy collapses.
\end{proof}

\begin{figure}
\includegraphics{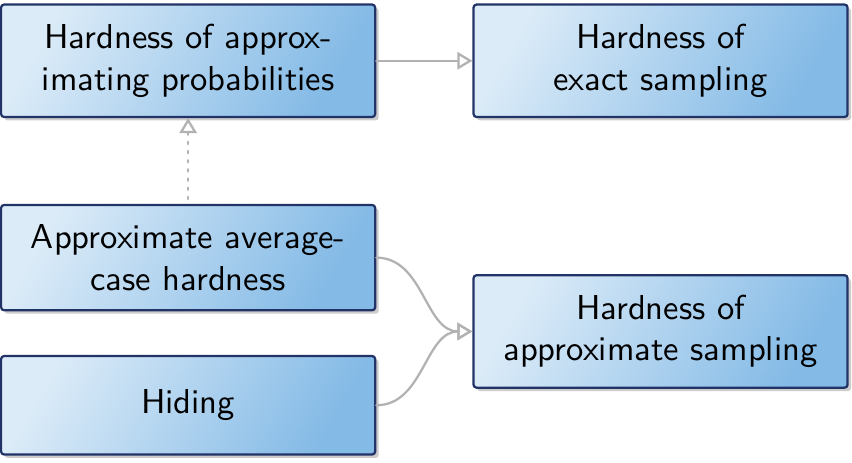}
  \caption{
  \label{fig:proof structure additive error} 
  While in the proof of exact sampling hardness, it was sufficient to build on the hardness of approximating the output probabilities of quantum circuits, in order to prove hardness of approximate sampling further properties of the circuit family $\mc C$ are required: approximate average-case hardness of computing the output probabilities and the hiding property. 
  } 
\end{figure}

We have proven approximate sampling hardness; see Fig.~\ref{fig:proof structure additive error}. 
\begin{theorem}[Additively robust sampling hardness]
\label{thm:approximate sampling hardness}
Consider a circuit family $\mc C$ that satisfies 
\begin{enumerate}[label = \arabic*.]
  \item \label{it:hiding} 
  the hiding property, and 
  \item \label{it:approx hardness} 
  approximate average-case hardness up to error \eqref{eq:stockmeyer error mixture} on any $(1-\delta)$ fraction of the instances. 
\end{enumerate}
Suppose there is an efficient classical sampling algorithm $\mc A$ that given $C \in \mc C $ drawn at random as an input, with success probability at least $1-\delta$ over $C$, outputs samples from an additive approximation $p$ to the outcome distribution $p(C)$ satisfying $\norm{p - p(C)}_{\rm TV} \leq \epsilon $.
Then the polynomial hierarchy collapses. 
\end{theorem}

We have walked a long route from the complexity-theoretic foundations of quantum speedups all the way to rigorous and approximate hardness-of-sampling arguments relevant to near-term quantum technology. 
The complexity-theoretic foundations of quantum speedups manifested themselves in the \gapp\ vs.\ \sharpP\ dichotomy:
while multiplicatively approximating the acceptance probabilities of classical circuits can be done in the third level of the polynomial hierarchy, this task remains \gapp-complete for certain quantum circuit families. 
We then saw how the at-first-sight different tasks of sampling from a probability distribution (weakly simulating it) and approximating its outcome probabilities (strongly simulating it) are related on a rigorous level: Stockmeyer's approximate counting algorithm and the concept of the polynomial hierarchy proved key to this question. 
Building on those methods, we could show that the task of sampling from the output distribution of certain random quantum computations cannot be achieved by an efficient classical algorithm. 
In a last step, we aimed at making this result robust to realistic errors, that is, additive errors in total-variation distance on the level of the output distributions. 
Making this leap involved stronger properties of the output distribution, however: approximate average-case hardness and the hiding property. 
The way we have formulated \cref{thm:approximate sampling hardness} provides a general framework for providing a hardness argument for approximately sampling from the output distributions of quantum circuit families. 
But of course, in order to complete the proof, crucially, the two properties---hiding and approximate average-case hardness---need to be shown for specific circuit families. 

We already hinted that the hiding property trivially holds for most circuit families: to show this, we merely need to show that $X$ gates at the end of the circuit do not alter the circuit family. 
The only instances of circuit families for which hiding is non-trivial, are boson sampling protocols. Let us briefly sketch the argument here. 

\paragraph{Hiding in boson sampling.}
\label{par:hiding boson sampling}
We have already seen above in Section~\ref{sec:sampling schemes} that the output probabilities of Fock boson sampling are given by permanents~\eqref{eq:bosonsamplingdistribution permanent} of submatrices of Haar-random unitaries. 
Conceivably, though, there is some structure in such submatrices. 
To see this, consider the case in which we obtain all bosons in a single mode as the outcome, i.e., $S = (n, 0, 0, \ldots) $. 
In this case, all columns of the submatrix $U_{S,1_n}$ are equal and, plausibly, this can be exploited to approximate $|\Perm(U_{S,1_n})|^2$, or in other words, because of the structure in the matrix, the specific outcome cannot be hidden. 
However, \textcite{aaronson_computational_2010} show that, under certain conditions, hiding holds in Fock boson sampling in virtue of the fact that the output probabilities of a random boson-sampling instance are determined by permanents of approximately Gauss-random and therefore highly unstructured matrices. 

In order to achieve this, \textcite{aaronson_computational_2010} consider the \emph{collision-free boson sampling distribution} $\pbos^*$.
The distribution $\pbos^*$ is obtained from $\pbos$ by discarding all output sequences $S$ with more than one boson per mode, i.e., all $S$ which are not in the set of \emph{collision-free} sequences
\begin{equation}\label{eq:collision-free subspace}
  \Phi^*_{m,n} = \Big\{ S \in \Phi_{m,n} : \forall s \in S : s \in \{0,1\} \Big\} .
\end{equation}
Why are collision-free outcomes advantageous when proving hardness? 
Intuitively, this is because for collision-free outcomes, the submatrix $U_{S,1_n}$ has much less structure than for outcomes with collisions because there are no repeated rows or columns. 
If moreover, the size of $U_{S,1_n}$ becomes sufficiently small compared to the full size of $U$, neither does there remain any of the structure in $U$ stemming from the orthogonality of its columns. 

The hiding property then follows from two facts.
First, we need to justify that restricting our attention to collision-free outcomes is valid. 
This is true if postselecting onto the collision-free subspace can be done efficiently in the sense that its probability weight is at least a constant, and Aaronson and Arkhipov prove that this is the case if $m$ grows sufficiently fast with $n$ and at least as $m \in \Omega(n^2)$ (see \textcite[Theorem 13.4]{aaronson_computational_2010} and \cite{arkhipov_bosonic_2011,jiang_how_2006}).
Second, they prove that if $m$ grows even faster, namely, as $m \in \Omega(n^5 \log(n)^2)$, 
the measure induced on $U \sim \mu_H$ by taking $n \times n$-submatrices of unitaries $U \in U(m)$ chosen with respect to the Haar measure $\mu_H$ is close to the complex Gaussian measure $\mu_G(\sigma)$ with mean zero and standard deviation $\sigma = 1/\sqrt m$ on $n \times n$-matrices.
Consequently, irrespective of which submatrix we choose, i.e., which collision-free outcome we obtain, the distribution of the submatrices is approximately Gaussian.

Conversely, \textcite[Lemma~5.8]{aaronson_computational_2010} prove that, given a Gauss-random instance $X \sim \mu_G(\sigma)$ as input, there is a $\bpp^\np$ algorithm\footnote{Like Stockmeyer's algorithm, this algorithm is therefore in the third level of \ph. } which, given $X$ hides this matrix in a large unitary matrix in the sense that it generates a Haar-random $U \in U(m)$ such that there is a uniformly random $S\in \Phi_{m,n}^*$ such that $X = U_{S,1_n}$. 
This provides the instance-generating algorithm. 
Hiding a Gauss-random instance $X$ is therefore possible by constructing a larger unitary matrix of which $X$ is a uniformly random submatrix, similarly to how we hid a qubit-circuit $C$ by appending uniformly random $X$-gates to it. 

A similar reasoning can be applied to Gaussian boson sampling, albeit with a slightly different distribution \cite{deshpande_quantum_2022}. 
Recall that the matrices of which the Hafnian is computed in Gaussian boson sampling with $k$ single-mode squeezed inputs and $n $ detected photons in $m$ modes are of the form $U_{S,1_k} U_{S,1_k}^T$, which, for collision-free outcomes, are outer products of random $n \times k$ submatrices of the linear optical unitary $U$. 
For those matrices, hiding plausibly holds with respect to symmetric Gaussian matrices $XX^T$, where $X \sim \mc G_{n,k}(0,1/m)$ is an $n \times k$ matrix drawn from the Gaussian distribution on $n \times k$ complex matrices. 
Indeed, this is provably true in two regimes \cite{deshpande_quantum_2022}:
first, for $m \in O(k^5\log^2 k)$ and $k=n$ the submatrices are individually Gaussian distributed by the result of \textcite{aaronson_computational_2010} and hence we can also bound the distance to the distribution of $XX^T$. 
Second, for $k=m$, \textcite{jiang_entries_2009} showed that whenever $n \in o(\sqrt{m}/\log m)$ the distribution of $n \times n$ submatrices of $UU^T$ for unitary $U$ converges asymptotically to the distribution of $XX^T$, where $X \sim \mc G_{n,m}(0,1/m)$ is an $n \times m$ complex Gaussian matrix.
For the intermediate regime $m^{1/5}< k < m$, there is numerical evidence that the hiding property remains true \cite{deshpande_quantum_2022}. 
The instance generation algorithm of \textcite[Lemma~5.8]{aaronson_computational_2010} will also work for this setting provided that the distributions of $U_{S,1_k}U_{S,1_k}^T$ for unitary $U$ and $XX^T$ for Gaussian $X \sim \mc G_{n,k}(0,1/m)$ are close not only in TVD, but also in a slightly stronger multiplicative sense.
This is because the instance-generating algorithm simply postselects on the matrix $XX^T$ appearing as a submatrix of $U\id_k U^T$ by making use of the \np\ oracle.

A very neat way of constructing a Gaussian boson sampling scheme that comes without the need of scaling $m \in \Omega(\poly(n))$ has been discovered by \textcite{grier_complexity_2022}. 
They observe that by programming a Gaussian boson sampling device in a bespoke way, it is possible to encode the permanent of an arbitrary matrix in the output probabilities.  
Specifically, they consider a bipartite system of $2m$ modes. 
The input state is given by a product of two-mode squeezed states on modes $i$ and $i+m$ for $i = 1, \ldots, m$ with squeezing parameters $r_1, \ldots, r_m$. 
In other words, the two halves of a two-mode squeezed state are associated with the two partitions, respectively. 
Now, a bipartite unitary mode transformation $U \otimes V$ is applied to the system, and all modes measured in the Fock basis. 
This gives rise to output probabilities which are proportional to a function of a submatrix of a matrix $C = U \diag(r) V^\dagger$, where $r = (r_1, \ldots, r_m)$ is the vector of squeezing values. 
Since this is nothing but a singular value decomposition, by choosing $r, U,V$ bespokely, $C$ can be programmed to be an arbitrary matrix, and in particular a Gaussian one which satisfies the hiding property by definition.  

\vspace{1ex}

Proving approximate average-case hardness is an entirely different story, however, and remains the central open theory problem in the context of quantum random sampling today.  
However, much work has been put into gathering evidence for the truth of approximate average-case hardness. 
In the following section, we discuss this evidence. 

\subsection{Approximate average-case hardness}
\label{ssec:approximate average-case hardness}

To do so, it is helpful to simplify the rather baroque error mixture \eqref{eq:stockmeyer error mixture} on any $(1-\delta)$ fraction to something more familiar---an exponentially small additive or a constant multiplicative error. 
Indeed, for those errors we already know that worst-case hardness of approximating the output probabilities and hence a necessary condition is true. 

\subsubsection{Reduction to additive or multiplicative average-case hardness}
\label{ssec:reduction additive/multiplicative average-case hardness}

To achieve this, we begin by observing that depending on which one of the two terms in \cref{eq:stockmeyer error mixture} is larger, the error will be relative or exponentially small additive, respectively. 
Hence, if we are able to get a handle on the comparative size of the two terms, we can reduce the error to a simpler form. 
Specifically if in the error bound~\eqref{eq:stockmeyer error mixture} the probability $p_0(C)$ is smaller than $\alpha/2^n$ for some constant $\alpha > 0 $ then \cref{eq:stockmeyer error mixture} can be upper-bounded in terms of an additive error $(2\epsilon/\delta + \alpha + o(1))/2^n$, if it is larger than $\alpha/2^n$, then \cref{eq:stockmeyer error mixture} can be upper bounded in terms of a relative error $2\epsilon/(\alpha \delta) + o(1)$. 

In order to reduce the error \eqref{eq:stockmeyer error mixture} to an exponentially small additive error, we can make use of concentration of the probabilities around their mean (given by $1/2^n$) using Markov's inequality
\begin{align}
\label{eq:markov bound stockmeyer}
  & \Pr_{C \sim \mc C} 
  \left[p_0(C) \geq \frac{1}{2^n \alpha} \right] 
  \leq \alpha, 
\end{align}
where the probability is taken over the choice of problem instances.
Since the probability in \cref{eq:markov bound stockmeyer} runs over the choice of random circuit, while in \cref{eq: markov error bound stockmeyer} it runs only over the uniformly random choice of outcome, the failure probabilities are independent from one another. 
Hence, both bounds are satisfied  with probability $(1-\delta)(1-\alpha)$ in which case \cref{eq:stockmeyer error mixture} is upper bounded by an exponentially small additive error
\begin{align}
   |q_0 - p_0(C)| 
   \leq \left[\frac{2 \epsilon }{ \delta} + \frac 1 {\poly(n)} \left( 1 + \frac 1 \alpha \right) \right] \frac 1 {2^n}. 
\end{align}

In order to reduce the error \eqref{eq:stockmeyer error mixture} to the arguably more ``natural'' case \cite[p.\ 61]{aaronson_computational_2010} of constant relative-error approximation we invoke a so-called \emph{anticoncentration property} introduced by \textcite{aaronson_computational_2010}. 
\begin{definition}[Anticoncentration]
\label{def:anticoncentration}
We say that a circuit family $\mc C$ anticoncentrates if for constant $\alpha> 0$ there exists $\gamma(\alpha) >0$ independent of $n$ such that 
\begin{align}
	\label{eq:anticoncentration}
	\Pr_{C \sim \mc C } \left [ p_0(C) \geq \frac \alpha {2^n} \right ] \geq \gamma(\alpha) . 
\end{align}
\end{definition}
Since the failure probabilities $\delta$ and $\gamma(\alpha)$ are independent, both bounds~\eqref{eq: markov error bound stockmeyer} and \eqref{eq:anticoncentration} are satisfied with probability at least $\gamma(\alpha) ( 1- \delta) $ in which case we obtain the relative-error bound bound 
\begin{align}
  \label{eq:relative stockmeyer bound} 
  |q_0 - p_0(C)| & \leq \left(\frac{2\epsilon }{ \delta \alpha } + \frac{1} {\poly(n)}\right) p_0(C). 
\end{align}

For the relative-error case, we can set $\alpha = 1/c$, $\epsilon = \gamma(\alpha)/4$ and $\delta = \gamma(\alpha)/2$ to obtain a $(c/2 + o(1))$-relative error approximation of $p_0(C)$ with probability at least $\gamma ( 1- \gamma/2)$ over the choice of instances.
For the additive-error case, we can set $2\epsilon/\delta = \kappa/2 $ and $\alpha$ constant to obtain a $(\kappa/2 + o(1))/2^n$-additive approximation of $p_0(C)$ with probability at least $4\epsilon \alpha/\kappa$ over the choice of instances.

We have reduced approximate average-case hardness (condition \ref{it:approx hardness} of \cref{thm:approximate sampling hardness}) to either of 
\begin{enumerate}[label = \arabic*a.]
\setcounter{enumi}{1}
	\item \label{it:additive approx hardness} 
	additive approximate average-case hardness up to an exponentially small additive error $O(2^{-n})$ on any $\gamma$ fraction, 
\end{enumerate}
or
\begin{enumerate}[label = \arabic*b.]
\setcounter{enumi}{1}
  \item \label{it:relative approx hardness} 
  relative approximate average-case hardness up to a relative error $1/4$ on any $\gamma(1 - \gamma/2)$ fraction, and
  \item \label{it:anticoncentration}
  anticoncentration for $\alpha=1 $ with constant $\gamma = \gamma(\alpha)$. 
\end{enumerate}

As of today, no proof of additive or relative approximate average-case hardness exists. 
But to see why a multiplicative-error
average-case hardness conjecture is plausibly true for \gapp-functions, consider again the argument above. 
For typical \sharpP\ functions the number of accepting paths is exponentially large and hence a multiplicative error is also of the same order of magnitude. 
In contrast, for typical \gapp\ functions, being differences of \sharpP\ functions, their number of accepting paths is a difference between two exponentially large numbers, which is often orders of magnitude smaller than each such number. 
This is why for \sharpP\ functions we often do not expect approximate average-case hardness, while for \gapp\ functions this conjecture seems reasonable. 

Another argument in favour of approximate average-case hardness makes use of universal quantities such as the Ising partition function~\eqref{eq:iqp ising partition function} 
\cite{bremner_classical_2010,goldberg_complexity_2014,bremner_averagecase_2016,boixo_characterizing_2016}, Tutte partition functions~\cite{goldberg_complexity_2014} or the Jones polynomial \cite{kuperberg_how_2015,mann_complexity_2017}. 
This argument observes that as we draw random instances of an Ising partition function $Z_W$ no additional structure is present as compared to a worst-case instance which a hypothetical approximation algorithm might be able to exploit.

While one might argue that these arguments are relatively weak, there have not been counterexamples to approximate average-case hardness in the standard settings either. 
In the following, we will see further and more substantial technical evidence towards the additive average-case hardness conjecture.

\subsubsection{Anticoncentration}
\label{sec:anticoncentration}

Let us begin with the anticoncentration property (\cref{def:anticoncentration}). 
The anticoncentration property allows us to reduce the baroque error \eqref{eq:stockmeyer error mixture} to a relative error, the arguably most natural error if we want to prove hardness of approximating the probabilities because \gapp\ naturally allows to reduce relative errors to exact computation. 
But anticoncentration can also serve as evidence for the additive approximate-average case hardness property to hold. 
By ruling out that almost all outcome probabilities are less than inverse exponentially small, anticoncentration rules out that an inverse-exponential additive error approximation is trivial: we cannot simply guess $0$ for all probabilities and be right almost always if anticoncentration holds. 

In this sense, a certain degree of anticoncentration is required to hold for approximate average-case hardness to be true.
Note, however, that anticoncentration is not a necessary property for hardness of sampling to hold---and neither is approximate average-case hardness. 
Both properties are merely used in the proof strategy we are describing in this section. 
But while approximate average-case hardness is sufficient for approximate hardness of sampling, anticoncentration is not.  

Notice that to prove anticoncentration we merely need to derive statistical properties of the respective random circuit families. 
To see this, we make use of the \emph{Paley-Zygmund inequality}~\cite{bremner_averagecase_2016},
a lower-bound analogue to Markov's inequality, which states that for a random variable $Z$ with $0 \leq Z \leq 1$
\begin{align}
  \label{eq:paley-zygmund}
    \Pr \left[ Z > \alpha \Eb[Z] \right] \geq (1- \alpha)^2 \frac{\Eb[Z]^2}{ \Eb[Z^2]}\ . 
\end{align}
Using the Paley-Zygmund inequality we can therefore  reduce the anticoncentration property  to the value of the second moments of the random circuit ensemble as 
\begin{align}
\label{eq:reduction to second moment}
    \Pr \left[ p_0(C) > \frac \alpha{2^n} \right] \geq (1- \alpha)^2 \frac{2^{-2n}}{ \Eb[p_0(C)^2]}\ . 
\end{align}
The normalized second moment $2^n \Eb[p_0(C)^2]$ is often also referred to as the \emph{average collision probability}.\footnote{The collision probability of a distribution $p$ is given by $\sum_x p(x)^2$. }
To prove anticoncentration for quantum random sampling, it is therefore sufficient to bound this  average collision probability as $O(2^{-n})$. 

This scaling of the average collision probability as $O(d^{-1})$ for quantum states in dimension $d$ is precisely the scaling that one obtains when drawing a quantum state $\ket \psi$ uniformly at random on the complex unit sphere $S(\mb C^d)$ and measuring it in the computational basis. 
Equivalently, we can draw a unitary $U \sim U(d)$ uniformly at random and apply it to a reference state as $U \ket 0$, giving rise to a uniformly distributed quantum state. 
The corresponding uniform measure $\mc U_{S(\mb C^d)}$ on the unit sphere is therefore invariant under the action of unitaries $U(d)$. 
For this measure, we can compute the $k^{\text{th}}$ moment projector as 
\begin{align}
\label{eq:moment operator states}
  M^k = \int_{S(\mb C^d)} (\proj \psi^{\otimes k}) d\mc U_{S(\mb C^d)}(\psi) = \frac {P_{[k]}}{D_{[k]}} , 
\end{align}
where $P_{[k]}$ is the projector on the symmetric subspace of $k$ tensor copies, and $D_{[k]} = \binom{d + k -1}{k}$ is the dimension of that subspace. 
See \cite[Sec.~I]{kliesch_theory_2021} for a pedagogical introduction to random unitaries and states. 

For uniformly random quantum states we can now compute the second moments of the output probabilities $|\braket x \psi|^2$ as 
\begin{align}
  \Eb[|\braket{x}{\psi}|^4] & = \mb E \left[\bra x^{\otimes 2} (\proj \psi)^{\otimes 2} \ket x^{\otimes 2}  \right]\\
  & = \bra x^{\otimes 2} \Eb\left[(\proj \psi)^{\otimes 2} \right] \ket x^{\otimes 2} \\
  & =  D_{[2]}^{-1} \bra x^{\otimes 2} P_{[2]} \ket x^{\otimes 2}\\
  & = D_{[2]}^{-1} = \frac2 {(d + 1)d},
\end{align}
where we have used that 
$\ket x^{\otimes 2}$ is in the symmetric subspace so that the projector $P_{[2]} = (\id + \mb S)/2$ with swap operator \begin{align}\mb S = \sum_{i,j =1}^{d^2} \ket i \ket j \bra j \bra i \end{align} acts trivially on it. 

For uniformly random quantum states, we therefore obtain from \cref{eq:reduction to second moment}  that the anticoncentration property holds with success probability at least $(1-\alpha^2)/2$. 
Proving anticoncentration of quantum circuit families can therefore be viewed as proving that the output probabilities of these families behave up to constant factors just like the output probabilities of uniformly random quantum states in terms of their average collision probability, their second moment. 
To prove bounds on the average collision probability, one may now proceed in various different ways. 
One can directly bound the average collision probability, or one can show that already the output states of circuits drawn from the family behave sufficiently similarly to uniformly random states. 
Let us briefly sketch the two most important ways via which anticoncentration can be proven for random quantum circuits: the so-called design property, and statistical-mechanics mappings.

\paragraph{Anticoncentration via spherical designs.}
While the circuit families proposed for quantum random sampling do not generate uniformly random quantum state  $C \ket 0 $, several families have the strong property that they mimic uniform randomness at the level of the second moment. 
A family of vectors $\Psi = \{\ket \psi_i\}_i$ which mimics uniform randomness for the $k^{\text{th}}$ moments in the sense that 
\begin{align}
M_\Psi^k = \frac 1 {|\Psi|} \sum_i (\proj {\psi_i})^{\otimes k} = \frac{P_{[k]}}{D_{[k]}} ,
\end{align}
forms a so-called complex (spherical) $k$-design.
We can slightly relax the notion of a $k$-design to approximations thereof and say that a family $\Psi$ is a (relative) $\epsilon$-approximate $k$-design if
\begin{align}
 (1-\epsilon) M_\Psi^k \leq \frac{P_{[k]}}{D_{[k]}} \leq  (1+\epsilon) M_\Psi^k. 
\end{align}
The proof of the following theorem then directly follows \cite{hangleiter_anticoncentration_2018}. 
\begin{lemma}[Anticoncentration of $2$-designs]
    \label{thm:anticoncentration} 
    Let $\Psi$ be a relative $\epsilon$-approximate $2$-design on $S(\mb C^d)$.  
    Then the output probabilities $\abs{\braket 0 \psi }^2$ of a randomly chosen $\ket \psi \in \Psi$ anticoncentrate in the sense that for $0 \leq \alpha \leq 1$
    \begin{equation}
        \Pr_{\substack{\ket \psi \sim \Psi}} \left( \abs{\braket 0 \psi }^2 > \frac{\alpha
(1-\epsilon) }{d} \right)  \geq  \frac{(1-\alpha)^2(1-\epsilon)^2}{2(1+\epsilon)} \ . 
    \label{eq:anticoncentration theorem} 
\end{equation} 
\end{lemma} 

Several circuit families considered for quantum random sampling approximately exhibit the $2$-design property when applied to a reference state. 
This holds in particular for universal random circuits in various settings. 
For random circuits, one can even prove a stronger property, namely that they are \emph{unitary designs}, mimicking uniform randomness on the unitary group as opposed to the complex sphere. 
Unitary designs by definition have the property that their columns form spherical designs and hence \cref{thm:anticoncentration} applies to them.
Historically, the first proof of the $2$-design property for random circuits is due to \textcite{harrow_random_2009}, albeit for a weaker (additive) notion of approximation than required for the proof of anticoncentration.

\textcite{brandao_local_2016} prove the stronger result that random circuits on $n$ qubits arranged in a linear chain form an $\epsilon$-approximate unitary $k$-design if they contain $ O(\poly(k) \cdot n (n + \log(1/\epsilon) ))$ many gates. 
The circuits they consider are composed of two-qubit gates that are applied either to random neighboring qubits or in alternating parallel ``brickwork'' configuration. 
The individual gates may be drawn either from a universal gate set containing its own inverses or uniformly (Haar) randomly. 
The key idea of the proof of \textcite{brandao_local_2016} is to map the design property to the gap of a local, frustration-free Hamiltonian, the local terms of which correspond to the individual two-qubit gates of the circuit and act on $4\cdot k$ many qubits, using the so-called \emph{detectability lemma}~\cite{aharonov_detectability_2009,anshu_simple_2016}. 
The gap of this Hamiltonian can then be bounded using a famous result due to \textcite{nachtergaele_spectral_1996}. 
\textcite{haferkamp_random_2022} has recently improved this result by showing a milder polynomial dependence in $k$,  providing an improved bound on the spectral gap. 
The same technique can also be applied to show the design property for other circuit families that encode universal quantum circuits, for example, random measurement-based quantum computations~\cite{haferkamp_closing_2020}.

Further examples of postselected-universal circuit families that exhibit the $2$-design property and therefore anticoncentration, are conjugated Clifford circuits \cite{bouland_complexity_2018}, Clifford circuits with magic-state inputs \cite{yoganathan_quantum_2018,hangleiter_anticoncentration_2018}, and diagonal quantum circuits applied to the state $\ket +^{\otimes n}$ \cite{nakata_generating_2014,hangleiter_anticoncentration_2018}.

Improving the result of \textcite{brandao_local_2016} to lattices of arbitrary dimension, \textcite{harrow_approximate_2018} prove that random universal circuits arranged on a lattice of dimension $D$ generate an approximate $k$-design using $\poly(k) \cdot n^{1+ 1/D}$ many gates. 
This result reflects the intuition that due to the fact that correlations in a parallel brickwork circuit spread ballistically, sufficiently random quantum states can only arise in a depth that scales linearly with the diameter of the system, and hence as $n^{1/D}$. 

\paragraph{Anticoncentration via computing the collision probability.}
 
While this intuition is presumably true for the design property of random circuits, it has recently been proven that anticoncentration already arises in logarithmic depth for nearest-neighbor random circuits in one dimension with uniformly random two-qubit gates \cite{barak_spoofing_2021}. 
To prove this result, \textcite{barak_spoofing_2021} directly bound the average collision probability, that is, the second moment $2^n \Eb[p_0(C)^2]$ using a mapping to a statistical-mechanics model due to \textcite{zhou_emergent_2019}. 
\textcite{dalzell_random_2020} show that this result is tight by complementing it with an $O(n \log(n))$ lower bound on the circuit size that holds for arbitrary geometries. 
For architectures with arbitrary connectivity, they further show that $5n \log(n)/6$
many gates are necessary and sufficient (up to subleading corrections) for exponentially small collision probability.
This in fact also holds directly for the anticoncentration property \cite{deshpande_tight_2022}.

Let us briefly sketch the idea of these proofs, following \textcite{hunter-jones_unitary_2019}. 
Again, the idea is to exploit the properties of the moment operator, albeit, now at the level of the individual quantum gates in the random circuit. 
For uniformly (Haar) random unitaries, we can, analogously to \cref{eq:moment operator states} define a moment operator $M_H^k$ on $U(d)$. 
This moment operator is characterized by so-called Weingarten functions $\mathrm{Wg}$ as \cite{brouwer_diagrammatic_1996,hunter-jones_unitary_2019}
\begin{equation}\label{eq:weingarten}
M_H^k = \Eb_{U\sim\mu_H} \left[ U^{\otimes k}\otimes \overline{U}^{\otimes k}\right] = \sum_{\sigma,\pi\in S_k} \mathrm{Wg}(\sigma^{-1}\pi,d)|\sigma\rangle\langle \pi|. 
\end{equation}
Here, $\ket\sigma = (\id \otimes r(\sigma)) \ket\Omega$, where $r$ is the representation of the symmetric group  $S_k$ on $(\mathbb C^{d})^{\otimes k}$ which permutes the vectors in the tensor product and $|\Omega\rangle=\sum_{j=1}^{d^t}\ket j \ket j $ is the maximally entangled state  up to normalization.
To evaluate formulae involving the moment operator \eqref{eq:weingarten}, it is useful to develop a graphical language for the moment operator. 
In this language, we can express the identity and the swap operator on two tensor copies, as well as the corresponding maximally entangled state as rewirings of single-copy identities as follows\footnote{See \cite{bridgeman_hand-waving_2017} for an introduction to the graphical representation.}. 
\begin{align}
  \id = \ipic[-.35]{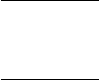}{.7}, 
  \hspace{1em}
  \mb S = \ipic[-.35]{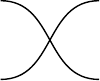}{.7}, 
  \hspace{1em} \ket \Omega = \ipic[-.35]{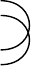}{1},\hspace{1em} \bra \Omega = \ipic[-.35]{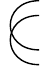}{1}. 
\end{align}
Hence, we can write 
\begin{align}
\ket{\mb S} \bra {\id}= \ipic[-.35]{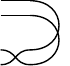}{1} \ipic[-.35]{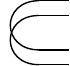}{1}= \ipic[-.35]{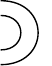}{1} \ipic[-.35]{bra_omega}{1}
.
\end{align}
For quantum circuits composed of Haar-random two-qubit unitaries, we can now evaluate the expectation value locally and the global moment operator is given by
\begin{align}
  \Eb_{U_1, \ldots, U_m \sim U(4)} \left[ U^{\otimes 2} \otimes \overline U^{\otimes 2} \right] = \prod_{i=1}^m \left(\Eb_{U_i \sim U(4)} \left[ U_i^{\otimes 2} \otimes \overline U_i^{\otimes 2} \right]\right), 
\end{align}
where $U = \prod_i U_i$ and in some abuse of notation we take the expectation over the individual quantum gates at their respective location in the quantum circuit.
Using Weingarten calculus, we can now evaluate the Weingarten formula for $k=2$, obtaining the result in graphical representation as
\begin{multline}
\label{eq:graphweingarten}
\Eb_{U\sim U(d) }\left[\ipic{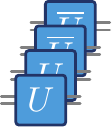}{1}\right]\\=\frac{1}{d^2-1}\left[\;\;\ipic{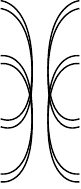}{0.8}-\frac{1}{d}\ipic{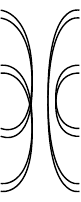}{0.8}-\frac{1}{d}\ipic{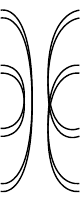}{0.8}+\ipic{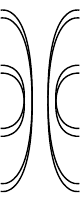}{0.8}\;\;\right]\;. 
\end{multline}
We can view the expectation value of a single two-qubit gate as an effective vertex 
\begin{align}
  \Eb_{U\sim U(d) }\left[\ipic{weingartenformula}{1}\right] \longrightarrow \ipic{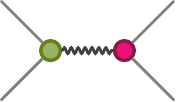}{1}, 
\end{align}
where the vertices can take one of two values $\id$ or $\mb S$ (corresponding to a spin up or down) that tell us how to contract each of the incoming or outgoing edges, and the curly edge between the vertices corresponds to a weight which is given by $-1/d/(d^2-1)$ for the configurations $\braket {\mb S}\id$ and $\braket {\id}{\mb S}$, and by $1/(d^2 -1)$ otherwise. The contractions themselves will pick up different values; for example, for a single contraction, we obtain $\braket{\mb S}{\id} = \braket{\id}{\mb S} = d$ and $\braket{\id}{\id} = \braket{\mb S}{\mb S}= d^2$. 
Computing the second moment $\Eb_U[|\bra x U \ket 0|^4$ now corresponds to computing a partition function over all local ``spin'' (aka.\ permutation) configurations with the corresponding weights and boundary conditions determined by $\ket x$ and $\ket 0$. 

\begin{equation}
\includegraphics{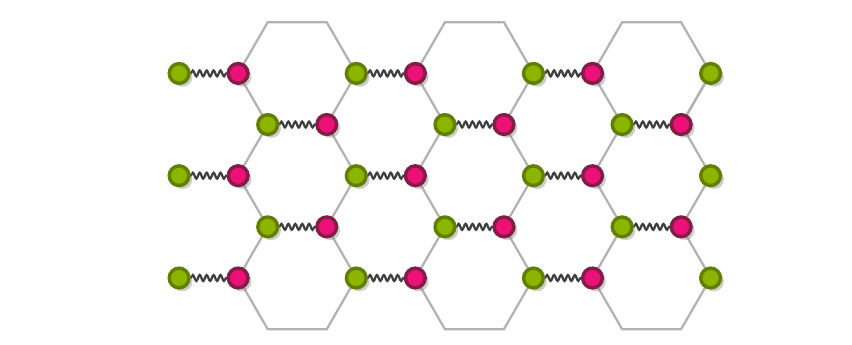}
\end{equation}
One can now sum over the pink vertices, giving rise to a new statistical mechanical model. 
This model is defined by terms that  with terms acting on the plaquettes of a triangular lattice. 
\begin{equation}
\includegraphics{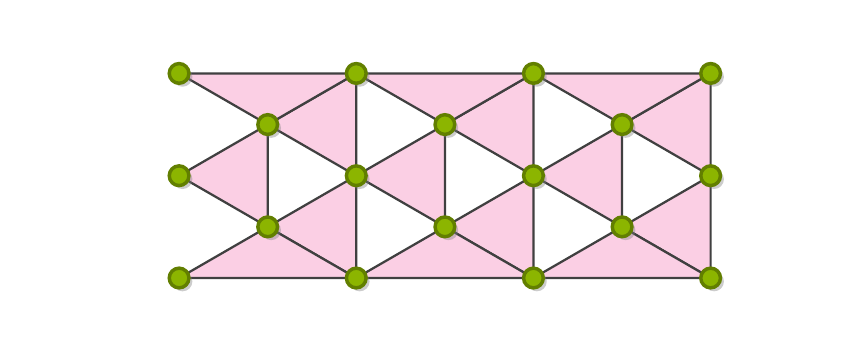}
\end{equation}
The plaquette terms are now just functions of permutations of the local spins with dimension $k$, which are nonzero only if the product of the permutations on a plaquette is the identity. 
For $k=2$, this allows one to perform simple domain-wall counting arguments in order to bound the value of the average collision probability.  

\paragraph{Further proofs of anticoncentration.}
An example of computing the second moments that makes use of the expression of the circuit amplitudes as partition functions is given by IQP circuits. 
For those circuits it is possible to directly compute the average collision probability, making use of the simple structure of the output probabilities as an Ising partition function (cf.\ \cref{eq:iqp ising partition function}) \cite{bremner_averagecase_2016}. 
Notably, there is also a direct proof of anticoncentration that does not rely on bounding second moments for the DQC1 model \cite{morimae_hardness_2017}.

The most important schemes for which anticoncentration has remained elusive are boson sampling protocols.
For Fock boson sampling, one can also compute the second moment of the output probabilities by making use of the hiding property so that the well-studied properties of Gaussian matrices can be exploited to compute $\Eb_{X \sim \mc G} [|\Perm(X)|^2]= n!$ and $\Eb_{X \sim \mc G} [|\Perm(X)|^4]/(n!)^2 = n+1$ \cite{aaronson_computational_2010}.
The value of the second-moments translates to a bound on the anticoncentration probability $\gamma$ in \cref{eq:anticoncentration} 
given by $1/(n+1)$ \cite{aaronson_computational_2010}. 
While numerical evidence suggests that anticoncentration 
is true for Fock boson sampling  \cite{aaronson_computational_2010}, second moments are therefore insufficient to prove this. 
Improving this bound, \textcite{tao_permanent_2008} prove that the permanent of $n\times n$ Bernoulli matrices is of order $n^{n( 1/2 -\epsilon)}$ with probability $1 - n^{-0.1}$, while a bound of order $n^{n(1/2 - O(\log(n)))}$ with inverse polynomial failure probability would be required for anticoncentration \cite[p.\ 75]{aaronson_computational_2010}.  
While this result may be extended to Gaussian distributions over $\mb C$, it is unclear how to further improve it \cite{tao_permanent_2008}. 
As a way around this, one might try to use higher moments of the Fock boson sampling distribution in order to obtain tighter bounds than provided by the Payley-Zygmund inequality. 
First steps in this direction have been taken by \textcite{nezami_permanent_2021}, who characterizes all moments of the distribution of Gaussian permanents and computes the lower ones, but concludes that only a closed formula for all moments may be sufficient to prove anticoncentration.  
For Gaussian boson sampling the situation remains even more elusive as, here, the distribution over which moments of the Hafnian \eqref{eq:Gaussianbosonsamplingdistribution} need to be computed is the so-called \emph{circular orthogonal ensemble} (COE), which is approximately given by symmetric Gaussian matrices of the form $XX^T$ with $X \sim \mc G$; see Paragraph \ref{par:hiding boson sampling}.

Be that as it may, recall that anticoncentration is merely a necessary condition for additively approximate average-case hardness, and a means to reduce this to relative-error approximations. 
The elephant in the room remains to prove the approximate average-case hardness conjecture in either its additive or its relative error version. This is the focus of the next section.  


\subsubsection{Average-case hardness: An overview}
\label{sec:average-case hardness}

Generally speaking, average-case complexity is a crucial question in cryptography, and comes with a number of intriguing peculiarities. 
Unfortunately, we have very few handles on average-case complexity and proofs of average-case hardness are only possible for very few complexity classes.
The question of average-case hardness has first been posed by \textcite{levin_average_1986} as a rigorous means to narrow down problem classes in which one can hope for simulation algorithms that work on average. 
What is the complexity of an instance drawn at random from some distribution $\mu$ over all possible problems? 
A key question in the context of average-case complexity is one posed already by \textcite{levin_average_1986}: 
how does the average-case complexity of a problem class depend on the distribution? 
Clearly, if one defines a probability measure to be supported on hard problem instances only, average-case complexity equals worst-case complexity. 
Intriguingly, there even exists a single so-called ``universal distribution'' for which the average-case complexity of any algorithm equals its worst-case complexity~\cite{li_average_1992}. 
The strong dependence on the distribution is part of the reason why average-case complexity under natural measures such as the uniform measure has remained largely elusive. 

Results that characterize average-case complexity of certain problems are only known for counting problems. 
The key conceptual idea underlying proofs of average-case hardness for such problems is the notion of \emph{random self-reducibility}. 
We say that a computational problem is randomly self-reducible if we can polynomially reduce the problem of evaluating any fixed instance $x$ to evaluating random instances $y_1, \ldots, y_k$ with a bounded probability that is independent of the input. 
Random self-reducibility is therefore a particular type of \emph{worst-to-average-case reduction}: 
We assume that there is a machine that solves random instances with probability bounded away from $1$ over a given distribution and then use this machine to try and efficiently solve an arbitrary fixed instance. 
If this is possible, then such a machine allows us to solve any instance in a time that is polynomially equivalent to the time it takes to solve a random instance. 
Hence, the problem must be as hard on average over this distribution as in the worst case.

A first step towards proving approximate average-case hardness of quantum output probabilities (that also constitutes a necessary condition) is to prove average-case hardness of near-exactly computing those output probabilities for the respective circuit family. 
Average-case complexity for near-exact computation has been pioneered by \textcite{lipton_permanent_1991} for the permanent as it prominently features in boson sampling~\cite{aaronson_computational_2010}. 
The key idea of Lipton's method is to use polynomial interpolation in order to interpolate from certain judiciously chosen random instances to an arbitrary, fixed instance.
This method is possible if the quantity in question can be written as a polynomial in the input parameters. 
While random quantum circuits lack this structure, the polynomial interpolation method of Lipton's can in fact be adapted to a broad class of quantum random sampling schemes~\cite{bouland_quantum_2018,movassagh_efficient_2018,movassagh_quantum_2020,kondo_fine-grained_2021,bouland_noise_2021,krovi_average-case_2022}.
In the following, we introduce and discuss these methods which eventually come close to proving approximate average case-hardness in that they tolerate an additive error of $O(2^{-O(m)})$ for random universal circuits, where $m$ is the number of gates in the circuit \cite{krovi_average-case_2022}. 
However, the step to inverse exponential $O(2^{-n})$ or relative error average-case complexity remains wide open and indeed remains the central open question in the field of quantum supremacy from a complexity-theoretic viewpoint.

\subsubsection{Random self-reducibility of the permanent}

\begin{figure*}
  \includegraphics{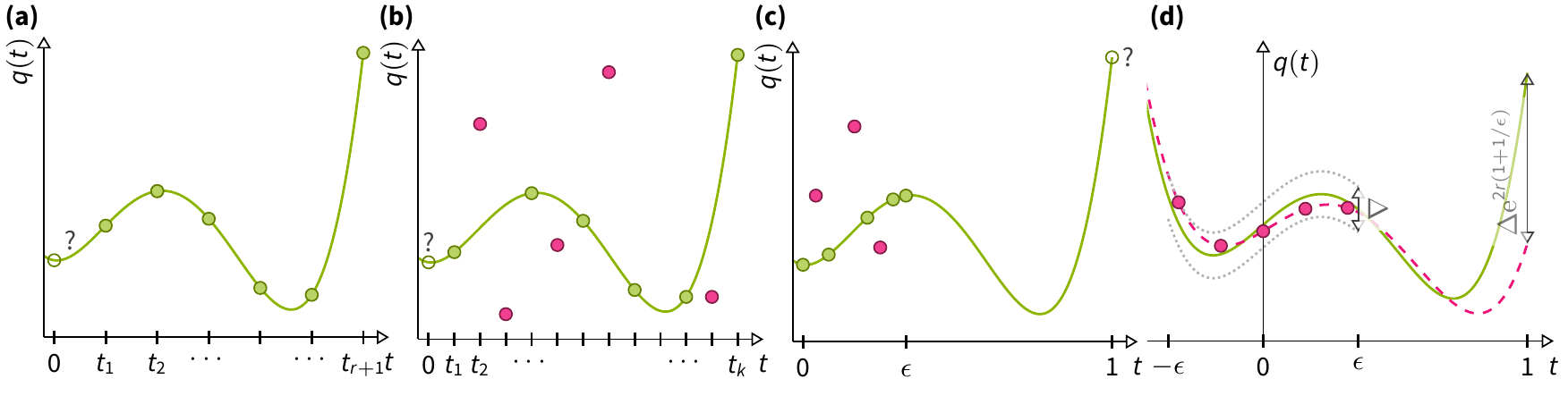}
  \caption{\label{fig:average-case hardness} \emph{(a)} From at least $r+1$ interpolation points $(t_i,q(t_i))$ one can efficiently interpolate a polynomial $q(t)$ of degree $r$.
  \emph{(b)} Using the Berlekamp-Welch decoding algorithm~\cite{welch_error_1986} for the Solomon-Reed code one can reconstruct a degree-$r$ polynomial from $k$ points $(t_i, y_i)$ if at least $ (k + r)/2$ of those points are correct. 
  \emph{(c)} When drawing instances from a distribution on the infinite field $\Cb$ as opposed to the uniform measure over a finite field, the interpolation points are chosen from the interval $[0,\epsilon]$ for $\epsilon = 1/\poly(n)$ so that the distribution of $G(t)$ in Eq.~\eqref{eq:g(t) worst-to-average case} does not deviate too far from the original distribution. 
  \emph{(d)} Using the result by \textcite[Theorem~\ref{thm:rakhmanov}]{rakhmanov_bounds_2007} one can bound the interpolation error of a degree-$r$ polynomial in the interval $(-\epsilon,\epsilon)$ when given evaluation points that are correct up to an error $\Delta$ (with inverse polynomial failure probability). 
  Using the Lemma by \textcite[Lemma~\ref{lem:paturi}]{paturi_degree_1992} one can then bound the extrapolation error when extrapolating to the hard problem instance at $t = 1$.
}
\end{figure*}

Let us start from the simplest and historically original proof of average-case hardness for \sharpP\---random self-reducibility of the permanent over a finite field $\mb F $ with respect to the uniform distribution over that field. 
Recall the definition of the permanent of an $n \times n $ matrix $X$ over $\mb F$ \eqref{eq:definitionpermanent}
\begin{align}
  \Perm(X) = \sum_{\sigma \in S_n} \prod_{j=1}^n x_{j,\sigma(j)}. 
\end{align}
The underlying structure in which the proof of random self-reducibility for the permanent is rooted is the algebraic fact that it is a degree-$n$ polynomial in the matrix entries of $X$ (and a degree-$2n$ polynomial in the case of $|\Perm(X)|^2$).
Concretely, the idea is the following: 
given an arbitrary instance $A \in \mb F^{n \times n}$,  draw a uniformly random matrix $B$ and for $t \in \mb F$ define the matrix 
\begin{align}
  \label{eq:average-case interpolation}
  E(t) = A + t B , 
\end{align}
for $t \in \mb F $. 
We think of $A$ as a `hard instance'.
Notice that for any fixed value of $t\neq 0$, $E(t)$ is distributed uniformly over $\mb F $.
This is in spite of the fact that, of course, $E(t)$ and $E(t')$ are correlated for values $t, t' \in \mb F $. 
As the permanent is a degree-$n$ polynomial in the matrix entries of an $n \times n$ matrix, the permanent of the matrix $E(t)$ is a degree-$n$ polynomial $q(t) = \Perm(E(t))$ in $t$. 

Let us now assume that there exists an efficient machine $\mc O$ that computes $\Perm(X)$ for uniformly random instances $X$ with failure probability $\delta$. 
Such an algorithm---while it may fail to evaluate $q(0) \equiv \Perm(A)$---will, by assumption, likely evaluate $q(t_i)$ correctly for some choice of evaluation points $t_i$. 
The idea is to infer $q(0)$ from the values of $q$ at the points $\{t_i\}_i$ using polynomial interpolation; see \cref{fig:average-case hardness}(a).

We can now query $\mc O$ on $n +1 $ distinct points $t_1, \ldots, t_{n+1} \neq 0 $ obtaining the values $q(t_i)$.\footnote{Notice that this requires the size of $\mb F $ to be at least $n+2$ and hence Lipton's proof does not work for the field $\mb F _2$, for instance. 
Indeed, for this case there are also known approximation schemes for the permanent~\cite{jerrum_polynomial-time_2004}. }
Applying a union bound, the probability that all of those values are correct is lower-bounded by $ 1 - (n+1)\delta $. 
Setting $\delta = 1/3n $ we thus obtain $n+1$ correct pairs $\{(t_i, q(t_i)), i \in [n+1]\}$ with probability at least $2/3 - 1/3n$. 
But $q$ is a degree-$n$ polynomial and hence those points uniquely determine $q$. 
We can now solve a linear system of equations to interpolate the polynomial $q$ and compute $q(0) = \Perm(A)$. 
Hence, an algorithm which solves random instances with probability at least $1-1/3n$ is able to solve arbitrary instances and computing the permanent over finite fields is average-case hard on any $1-1/3n$ fraction of the instances.  

\paragraph{Improving the success probability.}
Being correct on any $1-1/3n$ fraction of the instances is a rather strong requirement on the evaluation algorithm, however, and, by contraposition, requires only that at most a  $1/3n$-fraction of the instances \emph{need} indeed be \sharpP-hard to compute. 
Naturally, it is desirable to lower this requirement as far as possible to make stronger statements and assess average-case hardness as well as possible. 

Indeed, we can bring down the requirement on $\mc O $ to work correctly only for a constant $1/2 + 1/\poly(n) $ fraction of the instances \cite{gemmell_highly_1992,gemmell_self-testingcorrecting_1991}, \cite[see also][Section~8.7]{arora_computational_2009}. 
The idea is to use error-correction techniques for polynomial codes such as the Reed-Solomon code~\cite{reed_polynomial_1960}, where a string of $n$ symbols is identified with the coefficients of a degree-$(n-1)$ polynomial. 
Decoding algorithms for such codes output the correct polynomial even in the presence of some amount of errors. 

An error-correction algorithm for Reed-Solomon codes that will be extremely useful for our purposes is the algorithm by \textcite{welch_error_1986} as it works over \emph{arbitrary fields} and can even be extended to rational-function interpolation~\cite{movassagh_efficient_2018,movassagh_quantum_2020}. 
\begin{theorem}[Unique decoding for Reed-Solomon~\cite{welch_error_1986}]
  Let $q$ be a degree-$r$ polynomial over any field $\mb F $. 
  Suppose we are given $k$ pairs of elements $\{(t_i, y_i\})_{i \in [k]} $ with all $t_i $ distinct with the promise that $y_i = q(t_i)$ for at least $\max(r+1, (k + r )/2)$ points. 
  Then one can uniquely recover $q$ exactly in $\poly(k,r)$ deterministic time. 
\end{theorem}
We illustrate decoding with errors in \cref{fig:average-case hardness}(b).
Notice that for polynomially large $k$ the Berlekamp-Welch decoding algorithm tolerates an error rate that is arbitrarily close to a half. The Berlekamp-Welch algorithm is thus optimal in that as soon as less than half of the points are correct, no unique solution is guaranteed to exist. 

This issue is addressed by so-called \emph{list-decoding algorithms} which output a list of compatible solutions, observing that there cannot be too many such solutions~\cite[Sec.~19.5]{arora_computational_2009}. 
Such algorithms have been developed~\cite{beaver_hiding_1990,lipton_permanent_1991} for so-called 
\emph{Reed-Muller codes}~\cite{reed_class_1954,muller_application_1954}
over finite fields of which Reed-Solomon is a special case~\cite{sudan_decoding_1997}.
Using list-decoding algorithms, average-case hardness of the permanent over sufficiently large finite fields has been shown even for any inverse polynomial fraction of correct points \cite{cai_hardness_1999}; see \cite{guruswami_list_2006} for an overview of such approaches.

Let us illustrate the use of the Berlekamp-Welch algorithm to prove average-case hardness due to \textcite{gemmell_self-testingcorrecting_1991}: Using the Berlekamp-Welch algorithm, we can query the oracle $\mc O $ a number of times given by  $k> 2 (n+1)$ at distinct points $t_i$, obtaining pairs $(t_i, \mc O(t_i))$. 
We can then upper-bound the probability that less than $(k + n)/2$ of the obtained data points are correct as
\begin{align}
\label{eq:}
  \Pr\left[ \left| \{ i, \mc O(t_i) \neq q(t_i) \} \right| >  k - \frac { k + n} 2\right] < \frac{2\delta k }{k -n},
\end{align}
using Markov's inequality. 
This probability is at most $1/2$ if the failure probability of $\mc O $ satisfies
\begin{align}
  \delta < \frac 14 \left (  1 - \frac{k} {n}  \right)  . 
\end{align}
Hence the decoding procedure succeeds using $k$ samples as long as $\mc O$ works on a $3/4 + k/4n
= 3/4 + 1/\poly(n)$ fraction of the instances. 
Using an interpolation path to $A$ which is a polynomial in $k$. 
\textcite{gemmell_highly_1992} show that this can further be improved to a $1/2 + 1/\poly(n)$ fraction.

\paragraph{Distributions over infinite fields: the case of $\mb F = \Cb$.}
When considering the output probabilities of Fock boson sampling \eqref{eq:bosonsamplingdistribution} and Gaussian boson sampling \eqref{eq:Gaussianbosonsamplingdistribution}---and looking ahead also of quantum circuits---the matrices in question have entries not in a finite but an infinite field, the complex numbers $\mb F = \Cb$.
In this case, we are faced with two additional technical difficulties: 
first, there is no uniform or translation-invariant measure over the complex numbers. 
This means that when we construct the random matrix $E(t)$ as in Eq.~\eqref{eq:average-case interpolation} by drawing a random matrix $B$ from some distribution $\mu$, then $E(t)$ will be distributed according to some distribution $\mu'$ depending on the value of $t$ and the hard instance $A$. 
Assuming that we have found a solution to this problem, second, the polynomial interpolation and error-correction techniques that we have used above for the case of finite fields fail in case we only have a finite approximation of the values of $q(t_i)$. 
Numerically dealing with real numbers will, however, inevitably lead to finite-precision errors on the order of $2^{-\poly(n)}$. 

We can circumvent the first problem by choosing values of $t$ that are small such that the difference between $\mu'$ and $\mu$ in total-variation distance is small. 
As the total-variation distance upper-bounds the difference in probability that the two distributions assign to a specific event this difference translates to an additional contribution to the failure probability of~$\mc O $.   

The natural distribution over $\Cb$ which also appears in the Fock boson sampling problem is the complex normal distribution $\mc N_{\Cb}(\mu, \sigma)$ with mean $\mu$ and variance $\sigma^2$.
The following lemma, a variation of \textcite[Lemma~7.4]{aaronson_computational_2010}, bounds the total-variation distance between slightly shifted and squashed products Gaussian distributions with products of the standard distribution. 
\begin{lemma}[Autocorrelation of Gaussian distributions]
\label{lem:autocorrelation gaussian }
For the distributions 
\begin{align}
  \mc D_1 &= \mc N_{\Cb}(0,(1-\epsilon^2)\sigma)^M, \\
  \mc D_2 &= \prod_{i = 1}^M \mc N_{\Cb}(v_i,\sigma), 
\end{align}
with $v =(v_1,v_2, \ldots, v_M)\in \Cb^M$  and $\epsilon, \sigma > 0 $ it holds that 
\begin{align}
  \norm{\mc D_1 - \mc N_{\Cb}(0,\sigma)^M}_{\mathrm{TV}} & \leq 2 M \epsilon ,\\
  \norm{\mc D_2 - \mc N_{\Cb}(0,\sigma)^M}_{\mathrm{TV}} & \leq \frac 1 \sigma \norm{v}_{\ell_1}. 
  \end{align}
The same result holds for the uniform distribution $\mc U_{\Cb}(\mu,\sigma)$ centered around $\mu$ with cutoff $\sigma$. 
\end{lemma} 
For an arbitrary matrix $A=(a_{i,j})_{i,j}$ we now define the family of matrices 
\begin{align}
\label{eq:g(t) worst-to-average case}
  G(t) = t A + (1 - t) B 
\end{align}
similarly as above by drawing standard normal distributed instances $B \in \Cb^{n \times n}$. 
The matrix $E(t)$ is then distributed according to the new distribution 
\begin{align}
  \mc D = \prod_{i,j =1}^n \mc N_{\Cb}( t a_{i,j}, (1-t)^2).
\end{align}
Choosing equidistant values of $t_i$ in the interval $(0,\epsilon]$ for some cutoff $\epsilon > 0 $ will then result in a success probability of the algorithm $\mc O$ which has failure probability $\delta$ that is given by 
\begin{align}
  \Pr\left[\mc O(t_i) = q(t_i)\right] & \geq 1 - \delta - \norm{ \mc D - \mc G_{\Cb}(0,1)^{n^2}}_{TV} \\
  & \geq 1- \delta - 6n^2\epsilon. 
\end{align}
The remainder of the argument follows analogously as above by choosing $\epsilon = \delta/6n^2$. 
We illustrate the procedure in \cref{fig:average-case hardness}(c).

\paragraph{Robustness to finite-precision errors.}
The finite-precision problem requires somewhat more powerful machinery: 
using bounds on the stable extrapolation and interpolation of polynomials, we can recover the original proof using polynomial interpolation. 
This comes at the cost, however, that we cannot make use of the powerful error-correction techniques of Berlekamp and Welch anymore because those techniques require that some of the points are evaluated exactly. 

The two results that have been identified as being helpful to this effort by \textcite[Section~9.1]{aaronson_computational_2010} are a Lemma by \textcite{paturi_degree_1992} and a theorem by \textcite{rakhmanov_bounds_2007}.
\begin{lemma}[Stable extrapolation~\cite{paturi_degree_1992}]
\label{lem:paturi}
  Let $p: \Rb \rightarrow \Rb$ a polynomial of degree $r$ and suppose that $|p(x)| \leq \Delta$ for all $x $ such that $|x| \leq \epsilon$. 
  Then $|p(1)| \leq \Delta \ee^{2 r ( 1+ 1/\epsilon)}$. 
\end{lemma}
\begin{theorem}[Stable interpolation~\cite{rakhmanov_bounds_2007}]
\label{thm:rakhmanov}
  Let $E_k$ denote the set of $k$ equidistant points in $(-1,1)$. 
  Then for a polynomial $p: \Rb \rightarrow \Rb$ of degree $r$ such that $|p(x)| \leq 1$ for all $x \in E_k$, it holds that 
  \begin{align}
    |p(x)| \leq C \log \left( \frac \pi {\arctan \big(\frac kr \sqrt{\mc R^2 - x^2} \big)}\right) , 
  \end{align}
  for $|x| \leq \mc R = \sqrt{1 - r^2/k^2}$. 
\end{theorem}
We can now apply those results to the polynomial $p(t) = q(t) - q'(t)$, where $q'(t)$ is the polynomial defined by the slightly erroneous values $q'(t_i)$ of $q(t_i)$ satisfying $|q'(t_i) - q(t_i)| \leq 2^{-O(n^c)}$ for a sufficiently large $c$. 
Using Rakhmanov's result~\cite{rakhmanov_bounds_2007} we can bound the error between $q$ and $q'$ between the evaluation points; 
using Paturi's lemma~\cite{paturi_degree_1992}, we can then bound the error tolerance when extrapolating to $q(1)$; see \cref{fig:average-case hardness}(d).  
\vspace{1ex}

Let us note that exactly the same arguments apply to the output probabilities of Gaussian boson sampling, which are given by the squared Hafnian $|\Haf(XX^T)|^2$ for Gaussian $X \in \Cb^{2n\times 2k}$; recall \cref{GaussianBosonSamplingDistribution}. 
The squared Hafnian is a degree-$2n$ polynomial in its matrix entries (recall its relation \eqref{eq:permanent hafnian} to the permanent), and hence a degree-$4n$ polynomial in the matrix entries of the Gaussian-distributed matrix $X$. 

\subsubsection{Average-case hardness of quantum output probabilities}
\label{subsec:average-case hardness}

Let us now turn to average-case hardness of the output probabilities of quantum circuits. 
Firstly, let us observe, that there is a natural polynomial structure on the success probabilities of quantum circuits. 
For a quantum circuit $C = C_m \cdots C_2 C_1$ comprising $m$ gates $C_i$ acting on $n$ qubits, the output amplitudes can be expressed in terms of a path integral 
\begin{multline}
\label{eq:circuits average-case path integral}
  \bra 0 C \ket 0  =\\ \sum_{\lambda_1, \ldots \lambda_{m-1} \in \{0,1\}^n} \bra 0 C_m \ket {\lambda_{m-1}} \cdots \bra {\lambda_2 } C_2 \proj{\lambda_1} C_1 \ket 0 .
\end{multline} 
Consider that $C$ is drawn from some measure $\mu_{\mc C}$ that defines a circuit family $\mc C$. 
Some of the gates in $C$ might be randomly drawn from a gate set $\mc G$, others might be fixed across all $C \in \mc C $.

Now we are faced with a severe issue when trying to instantiate the idea of \textcite{lipton_permanent_1991}, however: 
when trying to construct an equivalent of $E(t)$ by choosing random instances $B$ for a fixed worst-case circuit $A$, the matrix given by $A + t B$ will not be unitary for $t \neq 0$ and therefore does not define a valid problem instance. 
Of course, this is because the unitary matrices do not form a group with respect to addition, but multiplication. 
How then can we perform a worst-to-average case reduction? 
A natural idea is to make use of the group structure by multiplying $A$ and $B$ in a gate-wise fashion in a way that is polynomial in an interpolation parameter, and then show that the distribution of the resulting instances does not deviate too much from the distribution of $B$. 
We can do so in different ways.

\begin{table*}
\centering
  \begin{tabular}{l >{\quad }c >{\quad }c>{\quad }c >{\quad}  c   }
  \toprule
     & Path & Interpolation method & Robustness & Instance fraction \\
  \midrule  
  \multirow{2}{*}{\cite{bouland_quantum_2018}*} & Truncated local Taylor series & Berlekamp-Welch (BW) & Exact & $3/4 + 1/\poly(n)$ \\
   & Truncated local Taylor series & Paturi + Rakhmanov & $2^{-\poly(n)}$ & $1 - 1/\poly(n)$ \\
  \cite{movassagh_efficient_2018} & Cayley paths & Rational BW & Exact & $3/4 + 1/\poly(n)$\\
  \cite{movassagh_quantum_2020} & Cayley paths & Paturi + Rakhmanov & $2^{-O(m^3)}$ & $1- 1/\poly(n)$\\
  \cite{bouland_noise_2021} & Cayley paths & Robust BW in $\bpp^\np$ & $2^{-O(m \log m)}$& $3/4+ 1/\poly(n)$\\
  \cite{kondo_fine-grained_2021} & Cayley paths & Lagrange interpolation + error bounds & $2^{-O(m \log m )}$ & $1 - 1/O(m)$\\ \cite{krovi_average-case_2022} &  Truncated global Taylor series & Robust BW in \bpp& $2^{-O(m)}$ & $> 3/4$\\
     \bottomrule
  \end{tabular}
  \caption{
\label{tab:average-case hardness} Comparison of the average-case hardness results for random quantum circuits on $n$ qubits with $m$ gates.\\ *Note that \textcite{bouland_complexity_2018}  prove average-case hardness for a non-unitary circuit whose output probabilities are $2^{-\poly(n)}$ close to the ideal output probabilities and the robustness we state is with respect to this non-unitary circuit, see the main text for a discussion of this point. 
  }
\end{table*}

\paragraph{Local Taylor series truncation \cite{bouland_quantum_2018}.}
On a high level, the first approach saves the polynomial structure of \cref{eq:average-case interpolation} by making use of Taylor expansion. 
We interpolate between a hard and a random instance as follows. 
For a hard instance of a circuit $C$ with random gates $C_1, \ldots, C_m$ drawn uniformly from a continuous subgroup $\mc G$ of the corresponding unitary group $ U(d)$ we define a new circuit by setting each gate  
\begin{align}
\label{eq:interpolation circuits bouland}
  C_i(t) = C_i H_i \ee^{- \ii  t h_i }, 
\end{align}
where $H_i$ is Haar-random in $\mc G$ and $h_i = - \ii \log H_i$ is its generator. 
Denote the resulting circuit as $C \circ H(t)$. 
$C_i(0)$ is Haar-random in $\mc G$, while for $t=1$ we recover the original gate $C_i$. 
Similarly to average-case hardness of  Gauss-random permanents, for tiny $t $ the gate $H_i \ee^{-\ii t h_i}$ looks almost Haar-random. 
One can therefore hope to follow the same procedure as above to extrapolate to $t = 1$, given values of $| \bra 0 C \circ H_t \ket 0 |^2$. 

However, the gates $C_i(t)$ and hence the output probability $| \bra 0 C \circ H_t \ket 0 |^2$ are no (low-degree) polynomial in $t$ so that polynomial interpolation cannot be applied. 
An easy way to circumvent this problem is to consider Taylor-approximations of the deformed gates $C_i(t)$. 
Let us define the $(t,K)$-truncated and perturbed Haar measure on the circuit family $\mc C$ by replacing each Haar-random gate $H_i$ in a circuit $C$ by 
\begin{align}
  G_i = H_i \cdot \left( \sum_{k = 0}^K \frac{(-\ii h_i t)^k}{k!}  \right) . 
\end{align}
We can now use the standard (Suzuki) bound on Taylor truncations 
\begin{align}
  | \bra \psi C_i G_i - C_i H_i \ee^{-\ii t h_i}  \ket \psi | \leq \frac \kappa {K!}, 
\end{align}
for a constant $\kappa > 0 $, set $K \in \poly(n)$, use an analogue of Lemma~\ref{lem:autocorrelation gaussian } 
to complete a worst-to-average case reduction for exactly computing the probabilities 
on any $3/4 + 1/\poly(n)$ fraction of the instances. 
Alternatively, as discussed above, we can apply the stability results by \textcite{rakhmanov_bounds_2007} and \textcite{paturi_degree_1992} to achieve robustness to additive errors $2^{-\poly(n)}$ on a $1- 1/\poly(n)$ fraction of the instances \cite[see also][Sec.~1.7 of the SM]{bouland_quantum_2018}.

A notable caveat of this approach is that in the reduction we have left the unitary group since the Taylor truncation of $\e^{-\ii t h_i}$ is non-unitary.
This means that average-case hardness is not achieved for exactly evaluating the circuit success probabilities, but only for exactly evaluating numbers $p_0(C)'$ which are $2^{-\poly(n)}$ additive approximations thereof and which do not correspond to success probabilities of valid quantum circuits. 
Nevertheless, average-case hardness of those numbers is a necessary requirement for the additive approximate average-case hardness property and hence serves as evidence for the conjecture. 
What is more, the additive approximate average-case hardness conjecture of the truncated distribution is equivalent to the additive approximate average-case hardness conjecture of the non-truncated distribution \cite[SM, Sec.~1.4]{bouland_quantum_2018}. For a more detailed discussion of this caveat, see \cite[App. D]{napp_efficient_2022} and \cite[Sec. 4.3]{movassagh_quantum_2020}.

\paragraph{Rational function interpolation \cite{movassagh_efficient_2018}.}
A more natural interpolation that remains within the unitary group and is much  more error-robust makes use of the \emph{Cayley function}
\begin{align}
  f(x)  = \frac { 1 + \ii x} { 1 - \ii x }, 
\end{align}
for $x \in \Rb$, defining $f(-\infty) = - 1$.
The Cayley function is a bijection between $\mb R \cup \{ - \infty \}$ and the complex unit circle. 
Observing that unitary matrices have eigenvalues on the complex unit circle, a Haar-random unitary matrix $H \in U(d)$ can therefore be uniquely represented as 
\begin{align}
  H = f(h), \quad h = h^\dagger, 
\end{align}
and $H^\dagger = f(-h)$. 
For each quantum gate $C_i \in U(d)$ we can then construct the path 
\begin{align}
\label{eq:cayley interpolation}
  C_i(t)  = C_i f(th_i) , 
\end{align}
with $h_i = f^{-1}(C_i^\dagger H_i)$ for Haar-random $H_i$ so that $C_i(0) = C_i$ and $C_i(1) = H_i$. 
The interpolated gate \eqref{eq:cayley interpolation} can be expressed as a fraction of two degree-$d$ polynomials using the spectral decomposition of $h = \sum_{\alpha =1}^d h_{i,\alpha} \proj{\psi_{i,\alpha}}$ as 
\begin{align}
  C_i(t) = \frac 1 {q_k(t)}\sum_{\alpha =1}^d p_{i,\alpha}(t) C_i \proj{\psi_{i,\alpha}}, 
 \end{align} 
 with 
 \begin{align}
  q_i(t) & = \prod_{\alpha = 1}^d ( 1 + \ii t h_{i,\alpha}), \\
  p_{i,\alpha}(t) & = f(h_{i,\alpha}) ( 1 - t h_{i,\alpha}) \prod_{\beta \in [d]\setminus \alpha} ( 1  + \ii t h_{i,\beta}) . 
 \end{align}
Denote the circuit resulting from this interpolation as $C \star H(t)$
Now, one can bound the total-variation distance for the distribution $\mc D_\epsilon$ on the circuit obtained when choosing $t = 1- \epsilon$ as $O(m\epsilon)$ \cite{movassagh_quantum_2020}.

However, while the techniques we have used so far were useful for polynomial interpolation, now we need to extrapolate a rational function. 
As a first step, it turns out one can generalize the Berlekamp-Welch algorithm to rational functions with degrees $k_1,k_2$ in the numerator and denominator, respectively~\cite{movassagh_efficient_2018,gemmell_highly_1992}. 
This algorithm requires that the number of evaluation points $t_i$ is at least $k_1 + k_2 + 2e$, where $e$ is the number of errors made by the evaluation algorithm $\mc O$.

A barrier to making this result robust lies in the fact that the results on stable interpolation~\cite{rakhmanov_bounds_2007} and extrapolation~\cite{paturi_degree_1992} of low-degree polynomials do not apply to rational functions.
\textcite{movassagh_quantum_2020} observes, however, that the output probabilities of the interpolated circuit can be reduced to a polynomial. 
To see this, observe that the output probabilities can be written as a fraction of two polynomials $Q(t) = \prod_{i=1}^m q_i(t)$ and $P(t) = \prod_{i=1}^m \sum_{\alpha} p_{i,\alpha}(t) C_i \proj{\psi_{i,\alpha}}$: 
\begin{align}
  |\bra 0 C \star H(t) \ket 0 |^2 = \frac {| \bra 0 P(t) \ket 0|^2}{|Q(t)|^2}. 
\end{align}
But as we can compute $Q(t)$ exactly in time $\Theta(m)$, we can reduce the rational function to a polynomial function by multiplying with $|\bra 0 C \star H(t) \ket 0 |^2$ with $|Q(t)|^2$. 
Now, one can show that $|Q(t)|^2 \leq 1 + O(m\epsilon)$ so that when choosing $\epsilon = 1/m$ the additional error incurred due to this multiplication is a multiplicative $O(1)$ error. 
The scaling of the extrapolation error in $|\bra 0 C \star H(t) \ket 0 |^2$ is therefore not disturbed when interpolating $|Q(t)|^2 \cdot |\bra 0 C \star H(t) \ket 0 |^2 $ instead. 

Now we can again resort to \cref{lem:paturi} and \cref{thm:rakhmanov} in order to compute the robustness as $2^{ - O( m/ \epsilon)} = 2^{ - O( m^2)}$ on any $1- 1/\poly(n)$ fraction of the instances~\cite[Theorem~3]{movassagh_quantum_2020} where $(0,\epsilon]$ defines the interval on which the success probabilities of $C(t)$ are evaluated.
\textcite{kondo_fine-grained_2021} observed that this can further be improved using the same strategy if Lagrange polynomials are used for the interpolation. 
For those polynomials, they find results analogous to \cref{lem:paturi} of \textcite{paturi_degree_1992}, and \cref{thm:rakhmanov} of \textcite{rakhmanov_bounds_2007} to obtain a robustness of $2^{-O(m \log m)}$ on any $1-1/O(m)$ fraction of the instances. 

The limitation of this approach, however, is that there is no error-correction procedure so that all results of the oracle need to be correct, giving rise to a very small tolerated failure probability because a union bound needs to be applied. 
Aiming to circumvent this issue, \textcite{bouland_noise_2021} observe that the failure probability can further be improved to $3/4 + 1/\poly(n)$ while retaining the same error scaling $2^{-O(m \log m )}$ by making use of an \np\ oracle.  
They achieve this by constructing more robust Berlekamp-Welch algorithm for polynomial interpolation over the real numbers. 
This algorithm makes use of the \np\ oracle in addition to randomness, and is therefore in the third level of the polynomial
hierarchy.

\paragraph{Global Taylor series truncation \cite{krovi_average-case_2022}.}
In a curious twist of events, \textcite{krovi_average-case_2022} recently observed that rather than performing a Taylor-series truncation on the level of individual gates, one can perform such truncation on the level of the global output distribution. 
The key observation of \textcite{krovi_average-case_2022} is that the output probabilities of circuits interpolated via \cref{eq:interpolation circuits bouland} can be expressed as a path integral
\begin{align}
  p(t)  = \sum_{r} e^{- i (t/m) \Delta \phi_r} A_r, 
\end{align}
in terms of $4^{2m}$ paths $r$, $|A_r| \leq 1$ and $|\Delta \phi_r|/m \in O(1)$. 
Here, the coefficients $A_r$ can be thought of as the path weights, and $\Delta \phi_r$ as their phases.  
Performing an appropriately chosen Taylor series truncation of $p(t)$, one finds that a degree-$O(m/\log m)$ polynomial is sufficient to achieve error robustness $2^{-O(m)}$ for circuits with Haar-random $2$-qubit gates and in fact robustness $2^{-O(n)}$ for IQP circuits. 
This result thus reduces the gap to the required robustness of $2^{-n}$ to constants in the exponent. 
By making use of recent results in polynomial interpolation that make use of specifically chosen points \cite{kane_robust_2017}, the success probability of the interpolation can further be improved to a constant without the need for an \np-oracle as in \cite{bouland_noise_2021}. 
We summarize the various average-case hardness results just discussed in \cref{tab:average-case hardness}.\footnote{Let us also note that a formulation of the above proof strategy using the language of representations of Lie groups is provided by \textcite{oszmaniec_fermion_2020}.}
\vspace{1ex}

A key issue to note in the worst-to-average case reductions on the unitary group is that the random gates in the circuit families need to be drawn from \emph{continuous subgroups} of the unitary group. 
Only if this is the case can one choose values of the interpolation parameter $t$ that are small enough such that the measure on the gate set is not perturbed too much in the interpolation step. 
In particular, this implies that the reduction does not apply to discrete gate sets and for some architectures the choice of random gates must be modified for the reduction to apply. 
For instance, to apply the average-case hardness results to IQP circuits family defined in \cref{eq:iqp circuit weights}, we need to choose the edge weights $w_{i,j}$ uniformly from the unit circle $S^1$ rather than from a discrete set of angles; see also~\cite{haferkamp_contracting_2020}.

One step in the direction of achieving an (exact) average-case hardness reduction for a discrete gate set has been taken by \textcite[Thm. 6]{dalzell_how_2018}. 
They consider the discrete family of IQP circuits whose output amplitudes are given by gaps of degree-3 Boolean polynomials (cf.\ Eqs.~\eqref{eq:iqp boolean polynomial} and \eqref{eq:iqp probabilities gap}). 
Specifically, they show a recursive reduction from the gap of a degree-3 polynomial with random degree-1 terms (but fixed degree-2 and degree-3 terms) to the gap of a worst-case polynomial (with the same degree-2 and degree-3 terms).
This translates to an exact average-case hardness result over a certain discrete family of IQP circuits. 
There are two issues with this approach, however. 
First, the family is very specific since it depends explicitly on the degree-2 and degree-3 terms of a worst-case instance. 
Second, it does not work for the output probabilities since these do not contain sign information about the gap anymore, which is crucial for the reduction---compare also the proof of approximate worst-case hardness of \gapp\ discussed in \cref{sec:approximating gapp}.
This strategy is still worth noting, however, since it is intrinsically distinct from the polynomial interpolation approaches discussed above and might yield another path to proving approximate average-case hardness.

\subsubsection{Discussion}
\label{ssec:discussion average-case hardness}

Using the techniques discussed above, we are currently able to prove approximate average-case hardness of universal random circuits with robustness $2^{-O(m)}$, where $m$ is the number of gates in the circuit.
This is further improved for IQP circuits to $2^{- O(n)}$, where $n$ is the number of qubits.   
To prove the approximate average-case hardness conjecture, we would need to improve this still to $O(2^{-n})$, however. 
Can we hope to prove such a result? 
The key technical obstacle on the way to addressing this question is the instability of polynomials with respect to variations in the interpolation points. 
Indeed, we saw in Paturi's lemma (Lemma~\ref{lem:paturi}) that the extrapolation error of a bounded error polynomial scales exponentially in the degree $r$ and size of the interval $\epsilon$ on which the bound holds, 
and the stronger version used by \textcite{kondo_fine-grained_2021} scales as an order-$d$ Chebyshev polynomial in $\epsilon$. 
As we have to make this interval inverse polynomially small to maintain closeness of the probability  distributions, this results in a strong increase of the Paturi bound, which can only be counter-weighted by an inversely scaling error bound on the interval $(-\epsilon, \epsilon)$. 
Small variations of a polynomial at a few points can thus lead to very large variations far away from those points. 

Random self-reducibility thus seems doomed when it comes to additive robustness of success probabilities on the order $2^{-n}$ as would be necessary for the quantum supremacy conjecture. 
Indeed, already \textcite[Section~9.2]{aaronson_computational_2010} argue that polynomial interpolation faces a significant barrier.
They argue that---in the presence of anticoncentration---the fact that polynomial interpolation is linear in the coefficients and hence linear with respect to additive errors prohibits it from allowing to prove approximate average-case hardness. 
Roughly speaking, this is because, even if two polynomials agree up to exponentially small error in an interval, they may exponentially disagree outside of that interval, while at the same time the target value of the polynomial might not be exponentially larger. Hence, constant relative-error approximations in the evaluation interval could translate to exponentially larger relative-error approximations at the target point.
The suggestion of \textcite[p.~91]{aaronson_computational_2010} is then to make use of a restricted class of polynomials that are not closed under addition, but which are at the same time able to capture the quantity of interest.

Making this argument somewhat quantitative, \textcite{bouland_noise_2021} investigate the applicability of random self-reducibility in the context of noisy circuits with error detection; see also \cite[Section 9.2]{aaronson_computational_2010}.
They show that even noisy, error-detectable probabilities which are conjectured to be $2^{-O(m)}$ close to uniform \cite{boixo_characterizing_2016} remain \sharpP-hard to compute up to error $2^{-16 m \log m- O(m)}$ in the average case via random self-reducibility.
But, so they argue, this implies that the average-case robustness of $2^{-O(m \log m)} $ is essentially optimal for this technique up to log-factors in the exponent.
The result of \textcite{krovi_average-case_2022} has further removed the log-factors in the exponent, providing a matching $2^{- O(m)}$ scaling of the robustness for universal random circuits and $2^{-O(n)}$ for IQP circuits. 
Similarly, for the case of Fock boson sampling on $m=O(n^c)$ many modes, they are able to show an even tighter error bound of $ e^{-(c+4)n \log n -O(n)}$, which is only constant factors in the exponent away from the $e^{-n \log n}$ robustness required to prove the approximate average-case hardness conjecture. 

Let us note that while the scaling conjecture of \textcite{boixo_characterizing_2016} has recently been shown in the low-noise limit \cite{dalzell_random_2021}, at high noise strengths, \textcite{deshpande_tight_2022} prove the expected convergence of the probabilities to uniform as $2^{-n - O(d)}$ \cite{deshpande_tight_2022}; see \cref{ssec:noisy distributions} for a discussion of these results. 
The latter result may lower the barrier for random circuits significantly. 

As another piece of evidence complicating a proof of approximate average-case hardness, for a (somewhat baroque) constant-depth universal circuit architecture in two dimensions, 
\textcite{napp_efficient_2022} proved that no approximate worst-to-average case reduction that enables the Stockmeyer argument is possible. 
Rather this architecture admits algorithms for both strong and weak simulation that are efficient on large fractions of the instances. 
But for the same architecture, strong simulation is classically intractable in the worst case unless \gapp\ admits a polynomial time algorithm and the polynomial hierarchy collapses to its third level, respectively. 
Moreover, they provide numerical evidence that random constant-depth universal circuits in 2D are efficiently simulable on average in practice. 
Strengthening this point, \textcite{deshpande_tight_2022} show that at sublogarithmic depth, almost all probabilities are subexponentially small for random universal circuits, so that anticoncentration does not hold. 
This implies that the trivial algorithm which always outputs $0$ is a good additive approximate average-case strong simulator for this case.
Note however, that this does not imply an average-case approximate sampling algorithm.    
Technically speaking, the upshot of these results is that any technique to prove approximate average-case hardness must be sensitive to the depth of the circuit, since we do not expect any technique to work at low depth.  
Moreover, while hardness of approximate sampling might hold for certain sublogarithmic depths, we are barred from proving it via the Stockmeyer argument. 

For an approximate worst-to-average case reduction we would require, it seems, quantum success probabilities that are extremely robust to noise in generic instances. 
Techniques such as quantum error correction~\cite{raussendorf_fault-tolerant_2006} might at first sight seem ideally suited for this task, but in such approaches errors need to be actively corrected. 
While in the framework of quantum sampling active correction can be bypassed using postselection~\cite{fujii_noise_2016,kapourniotis_nonadaptive_2019}, this means that only those probabilities corresponding to specific measurement outcomes on subsystems will be protected against errors. 
Since the postselection registers comprise at least a constant fraction of all registers the protected probabilities comprise only a $2^{-\Omega(n)}$ fraction of the instances. 
But by the hiding property every outcome probability is in one-to-one correspondence with the acceptance probability of a circuit from the family. 
So postselected fault-tolerance seems to be in conflict with average-case hardness.

To summarize, as it stands, we have very strong complexity-theoretic evidence of the hardness of \emph{exact sampling} from the output distributions of quantum random sampling schemes. 
This evidence is provided by the conjectured non-collapse of the polynomial hierarchy, which is a direct generalization of the unanimously believed $\classP \neq \np$ conjecture, whose failure would have extreme consequences on our widely tested view of the computational complexity of many different problems. 
Conversely, the evidence for the hardness of \emph{approximate sampling} is substantially weaker, since it is only based on the approximate average-case hardness conjecture.
The failure of this conjecture---while presumably unlikely---would not result in any meaningful consequences in complexity theory.
But while, as we have sketched in this section, there remain significant hurdles, proving this conjecture might still be possible in the not too far future.

\subsection{Fine-grained results}
\label{ssec:fine grained hardness}

The complexity-theoretic arguments we have discussed in quite some detail rule out an efficient classical simulation algorithm under the assumption of the non-collapse of the polynomial hierarchy and approximate average-case hardness of computing the respective output probabilities. 
However, they do not---and cannot---make any quantitative statements about lower bounds on the runtime of any classical simulation algorithm. 
But a convincing demonstration of quantum advantage not only require relies on asymptotic complexity-theoretic statements but also evidence that for the given finite size of the experiment there is no classical algorithm which can solve the problem using a reasonable amount of resources. 

This is the point at which so-called fine-grained complexity results continue to try and provide lower bounds on the runtime of classical simulation algorithms. 
The key idea of such results is to leverage versions of the so-called \emph{strong exponential time hypothesis} (SETH) which states that certain \np-complete problems cannot be solved in time faster than $2^{an}$ in the input size $n$ for some constant $a$ depending on the type of problem.
Such conjectures may then be leveraged to conjecture a fine-grained version of the collapse of the polynomial hierarchy.

Let us discuss this idea more concretely using the example of IQP circuits with output probabilities given by the squared gap of degree-3 polynomials (cf.\ \cref{eq:iqp boolean polynomial} and \eqref{eq:iqp probabilities gap}).
\textcite{dalzell_how_2018} provide a fine-grained hardness argument for this circuit family via a closely related problem, which they call \texttt{poly3-NONBALANCED}. 
The input to this problem is a degree-3 Boolean polynomial $f$, and the task is to decide whether $\gap(f) \neq 0$, i.e., whether the function $f$ has a different number of $0$ and $1$ outputs.
Since computing the gap of degree-3 Boolean polynomials is \sharpP-complete~\cite{montanaro_quantum_2017}, 
this problem is complete for a complexity class called \cocp. 
A language $L$ is contained in \cocp\ if there exists a polynomial-time algorithm $M$ such that for all $x \in \{0,1\}^*$
\begin{align}
  x \in L \Leftrightarrow \gap(M(x,\cdot)) \neq 0 , 
\end{align}
and is therefore closely related to the class \pp\ where the condition is 
\begin{align}
  x \in L \Leftrightarrow \gap(M(x,\cdot)) < 0 . 
\end{align}
\cocp\ is analogous to \pp\ and \sharpP\ in that an oracle to \cocp\ is sufficient to solve any problem in the polynomial hierarchy \cite{toda_counting_1992}, and conversely, an efficient algorithm for \cocp\ within the polynomial hierarchy would imply a collapse of \ph. 

The idea of fine-grained supremacy results is now analogously to the Stockmeyer argument to assume the existence of an efficient classical derandomizable sampling algorithm for the output distribution of an $n$-qubit IQP circuit $C_f$ up to a multiplicative error, using $g(n)$ gates and $t(n)$ time steps. 
This algorithm gives rise to a non-deterministic algorithm for \texttt{poly3-NONBALANCED} running in $s(n)$ steps in the sense that it accepts if and only if there is at least one computational path (i.e.\ input of randomness) giving rise to the all-zero sample. 
The fine-grained advantage result now relies on the following conjecture \cite{dalzell_how_2018}. 
\begin{conjecture}[\texttt{poly3-NSETH(a)}]
  Any non-deterministic classical algorithm that solves \texttt{poly3\-NONBALANCED} requires in the worst case $2^{an-1}$ time steps, where $n$ is the number of variables in the \texttt{poly3-NONBALANCED} instance.
\end{conjecture}
This conjecture directly yields a lower bound on the time complexity of the assumed classical sampling algorithm as $t(n) \geq 2^{an-1}$.
Omitting some fine print about the computational model in which this conjecture is phrased here\footnote{Since fine-grained complexity is about the concrete runtime, one has to fix the computational model. Typically, fine-grained complexity results are stated in terms of the so-called \emph{Word RAM model} \cite{williams_hardness_2015}.
},
the best known limit on $a$ is given by $a < 0.9965$ \cite{lokshtanov_beating_2017}. 

Analogously to the proof of additive-error sampling hardness via Stockmeyer's algorithm, 
fine-grained statements can be made for additive errors assuming an average-case lower bound on the runtime of a classical algorithm.
From this it is possible to estimate the number of qubits required to show a quantum advantage such that no classical computer will be able to reproduce the task. 
\textcite[Sec.\ 5.1]{dalzell_how_2018} estimate that IQP circuit sampling on roughly 200 qubits and $10^6$ many gates, would require at least a century using a classical-simulation algorithm running on state-of-the-art supercomputers.

Furthermore, statements can be made for different models by relating their simulation to well-studied problems such as \texttt{poly3-NONBALANCED}. 
In particular, this has been done for boson sampling \cite{dalzell_how_2018}, as well as the DQC1 model and Clifford+$T$ universal circuit sampling \cite{morimae_fine-grained_2019}.

\textcite{huang_explicit_2020} pursue a complementary approach on fine-grained results by considering strong simulation of quantum circuits via certain simulation algorithms. 
Specifically, they consider a subclass of classical simulation algorithms, which they call \emph{monotone simulators}. 
Roughly speaking, a monotone simulator is one that does not explicitly make use of the specific values of the nonzero matrix entries of the gates. 
A counterexample to a monotone method is therefore the simulator of \textcite{bravyi_improved_2016}, which explicitly uses the number of $T$ gates in the circuit. Note, however, that a $T$ gate does not differ from a simulable $Z$ or $S$ gate in terms of the locations of the non-zero matrix entries. 
Nonetheless, most tensor-network based methods (see \cref{sec:classical simulation} for details) are well captured by the monotone framework. 
They show an explicit lower bound of $\tilde O(2^{n-3})$ on the runtime of such monotone simulators. 
Furthermore, invoking the exponential-time hypothesis they  provide a $2^{n- o(n)}$ lower bound on strong simulation of quantum circuits.

\subsection{Complexity of sampling in the presence of noise}
\label{sec:noise complexity}

All of the complexity-theoretic analysis we have seen so far pertains to constant total-variation distance errors. 
While this is a meaningful notion of robustness, it is extremely challenging to achieve such errors in a scalable way: 
doing so requires local gate errors to scale at most inversely with the circuit size. 
Since local gate errors are the experimental bottleneck in any implementation of quantum random sampling, it is therefore natural to ask whether the sampling task remains hard in the presence of constant local gate errors. 
Constant gate errors tend to give rise to a TVD between the experimental output distribution and the target distribution that deviates from unity only by an inverse exponential. But it might still be the case that the sampling task remains difficult for a classical computer.

There are (at least) two ways of approaching this question. 
First, we can ask: Given certain local errors in a quantum random sampling scheme, what is the complexity of sampling from the output distribution? 
Second, we can ask: Is it possible to design a quantum random sampling scheme that is robust to constant local errors?  
While the first question requires an analysis of the noisy output distribution from the perspective of computational complexity, the second question might be solved by encoding a random sampling scheme in a fault-tolerant way. 
Let us briefly sketch some results along these directions in the following.

\subsubsection{Noisy output distributions}
\label{ssec:noisy distributions}
A natural noise model in the context of universal random circuits is given by single-qubit noise channels after each two-qubit gate in the circuit, since the fidelity of two-qubit gates is typically much worse than the single-qubit fidelity \cite{arute_quantum_2019}. 
Let us assume for simplicity that the noise channel is gate-independent, or that all two-qubit gates and the associated noise channel are the same, 
and that its average gate fidelity is given by $1 - \epsilon$. 
This model has been analyzed by \textcite{dalzell_random_2021} and \textcite{deshpande_tight_2022} in different regimes of the parameter $\epsilon$.

\textcite{dalzell_random_2021} compare the output distribution of the noisy circuit $p_{\text{noisy}}$ to the ``white-noise distribution'' with respect to an ideal distribution $p_{\rm ideal}$. 
Given a fidelity $F$, the white-noise distribution is defined as 
\begin{align}
  p_{\text{wn}} = F p_{\rm ideal}  + (1-F ) p_{\rm unif} , 
\end{align}
where $p_{\rm unif}$ is the uniform distribution. 
As it turns out, approximately sampling from the white noise distribution $p_{\text{wn}}$ with inverse polynomial fidelity $F$ within TVD error $\epsilon F$ is just as hard as approximately sampling from the ideal distribution $p_{\rm ideal}$ within TVD error $\epsilon$, given that $p_{\rm ideal}$ anticoncentrates in the sense that it has exponentially small second moments \cite[Theorem 4]{dalzell_random_2021}.
Notice that achieving an inverse polynomial fidelity would still require a local error rate of $\Theta(1/n)$ for circuits of size $O(n \log(n))$, which is the minimal size required for anticoncentration to hold, see \cref{sec:anticoncentration}.

\textcite{dalzell_random_2021} now show that the distance of the noisy distribution approaches the uniform distribution as $\ee^{-2m\epsilon + O(m \epsilon^2)}$, i.e., exponentially in the circuit size. 
At the same time, the distance to the white-noise distribution with fidelity parameter $F = \ee^{-2m\epsilon \pm O(m \epsilon^2)}$ scales as $O(F \epsilon \sqrt m)$, in the regime in which the noise parameter is small in the sense that $\epsilon n \log(n) \ll 1$ and the circuit  family satisfies the anticoncentration property, requiring $m \in \Omega(n \log(n)) $. 
Since the distance to the white noise distribution scales as a square root in the circuit size, their result shows a quadratic improvement in the required noise level for random quantum circuits as compared to the worst case for which the error would grow as $O(\epsilon m)$. 
To summarize, the average fidelity decay is exponential in $m$ and the typical distance to the corresponding white noise distribution grows slower than the worst case. 
Consequently, there is now an optimal scaling of $m$ with $n$ that achieves the minimal error to an appropriate white-noise distribution in terms of the circuit size.
It is in this regime that the XEB fidelity translates to a TVD bound and provides the best measure of quantum advantage; see also the discussion in \cref{ssec:xeb}.

Meanwhile, \textcite{deshpande_tight_2022} show that in the regime of large noise $\epsilon \in O(1)$, the expected total-variation distance to the uniform distribution is lower-bounded by $\exp(-O(d))$, where $d$ is the depth of the circuit. 
In certain regimes this result also holds for typical instances.
In light of the result of \textcite{dalzell_random_2021} which shows a fidelity decay in the circuit size $m = n \cdot d $, this is a surprisingly slow decay.

Notice that the respective bounds translate to a concentration bound on the distance of the individual probabilities to uniform as $2^{-O(m)-n}$ and $2^{-O(d)-n}$, respectively, by a Markov bound on the TVD. 
Let us also stress that the two results consider complementary regimes and their respective proof techniques fail beyond the considered regime. 
It remains an interesting open question to analyze the entire distribution of the TVD between the noisy distribution and the uniform distribution as well as its noise dependence in more detail. 

\subsubsection{Fault-tolerant random sampling}

As an alternative approach, one can consider the possibility of embedding quantum random sampling in a fault-tolerant encoding wherein error syndrome measurements are part of the sampling scheme. 
\textcite{fujii_noise_2016} observes that sampling from the entire distribution of such an encoding remains worst-case hard in the presence of noise. 
This is because one may postselect on the syndrome measurements returning the no-errors outcomes. 
In this case the conditional distribution on the sampling measurements is just given by the ideal distribution provided that the corresponding postselection probability is nonzero. 
Consequently, exact simulation of the noisy distribution remains computationally intractable in the worst case provided the local error rates are below the threshold for the used encoding. 

\textcite{kapourniotis_nonadaptive_2019} provide an explicit example of such an encoding in the measurement-based model of quantum computing, which also allows for an efficient verification scheme.
However, it is unclear to what extent the approximate average-case hardness conjecture required for this scheme is plausible, since it is based on the postselected success of magic-state distillation.
Building on ideas of \textcite{NoisyShallow}, \textcite{mezher_fault-tolerant_2020} develop high-dimensional and interactive measurement-based protocols in which this is achieved for every instance by appropriate classical postselection.


\section{Verification}
\label{sec:verification}

In the previous section, we discussed in detail the complexity-theoretic evidence for the classical intractability of quantum random sampling. 
But in order to demonstrate a quantum advantage via quantum random sampling the quantum implementation must be sufficiently accurate. 
It is therefore essential to verify that a claimed implementation of quantum random sampling in fact achieves the purported task. 

The verification task is extremely challenging, however. 
This is due to the difficulty of verifying sampling tasks in general, as well as the impossibility of efficiently simulating a sufficiently accurate implementation of quantum random sampling, or computing the corresponding output probabilities. 
In this section, we review different approaches to the verification problem, both inefficient and efficient ones. 

Clarifying the verification problem somewhat more formally  is a first nontrivial task since there are various distinct settings in which we can conceive of verification---we might allow for interaction between a skeptic and a quantum device which is claimed to produce samples from the correct distribution, or merely claimed to perform a task that is classically not efficiently solvable. 
We might ask to verify the device just from the samples it produces, or we might allow access to the quantum state of the device, i.e., by performing measurements in different bases. 

We begin by reviewing in some detail the reason why na\"ive verification from samples alone is impossible in the absence of assumptions on the device just because too many samples from the device would be required in \cref{ssec:hardness of verification}. 
We then move on to sample-efficient but computationally inefficient protocols for different verification settings that just use samples from the device in \cref{sec:sample-efficient verifiers}. 
Given the previous result, such protocols require assumptions on the device, or verify a weaker statement than the correctness of the samples. 
In \cref{sec:efficient quantum verification}, we then consider the setting in which we have direct access to the output state  $C \ket 0$ of the computation. As it turns out, this allows fully efficient and yet rigorous certification protocols for quantum sampling schemes that assume accurate quantum measurements in certain restricted bases.
Finally, we briefly discuss verification schemes that involve several rounds of interaction between a sceptic verifier and the quantum device under investigation in \cref{sec:efficient classical verification}.

\subsection{Hardness of verification from classical samples}
\label{ssec:hardness of verification}

In this section, we will see a simple argument why verifying the samples from quantum random sampling schemes typically requires exponentially many samples and is therefore infeasible---the quantum device would need to be run exponentially many times. 
To this end, one can invoke a strong result by \textcite{valiant_automatic_2017} on optimal identity testing and  properties of the output probability distribution of quantum random sampling \cite{hangleiter_sample_2019}.

\begin{theorem}[Optimal identity testing \cite{valiant_automatic_2017}]
  There exist constants $c_1, c_2 >0$ such that for any $\epsilon > 0 $ and any target distribution $P$, there exists a test which, given samples from a distribution $Q$, distinguishes whether $P=Q$ or $\tvd{P-Q} > \epsilon$, promised one is the case, given 
  \begin{align}
    c_1 \max\left\{ \frac 1 \epsilon, \frac 1 {\epsilon^2} \norm{P_{-\epsilon/16}^{-\max}}_{\ell_{2/3}} \right\} 
  \end{align}
  many samples. 
  On the other hand, there exists no such test from fewer than 
  \begin{align}
    c_2 \max\left\{ \frac 1 \epsilon, \frac 1 {\epsilon^2} \norm{P_{-2\epsilon}^{-\max}}_{\ell_{2/3}} \right\} 
  \end{align}
  samples. 
\end{theorem}
For a vector of non-negative numbers $P$ we here define $P^{-\max}$ to be the vector obtained from $P$ by setting the largest entry to zero, and $P_{-\epsilon}$ the vector obtained from $P$ by setting all of the smallest entries to zero such that the sum of the removed entries is less than $\epsilon$. 
Moreover, $\norm{P}_{\ell_{2/3}} = (\sum_x p_x^{2/3})^{3/2}$. 
The $\ell_{2/3}$ norm of ${P_{-\epsilon}^{-\max}}$ therefore completely characterizes the asymptotic complexity of identity testing up to constant factors in $\epsilon$. 
The intuition behind the  result of \textcite{valiant_automatic_2017} is that the largest probability as well as the tail of the distribution are easily detected in an identity test, because a constant deviation in these parts of the distribution will be visible in the samples obtained with high probability.
An important corollary of their result---which was known prior to it \cite[see e.g.][]{goldreich_introduction_2017}---is that the complexity of testing against the uniform distribution on a sample space $\Omega$ requires $O(\sqrt{|\Omega|})$ samples, while verification requires less samples for more peaked distributions.

Lower bounds on the certifiability of quantum random sampling, intuitively speaking, follow from the fact that the output distributions of the schemes are extremely flat with high probability. 
Technically speaking, we obtain the lower bounds from bounding the $\ell_{2/3}$ norm. 
It turns out that the second moments that were used to prove anticoncentration are just sufficient for that. 
To see this, following \textcite{hangleiter_sample_2019}, we first observe that the $\ell_{2/3}$ norm can be lower bounded in terms of the largest probability $p_0 $ of a distribution $P$ as 
\begin{align}
  \norm{P_{-\epsilon}^{-\max}}_{\ell_{2/3}} \geq p_0^{-1/2}(1- \epsilon - p_0)^{3/2}. 
\end{align}
Second, we observe that the R\'enyi-2 entropy $H_2(p) = - \log \sum_{x} p_x^2$ upper bounds the largest probability as 
\begin{align}
  \log p_0 \leq - \frac 1 2 H_2(P).  
\end{align}
But now we can use the fact that most quantum random sampling schemes given by a circuit family $\mc C$ have bounded average collision probabilities (cf.\ \cref{sec:anticoncentration}) and that they concentrate around the mean by Markov's inequality as 
\begin{align}
\sum_x \Eb_{C \sim \mc C }[p_x(C)^2] \leq O(2^{-n}/\delta), 
\end{align}
with probability at least $1-\delta$. 
This implies that the R\'enyi $2$-entropy is bounded as $H_2(P_C) \geq n + \log (O(\delta))$. 
Consequently, the largest probability is exponentially small with high probability, i.e., $\log p_0 \leq - (n + \log (O(\delta)))/2$. 
The sample complexity of certifying quantum random sampling from samples scales at least as $\Omega(2^{n/4 + O(\delta)})$ with probability $1-\delta$ over the choice of circuit instance.
Let us note that even though the second moments of the boson sampling probabilities are not sufficiently small to prove anticoncentration, they are small enough to prohibit sample-efficient verification \cite{hangleiter_sample_2019,gogolin_bosonsampling_2013}. 

While one might think that this result is actually not too bad in that few enough samples may be required for intermediate-scale instances of quantum random sampling, it turns out that the optimal identity test of \textcite{valiant_automatic_2017}, which employs a variant of the $
\chi^2$-test, is highly impractical in that the constants involved are much too large. 
What is more, the problem becomes more challenging when the test is also required to accept distributions not too far away from the ideal distribution. 
This is because this requirement poses an additional constraint on the testing protocol. 

\subsection{Sample-efficient classical verification via cross-entropy benchmarking}
\label{sec:sample-efficient verifiers}

To overcome the obstacle of exponential sample complexity, one may consider a weaker requirement than verifying the full total-variation distance. 
The most prominent approach that achieves this is a family of tests, which we subsume under the label \emph{cross-entropy benchmarking}. 
These tests have been introduced in a series of works~\cite{boixo_characterizing_2016,aaronson_complexity-theoretic_2017,arute_quantum_2019,neill_blueprint_2018}. 
The central idea is to use multiplicative measures of similarity between the implemented ``noisy'' distribution $Q$ and the ideal target distribution $P_C$ that 
measure the correlation between the two distributions. 
We can express those measures as follows.
\begin{definition}[Cross-entropy measures]
Let $f: [0,1] \rightarrow \Rb$ be a monotonously increasing function. Define
\begin{align}
\label{eq:cross entropy measures}
  F_f(Q,P_C) = \sum_{x \in \{0,1\}^n} Q(x) f(P_C(x)) 
\end{align}
as the cross-entropy measure corresponding to $f$. 
\end{definition}
The first observation that we can make is that  by a Chernoff bound the cross-entropy measures $F_f$ can be \emph{sample-efficiently} estimated from a number of samples that depends on the variance of $f(P_C(x))$ over $x$ and scales as $1/\epsilon^2$ in the estimation error. 
Notably, for exponentially small values of $F_f(Q,P_C)$, the error $\epsilon$ needs to scale inverse exponentially, too. 
Hence, sample efficiency is lost in that case.

The second observation is that estimating cross-entropy measures is \emph{computationally inefficient} for quantum advantage schemes since the probabilities $P_C(x)$ of the ideal distribution (or a function thereof) need to be computed for the observed outcomes. 
As we will see, this constitutes an important obstacle to their practical usage in verifying quantum random sampling in the quantum advantage regime.  

\begin{figure}
\includegraphics{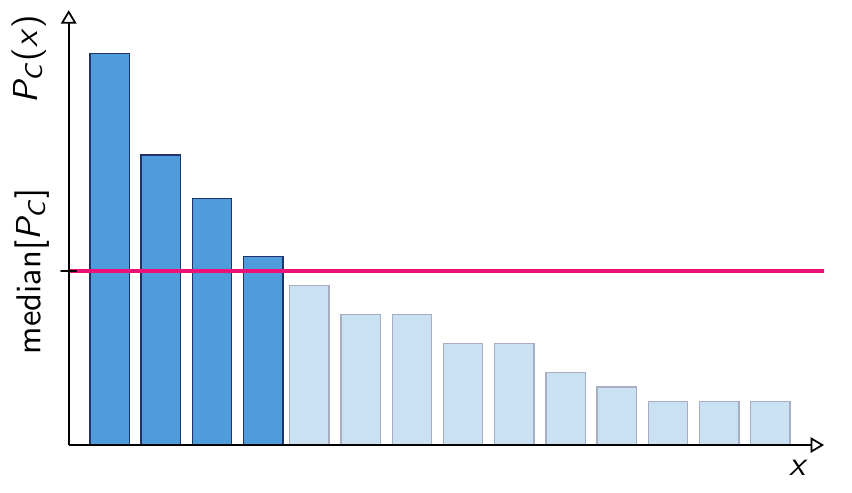}
\caption{\label{fig:hog}
In the task \emph{heavy outcome generation} (\hog) one is asked to output a list of strings $\{x_1, \ldots, x_k\} $ for which $P_C(x_i) \geq \median (P_C)$. 
}
\end{figure}

While different variants of the measure are interpreted differently, the intuition underlying all such measures is the following: 
those distributions which get the \emph{heavy outcomes} of a quantum computation correct will score well on cross-entropy measures because these outcomes dominate the measure~\cite{aaronson_complexity-theoretic_2017}. 
One can characterize ``heavy outcomes'' as those bit strings $x\in\{0,1\}^n$ for which the probability $P_C(x)$ of obtaining $x$ is large, for example, larger than the median of $P_C$, see Fig.~\ref{fig:hog}.

Before we introduce the most important measures---heavy outcome generation,
cross-entropy difference, and cross-entropy benchmarking fidelity---let us discuss in more detail the shape of the outcome distribution of random quantum circuits. 
Consider the success probability $p_U(0) = | \bra 0 U \ket 0 |^2$ of a Haar-random unitary $U \in U(d)$. 
Then the distribution of $p = p_U(0)$ over the choice of $U$ is given by the so-called \emph{Porter-Thomas distribution} \cite{porter_fluctuations_1956}
which is asymptotically exponentially distributed\footnote{See \cite[Chapter 4.9]{haake_quantum_2010} for the derivation.}
\begin{align}
  \label{eq:porter-thomas distribution}
  P_{\text{PT}}(p) = (d - 1)(1 - p)^d \xrightarrow{d \gg 1} d \exp(- d p). 
\end{align}
For $d \gg 1 $ one can now invoke Levy's lemma \footnote{Levy's lemma \cite{ledoux_concentration_2005} can be stated as follows.
Given a function $f : S^D \rightarrow \mb R$ defined on the 
$D$-dimensional hypersphere $S^D$ with zero mean and an
$x \in S^D$ chosen uniformly at random, then
\begin{align}
\Pr[|f(x)  | \geq \varepsilon] \leq 2 \exp
\left(-\frac{2C(D + 1)\varepsilon^2}{\eta^2}
\right),
\end{align}
where $\eta>0 $ is the Lipschitz constant of $f$ 
and $C>0$ is a constant. For normalized quantum 
state vectors of a complex vector space of dimension $d$,
$D=2d-1$.
The heuristic intuition developed here is that for random processes with an 
approximately constant Lipschitz constant, one would expect the fluctuation to scale
approximately as the inverse square root of the dimension $d=2^n$ of the underlying Hilbert space.} \cite{ledoux_concentration_2005} to see that the finite distribution of outcome probabilities of a fixed, Haar-randomly drawn unitary is expected to be $O(1/\sqrt{d})$ close to the Porter-Thomas distribution.
While exactly implementing Haar-random unitaries via a quantum circuit requires exponentially many gates, it has been numerically shown by \textcite{boixo_characterizing_2016} that the output distribution of universal random circuits quickly tends towards the Porter-Thomas distribution in terms of the lower moments of the distribution. 
This evidence serves as justification for the use of properties of the Porter-Thomas distribution as opposed to merely the second moments of the distribution in the analysis of cross-entropy measures.

\subsubsection{Heavy-outcome generation}
\label{ssec:hog}
The most basic cross-entropy measure that serves as intuition for the more involved measures we will see later, is based on the so-called \emph{heavy outcome generation} task or \hog\ as introduced by \textcite{aaronson_complexity-theoretic_2017}. 
\begin{problem}[\hog~\cite{aaronson_complexity-theoretic_2017}]
\label{prob:hog}
	Given as input a random quantum circuit $C \in \mc C $ from a family $\mc C $, generate distinct output strings $x_1, \ldots, x_k $, at least a $2/3$ fraction of which have a probability greater than the median $\median[P_C]$ of $P_C$. 
\end{problem}
\hog\ is equivalent to achieving a non-zero score in the \emph{\hog\ fidelity}
\begin{multline}
\label{eq:hog fidelity}
  F_{\text{\hog}}(Q,P_C) = \\
  \frac 2 { \ln 2} \sum_{x \in \{0,1\}^n} Q(x) \left( \theta(P_C(x) - \median[P_C]) - \frac 12 \right) , 
\end{multline}
defined in terms of the step function $\theta: \mb R \rightarrow \{0,1\}$ which is $0$ for $x < 0$ and $1$ otherwise.  

Because it is defined in terms of the bias of the target distribution, $F_\hog$ can be sample-efficiently estimated.  
First, the median can be estimated very efficiently up to a small error from few samples. 
Given $k$ samples $\{x_0, \ldots, x_k\}$ from a noisy distribution $Q$, we then need to compute the probabilities $P_C(x_i)$ and compare them to the median. 
By Hoeffding's inequality this can be achieved with error $O(1/\sqrt{k})$ with exponentially small failure probability. 

Let us briefly discuss the properties of $F_\hog$. 
If $Q$ is maximally noisy---that is, the uniform distribution---then 
\begin{multline}
F_{\text{\hog}}(Q,P_C) = \\
\frac 2 {\ln 2} \left(\frac 1{2^n} \left|\{ x: P_C(x) \geq \median[P_C] \}\right|  -\frac 12 \right)   = 0 ,
\end{multline}
as the median is defined as the largest number such that the sum of the output probabilities of $C$ exceeding that number is at least $1/2$. 
On the other hand, in an ideal implementation for which $ Q = P_C$, $F_{\text{\hog}}(Q,P_C) > 0 $ so long as $P_C$ is nonuniform. 
This is because, by definition, the probabilities above the median are larger than those below the median and hence the probability weight above the median is at least $1/2$.  
More specifically, if the outcome probabilities $P_C(x)$ are Porter-Thomas distributed, then $F_{\text{\hog}}(P_C,P_C) = 1$. 
To see this, observe that the median of the exponential distribution is given by $\ln 2 / 2^n $ and the total probability weight of $P_C$ above the median is then given by\footnote{See also \textcite[Footnote 3]{aaronson_complexity-theoretic_2017}.}
\begin{multline}
  \sum_{x \in \{0,1\}^n} P_C(x) \theta(P_C(x) - \ln 2/2^n) \\
  \approx \int_{\ln 2 / 2^n}^\infty 2^n \ee^{- 2^n p } \ \mathrm d p = \frac{ 1 + \ln 2}2 .     
\end{multline}

More generally, a distribution that scores well in terms of $F_{\text {\hog}}$ will therefore tend to be closer to an ideal implementation of $P_C$ in terms of total-variation distance. 
This is rigorously true in case the noisy distribution is a convex mixture
\begin{align}
\label{eq:uniformly noisy distribution}
  Q_\lambda(x) = (1 - \lambda) P_C(x) +  \lambda \frac 1 {2^n}, 
\end{align} 
of the ideal target distribution and the uniform distribution with $\lambda \in [0,1]$.

Clearly, there are also distributions, however, which score well on the \hog\ fidelity, but are far away from $P_C$. 
To see this, just take the distribution which is supported on $\{ x : P_C(x) \geq \median[P_C]|\}$. 
This distribution will have a \hog\ fidelity of $1/\ln 2 > 1$ even though its total variation-distance to $P_C$ is at least $ (1 - \ln 2)/2$. 
 
\paragraph{Computational hardness of HOG.}

It is presumably very hard to find a distribution which has high support on the heavy outcomes of the target distribution, though. 
Scoring well on $F_\hog$ may thus well be computationally hard even though it is a strictly easier task than approximately sampling from the target distribution. 
To see this observe that the ability to sample from the correct distribution implies the ability to score well on $F_\hog$, but not vice versa since $F_\hog$ does not quantify the TVD. 
\textcite{aaronson_complexity-theoretic_2017} conjecture precisely that: 
\hog\ is computationally intractable for random quantum circuits. 
To support this conjecture, they reduce it to the hardness of deciding whether $| \bra 0 C \ket 0|^2$ is larger than $\median[P_C]$ with probability at least $1/2 + \Omega(2^{-n})$ over the choice of $C$.  
The \emph{quantum threshold assumption} (\quath) states that this task is computationally intractable for classical computers. 
To reduce \quath\ to \hog\, we simply assume that there is an efficient routine solving \hog. 
Then, given a quantum circuit $C$ we can run that routine on the circuit $C'= \prod_i X_i^{z_i}$, where $z$ is a uniformly random string.
If $z$ is contained in the $k$ output samples, then we output ``Yes'', otherwise we output ``Yes'' with probability $1/2$ and ``No'' otherwise. 
This procedure decides whether $z$ is a heavy string for $C'$, or equivalently whether $0^n$ is heavy for $C$, with success probability at least $1/2+\Omega(1/2^n)$ since $z$ is uniformly random.

Conversely, \hog\ can be solved by a quantum algorithm for circuits with probability weight above the median greater than $2/3$ with probability at least $1 - \exp(-\Omega(k))$.  
\textcite[Lemma 8]{aaronson_complexity-theoretic_2017} provide a neat proof that this is indeed the case with high probability by showing that in expectation the probability weight above the median is lower-bounded by $5/8$. 

The \hog\ test, and the \hog-fidelity $F_\hog$ can therefore be considered benchmarks for quantum random sampling based on evidence independent of the argument presented in \cref{sec:hardness}. 
While \hog\ and \quath\ may be plausible conjectures, however, the level of complexity-theoretic evidence for either both \quath\ and the intractability of \hog\ is extremely weak.  
This is because we have no independent underpinning of those conjectures such as the non-collapse of the polynomial hierarchy which is independently grounded in significant evidence. 


\paragraph{Fine-graining \hog: Binned outcome generation.}
A natural way to connect the properties of the \hog\ fidelity with the TVD, is to bin probabilities in a more fine-grained fashion~\cite{bouland_quantum_2018}. 
This retains the complexity-theoretic intuition behind \hog\ that producing outcomes that are correlated with the ideal distribution is hard, and is also more directly supported by evidence for the intractability of simulating quantum random sampling within constant TVD. 
A natural starting point for such a more fine-grained measure is to observe that \hog\ effectively divides the probabilities into two bins---those that are larger than the median and those that are smaller. 
The \hog\ benchmark is then obtained from testing whether the empirically obtained samples satisfy certain properties expected from ideally distributed samples on the respective bins. 
The sample efficiency of computing this benchmark can be retained even when generalizing it to polynomially many bins and comparing the number of observed outcomes per bin with the number of expected outcomes. 

\begin{figure}
\includegraphics{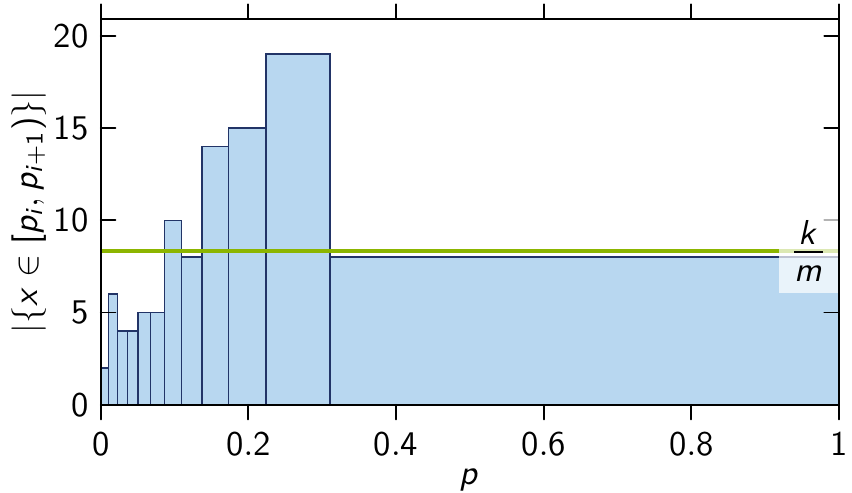}
  \caption{\label{fig:bog}
  The fine-grained generalization of heavy outcome generation is to bin the samples $x_1, \ldots, x_k$ from the noisy distribution $Q$ according to the probabilities $P_C(x_i)$. 
  This constitutes a coarse-grained estimator of the total-variation distance between $Q$ and $P_C$. 
  Since $P_C$ is (nearly) exponentially distributed for random circuits, a suitable choice of $m$ bins $[p_i, p_{i+1})$ is such that they are equifilled with a $1/m$ fraction of the ideal samples.
  This is shown in the figure for a noisy exponential distribution on a $n = 3$-qubit sample space, with $m = 12 $ bins and $k = 100 $ samples. 
  }
\end{figure}

Given that the distribution of outcome probabilities is expected to be an exponential distribution, the natural way to bin is to choose a larger number of bins. 
Concretely, we can choose $m$ equifilled bins $[p_i, p_i+1)$  satisfying 
\begin{equation}
    \int_{p_i}^{p_{i+1}}2^n \ee^{- 2^n p} \  \mathrm{d} p  = \frac{1}{m} \  ,
\end{equation}
for $i = 1 ,\ldots, m $ and $ p_0 = 0, \ p_m = 1$. 
Define $\Omega = \{ [p_i,p_i+1) \}_{i \in [m]} $. 
The task of \emph{binned outcome generation} \bog\ \cite{bouland_quantum_2018} is to obtain a good, i.e., low value of the binned distance 
\begin{align}
  d&_{\text{\bog}}(Q,P_C) \nonumber \\
  & = \sum_{X \in \Omega} \left | \frac 1 { 2^n} \sum_{ x\in \{0,1\}^n } (Q(x) - P_C(x)) \delta(P_C(x) \in X) \right| \\
  & =  \sum_{X \in \Omega} \left | \frac 1 { 2^n} \sum_{ x\in \{0,1\}^n } Q(x)  \delta(P_C(x) \in X) - \frac 1 m \right|,
\end{align}
where the last equality is true if $P_C$ is Porter-Thomas distributed. 
This is a discretized estimator of the total-variation distance of the outcome distribution and can be estimated from polynomially many samples; see Fig.~\ref{fig:bog}. 
Indeed, for $Q = P_C$ this measure is $0 $, while for any $ Q \neq P_C$ it converges to  $\tvd{Q - P_C}$ as $m \rightarrow \infty$. 
\textcite{canonne_testing_2020} prove that such \emph{binned identity testing} with $k$ bins up to error $\epsilon$ is possible using $O(k/\epsilon^2)$ many samples, and moreover, that this is asymptotically optimal.

\subsubsection{Cross-entropy difference}
\label{ssec:cross-entropy difference}
While \hog\ and its variants are conceptually intuitive, in practice, we want to capture as much about the distribution as possible, given the available samples. 
To capture correlations between the distribution $Q $ and $P_C$ as well as possible an appealing measure is the \emph{cross-entropy}~\cite{boixo_characterizing_2016}
\begin{align}
\label{eq:cross entropy}
\ce(Q,P_C)  = - \sum_x Q(x) \log P_C(x).
\end{align}
The cross-entropy is a well-known statistical measure of similarity between two distributions and measures correlations between the two distributions \cite{Murphy}.
It also gives rise to a distance measure between $Q$ and $P_C$, known as the 
\emph{cross-entropy difference}\footnote{Note that \textcite{boixo_characterizing_2016} define the cross-entropy difference in terms of \cref{eq:cross entropy measures} as the deviation of cross-entropy between $Q$ and $P_C$ from the cross-entropy between the uniform distribution and $P_C$
\begin{align}
  F_{\text{XE}}(Q,P_C) = \ce(1/2^n,P_C) - \ce(Q,P_C).
\end{align}
}
\begin{align}\label{eq:cross entropy difference}
	d_{\ce}(Q,P_C) & = \ce(Q,P_C) - H(P_C)  \\
	& =  \sum_{x \in \{0,1\}^n} (Q(x)  - P_C(x))  \log \frac 1 {P_C(x)}
\end{align}
where $H$ denotes the Shannon entropy.

But how does the cross-entropy difference fare when applied to the task of verifying quantum supremacy distributions? 
Again, using the assumption that the ideal probabilities are exponentially distributed, we observe that it constitutes a good measure for distributions of the form $Q_\lambda$ in \cref{eq:uniformly noisy distribution} \cite[Section~2 of the SM]{boixo_characterizing_2016}:
\begin{align}
\label{eq:ce linear interpolation}
	d&_\ce(Q_\lambda ,P_C)\nonumber \\
  &  = (1- \lambda) d_\ce(P_C,P_C)  + \lambda d_\ce(1/2^n,P_C) \\
	& \approx (1- \lambda ) \cdot 0 + \lambda \cdot 1  = \lambda. 
\end{align}
To see why this is the case, we can compute the expectation value of $H(P_C)$ over the random choice of $C$ as~\cite{boixo_characterizing_2016}
\begin{align}
	\Eb_C\left[H(P_C)\right] & = - \sum_{x } \Eb_C\left[P_C(x) \log P_C(x) \right]\\
	& = - 2^n \int_{0}^\infty 2^n \ee^{-2^n p }p \log p  \ \mathrm d p \\
	& = n - 1 + \gamma,  
\end{align}
where $\gamma \approx 0.5774 $ is the Euler constant. 
Likewise, the cross-entropy between $P_C$ and the uniform distribution is in expectation given by 
\begin{align}
	\Eb_C\left[ \ce(1/2^n, P_C)\right]  & = - \frac 1 { 2^n} \sum_{x \in \{0,1\}^n} \Eb_C \left[\log P_C(x) \right]\\
	& =  - \int_{0}^\infty 2^n \ee^{-2^n p } \log p  \ \mathrm d p\\
	& = n + \gamma. 
\end{align}
From this we obtain $\Eb_C[d_\ce(1/2^n,P_C)] = 1 $. 

By the above argument that the  probabilities $P_C(x)$ for a given (Haar) random and large enough unitary $C$ are (pairwise) independently identically distributed according to the Porter-Thomas distribution, with high probability over the choice of $C$, $d_\ce(1/2^n,P_C) = 1$ for a fixed circuit. 
Conversely, as the cross-entropy reduces to the Shannon entropy for $Q = P_C$ we trivially have $d_\ce(P_C,P_C) = 0 $. 
To summarize, the cross-entropy difference attains the value $1$ for the uniform distribution and vanishes for the ideal distribution, giving rise to linear interpolation \eqref{eq:ce linear interpolation} for states of the form $Q_\lambda$. 
Notice that this is equally true for any noisy distribution 
\begin{align}
\label{eq:unco{}rrelated Q}
 	Q'_\lambda = (1- \lambda)P_C + \lambda Q',
 \end{align} 
in which the uniform distribution is replaced by a distribution $Q'$ that is uncorrelated with $P_C$, i.e., $\Eb_C[\ce(Q',P_C)] = - \sum_x Q'(x) \Eb_C[ \log P_C(x) ] $.  

Under certain conditions the cross-entropy difference in fact bounds the total variation distance~\cite{bouland_quantum_2018}. 
To see this, notice that the definition of the cross-entropy difference is similar to that of the \emph{Kullback-Leibler divergence}
\begin{align}
	\kl(Q \Vert P_C ) = \ce (Q,P_C) - H(Q) , 
\end{align}
which bounds the total-variation distance by Pinsker's inequality as 
\begin{align}
	\tvd{Q- P_C } \leq \sqrt {\kl(Q \Vert P_C )/2 }. 
\end{align}
Hence, if the cross-entropy difference satisfies $d_\ce(Q,P_C) \leq \varepsilon$ and the noise is entropy-increasing such that $H(Q) \geq H(P_C)$, we have 
\begin{align}
\label{eq:tvd ced}
	\tvd{Q- P_C } & \leq \sqrt {\kl(Q \Vert P_C )/2 } \\
  & \leq \sqrt{ d_\ce(Q,P_C)/2} \leq \sqrt {\epsilon/2}.
\end{align}

The condition $H(Q) \geq H(P_C)$ is a fairly general condition on the type of noise under which the total-variation distance bound \eqref{eq:tvd ced} holds. 
But it is also a condition that cannot be checked from fewer than exponentially many samples from $Q$. 
Moreover, one can easily construct examples of distributions that violate the inequality~\eqref{eq:tvd ced}~\cite{bouland_quantum_2018}: 
those examples fare well on the cross-entropy difference, but are far from the ideal target distribution. 

The cross-entropy difference can be efficiently estimated up to accuracy $\epsilon$ with failure probability $\alpha$ from 
\begin{align}
\label{eq:sample complexity cross entropy}
  m \geq \frac{(n + O(\log n))^2 }{2\epsilon^2} \log(2/\alpha)  
\end{align}
many independently identically distributed (iid.)~samples from $Q$. 
To derive Eq.~\eqref{eq:sample complexity cross entropy}, we apply Hoeffding's inequality and assume that the probabilities $P_C(x)$ are Porter-Thomas distributed. 
We obtain that with probability at least $ 1- 1/f(n)$ over the choice of $U$, the probabilities $P_C(x)$ satisfy 
\begin{align}
\label{eq:bounds haar random probabilities}
	 2^{-2n}/f(n)\leq P_C(x )\leq  (n + \log f(n))2^{-n}, 
\end{align}
so that their logarithms $\log P_C(x) $ differ only by a constant factor of $\sim ( 2 +  O(\log(f(n))$ from $-n$. 

%

\subsubsection{Linear cross-entropy benchmarking (XEB) fidelity}
\label{ssec:xeb}

The most widely used cross-entropy benchmark is the \emph{linear cross-entropy benchmarking (XEB) fidelity} introduced by \textcite{arute_quantum_2019}. 
This measure simply chooses $f$ to be the identity function, up to rescaling and shifting, $f_\xeb(x) = 2^n x -1$, so that 
\begin{align}
\label{eq:xeb fidelity}
  F_\xeb(Q,P_C)  = \sum_{x \in \{ 0,1\}^n} Q(x) ( 2^n P_C(x) -1 ) . 
\end{align}
The XEB fidelity has the virtue that it can meaningfully be applied in two variants: 
in the first variant, it is a variant of a randomized benchmarking protocol with the goal of obtaining a fidelity measure averaged over random sequences of quantum gates. 
This variant is a special instance of randomized benchmarking 
\cite{helsen_new_2019,helsen_general_2020,liu_benchmarking_2021} and can be applied to gates acting on few qubits \cite[SM]{arute_quantum_2019}.
In its second reading, it can be used as a verification protocol for single instances of quantum random sampling.
By making use of a typicality argument based on Levy's lemma, guarantees for the average randomized-benchmarking behaviour can be transferred to the single-instance application.  
Therefore, the XEB fidelity unifies the idea of benchmarking a quantum processor by running random computations on it, and the idea of demonstrating a quantum advantage via sampling from the output distribution of such circuits. 

Even though it may serve as a measure of the fidelity of a single circuit instance for a large number of qubits, the XEB fidelity is intrinsically an average-case measure, and its ability to verify single instances is only derived from the fact that these instances are typical.  
Given the choice of rescaling and shifting, the \emph{average XEB fidelity} for a family of quantum circuits $\mc C$ that gives rise to a spherical $2$-design (recall \cref{sec:anticoncentration}) indeed gives rise to a meaningful measure of quantum advantage in the sense that 
\begin{multline}
\label{eq:xeb expectation}
  \Eb_{C \sim \mc C} [F_\xeb(Q_C, P_C)] \\ 
  =  \begin{cases}
    \sum_x 2^n \Eb_C[P_C(x)^2]  - 1 \approx 1  & Q_C = P_C, \\
    \sum_x P_C(x)  - 1 =  0 & Q_C = \frac1 {2^n}, 
  \end{cases}
\end{multline}
in the extreme cases in which for every $C$, $Q_C $ is the ideal target distribution and the uniform distribution, respectively. 
In the following, we discuss in more detail these interpretations of the XEB fidelity, 
and the extent to which the XEB provides a meaningful measure of quantum advantage.

\paragraph{Sample complexity of estimating the XEB fidelity.}
For Haar-random unitaries, $F_\xeb$ can be estimated up to error $\epsilon$ with probability at least $1-\delta$ from 
\begin{align}
  \ell
  \geq \frac {e^2}{2\epsilon^2} \ln^2\left (\frac{2} {2d}\right) \ln\left(\frac 2 \delta \right) 
\end{align}
many samples \cite{hangleiter_sampling_2021,kliesch_theory_2021}. 
Moreover, using the bounds~\eqref{eq:bounds haar random probabilities} on the size of the probabilities $P_C(x)$ we can estimate the average XEB fidelity
$\Eb_C[
F_\xeb(Q, P_C)] $ up to error $2 \epsilon $ with failure probability $\delta $ from 
\begin{align}
  \ell_C
   \geq \frac 1 {2\epsilon^2} \log \frac 2 \delta 
\end{align}
many distinct random circuits and 
\begin{align}
\label{eq:sample complexity xeb}
 \ell
 \geq \frac{(n + O(\log n))^2 }{2\epsilon^2/\ell_C^2} \log(2/\delta) 
\end{align}
many samples per circuit \cite{hangleiter_sampling_2021}. 
In fact, an $O(1)$ bound on the variance of $\mb E[F_{\xeb}]$ is true even if only the third moments of the circuit are close to the Haar-random value and the noise is gate-independent \cite{helsen_general_2020}. 

\paragraph{Benchmarking via XEB fidelity.}
\label{par:benchmarking xeb}
Let us now briefly sketch how XEB can be used to benchmark a quantum device. 
For instance \textcite[Sec. IV A, SM]{arute_quantum_2019} analyse how to estimate the depolarization error $p_c$ per cycle of the computation using the XEB fidelity.
Let us follow their argument. 
Consider the noisy quantum state 
\begin{align}\label{eq:noisy state average xeb}
  \rho_C = \epsilon_d C \proj 0   C^\dagger + ( 1- \epsilon_d) \chi_C, 
\end{align}
after applying a random circuit $C$ with $d$ gate layers (cf.\ \cref{eq:stockmeyer error mixture}). 
Here, $\epsilon_d$ describes the effect of errors on the state, and in the case of $\chi_C = \id/2^n$ is interpreted as the depolarization fidelity.  
We assume now that the erroneous state $\chi_C$ is uncorrelated with $C$ in the sense that the probabilities of a computational basis measurement are uncorrelated as $\Eb_C[\bra x \chi_C \proj x C \proj 0 C^\dagger \ket x] = \Eb_C[ \bra x \chi_C \ket x ]\Eb_C[\bra x  C \proj 0 C^\dagger \ket x] $. 

When averaging or ``twirling'' over random unitaries that form a unitary design we would then expect to obtain a fully mixed state
\begin{align}
  \Eb_C\left [  C^\dagger \chi_C C \right] = \frac \id {2^n} , 
\end{align}
so that one might expect 
\begin{align}
\label{eq:depolarizing error xeb}
  \Eb_C\left[C^\dagger \rho_C C \right] = \overline {\epsilon_d} \proj 0 + (1 - \overline {\epsilon_d} ) \frac \id {2^n} ,  
\end{align}
where $\overline{\epsilon_d}$ denotes the average of the individual values of $\epsilon_d$ over the random choice of unitaries. 
\cref{eq:depolarizing error xeb} precisely describes the effect of a depolarizing channel acting in each cycle of the computation with depolarization fidelity $p_c$ such that $p_c^d = \overline{\epsilon_d}$. 

We obtain an expression of the circuit-averaged XEB fidelity in terms of the depolarization fidelity
\begin{align}
  \label{eq:xeb fidelity estimation}
  \Eb_C\left[F_\xeb(Q,P_C) \right] = p_c^d \left ( 2^n\sum_{x} \Eb_C [ P_C(x)^2] - 1 \right), 
\end{align}
where $Q$ is the output distribution of the noisty state $\rho_C$ and $P_C$ is as usual the output distribution of $C \ket 0 $. 
We can now use this expression in order to estimate $p_c$ from $F_\xeb(Q,P_C)$. 
To do this, we classically estimate the quantity in brackets in Eq.~\eqref{eq:xeb fidelity estimation} and obtain 
\begin{align}
  p_c^d \eqsim \frac{\overline{\widehat{F_\xeb}(Q,P_C)} }{2^n \sum_{x} \Eb_U [ P_C(x)^2] - 1 }, 
\end{align}
where $\widehat{F_\xeb}(Q,P_C)$ denotes the empirical estimate of $F_\xeb(Q,P_C)$ for a fixed circuit and $\overline {F_\xeb(Q,P_C)} $ denotes the empirical average over random circuits. 
From an exponential fit of $p_c^d$ for various values of $d$ one can now estimate $p_c$. 

Notice that in writing \cref{eq:depolarizing error xeb}, we have used the average XEB fidelity $\overline{F_\xeb}$ as a proxy for the average fidelity $\overline{F}$ of the quantum state. 
Arguments for why the assumption that the noise is uncorrelated from the circuit should be true are essential to substantiate that connection. 

\textcite{liu_benchmarking_2021} provide further credence to the connection between average fidelity and average XEB fidelity by performing numerical simulations. 
They also further substantiate the claim that the model of \textcite{arute_quantum_2019} is valid---even in certain cases in which their uncorrelated noise assumption does not hold. 
To this end, they consider ``RCS benchmarking'' in the spirit of randomized benchmarking. 
Specifically, they formalize the protocol of \textcite{arute_quantum_2019} as estimating the average \emph{quantum fidelity} $\overline{F_d}$ of quantum circuits of increasing depth $d = 1, \ldots, D$, and finally performing an exponential fit $F = Ae^{-\lambda d}$. 
If (a) the average fidelity is in fact well fitted by a single exponential decay, and (b) the average XEB fidelity is a good proxy of the average quantum fidelity, then this model matches XEB benchmarking as performed by \textcite{arute_quantum_2019}. 

\textcite{liu_benchmarking_2021} make the connection (a) by proving the following: 
Consider random circuits that are comprised of layers of arbitrary non-Clifford gates (say, the two-qubit iSWAP$^*$ gates) and single-qubit Haar-random gates.\footnote{
This is the setup of a cycle benchmarking protocol \cite{erhard_characterizing_2019}.}
Now suppose that every layer of non-Clifford (two-qubit) gates comes with a Pauli noise channel $\mc N (\rho) = \sum_{\alpha \in \{0,1,2,3\}^n} p_\alpha \sigma_\alpha \rho \sigma_\alpha $, where $\sigma_\alpha$ denote the $n$-qubit Pauli matrices and $p_\alpha$ are their coefficients. 
Then the average fidelity $\mb E F_d$ of depth-$d$ circuits does in fact decay exponentially in the total error $\lambda = \sum_{\alpha \neq 0} p_\alpha$ in the sense that $e^{-\lambda d} \leq \mb E F_d  \leq e^{- \lambda d} (1 + K\lambda)$ for $d \ll 2^n$ up to a first order approximation in $\lambda$; see also the related discussion of \textcite[Section~IX]{helsen_general_2020}

For the second connection (b), they perform numerical simulations for various noise models. 
To this end, they make use of a somewhat more versatile fidelity estimator that is closely related to the XEB fidelity, which has been introduced by \textcite{rinott_statistical_2022}.\footnote{Unbiased estimators for other scenarios are discussed by \textcite{choi_emergent_2021,liu_benchmarking_2021}.  } 
Intuitively speaking, in this ``unbiased XEB'' estimator, instead of multiplying the ideal probability by $1/2^n$, it is multiplied by the inverse second moments of the ideal output distributions
\begin{align}
\label{eq:unbiased xeb}
  F_{\xeb,u}(Q,P_C) = \frac {F_{\xeb}(Q,P_C)} { \mb E_C [F_{\xeb}(Q,P_C)]}. 
\end{align}
This means that it is normalized on average to unity not only for deep quantum circuits which have design-like moments (recall \cref{eq:xeb expectation}), but also for more shallow circuits with differing second moments. 
\textcite{liu_benchmarking_2021} find good agreement between the fidelity and their unbiased XEB fidelity for various correlated noise models and more over show that the variance of the XEB fidelity scales as $O(1/\ell + \lambda^2 (\mb E F)^2)$ in the number of samples $\ell$ collected per circuit. 
The unbiased estimator \eqref{eq:unbiased xeb} has recently been tested as a measure of fidelity in small instances of measurement-based quantum random sampling \cite{IonSampling}.   

Let us also note that the \emph{maximum-likelihood estimator} (MLE) for the fidelity has been analyzed by \textcite{rinott_statistical_2022}. 
They find that the MLE has smaller bias and variance than the linear XEB estimator and---like the unbiased XEB estimator---is therefore a better fidelity estimator. 
They also find---as was already noted by \textcite[SM]{arute_quantum_2019}---that in the regime of small depolarization fidelity $\epsilon_d \ll 1$, the XEB fidelity estimator converges to the MLE of the fidelity.

\paragraph{Single-instance verification.}
When the number of qubits is large, and the unitary $C$ is drawn Haar-randomly, Levy's lemma implies that the fluctuations around the expectation value over $C$ \eqref{eq:xeb expectation} are expected to be on the order of $O(1/\sqrt{2^n})$. 
Consequently, for a large number of qubits, the fidelity concentrates around its expected value over the choice of random circuits \cite{arute_quantum_2019}.

For a large number of qubits, following \textcite[SM~IV.B]{arute_quantum_2019} we write again the noisy implementation of the quantum state  $C \ket 0 $ as 
\begin{align}
  \rho_C = F \, C \proj 0 C^\dagger + (1-F) \chi_C, 
\end{align}
where the mixed state $\chi_C$ describes the effect of noise and $F = \bra 0 C^\dagger \rho_C C \ket 0 $ is the fidelity of $\rho_C$ and the target state  $C \ket 0 $. We can now make the assumption that $\chi_C$ is uncorrelated from $C \ket 0 $ in the sense that \cite[Eq.~(25) in the SM]{arute_quantum_2019}
\begin{align}
\label{eq:uncorrelated noise definition}
  \sum_x \bra x \chi_C \ket x f(p_C(x)) = \frac 1 {2^n} \sum_x f(p_C(x)) + \epsilon ,
\end{align}
for $\epsilon \ll F$. 
By the Levy's Lemma argument, \textcite{arute_quantum_2019} expect a typical fluctuation $\epsilon \in O(1/\sqrt{2^n})$.

Large parts of the analysis of the theoretical proposal of random circuit sampling~\cite{boixo_characterizing_2016} and the experimental realization thereof~\cite[Section~IV B of the SM]{arute_quantum_2019} are indeed dedicated to validating the assumption of uncorrelated noise. 
This can be done, for example, by numerically studying realistic error models such as random Pauli errors. 
To summarize, given the arguments sketched above hold, the XEB fidelity quantifies the fidelity $F_\xeb(Q,P_C) = F$ up to a deviation of order $1/\sqrt{2^n}$. 

\paragraph{Hardness of achieving a nontrivial XEB fidelity.}
Similarly to \hog, we expect that achieving an exponentially small score in the XEB fidelity $b/2^n$ for constant $b>1$, formalized as the task \xhog, is computationally hard. 
This is because, intuitively, \xhog\ is a refined version of \hog\ in which the outcomes have to be produced according to their actual weight.  
Analogously to the argument reducing \hog\ to \quath\ \cite{aaronson_complexity-theoretic_2017}, 
\xhog\ can be reduced to an analogous conjecture \xquath\ \cite{aaronson_classical_2019}. 
\xquath\ states that given a circuit $C \sim \mc C $, there is no efficient classical algorithm that produces an estimate $p$ of $p_C(0)$ such that 
\begin{align}
  \Eb[(p_C(0) -  p)^2] = \Eb[(p_C(0) - 2^{-n})^2] - \Omega(2^{-3n})
\end{align}
where the expectation is taken over the choice of random circuit and the algorithm's internal randomness.

\paragraph{Spoofing the linear XEB fidelity.}
Summarizing the discussion above, the XEB fidelity serves two distinct functions \cite{gao_limitations_2021}. 
First, the argument of \textcite{aaronson_classical_2019} suggests that achieving a nontrivial XEB value is a computationally intractable task for random quantum circuits. 
Second, the XEB fidelity serves as a proxy for the quantum fidelity \cite{arute_quantum_2019,choi_emergent_2021,liu_benchmarking_2021}.

\textcite{zhou_what_2020-1,gao_limitations_2021} observe, however, that the XEB fidelity in fact overestimates the quantum fidelity in certain settings, leading to weaknesses that can be exploited by an adversarial classical simulator. 
More concretely, \textcite{gao_limitations_2021} characterize the conditions under which the XEB fidelity serves as a good proxy of the quantum fidelity when comparing a noisy quantum device to an ideal circuit. 
Based on these conditions they then demonstrate that the XEB fidelity is not a reliable measure of quantum advantage in an ``adversarial setting'' in which these conditions can be violated. 
\footnote{A similar overestimation of the fidelity has also been observed in the literature on randomized benchmarking \cite{boone_randomized_2019}. }
The explicit argument of \textcite{gao_limitations_2021,zhou_what_2020-1} is based on three properties of the XEB fidelity that make it distinct from the fidelity. 

First, the fidelity and the XEB fidelity exhibit different scaling behavior as multiple quantum systems are combined into a larger one: 
whereas the quantum fidelity generally decreases exponentially in the number of combined systems, the XEB fidelity generally increases. 
To see this, consider $k$ disjoint $n$-qubit quantum systems with XEB fidelity values $\chi_i = 2^n \sum_x q_i(x)p_i(x) -1 $ and fidelities $F_i$ for $i=1, \ldots, k$, where $q_i$ and $p_i$ are the output probabilities corresponding to the respective noisy and ideal circuits.
The fidelity then scales multiplicatively as $F = \prod_i F_i$, whereas the total XEB fidelity scales as 
\begin{align}
\label{eq:scaling xeb}
    \chi & = 2^{kn} \sum_{x_i} \prod_i p_i(x)q_i(x) - 1\\
    & = \prod_i (\chi_i + 1) -1 \approx \sum_i \chi_i, 
\end{align}  
assuming $\chi_i \ll 1$.
This difference in scaling behavior is fundamental to the fact that the first term of the XEB fidelity tends toward a nonzero value (namely unity) as $p$ and $q$ become uncorrelated from one another, which is explicitly subtracted. 

Second, their values may be distinct for highly correlated errors. 
To see this intuitively, consider a noisy quantum circuit with $m$ gates and independently and homogeneously distributed random errors across the circuit at rate $\epsilon$. 
The probability that no error occurs is then given by $(1-\epsilon)^m$.
If the presence of a single or more errors leads to vanishing contributions to the XEB or the fidelity, then both will be equal to $(1-\epsilon)^m$. 
However, outside of some limiting cases, there are non-zero correction terms for finite-size systems. 
Consider a single bit-flip error at depth $t$ in a 1D random circuit. 
In the Heisenberg picture, we can propagate $X(t)$ backward in time and consider its effect on the initial state  $\ket {0^n}$. 
If the dynamics are chaotic, then $X(t)$ becomes a linear combination of $4^{|s|}$ Pauli strings the support of which grows linearly as $|s| \approx 2 c t$ with an effective ``scrambling velocity'' $c$. 
But out of those operators $\sim 2^{|s|} $ are products of $\id$ and $Z$ and hence they do not cause an error on the input state $\ket {0^n}$. 
Consequently, a single error contributes $O(2^{-2ct})$ to the XEB fidelity and quantum fidelity alike.
Conversely, we can forward-propagate the error, but now the argument only holds for the XEB fidelity because measurements are performed in the $Z$ basis, while all terms contribute to the quantum fidelity, leading to a distinct behavior. 
\textcite{gao_limitations_2021} further argue that this difference can be amplified when considering specific spatial error patterns, and provide a lower bound on the total correction. 

In the complementary ``benign setting'' of error distributed independently and homogeneously across the system, they find necessary and sufficient conditions for the XEB fidelity and the quantum fidelity to agree, namely, that $n \epsilon f(c) \ll 1$, where $f(c) \in O(1)$ is a decreasing function depending on the architecture details. 
Via a mapping to a statistical mechanics model analogous to the one we have introduced in \cref{sec:anticoncentration} they derive a diffusion-reaction for how errors evolve in the circuit and analyse it for different ensembles of random gates. 
Using this model, they 
explore the intuition just described quantitatively, finding that the XEB fidelity starts to deviate from the fidelity for strong noise. 

Thirdly, because the XEB fidelity quantifies the correlations between the distribution $q$ and $p$, complete knowledge of $p$ allows to amplify those correlations by choosing $q$ adversarially. 

Building on those insights as well as a spoofing algorithm of the XEB fidelity for low-depth quantum circuits \cite{barak_spoofing_2021}, \textcite{gao_limitations_2021} construct an algorithm which achieves high scores for large quantum circuits. 
The key idea of this algorithm is to approximate the ideal circuit with a circuit that is given by a product over smaller subsystems each of which can be simulated on a classical computer. 
To achieve this, given a number of subsystems to divide the circuit in, they remove entangling gates between those subsystems. 
Using the algorithm they achieve a score of $1.85 \cdot 10^{-4}$ in $0.6$s on a single GPU, while the experiment at Google by \textcite{arute_quantum_2019} achieved $2.24 \cdot 10^{-3}$, and achieve a similar ratio for the larger follow-up experiments at USTC \cite{wu_strong_2021,zhu_quantum_2022}. 
They find, however, that for small system sizes, the ratio between the performance of their algorithm and the experimental score increases and conjecture that their algorithm will achieve an advantage over the quantum value of the XEB fidelity. 
Relating to the hardness argument of \textcite{aaronson_classical_2019}, their algorithm seems to refute the \xquath\ conjecture. 
More concretely, \textcite{gao_limitations_2021} show that for 1D circuits, their algorithm achieves an XEB fidelity that scales inverse exponentially $e^{-O(d)}$ in the circuit depth\footnote{A slightly weaker statement holds also for 2D circuits.}.
On the other hand, they show that the \xeb\ score of a variant of their algorithm precisely reflects the statement of the \xquath\ conjecture in terms of probability estimation on average.
Consequently, their results refute the \xquath\ conjecture for circuits of sublinear depth $d \in o(n)$.

Given this discussion, achieving a quantum advantage in terms of cross-entropy benchmarking via quantum random sampling boils down to the question whether the inverse exponential scaling of the quantum score of the linear XEB fidelity can be beaten by another inverse exponential scaling of a classical algorithm.   
And it seems that it can. 
A different way of benchmarking quantum advantage experiments from the linear XEB fidelity thus seems to be necessary to demonstrate that quantum devices are in fact able to scalably outperform classical algorithms in an adversarial setting. 
To this end, note that the spoofing algorithm of \textcite{gao_limitations_2021} and \textcite{zhou_what_2020-1} presumably do not work for the cross-entropy difference as it intrinsically builds on the linearity of the linear XEB fidelity. 
In spite of the results for the linear XEB, the cross-entropy difference remains to be a potential valid means of benchmarking quantum advantage.

Let us stress again, however, that while XEB measures may be estimated from few samples, 
all variants of XEB suffer from the problem that their evaluation is computationally inefficient. 
This limits their practical usage to a regime just below the quantum advantage threshold in which classically computing the output probabilities is still possible, but at a high cost. 
Alternatively, as we will see in \cref{sec:experimental implementations}, other quantities might be used in order to feasibly obtain an estimate of XEB measures. 
In the following, we discuss an alternative approach that does not suffer from the conceptual---in terms of quantifying quantum advantage---and computational---in terms of its efficient evaluation---disadvantages of XEB.  

%

\subsection{Efficient quantum verification}
\label{sec:efficient quantum verification}

An approach that is both natural in an experimental setting and a direct follow-up of the previous discussion regarding the relation between XEB fidelity and quantum fidelity, is to verify the sampling task directly on the level of the quantum state. 
This is reasonable: In an experimental setting, we know that there is a quantum state on which measurements are performed. 
Therefore we can exploit access to that quantum state in order to circumvent the no-go result of \cref{ssec:hardness of verification} and potentially achieve fully efficient verification of the TVD between the experimental and the target distribution, assuming that the measurements are carried out correctly. 

Of course, verification of a quantum state is possible if we have access to an ideal state preparation via a swap test, or by verification protocols that use measurements along the direction of the target state \cite{pallister_optimal_2018}. 
But assuming this capacity would already assume the ability to prepare the ideal target state. 
A reasonable quantum protocol for verifying quantum random sampling schemes should therefore make use of only restricted quantum capacities, such as the ability to implement single-qubit measurements, or to prepare single-qubit states reliably. 
Experimentally, such assumptions are extremely well justified: 
in most platforms single-qubit gate fidelities are orders of magnitude better than entangling-gate fidelities. 
It is also entirely different in kind when compared to assumptions on the \emph{global effect} of the noise on the outcome probability distribution $P_C$ such as the assumption $H(Q) \geq H(P_C)$ that was necessary for a cross-entropy based test to yield bounds on the total-variation distance: 
it is an assumption on single-qubit measurements and therefore \emph{local}. 
This means that it can be verified to the same degree that one can characterize those measurement apparata. 
For single- or two-qubit measurements this is possible using tools such as gate set tomography~\cite{MerGamSmo13,BluKinNie13,Gre15,BluGamNie17,cerfontaine_self-consistent_2019,EstimatingGateSetProperties,brieger_compressive_2022} or the device-independent verification of quantum processes and instruments~\cite{Sekatski:2018eo}. 

In contrast to classical verification from samples where we were given classical samples from an a-priori untrusted device, we can conceptualize quantum verification as the task to verify the preparation of a certain quantum state by a deep circuit using components of the device that are well characterized and known to work correctly. 

In the following we will see protocols that are able to verify or estimate the quantum fidelity between two quantum states $\sigma$ and $\proj \psi $ 
\begin{align}
\label{eq:fidelity def}
  F(\sigma,\proj \psi ) = \bra \psi \sigma \ket \psi . 
\end{align}
Via the Fuchs-van de  Graaf inequality, the fidelity bounds the TVD via the trace distance 
\begin{align}
  \label{eq:bounds trace fidelity}
 \tvd{ p_\sigma - p_\psi} \leq \norm{\sigma - \proj \psi}_{\tr} \leq \sqrt{1 - F(\sigma,\proj \psi)} ,  
\end{align}
where $p_\sigma$ and $p_\psi$ are the output distributions of $\sigma$ and $\proj \psi$ in the standard basis, respectively. 

Generally, we can think about such protocols in terms of their information gain versus their complexity in terms of number of measurements and distinct measurement settings as well as assumptions made in the derivation of the protocol \cite{BenchmarkingReview}. 
While protocols with low complexity tend to yield little information about an underlying quantum state, protocols with higher complexity can reveal more information about that state.  
In the following, we discuss two types of protocols to verify the output states of quantum random sampling via the fidelity---fidelity witnessing and fidelity estimation.

\subsubsection{Fidelity witnessing}
\label{ssec:fidelity witness}

\begin{figure}
\includegraphics{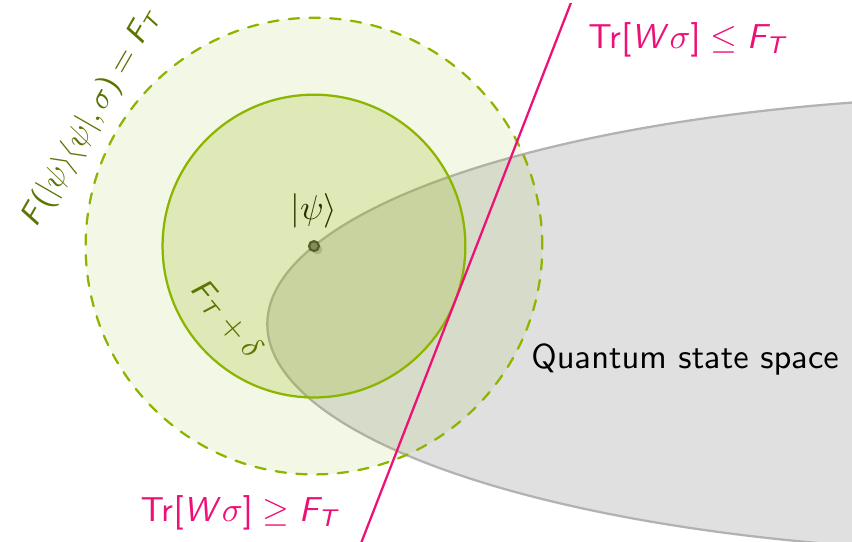}
\caption{\label{fig:fidelity witness}
Given a target state $\rho = \proj{\psi} $, a fidelity witness $W$ for $\rho$ provides a lower bound on the fidelity $F(\rho, \sigma) \geq \tr[W\sigma]$ so that, in particular, all states $\sigma$ such that $F(\rho,\sigma)\leq F_T$ it also holds that $\tr[W\sigma] \leq F_T$. 
Conversely, all states $\sigma$ satisfying $\tr[W \sigma]\geq F_T$ will also satisfy $F(\rho, \sigma)\geq F_T$. 
There is a gap $\delta \geq 1 - F_T$ such that all states $\sigma$ with 
fidelity $F(\rho, \sigma)\geq F_T + \delta$ lie on the left side of the witness. 
}
\end{figure}

We call an observable $W$ a \emph{fidelity witness} for a target state $\rho$, if \cite{gluza_fidelity_2018}
  \begin{enumerate}[label = \roman*.]
    \item $\tr[\sigma W] =1$  \emph{iff} $\rho = \sigma$,
    \item $\tr[\sigma W] \leq F(\rho,\sigma)$. 
  \end{enumerate}
Conceptually speaking, fidelity witnesses are very much like \emph{entanglement witnesses}~\cite{guehne_entanglement_2009} in that they cut a hyperplane through quantum state space, which detects a property of quantum states:  
Those states that lie on the left of the hyperplane defined by $\tr[W \sigma] \geq F_T$ are guaranteed to have a high fidelity of at least $ F_T$ since $\tr[W\sigma]$ lower-bounds $F(\rho,\sigma)$. 
For those states on the right of the hyperplane---satisfying $\tr[W\sigma] < F_T$ we cannot make a statement about their fidelity.  
Conversely though, all states $\sigma$ with low fidelity $F(\rho, \sigma) \leq F_T$ are guaranteed to lie to the right of the hyperplane as $\tr[W\sigma] \le F(\rho, \sigma) \le F_T$. 
We illustrate the idea of a fidelity witness in Fig.~\ref{fig:fidelity witness}. 

\paragraph{Fidelity witnessing via parent Hamiltonians.}
A simple fidelity witness $W_H = \id - H/\Delta$ can be constructed for the ground state of a Hamiltonian $H$ with gap $\Delta$ \cite{cramer_efficient_2010,hangleiter_direct_2017}. 
To see this, one can simply expand the Hamiltonian with ground state energy set to $0$ in its eigenbasis $\ket i$ with eigenvalues $\lambda_i $ in order to bound the fidelity between the ground state $\proj 0 $ and a state preparation $\rho$ using that  
\begin{multline}
 \tr (H \sigma) 
 = \sum_{i=1}^d \lambda_i \tr (\proj i \sigma) 
 \geq \Delta \sum_{i=1}^d  \tr (\proj i \sigma) \\
 = \Delta ( 1- \tr (\proj 0 \sigma))= \Delta ( 1- F(\proj 0, \sigma)). 
\end{multline}
To apply this witness, it is required to have knowledge of both the ground state energy and the gap of the Hamiltonian in question.
Applying this fidelity witness to quantum random sampling, it was observed by \textcite{hangleiter_direct_2017} that arbitrary quantum computations and in particular those required for quantum random sampling can be embedded in the ground state of a frustration-free, local Hamiltonian via the Feynman-Kitaev history state construction. 
This protocol finds a particularly natural application in the measurement-based model of quantum computation, which is universal for quantum computation~\cite{raussendorf_one-way_2001,raussendorf_measurement-based_2003}.
Since the prepared quantum state in measurement-based quantum computing is a stabilizer state, it is the ground state of a local, commuting Hamiltonian with gap $2$ comprised of the stabilizers, which are product operators. 
The state preparations of quantum random sampling schemes in the measurement-based model can therefore be verified via fidelity witnessing using only trusted single-qubit measurements \cite{gao_quantum_2017,bermejo-vega_architectures_2018}.

Let us briefly illustrate this point and define the cluster state  on $N$ qubits on a lattice as 
\begin{align}
  \label{eq:cluster state}
  \ket{\text{CS}} = \left( \prod_{\langle i,j\rangle} CZ_{i,j} \right) H^{\otimes N} \ket {0^N} ,  
\end{align}
where the symbol $\langle i,j\rangle$ denotes nearest neighbors on a lattice. 
Arbitrary quantum computations can be driven by single-qubit operations on that state---adaptive measurements at the correct angles in the $X$--$Y$ plane (multiples of $\pi/8$ suffice) \cite{mantri_universality_2017}. 
Assuming highly accurate single-qubit operations and measurements, we can now use the fidelity witness in order to verify the pre-measurement quantum state  $\ket{\text{CS}}$. 

To do so, we need to derive a ``parent Hamiltonian'' which has $\ket{\text{CS}}$ as its ground state. 
This can be done easily by observing that the diagonal Hamiltonian 
\begin{align}
  H_0 = -  \sum_{i = 1}^{N } Z_i \ , 
\end{align}
has the all-zero state  $\ket{0^N}$ as its ground state  with ground state energy $ E_0  = - N$ and gap $\Delta = 2$. 
Our strategy to derive a parent Hamiltonian $H$ of $\ket{\text{CS}}$ is based on the observation that conjugation by unitary transformations $U$ preserves the eigenvalues so that $ U \ket{0^N}$ is a ground state of $U H_0 U^\dagger$ with ground state energy $E_0$ and gap $\Delta$. 
Inserting $U = ( \prod_{\langle i , j \rangle}CZ_{i,j})  H^{\otimes N} $ and using the relation $CZ (X \otimes \id ) CZ = X \otimes Z $ we obtain that the Hamiltonian 
\begin{align}
  H = - \sum_{i = 1}^N \left( X_i \cdot \prod_{j \in \partial i } Z_j \right) 
  =
  - \sum_{i=1}^N S_i, 
\end{align}
is a parent of $\ket{ \text{CS}}$ with ground state energy $E_0 =  - N $ and gap $\Delta = 2$. 
Here, $\partial i = \{ j \in V: (i,j) \in E\} $ denotes the neighborhood of site $i$ on a graph $G = (V,E)$.
The operators $S_i = X_i \cdot \sum_{j \in \partial i} Z_j $ are often called \emph{stabilizers} of 
$\ket{\text{CS}}$.  
Of course, the same applies if we rotate the cluster state locally prior to a computational-basis measurement.

The fidelity witness has also been applied to the verification of IQP circuits the diagonal part of which comprises $Z$, $CZ$ and the non-Clifford $CCZ$ gate defined in \cref{eq:iqp boolean polynomial} \cite{miller_quantum_2017}. 
While the resulting non-local stabilizers $h_i$ are not directly products of Pauli operators in the same way as we obtained $CZ ( X \otimes \id) CZ = X \otimes Z $, \textcite{miller_quantum_2017} show
that single-qubit Pauli-$X$ and $Z$ measurements suffice to measure those stabilizers. 
More precisely, a measurement of the stabilizer $h_i$ can be achieved by measuring $X_i \cdot \prod_{j \neq i} Z_j$ with outcome $v = (v_1, \ldots, v_n)$ and returning $(-1)^{\partial_i f(v) + v_i}$, where $\partial_i f(x) = f(x_1, \ldots, x_i + 1, \ldots, x_n ) - f(x_1, \ldots, x_i, \ldots, x_n)$.

\paragraph{Fidelity witnesses for weighted graph states.}

Efficient fidelity estimation protocol for arbitrary weighted graph states as they are generated by the IQP circuit $C_W$ with arbitrary weights $w_{i,j}$ have been developed by \textcite{morimae_verification_2017,zhu_efficient_2019,hayashi_verifying_2019}. 
Those circuits can be seen to give rise to graph states in which not only vertices (as in the example above) but also edges can have arbitrary weights, so-called \emph{weighted graph states}.

\paragraph{Fidelity witnesses for quantum optical states.}

Another approach to constructing fidelity witnesses has been discovered by \textcite{chabaud_efficient_2021} in the context of linear-optical state preparations as a means to verify the output state of the boson-sampling protocol given by $\varphi(U) \ket {1_n}$ (\cref{eq:bosonsamplingdistribution}), where $U$ is a Haar-random linear-optical unitary.
They observe that if certain Gaussian measurements are performed on the state  $\varphi(U) \ket {1_n}$, then one can efficiently simulate the effect of the linear-optical unitary in the postprocessing. 
Specifically, consider a single-mode heterodyne measurement with POVM elements $\proj \alpha$, where $\ket \alpha = \e^{\alpha a^\dagger - \overline \alpha a} \ket 0 $ is a coherent state.   
Then the effect of a linear optical unitary multi-mode heterodyne POVM element $\pi^{-m} \prod_i \proj{\alpha_i}$ is to transform it into another element $\pi^{-m} \prod_i \proj {\beta_i}$, where $\varphi(U) \prod_i \ket{\alpha_i} = \prod_i \ket{\beta_i}$ and the values of $\beta_i$ are efficiently computable. 
This idea can be used to verify a noisy state preparation $\sigma$ of $\varphi(U) \ket{1_n}$ by performing heterodyne measurements, obtaining outcomes $\alpha_i$ and reinterpreting the outcomes as $\beta_i$. 
Now we observe that the fidelity of the quantum state $\sigma$ with a pure product state  $\psi = \prod_i \proj{\psi_i}$ can be bounded as 
\begin{align}
  \label{eq:fidelity bound ulysse}
  F\left(\psi, \sigma \right) \geq 1 - \sum_{i=1}^m (1 - F(\psi_i,\sigma_i)) \geq 1 - m (1- F(\psi, \sigma)) , 
\end{align} 
where $\rho_i = \tr_{{1,\ldots,m}\setminus \{i\}} \rho$ is the reduced state of $\rho$ on the $i$th mode.  
This reduces the verification problem to estimating the single-mode fidelities $F(\psi_i, \sigma_i)$. 
\textcite{chabaud_efficient_2021} show that this is possible using only heterodyne measurements on the state $\sigma_i$ if $\psi_i$ has bounded support in the Fock basis. 
The second inequality in \cref{eq:fidelity bound ulysse} moreover shows that the witness has a certain robustness to noise.

A similar protocol for Gaussian states and hence Gaussian boson sampling has been developed by \textcite{aolita_reliable_2015}. 
In this protocol, a witness is constructed directly on the level of the $m$-mode quantum state preparation $\sigma_p$, again, observing that the time evolution can be inverted classically for Gaussian measurements. 
More precisely, observe that 
\begin{align}
W  = 1 - \sum_{i=1}^m n_i,
\end{align}
witnesses the vacuum state  $\ket{0^m}$, and hence $\tilde W = 1 - \sum_i \tilde n_i$ with $\tilde n_i = U n_i U^\dagger$ witnesses the state  $U \ket{0^m}$. 
Since the number operator can be measured using homodyne ($x$ and $p$) measurements which can be seen through the equality $n_i = x_i^2 + p_i^2 - 1/2$, and since the action of a Gaussian unitary $U$ on those operators can be computed efficiently: 
Defining $r_{2i-1} = x_i$ and $r_{2i} = p_i$, the vector $r$ is transformed as 
\begin{align}
   U^\dagger r U = S r + d = \tilde r, 
 \end{align} 
where $S$ is a symplectic matrix corresponding to $U$ and $d \in \Rb^{2m}$.
Measuring all elements of $\tilde r ^2$, i.e., certain linear combinations of $x_i p_j, x_i x_j$ and $p_i p_j$ thus allows one to estimate $\sum_i \tilde n_i$ and hence the witness of $U \ket{0^m}$ for any Gaussian state.

All of the fidelity witnesses that we have seen in this section can be written in the form $W= 1 - \sum_{i=1}^k w_i$ with operators $w_i$ that we need to measure in an experiment. 
The sample complexity to achieve an overall estimation error $\epsilon$ thus scales as $O(k (\epsilon/k)^{-2})= O(k^3/\epsilon^2)$ since the error of every individual term needs to scale as $\epsilon/k$. 

A downside of fidelity witnesses is that while they provide a bound on the fidelity and are therefore well suited to verify state preparations that are very close to the ideal target state, the bound provided by the witness typically becomes loose rather quickly and hence the value of the witness becomes trivial even while the fidelity is still reasonably high. 
This motivates to directly estimate the fidelity, which, while potentially more difficult, yields much more detailed information regarding the state preparation.

\subsubsection{Fidelity estimation}
\label{sssec:fidelity estimation}

As it turns out, in certain settings, fidelity estimation is possible with a constant number of samples via the so-called \emph{direct fidelity estimation} protocol due to \textcite{flammia_direct_2011} and similar to the protocols proposed in \cite{bourennane_experimental_2004,kiesel_experimental_2005,toth_entanglement_2005,pallister_optimal_2018}. 
Using direct fidelity estimation we can estimate the fidelity of imperfect state preparations $\sigma$ with pure target states of the form 
\begin{align}
  \label{eq:rho decomposition}
  \rho = \sum_{\lambda \in \Lambda} p_\lambda A_\lambda , 
\end{align}
in terms of normal operators $\{A_\lambda \}_{\lambda \in \Lambda}$ weighted by probabilities $p_\lambda$. 

The idea is the following: Decompose $A_\lambda = \sum_{a \in \mathrm{spec}(A_\lambda)} a \pi^a_{\lambda}$ in terms of its eigenprojectors $\pi_\lambda^a$. 
The fidelity can then be written as 
\begin{align}
  F(\rho, \sigma)  = \sum_\lambda \sum_{a 
\in \mathrm{spec}(A_\lambda)} p_\lambda \tr[\pi^a_\lambda \sigma] \cdot a, 
\end{align}
and hence it can be estimated by sampling $\lambda \leftarrow p_\lambda$ and measuring $A_\lambda$ on the state preparation $\sigma$, obtaining outcome $a$ with probability $\tr[\sigma \pi^a_\lambda]$. 
Given $k$ samples $a_i$ obtained in this way, the fidelity can then be estimated as $\hat F(\rho, \sigma) = \frac 1 m \sum_{i =1}^m a_i$ with error $\epsilon$ using $O(1/\epsilon^2)$ many samples. 

Of course, for the protocol to be efficiently possible in practice, some requirements are necessary:  
\begin{enumerate}[label = \roman*.]
  \item For each $\lambda \in \Lambda$, $A_\lambda$ can be efficiently measured. 
  In particular, this is the case if $A_\lambda = A_{\lambda_1} \otimes \cdots \otimes A_{\lambda_n}$ with $\lambda = (\lambda_1, \ldots, \lambda_n)$ is a product of single-qubit operators~$A_{\lambda_i}$. 

  \item For each $\lambda \in \Lambda$, $\mathrm{spec}( A_\lambda )\subset [a_\lambda,b_\lambda]$ for constants $a_\lambda,b_\lambda \in \mb R$.

  \item The probability distribution $p = (p_\lambda)_{\lambda \in \Lambda}$ can be (classically) sampled efficiently.  
\end{enumerate}

A particularly simple application of the protocol is its application to stabilizer states such as the (locally rotated) cluster state   $\ket{\text{CS}}$, since such a state is in the joint $+1$ eigenspace of the stabilizer operators \cite{flammia_direct_2011}. 
A state $\ket \psi$ stabilized by $n$ operators $S_i$ with eigenvalues $\pm 1$ can therefore be expressed as $\proj \psi = \prod_i (\id - S_i)/2 = 2^{-n} \sum_{\lambda \in \mc S} s_\lambda$, where $\mc S$ denotes the stabilizer group of $\ket \psi$ which is generated by the $n$ operators $S_i$. 
Thus, it can be efficiently applied to quantum random sampling architectures which are based on state preparations that are locally equivalent to stabilizer states, in particular, ones based on measurement-based computation \cite{hangleiter_sampling_2021,IonSampling}.
Notice, though, that universal random circuits are not of this type.

A potential drawback of the direct fidelity estimation protocol as opposed to fidelity witnesses is that it in principle requires a different \emph{measurement setting} in each run of the experiment. 
In contrast, to evaluate the fidelity witness only two distinct measurement settings are repeated many times. 
So while the overall quantum sample complexity is dramatically reduced from $O(n^3)$ to $O(1)$ in the number of qubits, the measurement setting complexity is increased from $O(1)$ to $O(1/\epsilon^2)$ in the estimation error. 
Depending on the experimental setting at hand there may well be a trade-off between the time required to switch between settings and the time required for many repetitions of the same measurement setting, see for example \cite{IonSampling}. 
It has also been noted that, when restricting the operators $A_\lambda$ to Pauli operators, the sample complexity of verification scales exponentially in the number of non-Clifford gates in the circuit \cite{leone_magic_2022}. 

A closely related fidelity estimation protocol is so-called shadow fidelity estimation \cite{huang_predicting_2020}. 
In this protocol, measurements are performed in a random Clifford basis, see \cite[Sec.\ II.J]{kliesch_theory_2021} for an explanation. 
The sample complexity of shadow fidelity is also constant, but it is computationally inefficient for non-Clifford states, since overlaps between the target state and an arbitrary stabilizer state need to be computed.
Another fidelity estimation protocol that can be applied to quantum random sampling schemes is the adaptive protocol of \textcite{VerificationAnticoncentrated}, which requires two auxiliary qubits and entangling gates between the unknown state preparation and those auxiliary
qubits and on-the-fly classical computation. 
Interestingly, this scheme is sample-efficient precisely for  anticoncentrating distributions with exponentially small collision probability. 
To even further reduce the experimental effort of verification as compared to direct fidelity estimation one would need to improve the scaling in the tolerated estimation error $\epsilon$. 
For stabilizer states this has recently been studied by \textcite{kalev_validating_2019}.

\subsection{Efficient classical verification}
\label{sec:efficient classical verification}

In the previous sections, we have on the one hand seen classical verification methods that are sample efficient in that they require only few (polynomially many) samples from the quantum device, but require exponential computational runtime. 
On the other hand, we have seen quantum verification tools that are fully efficient but require trust in an experimental quantum measurement and are experimentally more demanding since they require measurements in different (local) bases. 
We conclude our discussion of verification protocols with classical verification protocols that are fully efficient but make other types of assumptions than experimental ones, or yield less information about the implemented distribution.

\subsubsection{State discrimination}
\label{subsec:state discrimination}
Rather than trying to certify the full target distribution in TVD we can alternatively discriminate the experimentally implemented distribution from our best guess of what a very noisy distribution or a closeby classically simulable distribution could be.  
One can see the full verification task in this mindset as distinguishing the imperfect preparation against \emph{all possible} distributions that are at least $\epsilon$-far away from the target distribution. 

The discrimination task is considered by \textcite{gogolin_bosonsampling_2013} in a setting of a highly restricted client aiming to verify a boson sampler just from the histogram of outcomes without using the information about which outcome has been obtained. 
They show that in this setting, a boson sampling distribution cannot be distinguished from uniform and prompted the development of a fully efficient and simple state discrimination test that makes use of the actual outcomes~\cite{aaronson_bosonsampling_2013}. 
To date, state discrimination remains the most convincing way of validating boson sampling experiments as it is unclear whether the XEB fidelity yield a meaningful benchmark of boson sampling experiments. 

Let us illustrate the idea by means of the test of \textcite{aaronson_bosonsampling_2013} for discriminating the Fock boson sampling distribution from the uniform distribution. 
The idea is to use the so-called 
\emph{row-norm estimator} for a matrix $X \in \mb C(n \times n) $
\begin{align}
\label{eq:row norm estimator}
  R^*(X) = \frac{1}{n^n} \prod_{i= 1}^n R_i(X), 
\end{align}
where $R_i(X)= \norm{x_{i}}_2^2 = |x_{i,1}|^2 + \cdots + |x_{i,n}|^2$ is the norm-squared of the $i$th row of $X$. 
Indeed, for a Gaussian normal matrix $X \sim \mc N \equiv \mc N_{\mb C}(0,1)^{n \times n} $ one expects $\Eb_{X \sim \mc N}[R^*(X)] = 1$. 
The fluctuations around this value depend on whether experimental samples are chosen from the boson sampling distribution or a uniform distribution and can be exploited to discriminate a device from uniform. 
In order to discriminate a distribution from uniform, we compute $R^*(U_{S,1_n})$ for a few samples $S$ and comparing the outcome to one's expectation. 
To see why this achieves the task, let $\mc H$ be the distribution $\mc N$ with distribution function $p_{\mc N}(X)$ scaled by the probability of obtaining the corresponding outcome, i.e., $p_{\mc H }(X)= p_{\mc N}(X) P(X)$. 
When specializing to boson sampling, the matrix $X$ will be an approximately Gaussian-distributed submatrix $U_{S,1_n}$ of the linear-optical unitary $U$. 
Remember that the probability of obtaining this matrix, corresponding to the outcome $S$, is given by $P_U(S) = |\Perm(U_{S,1_n})|^2/n!$ (cf.\ \cref{eq:bosonsamplingdistribution permanent}). 
One finds that \cite[Corollary 18]{aaronson_bosonsampling_2013}
\begin{multline}
  \Pr_{\mc H}[R^* \geq 1 ] - \Pr_{\mc N}[R^* \geq 1 ]\\
   = \frac 12 \Eb_{\mc N}\left[| R^* - 1| \right]
   \geq 0.146 - O\left(\frac 1 {\sqrt{n}} \right).
\end{multline}
In other words, the row norm estimator $R^*(X)$ is ever so slightly correlated with $\Perm(X)$. 
An intuitive reason for this is  that multiplying every row of $X$ by the same scalar $c$ also multiplies $\Perm(X)$ by $c$ \cite{aaronson_bosonsampling_2013}.
At the same time, it can be computed in time $O(n^2)$.
To discriminate a boson sampler from the uniform distribution, one therefore needs to simply collect $k$ samples $S_1, \ldots, S_k$ from a device claimed to realize a boson sampler and compute $\frac 1 k \sum_{i = 1}^k \left|R^*(U_{S_k}) - 1 \right|$ 
up to sufficiently high precision so as to confidently distinguish the resulting value from $0$.\footnote{
This may be done in a Bayesian framework~\cite{carolan_experimental_2014}.}

In the same framework, one can distinguish a boson sampler against other---some	what more informed---distributions such as a distribution of distinguishable particles that are sent through the linear-optical network \cite{carolan_experimental_2014,spagnolo_efficient_2014}.  
In the experiments of \textcite{zhong_quantum_2020,zhong_phase-programmable_2021}, the output distribution
is additionally distinguished from a thermal distribution. 
To distinguish from any classically efficient distribution, they use the Bayesian likelihood ratio estimator 
\begin{align}
  c =  \frac{\Pr(\{x_1, \ldots, x_S\}| P_0)}{\Pr(\{x_1, \ldots, x_S\}|P_0) + \Pr(\{x_1, \ldots, x_S\}|Q)}
\end{align}
where the likelihood of obtaining the experimental samples $x_1, \ldots, x_S$ is evaluated both with respect to the ideal target distribution $P_0$ and a distribution $Q$ that we want want to distinguish from $P_0$. 

An additional experimentally motivated test ruling out spoofing distributions that makes use of low-order marginal probabilities \cite{villalonga_efficient_2021}, performed by \textcite{zhong_phase-programmable_2021}, is to measure these marginals. 
One can then compare them to the theoretical predictions, thereby ruling out that a distribution which only agrees on the first two or three marginals is a good spoofing distribution.

The efficient state discrimination tests for boson sampling highlight a key difference between the output distributions of variants of boson sampling and universal circuit sampling: 
for universal circuit sampling we expect the output distribution not even to be efficiently distinguishable from the uniform distribution. 
This expectation can be understood in various readings. 
First, it can be viewed from the perspective of \hog-like tests, since high performance on a \hog-like test serves as a discriminator against the uniform distribution. 
Conversely, if \hog\ is indeed a computationally difficult task, then this provides evidence that discriminating against uniform is also a difficult task. 
Indeed, it is difficult to imagine a way of discriminating against uniform that does not make use of a \hog-like estimator.   
\textcite{franca_game_2020} make this intuition more rigorous. 
They show that if there exist functions defining a cross-entropy measure \eqref{eq:cross entropy measures} that gives rise to a sample-efficient state discrimination test, then full verification of the total-variation distance will be sample-efficiently possible in a multi-round scheme. 
Since we do not believe the latter to be possible, the result of \textcite{franca_game_2020} serves as more formal evidence against the possibility of efficient state discrimination for random quantum circuits.

\subsubsection{Cryptographic tests}
\label{sec:cryptographic tests}

A completely orthogonal but promising avenue of verifying sampling schemes has been pioneered by \textcite{shepherd_temporally_2009}:
By allowing the certifier to choose the classical input to the sampling device rather than drawing it fully at random, it may be possible to efficiently certify that a quantum device has performed a task that no classical device could have solved under cryptographic assumptions on the hardness of certain tasks.
This could be facilitated by checking a previously hidden bias in the obtained samples for a certain family of IQP circuits \cite{shepherd_temporally_2009}.

It is instructive to understand the idea behind such a test of computational quantumness. 
The protocol of \textcite{shepherd_temporally_2009} is formulated for a certain family of IQP circuits, called $X$-programs.
An $X$-program acting on $n$ qubits is defined by a list of pairs $(\theta_p, p ) \in [0,2\pi]\times \{0,1\}^n$ and acts as 
\begin{align}
	\ket 0 \mapsto \exp\left( \ii \sum_{p} \theta_p \prod_{j =1}^n X_j^{p_j}\right) \ket 0 .
\end{align}
For the purposes of the quantumness test it is sufficient to choose a constant value of $\theta$ that is the same for every nonvanishing term in the Hamiltonian. 
In this case, an $X$-program with $k$ nonvanishing Hamiltonian terms acting on $n$ qubits can be represented by a $0/1$ matrix $P \in \{0,1\}^{k\times n} $. 
Each row of this matrix specifies a Hamiltonian term and it is easy to see that the output distribution of such an $X$-program is given by 
\begin{align}
	P_P(x) = \left | \sum_{a \in \{0,1\}^k :\ P^T\cdot a = x } \cos(\theta)^{k - wt(a)} \sin(\theta)^{wt(a)} \right| ^2, 
\end{align}
where $wt(a) = |\{ l \in [k] : a_l = 1 \} |  $ is the \emph{Hamming weight} of the binary string $a \in \{0,1\}^k$. 

For a random variable $X$ taking values in $\{0,1\}^n$, and $s \in \{0,1\}^n$, the \emph{bias} of $X$ in the direction of $s$ is just the probability that a sample $x \sim P_P$ is orthogonal to $s$, i.e., that $x^T s = 0 $. 
The key idea of the test of computational quantumness is to hide a string $s$ the output probability distribution of an $X$-program in such a way that this string $s$ cannot be determined efficiently, but at the same time the bias  of the output distribution of the $X$-program in direction $s$ is significantly larger than the bias of any cheating distribution that can efficiently be obtained using classical computing resources. 
In particular, the bias of the output distribution $P_P$ of the $X$-program defined by a matrix $P \in \{0,1\}^{k \times n }$ and angle $\theta$ is given by 
\begin{align}
	\Pr_{x \sim P_P} [x^T s = 0 ] =   \sum_{x : \ x^T \cdot s = 0 } P_P(x).
	\label{eq:bias iqp}
\end{align}
To achieve this, \textcite{shepherd_temporally_2009} notice that the matrix $P$ can be viewed as the \emph{generator matrix} of a linear code. 
That is, the columns of $P$ span the code space $\mc C = \{ P \cdot d: \ d \in \{0,1\}^n\} $. 
If we let $P_s$ be the $n_s \times n $ submatrix of $P$ obtained by deleting all rows $p$ for which $p^T s = 0 $,\footnote{Of course, this leaves only rows for which $p^T s = 1$. } and $\mc C_s$ be the code generated by $P_s$, then we can rewrite the bias \eqref{eq:bias iqp} of $P_p$ as \cite[Thm.\ 2.7]{shepherd_temporally_2009}
\begin{align}
	\Pr_{x \sim P_P} [x^T s = 0 ] = \Eb_{c \sim \mc C_s}\left[ \cos^2 \left(\theta (n_s - 2 \cdot  wt(c)\right) \right ] .
	\label{eq:bias iqp 2}
\end{align}

We can now set a quantum challenge that is intrinsically verifiable in the following way. 
We choose a code $\mc C_s$ and a value of $\theta$ in such a way that both the bias \eqref{eq:bias iqp 2} is strictly larger than $1/2$, and that any classical strategy can only achieve a bias that is significantly lower, say, by a constant. 
We then choose a generating matrix $P_s$ for $\mc C_s$ such that $s$ is not orthogonal to any of the rows of $P_s$. 
Finally, we obfuscate this matrix by adding rows that are orthogonal to $s$, permuting all rows and potentially performing reversible column operations, giving rise to a matrix $P$. 
Given samples from the distribution $P_P$, we can now distinguish the hypothesis that the sampling device has quantum capacities from the hypothesis that it is cheating by comparing the frequencies of outcomes that are orthogonal to the hidden string $s$.
Notice that this protocol \emph{does not} certify that the samples are distributed according to the correct distribution. Therefore, it does not constitute a workaround to the no-go theorem of \cref{ssec:hardness of verification} based on cryptographic assumptions. 
Similarly to the \hog\ test (\cref{prob:hog}), this cryptographic test of quantumness merely certifies that the device has the capacity to do something that presumably---under assumptions---no classical computing device could have achieved. 

The particular suggestion of \textcite{shepherd_temporally_2009} is to use \emph{quadratic residue codes} and a particular obfuscation procedure that exploits specific properties of such codes (such as that the full-weight vector is always a codeword).
They conjecture that recovering the matrix $P_s$ from the obfuscated matrix $P$ is \np-complete. 
Choosing $\theta = \pi/8$, this construction gives rise to a bias that serendipitously matches that of the Bell inequality: $\cos^2(\pi/8) \approx 0.854$ for the quantum value, and $3/4$ for the best classical strategy discussed by \textcite{bremner_classical_2010}.\footnote{There is no proof that this $3/4$ is the optimal classical value.
}

Note also that besides the security assumption on the obfuscation procedure, additional conjectures need to be made \cite[Conjectures~4.2 and~4.3]{shepherd_temporally_2009} for such a test to achieve its goal: 
First, the distribution $P_P$ of a randomly selected $X$-program with constant $\theta = \pi/8$ should be hard to sample from so that only a quantum device can perform this task. 
Second, the output distribution should be sufficiently flat in the sense that its R\'enyi $2$-entropy or \emph{collision entropy} is close to maximal, i.e., $H_2(P_P) = \Omega(n)$ so that cheating becomes more difficult. 

Iterating the importance of extensively testing cryptographic assumptions for their security, \textcite{kahanamoku-meyer_forging_2019} has  developed a classical cheating strategy for the protocol proposed by \textcite{shepherd_temporally_2009}. 
Given a description of an $X$-program in the form of the matrix $P$, the cheating strategy extracts the secret vector $s$ with probability arbitrarily close to unity in an (empirically observed) average runtime of $O(n^3)$. 

In a similar mindset, albeit without restricting to sampling tasks for which there is strong complexity-theoretic evidence for hardness, cryptographic tests of quantumness have been devised by \textcite{brakerski_cryptographic_2018,brakerski_simpler_2020}. 
There, the authors made use of so-called \emph{trapdoor claw-free functions} to delegate a simple task that no classical device can efficiently solve, but a quantum device succeeds with higher probability. 
A trapdoor claw-free function is a 2-to-1 efficiently computable function $f$ such that it is difficult to find a \emph{claw} $x,x'$ for which $f(x) = f(x')$, but it becomes easy when given access to the trapdoor. 
So while a classical algorithm can only every hold $y = f(x) $ and $x$ but not at the same time $x'$, a quantum algorithm can compute $f$ in superposition and therefore hold $y$ as well as a superposition $\ket x + \ket {x'}$. 
The idea of the proof is to exploit this superposition: we can ask the device to perform a measurement in the computational basis, obtaining $x$ or $x'$, or in the Hadamard basis, obtaining $d$ for which $d \cdot ( x \oplus x') = 0$. 
This reveals some information about $x$ and $x'$ that is not accessible to a classical device. 
Such protocols were recently improved to much simpler functions \cite{kahanamoku-meyer_classically-verifiable_2021} and low-depth implementations \cite{hirahara_test_2021,liu_depth-efficient_2021}, bringing their experimental demonstration within closer reach \cite{zhu_interactive_2021}. 

Using such trapdoor claw-free functions, it is also possible to classically delegate a \bqp\ computation to a fully untrusted quantum server \cite{mahadev_classical_2018}, 
and even to verify sampling problems \cite{chung_constant-round_2020}. 
A drawback of the protocol of \textcite{chung_constant-round_2020}, however, is that it only has inverse polynomially large soundness, so that it cannot be used as a subroutine in secure computation problems. 
More severely, for an application to verifying quantum random sampling, the overhead is unfeasibly large.

\subsection{Further approaches to the verification of quantum samplers}
\label{sec:further approaches verification}

Another approach to verification of quantum states from measurements---blind verified quantum computation---has been developed by \textcite{broadbent_universal_2009} and \textcite{fitzsimons_unconditionally_2017}. 
While the protocols discussed in \cref{ssec:fidelity witness} make use of the ability of the experimenter to measure single qubits with high fidelity, blind verified quantum computing presupposes the ability to accurately prepare single qubits.
And indeed, blind verified quantum computing also applies measurement-based computation using cluster states, exploiting the property that single-qubit phase gates commute through the state preparation. 
While in our approaches, the imperfect state preparation is directly verified, however, in verified blind quantum computing so-called \emph{trap qubits} are made use of. 
The outcome of measurements on those qubits is deterministic and can thus be checked to build confidence in the correct functioning of an untrusted quantum server. 
By turning blind quantum computing upside down, a ``post hoc verification protocol'' for quantum computations has been developed by \textcite{fitzsimons_post_2018}. 

In order to build trust in the correct functioning of a sampling device one may also resort to weaker types of verification than direct verification of the quantum state or output distribution. 
For instance, instead of directly running a randomly chosen unitary circuit, one can run specific computations on the device, the output distribution of which is highly structured, such as the 
quantum Fourier transform \cite{TicMayBuc14}. 
Finally, one can build trust in the device from certain efficiently computable benchmarks such as two-point correlation functions~\cite{phillips_certification_2019}, higher correlation functions \cite{zhong_phase-programmable_2021}, the click-number distribution in boson sampling with threshold detection \cite{drummond_simulating_2022}, or comparison to a coarse-grained distribution~\cite{wang_certification_2016}. 

Let us also  note that with the exception of the classical verification protocol due to \textcite{mahadev_classical_2018,chung_constant-round_2020} most of the verification protocols considered here require independently identically distributed (iid.) state preparations which is an additional assumption---albeit a very realistic one.  
In order to relax this assumption to the non-iid.\ case, one can make use of de-Finetti arguments~\cite{finetti_prevision_1937,hudson_locally_1976,caves_unknown_2002,konig_finetti_2005}. 
This has been done by \textcite{takeuchi_verification_2018} and optimized for an application to graph states by \textcite{takeuchi_resource-efficient_2019,markham_simple_2020} and to bosonic states by \textcite{chabaud_efficient_2021,chabaud_building_2020}.


\section{Experimental implementations}
\label{sec:experimental implementations}

It is the comparative simplicity of quantum random sampling schemes that renders them 
particularly compelling for an implementation on current-day devices. 
In contrast to other proposals for quantum advantage they do precisely not require interactive or multi-round feedback. 
Moreover, comparably small circuit sizes are required so that it might be possible to implement the circuits with non-negligible fidelity without full-fledged quantum error correction. 
This makes quantum random sampling schemes attractive as proofs of quantum advantage from an experimental point of view. 
Experimental implementations of quantum random sampling start with the first proof-of-principle demonstrations of boson sampling \cite{spring_boson_2013,crespi_integrated_2013,tillmann_experimental_2013,broome_photonic_2013} and universal circuit sampling \cite{neill_blueprint_2018}, and culminate in the recent large-scale implementations of universal circuit sampling \cite{arute_quantum_2019,zhu_quantum_2022,wu_strong_2021} and Gaussian boson sampling \cite{zhong_quantum_2020,zhong_phase-programmable_2021,madsen_quantum_2022}, which are arguably in the classically intractable regime.  
In this section, we summarize the most important technological developments and experimental subleties of quantum random sampling implementations with a focus on universal circuit sampling.

\subsection{Universal circuit sampling with superconducting circuits}
\label{subsec:arute experiment}

At the current state of the art, universal circuit sampling is most feasibly implemented using superconducting 
transmon devices. 
The first large-scale experiment aimed at reaching a quantum advantage was performed in such an architecture \cite{arute_quantum_2019}.
This experiment is a landmark experiment which has arguably first reached the regime of a quantum advantage over the capabilities of classical supercomputers and hence the ``quantum supremacy'' regime. 
We therefore dedicate slightly more detail to the discussion of this experiment, as an exemplary discussion \emph{pars pro toto}.
The experiment implemented a random circuit consisting of up to 20 layers of the universal random circuits introduced in \cref{subsec:universalcircuitsampling} acting on 53 qubits.

\subsubsection{Design of the experiment}

The experiment of \textcite{arute_quantum_2019} was performed on a \emph{transmon} superconducting chip referred to as \emph{Sycamore} chip. 
Transmons are superconducting charge qubits that have been designed to be less sensitive to charge noise than is common in other settings, a feature that renders them particularly
attractive for the use in quantum computational schemes. 
Generally speaking, in a superconducting circuit,
currents and voltages behave quantum mechanically, as conduction electrons condense into a macroscopic
quantum state. 
For this to be possible and to ensure that the ambient thermal energy is reduced to well below the native energy scales of the qubits, cryogenic temperatures are required. 
The extremely low temperatures of $\sim 20$ mK required for the experiments are to date only accessible in  dilution refrigerators. 
Each of the qubits can be seen as a non-linear superconducting resonator operating at $5$--$7$ GHz. 
These qubits can be tuned by resorting to two degrees of freedom.
On the one hand, there is a microwave drive that allows to drive Rabi oscillations of the qubit. 
On the other hand, there
is a magnetic flux control that allows to tune the frequency.  

During a quantum circuit, the qubits are tuned to three different frequencies: first, there is the qubit idle frequency at which single qubit gates are performed. 
Second, there is an interaction frequency to which neighbouring qubits are tuned in order to interact. 
The idle frequency is chosen such that there is as little as possible crosstalk during single-qubit gates, while at the same time minimizing the frequency distance required for interaction with its neighbours. 
Finally, the qubits are tuned to a readout frequency.
When selecting those frequencies, there are trade-offs to be accounted for between energy-relaxation, dephasing, leakage, and control imperfections \cite[SM, VI.A.4]{arute_quantum_2019}. 
At the idle frequency, single-qubit gates are implemented by driving the qubits with 25 ns microwave pulses.

In the Sycamore superconducting-qubit architecture, two-qubit gates are implemented using adjustable couplers. 
Since the qubits are arranged in a planar two-dimensional architecture, these couplers are naturally placed between nearest neighbors on a lattice.\footnote{
This architecture has also been chosen to be forward-compatible with the realization of a surface code for
quantum error correction.} 
The couplers allow to quickly switch on and off a coupling of up to 40 MHz by tuning the frequency of the coupler qubits. 
Specifically, the coupling is achieved 
by tuning neighboring qubits' frequencies on-resonance and turning 
on a 20 MHz coupling for 12 ns. 
The coupling in the system natively gives rise to the two-qubit gate
\begin{align}
\label{eq:fsim gate}
  \text{fSim}(\theta,\phi) = \begin{pmatrix}    
    1 & 0 & 0 & 0 \\ 
    0 & \cos(\theta) & - \ii \sin(\theta) & 0\\
    0 & - \ii \sin(\theta) &\cos(\theta) & 0 \\
    0 & 0 & 0 & \e^{-\ii \phi} 
  \end{pmatrix}  ,   
\end{align}
with tunable angles $\theta,\phi$.
Here, the angle $\theta$ is interpreted as the swap angle and the angle $\phi$ is a conditional phase. 
The fSim gate captures a wide range of entangling gates, including the iSWAP gate with $\theta=\pi/2$ and $\phi = 0$, the CZ gate with $\theta = 0$ and $\phi = \pi$.

In the experiment of \textcite{arute_quantum_2019}, iSWAP-like fSim gates close to iSWAP$^*$ \eqref{eq:iswapstar} with $\theta \approx \pi/2$ and a conditional phase $\phi \approx \pi/6$ were performed.
Notably, the specific phases of each two-qubit gate, corresponding to a particular physical coupler between two qubits varied around their ideal values. 
\textcite{arute_quantum_2019} were able to measure the precise angle, thus ensuring higher accuracy of the resulting computational task. 
The uncertainty in the actual angles implemented in the circuit can be viewed as limited programmability of the device: 
some parameters of the circuit are only determined contingently on the specific physical implementation. 
More recently, progress has been made towards achieving full programmability of the angles in the fSim gate \cite{foxen_demonstrating_2020}.  

Every qubit can be read out by means of a linear resonator.
To this end, the qubit frequency is tuned to its readout value and coupled to a far-detuned resonator via a neighbouring coupler \cite{blais_cavity_2004,bultink_general_2018,gambetta_qubit-photon_2006}.
As the qubit state changes from $\ket 0 $ to $\ket 1$, there occurs a frequency shift in the resonator, which can be read out via the phase shift incurred by a microwave probe signal applied to the resonator \cite[SM~III.B]{arute_quantum_2019}.
On the chip, the qubits are divided into groups of six qubits which are each coupled to their own resonator, but resonators within a group are simultaneously read out via frequency multiplexing.
Overall, the architecture consists of 53 such transmon qubits, each of which is connected to a read-out device,
with 86 couplers connecting nearest-neighbor qubits.

\subsubsection{Benchmarking of the components}

Naturally, substantial efforts have been made to carefully benchmark the experiment. 

\paragraph{Benchmarking of single-qubit gates.} 
At the lowest level, benchmarking of the experiment was performed on the level of the individual components of the device. 
For the individual components, the single-qubit operations, the entangling gates and the read-out were benchmarked individually. For both single-qubit gates and  two-qubit gates as well as the benchmarking of the entire device, 
\textcite{arute_quantum_2019} make use of linear XEB.
Here, we pick up the line of thought developed in   \cref{ssec:xeb} and put it into the context of experimental findings. 
As developed there, XEB provides a unified picture for average-case benchmarking of small-scale operations in the sense of randomized benchmarking  on the one hand \cite{arute_quantum_2019,liu_benchmarking_2021}, and single-instance benchmarking of typical large-scale quantum states on the other hand \cite{arute_quantum_2019}.
Again, the use of linear XEB is attractive in this context as this procedure does not require the classical computation of all possible events, but the classical simulations only need to compute the likelihood of the set of bit strings obtained in an experiment. 

For the benchmarking
of single-qubit gates, linear XEB benchmarking has been used to estimate the probability
of an error occurring on the single-qubit level.
For each qubit, a sequence of a variable number of randomly selected gates is applied and $F_\xeb(Q,P_C)$ as defined in \cref{eq:xeb fidelity} and discussed in \cref{ssec:xeb} is estimated. 
Notably, the resulting scheme can be seen as a randomized
benchmarking protocol \cite{EstimatingGateSetProperties,liu_benchmarking_2021}.
One finds a decay of the signal in the length $\ell$ of the sequence that is well
described by an exponential  dependence of the form $(1-3 e_1/4)^\ell$,
where $e_1\in[0,1]$ is the single-qubit Pauli error probability. 
The single-qubit errors $e_1$ over the various qubits follows a distribution that is estimated by suitable histograms. 
From these histograms, one can then estimate an average of about $e_1 = 0.16$\% in simultaneous operation of the qubits on the chip.

\paragraph{Benchmarking of two-qubit gates.} 
For the linear XEB benchmarking of the two-qubit gates, as in the single-qubit case, sequences of cycles are employed. 
Now, each cycle consists of randomly chosen single-qubit gates followed by the iSWAP$^*$ two-qubit gate. 
This gives rise to an interleaved randomized benchmarking scheme \cite{arute_quantum_2019}, in which the same logic as for single-qubit gate benchmarking is applied: 
an exponential curve is fitted to the decay and one can estimate
the two-qubit error rate $e_2$ by subtracting the single-qubit error rate $e_1$. 
After appropriate
corrections for dispersive shifts and cross-talk, an average of about $e_2= 0.62$\% is
found when operating gates simultaneously on the chip. 
Finally, the combined single- and two-qubit error rate $e_{2C}$ which characterizes a single layer of a gate cycle is measured to be $e_{2C} = 0.93$\% on average.

\paragraph{Characterization of single-qubit measurements.}
Measurement errors of single-qubit readout are obtained by preparing $\ket 0$ and $\ket 1$ and performing a measurement of the state. 
The identification error is taken to be the probability that the qubit was read out in a state other than intended, giving rise to a median identification error of $0.97$\% for the $\ket 0$ state 
and $4.5$\% for the $\ket 1$ state \cite[SM~VI.D.3]{arute_quantum_2019}. 
The fact that the state preparation fidelity is much higher than the measurement fidelity justifies this procedure. 

In a second step, multi-qubit readout is characterized by preparing and measuring $150$ random classical bit strings with $53$ qubits and repeating each measurement $3000$ times,  resulting in a $13.6$\% probability of correctly identifying the state.
This can be decomposed to a median error for simultaneous single-qubit readout of $1.8$\% for 
 $\ket 0$ and~$5.1$\% for $\ket 1 $, giving an overall simultaneous readout error of about~$3.8$\%.

\subsubsection{Verifying the sampling task}

The entire setup of the experiment has been tailored to achieving a quantum computational advantage.
Benchmarking the individual components builds trust in the functioning of the 53 qubit device as a whole, but does not yet constitute a test of quantum advantage as outlined in the introduction (\cref{sec:intro}). 
In light of the hardness of rigorous verification of the sampling task using the samples as explained in \cref{ssec:hardness of verification}, the entire scheme has been 
benchmarked via linear XEB, but now applied to typical instances of high-dimensional quantum states as discussed in \ref{ssec:xeb}. 
As discussed there, while the linear XEB fidelity does not yield a rigorous certificate for the sampling task, achieving a nontrivial XEB value might be a computationally difficult task in itself. 
Having said that, the claim of \textcite{arute_quantum_2019} is indeed to have performed the \emph{sampling task} to nontrivial precision. 

In order to estimate the XEB fidelity, the probability of each bit string obtained in the experiment needs to be computed. 
As laid out in more detail in the following section (\cref{sec:classical simulation}), for the 
full random quantum circuit this is  beyond the reach of classical computers. 
This is why proxy methods need to be used in order to reduce the complexity of computing the output probabilities of the  implemented quantum circuits. 
Specifically, \textcite{arute_quantum_2019} make use of three different simulation strategies. 

\begin{figure*}
  \includegraphics[width=\textwidth]{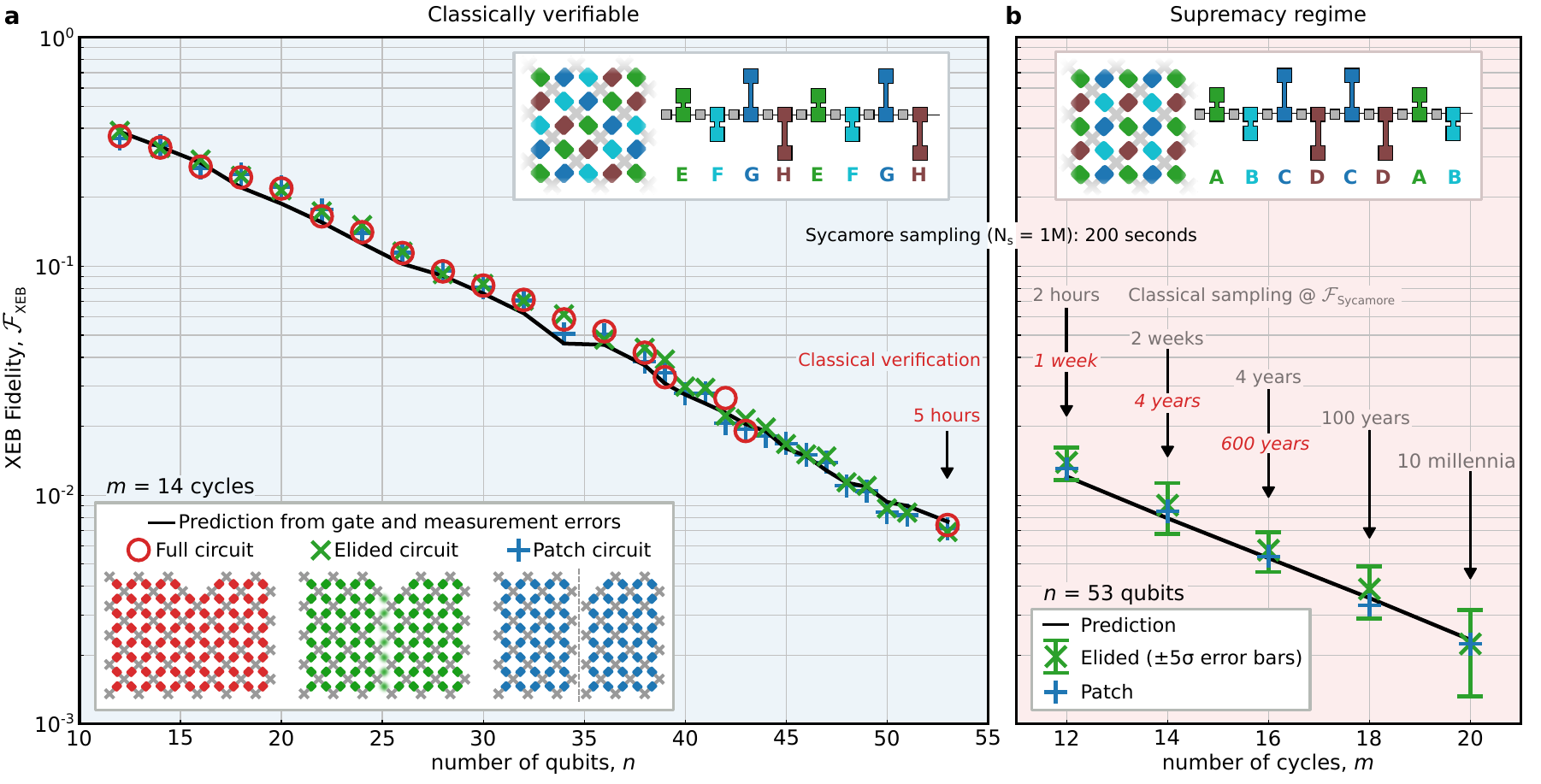}
  \caption{\label{fig:arute_fig3}
  \cite[Fig.~4]{arute_quantum_2019}
  \emph{(a)} To build trust in the proxy methods (elided and patch circuits) for full circuit simulation used to estimate the XEB fidelity, sampling on the quantum processor is performed using circuits with the same gate count as the ``Supremacy circuits'', but in a simplifiable pattern and with depth $14$.   
  Each data point is an average of the XEB fidelity of $10$ circuit instances with $0.5 - 2.5 \cdot 10^6$ many samples per instance. 
  The solid line represents the predicted value of the XEB fidelity, given the error model. 
  \emph{(b)} In the ``supremacy regime'' of $53 $ qubits and depth up to $20$, the elided and patch methods are used to estimate the XEB fidelity, and the classical simulation time for verification and sampling is extrapolated. 
  }
\end{figure*}

In a full circuit simulation, the exact output probabilities of a given quantum circuit are computed.
In ``patch circuits'', one removes all two-qubit gates along a slice through the 2D qubit array, so that the
circuit is split into two unconnected parts and the overall fidelity is nothing but the product of two fidelities. 
In ``elided circuits'' one removes a fraction of two-qubit gates between the two partitions of the qubits, so that 
the parts are coupled, but  less entanglement is being generated.

In order to benchmark the patch circuit and elided circuit method against the full circuit method as a means to estimate the XEB fidelity, \textcite{arute_quantum_2019} perform what they call ``verification circuits''. 
Those circuits are chosen in such a way that a full circuit simulation is still possible. Specifically, two-qubit gates are arranged in a simplifiable tiling so that circuits with exactly the
same gate count as in the full experiment are easier to classically simulate.
For circuits with 14 cycles on up to 53 qubits, this allows for the comparison of the three different methods of estimating the XEB fidelity, see \cref{fig:arute_fig3}(a), showing that all methods yield roughly the same value of the XEB fidelity. 

For full circuit simulation of up to $43$ qubits,
a ``Schr{\"o}dinger type'' simulation algorithm is run for the
simulation of the full quantum state, making use of 100.000 cores and
250 Terabytes memory. 
For larger qubit sizes, a hybrid  ``Schr{\"o}dinger-Heisenberg type'' simulation algorithm is run.

Providing further justification \textcite{arute_quantum_2019} provide a model for how the
XEB fidelity $F_\xeb(Q,P_C)$ scales 
given the errors obtained for the individual circuit components, yielding good agreement with the predictions obtained via the various simulation methods.   
Altogether, these tests constitute the justification for the use of the ``elided'' and ``patch'' methods as a substitute of full circuit simulation when computing the XEB benchmark. 

In the ``supremacy regime'' of 53 qubits and depth 20, elided and patch circuit method remain close to the error model (\cref{fig:arute_fig3}) yielding a value of $F_\xeb(Q,P_C)\approx
(2.24 \pm 0.21)\times 10^{-3}$ averaged over 10 circuit instances. 
Here, the error bar is a $\sigma $ interval, where $\sigma$ combines statistical errors of the finite-sample XEB benchmark and systematic errors due to the elided simulation method. 
This shows that the XEB value is larger than $10^{-3}$ with $5\sigma$ significance. 
Importantly, since the \xeb-fidelity scales inverse exponentially, the number of samples required to obtain the required significance scales exponentially.

In order to further substantiate the claim that, on the quantum device, the sampling task has in fact been achieved to nontrivial accuracy, \textcite[SM~VIII]{arute_quantum_2019} perform further tests. 
First, they compare the values of the linear XEB with the logarithmic XEB or cross-entropy difference \eqref{eq:cross entropy difference}, see \cref{ssec:cross-entropy difference} for a discussion. 
This measure is expected to have a larger variance than the linear XEB fidelity as it puts more weight on the tail of the distribution but at the same time relates more closely to the actual total-variation distance \cite{bouland_complexity_2018}.
\textcite{arute_quantum_2019} argue that both measures can serve as a proxy for the quantum fidelity as discussed in \cref{ssec:xeb}. 
Second, they analyse in more detail the distribution of bit string probabilities obtained in the experiment. 
They find an excellent fit with the expected (Porter-Thomas) distribution of the outcome probabilities, and perform hypothesis tests to reject the hypothesis that the samples stem from a uniform distribution.

The claim of having achieved quantum computational advantage in a practical sense is substantiated by extrapolating the computational effort to estimate the computational cost of the quantum advantage circuits to larger system sizes.
\textcite{arute_quantum_2019} estimate that for $n=53$ and $d=14$, sampling of three million bit strings with $0.01$ fidelity would take about a year. 
By extrapolation, they then argue that for the full  $n=53$ and $d=20$, obtaining a million samples on the quantum processor  takes about 200 s, while sampling to a comparable fidelity classically would take 10\,000 years on a million cores, and the verification of the fidelity would require millions of years.
These claims have naturally been challenged by new improved classical simulation methods  explained in \cref{sec:classical simulation}.

\subsubsection{Follow-up work}

\textcite{wu_strong_2021,zhu_quantum_2022} follow up on the landmark experiment of \textcite{arute_quantum_2019},
presenting comprehensive, and qualitatively similar data from a superconducting platform, but with larger number of qubits and circuit sizes. 
The superconducting processor of \textcite{wu_strong_2021,zhu_quantum_2022} of $n=66$ transmon qubits, which are coupled by $110$ tunable nearest neighbor couplers. 

However, quantitatively, the experiment improves in several ways over the experiment of \textcite{arute_quantum_2019}. 
\textcite{wu_strong_2021} benchmark the device using $56$-qubit, depth-$20$ random Sycamore circuits (i.e., in the same scheme as \textcite{arute_quantum_2019})  and achieve comparable error rates.
They find an XEB fidelity of $0.0662$\% for roughly 10 million bit strings observed in the experiment.
\textcite{zhu_quantum_2022} improve upon this and measure an XEB fidelity of $0.0758$\% for 60-qubit, 22-cycle circuits, and $(3.66 \pm 0.345)\cdot 10^{-4}$ for 60-qubit, 24-cycle circuits. 
Their experiment improves over that of \textcite{arute_quantum_2019} especially when it comes to readout fidelity, for which they achieve an average fidelity of $2.26$\%. 
\textcite{zhu_quantum_2022} estimate
that the sampling task  requires about four orders of magnitude more 
resources compared to the sampling task considered by \textcite{arute_quantum_2019}.\\

To summarize this discussion, both \textcite{arute_quantum_2019} and \textcite{zhu_quantum_2022,wu_strong_2021} claim a significant advantage of their respective quantum device over all possible classical algorithms applied to the same task.
In a nutshell, the advantage claim of those experiments is based on a placeholder for the linear XEB fidelity which can be computed in the advantage regime, as well as empirical and numerical evidence for the validity of this estimator.
In \cref{sec:classical simulation}, we discuss in detail how and to what extent this quantum advantage claim is challenged by tailored classical simulation algorithms, as well as how the particular choice of benchmark affects the claim. 

\subsection{Photonic implementations}
\label{subsec:photonic implementations}

Historically preceding implementations using superconducting quantum circuits, photonic implementations of variants of boson sampling have developed significantly over the past ten years. 
These fall into implementations of the original proposal of \textcite{aaronson_computational_2010} to make use of initial Fock state preparations, and implementations of the Gaussian boson sampling protocol initially proposed by \textcite{lund_boson_2014}, and refined by \textcite{hamilton_gaussian_2017,kruse_detailed_2019}.

\subsubsection{Fock boson sampling}
\label{subsec:experimental fock boson sampling}

Soon after the proposal of boson sampling had become available \cite{aaronson_computational_2010}, 
first experiments with photonic systems 
were 
conducted, all around the same time \cite{crespi_integrated_2013,spring_boson_2013,broome_photonic_2013,tillmann_experimental_2013}. 
These first implementations involve a comparably small number of 
modes and photons, even though already these early experiments were often performed on photonic chips in integrated optics. 
\textcite{spring_boson_2013}
present data from an experiment involving $m=6$ modes and $n=3$ and $n=4$ photons, resorting to silica-on-silicon integrated waveguide circuits. 
In such waveguide circuits fabricated by ultra-violet writing, evanescent waves overlap, giving rise to effective beam-splitter arrays.
In this experiment, two parametric down-conversion pair sources are used to inject up to four photons into a photonic circuit. 
That is to say, the sources are not used in a heralded mode, where one port provides a classical
signal for the presence of a photon in the other port, but both output ports of the sources are fed into the device. 
The dominant sources of inaccuracy in this type of sampling are 
consequently multi-photon emission as well as partial distinguishability of our
photon sources.

In fact, limitations of single-photon sources to date still constitute a key limitation in the way of  large-scale implementations of Fock boson sampling experiments.
Postselection is made use of to ensure that higher photon numbers
which are intrinsically also produced in the process do not contribute substantially.  
To build trust in the functioning of the device,
the measured relative frequencies of outcomes in which the photons are detected in
distinct modes are compared with expected numbers. 
This is possible as up to these system sizes, the relevant probabilities can still be classically computed. 

The experiment of \textcite{crespi_integrated_2013} also shows three-photon interference in an integrated interferometer, here involving $m=5$ optical modes. Similarly, \textcite{tillmann_experimental_2013}
present data from a three photon in a $m=5$ mode integrated optical interferometer.
In each case, single photons were created using parametric down-conversion.
\textcite{broome_photonic_2013} perform boson sampling in a tunable architecture on $m = 6$ modes with  $n = 2$ and $n = 3$ photons. 
Here, polarization controllers at the inputs and outputs can be used to perform different unitary evolutions.

The next step in implementation sized up the instances slightly to $n=3$ photons in $m=9$ modes \cite{carolan_experimental_2014,spagnolo_efficient_2014}. 
More significantly, both \textcite{carolan_experimental_2014} and \textcite{spagnolo_efficient_2014} perform the efficient state discrimination test proposed by \textcite{aaronson_bosonsampling_2013} in order to distinguish the experimental samples from a uniform distribution. 
\textcite{carolan_experimental_2014} furthermore distinguishes the samples from a distribution obtained if the bosons were distinguishable, making use of a technique called \emph{bosonic clouding}.
More recently, \textcite{giordani_experimental_2018} experimentally demonstrated a way to efficiently witness multi-photon interference in a $n=3$ photon experiment, following a proposal of \textcite{walschaers_statistical_2016}. 

These small-scale experimental results have more recently been brought to a new level in terms of large-scale
photonic implementations. 
This advance has been made possible by substantial technological development \cite{wang_high-efficiency_2017,loredo_boson_2017}. 
On the one hand, solid-state sources of highly efficient, pure, and indistinguishable single photons have been developed. 
Such quantum dot-micropillar systems allow for the deterministic generation of indistinguishable single photons with high sample rate.  
On the other hand, the transmissivity of the linear-optical circuits has been dramatically improved with the development of ultralow-loss optical circuits. 
These developments allowed \textcite{wang_high-efficiency_2017} to implement a $n=5$-photon, $m=9$-mode boson sampler with high sample rate. 
Improving those components even further, importantly, by integrating the optical circuit in a three-dimensional architecture, \textcite{PanSampling} have performed a boson-sampling experiment with $n=20$ photons and $m=60$ modes---the largest implementation of Fock boson sampling to date. 
For a detailed discussion of the early photonic implementations of boson sampling, we refer the reader to the review of \textcite{brod_photonic_2019}.

\subsubsection{Gaussian boson sampling}
\label{subsec:experimental gbs}

Gaussian boson sampling allows for even larger system sizes, given the comparably 
easier availability of suitable sources.
Recall that in Gaussian boson sampling, single-mode squeezed states are prepared at the input, whereas in Fock boson sampling, single-qubit Fock states need to be prepared---a much more challenging task. 

After early demonstrations  of the so-called scattershot boson sampling variant of GBS \cite{bentivegna_experimental_2015,zhong_12-photon_2018,paesani_generation_2019}, 
\textcite{zhong_quantum_2020} performed a large-scale GBS experiment that involved $50$ input single-mode squeezed states featuring high indistinguishability and squeezing parameters. 
The resource states are fed into a large-scale bulk optical (and hence not integrated) interferometer 
with full connectivity among $m=100$ modes that implements a random transformation with low loss. 
Notably, some randomness in this interferometer is physical: the interferometer is fabricated to implement a certain unitary transformation but imperfections of the process alter the targeted unitary.
To obtain an accurate description of the unitary, the interferometer is characterized post-hoc via tomography. 
Strictly speaking, the boson sampling device used in this (and all previous) experiment is therefore not a programmable device. 
Rather, it is designed to implement a specific transformation which is slightly altered in the fabrication process. 
The output of the interferometer is then sampled from making use of high-efficiency single-photon detectors. In this experiment, up to
$n= 76$ output photon-clicks have been detected. 

This scheme has yet been improved by \textcite{zhong_phase-programmable_2021} in two ways.
First, a restricted programmability of the boson sampling device has been achieved by making use of the capacity to vary the phase of the input squeezed states. This can also be viewed as introducing programmable phases in the random unitary transformation. 
Moreover, the experiment has been pushed further to detecting
$n=113$ photon events at the output of a photonic circuit comprising $m= 144$ optical modes. 
Key to this latter improvement is the availability of a high-brightness and scalable quantum light source that has been developed for this purpose. 
This source builds on methods of the stimulated emission of squeezed photons, which  are improved to achieve near-unity purity and high efficiency. 

In principle, such experiments can be efficiently verified in their functioning using quantum measurements \cite{chabaud_efficient_2021}, see \cref{ssec:fidelity witness}. 
While such tests have been performed in this experiment, subsystem properties---namely, low-order mode marginals---were used by \textcite{zhong_phase-programmable_2021} to efficiently distinguish from classically simulable distributions such as distinguishable photons and thermal states, see \cref{subsec:state discrimination}. 
To this end, they use a variant of Bayesian likelihood ratio estimators, which can be recast as a ratio of cross-entropy scores. 
In a similar vein, \textcite{drummond_simulating_2022} found good agreement between the distribution of total number of clicks of the treshold detectors observed by \textcite{zhong_phase-programmable_2021} with the theoretical click number distribution, including some decoherence effects. 

Very recently, \textcite{madsen_quantum_2022} have performed Gaussian boson sampling using time multiplexing in order to implement low-depth, but high-dimensional unitary mode transformations, as proposed by 
\textcite{deshpande_quantum_2022}.
The lower depth of the unitary transformation allows reaching larger system sizes since the loss does not contribute as much.
At the same time, classical simulation may become easier, but \textcite{deshpande_quantum_2022} provide numerical evidence that low-depth, high-dimensional transformations remain computationally intractable in practice.
The experiment uses $m=216$ single-mode squeezed input states, a linear-optical transformation with three-dimensional connectivity, and photon-number resolving detectors.
The average number of detected photons is 125. 
In order to benchmark the experiment, \textcite{madsen_quantum_2022} apply a number of tests. 
For the events  with very low photon number of $n\leq 6$ and $m=16$, they compute the TVD between the experimental and the target distribution. 
In the intermediate regime of photon numbers $n\leq 26$ and $m = 216$ modes, they estimate the cross-entropy difference as well as the Bayesian estimator of \textcite{zhong_phase-programmable_2021} in order to compare to potential classical spoofing algorithms as discussed in \cref{subsec:state discrimination}. 
Finally, in the classically intractable regime, they compute first and second order cumulants of the experimental distribution. 

Let us close this section by mentioning that a variant of the original scheme of Gaussian boson sampling has been implemented by \textcite{thekkadath_experimental_2022} on an interferometer comprising $m=15$ modes.
This scheme allows for shifts of the input squeezed states in phase space. 
Such displacements are useful when anticipating applications of Gaussian
boson sampling as sketched in \cref{sec:perspectives}.
A direct implementation of a scheme of approximating vibronic spectroscopy 
with imperfect quantum optics as a variant of boson sampling has been reported by
\textcite{clements_approximating_2018}.

\subsection{Further implementations of quantum random sampling}
\label{sec:further implementations}

The schemes of quantum random sampling discussed above are by far the most common
schemes that have been implemented experimentally. 
That said, other platforms different from superconducting or photonic architectures have also been considered, 
sometimes even leading to an actual
experimental realization. 
\textcite{wang_efficient_2020}, discussed also below, suggests to overcome the challenge of preparing and detecting bosonic quantum states in photonic implementations and implements a boson sampling protocol in a two-mode superconducting device, deviating from the common implementations of boson sampling on photonic platforms. 
This is used for simulating molecular vibronic spectra as suggested by
\textcite{huh_boson_2015}.

Quantum random sampling in the measurement-based model of quantum computing has recently been demonstrated on small scales by \textcite{IonSampling}. 
The advantage of this approach over gate-based circuits is that---in principle---significantly less device control is required. 
This is because all entangling gates are fixed and can be applied in a single layer, and only a single layer of random $Z$-type rotations are required \cite{bermejo-vega_architectures_2018,haferkamp_closing_2020}. 
The trade-off of this approach compared to a gate-based one is therefore one between depth of the circuit and space: 
in order to achieve a hard-to-simulate circuit comparable to the one of \textcite{arute_quantum_2019}, presumably $2\,500-10\,000$ qubits are required. 
\textcite{IonSampling} make this trade-off explicit: 
By `recycling'---i.e., measuring and re-preparing---certain qubits during the computation while keeping the remaining qubits coherent, depth of the physically implemented circuit can be traded with the number of qubits available in the device. 
A major advantage of the measurement-based approach to quantum random sampling is that it is possible to efficiently witness and measure the quantum fidelity using single-qubit measurements as discussed in \cref{ssec:fidelity witness,sssec:fidelity estimation} \cite{bermejo-vega_architectures_2018,hangleiter_direct_2017,hangleiter_sampling_2021}. 
This allows to perform the benchmarking and verification methods discussed in \cref{ssec:xeb} using the quantum fidelity, and thereby circumvent important caveats of the XEB fidelity. 

Along similar lines, to lessen the burden of actually explicitly implementing random circuits in a gate-based approach, a number of schemes have been suggested that would in effect give rise to such circuits, but based on physical interaction mechanisms. 
For example, \textcite{Deutsch}
considers the complexity of a probability distribution associated with an ensemble of non-interacting massive bosons undergoing a quantum random walk on a one-dimensional lattice. 
Such settings are potentially more feasible to implement in cold atomic systems.
In fact, the coherent cold collisions that have already been experimentally
implemented \cite{ColdCollisions} in systems of neutral ultra-cold atoms in optical lattices gives rise to precisely the interaction required for the implementation of the scheme of \textcite{bermejo-vega_architectures_2018,haferkamp_closing_2020}, which allows for efficient quantum verification.


\section{Classically simulating quantum random sampling schemes}
\label{sec:classical simulation}

Random quantum sampling schemes are set up to showcase the computational power of quantum devices, to
demonstrate that there are computational advantages of paradigmatic quantum computers over classical computers.
The rigorous statements discussed in \cref{sec:hardness} always involve a 
separation in the \emph{scaling} of classical versus quantum computations. 
Such statements show that, as systems are scaled up, the speed of the respective quantum computations will at some point certainly surpass that of every classical algorithm. 
But how large does one actually have to make a quantum sampler such that it cannot be simulated classically? 
In other words, what is the \emph{finite-size} behavior of the complexity of simulating quantum random sampling?

This question can only explicitly be answered for specific classical algorithms at a time.\footnote{Alternatively, one can invoke fine-grained complexity assumptions as discussed in \cref{ssec:fine grained hardness}.} 
The effort of devising such specific algorithms constitutes 
a crucial part in the quest of demonstrating a quantum advantage and thereby violating the extended Church-Turing thesis: 
One has to not only demonstrate that the scaling is possible in principle,  
but also that the frontier determined by the best available classical algorithm run 
on the fastest available supercomputers can be surpassed using actual quantum devices.

We can conceive of this situation as a competition between classical algorithms with an unfavourable scaling of the complexity, but run on extremely large supercomputers, and small but extremely noisy quantum devices. 
In the absence of quantum error correction, both competitors will hit a ceiling sooner or later, and the competition between classical and quantum devices is determined by which ceiling is more favourable: 
Roughly speaking, the quantum device, which is constrained by the noise present in current-day experiments will hit the simulation barrier as the circuit size reaches the tolerated error divided by the local gate error. 
Conversely, the classical algorithm, which is constrained by the scaling of the simulation task will hit a barrier once the time or space complexity reaches the tolerable limit determined by the speed and memory size of current-day supercomputers. 

Of course, what is and what is not possible in this situation depends heavily on the precise setting considered: 
is the goal to exactly simulate the sampling task, to sample from a distribution close in TVD, or to simulate a quantum experiment while including realistic amounts and sources of noise? 
Or is it to score at least as high as the quantum device on a given benchmark, potentially via other means than simulating the sampling task?  
Depending on the task at hand, a classical simulation algorithm may be able to exploit weaknesses in the benchmark, or optimally exploit the available time and space resources to beat the performance of a noisy quantum device.\footnote{A quantitative analysis of the competing scalings for the task of sampling from the exact or noisy distribution as measured by the linear XEB fidelity has been made by \textcite{zlokapa_boundaries_2020}.}

In this section, we provide a brief overview of classical simulation algorithms for different tasks related to quantum random sampling. 
We categorize those tasks into two categories: 
first, 
computing the output probabilities---a task that inevitably requires exponential precision for random instances as almost all probabilities are exponentially small (recall \cref{ssec:hardness of verification})---and second, simulating the sampling task.  
Computing the output probabilities is first and foremost required as a subroutine of most sampling algorithms, and also for the estimation of the XEB fidelity of an experimental system. 
In contrast, the goals of simulating the sampling task are manifold: 
the goal can be to sample from the ideal output distribution or a distribution close to that, it can be to simulate a noisy quantum experiment as well as possible, or it may be to achieve high scores on a given quantum advantage benchmark such as the XEB fidelity.

\subsection{Sampling versus computing output probabilities}
\label{sec:computing probabilities}

Computing the output probabilities of a (random) quantum computation, or \emph{strongly simulating} it, involves computing the output amplitudes of a quantum circuit.  
On a high level, most classical  algorithms for computing probabilities can be broadly 
categorized into ``Feynman'' type algorithms and ``Schr\"odinger'' type algorithms \cite{aaronson_complexity-theoretic_2017}.
Consider a quantum circuit with $m$ gates acting on $n$ qubits.
A Schr\"odinger algorithm stores and consecutively updates the entire state using $\sim 2^n$ space and $\sim m 2^n$ time.
A Feynman algorithm, in contrast, makes use of a path-integral formulation of the output amplitudes (recall \cref{eq:path integral unitary time evolution}) that expresses them as a sum of $\sim 4^{m}$ many products of $m$ matrix entries of the quantum gates in the circuit. 
Such an algorithm computes each term and sums all term up consecutively, and therefore requires merely $\sim m +n$ space.  
In fact, \textcite{aaronson_complexity-theoretic_2017} show that Feynman algorithms can have a much reduced runtime for local circuits that can be decomposed into $d$ layers of $m/d$ gates, by recursively computing sums over paths over portions of the circuit. 
This gives rise to a runtime scaling of $O(n (2d)^{n+1})$ for general circuits and $2^{O(d\sqrt n)}$ for circuits on a two-dimensional grid, while the space consumption scales as $n\log n$. 
Typically $m \gg n$ and hence, depending on the setting at hand, space or time may be the limiting factor and determine the choice of simulation algorithm. 

Of course, in practice, more intricate algorithms are used, but the basic idea remains often the same. 
For qubit-based architectures, most importantly, universal random circuits, hybrid Schr\"odinger-Feynman type algorithms turn out to be most efficient in practice. 
The most important tool here are so-called tensor-network algorithms (see \cite{bridgeman_hand-waving_2017} for an introduction), which allow the exploitation of locality structure in quantum circuits. 
For boson sampling schemes, Feynman algorithms are natural since the output probabilities are expressed in terms of matrix polynomials in  the entries of an $n \times n$ linear-optical unitary, albeit with exponentially many terms. 
Locality can (in most instances) not be meaningfully exploited for those systems. 

Let us now turn to the task of sampling from the output distribution of a quantum circuit, or \emph{weakly simulating} it. 
Computing the probabilities is not sufficient for sampling from a given distribution, and in fact not even necessary, however.
Having said that, computing the output probabilities is often the key subroutine of sampling algorithms, and all methods for sampling that we are aware of make use of that subroutine. 
Let us sketch the most important ideas for how to sample from a given distribution that are used in simulations of quantum random sampling. 

First, there are \emph{ancestral sampling} techniques. Here, the idea is that in order to sample from a multivariate distribution, say a distribution $p$ over length-$n$ bit strings with probabilities $p(x_1, \ldots, x_n)$, we can iteratively sample from the marginal distributions of larger and larger portions of the bit string. 
In the first step of such an algorithm we sample a bit $y_1$ from the marginal distribution $p_1 = \sum_{x_2, \ldots,x_n} p(\,\cdot\,, x_2, \ldots, x_n)$, in the second step we sample from the conditional distribution $p(\,\cdot\, | y_1)$, etc. 
The key obstacle to notice for this approach is that it requires an algorithm not only for individual probabilities but also for all marginals of the distribution, a potentially considerably more difficult task as it na\"ively requires summing over exponentially many probabilities. 

Second, there are \emph{rejection-sampling} techniques. 
The idea of rejection sampling is to generate a sample $y$ from a distribution $q$, as well as a uniformly random number $u \in [0,1]$ in the first step.  
The distribution $q$ should be such that we can efficiently sample from it and it must satisfy $p(x) \leq c q(x) $ for some number $c$ and all $x$.
In the second step, the sample $x$ is accepted if $ucq(x) \leq p(x)$ and rejected otherwise. 
If it is rejected, the procedure is repeated. 
The expected number of probabilities that need to be computed per sample is given by $c$.  
Rejection sampling has a natural geometrical intuition: 
Suppose that $q$ is the uniform distribution over length-$n$ bit strings and $c =2^n$. 
Then we sample uniformly random points in the rectangle $\{0,1\}^n \times [0,1]$ and accept a sample if it lies within the histogram of the distribution $p$. 

Then, there are so-called \emph{Markov-chain Monte Carlo} techniques. 
Here, the idea is to set up a Markov chain of bit strings $x_1 \rightarrow x_2 \rightarrow \cdots \rightarrow x_m$ that converges to the target distribution $p$ as its stationary distribution. 
This Markov chain is specified by the probability $P_t(x)$ of being in state $x$ at time step $t$, and rates $W_{x\rightarrow x'}$ for the transition $x\rightarrow x'$ that determine the probability of moving from state $x$ to state $x'  $.
The overall idea is to construct the Markov chain based on a proposal distribution $q$. 
The proposal distribution determines the probability $q(x'|x)$ of moving to state $x'$ given that the Markov chain is in state $x$. 
The transition probabilities are then given by 
\begin{align}
  W_{x \rightarrow x'}  = \Pr[\texttt{Accept}|(x'|x)] q(x'|x) . 
\end{align}
A simple choice of the acceptance probability is the Metropolis choice
\begin{align}
  \label{eq:metropolis acceptance}
  \Pr[\texttt{Accept}|(x'|x)] &= \min\left\{ \frac{p(x')q(x|x')}{p(x) q(x'|x)} , 1 \right\} \, . 
\end{align}
This choice has the favourable property that it only depends on the ratio $p(x')/p(x)$. 
This means that one need not be able to compute those probabilities directly but only a function $f \propto p$. 

\subsection{Simulating universal circuit sampling}

The best studied family of quantum circuits are universal random circuits, and in particular, the circuits implemented by \textcite{arute_quantum_2019} and subsequently by \textcite{wu_strong_2021,zhu_quantum_2022}.
Recall that these circuits comprise single-qubit gates $\sqrt X, \sqrt Y , \sqrt W$ and the two-qubit gate $\text{iSWAP}^* = \text{fSim}(\pi/2, \pi/6)$.
The goal of large-scale simulations performed for this task has been to compare the performance of inefficient classical algorithms potentially including approximations and the noisy quantum devices in the lab. 
The methods devised for this task have similar complexity for the task of computing the probabilities and simulating the experiment since amplitude estimation dominates the computational cost.
Nonetheless, the number of amplitudes required for sampling typically scales linearly in the number of samples and hence producing millions of samples can be prohibitively costly while computing a single amplitude is doable.

The figure of merit in terms of which the success of these simulations is measured is 
either the fidelity of the classical representation of an approximate quantum state in cases in which such a representation exists, or the XEB fidelity of the produced classical samples as a classical benchmark that acts as a placeholder for the circuit fidelity. 

\subsubsection{Using tensor networks to simulate quantum circuits}

\begin{figure}
  \includegraphics{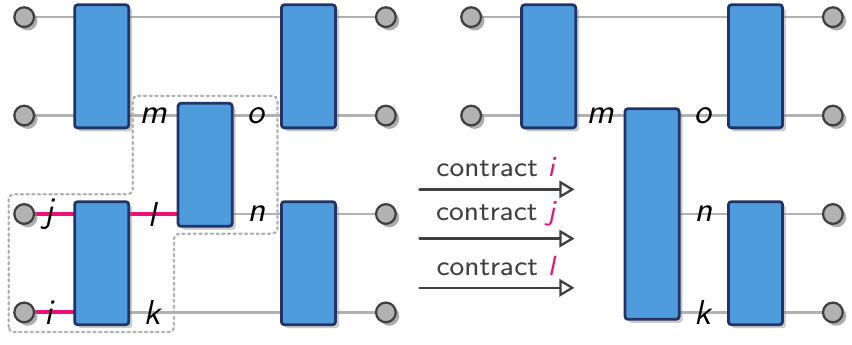}
  \caption{\label{fig:tensor network}
  In a \emph{tensor network} every edge corresponds to a rule to sum over the corresponding index of the neighbouring tensors. 
  In a quantum circuit two-qubit gates are represented as four-index tensors (boxes) and single-qubit computational basis states (vertices) are single index tensors or vectors. 
  Contracting an edge with a neighbouring computational basis state (indices $i$ and $j$) corresponds to selecting a slice of the neighbouring tensor. 
  Contracting an edge between two arbitrary tensors (index $l$) corresponds to summing over the entries of the neighbouring tensors over that index, resulting in a new, larger tensor. 
  }
\end{figure}

The most important tool for the simulation of universal random circuits are \emph{tensor networks} \cite{markov_simulating_2008,boixo_simulation_2017}. 
The basic idea of a tensor network is to express a quantity of interest in terms of a network of multi-index tensors in which the edges correspond to a prescription to sum over the corresponding index. 
Amplitudes of quantum circuits are therefore naturally tensor networks since two-qubit gates are rank-four tensors, single-qubit gates are rank-two tensors and a product state is just a product of vectors (rank-one tensors). The circuit description is just a rule specifying how to connect those tensors. 
In order to compute the quantity of interest, one then needs to contract the tensors across their edges, i.e., perform tensor multiplication by summing over the corresponding index, see \cref{fig:tensor network}.
The contraction complexity is determined by the largest dimension of an index that appears in a particular contraction scheme, which is roughly determined by the treewidth of the underlying graph~\cite{markov_simulating_2008}.

While the properties of one-dimensional efficient tensor networks can be computed efficiently in the dimensions and size of the tensor network, this no longer holds generally true for
higher-dimensional geometries that do not admit a linear contraction scheme ~\cite{schuch_computational_2007,haferkamp_contracting_2020}. 
Nonetheless, it often remains possible to find contraction schemes that scale much better than the worst-case runtime in practice. 

Tensor networks admit various sampling algorithms. 
Very naturally, one can make use of ancestral sampling because the data structure of a tensor network naturally admits the computation of marginals at a cost that is similar to the cost of computing an individual output amplitude 
\cite{PhysRevB.85.165146}. 
Still, this method is rather costly since every sample requires $n$ different contractions of the circuit tensor network. 

\begin{figure}
  \includegraphics{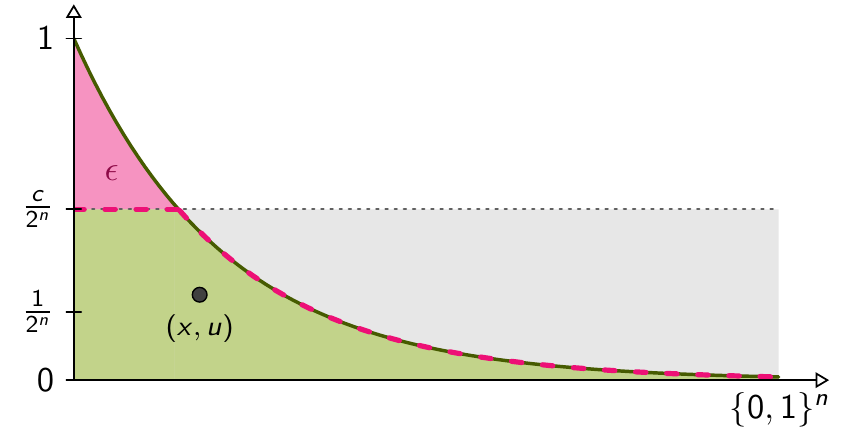}
  \caption{\label{fig:frugal rejection}
  In frugal rejection sampling, we sample a point $(x,u)$ uniformly at random over the area $\{0,1\}^n \times [0,c/2^n]$. 
  A sample is accepted if $u\leq p(x)$ (green area) and rejected otherwise. 
  For Porter-Thomas (exponentially) distributed outcome probabilities (solid (green) line), this will result in a TVD error $\epsilon = 2 \exp(-c/(1-e^{-c}))$ of the actually sampled distribution (dashed (pink) line) compared to the target distribution. 
  }
\end{figure}

For the output distributions of random universal circuits it turns out that variants of rejection sampling are much more efficient however. 
This is because the output distribution of random universal circuits is exponentially (Porter-Thomas) distributed, which implies that the largest probability is exponentially small with inverse polynomial failure probability over the choice of the random circuit (recall \cref{eq:bounds haar random probabilities}). 
Choosing the uniform proposal distribution and the bound $c=\log(2^n/\epsilon)$ in the rejection sampling algorithm (see \cref{sec:computing probabilities}) one can therefore simulate Porter-Thomas distributed probability distributions for $n=49$ up to error $\epsilon=10^{-3}$ using $41$ probabilities per bit string on average \cite{markov_quantum_2018}.
Further improving this, 
\textcite{markov_quantum_2018} introduce a ``frugal'' sampling scheme that reduces the fraction of rejected strings. 
To do so, frugal rejection sampling chooses $c$ such that the upper tail of the distribution with probabilities $> c/2^n$ has fixed weight $\epsilon$ and accepts all proposed strings $x_j$ with unit probability if their probability is larger than $2^n p(x_j)/c$, see \cref{fig:frugal rejection}. 
This effectively reduces the probability of such outcomes to $c/2^n$ and improves the average number of probabilities required per sample and makes them independent of $n$. 
At the same time, it introduces an error of the sampled distribution compared to the target distribution. 
Quantitatively, this error is given by $2 \exp(-c/(1-e^{-c}))$ as measured by the TVD of the sampled distribution to the ideal one assuming exponentially distributed probabilities. 
For instance, for $c=10$, it is given by $\sim 10^{-4}$. 

A further important advantage of rejection sampling over ancestral sampling is that all probabilities can be pre-computed. 
This is crucial because it allows a more efficient use of the contractions of a tensor-network. 
Instead of contracting a new tensor network for each amplitude, the desired output strings are stored in a large tensor so that only a single, albeit slightly more complex, contraction is required for the entire batch.
For instance in the algorithm of \textcite{markov_quantum_2018} saving $10^7$ amplitudes instead of a single one leads to only a $2.76$-fold slowdown of the simulation.

\subsubsection{Simulation of random quantum circuits}

The question of how to approximately simulate random quantum circuits thus boils down to the question how to best contract the tensor network representing the quantum circuit. 
In the following, we discuss various techniques for that in more detail.

\paragraph{State-vector simulation.}
The in some sense simplest algorithms for the simulation of quantum circuits just store the entire quantum state and time evolve that state
\cite{de_raedt_massively_2007,smelyanskiy_qhipster_2016,haner_05_2017,de_raedt_massively_2019,pednault_leveraging_2019}.
Here, a key challenge is to exploit all the available storage on a large computer. 
For this, the state-vector simulation needs to be distributed among the different parts of the storage. 
To our knowledge, the largest such simulation runs on $49$ ($7 \times 7 )$ qubits \cite{li_quantum_2018,pednault_pareto-efficient_2017}. 

An alternative and arguably more natural way of storing multipartite quantum states is given by a tensor network as discussed above.  
In order to compute all amplitudes of the output state of the quantum circuit, one can contract the tensor network along the time dimension, giving rise to a tensor-network representation of the output state.
The description complexity of a tensor network state is bounded by $\sim n 2^n$, while the simulation time scales as $\sim m 2^n$ in the worst case. 
This approach has been pursued by a number of works \cite{guo_general-purpose_2019,pan_contracting_2020,zhou_what_2020-1,mccaskey_validating_2018}, building on work in the simulation of quantum many-body systems \cite{schollwock_density-matrix_2005,VerstraeteBig}. 
While tensor networks can efficiently  approximate states with low entanglement \cite{schollwock_density-matrix_2005}, 
this is not the case for random quantum circuits which have high entanglement by construction.

Indeed, an important feature of tensor network algorithms is that they allow for a natural way of relaxing the precision of the simulation. 
When contracting a two-qubit gate into a two-site tensor network, the dimension of the tensors are multiplied. 
In order to keep the storage effort constant, the usual approach is to perform a singular-value decomposition (SVD) of the new, larger tensor and then to truncate the smallest singular values. 
Thus, the tensor size is kept fixed. 
For quantum states with low entanglement the singular value distribution will be nontrivial, allowing for an efficient approximation scheme. 
For random quantum circuits, however, the singular value distribution tends to be very flat so that a reduction of the bond dimension results in large errors \cite{guo_general-purpose_2019,markov_simulating_2008}. 

The introduced error rate due to such truncation can be viewed as analogous to a finite gate fidelity in a real quantum quantum circuit.
Such a sequential compression has been pursued for one and two-dimensional
quantum circuits by \textcite{zhou_what_2020-1}.
Using this approach, fidelities of the output state on the same order of magnitude as seen in the experiment by \textcite{arute_quantum_2019} can be reached for a
two-dimensional circuit with $CZ$ entangling gates acting on $54$ qubits.
These simulations could be carried out on a laptop in a few hours. 
For 20 qubits, the linear XEB fidelity of the resulting state classically can be computed using the exact probabilities that are obtained from an untruncated tensor network contraction.
Note that this approach does not (yet) achieve the advantage regime of $F_\xeb \approx 0.002$ for the iSWAP$^*$ entangling gate which is considerably more difficult to simulate. 
\textcite{zhou_what_2020-1} estimate that this would require a bond dimension of roughly $10^4$ which is an order of magnitude above what is needed for the $CZ$ gate.

Even for algorithms that do not involve approximations, a clever choice of contraction order can yield better runtimes, however. 
For instance, \textcite{guo_general-purpose_2019} provide a simulator for quantum circuits  acting on a two-dimensional lattice based on specific contraction strategies of the tensor-network representation of the quantum state after the circuit has been applied. 
For a lattice of side length $L$ and a circuit of depth $d$ their most generic contraction scheme achieves space and time complexities of the resulting algorithm that scale as $2^{d(L+1)/8 }$ and $L^2 2^{d(L+1)/8 }$, respectively. 
This allows for the computation of a single output probability of a random quantum circuit with $CZ$ entangling gates of depth $26$ on $10\times 10$ qubits on a supercomputer in 9 minutes and a circuit of depth $40$ on $7 \times 7$ qubits in 31 minutes and 92.51 TB memory usage.

\paragraph{Hybrid algorithms.}
Implementing the idea of \textcite{aaronson_complexity-theoretic_2017} to balance memory consumption and computation time in a Schr\"odinger-Feynman hybrid algorithm, \textcite{chen_64-qubit_2018,li_quantum_2018,chen_classical_2018-1} introduced ``slicing algorithms'' in which the system is sliced into smaller subcircuits that are independently simulated. 
Every time an entangling gate occurs between those sub-circuits the number of independent circuits to be simulated is multiplied by the Schmidt (product) rank of the entangling gate. 
By judiciously choosing the slices, one can thus optimally balance the memory consumption and computation time.

Importantly, all of the simulations above used $CZ$ entangling gates. 
In the universal circuit sampling experiments \cite{arute_quantum_2019,wu_strong_2021,zhu_quantum_2022}, the entangling gates are ones that are close to the iSWAP$^*$ gate, however.
This gate is significantly more challenging to simulate. 
This is because while the $CZ$ gate can be decomposed into a sum of two equally weighted product operators as $CZ = \proj 0 \otimes \id + \proj 1 \otimes Z$, the iSWAP$^*$ gate saturates the decomposition rank of four with equal magnitude weights as 
\begin{multline}
  \text{iSWAP}^* = \proj 0 \otimes \proj 0 + \ee^{-\ii \pi/6} \proj 1 \otimes \proj 1 \\- i \ket 0 \bra 1 \otimes \ket 1 \bra 0 -  i \ket 1 \bra 0 \otimes \ket 0 \bra 1 . 
\end{multline}
Roughly speaking, the effort of simulating circuits including  iSWAP-like gates will therefore be quadratically larger in the number of gates across partitions of the circuit compared to circuits with $CZ$ entangling operations. 

\textcite{markov_quantum_2018} exploit such a decomposition to match a given fidelity in a classical simulation. 
This makes use of the observation that all two-qubit gate paths have equal weight in absolute value, while the remainder of the circuit is chaotic, meaning that different paths contribute roughly equally to the final amplitude \cite{villalonga_establishing_2020}.
This implies that one may just estimate an output probability to a given fidelity by summing over a fraction of the paths given by the fidelity. 
This allows the simulator to produce a (correlated) sample of $M$ bit strings with target fidelity $f$ at the same cost as computing $f \cdot M$ noiseless amplitudes.

\textcite{villalonga_flexible_2019} further show that yet faster sampling can be achieved by ``recycling'' an initial tensor contraction to obtain contractions for nearby bit strings. 
The resulting simulation algorithm has been executed on one of the fastest supercomputers available to simulate with fidelity 0.5\% depth-$40$ $7 \times 7 $ random circuits with $CZ$ entangling gates in 2.44 h, and depth-$24$ $11 \times 11$ circuits in 0.28 h \cite{villalonga_establishing_2020}.

A number of works aim at finding the optimal way of contracting the corresponding tensor networks by finding good contraction paths that keep the tensors relatively small \cite{chen_classical_2018-1,chen_64-qubit_2018,guo_general-purpose_2019,huang_classical_2020,gray_hyper-optimized_2021,guo_verifying_2021,schutski_simple_2020,pan_contracting_2020}.
An approach that is closely related to tensor-network contraction has been introduced by \textcite{boixo_simulation_2017}. 
This approach makes use of undirected graphical models which are probabilistic models for which a graph expresses the conditional dependence structure between random variables \cite{Barber}.\footnote{As such they are closely related to tensor networks \cite{Expressive}.}
The key idea of this approach is the following:
 When representing  
the quantum circuit by a product of unitary
matrices acting at different clock cycles, 
expressions for probabilities can be viewed as a path integral with individual paths formed by a sequence of the computational
basis states. 
The dependencies can then be cast into the form of a probabilistic graphical model,
except that in contrast with actual probabilistic models,
the factors in general take complex values. 
To evaluate the resulting expressions, a new variant of a variable elimination algorithm \cite{Murphy} has been suggested. 
This algorithm allows to sample from the output distribution of circuits featuring a sufficiently small tree width, as well as to estimate the XEB benchmark.
\textcite{chen_classical_2018-1,huang_efficient_2021-1} have further improved this approach and combined it with tensor-network contraction techniques for an application in a parallelized architecture.

Other interesting variants of circuit contraction schemes have been proposed by \textcite{chen_quantum-teleportation-inspired_2020}, inspired by quantum teleportation, to swap space and time in order to take advantage of low-depth quantum circuits.
Finally, \textcite{kalachev_recursive_2021} devise a ``multi-tensor contraction scheme'' in which the tensor network contraction is performed by assigning a so-called \emph{contraction tree} with a recursive relation. 
In this relation, certain pre-computed sub-expressions are re-used as often as possible to speed up the overall computation.
In this way, they are able to compute individual probabilities of Sycamore circuits of depth up to 16.

\begin{table*}
\centering
  \begin{tabular}{l >{\quad }c >{\quad }c>{\quad }c<{\quad}  c  >{\quad}c }
  \toprule
    & Depth & \# bit strings & & Time complexity (\# FLOPs) & Space complexity \\
  \midrule  
  \cite{kalachev_recursive_2021} & $16$ &$2 \cdot 10^6 $ & uncorrelated & $ 1.1\cdot 10^{19}$ & ? \\
  \cite{gray_hyper-optimized_2021} & $20 $ &$1$ & uncorrelated & $ 3.1 \cdot 10^{22}$ & $2^{27}$\\
  \cite{huang_classical_2020} & $20 $ &$64$ & uncorrelated& $ 6.7 \cdot 10^{18}$ & $2^{29}$\\
  \cite{pan_simulation_2022} & $20 $ &$2 \cdot 10^6 $ & correlated &  $ 4.5\cdot 10^{18}$ & $2^{30}$\\
  \cite{pan_solving_2022} & $20$ &$2^{26} $ & uncorrelated \& correlated &  $ 3.5\cdot 10^{18}$ & $2^{30}$\\
  \cite{kalachev_classical_2021} & $20$ & $2^{25}$ & correlated &  $6.9 \cdot 10^{18}$ & ?\\
     \bottomrule
  \end{tabular}
  \caption{
\label{tab:computational cost} Comparison of the time and space complexity of computing bit string probabilities of Sycamore circuits with 53 qubits for selected simulation schemes in terms of the total number of floating point operations (FLOPs).
Sources: \cite{pan_simulation_2022,kalachev_recursive_2021,pan_solving_2022,kalachev_classical_2021}.  }
\end{table*}

\paragraph{Simulating the experiment of \textcite{arute_quantum_2019}.}
The previously mentioned methods have been used to approximately simulate different random circuits or different sizes as compared to the ones performed by \textcite{arute_quantum_2019,wu_strong_2021,zhu_quantum_2022}. 
In order to fairly compare the noisy experiment with a classical algorithm it is necessary to perform the same task or at least fairly comparable tasks in the first place, however. 
Unfortunately, it is not fully clear what exactly that task should be. 
Ideally, the task on which we compare quantum and classical algorithms is to produce samples from the correct probability distribution. 
However, this point of view has the issue that it is not possible to verify the distribution. 
\textcite{arute_quantum_2019} seem to have precisely this in mind when they argue that the linear XEB fidelity is a placeholder for the quantum fidelity, and perform further tests to corroborate this such as computing the logarithmic cross entropy and estimate the entropy of the sampled distribution.
Alternatively, we could think that the task on which the experiment has to be beaten is merely to score high on the XEB benchmark. 
This interpretation has the advantage that there is a clear-cut benchmark, which, while not efficiently computable, can at least be sample-efficiently estimated (see \cref{ssec:xeb}) and is well defined. 

The latter approach is taken by \textcite{pan_simulation_2022}. 
They devise a tensor-contraction method which allows them to exactly compute a certain subset of the probabilities. 
The basic idea of their ``big head'' algorithm is to identify a bottleneck in the contraction of the tensor network and split the tensor network into two parts across that edge, which is close to the output. 
This gives rise to a very large ``head part'' of the network and a ``tail part'' of the network.
The output qubits in the head-part of the network are projected onto a fixed bit string $s_1 =(0, \ldots, 0)$.  
The tail part of the network is much smaller than the head part and contains a subset of the output qubits. 
Thus, the head part of the network need only be contracted once, while the output amplitude of a bit string $(s_1,s_2)$ can be computed as the inner product of the vectors corresponding to the contractions of the head and the tail part. 
In this way, \textcite{pan_simulation_2022} are able to obtain $2^{21}$ correlated bit strings, and postselecting onto the $10^6$ largest ones gives an XEB score of $0.736$.
Importantly the probabilities computed by the algorithm are  \emph{exact}. 
\cite{liu_closing_2021-1} reduce the runtime of the algorithm to a few minutes by making use of a large supercomputer.

Now, arguably, this method does not produce \emph{uncorrelated} or independent samples from the correct distribution and does therefore not achieve the \emph{sampling task} associated with the universal circuit. 
\textcite{liu_redefining_2021} make use of the  algorithm of \textcite{pan_simulation_2022} in order to produce perfect samples from the target distribution. To this end, they leave fewer legs of the tensor network open (6 instead of 21) and use the outcomes to produce a single perfect sample of depth-20 circuits in 276 seconds on a supercomputer. 
To produce many samples with smaller (XEB) fidelity, these perfect samples can be diluted by uniform samples as proposed by \textcite{huang_classical_2020}. 
Producing one million samples with XEB fidelity $0.2 \%$ is therefore equivalent to producing 2000 perfect samples. 

\textcite{pan_solving_2022} pursue a different strategy and achieve the approximate sampling task by artificially introducing approximations, and using a sparse representation of the output state. Specifically, they ``drill holes'' into the tensor network by judiciously removing a few of its edges at various positions in the circuit.
To achieve this, they remove $k$ pairs of edges of the iSWAP$^*$ gate, which allows them to significantly reduce the complexity while decreasing the fidelity of the state by roughly a factor of $2^{-2k}$.
Second, they compute the output probabilities associated to $L$ uniformly random groups of $l$ correlated bit strings. 
By removing $2k = 8 $ edges from the tensor network they are able to compute $2^{26} $ uncorrelated batches of correlated probabilities. 
Using those they obtain $2^{20}$ independent samples from a state with fidelity, or equivalently, XEB fidelity of $\approx 0.37 \%$.
The simulation has a cost about 15 hours on a small cluster of 512 GPUs. 
Importantly, the samples produced in this way pass the same tests that were performed by \textcite{arute_quantum_2019} to validate the experimental samples. 

An analogous approach is pursued by \textcite{kalachev_classical_2021}, who devise a slicing procedure based on maximizing the norm of partially summed slices to match a targeted fidelity.
Similarly to the approach of \textcite{pan_solving_2022}, this gives rise to batches of correlated probabilities. 
\textcite{kalachev_classical_2021} further provide an optimized sampling procedure that minimizes the sampling overhead in terms of how many probabilities need to be computed to a factor of $2$.
They estimate that this would allow them to sample from a distribution with fidelity $0.2 \%$ using 15 months time on a single GPU. 
By postselecting on the largest amplitudes in a number of batches, they are able to spoof the linear XEB benchmark with a value of $0.47 \%$ in four hours on a single GPU. 

In a similar spirit, and as discussed in more detail above, 
\textcite{barak_spoofing_2021,zhou_what_2020-1,gao_limitations_2021} provide evidence that weaknesses in the linear XEB fidelity can be exploited in order to devise classical algorithms with a score comparable to the score noisy quantum devices achieve.  
Specifically, the algorithm of \textcite{gao_limitations_2021} scores only one order of magnitude below the experiment of \textcite{arute_quantum_2019} using a laptop. 
It is projected to keep roughly constant score for larger circuits, while the experimental score is expected to further decrease exponentially.
At the same time, the \emph{quantum} fidelity of the quantum state from which those samples were produced is presumably exponentially small. 
It may therefore remain to be difficult to sample from the output distribution of a state with comparably high fidelity.  

In \cref{tab:computational cost} we compare the most advanced algorithms for (approximately) computing the output probabilities of Sycamore circuits. 
These algorithms are used as the crucial subroutines in algorithms that approximately sample from the output distribution of those circuits, and algorithms which perform the weaker task of outputting samples with a high XEB score. 

Summarizing the discussion above, it seems fair to say that the experiment of \textcite{arute_quantum_2019} has been simulated on conventional computers, probably most convincingly by \textcite{pan_solving_2022}. 
However, all of the simulation methods mentioned here fail already for the slightly larger implementation of universal circuit sampling by \textcite{zhu_quantum_2022}. 
Let us stress that our discussion once again highlights how difficult it is to fairly compare different spoofing strategies to experimental samples either of which can only be validated by incomplete methods such as cross-entropy benchmarking. 
For example, one might argue that the bit strings produced by \textcite{pan_simulation_2022} have already outperformed the experimental samples in terms of the relevant benchmark---the XEB fidelity, since this is the benchmark which \textcite{arute_quantum_2019} have decided on as the central quantity characterizing the quality of their experiment (granting that they did perform further benchmarks). 
But one could equally well argue that actually the samples of \textcite{arute_quantum_2019} are approximately sampled from the targeted distribution. 
In this reading, a high XEB fidelity is only one of many features that those samples should have. 
In addition, they should be independent samples, have a high entropy, even be sampled from the output distribution of a quantum state which has high fidelity with the ideal target state. 
Evidence for all of those features features was collected in the experiment of \textcite{arute_quantum_2019} by means of various tests. 
In this reading only the samples of \textcite{pan_solving_2022} can actually be said to ``reproduce'' the experiment---as viewed through those tests.
This discussion highlights the importance of clearly identifying and stating reproducible criteria under which we will consider quantum random sampling to be successfully achieved---on a classical or a quantum computer and in the absence of an unambiguous and efficient means of verifying samples. 

\paragraph{Efficient algorithms.}
While the above simulation algorithms have exponential runtime or result in large errors, \textcite{napp_efficient_2022} provide both numerical and analytical evidence that shallow (depth-$3$) universal circuits in a two-dimensional brickwork architecture can be strongly simulated as well as efficiently weakly simulated within a constant total-variation distance error. 
They do so in a twofold approach: 
first, they numerically demonstrate approximate simulation of random universal circuits in a $400 \times 400$ brickwork architecture  using a tensor-network algorithm (which is worst-case hard to simulate strongly). 
They then provide analytical evidence for easiness using a mapping to a recently developed model consisting of alternating rounds of random unitaries and weak measurements~\cite{bao_theory_2019,jian_measurement-induced_2019}, see also \cref{ssec:discussion average-case hardness}.

\paragraph{Alternative simulation schemes.}
Yet another approach of entirely different type is to make use of the so-called \emph{stabilizer decomposition}
of quantum states. 
This method is based on the observation that stabilizer states, that is, states generated by Clifford circuits can be efficiently simulated both weakly and strongly \cite{gottesman_stabilizer_1997}. 
Circuits that comprise additional non-Clifford gates, can then be expressed as linear combinations of Clifford circuits \cite{aaronson_improved_2004,bravyi_improved_2016,qassim_improved_2021,bennink_unbiased_2017-1}. 
The complexity of this scheme grows exponentially in the stabilizer rank $\chi$ of a quantum state, that is, the number of stabilizer states in this decomposition. 
Since the number of non-Clifford gates typically grows much faster than the number of qubits, this approach is currently not practically useful, however.  

To summarize, the above efforts let us conclude that using sophisticated classical algorithms,
modern supercomputers can just so keep track of existing experimental schemes of
universal circuit sampling.

\subsubsection{Analysis of noise}

Intuitively speaking, noise should render the simulation of quantum random sampling schemes less computationally demanding. 
In an idealized scenario,
if local depolarizing noise with constant strength is applied at the end of a quantum circuit, the output distribution will be very close to the uniform distribution.
However, often it is a priori unclear how to exploit specific types of noise in a particular simulation algorithm, and more specifically which noise levels will be classically simulable. 
It has therefore been a subject of some research to pin down regions---determined by the type and strength of noise---in which quantum random sampling schemes are efficiently simulable via classical algorithms. 
Conversely, one can ask the question whether it is possible to mitigate certain forms of noise without resorting to quantum error correction techniques. 

Early on \textcite{aharonov_polynomial_1996} have already considered the effect of depolarizing noise on the complexity of quantum circuit simulation. 
They find a polynomial-time algorithm for noisy circuits whenever the depolarizing fidelity is higher than some threshold.
More specifically to the case of quantum random sampling, \textcite{bremner_achieving_2017} show that IQP circuits subject to local depolarizing noise at the end of the circuit are classically simulable for any constant noise strength, provided the ideal distribution in question is sufficiently anticoncentrated. 
The key idea of their simulation scheme is to make use of a simulation algorithm based on a sparse Fourier representation of the output distribution for sufficiently  anticoncentrated IQP distributions \cite{schwarz_simulating_2013,kushilevitz_learning_1993}.  
For measurement depolarizing noise with strength $\epsilon$ and distributions with collision probability $\leq \alpha/2^n$ their algorithm runs in time $O(n^{\log(\alpha/\delta)/\epsilon})$ to sample from the target distribution up to TVD $\delta$. 
This simulation scheme can be further extended to universal circuits  \cite{yung_can_2017} using a measurement-based embedding \cite{gao_quantum_2017} and then exploiting the algorithm of \textcite{bremner_achieving_2017} on individual branches of that embedding. 
It may not be clear, however, to what extent the considered type of noise channel (local depolarizing noise at the end of the circuit) is actually realistic and reflective of common physical sources of quantum noise \cite{boixo_characterizing_2016,boixo_fourier_2017}. 
It has also been noted that, asymptotically, such Fourier-based simulation algorithms are no more efficient than trivial algorithms \cite{boixo_fourier_2017}.  
Nonetheless, there may well be an intermediate regime in which an advantage can be gained by exploiting the specific structure of the Fourier coefficients.

Following up on this, \textcite{gao_efficient_2018} have proven convergence to the uniform distribution for local Pauli noise associated with single-qubit gates.
This convergence result has been refined recently by \textcite{deshpande_tight_2022,dalzell_random_2021}, who further delineate the regime in which we expect classical simulation algorithms to be feasible; recall \cref{ssec:noisy distributions}.
From the result of \textcite{deshpande_tight_2022}, it follows that random with a constant amount of noise are efficiently simulatable up to inverse polynomial TVD---by the trivial algorithm which just outputs uniform samples---whenever the depth grows as $\omega(\log n)$ since in this case the TVD between the noisy output distribution and the uniform distribution is smaller than any inverse polynomial. 

\textcite{gao_efficient_2018} make a significant step forward from this and give an average-case simulation algorithm for the output distributions of universal quantum circuits with a noiseless Clifford part and Pauli noise on non-Clifford gates with noise strength $\eta$. 
The output distribution of these noisy circuits is nontrivial in that it is far from uniform yet simulable up to TVD error $\epsilon$ with a runtime of $n^{O(\log 1/\epsilon)/\eta}$ and hence efficient for constant $\epsilon, \eta>0$. 
This algorithm only runs in quasi-polynomial time if the goal is to simulate the noisy circuit up to inverse polynomial TVD. 
This regime is significant since an algorithm that can only simulate a noisy experimental circuit up to constant TVD can be efficiently distinguished from the actual noisy experiment at polynomial overhead.
Building on this algorithm of \textcite{gao_efficient_2018}, \textcite{aharonov_polynomial-time_2022} close this gap and find an algorithm that can efficiently simulate a noisy universal random circuit with constant local depolarizing noise after every gate up to any inverse polynomial TVD whenever the output distribution anticoncentrates. 
Since anticoncentration requires at least logarithmic depth \cite{dalzell_random_2020}, the algorithm of \textcite{aharonov_polynomial-time_2022} is therefore nontrivial precisely in the regime of logarithmic depth. 
Given previous results, this is the regime in which one might have hoped for an asymptotic quantum advantage even for quantum circuits with a constant amount of noise \cite{deshpande_tight_2022}; see our discussion of these results in \cref{ssec:noisy distributions}. 
These results hence show that random quantum circuits of logarithmic depth do not offer a 
``sweet spot'' at which anticoncentration already sets in and yet constant noise levels are not yet overwhelming.

Let us briefly sketch the idea of the algorithm of \textcite{gao_efficient_2018,aharonov_polynomial-time_2022}, which draws its key ideas from the work of \textcite{bremner_achieving_2017}. 
The starting observation of the algorithm is that the (ideal) output distribution of a quantum circuit $C = U_d U_{d-1} \cdots U_1$ can be expressed as a \emph{Pauli path integral}
\begin{align}
\label{eq:pauli path integral}
  P_C(x) = &\sum_{s_0, \ldots, s_d \in \mathsf{P}_n} \tr[\proj x s_d] \tr[s_d U_d s_{d-1}U_d^\dagger] \cdots  \nonumber\\
  &\qquad\cdots \tr[s_1 U_1 s_{0}U_1^\dagger] \tr[s_0 \proj {0^n}]\\
  \eqqcolon& \sum_{s \in \mathsf{P}_{n}^{d+1}} f(C,s,x),
\end{align}
where $\mathsf{P}_n$ is the $n$-qubit Pauli group. 
This can be easily seen from the fact that the Pauli matrices form a complete operator basis and therefore $\tr[U\rho U^\dagger s] =\sum_{t \in \mathsf{P}_n} \tr[Ut U^\dagger s]  \tr[\rho t]$.
We can also think of the Pauli path integral as a Fourier decomposition of the output probabilities. 

In the Fourier representation, the effect of local depolarizing noise can be easily analyzed since it just acts as $\mc E(\rho) = (1- \epsilon) \rho + \epsilon \tr[\rho] \id/2^n$. 
The contribution of a Pauli path of a noisy quantum circuit to the total output probability thus decays with the number of non-identity Pauli operators in it (the Hamming weight of $s$) as\footnote{This reflects an analogous expression derived by \textcite{bremner_achieving_2017} for noisy IQP circuits.}
\begin{align}
  \tilde p_C(x) = \sum_{s \in \mathsf{P}_{n}^{d+1}} (1- \epsilon)^{|s|} f(C,s,x). 
\end{align}
\textcite{aharonov_polynomial-time_2022} now show that the sum can be approximated by including only path weights $f(C,s,x)$ with Hamming weight $|s| \leq \ell$ incurring a TVD error on the order of $2^{- \Omega(\ell)}$ on average.
Then, they show that the truncated sum can be calculated efficiently using that the low-weight Pauli paths are sparse in that most of them actually have weight $0$, using ideas very similar to those of \textcite{bremner_achieving_2017} and \textcite{kushilevitz_learning_1993} on computing quantities with sparse Fourier spectrum. 
This completes an algorithm for approximate strong simulation. 
The algorithm for approximating the probabilities can be straightforwardly extended to an algorithm that also approximates all marginals (over bits of the measurement outcome $x$) of the truncated noisy distribution.
Consequently the marginal sampling algorithm can be used to sample from the output distribution up to TVD $2^{- \Omega(\ell)}$ in time $2^{O(\ell)}$.

Interestingly, for IQP circuits it has also been shown that it is possible to classically protect against noise \cite{bremner_achieving_2017}: 
Using classical coding techniques one can encode a smaller IQP circuit $\mc C$ redundantly in a larger one $\mc C'$ such that even if local depolarizing noise is applied to the output of $\mc C'$ one can sample efficiently from a distribution arbitrarily close to the ideal output distribution of $\mc C$. 
Unfortunately, it is not at all clear however, how these coding techniques (similar to the ideas used in the idea to use cryptography to verify IQP circuits which we discussed in \cref{sec:cryptographic tests}) can be extended 
beyond IQP circuits.

\subsection{Simulating boson sampling protocols}
\label{sec:boson sampling classical simulation}

Classical simulation methods
for boson sampling naturally exploit the expression of the output probabilities in terms of the permanent or related matrix polynomials. 
The individual terms in those polynomials can be viewed as the weights of a Feynman path integral expansion of the polynomial, and hence Feynman-type algorithms are natural candidates for the simulation of those schemes. 

\subsubsection{Computing probabilities: permanents and Hafnians}

Computing the output probabilities of boson sampling amounts to computing the permanent \eqref{eq:definitionpermanent} for Fock input states, and the Hafnian \eqref{eq:definition hafnian} for Gaussian input states.
The na\"ive runtime of computing the permanent of an $n \times n$ matrix scales linearly in the number of all permutations of $n$ elements, given by $n!$, multiplied by the complexity of computing the product of $n$ numbers, given by $n^2$, while the space complexity is given by $O(n)$.
Similarly, we can express the Hafnian as a sum over all perfect matching permutations of $2n$ elements and hence the worst-case runtime is given by $n^2$ times the number of perfect matching permutations $| \mathrm{PMP}(2n)| = (2n -1)!! = 1 \cdot 3 \cdot 5\cdots (2n - 1)$ \cite{gupt_walrus_2019}.

These worst-case estimates can be significantly improved, however, via clever re-expressions of the permanent and Hafnian, respectively. 
Indeed, \textcite{ryser_combinatorial_1963-1} has found a way to re-express the permanent via the \emph{principle of inclusion and exclusion} as a sum of $2^n$ terms and hence the complexity of computing the permanent is reduced to $O(n^2 2^n)$ and further to $O(n 2^n)$ by using Gray codes. 
Alternative expressions for the permanent with the same number of terms and hence the same complexity have been found by \textcite{glynn_permanent_2010}, using the polarization identity for symmetric tensors and making use of partial derivatives.\footnote{See \cite{huh_fast_2022} for a use of the Glynn formula in a quantum algorithm for permanent estimation. 
}
Similarly, it turns out that the Hafnian of a $n \times n$ matrix can also be computed in time $O(n^3 2^{n/2})$ \cite{bjorklund_counting_2012}.
The Ryser formula for the permanent can be further reduced to incorporate collision events, reducing the number of terms from $O(n2^n)$ to $\prod_i (n_i +1)$, where $n_i$ is the number of photons observed in mode $i$ \cite{chin_generalized_2018-1,tichy_interference_2014,shchesnovich_asymptotic_2013}. 
Algorithms based on these re-expressions remain the fastest for permanents and Hafnians \cite{bjorklund_faster_2019-1,gupt_walrus_2019,wu_benchmark_2018}, allowing for the computation of matrix permanents of size up to $54 \times 54$ \cite{lundow_efficient_2022}. 
Their runtime can also be further improved by exploiting specific structures such as sparsity or the matrix bandwidth \cite{lundow_efficient_2022}. 

A natural way to exploit path-integral expressions for approximate computation of permanents and Hafnians is to randomly sample out paths and sum up their weights to construct a randomized estimator of the permanent or Hafnian.
\textcite{gurvits_classical_2003} does precisely that, making use of Ryser's or Glynn's formula, to obtain an algorithm that takes time $O(n^2/\epsilon^2)$ to achieve an additive error $\pm \epsilon \norm{A}^n$ estimate of the permanent of $A$.  
\textcite{aaronson_generalizing_2012} generalize the algorithm, obtaining an improved runtime for permanents with repeated rows and columns, corresponding to bunching events, and derandomizing the algorithm for non-negative matrices.
Furthermore, it can be extended to arbitrary input states \cite{yung_universal_2016}.

In specific instances one can also obtain multiplicative-error approximations in sub-exponential or even polynomial time. 
Such results delineate out regimes in which the permanent is in fact not \sharpP-hard to approximate and hence the sampling task will not be intractable, too.  
Specifically, for non-negative matrices \textcite{jerrum_polynomial-time_2004} give a Markov-chain Monte Carlo based randomized algorithm that is able to approximate the permanent up to multiplicative error $\epsilon$ in time $\poly(n,1/\epsilon)$ while, deterministically, only an approximation factor of $2^n$ is currently achievable \cite{gurvits_deterministic_2002,linial_deterministic_1998,barvinok_polynomial_1999}.
Using an ingenious method based on a Taylor-series approximation of the complex polynomial $f(z) = \ln [\Perm(J + z (A - J))]$, where $z \in \mb C$ and $J$ is the matrix filled with ones, \citeauthor{barvinok_combinatorics_2016} has identified certain regimes for which quasi-polynomial relative-error approximations of the permanent and the Hafnian are possible.
This is the case if the function $f(z)$ is holomorphic on the unit disc---for matrices with entries $a_{i,j}$ satisfying $|a_{i,j} - 1| \leq 0.19$ \cite{barvinok_computing_2016}, matrices with entries satisfying $\delta < a_{i,j} \leq 1$ \cite{barvinok_approximating_2017}, and diagonally dominant matrices \cite{barvinok_computing_2019}. 
An interesting case is that of positive semi-definite matrices, as it has been shown that exactly computing the permanent of such matrices remains \sharpP-hard \cite{grier_new_2018}, but multiplicative-error approximation algorithms in $\bpp^\np$ \cite{rahimi-keshari_what_2015} and with quasi-polynomial runtime \cite{barvinok_remark_2020,anari_simply_2017} exist in some circumstances.
Building on the approach of \citeauthor{barvinok_combinatorics_2016}, \textcite{eldar_approximating_2018} show that for random Gaussian matrices with non-zero but vanishing mean there is a quasi-polynomial time algorithm that approximates the permanent to within a multiplicative error.

Physically interesting cases include the case of low-rank matrices since, such matrices determine the probabilities of outcomes with collisions: 
for constant rank, the corresponding permanents can be computed efficiently in the matrix dimension \cite{barvinok_two_1996}.  
\textcite{quesada_franck-condon_2019,quesada_simulating_2019} analyze the complexity of computing of the output probabilities of Gaussian states with finite displacement. 
In this case, the probabilities correspond to so-called \emph{loop-Hafnians} \cite{bjorklund_faster_2019-1} which can be viewed as counting the perfect matching of a graph with self-loops.
Using similar techniques, \textcite{chabaud_classical_2021,chabaud_resources_2022} find efficient algorithms for states with polynomial \emph{stellar rank} and polynomial support over the Fock basis. 
Another physically relevant simplifying modification is to analyze the complexity of computing the outcome probabilities if the detectors can only distinguish between $0$ and at least $1$ photon, so-called threshold detectors. 
In this case, the output probabilities can be expressed in terms of what \textcite{quesada_gaussian_2018} called the ``Torontonian''.
The complexity of directly computing the Torontonian is given by $O (n^3 2^n)$ which is equivalent to the complexity of directly computing the Hafnian. 
It remains an open question whether threshold detectors significantly reduce the complexity of simulating Gaussian boson sampling experiments. 
Furthermore, the output probabilities of Gaussian boson sampling with local and shallow linear-optical circuits can be efficiently computed by making use of the banded structure of the adjacency matrix \cite{qi_efficient_2020}.

Exploiting the fact that for  realistic experiments the number of modes is not much larger than the number of observed photons as required by the proofs of hardness (see Section~\ref{par:hiding boson sampling}) and the fact that threshold detectors are used which only distinguish between $0$ and $\geq 1$ photon, 
\textcite{popova_cracking_2021} introduce an iterative series of approximations to the ideal outcome probabilities. 
To this end, they exploit that low-order moments $\sum_k k^j p_n(k)$ can be efficiently computed with complexity scaling exponential in $j$. Here, $p_n(k)$ is the probability that photons have been detected in $k$ of $m$ detectors, conditioned on a total number $n$ of photons. 
Then they solve an inverse-moment problem to estimate $p_n(k)$, projecting that using up to $j=4$th order, probabilities of a $m=100$ mode-device can be estimated with 50 \% relative accuracy.

\subsubsection{Simulating the sampling task}

Given the entirely different structure of the circuits in variants of boson sampling schemes, the sampling algorithms used are also different in type compared to simulations of universal circuit sampling.
An important further distinction in the quantitative comparison of classical simulation algorithms to actual experiments is the lack of a simple benchmark analogous to the XEB fidelity. 
As discussed in \cref{sec:efficient classical verification} the most important way of verifying boson sampling experiments are state discrimination schemes, as well as certain efficiently computable quantities such as low-order correlations of the respective distributions. 
Consequently, in boson sampling experiments it is also much less clear at which point quantum advantage has been reached experimentally.

A first competitive classical simulation algorithm for Fock state boson sampling uses the Markov-chain Monte Carlo method described above, giving rise to a much better runtime than the na\"ive worst-case complexity \cite{neville_classical_2017}. 
This algorithm takes into account noise in actual devices, in particular, photon loss, which is the dominant source of errors. 
This approximate sampling algorithm has been vastly improved by \textcite{clifford_classical_2018}, 
who provide an \emph{exact} boson sampling algorithm with the same  improved runtime of $O(n 2^n + \poly(m,n))$ as compared to the worst-case  runtime of  $O(\binom{m+n-1}{n} n2^n)$, 
where $n$ corresponds to the number of photons and $m$ is the number of output modes.
In follow-up work by the same authors, this algorithm has been even further improved, achieving an average-case time complexity that is much lower when $m$ is proportional to $n$ \cite{clifford_faster_2020}. 
When $m=n$, specifically, the algorithm runs in time  approximately $ O(n 1.69^n)$ on average.
The sampling algorithms by \citeauthor{clifford_faster_2020} are based on ancestral or marginal sampling. 
The key insight of their algorithms is an expression of the low-order photon marginals in terms of permanents of smaller and smaller matrices, so that the runtime of the algorithm is dominated by the final marginal, where a single permanent of the full $n \times n$ matrix needs to be computed which in the worst case is given by $O(n2^n)$.
Altogether, these results indicate that Fock boson samplers require at least $ \sim 40$ photons before one can hope to surpass the capabilities of currently available classical computers. 

An exact algorithm for Gaussian boson sampling with threshold detectors \cite{quesada_gaussian_2018} that has been implemented by \textcite{gupt_classical_2020-1} requires exponential space, since the entire probability distribution needs to be saved.
\textcite{quesada_exact_2020} improve this and devise an exponential-time exact sampling algorithm that uses only polynomial space and has a runtime $O(n^3 2^{n})$ for generating a single sample with $n$ photons.
To achieve a runtime scaling proportional to the runtime required for a single Hafnian computation
\textcite{quesada_quadratic_2022,bulmer_boundary_2022} 
give algorithms with a further quadratic improvement, achieving runtime $O(n^3 2^{n/2})$.
The key idea of \textcite{quesada_quadratic_2022} is to perform a virtual heterodyne measurement in all modes first. Such a measurement can be efficiently simulated. 
Then, one can iteratively replace the heterodyne outcomes with photon-number measurements and sample from the photon-number distribution conditioned on the heterodyne outcomes in the remaining modes. 
These probabilities are described by loop Hafnians of matrices with increasing size, similar to how the algorithm of \textcite{clifford_faster_2020} expressed probabilities in terms of smaller permanents for standard boson sampling. 
The idea of an algorithm for Gaussian boson sampling with threshold detectors by \textcite{bulmer_boundary_2022} is to simulate a photon-number resolving measurement, and then set all nonzero photon numbers in a sample to one.
In the dilute regime, this reduces the computation to a loop Hafnian of size $n \times n$, containing $2^{n/2}$ terms.
\textcite{bulmer_boundary_2022} then provide a construction that reduces sampling in the non-dilute regime to sampling in the dilute regime by artificially introducing `sub-modes' for each detector. 

\textcite{bulmer_boundary_2022} also present the most advanced implementation of (near-)exact sampling algorithms for Gaussian boson sampling with photon-number-resolving detectors.
To this end, they implement the ancestral sampling algorithm of \textcite{quesada_quadratic_2022} with a variety of improvements. 
Specifically, they reduce the runtime of computing loop Hafnians by making use of an inclusion/exclusion principle on pairs of photons and using a so-called finite difference sieve analogous to \citeauthor{glynn_permanent_2010}'s formula. 
Furthermore, they exploit threshold detectors explicitly in their sampling algorithm. 
For low photon density, it turns out that simulating photon-number resolving detectors and reducing collisions subsequently is advantageous to computing the Torontonians which exactly describe the output distribution with threshold detectors. 
Running on a $ \sim 100 000$ core supercomputer, they are able to simulate $m=60$ modes with up to 80 photons observed by photon-number resolving detectors with a mean time per sample of 3 seconds, and $m=100$ modes with up to 60 click events  with a mean time per sample of 8.4 seconds.
Finally, they generate a single 92-photon event in $m=100$ modes and photon-number resolving detectors in 82 minutes.

A complementary approach has been pursued by \textcite{villalonga_efficient_2021}. 
The idea is to sample from a distribution that reproduces the low-order mode marginals of the ideal target distribution. 
These low-order marginals can indeed be classically efficiently calculated, since they are just determined by a submatrix of the covariance matrix. 
The practical challenge is to efficiently sample from a distribution with the correct marginals. 
\textcite{villalonga_efficient_2021} present two heuristic approaches that are able to achieve this. 
The first heuristic employs a maximum-entropy principle, which corresponds to a Boltzmann machine, i.e., a distribution of the form $p(\mathbf z) = 1/Z \exp(\sum_i \lambda_i z_i + \sum_{i < j} \lambda_{i,j} z_i z_j + \ldots)$, where $Z $ is the partition function that normalizes the distribution. 
To find the correct parameters $\lambda_i, \lambda_{i,j}, \ldots $, a mean-field approximation is used for the second order and costly log-likelihood minimization for the higher orders. 
Another method makes use of a greedy algorithm to generate samples with the correct low order marginals with a cost exponential in the order of the marginal.
\textcite{villalonga_efficient_2021} implement their sampler using the ideal second- and third-order marginals and compare to the experiment of \textcite{zhong_phase-programmable_2021} using $m=144$ modes and squeezing values that give rise to an average photon number of up to $66.9$. 
The total-variation distance of the low-order marginal distributions of up to $14$ modes compared to the corresponding ideal distribution is lower than that of the experimental distribution. 
In a similar vein, first steps towards an approach analogous to that of \citeauthor{clifford_faster_2020} are taken by \textcite{renema_marginal_2020-1} who computes low-order marginals in terms of photons rather than modes, potentially offering a better approximation of the ideal distribution.

Another variant of a spoofing algorithm for Gaussian boson sampling has recently been proposed by \textcite{martinez-cifuentes_classical_2022} and exploits the fact that the quantum device is noisy. 
The idea is to replace the input squeezed states with so-called \emph{squashed states}, that is, coherent states with vacuum fluctuations in one quadrature and larger fluctuations in the other.
Linearly transformed squashed states are classical Gaussian states in that a photon-number measurement can be efficiently simulated classically.
The definition of squashed states is motivated by the fact that loss in the network can be incorporated by replacing the initial squeezed states with squeezed thermal states. 
Squashed states are indeed those Gaussian states which best approximate squeezed thermal states and are at the same time classically simulable in the photon-number basis. 
\textcite{martinez-cifuentes_classical_2022} find that while the experiment of \textcite{zhong_quantum_2020} can be spoofed by squashed-state Gaussian boson sampling in the sense that the  correlations in the distribution match the ideal correlations better than the experiment, the more recent experiment of \textcite{zhong_phase-programmable_2021} cannot.

Interestingly, such approaches cannot be applied to universal circuit sampling because the approximation that reproduces marginals and correlations up to a constant order would be exponentially close to the uniform distribution due to the highly entangled nature of the output distribution.

Let us also note that the complexity of Fock boson sampling has been considered under locality constraints \cite{deshpande_dynamical_2018,maskara_complexity_2019}. 
In certain setting, such structure renders the classical simulation of boson sampling efficient. 
\textcite{oh_classical_2022} follow up on these results and derive general algorithms for Fock boson sampling and Gaussian boson sampling that exploit the graph structure of a linear-optical circuit.
For a sufficiently small tree-width of the interaction graph, i.e., in particular, for low-depth, geometrically local linear-optical circuits this exact sampling is efficient. 

\subsubsection{Analysis of noise} 
\label{subsec:noisy boson sampling}

In boson sampling experiments with photons, the dominant sources of noise are losses of photons due to finite transmittivity of wave-guides and other optical elements, finite distinguishability of the photons due to imperfect time or frequency synchronization between the single-photon sources, so-called \emph{mode mismatch}.
The asymptotic effects of these 
noise types has been studied extensively---for photon loss and detector noise (dark counts)
\cite{oszmaniec_classical_2018,garcia-patron_simulating_2019,renema_classical_2019,moylett_classically_2019,qi_regimes_2020,oh_classical_2021-2,rahimi-keshari_sufficient_2016} and 
partial distinguishability of the photons \cite{tichy_sampling_2015,renema_efficient_2018,moylett_classically_2019,renema_simulability_2020,shchesnovich_sufficient_2014,rahimi-keshari_sufficient_2016}. 
The overall observation of these studies is that already comparably low noise levels drive the output probability distribution closer to distributions that are simulable with less effort.

An interesting `toy' noise model for boson sampling has been considered by \textcite{kalai_gaussian_2014}.
In this model, additive Gaussian noise is applied to the random (Gaussian) submatrix of which the permanent is taken to compute the outcome probabilities of boson sampling, see \cref{eq:bosonsamplingdistribution permanent}.
\textcite{kalai_gaussian_2014} show that the (collision-free) output probabilities of boson sampling with a constant amount of such Gaussian noise can be approximated by sparse low-degree polynomials.
This gives an efficient approximation algorithm for the noisy xoutput probabilities with constant precision.
This noise model turns out to be very appealing: on the one hand it ``preserves the mathematical connection to random Gaussian matrices, used to establish hardness of boson sampling'' 
\cite[p.3]{shchesnovich_noise_2019}. 
As \textcite{shchesnovich_noise_2019} shows, on the other hand, this noise model is closely related to experimentally more relevant noise sources: it is equivalent to photon loss at the input of the interfermoeter and dark counts in the measurement that exactly compensate for the lost photons, as well as partial distinguishability of bosons.

While \textcite{kalai_gaussian_2014} do not provide a total-variation distance bound on the approximate noisy distributions (with approximations given by the low-degree polynomial), \textcite{shchesnovich_noise_2019} provides such a bound and shows that it can be made inverse polynomially small at a polynomial cost in the time it takes to compute the corresponding probabilities.  
Analogously to \textcite{renema_efficient_2018,renema_simulability_2020}, they then argue that a Metropolis Markov-chain Monte Carlo algorithm can be used to efficiently sample from this distribution.\footnote{Notice, though, that this falls short of an efficiency proof since none of those works actually bounds the mixing time of the corresponding Markov chain. }
The result of \textcite{shchesnovich_noise_2019} can thus be viewed as unifying several more specific previous results \cite{PhysRevA.92.062326,shchesnovich_sufficient_2014,PhysRevA.93.012335,renema_classical_2019,renema_efficient_2018,oszmaniec_classical_2018,garcia-patron_simulating_2019,leverrier_analysis_2015} on the easiness and hardness of noisy boson sampling in certain noise regimes, see \cite[Table I]{shchesnovich_noise_2019} for an overview, and gives rise to the following heuristic picture:
Boson sampling with noise strength on the order of $\Omega(1)$ can be simulated classically to total-variation distance error $\epsilon$ with polynomial effort in $n$ and $1/\epsilon$. 
Conversely, for a noise strength scaling as $O(1/n)$ the simulation complexity remains the same as that of ideal boson sampling. 
In other words: constant local noise renders boson sampling classically simulatable, while local noise scaling inversely with the number of photons presumably remains classically intractable, with the intermediate regime remaining open.

With the Fourier picture of \textcite{bremner_achieving_2017,gao_efficient_2018,aharonov_polynomial-time_2022} in mind, \textcite{oh_classical_2023} build upon those prior works and provide a fully provable algorithm for sampling from the output distribution of noisy Fock boson sampling.
Their sampling algorithm is based on the low-degree polynomial decomposition of \textcite{kalai_gaussian_2014}, which they show also works for the marginals of Fock boson sampling written in first quantization. 
This allows them to compute all marginals of the low-degree approximation to the noisy probabilities and hence provides an approximate sampling algorithm for the noisy distribution. 
At a constant noise rate, the total runtime of the algorithm is quasi-polynomial and given by $n^{O(\log n,\log (1/\epsilon), \log (1/\delta))}$ per sample to within a total-variation distance $\epsilon>0$ for a proportion $1-\delta$ of Haar-random unitaries, assuming the hiding condition $m \in \omega (n^5)$\footnote{To achieve a provable polynomial-time algorithm as for universal circuits,  the total noise rate has to scale
like $1-x^\gamma$ with $\gamma = \Omega(\log n)$ and a constant $x\in [0,1)$. The authors argue that this is also the fair comparison since a constant noise rate per gate results in an overall noise that scales with the number of gates.}.

A compelling physical picture for why noise renders classical simulations tractable has been developed by \textcite{renema_classical_2019,renema_simulability_2020}, who conceive of the boson sampling distribution as arising from interference processes with increasing order. 
As it turns out, the effect of noise in a precise way results in higher order interference terms to contribute exponentially less. 
This gives rise to a distribution that just arises from low-order interference. 
The resulting distribution can therefore be classically simulated efficiently.
A similar intuition, albeit on the level of individual modes, is followed by the simulation algorithm of \textcite{villalonga_efficient_2021}. 
\textcite{shchesnovich_boson_2022,shchesnovich_distinguishing_2021} shows, however, that the output data from classical simulation methods based on lower-order multi-boson interferences can be efficiently distinguished from a noisy boson sampling distribution since the higher-order correlations remain sufficiently significant. 
This matches the observation of \textcite{zhong_phase-programmable_2021}, who find that higher-order correlations remain present in experimental data.

To conclude, the fact that noise renders the
classical simulation of imperfect devices less computationally demanding adds a challenge to the
experimental realization of quantum random sampling schemes that show an unambiguous
quantum advantage.

\section{Perspectives}
\label{sec:perspectives}

The field of quantum random sampling has now reached a state at which the theoretical foundations are thoroughly explored, and important but extremely difficult open questions have been identified. 
It has reached a state at which we have seen first demonstrations on the verge of classical intractability and first pushbacks from classical algorithms. 
In this review, we have comprehensively discussed these theoretical and practical aspects of quantum random sampling.

But what is the road ahead?  
Some features of this road are quite clear; 
important technical questions such as approximate average-case hardness remain to be tackled, quantum devices and classical algorithms alike are going to be further improved.  
At the same time, the leap from  demonstrations of quantum computational advantage via quantum random sampling or other means to achieving a practically useful task with a quantum advantage seems enormous. 

This section is more than a mere outlook.
While we summarize the key open question, we also provide summaries of ideas to highlight exciting future directions of the field. 
We start by summarizing the key open questions in the field of quantum random sampling, most of which already appeared in earlier sections.  
We then move on to take a broader perspective on the field of quantum advantages in general and quantum random sampling schemes in particular, to see what questions have already been comprehensively settled, what is ahead, and what are reasonable next steps. 
In particular, we begin an outlook by drawing connections between quantum random sampling and other fields, for example, quantum simulation. 
Finally, we sketch some ideas that have been developed with the goal of practical applications of quantum random sampling in mind. 
These applications either make direct use of the randomness of quantum random sampling, or of the programmability of a quantum random sampler in order to solve a specific task.

\subsection{Open questions on quantum random sampling}
\label{ssec:open questions}

Throughout the course of this review, we have highlighted important open technical questions regarding our understanding of quantum random sampling.  
Let us summarize some of the most important ones here. 

\subsubsection{Understanding random quantum circuits better}

From the perspective of the computational complexity of quantum random sampling, the key open question is to prove approximate average-case hardness, as discussed in detail in \cref{subsec:average-case hardness,ssec:discussion average-case hardness}. 
As of now, approximate average-case hardness is a conjecture that is based solely on the lack of efficient classical simulation algorithms and the observation that random instances do not offer any additional structure that a classical simulation algorithm might exploit to perform better than in the worst case. 
While we have progressively moved forward on this question by making polynomial interpolation techniques more robust \cite{bouland_complexity_2018,movassagh_quantum_2020,bouland_noise_2021,kondo_fine-grained_2021,krovi_average-case_2022}, there remain fundamental barriers to improving this result to the required robustness $O(2^{-n})$ as discussed in detail in \cref{ssec:discussion average-case hardness}.
It seems that, from this point onward, polynomial interpolation alone will not be able to help us solve the question of approximate average-case hardness and new proof ideas are required. 
The development of new methods, while most pressing, is also elusive and constitutes a major challenge that reaches beyond the field of quantum random sampling schemes all the way into the midst of computational complexity theory.
For example, 
\textcite[p.~91]{aaronson_computational_2010} suggest to make use of a restricted class of polynomials that are not closed under addition which are at the same time able to capture the quantity of interest.

Let us zoom out from the details of the proof of robust sampling hardness and consider the task of sampling from the output distribution of a random quantum  circuit up to a constant total-variation distance error.  
To achieve such an overall constant additive error on the global distribution, the gate errors need to scale inversely as $1/(m+2n)$ with the total number $m$ of gate applications, and $n$ single-qubit state preparations and measurements, respectively. 
But in experiments the gate application, state-preparation and measurement errors typically do not scale in the size of the system, or circuit, but are rather fixed by physical details. 
This raises the question of what the optimal trade-off for achieving a quantum advantage is in terms of circuit-depth and system size. 
While on the one hand, random short depth circuits might be easy to sample from with a global error budget \cite{napp_efficient_2022}, too large a circuit will incur a devastating amount of errors that renders classical simulation trivial. 
This is why the detailed study of noise in random circuits and its effect on the output distribution is paramount to better understanding the computationally most difficult regimes.  
First steps towards this have recently been taken by the complementing approaches of \textcite{dalzell_random_2021} and \textcite{deshpande_tight_2022}. 
These two works study the convergence to the white-noise or uniform distribution in the regime of low gate noise $\epsilon \in \tilde O(1/n)$, and constant gate noise, respectively. 
A better understanding of how different types of experimentally relevant noise affect the output distribution of typical random quantum circuits is thus paramount to optimizing the parameters in a demonstration of quantum advantage. 

\subsubsection{Verification beyond XEB}

The issue of noise in random quantum circuits directly leads to the next open question. 
In \cref{sec:verification}, we discussed in detail in what sense samples from random quantum circuits can be verified. 
The standard measure of quantum advantage as of today is the linear XEB fidelity \eqref{eq:xeb fidelity}. 
On the one hand, this is because it  offers the best available compromise between being practically viable and providing a meaningful benchmark for achieving a nontrivial task on the quantum device. 
On the other hand, it is because it provides a unified view for the use of quantum random sampling as a benchmark of a quantum device and as a means to demonstrate a computational advantage. 
However, both interpretations of XEB fidelity are not fully understood.

Coming from the perspective of quantum advantage demonstrations, there is the question under which circumstances cross-entropy type measures---and in particular the XEB fidelity---can yield certificates for the global distribution. 
The logarithmic XEB fidelity, for instance, provides rigorous bounds on the total-variation distance only if the noise in the device is such that it increases the entropy of the ideal distribution~\cite{bouland_quantum_2018}. 
But estimating the entropy of the noisy distribution is an infeasible task in itself.

Coming from a practical perspective of device development and characterization, the question remains to identify in which settings random quantum circuits can be used to benchmark noise in the quantum circuit. 
As discussed in \cref{ssec:xeb}, \textcite{liu_benchmarking_2021} have made first steps towards understanding how a noise parameter can be extracted from the XEB fidelity for the case of global Pauli noise via a perturbative analysis in the noise parameter. 
Going beyond perturbative methods, further steps in this direction might make use of the framework of Fourier analysis for randomized benchmarking \cite{helsen_general_2020}. 
Ultimately, one would hope to analyse gate-dependent noise channels in a local quantum circuit. 

Finally, let us mention that the most important problem with the use of XEB to verify for quantum random sampling is the fact that evaluating XEB-like measures while sample-efficient, incur the exponential computational cost of estimating some of the target probabilities. 
Going beyond XEB, an interesting open problem is whether the no-go result prohibiting sample-efficient verification of flat distributions in \cref{ssec:hardness of verification} can be circumvented in random sampling schemes with larger second moments.
Is there any ``room in the middle'' between exponentially flat distributions that are hard to verify but anticoncentrate, and polynomially concentrated distributions which do not anticoncentrate but are sample-efficiently verifiable. 
If there is, then distributions could exist which we \emph{can} (sample-)efficiently verify from classical samples and which we can sample from efficiently on a quantum computer, but \emph{cannot} efficiently sample from on a classical computer \cite[see also][]{hangleiter_sample_2019}.
Works such as the one by \textcite{morimae_hardness_2017} proving anticoncentration of the DQC1 model without resorting to a second-moment bound might yield some leeway in this direction.

But one may also ask whether there are fully efficient ways of verifying that a classically intractable task has been achieved via quantum random sampling without resorting to directly verifying total-variation distance? 
We have mentioned ideas for the verification of quantum samplers that make use of cryptographic secret hiding \cite{shepherd_temporally_2009} in a delegated scheme. 
But such ideas remain prone to classical attacks \cite{kahanamoku-meyer_forging_2019}, or remain orthogonal to the spirit of quantum random sampling as a specifically simple, unstructured task that is executed on a given quantum device. 
Having said that for very simple tasks, proofs of quantumness that might not be too far realm of practical feasibility can be devised using such ideas \cite{kahanamoku-meyer_classically-verifiable_2021,zhu_interactive_2021,liu_depth-efficient_2021,hirahara_test_2021}. 
Interesting progress in the direction of merging these worlds with public verifiability of \np\ problems such as factoring has been made recently by \textcite{yamakawa_verifiable_2022}. 
It remains an exciting question to further explore the possibility of verifying quantum random sampling efficiently.

\subsection{Developing novel schemes}
\label{ssec:larger scales}

Going beyond better understanding the current schemes, there is the overarching question of how quantum random sampling schemes can be extended beyond their current realm of applicability. 
This regards both the extension from digital quantum devices to analog ones and to a larger error resilience.

\subsubsection{Improving error resilience}

Given all the strengths of the various approaches to quantum random sampling, it is also limited in its capacity to demonstrate quantum speedups. 
This is because these schemes do not allow for any type of error correction, making the region that is nontrivially accessible with finite errors limited. 
Going from relative to additive errors has been a tremendous technical achievement, matching complexity
theoretic arguments more closely with experimental desiderata, but it still falls short of capturing 
fully realistic errors. 
The central challenge one has to overcome when realizing quantum supremacy is thus to bring this barrier down as far as possible such that the computing capabilities of classical computers can be surpassed before the barrier is hit. 

Ultimately, one would hope to make the hardness of quantum sampling robust to constant local errors. 
This can indeed be achieved for universal computations using quantum error correction codes. 
However, quantum error correction is intrinsically based on the continuous measurement of error syndromes, giving information about which errors have occurred during one cycle of the computation. 
Those errors then need to be actively corrected, requiring an elaborate machinery that is again well outside the realm of what we envision the context of quantum random sampling to be. 
What is more, from a conceptual point of view, quantum random sampling is intrinsically based on a global property of the outcome state, namely, the full probability distribution. 
To make this global property robust to constant local errors will therefore likely require invoking a global error-detection machinery such as the one of \textcite{bremner_achieving_2017}.

One might think that coherent errors do not constitute a specifically grave problem for quantum random sampling schemes since, say, Pauli errors can often simply be absorbed in the random ensemble, giving rise to a different computation distributed according to the same ensemble.
Let us stress, however, that to maintain hardness of sampling, we actually \emph{need to know} how the circuit has changed due to the errors. 
In other words, the errors need to be ``heralded''.
But continuous measurements of syndromes complicate the computation significantly. 
Conversely, if the ongoing computation is not continuously measured in every gate cycle, it is not clear under which circumstances such a ``heralded noise model'' is actually realistic. 
Finding ways around this obstacle, possibly using error detection and post-hoc corrections, is the major challenge towards making quantum random sampling schemes robust to physical noise and thus scalable.

Further intuition on the resilience of quantum circuits to local errors is also provided 
by the analysis of constant-depth quantum circuits: \textcite{Shallow} show that constant-depth quantum circuits are more powerful than their classical counterparts. 
Any classical probabilistic circuit composed of bounded fan-in gates that solves what \textcite{Shallow} call the \emph{two-dimensional hidden linear function problem} with high probability must have depth at least logarithmic in the system size. 
In contrast, the same problem can be solved with certainty by a constant-depth quantum circuit that is composed of one- and two-qubit quantum gates acting that act on a two-dimensional lattice. 
It turns out that this scheme is robust to noise in that the above separation in computational power persists even when the shallow quantum circuits are restricted to three dimensions and are corrupted by noise \cite{NoisyShallow}. Technically, the argument supporting this conclusion is rooted in ideas on the  generation of a long-ranged entanglement in noisy three-dimensional cluster states
\cite{PhysRevA.71.062313}.

In a similar spirit, first ideas towards using nonadaptive error-correction by embedding a computation in an error correction code have been made \cite{fujii_noise_2016,kapourniotis_nonadaptive_2019} and, indeed, if experimental errors remain within the specific error model considered, sampling hardness remains. 
However, robustness may be lost since the distribution for which an approximate average-case hardness conjecture for the outcome probabilities holds has significantly changed compared to the non-error-corrected distribution.

\subsubsection{Relation to analogue quantum simulation}

Aside from the largely technical open questions discussed so far, another avenue for making progress en route to larger-scale implementations of quantum random sampling, is to connect it to the setting of analog quantum simulators. 
Such devices offer a limited amount of control, but often a large number of coherently and highly accurately controlled quantum degrees of freedom, which in several instances cannot be simulated by the best classical algorithms \cite{Trotzky,choi_exploring_2016,debnath_demonstration_2016,Emergence,ebadi_quantum_2021}. 

Along these lines a reasonable goal would be to prove a rigorous complexity-theoretic separation for a task that is natural in a physics mindset in general, and quantum simulations in particular. 
Quantities that first come to mind here are measurements of $k$-point correlation functions of the type $\langle b_i^\dagger b_j\rangle$. 
First steps towards this have been taken by \textcite{novo_quantum_2019} who have shown that one can run a Stockmeyer argument for the task of reproducing the statistics of an energy measurement of a local Hamiltonian. 
Deviating from the mindset of quantum random sampling \textcite{baez_dynamical_2020} have shown a quantum advantage for the estimation of dynamical structure factors, bringing the insight that performing measurements on quantum states arising from time evolution under local Hamiltonians is $\bqp$-complete \cite{Vollbrecht,nagaj_universal_2012,nagaj_hamiltonian_2008} closer to experimental reality.

These works are simply based on the assumption that quantum computers are more powerful than classical computers and therefore do not offer independent evidence for this separation. 
In technical terms, they show a much weaker complexity-theoretic consequence than a collapse of the polynomial hierarchy,  namely, that $\bpp = \bqp$. 
Coming from a complexity-theoretical perspective, they are thus begging the question as, from this perspective, one would like to precisely collect evidence that $\bpp \neq \bqp$. 
Accepting this, it remains not obvious whether one expects average-case hardness of the respective tasks for problems in \bqp. 
Coming from a more practically minded perspective, accepting $\bqp \subsetneq \bpp $ is a fair assumption. Such ideas may thus help to demonstrate quantum advantages for tasks that are more useful than sampling alone. From a technological perspective, it is interesting to see whether one can reach the regime in which quantum advantages in this sense are conceivable.

Another interesting perspective that has been considered in this context is the relation of sampling hardness to physical phenomena. 
For instance, phase transitions in sampling complexity of two-dimensional bosonic lattice systems have been considered by \textcite{deshpande_dynamical_2018,maskara_complexity_2019}. 
Here, the idea is to vary a physical parameter of the system, in this case, the spacing between bosons in the initial state and consider the complexity as a function of time when evolving the system. 
In a similar vein, \textcite{ehrenberg_simulation_2022} study transitions in the complexity of sampling from the output distribution of many-body-localizing time evolution. 
In such approaches, the hope is to narrow down and better understand the physical mechanisms underlying sampling complexity.

\subsection{Towards applications of quantum random sampling}
\label{ssec:applications}

What is next? 
On the road towards practically useful quantum computers, quantum random sampling schemes are an important stepping stone. 
But quantum random sampling has been conceived as a proof-of-principle task to show that quantum devices have the capability to computationally outperform classical computers and nothing more. 
It is therefore not set up to realize practically interesting applications in their own right. 
It goes without saying, however,  that a natural next question is whether one can exploit the provable speedup over classical sampling algorithms on the specific random sampling task for relevant practically motivated applications. 
Here, we discuss some of those first steps at identifying applications of quantum random sampling. 

Roughly speaking, these applications of quantum random sampling fall into two categories. 
On the one hand, there are applications that exploit the intrinsic quantum randomness of typical quantum circuits. 
Such applications make use of the fact that the output distributions of random quantum circuits are highly unstructured or, in technical terms, have a high min-entropy, as explained in detail in \cref{ssec:hardness of verification}. 
On the other hand, there are applications that take programmable quantum random sampling devices as their starting point and ask the question: 
What applications can those devices be used for? 
In such applications, structure of the output distributions is explicitly exploited to solve a computational task, or serve as a subroutine in an algorithm solving such a task.  
In the following we explain some of the ideas in this mindset with the goal of giving the reader a concrete idea about potentially interesting directions of study.  

\subsubsection{Exploiting randomness}

One of the most promising near-term applications of quantum devices is the generation of certified random numbers. 
In the classical world, bits that are perfectly random in that they are unpredictable not only to the user of the device, but to any observer, cannot be realized in principle because the laws of classical physics are deterministic. 
In practice one has to therefore rely on---albeit possibly extremely weak and plausible---hypotheses to design \emph{pseudo-random-number generators}. Going beyond this, so-called \emph{true random-number generators} exploit physical processes from the realm of classical physics that are hard to predict. 
Quantum random-number generators make use of the intrinsic randomness offered by quantum mechanics. 
The possibility of harnessing this randomness makes quantum technologies attractive as a means to generate certified random numbers \cite{acincertified2016} that cannot be predicted by any adversary. 

Given that the output distributions of random quantum circuits have a high min-entropy, statistically verified quantum random sampling would naturally give rise to a large number of intrinsically random bits.
In the absence of such statistical tests, 
\textcite{aaronson_aspects_QCrypt,aaronson_aspects_2019} proposed protocols for certified randomness that use universal circuit sampling and the XEB benchmark. 
The proof of security of the proposed protocols is based on a strong and highly nonstandard complexity-theoretic conjecture on the hardness of what \textcite{aaronson_aspects_2019} calls the \emph{long list quantum sample verification (LLQSV)} problem.
This problem asks to distinguish exponentially many output bit strings from a quantum random sampler, given by an oracle, from uniformly random numbers.  
More specifically, the conjecture is that LLQSV is not in a complexity class called \qcam\ which contains \am\ and \bqp\ and more.

\textcite{bassirian_certified_2021} provides complexity-theoretic evidence in support of the classical intractability of this problem. 
This support lives to the same standard as the evidence for computational hardness of achieving a high XEB score via \xhog\ and \hog. 
They prove two statements regarding the hardness of LLQSV, or in other words, the hardness of distinguishing the high min-entropy samples from the quantum device from uniformly random samples. 
They do so in the black-box model in which query access to a random Boolean function is granted, instead of a random circuit. 
First,  \textcite{bassirian_certified_2021} prove an average-case linear min-entropy bound for quantum algorithms that pass an XEB-like test. 
Second, they show that no \bqp\ or \ph\ algorithm can solve the LLQSV\ problem, thereby individually showing separations from major classes contained in \qcam.
To do so, they reduce it to a variant of the so-called \emph{forrelation} problem introduced by \textcite{aaronson_bqp_2010}.
These results imply that if one believes that quantum circuits viewed as random functions are sufficiently unstructured, then quantum random sampling can generate random samples that are certified by an XEB-like test.

In a different vein, the fact that quantum states prepared by random quantum circuits are highly entangled might be useful in \emph{quantum metrology}. 
In this context, it is less the flatness of the classical output distribution that is exploited, but rather the full quantum state. 
Along these lines, \textcite{PhysRevX.6.041044} study how useful random bosonic states are for quantum metrology. 
They indeed find that a close to 
optimal (Heisenberg) scaling is typically achieved.
\textcite{PhysRevA.103.032613}
explores the phase sensitivity of generic linear interferometric schemes using Gaussian resources and measurements, in what could be called boson-sampling-inspired strategies. 
Multimode metrology via a variant of Gaussian boson sampling was  studied by\textcite{PhysRevA.104.032607}.
Finally, it has been suggested that the high min-entropy of the output distributions can be exploited to devise cryptographic schemes \cite{nikolopoulos_cryptographic_2019,huang_boson_2019,huang_protecting_2020}.

\subsubsection{Exploiting structure}

Rather than exploiting the randomness of quantum random samplers, one may alternatively also program such devices in a bespoke way in order to solve computational problems. 
Such applications make use of structure in the output probability distributions, an idea which has to date been most developed for variants of Gaussian boson sampling.
Let us give two examples that argue along these lines. 
While the first example makes specific use of \emph{samples}, the second example uses samples in order to estimate \emph{probabilities}.  

\paragraph{Using samples to solve graph problems.}
A natural class of problems that can be studied in the context of Gaussian boson sampling are graph problems. 
This is because the Hafnian \eqref{eq:definition hafnian} of an adjacency matrix of a graph equals the number of \emph{perfect matchings} of that graph, that is, the number of disjoint sets of edges in which every vertex of the graph is connected to exactly one edge. 

As an example, consider the so-called \emph{densest $k$-subgraph problem} \cite{arrazola_using_2018}.
This problem asks, given a graph $G$ with $n$ vertices to find the subgraph with $k<n$ vertices that has the largest number of edges.
Recall that the probability $\Gpbos(S)$ \eqref{GaussianBosonSamplingDistribution} of obtaining a collision-free output pattern $S$ in Gaussian boson sampling is determined by the Hafnian of a submatrix $M_S$ of a certain matrix $M$ \eqref{eq:M GBS} that depends on the covariance matrix of the input state. 
Given the adjacency matrix $A\in \{0,1\}^{m\times m}$ of $G$ we can now choose the squeezing parameters and linear-optical unitary in order to ``program'' that matrix to be  
\begin{equation}
M = c(A\oplus A),
\end{equation}
where $c< \lambda^{-1}$ and $\lambda$ is the largest eigenvalue of $A$.
It turns out that the corresponding Gaussian state is pure and hence a valid state that can be prepared in Gaussian boson sampling.  
The output probabilities postselected on the collision-free subspace will then be proportional to $|\Haf(A_S)|^2$, where $A_S$ is a submatrix of $A$ determined by the outcome $S$, or equivalently, the adjacency matrix of a subgraph of $G$ with vertices selected by $S$. 
Since the  Hafnian of an adjacency matrix equals the number of perfect matchings of the corresponding graph, the  larger the number of perfect matchings in a subgraph, the more likely its corresponding sample is obtained as an output in Gaussian boson sampling.

The next step is to establish a connection between the number of perfect matchings in a graph and its density. 
On an intuitive level, the number of perfect matchings corresponds to the density of a graph since a graph with many perfect matchings will have many edges. 
Indeed, the number of perfect matchings provides a lower bound to the number of edges in the graph \cite{aaghabali_upper_2015}.
Consequently, by programming the quantum device in an appropriate way, one can sample from a distribution that has a 
bias in favor of dense subgraphs. 
For this reason, stochastic algorithms \cite{GraphData} for the densest $k$-subgraph problem that make use of uniform randomness can be enhanced by having access to samples drawn from the output distribution of Gaussian boson sampling.
In a proof-of-principle experiment using time-bin encoded GBS, this has been demonstrated recently \cite{sempere-llagostera_experimentally_2022}.

\textcite{arrazola_quantum_2018} follow a similar line of thought by introducing an \np-hard problem referred to as \textsf{Max-Haf}. 
They show that access to samples from the Gaussian boson sampling distribution defined by the probabilities
$\Gpbos(S) $ of obtaining the output pattern $S$ can enhance classical stochastic algorithms for this problem. 
This work not only presents the idea and compares the performance of this algorithm with classical algorithms based on uniform randomness, but also numerical data from use cases. 
\textcite{bradler_gaussian_2018} discuss
the problem of actually finding perfect matchings of arbitrary graphs enhanced by having access to samples from Gaussian boson sampling. 

Coming from a perspective of quantum machine learning, \textcite{PhysRevE.101.022134} propose an application of quantum random sampling to statistical modelling. 
\textcite{havlicek_supervised_2019} show how minimally enhanced IQP circuits might be used to enhance the feature space of machine-learning algorithms for supervised-learning.
More concretely, samples from Gaussian boson samplers can be made use of to construct feature vectors of graphs that give rise a natural measure of similarity between graphs \cite{schuld_quantum_2020}. 
The connection to quantum-enhanced machine learning is made even more explicit by \textcite{PhysRevA.102.012417}, who show how  Gaussian boson sampling devices can be trained in the following sense: 
Analytical gradient formulae for the GBS distribution can be exploited when training devices using gradient-descent based methods. 
Finally, \textcite{chabaud_quantum_2021} study supervised learning using minimal extensions of Fock boson sampling. 

\paragraph{Estimating physical quantities using Gaussian boson samplers.}
Using the samples from a quantum device in order to estimate outcome probabilities is the basis of a line of thought initiated by \textcite{huh_boson_2015}. 
As it turns out, when preparing \emph{displaced squeezed states} at the input of a linear optical device, the output probabilities of a Gaussian boson sampler can be used to estimate so-called \emph{Franck-Condon factors}, which represent the transition frequencies of molecular vibronic spectra. 
This is a problem for which no efficient classical algorithm is currently known. 
In this way, Franck-Condon factors can be estimated from Gaussian boson sampling data. 
Following up on this, \textcite{AnalogSimulation} suggest an analog quantum simulation of molecular vibronic spectra based
on boson-sampling like schemes, incorporating the non-Condon scattering operation with a 
quadratically small truncation error. 
Pursuing a similar aim, \emph{molecular docking} is studied by \textcite{MolecularDocking}, who suggest that Gaussian boson samplers provide insights into molecular docking configurations, 
which are spatial orientations that molecules assume when they bind to larger proteins.
Connecting these ideas to the (loop) Hafnian picture of the output probabilities, \textcite{quesada_franck-condon_2019} suggests to estimate Franck-Condon factors by counting perfect matchings 
of graphs with loops. 
To this end, he shows that the Franck-Condon factor associated with a transition between initial and 
final vibrational states in two different potential energy surfaces can be reduced to the number of perfect matchings of a suitable weighted graph with loops. 
\textcite{clements_approximating_2018} explores the impact of experimental imperfections on the performance of the protocol of \textcite{huh_boson_2015} for performing 
quantum simulations of vibronic spectroscopy, providing stringent benchmarks that have to be met
by experiments. 
This work also discusses practically meaningful examples such as 
Franck-Condon factors for vibronic transitions in molecules such as tropolone.
Departing from the above prescriptions in a different way, \textcite{wang_efficient_2020}
implement a small-scale instance of the protocol of \textcite{huh_boson_2015} in a two-mode superconducting device.
    
Known classical simulation methods for boson sampling with sparse outputs as they have been 
presented by \textcite{RogaSimulation,oh_classical_2022} have challenged these results, in that they argue that the
instances considered when sampling from Franck-Condon factors are often sparse in the
appropriate sense. 
Technically, this work demonstrates that the computationally costly
support detection step, i.e., the localization of the largest element from a long list, can be reduced to solving an Ising model that can be solved in polynomial time under suitable 
conditions. 
\textcite{oh_quantum-inspired_2022} have followed up on this line of thought by presenting a quantum-inspired classical algorithm for molecular vibronic spectra. Technically, they find an exact solution of the Fourier components of molecular vibronic spectra at zero temperature using a positive $P$-representation method. 
The resulting algorithm resembles that of \textcite{baiardi_general_2013}.

Both of the just discussed lines of work are contributions that show the potential of achieving computational advantages in practically motivated problems by using Gaussian boson sampling devices. 
At the same time, as the classical algorithms by \textcite{oh_quantum-inspired_2022,oh_classical_2022,RogaSimulation} show, it may be possible to find classical algorithms that are efficient for those instances of Gaussian boson sampling that are used to solve a specific computational problem.
For these instances there is certainly no complexity-theoretic reason analogous to the polynomial hierarchy collapse to believe in a quantum speedup. 
Rather, now we are moving into the realm of comparing quantum algorithms with the best classical algorithm for specific problems---as one would also expect when considering practically relevant problems.

\subsection{Concluding thoughts}
\label{ssec:conclusion}

In this review, we have aspired to provide a comprehensive overview of the efforts aimed at understanding in theory and demonstrating in practice the computational advantage of quantum random sampling over classical computation. 
Quantum random sampling schemes are particularly attractive as they are extremely simple conceptually and come along with comparably small experimental desiderata. 
On the highest level, there seem to be two main lessons that can be drawn from the exciting research efforts that are the focus of this review. 

One of those lessons is of a \emph{foundational nature}. 
Ultimately, the questions asked in endeavours to show quantum advantages with quantum random sampling schemes follow up on the thoughts of Turing about the intertwinement of the complexity of processes in nature and what can be computed using the mechanisms allowed by natural laws. 
Boldly stated, the question on the desk is: What is, after all, the computational nature of Nature? 
Put in more elaborate words:
Can all naturally feasible computations be efficiently described within a classical Turing machine model? 
The extended Church-Turing thesis asserts that this is indeed the case, but it is challenged by the onset of physical quantum computers. 
We have walked a long route along this path, starting from theoretical arguments against the validity of the extended Church-Turing thesis to the question of how to verify those claims 
experimentally. 
Further efforts on realizing sampling schemes will shine light onto this matter.

The other lesson relates to \emph{technological issues}. 
The present efforts towards realizing quantum advantage schemes can hardly be underestimated in their importance of providing guidance for the next steps to be taken in the development of quantum technologies. 
The experimental demonstration of quantum random sampling schemes provide impetus for achieving unprecedented control in experiments, of pursuing large scale quantum computations. 
Next natural steps are to pursue practically motivated quantum algorithms on such quantum devices, a process that is well under way. 
Some schemes can be seen as variations of quantum random sampling schemes, addressing pragmatically motivated questions. 
This applies, e.g., to photonic experiments that explore vibronic
spectra \cite{clements_approximating_2018,wang_efficient_2020}, implement variational schemes \cite{Peruzzo},
or quantum simulations of processes in statistical physics \cite{QuantumPhotoThermodynamics}. 
Then, the layout of the superconducting quantum advantage experiment of \textcite{arute_quantum_2019} has been made to be forward compatible with realizing the surface code \cite{satzinger_realizing_2021}. 
Indeed, arguably the most 
substantial next step will be to achieve fault tolerance in quantum computing, a step that may still be relatively
far away. 
The efforts on quantum random sampling schemes can be seen as a first milestone along this way.

In a similar way, the questions of ``what next'' apply to theoretical research. 
Steps have been taken towards developing protocols that show a more practically minded quantum advantage. 
Quantum approximate optimization algorithms in their various variants suggest to address questions of combinatoric optimization \cite{QAOA,PhysRevX.10.021067}, variational quantum eigensolvers may solve variational principles beyond the
capabilities of classical efficient variational methods  \cite{McClean_2016}. 
These applications are thought to be pursued without quantum error correction---but the key question remains open of what noise levels quantum devices may ultimately tolerate while maintaining a quantum advantage \cite{stilck_franca_limitations_2021}. 
The efforts towards achieving quantum advantages can be seen as a first stepping stone en route to building useful quantum computers and an invitation to master the next hurdle along that route.


\section*{Acknowledgements}

We would like to thank 
Scott Aaronson,
Dorit Aharonov,
Juani Bermejo-Vega,
Adam Bouland,
Ulysse Chabaud,
Abhinav Deshpande,
Ish Dhand,
Adam Ehrenberg,
Bill Fefferman,
Ra\'ul Garc\'ia-Patr\'on,
Christian Gogolin,
Alexey Gorshkov,
Jonas Haferkamp,
Aram Harrow,
Marcel Hinsche,
Marios Ioannou,
Martin Kliesch,
Austin Lund,
Arthur Mehta,
Ashley Montanaro,
Tomoyuki Morimae,
Hakop Pashayan,
Jelmer Renema,
Nicol\'as Quesada,
Robert Raussendorf,
Ingo Roth, 
Martin Schwarz,
and
Barbara Terhal
for numerous exciting and enlightening discussions on quantum random sampling in the past years. 
We would also like to thank 
Sergio Boixo,
Bill Fefferman,
Jonas Haferkamp,
Chao-Yang Lu,
Ingo Roth,
Pedram Roushan, 
and especially 
Abhinav Deshpande and Ulysse Chabaud
for very helpful and extensive comments on drafts of this work.
 
D.~H.\ is specifically grateful to 
Marcel Hinsche for insights into Gaussian boson sampling, 
to Marcel Hinsche and Scott Aaronson for discussions on the hiding problem, 
to Abhinav Deshpande for discussions surrounding the comparison of near-term quantum and large-scale classical algorithms, 
to Ingo Roth and Jonas Helsen for discussions about the XEB fidelity, 
and to 
Bill Fefferman for discussions regarding anticoncentration and the relevance of the total-variation distance. 

D.~H.\ acknowledges financial support from the U.S. Department of Defense through a QuICS Hartree Fellowship.
This work was completed while D.~H.\ was visiting the Simons Institute for the Theory of Computing.
J.~E.\ acknowledges funding from the DFG (CRC 183, EI 519/21-1), the BMBF (PhoQuant, QPIC-1, DAQC, HYBRID,  
MUNIQC-ATOMS, FermiQP, QSolid), the BMWK (EniQmA, PlanQK), the QuantERA (HQCC), the Munich Quantum Valley (MQV-K8), 
and the Einstein Foundation (Einstein Research Unit on quantum devices).
He has also received funding from the EU's Horizon 2020 research and innovation programme under grant agreement No.~817482 (PASQuanS). \\

\emph{Note: Sections \ref{sec:sampling schemes}--\ref{sec:verification} build on previously unpublished parts of \textcite[Chpts.~1, 2, 3, 7, 8]{hangleiter_sampling_2021}.}

\bibliography{CleanReferences}

\begin{thebibliography}{451}%
\makeatletter
\providecommand \@ifxundefined [1]{%
 \@ifx{#1\undefined}
}%
\providecommand \@ifnum [1]{%
 \ifnum #1\expandafter \@firstoftwo
 \else \expandafter \@secondoftwo
 \fi
}%
\providecommand \@ifx [1]{%
 \ifx #1\expandafter \@firstoftwo
 \else \expandafter \@secondoftwo
 \fi
}%
\providecommand \natexlab [1]{#1}%
\providecommand \enquote  [1]{``#1''}%
\providecommand \bibnamefont  [1]{#1}%
\providecommand \bibfnamefont [1]{#1}%
\providecommand \citenamefont [1]{#1}%
\providecommand \href@noop [0]{\@secondoftwo}%
\providecommand \href [0]{\begingroup \@sanitize@url \@href}%
\providecommand \@href[1]{\@@startlink{#1}\@@href}%
\providecommand \@@href[1]{\endgroup#1\@@endlink}%
\providecommand \@sanitize@url [0]{\catcode `\\12\catcode `\$12\catcode
  `\&12\catcode `\#12\catcode `\^12\catcode `\_12\catcode `\%12\relax}%
\providecommand \@@startlink[1]{}%
\providecommand \@@endlink[0]{}%
\providecommand \url  [0]{\begingroup\@sanitize@url \@url }%
\providecommand \@url [1]{\endgroup\@href {#1}{\urlprefix }}%
\providecommand \urlprefix  [0]{URL }%
\providecommand \Eprint [0]{\href }%
\providecommand \doibase [0]{https://doi.org/}%
\providecommand \selectlanguage [0]{\@gobble}%
\providecommand \bibinfo  [0]{\@secondoftwo}%
\providecommand \bibfield  [0]{\@secondoftwo}%
\providecommand \translation [1]{[#1]}%
\providecommand \BibitemOpen [0]{}%
\providecommand \bibitemStop [0]{}%
\providecommand \bibitemNoStop [0]{.\EOS\space}%
\providecommand \EOS [0]{\spacefactor3000\relax}%
\providecommand \BibitemShut  [1]{\csname bibitem#1\endcsname}%
\let\auto@bib@innerbib\@empty
\bibitem [{\citenamefont {Aaghabali}\ \emph {et~al.}(2015)\citenamefont
  {Aaghabali}, \citenamefont {Akbari}, \citenamefont {Friedland}, \citenamefont
  {Markstr{\"o}m},\ and\ \citenamefont {Tajfirouz}}]{aaghabali_upper_2015}%
  \BibitemOpen
  \bibfield  {author} {\bibinfo {author} {\bibnamefont {Aaghabali},
  \bibfnamefont {M}}, \bibinfo {author} {\bibfnamefont {S.}~\bibnamefont
  {Akbari}}, \bibinfo {author} {\bibfnamefont {S.}~\bibnamefont {Friedland}},
  \bibinfo {author} {\bibfnamefont {K.}~\bibnamefont {Markstr{\"o}m}}, and\
  \bibinfo {author} {\bibfnamefont {Z.}~\bibnamefont {Tajfirouz}}} (\bibinfo
  {year} {2015}),\ \bibfield  {title} {\enquote {\bibinfo {title} {Upper bounds
  on the number of perfect matchings and directed 2-factors in graphs with
  given number of vertices and edges},}\ }\href
  {https://doi.org/10.1016/j.ejc.2014.11.001} {\bibfield  {journal} {\bibinfo
  {journal} {Europ. J. Comb.}\ }\textbf {\bibinfo {volume} {45}},\ \bibinfo
  {pages} {132--144}}\BibitemShut {NoStop}%
\bibitem [{\citenamefont {Aaronson}(2005)}]{Aaronson-ProcRS-2005}%
  \BibitemOpen
  \bibfield  {author} {\bibinfo {author} {\bibnamefont {Aaronson},
  \bibfnamefont {S}}} (\bibinfo {year} {2005}),\ \bibfield  {title} {\enquote
  {\bibinfo {title} {Quantum computing, post-selection, and probabilistic
  polynomial-time},}\ }\href {https://doi.org/10.1098/rspa.2005.1546}
  {\bibfield  {journal} {\bibinfo  {journal} {Proc. Roy. Soc. A}\ }\textbf
  {\bibinfo {volume} {461}},\ \bibinfo {pages} {3473--3482}}\BibitemShut
  {NoStop}%
\bibitem [{\citenamefont {Aaronson}(2010)}]{aaronson_bqp_2010}%
  \BibitemOpen
  \bibfield  {author} {\bibinfo {author} {\bibnamefont {Aaronson},
  \bibfnamefont {S}}} (\bibinfo {year} {2010}),\ \bibfield  {title} {\enquote
  {\bibinfo {title} {{{BQP}} and the polynomial hierarchy},}\ }in\ \href
  {https://doi.org/10.1145/1806689.1806711} {\emph {\bibinfo {booktitle} {Proc.
  42nd {{ACM}} Symp. {{Th.}} Comp.}}},\ \bibinfo {series and number} {{{STOC}}
  '10}\ (\bibinfo  {publisher} {{Association for Computing Machinery}},\
  \bibinfo {address} {{New York, NY, USA}})\ pp.\ \bibinfo {pages}
  {141--150}\BibitemShut {NoStop}%
\bibitem [{\citenamefont {Aaronson}(2018)}]{aaronson_aspects_QCrypt}%
  \BibitemOpen
  \bibfield  {author} {\bibinfo {author} {\bibnamefont {Aaronson},
  \bibfnamefont {S}}} (\bibinfo {year} {2018}),\ \href
  {https://www.youtube.com/watch?v=hf7-Elx1Y4w} {\enquote {\bibinfo {title}
  {Certified randomness from quantum supremacy},}\ }\bibinfo {note}
  {{PowerPoint} presentation, accessed on 2022-05-14}\BibitemShut {NoStop}%
\bibitem [{\citenamefont {Aaronson}(2019)}]{aaronson_aspects_2019}%
  \BibitemOpen
  \bibfield  {author} {\bibinfo {author} {\bibnamefont {Aaronson},
  \bibfnamefont {S}}} (\bibinfo {year} {2019}),\ \href
  {https://www.scottaaronson.com/talks/certrand2.ppt} {\enquote {\bibinfo
  {title} {Aspects of certified randomness from quantum supremacy},}\ }\bibinfo
  {note} {{PowerPoint} presentation, accessed on 2020-09-04}\BibitemShut
  {NoStop}%
\bibitem [{\citenamefont {Aaronson}\ and\ \citenamefont
  {Arkhipov}(2013)}]{aaronson_computational_2010}%
  \BibitemOpen
  \bibfield  {author} {\bibinfo {author} {\bibnamefont {Aaronson},
  \bibfnamefont {S}}, and\ \bibinfo {author} {\bibfnamefont {A.}~\bibnamefont
  {Arkhipov}}} (\bibinfo {year} {2013}),\ \bibfield  {title} {\enquote
  {\bibinfo {title} {The computational complexity of linear optics},}\ }\href
  {https://doi.org/10.4086/toc.2013.v009a004} {\bibfield  {journal} {\bibinfo
  {journal} {Th. Comp.}\ }\textbf {\bibinfo {volume} {9}},\ \bibinfo {pages}
  {143--252}}\BibitemShut {NoStop}%
\bibitem [{\citenamefont {Aaronson}\ and\ \citenamefont
  {Arkhipov}(2014)}]{aaronson_bosonsampling_2013}%
  \BibitemOpen
  \bibfield  {author} {\bibinfo {author} {\bibnamefont {Aaronson},
  \bibfnamefont {S}}, and\ \bibinfo {author} {\bibfnamefont {A.}~\bibnamefont
  {Arkhipov}}} (\bibinfo {year} {2014}),\ \bibfield  {title} {\enquote
  {\bibinfo {title} {Bosonsampling is far from uniform},}\ }\href
  {https://doi.org/10.26421/QIC14.15-16-7} {\bibfield  {journal} {\bibinfo
  {journal} {Quantum Inf. Comp.}\ }\textbf {\bibinfo {volume} {14}},\ \bibinfo
  {pages} {1383--1423}},\ \Eprint {https://arxiv.org/abs/1309.7460}
  {arxiv:1309.7460} \BibitemShut {NoStop}%
\bibitem [{\citenamefont {Aaronson}\ \emph {et~al.}(2016)\citenamefont
  {Aaronson}, \citenamefont {Bouland}, \citenamefont {Kuperberg},\ and\
  \citenamefont {Mehraban}}]{aaronson_computational_2016}%
  \BibitemOpen
  \bibfield  {author} {\bibinfo {author} {\bibnamefont {Aaronson},
  \bibfnamefont {S}}, \bibinfo {author} {\bibfnamefont {A.}~\bibnamefont
  {Bouland}}, \bibinfo {author} {\bibfnamefont {G.}~\bibnamefont {Kuperberg}},
  and\ \bibinfo {author} {\bibfnamefont {S.}~\bibnamefont {Mehraban}}}
  (\bibinfo {year} {2016}),\ \bibfield  {title} {\enquote {\bibinfo {title}
  {The computational complexity of ball permutations},}\ }\href
  {http://arxiv.org/abs/1610.06646} {\ }\Eprint
  {https://arxiv.org/abs/1610.06646} {arxiv:1610.06646} \BibitemShut {NoStop}%
\bibitem [{\citenamefont {Aaronson}\ and\ \citenamefont
  {Brod}(2016)}]{PhysRevA.93.012335}%
  \BibitemOpen
  \bibfield  {author} {\bibinfo {author} {\bibnamefont {Aaronson},
  \bibfnamefont {S}}, and\ \bibinfo {author} {\bibfnamefont {D.~J.}\
  \bibnamefont {Brod}}} (\bibinfo {year} {2016}),\ \bibfield  {title} {\enquote
  {\bibinfo {title} {Bosonsampling with lost photons},}\ }\href
  {https://doi.org/10.1103/PhysRevA.93.012335} {\bibfield  {journal} {\bibinfo
  {journal} {Phys. Rev. A}\ }\textbf {\bibinfo {volume} {93}},\ \bibinfo
  {pages} {012335}}\BibitemShut {NoStop}%
\bibitem [{\citenamefont {Aaronson}\ and\ \citenamefont
  {Chen}(2017)}]{aaronson_complexity-theoretic_2017}%
  \BibitemOpen
  \bibfield  {author} {\bibinfo {author} {\bibnamefont {Aaronson},
  \bibfnamefont {S}}, and\ \bibinfo {author} {\bibfnamefont {L.}~\bibnamefont
  {Chen}}} (\bibinfo {year} {2017}),\ \bibfield  {title} {\enquote {\bibinfo
  {title} {Complexity-{theoretic} {foundations} of {quantum} {supremacy}
  {experiments}},}\ }in\ \href {https://doi.org/10.4230/LIPIcs.CCC.2017.22}
  {\emph {\bibinfo {booktitle} {32nd {Computational} {Complexity} {Conference}
  ({CCC} 2017)}}},\ \bibinfo {series} {Leibniz {International} {Proceedings} in
  {Informatics} ({LIPIcs})}, Vol.~\bibinfo {volume} {79},\ \bibinfo {editor}
  {edited by\ \bibinfo {editor} {\bibfnamefont {Ryan}\ \bibnamefont
  {O'Donnell}}}\ (\bibinfo  {publisher} {Schloss Dagstuhl-Leibniz-Zentrum fuer
  Informatik},\ \bibinfo {address} {Dagstuhl, Germany})\ pp.\ \bibinfo {pages}
  {22:1--22:67},\ \Eprint {https://arxiv.org/abs/1612.05903} {arXiv:1612.05903}
  \BibitemShut {NoStop}%
\bibitem [{\citenamefont {Aaronson}\ and\ \citenamefont
  {Gottesman}(2004)}]{aaronson_improved_2004}%
  \BibitemOpen
  \bibfield  {author} {\bibinfo {author} {\bibnamefont {Aaronson},
  \bibfnamefont {S}}, and\ \bibinfo {author} {\bibfnamefont {D.}~\bibnamefont
  {Gottesman}}} (\bibinfo {year} {2004}),\ \bibfield  {title} {\enquote
  {\bibinfo {title} {Improved simulation of stabilizer circuits},}\ }\href
  {https://doi.org/10.1103/PhysRevA.70.052328} {\bibfield  {journal} {\bibinfo
  {journal} {Phys. Rev. A}\ }\textbf {\bibinfo {volume} {70}},\ \bibinfo
  {pages} {052328}}\BibitemShut {NoStop}%
\bibitem [{\citenamefont {Aaronson}\ and\ \citenamefont
  {Gunn}(2019)}]{aaronson_classical_2019}%
  \BibitemOpen
  \bibfield  {author} {\bibinfo {author} {\bibnamefont {Aaronson},
  \bibfnamefont {S}}, and\ \bibinfo {author} {\bibfnamefont {S.}~\bibnamefont
  {Gunn}}} (\bibinfo {year} {2019}),\ \bibfield  {title} {\enquote {\bibinfo
  {title} {On the classical hardness of spoofing linear cross-entropy
  benchmarking},}\ }\href {https://arxiv.org/abs/1910.12085} {\ }\Eprint
  {https://arxiv.org/abs/1910.12085} {arXiv:1910.12085} \BibitemShut {NoStop}%
\bibitem [{\citenamefont {Aaronson}\ and\ \citenamefont
  {Hance}(2012)}]{aaronson_generalizing_2012}%
  \BibitemOpen
  \bibfield  {author} {\bibinfo {author} {\bibnamefont {Aaronson},
  \bibfnamefont {S}}, and\ \bibinfo {author} {\bibfnamefont {T.}~\bibnamefont
  {Hance}}} (\bibinfo {year} {2012}),\ \bibfield  {title} {\enquote {\bibinfo
  {title} {Generalizing and derandomizing {{Gurvits}}'s approximation algorithm
  for the permanent},}\ }\href@noop {} {\ }\Eprint
  {https://arxiv.org/abs/1212.0025} {arXiv:1212.0025} \BibitemShut {NoStop}%
\bibitem [{\citenamefont {Acharya}\ \emph {et~al.}(2022)\citenamefont
  {Acharya}, \citenamefont {Aleiner}, \citenamefont {Allen}, \citenamefont
  {Andersen}, \citenamefont {Ansmann}, \citenamefont {Arute}, \citenamefont
  {Arya}, \citenamefont {Asfaw}, \citenamefont {Atalaya}, \citenamefont
  {Babbush}, \citenamefont {Bacon}, \citenamefont {Bardin}, \citenamefont
  {Basso}, \citenamefont {Bengtsson}, \citenamefont {Boixo}, \citenamefont
  {Bortoli}, \citenamefont {Bourassa}, \citenamefont {Bovaird}, \citenamefont
  {Brill}, \citenamefont {Broughton}, \citenamefont {Buckley}, \citenamefont
  {Buell}, \citenamefont {Burger}, \citenamefont {Burkett}, \citenamefont
  {Bushnell}, \citenamefont {Chen}, \citenamefont {Chen}, \citenamefont
  {Chiaro}, \citenamefont {Cogan}, \citenamefont {Collins}, \citenamefont
  {Conner}, \citenamefont {Courtney}, \citenamefont {Crook}, \citenamefont
  {Curtin}, \citenamefont {Debroy}, \citenamefont {Barba}, \citenamefont
  {Demura}, \citenamefont {Dunsworth}, \citenamefont {Eppens}, \citenamefont
  {Erickson}, \citenamefont {Faoro}, \citenamefont {Farhi}, \citenamefont
  {Fatemi}, \citenamefont {Burgos}, \citenamefont {Forati}, \citenamefont
  {Fowler}, \citenamefont {Foxen}, \citenamefont {Giang}, \citenamefont
  {Gidney}, \citenamefont {Gilboa}, \citenamefont {Giustina}, \citenamefont
  {Dau}, \citenamefont {Gross}, \citenamefont {Habegger}, \citenamefont
  {Hamilton}, \citenamefont {Harrigan}, \citenamefont {Harrington},
  \citenamefont {Higgott}, \citenamefont {Hilton}, \citenamefont {Hoffmann},
  \citenamefont {Hong}, \citenamefont {Huang}, \citenamefont {Huff},
  \citenamefont {Huggins}, \citenamefont {Ioffe}, \citenamefont {Isakov},
  \citenamefont {Iveland}, \citenamefont {Jeffrey}, \citenamefont {Jiang},
  \citenamefont {Jones}, \citenamefont {Juhas}, \citenamefont {Kafri},
  \citenamefont {Kechedzhi}, \citenamefont {Kelly}, \citenamefont {Khattar},
  \citenamefont {Khezri}, \citenamefont {Kieferov{\'a}}, \citenamefont {Kim},
  \citenamefont {Kitaev}, \citenamefont {Klimov}, \citenamefont {Klots},
  \citenamefont {Korotkov}, \citenamefont {Kostritsa}, \citenamefont
  {Kreikebaum}, \citenamefont {Landhuis}, \citenamefont {Laptev}, \citenamefont
  {Lau}, \citenamefont {Laws}, \citenamefont {Lee}, \citenamefont {Lee},
  \citenamefont {Lester}, \citenamefont {Lill}, \citenamefont {Liu},
  \citenamefont {Locharla}, \citenamefont {Lucero}, \citenamefont {Malone},
  \citenamefont {Marshall}, \citenamefont {Martin}, \citenamefont {McClean},
  \citenamefont {Mccourt}, \citenamefont {McEwen}, \citenamefont {Megrant},
  \citenamefont {Costa}, \citenamefont {Mi}, \citenamefont {Miao},
  \citenamefont {Mohseni}, \citenamefont {Montazeri}, \citenamefont {Morvan},
  \citenamefont {Mount}, \citenamefont {Mruczkiewicz}, \citenamefont {Naaman},
  \citenamefont {Neeley}, \citenamefont {Neill}, \citenamefont {Nersisyan},
  \citenamefont {Neven}, \citenamefont {Newman}, \citenamefont {Ng},
  \citenamefont {Nguyen}, \citenamefont {Nguyen}, \citenamefont {Niu},
  \citenamefont {O'Brien}, \citenamefont {Opremcak}, \citenamefont {Platt},
  \citenamefont {Petukhov}, \citenamefont {Potter}, \citenamefont {Pryadko},
  \citenamefont {Quintana}, \citenamefont {Roushan}, \citenamefont {Rubin},
  \citenamefont {Saei}, \citenamefont {Sank}, \citenamefont {Sankaragomathi},
  \citenamefont {Satzinger}, \citenamefont {Schurkus}, \citenamefont
  {Schuster}, \citenamefont {Shearn}, \citenamefont {Shorter}, \citenamefont
  {Shvarts}, \citenamefont {Skruzny}, \citenamefont {Smelyanskiy},
  \citenamefont {Smith}, \citenamefont {Sterling}, \citenamefont {Strain},
  \citenamefont {Szalay}, \citenamefont {Torres}, \citenamefont {Vidal},
  \citenamefont {Villalonga}, \citenamefont {Heidweiller}, \citenamefont
  {White}, \citenamefont {Xing}, \citenamefont {Yao}, \citenamefont {Yeh},
  \citenamefont {Yoo}, \citenamefont {Young}, \citenamefont {Zalcman},
  \citenamefont {Zhang},\ and\ \citenamefont {Zhu}}]{acharya_suppressing_2022}%
  \BibitemOpen
  \bibfield  {author} {\bibinfo {author} {\bibnamefont {Acharya}, \bibfnamefont
  {R}}, \bibinfo {author} {\bibfnamefont {I.}~\bibnamefont {Aleiner}}, \bibinfo
  {author} {\bibfnamefont {R.}~\bibnamefont {Allen}}, \bibinfo {author}
  {\bibfnamefont {T.~I.}\ \bibnamefont {Andersen}}, \bibinfo {author}
  {\bibfnamefont {M.}~\bibnamefont {Ansmann}}, \bibinfo {author} {\bibfnamefont
  {F.}~\bibnamefont {Arute}}, \bibinfo {author} {\bibfnamefont
  {K.}~\bibnamefont {Arya}}, \bibinfo {author} {\bibfnamefont {A.}~\bibnamefont
  {Asfaw}}, \bibinfo {author} {\bibfnamefont {J.}~\bibnamefont {Atalaya}},
  \bibinfo {author} {\bibfnamefont {R.}~\bibnamefont {Babbush}}, \bibinfo
  {author} {\bibfnamefont {D.}~\bibnamefont {Bacon}}, \bibinfo {author}
  {\bibfnamefont {J.~C.}\ \bibnamefont {Bardin}}, \bibinfo {author}
  {\bibfnamefont {J.}~\bibnamefont {Basso}}, \bibinfo {author} {\bibfnamefont
  {A.}~\bibnamefont {Bengtsson}}, \bibinfo {author} {\bibfnamefont
  {S.}~\bibnamefont {Boixo}}, \bibinfo {author} {\bibfnamefont
  {G.}~\bibnamefont {Bortoli}}, \bibinfo {author} {\bibfnamefont
  {A.}~\bibnamefont {Bourassa}}, \bibinfo {author} {\bibfnamefont
  {J.}~\bibnamefont {Bovaird}}, \bibinfo {author} {\bibfnamefont
  {L.}~\bibnamefont {Brill}}, \bibinfo {author} {\bibfnamefont
  {M.}~\bibnamefont {Broughton}}, \bibinfo {author} {\bibfnamefont {B.~B.}\
  \bibnamefont {Buckley}}, \bibinfo {author} {\bibfnamefont {D.~A.}\
  \bibnamefont {Buell}}, \bibinfo {author} {\bibfnamefont {T.}~\bibnamefont
  {Burger}}, \bibinfo {author} {\bibfnamefont {B.}~\bibnamefont {Burkett}},
  \bibinfo {author} {\bibfnamefont {N.}~\bibnamefont {Bushnell}}, \bibinfo
  {author} {\bibfnamefont {Y.}~\bibnamefont {Chen}}, \bibinfo {author}
  {\bibfnamefont {Z.}~\bibnamefont {Chen}}, \bibinfo {author} {\bibfnamefont
  {B.}~\bibnamefont {Chiaro}}, \bibinfo {author} {\bibfnamefont
  {J.}~\bibnamefont {Cogan}}, \bibinfo {author} {\bibfnamefont
  {R.}~\bibnamefont {Collins}}, \bibinfo {author} {\bibfnamefont
  {P.}~\bibnamefont {Conner}}, \bibinfo {author} {\bibfnamefont
  {W.}~\bibnamefont {Courtney}}, \bibinfo {author} {\bibfnamefont {A.~L.}\
  \bibnamefont {Crook}}, \bibinfo {author} {\bibfnamefont {B.}~\bibnamefont
  {Curtin}}, \bibinfo {author} {\bibfnamefont {D.~M.}\ \bibnamefont {Debroy}},
  \bibinfo {author} {\bibfnamefont {A.~Del~Toro}\ \bibnamefont {Barba}},
  \bibinfo {author} {\bibfnamefont {S.}~\bibnamefont {Demura}}, \bibinfo
  {author} {\bibfnamefont {A.}~\bibnamefont {Dunsworth}}, \bibinfo {author}
  {\bibfnamefont {D.}~\bibnamefont {Eppens}}, \bibinfo {author} {\bibfnamefont
  {C.}~\bibnamefont {Erickson}}, \bibinfo {author} {\bibfnamefont
  {L.}~\bibnamefont {Faoro}}, \bibinfo {author} {\bibfnamefont
  {E.}~\bibnamefont {Farhi}}, \bibinfo {author} {\bibfnamefont
  {R.}~\bibnamefont {Fatemi}}, \bibinfo {author} {\bibfnamefont {L.~F.}\
  \bibnamefont {Burgos}}, \bibinfo {author} {\bibfnamefont {E.}~\bibnamefont
  {Forati}}, \bibinfo {author} {\bibfnamefont {A.~G.}\ \bibnamefont {Fowler}},
  \bibinfo {author} {\bibfnamefont {B.}~\bibnamefont {Foxen}}, \bibinfo
  {author} {\bibfnamefont {W.}~\bibnamefont {Giang}}, \bibinfo {author}
  {\bibfnamefont {C.}~\bibnamefont {Gidney}}, \bibinfo {author} {\bibfnamefont
  {D.}~\bibnamefont {Gilboa}}, \bibinfo {author} {\bibfnamefont
  {M.}~\bibnamefont {Giustina}}, \bibinfo {author} {\bibfnamefont {A.~G.}\
  \bibnamefont {Dau}}, \bibinfo {author} {\bibfnamefont {J.~A.}\ \bibnamefont
  {Gross}}, \bibinfo {author} {\bibfnamefont {S.}~\bibnamefont {Habegger}},
  \bibinfo {author} {\bibfnamefont {M.~C.}\ \bibnamefont {Hamilton}}, \bibinfo
  {author} {\bibfnamefont {M.~P.}\ \bibnamefont {Harrigan}}, \bibinfo {author}
  {\bibfnamefont {S.~D.}\ \bibnamefont {Harrington}}, \bibinfo {author}
  {\bibfnamefont {O.}~\bibnamefont {Higgott}}, \bibinfo {author} {\bibfnamefont
  {J.}~\bibnamefont {Hilton}}, \bibinfo {author} {\bibfnamefont
  {M.}~\bibnamefont {Hoffmann}}, \bibinfo {author} {\bibfnamefont
  {S.}~\bibnamefont {Hong}}, \bibinfo {author} {\bibfnamefont {T.}~\bibnamefont
  {Huang}}, \bibinfo {author} {\bibfnamefont {A.}~\bibnamefont {Huff}},
  \bibinfo {author} {\bibfnamefont {W.~J.}\ \bibnamefont {Huggins}}, \bibinfo
  {author} {\bibfnamefont {L.~B.}\ \bibnamefont {Ioffe}}, \bibinfo {author}
  {\bibfnamefont {S.~V.}\ \bibnamefont {Isakov}}, \bibinfo {author}
  {\bibfnamefont {J.}~\bibnamefont {Iveland}}, \bibinfo {author} {\bibfnamefont
  {E.}~\bibnamefont {Jeffrey}}, \bibinfo {author} {\bibfnamefont
  {Z.}~\bibnamefont {Jiang}}, \bibinfo {author} {\bibfnamefont
  {C.}~\bibnamefont {Jones}}, \bibinfo {author} {\bibfnamefont
  {P.}~\bibnamefont {Juhas}}, \bibinfo {author} {\bibfnamefont
  {D.}~\bibnamefont {Kafri}}, \bibinfo {author} {\bibfnamefont
  {K.}~\bibnamefont {Kechedzhi}}, \bibinfo {author} {\bibfnamefont
  {J.}~\bibnamefont {Kelly}}, \bibinfo {author} {\bibfnamefont
  {T.}~\bibnamefont {Khattar}}, \bibinfo {author} {\bibfnamefont
  {M.}~\bibnamefont {Khezri}}, \bibinfo {author} {\bibfnamefont
  {M.}~\bibnamefont {Kieferov{\'a}}}, \bibinfo {author} {\bibfnamefont
  {S.}~\bibnamefont {Kim}}, \bibinfo {author} {\bibfnamefont {A.}~\bibnamefont
  {Kitaev}}, \bibinfo {author} {\bibfnamefont {P.~V.}\ \bibnamefont {Klimov}},
  \bibinfo {author} {\bibfnamefont {A.~R.}\ \bibnamefont {Klots}}, \bibinfo
  {author} {\bibfnamefont {A.~N.}\ \bibnamefont {Korotkov}}, \bibinfo {author}
  {\bibfnamefont {F.}~\bibnamefont {Kostritsa}}, \bibinfo {author}
  {\bibfnamefont {J.~Mark}\ \bibnamefont {Kreikebaum}}, \bibinfo {author}
  {\bibfnamefont {D.}~\bibnamefont {Landhuis}}, \bibinfo {author}
  {\bibfnamefont {P.}~\bibnamefont {Laptev}}, \bibinfo {author} {\bibfnamefont
  {K.-M.}\ \bibnamefont {Lau}}, \bibinfo {author} {\bibfnamefont
  {L.}~\bibnamefont {Laws}}, \bibinfo {author} {\bibfnamefont {J.}~\bibnamefont
  {Lee}}, \bibinfo {author} {\bibfnamefont {K.}~\bibnamefont {Lee}}, \bibinfo
  {author} {\bibfnamefont {B.~J.}\ \bibnamefont {Lester}}, \bibinfo {author}
  {\bibfnamefont {A.}~\bibnamefont {Lill}}, \bibinfo {author} {\bibfnamefont
  {W.}~\bibnamefont {Liu}}, \bibinfo {author} {\bibfnamefont {A.}~\bibnamefont
  {Locharla}}, \bibinfo {author} {\bibfnamefont {E.}~\bibnamefont {Lucero}},
  \bibinfo {author} {\bibfnamefont {F.~D.}\ \bibnamefont {Malone}}, \bibinfo
  {author} {\bibfnamefont {J.}~\bibnamefont {Marshall}}, \bibinfo {author}
  {\bibfnamefont {O.}~\bibnamefont {Martin}}, \bibinfo {author} {\bibfnamefont
  {J.~R.}\ \bibnamefont {McClean}}, \bibinfo {author} {\bibfnamefont
  {T.}~\bibnamefont {Mccourt}}, \bibinfo {author} {\bibfnamefont
  {M.}~\bibnamefont {McEwen}}, \bibinfo {author} {\bibfnamefont
  {A.}~\bibnamefont {Megrant}}, \bibinfo {author} {\bibfnamefont {B.~Meurer}\
  \bibnamefont {Costa}}, \bibinfo {author} {\bibfnamefont {X.}~\bibnamefont
  {Mi}}, \bibinfo {author} {\bibfnamefont {K.~C.}\ \bibnamefont {Miao}},
  \bibinfo {author} {\bibfnamefont {M.}~\bibnamefont {Mohseni}}, \bibinfo
  {author} {\bibfnamefont {S.}~\bibnamefont {Montazeri}}, \bibinfo {author}
  {\bibfnamefont {A.}~\bibnamefont {Morvan}}, \bibinfo {author} {\bibfnamefont
  {E.}~\bibnamefont {Mount}}, \bibinfo {author} {\bibfnamefont
  {W.}~\bibnamefont {Mruczkiewicz}}, \bibinfo {author} {\bibfnamefont
  {O.}~\bibnamefont {Naaman}}, \bibinfo {author} {\bibfnamefont
  {M.}~\bibnamefont {Neeley}}, \bibinfo {author} {\bibfnamefont
  {C.}~\bibnamefont {Neill}}, \bibinfo {author} {\bibfnamefont
  {A.}~\bibnamefont {Nersisyan}}, \bibinfo {author} {\bibfnamefont
  {H.}~\bibnamefont {Neven}}, \bibinfo {author} {\bibfnamefont
  {M.}~\bibnamefont {Newman}}, \bibinfo {author} {\bibfnamefont {J.~H.}\
  \bibnamefont {Ng}}, \bibinfo {author} {\bibfnamefont {A.}~\bibnamefont
  {Nguyen}}, \bibinfo {author} {\bibfnamefont {M.}~\bibnamefont {Nguyen}},
  \bibinfo {author} {\bibfnamefont {M.~Y.}\ \bibnamefont {Niu}}, \bibinfo
  {author} {\bibfnamefont {T.~E.}\ \bibnamefont {O'Brien}}, \bibinfo {author}
  {\bibfnamefont {A.}~\bibnamefont {Opremcak}}, \bibinfo {author}
  {\bibfnamefont {J.}~\bibnamefont {Platt}}, \bibinfo {author} {\bibfnamefont
  {A.}~\bibnamefont {Petukhov}}, \bibinfo {author} {\bibfnamefont
  {R.}~\bibnamefont {Potter}}, \bibinfo {author} {\bibfnamefont {L.~P.}\
  \bibnamefont {Pryadko}}, \bibinfo {author} {\bibfnamefont {C.}~\bibnamefont
  {Quintana}}, \bibinfo {author} {\bibfnamefont {P.}~\bibnamefont {Roushan}},
  \bibinfo {author} {\bibfnamefont {N.~C.}\ \bibnamefont {Rubin}}, \bibinfo
  {author} {\bibfnamefont {N.}~\bibnamefont {Saei}}, \bibinfo {author}
  {\bibfnamefont {D.}~\bibnamefont {Sank}}, \bibinfo {author} {\bibfnamefont
  {K.}~\bibnamefont {Sankaragomathi}}, \bibinfo {author} {\bibfnamefont
  {K.~J.}\ \bibnamefont {Satzinger}}, \bibinfo {author} {\bibfnamefont {H.~F.}\
  \bibnamefont {Schurkus}}, \bibinfo {author} {\bibfnamefont {C.}~\bibnamefont
  {Schuster}}, \bibinfo {author} {\bibfnamefont {M.~J.}\ \bibnamefont
  {Shearn}}, \bibinfo {author} {\bibfnamefont {A.}~\bibnamefont {Shorter}},
  \bibinfo {author} {\bibfnamefont {V.}~\bibnamefont {Shvarts}}, \bibinfo
  {author} {\bibfnamefont {J.}~\bibnamefont {Skruzny}}, \bibinfo {author}
  {\bibfnamefont {V.}~\bibnamefont {Smelyanskiy}}, \bibinfo {author}
  {\bibfnamefont {W.~C.}\ \bibnamefont {Smith}}, \bibinfo {author}
  {\bibfnamefont {G.}~\bibnamefont {Sterling}}, \bibinfo {author}
  {\bibfnamefont {D.}~\bibnamefont {Strain}}, \bibinfo {author} {\bibfnamefont
  {M.}~\bibnamefont {Szalay}}, \bibinfo {author} {\bibfnamefont
  {A.}~\bibnamefont {Torres}}, \bibinfo {author} {\bibfnamefont
  {G.}~\bibnamefont {Vidal}}, \bibinfo {author} {\bibfnamefont
  {B.}~\bibnamefont {Villalonga}}, \bibinfo {author} {\bibfnamefont {C.~V.}\
  \bibnamefont {Heidweiller}}, \bibinfo {author} {\bibfnamefont
  {T.}~\bibnamefont {White}}, \bibinfo {author} {\bibfnamefont
  {C.}~\bibnamefont {Xing}}, \bibinfo {author} {\bibfnamefont {Z.~J.}\
  \bibnamefont {Yao}}, \bibinfo {author} {\bibfnamefont {P.}~\bibnamefont
  {Yeh}}, \bibinfo {author} {\bibfnamefont {J.}~\bibnamefont {Yoo}}, \bibinfo
  {author} {\bibfnamefont {G.}~\bibnamefont {Young}}, \bibinfo {author}
  {\bibfnamefont {A.}~\bibnamefont {Zalcman}}, \bibinfo {author} {\bibfnamefont
  {Y.}~\bibnamefont {Zhang}}, and\ \bibinfo {author} {\bibfnamefont
  {N.}~\bibnamefont {Zhu}}} (\bibinfo {year} {2022}),\ \bibfield  {title}
  {\enquote {\bibinfo {title} {Suppressing quantum errors by scaling a surface
  code logical qubit},}\ }\href@noop {} {\ }\Eprint
  {https://arxiv.org/abs/2207.06431} {arXiv:2207.06431} \BibitemShut {NoStop}%
\bibitem [{\citenamefont {Ac\'in}\ and\ \citenamefont
  {Masanes}(2016)}]{acincertified2016}%
  \BibitemOpen
  \bibfield  {author} {\bibinfo {author} {\bibnamefont {Ac\'in}, \bibfnamefont
  {A}}, and\ \bibinfo {author} {\bibfnamefont {L.}~\bibnamefont {Masanes}}}
  (\bibinfo {year} {2016}),\ \bibfield  {title} {\enquote {\bibinfo {title}
  {Certified randomness in quantum physics},}\ }\href
  {https://doi.org/10.1038/nature20119} {\bibfield  {journal} {\bibinfo
  {journal} {Nature}\ }\textbf {\bibinfo {volume} {540}},\ \bibinfo {pages}
  {213--219}}\BibitemShut {NoStop}%
\bibitem [{\citenamefont {Aharonov}(2003)}]{aharonov_simple_2003}%
  \BibitemOpen
  \bibfield  {author} {\bibinfo {author} {\bibnamefont {Aharonov},
  \bibfnamefont {D}}} (\bibinfo {year} {2003}),\ \bibfield  {title} {\enquote
  {\bibinfo {title} {{A simple proof that Toffoli and Hadamard are quantum
  universal}},}\ }\href@noop {} {\ }\Eprint
  {https://arxiv.org/abs/quant-ph/0301040} {arXiv:quant-ph/0301040}
  \BibitemShut {NoStop}%
\bibitem [{\citenamefont {Aharonov}\ \emph {et~al.}(2009)\citenamefont
  {Aharonov}, \citenamefont {Arad}, \citenamefont {Landau},\ and\ \citenamefont
  {Vazirani}}]{aharonov_detectability_2009}%
  \BibitemOpen
  \bibfield  {author} {\bibinfo {author} {\bibnamefont {Aharonov},
  \bibfnamefont {D}}, \bibinfo {author} {\bibfnamefont {I.}~\bibnamefont
  {Arad}}, \bibinfo {author} {\bibfnamefont {Z.}~\bibnamefont {Landau}}, and\
  \bibinfo {author} {\bibfnamefont {U.}~\bibnamefont {Vazirani}}} (\bibinfo
  {year} {2009}),\ \bibfield  {title} {\enquote {\bibinfo {title} {The
  detectability lemma and quantum gap amplification},}\ }in\ \href
  {https://doi.org/10.1145/1536414.1536472} {\emph {\bibinfo {booktitle} {Proc.
  41st Ann. {{ACM}} Symp. {{Th.}} Comp.}}},\ \bibinfo {series and number}
  {{{STOC}} '09}\ (\bibinfo  {publisher} {{Association for Computing
  Machinery}},\ \bibinfo {address} {{New York, NY, USA}})\ pp.\ \bibinfo
  {pages} {417--426},\ \Eprint {https://arxiv.org/abs/0811.3412}
  {arXiv:0811.3412} \BibitemShut {NoStop}%
\bibitem [{\citenamefont {Aharonov}\ and\ \citenamefont
  {{Ben-Or}}(1996)}]{aharonov_polynomial_1996}%
  \BibitemOpen
  \bibfield  {author} {\bibinfo {author} {\bibnamefont {Aharonov},
  \bibfnamefont {D}}, and\ \bibinfo {author} {\bibfnamefont {M.}~\bibnamefont
  {{Ben-Or}}}} (\bibinfo {year} {1996}),\ \bibfield  {title} {\enquote
  {\bibinfo {title} {Polynomial simulations of decohered quantum computers},}\
  }in\ \href {https://doi.org/10.1109/SFCS.1996.548463} {\emph {\bibinfo
  {booktitle} {Proceedings of 37th {{Conference}} on {{Foundations}} of
  {{Computer Science}}}}},\ pp.\ \bibinfo {pages} {46--55}\BibitemShut
  {NoStop}%
\bibitem [{\citenamefont {Aharonov}\ and\ \citenamefont
  {{Ben-Or}}(1997)}]{aharonov_fault-tolerant_1997}%
  \BibitemOpen
  \bibfield  {author} {\bibinfo {author} {\bibnamefont {Aharonov},
  \bibfnamefont {D}}, and\ \bibinfo {author} {\bibfnamefont {M.}~\bibnamefont
  {{Ben-Or}}}} (\bibinfo {year} {1997}),\ \bibfield  {title} {\enquote
  {\bibinfo {title} {Fault-tolerant quantum computation with constant error},}\
  }in\ \href {https://doi.org/10.1145/258533.258579} {\emph {\bibinfo
  {booktitle} {Proc. 29th Ann. {{ACM}} Symp. {{Th.}} Comp.}}},\ \bibinfo
  {series and number} {{{STOC}} '97}\ (\bibinfo  {publisher} {{Association for
  Computing Machinery}},\ \bibinfo {address} {{New York, NY, USA}})\ pp.\
  \bibinfo {pages} {176--188}\BibitemShut {NoStop}%
\bibitem [{\citenamefont {Aharonov}\ \emph {et~al.}(2022)\citenamefont
  {Aharonov}, \citenamefont {Gao}, \citenamefont {Landau}, \citenamefont
  {Liu},\ and\ \citenamefont {Vazirani}}]{aharonov_polynomial-time_2022}%
  \BibitemOpen
  \bibfield  {author} {\bibinfo {author} {\bibnamefont {Aharonov},
  \bibfnamefont {D}}, \bibinfo {author} {\bibfnamefont {X.}~\bibnamefont
  {Gao}}, \bibinfo {author} {\bibfnamefont {Z.}~\bibnamefont {Landau}},
  \bibinfo {author} {\bibfnamefont {Y.}~\bibnamefont {Liu}}, and\ \bibinfo
  {author} {\bibfnamefont {U.}~\bibnamefont {Vazirani}}} (\bibinfo {year}
  {2022}),\ \bibfield  {title} {\enquote {\bibinfo {title} {A polynomial-time
  classical algorithm for noisy random circuit sampling},}\ }\href@noop {} {\
  }\Eprint {https://arxiv.org/abs/2211.03999} {arXiv:2211.03999} \BibitemShut
  {NoStop}%
\bibitem [{\citenamefont {Anari}\ \emph {et~al.}(2017)\citenamefont {Anari},
  \citenamefont {Gurvits}, \citenamefont {Gharan},\ and\ \citenamefont
  {Saberi}}]{anari_simply_2017}%
  \BibitemOpen
  \bibfield  {author} {\bibinfo {author} {\bibnamefont {Anari}, \bibfnamefont
  {N}}, \bibinfo {author} {\bibfnamefont {L.}~\bibnamefont {Gurvits}}, \bibinfo
  {author} {\bibfnamefont {S.~O.}\ \bibnamefont {Gharan}}, and\ \bibinfo
  {author} {\bibfnamefont {A.}~\bibnamefont {Saberi}}} (\bibinfo {year}
  {2017}),\ \bibfield  {title} {\enquote {\bibinfo {title} {Simply
  {{exponential approximation}} of the {{permanent}} of {{positive semidefinite
  matrices}}},}\ }\href@noop {} {\ }\Eprint {https://arxiv.org/abs/1704.03486}
  {arXiv:1704.03486} \BibitemShut {NoStop}%
\bibitem [{\citenamefont {Anshu}\ \emph {et~al.}(2016)\citenamefont {Anshu},
  \citenamefont {Arad},\ and\ \citenamefont {Vidick}}]{anshu_simple_2016}%
  \BibitemOpen
  \bibfield  {author} {\bibinfo {author} {\bibnamefont {Anshu}, \bibfnamefont
  {A}}, \bibinfo {author} {\bibfnamefont {I.}~\bibnamefont {Arad}}, and\
  \bibinfo {author} {\bibfnamefont {T.}~\bibnamefont {Vidick}}} (\bibinfo
  {year} {2016}),\ \bibfield  {title} {\enquote {\bibinfo {title} {Simple proof
  of the detectability lemma and spectral gap amplification},}\ }\href
  {https://doi.org/10.1103/PhysRevB.93.205142} {\bibfield  {journal} {\bibinfo
  {journal} {Phys. Rev. B}\ }\textbf {\bibinfo {volume} {93}},\ \bibinfo
  {pages} {205142}}\BibitemShut {NoStop}%
\bibitem [{\citenamefont {Aolita}\ \emph {et~al.}(2015)\citenamefont {Aolita},
  \citenamefont {Gogolin}, \citenamefont {Kliesch},\ and\ \citenamefont
  {Eisert}}]{aolita_reliable_2015}%
  \BibitemOpen
  \bibfield  {author} {\bibinfo {author} {\bibnamefont {Aolita}, \bibfnamefont
  {L}}, \bibinfo {author} {\bibfnamefont {C.}~\bibnamefont {Gogolin}}, \bibinfo
  {author} {\bibfnamefont {M.}~\bibnamefont {Kliesch}}, and\ \bibinfo {author}
  {\bibfnamefont {J.}~\bibnamefont {Eisert}}} (\bibinfo {year} {2015}),\
  \bibfield  {title} {\enquote {\bibinfo {title} {Reliable quantum
  certification of photonic state preparations},}\ }\href
  {https://doi.org/10.1038/ncomms9498} {\bibfield  {journal} {\bibinfo
  {journal} {Nature Comm.}\ }\textbf {\bibinfo {volume} {6}},\ \bibinfo {pages}
  {8498}}\BibitemShut {NoStop}%
\bibitem [{\citenamefont {Arkhipov}(2015)}]{PhysRevA.92.062326}%
  \BibitemOpen
  \bibfield  {author} {\bibinfo {author} {\bibnamefont {Arkhipov},
  \bibfnamefont {A}}} (\bibinfo {year} {2015}),\ \bibfield  {title} {\enquote
  {\bibinfo {title} {Bosonsampling is robust against small errors in the
  network matrix},}\ }\href {https://doi.org/10.1103/PhysRevA.92.062326}
  {\bibfield  {journal} {\bibinfo  {journal} {Phys. Rev. A}\ }\textbf {\bibinfo
  {volume} {92}},\ \bibinfo {pages} {062326}}\BibitemShut {NoStop}%
\bibitem [{\citenamefont {Arkhipov}\ and\ \citenamefont
  {Kuperberg}(2011)}]{arkhipov_bosonic_2011}%
  \BibitemOpen
  \bibfield  {author} {\bibinfo {author} {\bibnamefont {Arkhipov},
  \bibfnamefont {A}}, and\ \bibinfo {author} {\bibfnamefont {G.}~\bibnamefont
  {Kuperberg}}} (\bibinfo {year} {2011}),\ \bibfield  {title} {\enquote
  {\bibinfo {title} {The bosonic birthday paradox},}\ }\href@noop {} {\
  }\Eprint {https://arxiv.org/abs/1106.0849} {arXiv:1106.0849} \BibitemShut
  {NoStop}%
\bibitem [{\citenamefont {Arora}\ and\ \citenamefont
  {Barak}(2009)}]{arora_computational_2009}%
  \BibitemOpen
  \bibfield  {author} {\bibinfo {author} {\bibnamefont {Arora}, \bibfnamefont
  {S}}, and\ \bibinfo {author} {\bibfnamefont {B.}~\bibnamefont {Barak}}}
  (\bibinfo {year} {2009}),\ \href@noop {} {\emph {\bibinfo {title}
  {Computational complexity a modern approach}}}\ (\bibinfo  {publisher}
  {Cambridge University Press})\BibitemShut {NoStop}%
\bibitem [{\citenamefont {Arrazola}\ and\ \citenamefont
  {Bromley}(2018)}]{arrazola_using_2018}%
  \BibitemOpen
  \bibfield  {author} {\bibinfo {author} {\bibnamefont {Arrazola},
  \bibfnamefont {J~M}}, and\ \bibinfo {author} {\bibfnamefont {T.~R.}\
  \bibnamefont {Bromley}}} (\bibinfo {year} {2018}),\ \bibfield  {title}
  {\enquote {\bibinfo {title} {{Using Gaussian boson sampling to find dense
  subgraphs}},}\ }\href {https://doi.org/10.1103/PhysRevLett.121.030503}
  {\bibfield  {journal} {\bibinfo  {journal} {Phys. Rev. Lett.}\ }\textbf
  {\bibinfo {volume} {121}},\ \bibinfo {pages} {030503}}\BibitemShut {NoStop}%
\bibitem [{\citenamefont {Arrazola}\ \emph {et~al.}(2018)\citenamefont
  {Arrazola}, \citenamefont {Bromley},\ and\ \citenamefont
  {Rebentrost}}]{arrazola_quantum_2018}%
  \BibitemOpen
  \bibfield  {author} {\bibinfo {author} {\bibnamefont {Arrazola},
  \bibfnamefont {J~M}}, \bibinfo {author} {\bibfnamefont {T.~R.}\ \bibnamefont
  {Bromley}}, and\ \bibinfo {author} {\bibfnamefont {P.}~\bibnamefont
  {Rebentrost}}} (\bibinfo {year} {2018}),\ \bibfield  {title} {\enquote
  {\bibinfo {title} {{Quantum approximate optimization with Gaussian boson
  sampling}},}\ }\href {https://doi.org/10.1103/PhysRevA.98.012322} {\bibfield
  {journal} {\bibinfo  {journal} {Phys. Rev. A}\ }\textbf {\bibinfo {volume}
  {98}},\ \bibinfo {pages} {012322}}\BibitemShut {NoStop}%
\bibitem [{\citenamefont {Arute}\ \emph {et~al.}(2019)\citenamefont {Arute},
  \citenamefont {Arya}, \citenamefont {Babbush}, \citenamefont {Bacon},
  \citenamefont {Bardin}, \citenamefont {Barends}, \citenamefont {Biswas},
  \citenamefont {Boixo}, \citenamefont {Brand{\~a}o}, \citenamefont {Buell},
  \citenamefont {Burkett}, \citenamefont {Chen}, \citenamefont {Chen},
  \citenamefont {Chiaro}, \citenamefont {Collins}, \citenamefont {Courtney},
  \citenamefont {Dunsworth}, \citenamefont {Farhi}, \citenamefont {Foxen},
  \citenamefont {Fowler}, \citenamefont {Gidney}, \citenamefont {Giustina},
  \citenamefont {Graff}, \citenamefont {Guerin}, \citenamefont {Habegger},
  \citenamefont {Harrigan}, \citenamefont {Hartmann}, \citenamefont {Ho},
  \citenamefont {Hoffmann}, \citenamefont {Huang}, \citenamefont {Humble},
  \citenamefont {Isakov}, \citenamefont {Jeffrey}, \citenamefont {Jiang},
  \citenamefont {Kafri}, \citenamefont {Kechedzhi}, \citenamefont {Kelly},
  \citenamefont {Klimov}, \citenamefont {Knysh}, \citenamefont {Korotkov},
  \citenamefont {Kostritsa}, \citenamefont {Landhuis}, \citenamefont
  {Lindmark}, \citenamefont {Lucero}, \citenamefont {Lyakh}, \citenamefont
  {Mandra}, \citenamefont {{McClean}}, \citenamefont {{McEwen}}, \citenamefont
  {Megrant}, \citenamefont {Mi}, \citenamefont {Michielsen}, \citenamefont
  {Mohseni}, \citenamefont {Mutus}, \citenamefont {Naaman}, \citenamefont
  {Neeley}, \citenamefont {Neill}, \citenamefont {Niu}, \citenamefont {Ostby},
  \citenamefont {Petukhov}, \citenamefont {Platt}, \citenamefont {Quintana},
  \citenamefont {Rieffel}, \citenamefont {Roushan}, \citenamefont {Rubin},
  \citenamefont {Sank}, \citenamefont {Satzinger}, \citenamefont {Smelyanskiy},
  \citenamefont {Sung}, \citenamefont {Trevithick}, \citenamefont
  {Vainsencher}, \citenamefont {Villalonga}, \citenamefont {White},
  \citenamefont {Yao}, \citenamefont {Yeh}, \citenamefont {Zalcman},
  \citenamefont {Neven},\ and\ \citenamefont {Martinis}}]{arute_quantum_2019}%
  \BibitemOpen
  \bibfield  {author} {\bibinfo {author} {\bibnamefont {Arute}, \bibfnamefont
  {F}}, \bibinfo {author} {\bibfnamefont {K.}~\bibnamefont {Arya}}, \bibinfo
  {author} {\bibfnamefont {R.}~\bibnamefont {Babbush}}, \bibinfo {author}
  {\bibfnamefont {D.}~\bibnamefont {Bacon}}, \bibinfo {author} {\bibfnamefont
  {J.~C.}\ \bibnamefont {Bardin}}, \bibinfo {author} {\bibfnamefont
  {R.}~\bibnamefont {Barends}}, \bibinfo {author} {\bibfnamefont
  {R.}~\bibnamefont {Biswas}}, \bibinfo {author} {\bibfnamefont
  {S.}~\bibnamefont {Boixo}}, \bibinfo {author} {\bibfnamefont {F.~G. S.~L.}\
  \bibnamefont {Brand{\~a}o}}, \bibinfo {author} {\bibfnamefont {D.~A.}\
  \bibnamefont {Buell}}, \bibinfo {author} {\bibfnamefont {B.}~\bibnamefont
  {Burkett}}, \bibinfo {author} {\bibfnamefont {Y.}~\bibnamefont {Chen}},
  \bibinfo {author} {\bibfnamefont {Z.}~\bibnamefont {Chen}}, \bibinfo {author}
  {\bibfnamefont {B.}~\bibnamefont {Chiaro}}, \bibinfo {author} {\bibfnamefont
  {R.}~\bibnamefont {Collins}}, \bibinfo {author} {\bibfnamefont
  {W.}~\bibnamefont {Courtney}}, \bibinfo {author} {\bibfnamefont
  {A.}~\bibnamefont {Dunsworth}}, \bibinfo {author} {\bibfnamefont
  {E.}~\bibnamefont {Farhi}}, \bibinfo {author} {\bibfnamefont
  {B.}~\bibnamefont {Foxen}}, \bibinfo {author} {\bibfnamefont
  {A.}~\bibnamefont {Fowler}}, \bibinfo {author} {\bibfnamefont
  {C.}~\bibnamefont {Gidney}}, \bibinfo {author} {\bibfnamefont
  {M.}~\bibnamefont {Giustina}}, \bibinfo {author} {\bibfnamefont
  {R.}~\bibnamefont {Graff}}, \bibinfo {author} {\bibfnamefont
  {K.}~\bibnamefont {Guerin}}, \bibinfo {author} {\bibfnamefont
  {S.}~\bibnamefont {Habegger}}, \bibinfo {author} {\bibfnamefont {M.~P.}\
  \bibnamefont {Harrigan}}, \bibinfo {author} {\bibfnamefont {M.~J.}\
  \bibnamefont {Hartmann}}, \bibinfo {author} {\bibfnamefont {A.}~\bibnamefont
  {Ho}}, \bibinfo {author} {\bibfnamefont {M.}~\bibnamefont {Hoffmann}},
  \bibinfo {author} {\bibfnamefont {T.}~\bibnamefont {Huang}}, \bibinfo
  {author} {\bibfnamefont {T.~S.}\ \bibnamefont {Humble}}, \bibinfo {author}
  {\bibfnamefont {S.~V.}\ \bibnamefont {Isakov}}, \bibinfo {author}
  {\bibfnamefont {E.}~\bibnamefont {Jeffrey}}, \bibinfo {author} {\bibfnamefont
  {Z.}~\bibnamefont {Jiang}}, \bibinfo {author} {\bibfnamefont
  {D.}~\bibnamefont {Kafri}}, \bibinfo {author} {\bibfnamefont
  {K.}~\bibnamefont {Kechedzhi}}, \bibinfo {author} {\bibfnamefont
  {J.}~\bibnamefont {Kelly}}, \bibinfo {author} {\bibfnamefont {P.~V.}\
  \bibnamefont {Klimov}}, \bibinfo {author} {\bibfnamefont {S.}~\bibnamefont
  {Knysh}}, \bibinfo {author} {\bibfnamefont {A.}~\bibnamefont {Korotkov}},
  \bibinfo {author} {\bibfnamefont {F.}~\bibnamefont {Kostritsa}}, \bibinfo
  {author} {\bibfnamefont {D.}~\bibnamefont {Landhuis}}, \bibinfo {author}
  {\bibfnamefont {M.}~\bibnamefont {Lindmark}}, \bibinfo {author}
  {\bibfnamefont {E.}~\bibnamefont {Lucero}}, \bibinfo {author} {\bibfnamefont
  {D.}~\bibnamefont {Lyakh}}, \bibinfo {author} {\bibfnamefont
  {S.}~\bibnamefont {Mandra}}, \bibinfo {author} {\bibfnamefont {J.~R.}\
  \bibnamefont {{McClean}}}, \bibinfo {author} {\bibfnamefont {M.}~\bibnamefont
  {{McEwen}}}, \bibinfo {author} {\bibfnamefont {A.}~\bibnamefont {Megrant}},
  \bibinfo {author} {\bibfnamefont {X.}~\bibnamefont {Mi}}, \bibinfo {author}
  {\bibfnamefont {K.}~\bibnamefont {Michielsen}}, \bibinfo {author}
  {\bibfnamefont {M.}~\bibnamefont {Mohseni}}, \bibinfo {author} {\bibfnamefont
  {J.}~\bibnamefont {Mutus}}, \bibinfo {author} {\bibfnamefont
  {O.}~\bibnamefont {Naaman}}, \bibinfo {author} {\bibfnamefont
  {M.}~\bibnamefont {Neeley}}, \bibinfo {author} {\bibfnamefont
  {C.}~\bibnamefont {Neill}}, \bibinfo {author} {\bibfnamefont {M.~Y.}\
  \bibnamefont {Niu}}, \bibinfo {author} {\bibfnamefont {E.}~\bibnamefont
  {Ostby}}, \bibinfo {author} {\bibfnamefont {A.}~\bibnamefont {Petukhov}},
  \bibinfo {author} {\bibfnamefont {J.~C.}\ \bibnamefont {Platt}}, \bibinfo
  {author} {\bibfnamefont {C.}~\bibnamefont {Quintana}}, \bibinfo {author}
  {\bibfnamefont {E.~G.}\ \bibnamefont {Rieffel}}, \bibinfo {author}
  {\bibfnamefont {P.}~\bibnamefont {Roushan}}, \bibinfo {author} {\bibfnamefont
  {N.~C.}\ \bibnamefont {Rubin}}, \bibinfo {author} {\bibfnamefont
  {D.}~\bibnamefont {Sank}}, \bibinfo {author} {\bibfnamefont {K.~J.}\
  \bibnamefont {Satzinger}}, \bibinfo {author} {\bibfnamefont {V.}~\bibnamefont
  {Smelyanskiy}}, \bibinfo {author} {\bibfnamefont {K.~J.}\ \bibnamefont
  {Sung}}, \bibinfo {author} {\bibfnamefont {M.~D.}\ \bibnamefont
  {Trevithick}}, \bibinfo {author} {\bibfnamefont {A.}~\bibnamefont
  {Vainsencher}}, \bibinfo {author} {\bibfnamefont {B.}~\bibnamefont
  {Villalonga}}, \bibinfo {author} {\bibfnamefont {T.}~\bibnamefont {White}},
  \bibinfo {author} {\bibfnamefont {Z.~J.}\ \bibnamefont {Yao}}, \bibinfo
  {author} {\bibfnamefont {P.}~\bibnamefont {Yeh}}, \bibinfo {author}
  {\bibfnamefont {A.}~\bibnamefont {Zalcman}}, \bibinfo {author} {\bibfnamefont
  {H.}~\bibnamefont {Neven}}, and\ \bibinfo {author} {\bibfnamefont {J.~M.}\
  \bibnamefont {Martinis}}} (\bibinfo {year} {2019}),\ \bibfield  {title}
  {\enquote {\bibinfo {title} {Quantum supremacy using a programmable
  superconducting processor},}\ }\href
  {https://doi.org/10.1038/s41586-019-1666-5} {\bibfield  {journal} {\bibinfo
  {journal} {Nature}\ }\textbf {\bibinfo {volume} {574}},\ \bibinfo {pages}
  {505--510}}\BibitemShut {NoStop}%
\bibitem [{\citenamefont {Aspect}\ \emph
  {et~al.}(1982{\natexlab{a}})\citenamefont {Aspect}, \citenamefont
  {Dalibard},\ and\ \citenamefont {Roger}}]{aspect_experimental_1982}%
  \BibitemOpen
  \bibfield  {author} {\bibinfo {author} {\bibnamefont {Aspect}, \bibfnamefont
  {A}}, \bibinfo {author} {\bibfnamefont {J.}~\bibnamefont {Dalibard}}, and\
  \bibinfo {author} {\bibfnamefont {G.}~\bibnamefont {Roger}}} (\bibinfo {year}
  {1982}{\natexlab{a}}),\ \bibfield  {title} {\enquote {\bibinfo {title}
  {{Experimental test of Bell's inequalities using time-varying analyzers}},}\
  }\href {https://doi.org/10.1103/PhysRevLett.49.1804} {\bibfield  {journal}
  {\bibinfo  {journal} {Phys. Rev. Lett.}\ }\textbf {\bibinfo {volume} {49}},\
  \bibinfo {pages} {1804--1807}}\BibitemShut {NoStop}%
\bibitem [{\citenamefont {Aspect}\ \emph
  {et~al.}(1982{\natexlab{b}})\citenamefont {Aspect}, \citenamefont
  {Grangier},\ and\ \citenamefont {Roger}}]{aspect_experimental_1982-1}%
  \BibitemOpen
  \bibfield  {author} {\bibinfo {author} {\bibnamefont {Aspect}, \bibfnamefont
  {A}}, \bibinfo {author} {\bibfnamefont {P.}~\bibnamefont {Grangier}}, and\
  \bibinfo {author} {\bibfnamefont {G.}~\bibnamefont {Roger}}} (\bibinfo {year}
  {1982}{\natexlab{b}}),\ \bibfield  {title} {\enquote {\bibinfo {title}
  {{Experimental realization of Einstein-Podolsky-Rosen-Bohm
  gedankenexperiment: A new violation of Bell's inequalities}},}\ }\href
  {https://doi.org/10.1103/PhysRevLett.49.91} {\bibfield  {journal} {\bibinfo
  {journal} {Phys. Rev. Lett.}\ }\textbf {\bibinfo {volume} {49}},\ \bibinfo
  {pages} {91--94}}\BibitemShut {NoStop}%
\bibitem [{\citenamefont {Baez}\ \emph {et~al.}(2020)\citenamefont {Baez},
  \citenamefont {Goihl}, \citenamefont {Haferkamp}, \citenamefont
  {Bermejo-Vega}, \citenamefont {Gluza},\ and\ \citenamefont
  {Eisert}}]{baez_dynamical_2020}%
  \BibitemOpen
  \bibfield  {author} {\bibinfo {author} {\bibnamefont {Baez}, \bibfnamefont
  {M~L}}, \bibinfo {author} {\bibfnamefont {M.}~\bibnamefont {Goihl}}, \bibinfo
  {author} {\bibfnamefont {J.}~\bibnamefont {Haferkamp}}, \bibinfo {author}
  {\bibfnamefont {J.}~\bibnamefont {Bermejo-Vega}}, \bibinfo {author}
  {\bibfnamefont {M.}~\bibnamefont {Gluza}}, and\ \bibinfo {author}
  {\bibfnamefont {J.}~\bibnamefont {Eisert}}} (\bibinfo {year} {2020}),\
  \bibfield  {title} {\enquote {\bibinfo {title} {Dynamical structure factors
  of dynamical quantum simulators},}\ }\href
  {https://doi.org/10.1073/pnas.2006103117} {\bibfield  {journal} {\bibinfo
  {journal} {{PNAS}}\ }\textbf {\bibinfo {volume} {117}},\ \bibinfo {pages}
  {26123--26134}}\BibitemShut {NoStop}%
\bibitem [{\citenamefont {Baiardi}\ \emph {et~al.}(2013)\citenamefont
  {Baiardi}, \citenamefont {Bloino},\ and\ \citenamefont
  {Barone}}]{baiardi_general_2013}%
  \BibitemOpen
  \bibfield  {author} {\bibinfo {author} {\bibnamefont {Baiardi}, \bibfnamefont
  {A}}, \bibinfo {author} {\bibfnamefont {J.}~\bibnamefont {Bloino}}, and\
  \bibinfo {author} {\bibfnamefont {V.}~\bibnamefont {Barone}}} (\bibinfo
  {year} {2013}),\ \bibfield  {title} {\enquote {\bibinfo {title} {General
  {{time dependent approach}} to {{vibronic spectroscopy including
  Franck}}\textendash{{Condon}}, {{Herzberg}}\textendash{{Teller}}, and
  {{Duschinsky effects}}},}\ }\href {https://doi.org/10.1021/ct400450k}
  {\bibfield  {journal} {\bibinfo  {journal} {J. Chem. Theory Comput.}\
  }\textbf {\bibinfo {volume} {9}},\ \bibinfo {pages} {4097--4115}}\BibitemShut
  {NoStop}%
\bibitem [{\citenamefont {Banchi}\ \emph
  {et~al.}(2020{\natexlab{a}})\citenamefont {Banchi}, \citenamefont
  {Fingerhuth}, \citenamefont {Babej}, \citenamefont {Ing},\ and\ \citenamefont
  {Arrazola}}]{MolecularDocking}%
  \BibitemOpen
  \bibfield  {author} {\bibinfo {author} {\bibnamefont {Banchi}, \bibfnamefont
  {L}}, \bibinfo {author} {\bibfnamefont {M.}~\bibnamefont {Fingerhuth}},
  \bibinfo {author} {\bibfnamefont {T.}~\bibnamefont {Babej}}, \bibinfo
  {author} {\bibfnamefont {C.}~\bibnamefont {Ing}}, and\ \bibinfo {author}
  {\bibfnamefont {J.~M.}\ \bibnamefont {Arrazola}}} (\bibinfo {year}
  {2020}{\natexlab{a}}),\ \bibfield  {title} {\enquote {\bibinfo {title}
  {{Molecular docking with Gaussian boson sampling}},}\ }\href
  {https://doi.org/10.1126/sciadv.aax1950} {\bibfield  {journal} {\bibinfo
  {journal} {Science Adv.}\ }\textbf {\bibinfo {volume} {6}},\ \bibinfo {pages}
  {eaax1950}}\BibitemShut {NoStop}%
\bibitem [{\citenamefont {Banchi}\ \emph
  {et~al.}(2020{\natexlab{b}})\citenamefont {Banchi}, \citenamefont {Quesada},\
  and\ \citenamefont {Arrazola}}]{PhysRevA.102.012417}%
  \BibitemOpen
  \bibfield  {author} {\bibinfo {author} {\bibnamefont {Banchi}, \bibfnamefont
  {L}}, \bibinfo {author} {\bibfnamefont {N.}~\bibnamefont {Quesada}}, and\
  \bibinfo {author} {\bibfnamefont {J.~M.}\ \bibnamefont {Arrazola}}} (\bibinfo
  {year} {2020}{\natexlab{b}}),\ \bibfield  {title} {\enquote {\bibinfo {title}
  {{Training Gaussian boson sampling distributions}},}\ }\href
  {https://doi.org/10.1103/PhysRevA.102.012417} {\bibfield  {journal} {\bibinfo
   {journal} {Phys. Rev. A}\ }\textbf {\bibinfo {volume} {102}},\ \bibinfo
  {pages} {012417}}\BibitemShut {NoStop}%
\bibitem [{\citenamefont {Bao}\ \emph {et~al.}(2020)\citenamefont {Bao},
  \citenamefont {Choi},\ and\ \citenamefont {Altman}}]{bao_theory_2019}%
  \BibitemOpen
  \bibfield  {author} {\bibinfo {author} {\bibnamefont {Bao}, \bibfnamefont
  {Y}}, \bibinfo {author} {\bibfnamefont {S.}~\bibnamefont {Choi}}, and\
  \bibinfo {author} {\bibfnamefont {E.}~\bibnamefont {Altman}}} (\bibinfo
  {year} {2020}),\ \bibfield  {title} {\enquote {\bibinfo {title} {Theory of
  the phase transition in random unitary circuits with measurements},}\ }\href
  {https://doi.org/10.1103/PhysRevB.101.104301} {\bibfield  {journal} {\bibinfo
   {journal} {Phys. Rev. B}\ }\textbf {\bibinfo {volume} {101}},\ \bibinfo
  {pages} {104301}}\BibitemShut {NoStop}%
\bibitem [{\citenamefont {Barak}\ \emph {et~al.}(2021)\citenamefont {Barak},
  \citenamefont {Chou},\ and\ \citenamefont {Gao}}]{barak_spoofing_2021}%
  \BibitemOpen
  \bibfield  {author} {\bibinfo {author} {\bibnamefont {Barak}, \bibfnamefont
  {B}}, \bibinfo {author} {\bibfnamefont {C.-N.}\ \bibnamefont {Chou}}, and\
  \bibinfo {author} {\bibfnamefont {X.}~\bibnamefont {Gao}}} (\bibinfo {year}
  {2021}),\ \bibfield  {title} {\enquote {\bibinfo {title} {Spoofing {{linear
  cross-entropy benchmarking}} in {{shallow quantum circuits}}},}\ }in\ \href
  {https://doi.org/10.4230/LIPIcs.ITCS.2021.30} {\emph {\bibinfo {booktitle}
  {12th {{Innovations}} in {{Theoretical Computer Science Conference}}
  ({{ITCS}} 2021)}}},\ \bibinfo {series} {Leibniz {{International Proceedings}}
  in {{Informatics}} ({{LIPIcs}})}, Vol.\ \bibinfo {volume} {185},\ \bibinfo
  {editor} {edited by\ \bibinfo {editor} {\bibfnamefont {James~R.}\
  \bibnamefont {Lee}}}\ (\bibinfo  {publisher} {{Schloss Dagstuhl\textendash
  Leibniz-Zentrum f\"ur Informatik}},\ \bibinfo {address} {{Dagstuhl,
  Germany}})\ pp.\ \bibinfo {pages} {30:1--30:20},\ \Eprint
  {https://arxiv.org/abs/2005.02421} {arXiv:2005.02421} \BibitemShut {NoStop}%
\bibitem [{\citenamefont {Barber}(2012)}]{Barber}%
  \BibitemOpen
  \bibfield  {author} {\bibinfo {author} {\bibnamefont {Barber}, \bibfnamefont
  {D}}} (\bibinfo {year} {2012}),\ \href
  {https://doi.org/10.1017/CBO9780511804779} {\emph {\bibinfo {title} {Bayesian
  reasoning and machine learning}}}\ (\bibinfo {address}
  {Cambridge})\BibitemShut {NoStop}%
\bibitem [{\citenamefont {Barends}\ \emph {et~al.}(2014)\citenamefont
  {Barends}, \citenamefont {Kelly}, \citenamefont {Megrant}, \citenamefont
  {Veitia}, \citenamefont {Sank}, \citenamefont {Jeffrey}, \citenamefont
  {White}, \citenamefont {Mutus}, \citenamefont {Fowler}, \citenamefont
  {Campbell}, \citenamefont {Chen}, \citenamefont {Chen}, \citenamefont
  {Chiaro}, \citenamefont {Dunsworth}, \citenamefont {Neill}, \citenamefont
  {O’Malley}, \citenamefont {Roushan}, \citenamefont {Vainsencher},
  \citenamefont {Wenner}, \citenamefont {Korotkov}, \citenamefont {Cleland},\
  and\ \citenamefont {Martinis}}]{barends_superconducting_2014}%
  \BibitemOpen
  \bibfield  {author} {\bibinfo {author} {\bibnamefont {Barends}, \bibfnamefont
  {R}}, \bibinfo {author} {\bibfnamefont {J.}~\bibnamefont {Kelly}}, \bibinfo
  {author} {\bibfnamefont {A.}~\bibnamefont {Megrant}}, \bibinfo {author}
  {\bibfnamefont {A.}~\bibnamefont {Veitia}}, \bibinfo {author} {\bibfnamefont
  {D.}~\bibnamefont {Sank}}, \bibinfo {author} {\bibfnamefont {E.}~\bibnamefont
  {Jeffrey}}, \bibinfo {author} {\bibfnamefont {T.~C.}\ \bibnamefont {White}},
  \bibinfo {author} {\bibfnamefont {J.}~\bibnamefont {Mutus}}, \bibinfo
  {author} {\bibfnamefont {A.~G.}\ \bibnamefont {Fowler}}, \bibinfo {author}
  {\bibfnamefont {B.}~\bibnamefont {Campbell}}, \bibinfo {author}
  {\bibfnamefont {Y.}~\bibnamefont {Chen}}, \bibinfo {author} {\bibfnamefont
  {Z.}~\bibnamefont {Chen}}, \bibinfo {author} {\bibfnamefont {B.}~\bibnamefont
  {Chiaro}}, \bibinfo {author} {\bibfnamefont {A.}~\bibnamefont {Dunsworth}},
  \bibinfo {author} {\bibfnamefont {C.}~\bibnamefont {Neill}}, \bibinfo
  {author} {\bibfnamefont {P.}~\bibnamefont {O’Malley}}, \bibinfo {author}
  {\bibfnamefont {P.}~\bibnamefont {Roushan}}, \bibinfo {author} {\bibfnamefont
  {A.}~\bibnamefont {Vainsencher}}, \bibinfo {author} {\bibfnamefont
  {J.}~\bibnamefont {Wenner}}, \bibinfo {author} {\bibfnamefont {A.~N.}\
  \bibnamefont {Korotkov}}, \bibinfo {author} {\bibfnamefont {A.~N.}\
  \bibnamefont {Cleland}}, and\ \bibinfo {author} {\bibfnamefont {J.~M.}\
  \bibnamefont {Martinis}}} (\bibinfo {year} {2014}),\ \bibfield  {title}
  {\enquote {\bibinfo {title} {Superconducting quantum circuits at the surface
  code threshold for fault tolerance},}\ }\href
  {https://doi.org/10.1038/nature13171} {\bibfield  {journal} {\bibinfo
  {journal} {Nature}\ }\textbf {\bibinfo {volume} {508}},\ \bibinfo {pages}
  {500--503}}\BibitemShut {NoStop}%
\bibitem [{\citenamefont {Bartolucci}\ \emph {et~al.}(2021)\citenamefont
  {Bartolucci}, \citenamefont {Birchall}, \citenamefont {Bombin}, \citenamefont
  {Cable}, \citenamefont {Dawson}, \citenamefont {Gimeno-Segovia},
  \citenamefont {Johnston}, \citenamefont {Kieling}, \citenamefont {Nickerson},
  \citenamefont {Pant}, \citenamefont {Pastawski}, \citenamefont {Rudolph},\
  and\ \citenamefont {Sparrow}}]{Fusion}%
  \BibitemOpen
  \bibfield  {author} {\bibinfo {author} {\bibnamefont {Bartolucci},
  \bibfnamefont {S}}, \bibinfo {author} {\bibfnamefont {P.}~\bibnamefont
  {Birchall}}, \bibinfo {author} {\bibfnamefont {H.}~\bibnamefont {Bombin}},
  \bibinfo {author} {\bibfnamefont {H.}~\bibnamefont {Cable}}, \bibinfo
  {author} {\bibfnamefont {C.}~\bibnamefont {Dawson}}, \bibinfo {author}
  {\bibfnamefont {M.}~\bibnamefont {Gimeno-Segovia}}, \bibinfo {author}
  {\bibfnamefont {E.}~\bibnamefont {Johnston}}, \bibinfo {author}
  {\bibfnamefont {K.}~\bibnamefont {Kieling}}, \bibinfo {author} {\bibfnamefont
  {N.}~\bibnamefont {Nickerson}}, \bibinfo {author} {\bibfnamefont
  {M.}~\bibnamefont {Pant}}, \bibinfo {author} {\bibfnamefont {F.}~\bibnamefont
  {Pastawski}}, \bibinfo {author} {\bibfnamefont {T.}~\bibnamefont {Rudolph}},
  and\ \bibinfo {author} {\bibfnamefont {C.}~\bibnamefont {Sparrow}}} (\bibinfo
  {year} {2021}),\ \bibfield  {title} {\enquote {\bibinfo {title} {Fusion-based
  quantum computation},}\ }\href@noop {} {\ }\Eprint
  {https://arxiv.org/abs/2101.09310} {arXiv:2101.09310} \BibitemShut {NoStop}%
\bibitem [{\citenamefont {Barvinok}(1999)}]{barvinok_polynomial_1999}%
  \BibitemOpen
  \bibfield  {author} {\bibinfo {author} {\bibnamefont {Barvinok},
  \bibfnamefont {A}}} (\bibinfo {year} {1999}),\ \bibfield  {title} {\enquote
  {\bibinfo {title} {Polynomial {{time algorithms}} to {{approximate
  permanents}} and {{mixed discriminants within}} a {{simply exponential
  factor}}},}\ }\href
  {https://doi.org/10.1002/(SICI)1098-2418(1999010)14:1<29::AID-RSA2>3.0.CO;2-X}
  {\bibfield  {journal} {\bibinfo  {journal} {Rand. Struc. Alg.}\ }\textbf
  {\bibinfo {volume} {14}},\ \bibinfo {pages} {29--61}}\BibitemShut {NoStop}%
\bibitem [{\citenamefont
  {Barvinok}(2016{\natexlab{a}})}]{barvinok_combinatorics_2016}%
  \BibitemOpen
  \bibfield  {author} {\bibinfo {author} {\bibnamefont {Barvinok},
  \bibfnamefont {A}}} (\bibinfo {year} {2016}{\natexlab{a}}),\ \href
  {https://doi.org/10.1007/978-3-319-51829-9} {\emph {\bibinfo {title}
  {Combinatorics and {{complexity}} of {{partition functions}}}}},\ Algorithms
  and {{Combinatorics}}\ (\bibinfo  {publisher} {{Springer International
  Publishing}},\ \bibinfo {address} {{Cham}})\BibitemShut {NoStop}%
\bibitem [{\citenamefont
  {Barvinok}(2016{\natexlab{b}})}]{barvinok_computing_2016}%
  \BibitemOpen
  \bibfield  {author} {\bibinfo {author} {\bibnamefont {Barvinok},
  \bibfnamefont {A}}} (\bibinfo {year} {2016}{\natexlab{b}}),\ \bibfield
  {title} {\enquote {\bibinfo {title} {Computing the {{permanent}} of
  ({{some}}) {{complex matrices}}},}\ }\href
  {https://doi.org/10.1007/s10208-014-9243-7} {\bibfield  {journal} {\bibinfo
  {journal} {Found Comput Math}\ }\textbf {\bibinfo {volume} {16}},\ \bibinfo
  {pages} {329--342}}\BibitemShut {NoStop}%
\bibitem [{\citenamefont {Barvinok}(2017)}]{barvinok_approximating_2017}%
  \BibitemOpen
  \bibfield  {author} {\bibinfo {author} {\bibnamefont {Barvinok},
  \bibfnamefont {A}}} (\bibinfo {year} {2017}),\ \bibfield  {title} {\enquote
  {\bibinfo {title} {{Approximating permanents and Hafnians}},}\ }\href
  {https://doi.org/10.19086/da.1244} {\bibfield  {journal} {\bibinfo  {journal}
  {Discrete Ana.}\ }\textbf {\bibinfo {volume} {2}},\ \bibinfo {pages}
  {34}}\BibitemShut {NoStop}%
\bibitem [{\citenamefont {Barvinok}(2019)}]{barvinok_computing_2019}%
  \BibitemOpen
  \bibfield  {author} {\bibinfo {author} {\bibnamefont {Barvinok},
  \bibfnamefont {A}}} (\bibinfo {year} {2019}),\ \bibfield  {title} {\enquote
  {\bibinfo {title} {Computing permanents of complex diagonally dominant
  matrices and tensors},}\ }\href {https://doi.org/10.1007/s11856-019-1896-0}
  {\bibfield  {journal} {\bibinfo  {journal} {Isr. J. Math.}\ }\textbf
  {\bibinfo {volume} {232}},\ \bibinfo {pages} {931--945}}\BibitemShut
  {NoStop}%
\bibitem [{\citenamefont {Barvinok}(2020)}]{barvinok_remark_2020}%
  \BibitemOpen
  \bibfield  {author} {\bibinfo {author} {\bibnamefont {Barvinok},
  \bibfnamefont {A}}} (\bibinfo {year} {2020}),\ \bibfield  {title} {\enquote
  {\bibinfo {title} {A remark on approximating permanents of positive definite
  matrices},}\ }\href@noop {} {\ }\Eprint {https://arxiv.org/abs/2005.06344}
  {arXiv:2005.06344} \BibitemShut {NoStop}%
\bibitem [{\citenamefont {Barvinok}(1996)}]{barvinok_two_1996}%
  \BibitemOpen
  \bibfield  {author} {\bibinfo {author} {\bibnamefont {Barvinok},
  \bibfnamefont {A~I}}} (\bibinfo {year} {1996}),\ \bibfield  {title} {\enquote
  {\bibinfo {title} {Two {{algorithmic results}} for the {{traveling salesman
  problem}}},}\ }\href {https://doi.org/10.1287/moor.21.1.65} {\bibfield
  {journal} {\bibinfo  {journal} {Math. OR}\ }\textbf {\bibinfo {volume}
  {21}},\ \bibinfo {pages} {65--84}}\BibitemShut {NoStop}%
\bibitem [{\citenamefont {Bassirian}\ \emph {et~al.}(2021)\citenamefont
  {Bassirian}, \citenamefont {Bouland}, \citenamefont {Fefferman},
  \citenamefont {Gunn},\ and\ \citenamefont {Tal}}]{bassirian_certified_2021}%
  \BibitemOpen
  \bibfield  {author} {\bibinfo {author} {\bibnamefont {Bassirian},
  \bibfnamefont {R}}, \bibinfo {author} {\bibfnamefont {A.}~\bibnamefont
  {Bouland}}, \bibinfo {author} {\bibfnamefont {B.}~\bibnamefont {Fefferman}},
  \bibinfo {author} {\bibfnamefont {S.}~\bibnamefont {Gunn}}, and\ \bibinfo
  {author} {\bibfnamefont {A.}~\bibnamefont {Tal}}} (\bibinfo {year} {2021}),\
  \bibfield  {title} {\enquote {\bibinfo {title} {On {{certified randomness}}
  from {{quantum advantage experiments}}},}\ }\href@noop {} {\ }\Eprint
  {https://arxiv.org/abs/2111.14846} {arXiv:2111.14846} \BibitemShut {NoStop}%
\bibitem [{\citenamefont {Beaver}\ and\ \citenamefont
  {Feigenbaum}(1990)}]{beaver_hiding_1990}%
  \BibitemOpen
  \bibfield  {author} {\bibinfo {author} {\bibnamefont {Beaver}, \bibfnamefont
  {D}}, and\ \bibinfo {author} {\bibfnamefont {J.}~\bibnamefont {Feigenbaum}}}
  (\bibinfo {year} {1990}),\ \bibfield  {title} {\enquote {\bibinfo {title}
  {Hiding instances in multi-oracle queries},}\ }in\ \href
  {https://doi.org/10.1007/3-540-52282-4_30} {\emph {\bibinfo {booktitle}
  {{STACS} 90}}},\ \bibinfo {series and number} {Lecture Notes in Computer
  Science},\ \bibinfo {editor} {edited by\ \bibinfo {editor} {\bibfnamefont
  {C.}~\bibnamefont {Choffrut}}\ and\ \bibinfo {editor} {\bibfnamefont
  {T.}~\bibnamefont {Lengauer}}}\ (\bibinfo  {publisher} {Springer})\ pp.\
  \bibinfo {pages} {37--48}\BibitemShut {NoStop}%
\bibitem [{\citenamefont {Bell}(1964)}]{bell_einstein_1964}%
  \BibitemOpen
  \bibfield  {author} {\bibinfo {author} {\bibnamefont {Bell}, \bibfnamefont
  {J~S}}} (\bibinfo {year} {1964}),\ \bibfield  {title} {\enquote {\bibinfo
  {title} {{On the Einstein Podolsky Rosen paradox}},}\ }\href
  {https://doi.org/10.1103/PhysicsPhysiqueFizika.1.195} {\bibfield  {journal}
  {\bibinfo  {journal} {Physics}\ }\textbf {\bibinfo {volume} {1}},\ \bibinfo
  {pages} {195--200}}\BibitemShut {NoStop}%
\bibitem [{\citenamefont {Benioff}(1980)}]{benioff_computer_1980}%
  \BibitemOpen
  \bibfield  {author} {\bibinfo {author} {\bibnamefont {Benioff}, \bibfnamefont
  {P}}} (\bibinfo {year} {1980}),\ \bibfield  {title} {\enquote {\bibinfo
  {title} {The computer as a physical system: {{A}} microscopic quantum
  mechanical {{Hamiltonian}} model of computers as represented by {{Turing}}
  machines},}\ }\href {https://doi.org/10.1007/BF01011339} {\bibfield
  {journal} {\bibinfo  {journal} {J. Stat. Phys.}\ }\textbf {\bibinfo {volume}
  {22}},\ \bibinfo {pages} {563--591}}\BibitemShut {NoStop}%
\bibitem [{\citenamefont {Bennink}(2021)}]{VerificationAnticoncentrated}%
  \BibitemOpen
  \bibfield  {author} {\bibinfo {author} {\bibnamefont {Bennink}, \bibfnamefont
  {R~S}}} (\bibinfo {year} {2021}),\ \bibfield  {title} {\enquote {\bibinfo
  {title} {Efficient verification of anticoncentrated quantum states},}\ }\href
  {https://doi.org/10.1038/s41534-021-00455-6} {\bibfield  {journal} {\bibinfo
  {journal} {npj Quant. Inf.}\ }\textbf {\bibinfo {volume} {7}},\ \bibinfo
  {pages} {127}}\BibitemShut {NoStop}%
\bibitem [{\citenamefont {Bennink}\ \emph {et~al.}(2017)\citenamefont
  {Bennink}, \citenamefont {Ferragut}, \citenamefont {Humble}, \citenamefont
  {Laska}, \citenamefont {Nutaro}, \citenamefont {Pleszkoch},\ and\
  \citenamefont {Pooser}}]{bennink_unbiased_2017-1}%
  \BibitemOpen
  \bibfield  {author} {\bibinfo {author} {\bibnamefont {Bennink}, \bibfnamefont
  {R~S}}, \bibinfo {author} {\bibfnamefont {E.~M.}\ \bibnamefont {Ferragut}},
  \bibinfo {author} {\bibfnamefont {T.~S.}\ \bibnamefont {Humble}}, \bibinfo
  {author} {\bibfnamefont {J.~A.}\ \bibnamefont {Laska}}, \bibinfo {author}
  {\bibfnamefont {J.~J.}\ \bibnamefont {Nutaro}}, \bibinfo {author}
  {\bibfnamefont {M.~G.}\ \bibnamefont {Pleszkoch}}, and\ \bibinfo {author}
  {\bibfnamefont {R.~C.}\ \bibnamefont {Pooser}}} (\bibinfo {year} {2017}),\
  \bibfield  {title} {\enquote {\bibinfo {title} {Unbiased simulation of
  near-{{Clifford}} quantum circuits},}\ }\href
  {https://doi.org/10.1103/PhysRevA.95.062337} {\bibfield  {journal} {\bibinfo
  {journal} {Phys. Rev. A}\ }\textbf {\bibinfo {volume} {95}},\ \bibinfo
  {pages} {062337}}\BibitemShut {NoStop}%
\bibitem [{\citenamefont {Bentivegna}\ \emph {et~al.}(2015)\citenamefont
  {Bentivegna}, \citenamefont {Spagnolo}, \citenamefont {Vitelli},
  \citenamefont {Flamini}, \citenamefont {Viggianiello}, \citenamefont
  {Latmiral}, \citenamefont {Mataloni}, \citenamefont {Brod}, \citenamefont
  {Galv{\~a}o}, \citenamefont {Crespi}, \citenamefont {Ramponi}, \citenamefont
  {Osellame},\ and\ \citenamefont {Sciarrino}}]{bentivegna_experimental_2015}%
  \BibitemOpen
  \bibfield  {author} {\bibinfo {author} {\bibnamefont {Bentivegna},
  \bibfnamefont {M}}, \bibinfo {author} {\bibfnamefont {N.}~\bibnamefont
  {Spagnolo}}, \bibinfo {author} {\bibfnamefont {C.}~\bibnamefont {Vitelli}},
  \bibinfo {author} {\bibfnamefont {F.}~\bibnamefont {Flamini}}, \bibinfo
  {author} {\bibfnamefont {N.}~\bibnamefont {Viggianiello}}, \bibinfo {author}
  {\bibfnamefont {L.}~\bibnamefont {Latmiral}}, \bibinfo {author}
  {\bibfnamefont {P.}~\bibnamefont {Mataloni}}, \bibinfo {author}
  {\bibfnamefont {D.~J.}\ \bibnamefont {Brod}}, \bibinfo {author}
  {\bibfnamefont {E.~F.}\ \bibnamefont {Galv{\~a}o}}, \bibinfo {author}
  {\bibfnamefont {A.}~\bibnamefont {Crespi}}, \bibinfo {author} {\bibfnamefont
  {R.}~\bibnamefont {Ramponi}}, \bibinfo {author} {\bibfnamefont
  {R.}~\bibnamefont {Osellame}}, and\ \bibinfo {author} {\bibfnamefont
  {F.}~\bibnamefont {Sciarrino}}} (\bibinfo {year} {2015}),\ \bibfield  {title}
  {\enquote {\bibinfo {title} {Experimental scattershot boson sampling},}\
  }\href {https://doi.org/10.1126/sciadv.1400255} {\bibfield  {journal}
  {\bibinfo  {journal} {Science Adv.}\ }\textbf {\bibinfo {volume} {1}},\
  \bibinfo {pages} {e1400255}}\BibitemShut {NoStop}%
\bibitem [{\citenamefont {Bermejo-Vega}\ \emph {et~al.}(2018)\citenamefont
  {Bermejo-Vega}, \citenamefont {Hangleiter}, \citenamefont {Schwarz},
  \citenamefont {Raussendorf},\ and\ \citenamefont
  {Eisert}}]{bermejo-vega_architectures_2018}%
  \BibitemOpen
  \bibfield  {author} {\bibinfo {author} {\bibnamefont {Bermejo-Vega},
  \bibfnamefont {J}}, \bibinfo {author} {\bibfnamefont {D.}~\bibnamefont
  {Hangleiter}}, \bibinfo {author} {\bibfnamefont {M.}~\bibnamefont {Schwarz}},
  \bibinfo {author} {\bibfnamefont {R.}~\bibnamefont {Raussendorf}}, and\
  \bibinfo {author} {\bibfnamefont {J.}~\bibnamefont {Eisert}}} (\bibinfo
  {year} {2018}),\ \bibfield  {title} {\enquote {\bibinfo {title}
  {Architectures for {quantum} {simulation} {showing} a {quantum} {speedup}},}\
  }\href {https://doi.org/10.1103/PhysRevX.8.021010} {\bibfield  {journal}
  {\bibinfo  {journal} {Phys. Rev. X}\ }\textbf {\bibinfo {volume} {8}},\
  \bibinfo {pages} {021010}}\BibitemShut {NoStop}%
\bibitem [{\citenamefont {Bernien}\ \emph {et~al.}(2017)\citenamefont
  {Bernien}, \citenamefont {Schwartz}, \citenamefont {Keesling}, \citenamefont
  {Levine}, \citenamefont {Omran}, \citenamefont {Pichler}, \citenamefont
  {Choi}, \citenamefont {Zibrov}, \citenamefont {Endres}, \citenamefont
  {Greiner}, \citenamefont {Vuleti{\'c}},\ and\ \citenamefont
  {Lukin}}]{bernien_probing_2017}%
  \BibitemOpen
  \bibfield  {author} {\bibinfo {author} {\bibnamefont {Bernien}, \bibfnamefont
  {H}}, \bibinfo {author} {\bibfnamefont {S.}~\bibnamefont {Schwartz}},
  \bibinfo {author} {\bibfnamefont {A.}~\bibnamefont {Keesling}}, \bibinfo
  {author} {\bibfnamefont {H.}~\bibnamefont {Levine}}, \bibinfo {author}
  {\bibfnamefont {A.}~\bibnamefont {Omran}}, \bibinfo {author} {\bibfnamefont
  {H.}~\bibnamefont {Pichler}}, \bibinfo {author} {\bibfnamefont
  {S.}~\bibnamefont {Choi}}, \bibinfo {author} {\bibfnamefont {A.~S.}\
  \bibnamefont {Zibrov}}, \bibinfo {author} {\bibfnamefont {M.}~\bibnamefont
  {Endres}}, \bibinfo {author} {\bibfnamefont {M.}~\bibnamefont {Greiner}},
  \bibinfo {author} {\bibfnamefont {V.}~\bibnamefont {Vuleti{\'c}}}, and\
  \bibinfo {author} {\bibfnamefont {M.~D.}\ \bibnamefont {Lukin}}} (\bibinfo
  {year} {2017}),\ \bibfield  {title} {\enquote {\bibinfo {title} {Probing
  many-body dynamics on a 51-atom quantum simulator},}\ }\href
  {https://doi.org/10.1038/nature24622} {\bibfield  {journal} {\bibinfo
  {journal} {Nature}\ }\textbf {\bibinfo {volume} {551}},\ \bibinfo {pages}
  {579--584}}\BibitemShut {NoStop}%
\bibitem [{\citenamefont {Bernstein}\ and\ \citenamefont
  {Vazirani}(1993)}]{bernstein_quantum_1993}%
  \BibitemOpen
  \bibfield  {author} {\bibinfo {author} {\bibnamefont {Bernstein},
  \bibfnamefont {E}}, and\ \bibinfo {author} {\bibfnamefont {U.}~\bibnamefont
  {Vazirani}}} (\bibinfo {year} {1993}),\ \bibfield  {title} {\enquote
  {\bibinfo {title} {Quantum complexity theory},}\ }in\ \href
  {https://doi.org/10.1145/167088.167097} {\emph {\bibinfo {booktitle} {Proc.
  25th Ann. {{ACM}} Symp. {{Th.}} {{Comp.}}}}},\ \bibinfo {series and number}
  {{{STOC}} '93}\ (\bibinfo  {publisher} {{Association for Computing
  Machinery}},\ \bibinfo {address} {{New York, NY, USA}})\ pp.\ \bibinfo
  {pages} {11--20}\BibitemShut {NoStop}%
\bibitem [{\citenamefont {Bernstein}\ and\ \citenamefont
  {Vazirani}(1997)}]{bernstein_quantum_1997}%
  \BibitemOpen
  \bibfield  {author} {\bibinfo {author} {\bibnamefont {Bernstein},
  \bibfnamefont {E}}, and\ \bibinfo {author} {\bibfnamefont {U.}~\bibnamefont
  {Vazirani}}} (\bibinfo {year} {1997}),\ \bibfield  {title} {\enquote
  {\bibinfo {title} {Quantum complexity theory},}\ }\href
  {https://doi.org/10.1137/S0097539796300921} {\bibfield  {journal} {\bibinfo
  {journal} {SIAM J. Comp.}\ }\textbf {\bibinfo {volume} {26}},\ \bibinfo
  {pages} {1411--1473}}\BibitemShut {NoStop}%
\bibitem [{\citenamefont {Bj{\"o}rklund}(2012)}]{bjorklund_counting_2012}%
  \BibitemOpen
  \bibfield  {author} {\bibinfo {author} {\bibnamefont {Bj{\"o}rklund},
  \bibfnamefont {A}}} (\bibinfo {year} {2012}),\ \bibfield  {title} {\enquote
  {\bibinfo {title} {Counting {{perfect matchings}} as {{fast}} as
  {{ryser}}},}\ }in\ \href {https://doi.org/10.1137/1.9781611973099.73} {\emph
  {\bibinfo {booktitle} {Proc. 2012 {{Ann. ACM-SIAM Symp.}} {{Disc. Alg.}}
  ({{SODA}})}}},\ \bibinfo {series and number} {Proceedings}\ (\bibinfo
  {publisher} {{Society for Industrial and Applied Mathematics}})\ pp.\
  \bibinfo {pages} {914--921}\BibitemShut {NoStop}%
\bibitem [{\citenamefont {Bj{\"o}rklund}\ \emph {et~al.}(2019)\citenamefont
  {Bj{\"o}rklund}, \citenamefont {Gupt},\ and\ \citenamefont
  {Quesada}}]{bjorklund_faster_2019-1}%
  \BibitemOpen
  \bibfield  {author} {\bibinfo {author} {\bibnamefont {Bj{\"o}rklund},
  \bibfnamefont {A}}, \bibinfo {author} {\bibfnamefont {B.}~\bibnamefont
  {Gupt}}, and\ \bibinfo {author} {\bibfnamefont {N.}~\bibnamefont {Quesada}}}
  (\bibinfo {year} {2019}),\ \bibfield  {title} {\enquote {\bibinfo {title} {A
  {{faster Hafnian formula}} for {{complex matrices}} and {{its benchmarking}}
  on a {{supercomputer}}},}\ }\href {https://doi.org/10.1145/3325111}
  {\bibfield  {journal} {\bibinfo  {journal} {ACM J. Exp. Algorithmics}\
  }\textbf {\bibinfo {volume} {24}},\ \bibinfo {pages} {1.11:1}}\BibitemShut
  {NoStop}%
\bibitem [{\citenamefont {Blais}\ \emph {et~al.}(2004)\citenamefont {Blais},
  \citenamefont {Huang}, \citenamefont {Wallraff}, \citenamefont {Girvin},\
  and\ \citenamefont {Schoelkopf}}]{blais_cavity_2004}%
  \BibitemOpen
  \bibfield  {author} {\bibinfo {author} {\bibnamefont {Blais}, \bibfnamefont
  {A}}, \bibinfo {author} {\bibfnamefont {R.-S.}\ \bibnamefont {Huang}},
  \bibinfo {author} {\bibfnamefont {A.}~\bibnamefont {Wallraff}}, \bibinfo
  {author} {\bibfnamefont {S.~M.}\ \bibnamefont {Girvin}}, and\ \bibinfo
  {author} {\bibfnamefont {R.~J.}\ \bibnamefont {Schoelkopf}}} (\bibinfo {year}
  {2004}),\ \bibfield  {title} {\enquote {\bibinfo {title} {Cavity quantum
  electrodynamics for superconducting electrical circuits: {{An}} architecture
  for quantum computation},}\ }\href
  {https://doi.org/10.1103/PhysRevA.69.062320} {\bibfield  {journal} {\bibinfo
  {journal} {Phys. Rev. A}\ }\textbf {\bibinfo {volume} {69}},\ \bibinfo
  {pages} {062320}}\BibitemShut {NoStop}%
\bibitem [{\citenamefont {Blatt}\ and\ \citenamefont
  {Roos}(2012)}]{blatt_quantum_2012}%
  \BibitemOpen
  \bibfield  {author} {\bibinfo {author} {\bibnamefont {Blatt}, \bibfnamefont
  {R}}, and\ \bibinfo {author} {\bibfnamefont {C.~F.}\ \bibnamefont {Roos}}}
  (\bibinfo {year} {2012}),\ \bibfield  {title} {\enquote {\bibinfo {title}
  {Quantum simulations with trapped ions},}\ }\href
  {https://doi.org/10.1038/nphys2252} {\bibfield  {journal} {\bibinfo
  {journal} {Nature Phys.}\ }\textbf {\bibinfo {volume} {8}},\ \bibinfo {pages}
  {277--284}}\BibitemShut {NoStop}%
\bibitem [{\citenamefont {Bloch}\ \emph {et~al.}(2008)\citenamefont {Bloch},
  \citenamefont {Dalibard},\ and\ \citenamefont
  {Zwerger}}]{bloch_many-body_2008}%
  \BibitemOpen
  \bibfield  {author} {\bibinfo {author} {\bibnamefont {Bloch}, \bibfnamefont
  {I}}, \bibinfo {author} {\bibfnamefont {J.}~\bibnamefont {Dalibard}}, and\
  \bibinfo {author} {\bibfnamefont {W.}~\bibnamefont {Zwerger}}} (\bibinfo
  {year} {2008}),\ \bibfield  {title} {\enquote {\bibinfo {title} {Many-body
  physics with ultracold gases},}\ }\href
  {https://doi.org/10.1103/RevModPhys.80.885} {\bibfield  {journal} {\bibinfo
  {journal} {Rev. Mod. Phys.}\ }\textbf {\bibinfo {volume} {80}},\ \bibinfo
  {pages} {885--964}}\BibitemShut {NoStop}%
\bibitem [{\citenamefont {{Blume-Kohout}}\ \emph {et~al.}(2017)\citenamefont
  {{Blume-Kohout}}, \citenamefont {{Gamble}}, \citenamefont {{Nielsen}},
  \citenamefont {{Rudinger}}, \citenamefont {{Mizrahi}}, \citenamefont
  {{Fortier}},\ and\ \citenamefont {{Maunz}}}]{BluGamNie17}%
  \BibitemOpen
  \bibfield  {author} {\bibinfo {author} {\bibnamefont {{Blume-Kohout}},
  \bibfnamefont {R}}, \bibinfo {author} {\bibfnamefont {J.~K.}\ \bibnamefont
  {{Gamble}}}, \bibinfo {author} {\bibfnamefont {E.}~\bibnamefont {{Nielsen}}},
  \bibinfo {author} {\bibfnamefont {K.}~\bibnamefont {{Rudinger}}}, \bibinfo
  {author} {\bibfnamefont {J.}~\bibnamefont {{Mizrahi}}}, \bibinfo {author}
  {\bibfnamefont {K.}~\bibnamefont {{Fortier}}}, and\ \bibinfo {author}
  {\bibfnamefont {P.}~\bibnamefont {{Maunz}}}} (\bibinfo {year} {2017}),\
  \bibfield  {title} {\enquote {\bibinfo {title} {{Demonstration of qubit
  operations below a rigorous fault tolerance threshold with gate set
  tomography}},}\ }\href {https://doi.org/10.1038/ncomms14485} {\bibfield
  {journal} {\bibinfo  {journal} {Nature Comm.}\ }\textbf {\bibinfo {volume}
  {8}},\ \bibinfo {pages} {14485}}\BibitemShut {NoStop}%
\bibitem [{\citenamefont {{Blume-Kohout}}\ \emph {et~al.}(2013)\citenamefont
  {{Blume-Kohout}}, \citenamefont {{King Gamble}}, \citenamefont {{Nielsen}},
  \citenamefont {{Mizrahi}}, \citenamefont {{Sterk}},\ and\ \citenamefont
  {{Maunz}}}]{BluKinNie13}%
  \BibitemOpen
  \bibfield  {author} {\bibinfo {author} {\bibnamefont {{Blume-Kohout}},
  \bibfnamefont {R}}, \bibinfo {author} {\bibfnamefont {J.}~\bibnamefont {{King
  Gamble}}}, \bibinfo {author} {\bibfnamefont {E.}~\bibnamefont {{Nielsen}}},
  \bibinfo {author} {\bibfnamefont {J.}~\bibnamefont {{Mizrahi}}}, \bibinfo
  {author} {\bibfnamefont {J.~D.}\ \bibnamefont {{Sterk}}}, and\ \bibinfo
  {author} {\bibfnamefont {P.}~\bibnamefont {{Maunz}}}} (\bibinfo {year}
  {2013}),\ \bibfield  {title} {\enquote {\bibinfo {title} {{Robust,
  self-consistent, closed-form tomography of quantum logic gates on a trapped
  ion qubit}},}\ }\href@noop {} {\ }\Eprint {https://arxiv.org/abs/1310.4492}
  {arXiv:1310.4492} \BibitemShut {NoStop}%
\bibitem [{\citenamefont {Boixo}\ \emph {et~al.}(2018)\citenamefont {Boixo},
  \citenamefont {Isakov}, \citenamefont {Smelyanskiy}, \citenamefont {Babbush},
  \citenamefont {Ding}, \citenamefont {Jiang}, \citenamefont {Bremner},
  \citenamefont {Martinis},\ and\ \citenamefont
  {Neven}}]{boixo_characterizing_2016}%
  \BibitemOpen
  \bibfield  {author} {\bibinfo {author} {\bibnamefont {Boixo}, \bibfnamefont
  {S}}, \bibinfo {author} {\bibfnamefont {S.~V.}\ \bibnamefont {Isakov}},
  \bibinfo {author} {\bibfnamefont {V.~N.}\ \bibnamefont {Smelyanskiy}},
  \bibinfo {author} {\bibfnamefont {R.}~\bibnamefont {Babbush}}, \bibinfo
  {author} {\bibfnamefont {N.}~\bibnamefont {Ding}}, \bibinfo {author}
  {\bibfnamefont {Z.}~\bibnamefont {Jiang}}, \bibinfo {author} {\bibfnamefont
  {M.~J.}\ \bibnamefont {Bremner}}, \bibinfo {author} {\bibfnamefont {J.~M.}\
  \bibnamefont {Martinis}}, and\ \bibinfo {author} {\bibfnamefont
  {H.}~\bibnamefont {Neven}}} (\bibinfo {year} {2018}),\ \bibfield  {title}
  {\enquote {\bibinfo {title} {Characterizing quantum supremacy in near-term
  devices},}\ }\href {https://doi.org/10.1038/s41567-018-0124-x} {\bibfield
  {journal} {\bibinfo  {journal} {Nature Phys.}\ }\textbf {\bibinfo {volume}
  {14}},\ \bibinfo {pages} {595--600}}\BibitemShut {NoStop}%
\bibitem [{\citenamefont {Boixo}\ \emph
  {et~al.}(2017{\natexlab{a}})\citenamefont {Boixo}, \citenamefont {Isakov},
  \citenamefont {Smelyanskiy},\ and\ \citenamefont
  {Neven}}]{boixo_simulation_2017}%
  \BibitemOpen
  \bibfield  {author} {\bibinfo {author} {\bibnamefont {Boixo}, \bibfnamefont
  {S}}, \bibinfo {author} {\bibfnamefont {S.~V.}\ \bibnamefont {Isakov}},
  \bibinfo {author} {\bibfnamefont {V.~N.}\ \bibnamefont {Smelyanskiy}}, and\
  \bibinfo {author} {\bibfnamefont {H.}~\bibnamefont {Neven}}} (\bibinfo {year}
  {2017}{\natexlab{a}}),\ \bibfield  {title} {\enquote {\bibinfo {title}
  {Simulation of low-depth quantum circuits as complex undirected graphical
  models},}\ }\href@noop {} {\ }\Eprint {https://arxiv.org/abs/1712.05384}
  {arXiv:1712.05384} \BibitemShut {NoStop}%
\bibitem [{\citenamefont {Boixo}\ \emph
  {et~al.}(2017{\natexlab{b}})\citenamefont {Boixo}, \citenamefont
  {Smelyanskiy},\ and\ \citenamefont {Neven}}]{boixo_fourier_2017}%
  \BibitemOpen
  \bibfield  {author} {\bibinfo {author} {\bibnamefont {Boixo}, \bibfnamefont
  {S}}, \bibinfo {author} {\bibfnamefont {V.~N.}\ \bibnamefont {Smelyanskiy}},
  and\ \bibinfo {author} {\bibfnamefont {H.}~\bibnamefont {Neven}}} (\bibinfo
  {year} {2017}{\natexlab{b}}),\ \bibfield  {title} {\enquote {\bibinfo {title}
  {Fourier analysis of sampling from noisy chaotic quantum circuits},}\
  }\href@noop {} {\ }\Eprint {https://arxiv.org/abs/1708.01875}
  {arxiv:1708.01875} \BibitemShut {NoStop}%
\bibitem [{\citenamefont {Boone}\ \emph {et~al.}(2019)\citenamefont {Boone},
  \citenamefont {{Carignan-Dugas}}, \citenamefont {Wallman},\ and\
  \citenamefont {Emerson}}]{boone_randomized_2019}%
  \BibitemOpen
  \bibfield  {author} {\bibinfo {author} {\bibnamefont {Boone}, \bibfnamefont
  {K}}, \bibinfo {author} {\bibfnamefont {A.}~\bibnamefont {{Carignan-Dugas}}},
  \bibinfo {author} {\bibfnamefont {J.~J.}\ \bibnamefont {Wallman}}, and\
  \bibinfo {author} {\bibfnamefont {J.}~\bibnamefont {Emerson}}} (\bibinfo
  {year} {2019}),\ \bibfield  {title} {\enquote {\bibinfo {title} {Randomized
  {{benchmarking}} under {{different gatesets}}},}\ }\href
  {https://doi.org/10.1103/PhysRevA.99.032329} {\bibfield  {journal} {\bibinfo
  {journal} {Phys. Rev. A}\ }\textbf {\bibinfo {volume} {99}},\ \bibinfo
  {pages} {032329}}\BibitemShut {NoStop}%
\bibitem [{\citenamefont {Bouland}\ \emph {et~al.}(2022)\citenamefont
  {Bouland}, \citenamefont {Fefferman}, \citenamefont {Landau},\ and\
  \citenamefont {Liu}}]{bouland_noise_2021}%
  \BibitemOpen
  \bibfield  {author} {\bibinfo {author} {\bibnamefont {Bouland}, \bibfnamefont
  {A}}, \bibinfo {author} {\bibfnamefont {B.}~\bibnamefont {Fefferman}},
  \bibinfo {author} {\bibfnamefont {Z.}~\bibnamefont {Landau}}, and\ \bibinfo
  {author} {\bibfnamefont {Y.}~\bibnamefont {Liu}}} (\bibinfo {year} {2022}),\
  \bibfield  {title} {\enquote {\bibinfo {title} {Noise and the {{frontier}} of
  {{quantum supremacy}}},}\ }\bibfield  {booktitle} {\emph {\bibinfo
  {booktitle} {2021 {{IEEE}} 62nd {{Annual Symposium}} on {{Foundations}} of
  {{Computer Science}} ({{FOCS}})}},\ }\href
  {https://doi.org/10.1109/FOCS52979.2021.00127} {\ ,\ \bibinfo {pages}
  {1308--1317}}\Eprint {https://arxiv.org/abs/2102.01738} {arXiv:2102.01738}
  \BibitemShut {NoStop}%
\bibitem [{\citenamefont {Bouland}\ \emph {et~al.}(2019)\citenamefont
  {Bouland}, \citenamefont {Fefferman}, \citenamefont {Nirkhe},\ and\
  \citenamefont {Vazirani}}]{bouland_quantum_2018}%
  \BibitemOpen
  \bibfield  {author} {\bibinfo {author} {\bibnamefont {Bouland}, \bibfnamefont
  {A}}, \bibinfo {author} {\bibfnamefont {B.}~\bibnamefont {Fefferman}},
  \bibinfo {author} {\bibfnamefont {C.}~\bibnamefont {Nirkhe}}, and\ \bibinfo
  {author} {\bibfnamefont {U.}~\bibnamefont {Vazirani}}} (\bibinfo {year}
  {2019}),\ \bibfield  {title} {\enquote {\bibinfo {title} {On the complexity
  and verification of quantum random circuit sampling},}\ }\href
  {https://doi.org/10.1038/s41567-018-0318-2} {\bibfield  {journal} {\bibinfo
  {journal} {Nature Phys.}\ }\textbf {\bibinfo {volume} {15}},\ \bibinfo
  {pages} {159--163}}\BibitemShut {NoStop}%
\bibitem [{\citenamefont {Bouland}\ \emph {et~al.}(2018)\citenamefont
  {Bouland}, \citenamefont {Fitzsimons},\ and\ \citenamefont
  {Koh}}]{bouland_complexity_2018}%
  \BibitemOpen
  \bibfield  {author} {\bibinfo {author} {\bibnamefont {Bouland}, \bibfnamefont
  {A}}, \bibinfo {author} {\bibfnamefont {J.~F.}\ \bibnamefont {Fitzsimons}},
  and\ \bibinfo {author} {\bibfnamefont {D.~E.}\ \bibnamefont {Koh}}} (\bibinfo
  {year} {2018}),\ \bibfield  {title} {\enquote {\bibinfo {title} {Complexity
  {classification} of {conjugated} {Clifford} {circuits}},}\ }in\ \href
  {https://doi.org/10.4230/LIPIcs.CCC.2018.21} {\emph {\bibinfo {booktitle}
  {33rd {Computational} {Complexity} {Conference} ({CCC} 2018)}}},\ \bibinfo
  {series} {Leibniz {International} {Proceedings} in {Informatics} ({LIPIcs})},
  Vol.\ \bibinfo {volume} {102},\ \bibinfo {editor} {edited by\ \bibinfo
  {editor} {\bibfnamefont {Rocco~A.}\ \bibnamefont {Servedio}}}\ (\bibinfo
  {publisher} {Schloss Dagstuhl, Leibniz-Zentrum f{\"u}r Informatik},\ \bibinfo
  {address} {Dagstuhl, Germany})\ pp.\ \bibinfo {pages} {21:1--21:25},\ \Eprint
  {https://arxiv.org/abs/1709.01805} {arXiv:1709.01805} \BibitemShut {NoStop}%
\bibitem [{\citenamefont {Bourennane}\ \emph {et~al.}(2004)\citenamefont
  {Bourennane}, \citenamefont {Eibl}, \citenamefont {Kurtsiefer}, \citenamefont
  {Gaertner}, \citenamefont {Weinfurter}, \citenamefont {G{\"u}hne},
  \citenamefont {Hyllus}, \citenamefont {Bru{\ss}}, \citenamefont
  {Lewenstein},\ and\ \citenamefont {Sanpera}}]{bourennane_experimental_2004}%
  \BibitemOpen
  \bibfield  {author} {\bibinfo {author} {\bibnamefont {Bourennane},
  \bibfnamefont {M}}, \bibinfo {author} {\bibfnamefont {M.}~\bibnamefont
  {Eibl}}, \bibinfo {author} {\bibfnamefont {C.}~\bibnamefont {Kurtsiefer}},
  \bibinfo {author} {\bibfnamefont {S.}~\bibnamefont {Gaertner}}, \bibinfo
  {author} {\bibfnamefont {H.}~\bibnamefont {Weinfurter}}, \bibinfo {author}
  {\bibfnamefont {O.}~\bibnamefont {G{\"u}hne}}, \bibinfo {author}
  {\bibfnamefont {P.}~\bibnamefont {Hyllus}}, \bibinfo {author} {\bibfnamefont
  {D.}~\bibnamefont {Bru{\ss}}}, \bibinfo {author} {\bibfnamefont
  {M.}~\bibnamefont {Lewenstein}}, and\ \bibinfo {author} {\bibfnamefont
  {A.}~\bibnamefont {Sanpera}}} (\bibinfo {year} {2004}),\ \bibfield  {title}
  {\enquote {\bibinfo {title} {Experimental {detection} of {multipartite}
  {entanglement} using {witness} {operators}},}\ }\href
  {https://doi.org/10.1103/PhysRevLett.92.087902} {\bibfield  {journal}
  {\bibinfo  {journal} {Phys. Rev. Lett.}\ }\textbf {\bibinfo {volume} {92}},\
  \bibinfo {pages} {087902}}\BibitemShut {NoStop}%
\bibitem [{\citenamefont {Br{\'a}dler}\ \emph {et~al.}(2018)\citenamefont
  {Br{\'a}dler}, \citenamefont {Dallaire-Demers}, \citenamefont {Rebentrost},
  \citenamefont {Su},\ and\ \citenamefont {Weedbrook}}]{bradler_gaussian_2018}%
  \BibitemOpen
  \bibfield  {author} {\bibinfo {author} {\bibnamefont {Br{\'a}dler},
  \bibfnamefont {K}}, \bibinfo {author} {\bibfnamefont {P.-L.}\ \bibnamefont
  {Dallaire-Demers}}, \bibinfo {author} {\bibfnamefont {P.}~\bibnamefont
  {Rebentrost}}, \bibinfo {author} {\bibfnamefont {D.}~\bibnamefont {Su}}, and\
  \bibinfo {author} {\bibfnamefont {C.}~\bibnamefont {Weedbrook}}} (\bibinfo
  {year} {2018}),\ \bibfield  {title} {\enquote {\bibinfo {title} {Gaussian
  boson sampling for perfect matchings of arbitrary graphs},}\ }\href
  {https://doi.org/10.1103/PhysRevA.98.032310} {\bibfield  {journal} {\bibinfo
  {journal} {Phys. Rev. A}\ }\textbf {\bibinfo {volume} {98}},\ \bibinfo
  {pages} {032310}}\BibitemShut {NoStop}%
\bibitem [{\citenamefont {Brakerski}\ \emph {et~al.}(2018)\citenamefont
  {Brakerski}, \citenamefont {Christiano}, \citenamefont {Mahadev},
  \citenamefont {Vazirani},\ and\ \citenamefont
  {Vidick}}]{brakerski_cryptographic_2018}%
  \BibitemOpen
  \bibfield  {author} {\bibinfo {author} {\bibnamefont {Brakerski},
  \bibfnamefont {Z}}, \bibinfo {author} {\bibfnamefont {P.}~\bibnamefont
  {Christiano}}, \bibinfo {author} {\bibfnamefont {U.}~\bibnamefont {Mahadev}},
  \bibinfo {author} {\bibfnamefont {U.}~\bibnamefont {Vazirani}}, and\ \bibinfo
  {author} {\bibfnamefont {T.}~\bibnamefont {Vidick}}} (\bibinfo {year}
  {2018}),\ \bibfield  {title} {\enquote {\bibinfo {title} {A cryptographic
  test of quantumness and certifiable randomness from a single quantum
  device},}\ }in\ \href {https://doi.org/10.1109/FOCS.2018.00038} {\emph
  {\bibinfo {booktitle} {2018 {IEEE} 59th Ann. Symp. Found. Comp. Sc.
  ({FOCS})}}},\ pp.\ \bibinfo {pages} {320--331},\ \Eprint
  {https://arxiv.org/abs/1804.00640} {arxiv:1804.00640} \BibitemShut {NoStop}%
\bibitem [{\citenamefont {Brakerski}\ \emph {et~al.}(2020)\citenamefont
  {Brakerski}, \citenamefont {Koppula}, \citenamefont {Vazirani},\ and\
  \citenamefont {Vidick}}]{brakerski_simpler_2020}%
  \BibitemOpen
  \bibfield  {author} {\bibinfo {author} {\bibnamefont {Brakerski},
  \bibfnamefont {Z}}, \bibinfo {author} {\bibfnamefont {V.}~\bibnamefont
  {Koppula}}, \bibinfo {author} {\bibfnamefont {U.}~\bibnamefont {Vazirani}},
  and\ \bibinfo {author} {\bibfnamefont {T.}~\bibnamefont {Vidick}}} (\bibinfo
  {year} {2020}),\ \bibfield  {title} {\enquote {\bibinfo {title} {Simpler
  proofs of quantumness},}\ }\href {http://arxiv.org/abs/2005.04826} {\
  }\Eprint {https://arxiv.org/abs/2005.04826} {arxiv:2005.04826} \BibitemShut
  {NoStop}%
\bibitem [{\citenamefont {Brand{\~a}o}\ \emph {et~al.}(2016)\citenamefont
  {Brand{\~a}o}, \citenamefont {Harrow},\ and\ \citenamefont
  {Horodecki}}]{brandao_local_2016}%
  \BibitemOpen
  \bibfield  {author} {\bibinfo {author} {\bibnamefont {Brand{\~a}o},
  \bibfnamefont {F~G S~L}}, \bibinfo {author} {\bibfnamefont {A.~W.}\
  \bibnamefont {Harrow}}, and\ \bibinfo {author} {\bibfnamefont
  {M.}~\bibnamefont {Horodecki}}} (\bibinfo {year} {2016}),\ \bibfield  {title}
  {\enquote {\bibinfo {title} {Local {random} {quantum} {circuits} are
  {approximate} {polynomial}-{designs}},}\ }\href
  {https://doi.org/10.1007/s00220-016-2706-8} {\bibfield  {journal} {\bibinfo
  {journal} {Commun. Math. Phys.}\ }\textbf {\bibinfo {volume} {346}},\
  \bibinfo {pages} {397--434}}\BibitemShut {NoStop}%
\bibitem [{\citenamefont {Brand{\~a}o}\ and\ \citenamefont
  {Svore}(2016)}]{QuantumSDP}%
  \BibitemOpen
  \bibfield  {author} {\bibinfo {author} {\bibnamefont {Brand{\~a}o},
  \bibfnamefont {F~G S~L}}, and\ \bibinfo {author} {\bibfnamefont
  {K.}~\bibnamefont {Svore}}} (\bibinfo {year} {2016}),\ \bibfield  {title}
  {\enquote {\bibinfo {title} {Quantum speed-ups for semidefinite
  programming},}\ }\href@noop {} {\ }\Eprint {https://arxiv.org/abs/1609.05537}
  {arXiv:1609.05537} \BibitemShut {NoStop}%
\bibitem [{\citenamefont {Braun}\ \emph {et~al.}(2015)\citenamefont {Braun},
  \citenamefont {Friesdorf}, \citenamefont {Hodgman}, \citenamefont
  {Schreiber}, \citenamefont {Ronzheimer}, \citenamefont {Riera}, \citenamefont
  {del Rey}, \citenamefont {Bloch}, \citenamefont {Eisert},\ and\ \citenamefont
  {Schneider}}]{Emergence}%
  \BibitemOpen
  \bibfield  {author} {\bibinfo {author} {\bibnamefont {Braun}, \bibfnamefont
  {S}}, \bibinfo {author} {\bibfnamefont {M.}~\bibnamefont {Friesdorf}},
  \bibinfo {author} {\bibfnamefont {S.~S.}\ \bibnamefont {Hodgman}}, \bibinfo
  {author} {\bibfnamefont {M.}~\bibnamefont {Schreiber}}, \bibinfo {author}
  {\bibfnamefont {J.~P.}\ \bibnamefont {Ronzheimer}}, \bibinfo {author}
  {\bibfnamefont {A.}~\bibnamefont {Riera}}, \bibinfo {author} {\bibfnamefont
  {M.}~\bibnamefont {del Rey}}, \bibinfo {author} {\bibfnamefont
  {I.}~\bibnamefont {Bloch}}, \bibinfo {author} {\bibfnamefont
  {J.}~\bibnamefont {Eisert}}, and\ \bibinfo {author} {\bibfnamefont
  {U.}~\bibnamefont {Schneider}}} (\bibinfo {year} {2015}),\ \bibfield  {title}
  {\enquote {\bibinfo {title} {Emergence of coherence and the dynamics of
  quantum phase transitions},}\ }\href
  {https://doi.org/10.1073/pnas.1408861112} {\bibfield  {journal} {\bibinfo
  {journal} {PNAS}\ }\textbf {\bibinfo {volume} {112}},\ \bibinfo {pages}
  {3641}}\BibitemShut {NoStop}%
\bibitem [{\citenamefont {Bravyi}\ \emph {et~al.}(2018)\citenamefont {Bravyi},
  \citenamefont {Gosset},\ and\ \citenamefont {K{\"o}nig}}]{Shallow}%
  \BibitemOpen
  \bibfield  {author} {\bibinfo {author} {\bibnamefont {Bravyi}, \bibfnamefont
  {S}}, \bibinfo {author} {\bibnamefont {Gosset}}, and\ \bibinfo {author}
  {\bibfnamefont {R.}~\bibnamefont {K{\"o}nig}}} (\bibinfo {year} {2018}),\
  \bibfield  {title} {\enquote {\bibinfo {title} {Quantum advantage with
  shallow circuits},}\ }\href {https://doi.org/10.1126/science.aar3106}
  {\bibfield  {journal} {\bibinfo  {journal} {Science}\ }\textbf {\bibinfo
  {volume} {362}},\ \bibinfo {pages} {308}}\BibitemShut {NoStop}%
\bibitem [{\citenamefont {Bravyi}\ and\ \citenamefont
  {Gosset}(2016)}]{bravyi_improved_2016}%
  \BibitemOpen
  \bibfield  {author} {\bibinfo {author} {\bibnamefont {Bravyi}, \bibfnamefont
  {S}}, and\ \bibinfo {author} {\bibfnamefont {D.}~\bibnamefont {Gosset}}}
  (\bibinfo {year} {2016}),\ \bibfield  {title} {\enquote {\bibinfo {title}
  {Improved classical simulation of quantum circuits dominated by {{Clifford}}
  gates},}\ }\href {https://doi.org/10.1103/PhysRevLett.116.250501} {\bibfield
  {journal} {\bibinfo  {journal} {Phys. Rev. Lett.}\ }\textbf {\bibinfo
  {volume} {116}},\ \bibinfo {pages} {250501}}\BibitemShut {NoStop}%
\bibitem [{\citenamefont {Bravyi}\ \emph {et~al.}(2020)\citenamefont {Bravyi},
  \citenamefont {Gosset}, \citenamefont {K{\"o}nig},\ and\ \citenamefont
  {Tomamichel}}]{NoisyShallow}%
  \BibitemOpen
  \bibfield  {author} {\bibinfo {author} {\bibnamefont {Bravyi}, \bibfnamefont
  {S}}, \bibinfo {author} {\bibfnamefont {D.}~\bibnamefont {Gosset}}, \bibinfo
  {author} {\bibfnamefont {R.}~\bibnamefont {K{\"o}nig}}, and\ \bibinfo
  {author} {\bibfnamefont {M.}~\bibnamefont {Tomamichel}}} (\bibinfo {year}
  {2020}),\ \bibfield  {title} {\enquote {\bibinfo {title} {Quantum advantage
  with noisy shallow circuits},}\ }\href
  {https://doi.org/10.1038/s41567-020-0948-z} {\bibfield  {journal} {\bibinfo
  {journal} {Nature Phys.}\ }\textbf {\bibinfo {volume} {16}},\ \bibinfo
  {pages} {1040--1045}}\BibitemShut {NoStop}%
\bibitem [{\citenamefont {Bremner}\ \emph {et~al.}(2010)\citenamefont
  {Bremner}, \citenamefont {Jozsa},\ and\ \citenamefont
  {Shepherd}}]{bremner_classical_2010}%
  \BibitemOpen
  \bibfield  {author} {\bibinfo {author} {\bibnamefont {Bremner}, \bibfnamefont
  {M~J}}, \bibinfo {author} {\bibfnamefont {R.}~\bibnamefont {Jozsa}}, and\
  \bibinfo {author} {\bibfnamefont {D.~J.}\ \bibnamefont {Shepherd}}} (\bibinfo
  {year} {2010}),\ \bibfield  {title} {\enquote {\bibinfo {title} {Classical
  simulation of commuting quantum computations implies collapse of the
  polynomial hierarchy},}\ }\href {https://doi.org/10.1098/rspa.2010.0301}
  {\bibfield  {journal} {\bibinfo  {journal} {Proc. Roy. Soc. A}\ }\textbf
  {\bibinfo {volume} {467}},\ \bibinfo {pages} {459--472}}\BibitemShut
  {NoStop}%
\bibitem [{\citenamefont {Bremner}\ \emph {et~al.}(2016)\citenamefont
  {Bremner}, \citenamefont {Montanaro},\ and\ \citenamefont
  {Shepherd}}]{bremner_averagecase_2016}%
  \BibitemOpen
  \bibfield  {author} {\bibinfo {author} {\bibnamefont {Bremner}, \bibfnamefont
  {M~J}}, \bibinfo {author} {\bibfnamefont {A.}~\bibnamefont {Montanaro}}, and\
  \bibinfo {author} {\bibfnamefont {D.~J.}\ \bibnamefont {Shepherd}}} (\bibinfo
  {year} {2016}),\ \bibfield  {title} {\enquote {\bibinfo {title} {Average-case
  complexity versus approximate simulation of commuting quantum
  computations},}\ }\href {https://doi.org/10.1103/PhysRevLett.117.080501}
  {\bibfield  {journal} {\bibinfo  {journal} {Phys. Rev. Lett.}\ }\textbf
  {\bibinfo {volume} {117}},\ \bibinfo {pages} {080501}}\BibitemShut {NoStop}%
\bibitem [{\citenamefont {Bremner}\ \emph {et~al.}(2017)\citenamefont
  {Bremner}, \citenamefont {Montanaro},\ and\ \citenamefont
  {Shepherd}}]{bremner_achieving_2017}%
  \BibitemOpen
  \bibfield  {author} {\bibinfo {author} {\bibnamefont {Bremner}, \bibfnamefont
  {M~J}}, \bibinfo {author} {\bibfnamefont {A.}~\bibnamefont {Montanaro}}, and\
  \bibinfo {author} {\bibfnamefont {D.~J.}\ \bibnamefont {Shepherd}}} (\bibinfo
  {year} {2017}),\ \bibfield  {title} {\enquote {\bibinfo {title} {Achieving
  quantum supremacy with sparse and noisy commuting quantum computations},}\
  }\href {https://doi.org/10.22331/q-2017-04-25-8} {\bibfield  {journal}
  {\bibinfo  {journal} {Quantum}\ }\textbf {\bibinfo {volume} {1}},\ \bibinfo
  {pages} {8}}\BibitemShut {NoStop}%
\bibitem [{\citenamefont {Bridgeman}\ and\ \citenamefont
  {Chubb}(2017)}]{bridgeman_hand-waving_2017}%
  \BibitemOpen
  \bibfield  {author} {\bibinfo {author} {\bibnamefont {Bridgeman},
  \bibfnamefont {J~C}}, and\ \bibinfo {author} {\bibfnamefont {C.~T.}\
  \bibnamefont {Chubb}}} (\bibinfo {year} {2017}),\ \bibfield  {title}
  {\enquote {\bibinfo {title} {Hand-waving and interpretive dance: An
  introductory course on tensor networks},}\ }\href
  {https://doi.org/10.1088/1751-8121/aa6dc3} {\bibfield  {journal} {\bibinfo
  {journal} {J. Phys. A}\ }\textbf {\bibinfo {volume} {50}},\ \bibinfo {pages}
  {223001}}\BibitemShut {NoStop}%
\bibitem [{\citenamefont {Brieger}\ \emph {et~al.}(2022)\citenamefont
  {Brieger}, \citenamefont {Roth},\ and\ \citenamefont
  {Kliesch}}]{brieger_compressive_2022}%
  \BibitemOpen
  \bibfield  {author} {\bibinfo {author} {\bibnamefont {Brieger}, \bibfnamefont
  {R}}, \bibinfo {author} {\bibfnamefont {I.}~\bibnamefont {Roth}}, and\
  \bibinfo {author} {\bibfnamefont {M.}~\bibnamefont {Kliesch}}} (\bibinfo
  {year} {2022}),\ \bibfield  {title} {\enquote {\bibinfo {title} {Compressive
  gate set tomography},}\ }\href@noop {} {\ }\Eprint
  {https://arxiv.org/abs/2112.05176} {arXiv:2112.05176} \BibitemShut {NoStop}%
\bibitem [{\citenamefont {Broadbent}\ \emph {et~al.}(2009)\citenamefont
  {Broadbent}, \citenamefont {Fitzsimons},\ and\ \citenamefont
  {Kashefi}}]{broadbent_universal_2009}%
  \BibitemOpen
  \bibfield  {author} {\bibinfo {author} {\bibnamefont {Broadbent},
  \bibfnamefont {A}}, \bibinfo {author} {\bibfnamefont {J.}~\bibnamefont
  {Fitzsimons}}, and\ \bibinfo {author} {\bibfnamefont {E.}~\bibnamefont
  {Kashefi}}} (\bibinfo {year} {2009}),\ \bibfield  {title} {\enquote {\bibinfo
  {title} {Universal blind quantum computation},}\ }in\ \href
  {https://doi.org/10.1109/FOCS.2009.36} {\emph {\bibinfo {booktitle} {2009
  50th Annual {IEEE} Symposium on Foundations of Computer Science (FOCS)}}},\
  pp.\ \bibinfo {pages} {517--526},\ \Eprint {https://arxiv.org/abs/0807.4154}
  {arxiv:0807.4154} \BibitemShut {NoStop}%
\bibitem [{\citenamefont {Brod}\ \emph {et~al.}(2019)\citenamefont {Brod},
  \citenamefont {Galv{\~a}o}, \citenamefont {Crespi}, \citenamefont {Osellame},
  \citenamefont {Spagnolo},\ and\ \citenamefont
  {Sciarrino}}]{brod_photonic_2019}%
  \BibitemOpen
  \bibfield  {author} {\bibinfo {author} {\bibnamefont {Brod}, \bibfnamefont
  {D~J}}, \bibinfo {author} {\bibfnamefont {E.~F.}\ \bibnamefont {Galv{\~a}o}},
  \bibinfo {author} {\bibfnamefont {A.}~\bibnamefont {Crespi}}, \bibinfo
  {author} {\bibfnamefont {R.}~\bibnamefont {Osellame}}, \bibinfo {author}
  {\bibfnamefont {N.}~\bibnamefont {Spagnolo}}, and\ \bibinfo {author}
  {\bibfnamefont {F.}~\bibnamefont {Sciarrino}}} (\bibinfo {year} {2019}),\
  \bibfield  {title} {\enquote {\bibinfo {title} {Photonic implementation of
  boson sampling: A review},}\ }\href {https://doi.org/10.1117/1.AP.1.3.034001}
  {\bibfield  {journal} {\bibinfo  {journal} {Adv. Phot.}\ }\textbf {\bibinfo
  {volume} {1}},\ \bibinfo {pages} {034001}}\BibitemShut {NoStop}%
\bibitem [{\citenamefont {Broome}\ \emph {et~al.}(2013)\citenamefont {Broome},
  \citenamefont {Fedrizzi}, \citenamefont {{Rahimi-Keshari}}, \citenamefont
  {Dove}, \citenamefont {Aaronson}, \citenamefont {Ralph},\ and\ \citenamefont
  {White}}]{broome_photonic_2013}%
  \BibitemOpen
  \bibfield  {author} {\bibinfo {author} {\bibnamefont {Broome}, \bibfnamefont
  {M~A}}, \bibinfo {author} {\bibfnamefont {A.}~\bibnamefont {Fedrizzi}},
  \bibinfo {author} {\bibfnamefont {S.}~\bibnamefont {{Rahimi-Keshari}}},
  \bibinfo {author} {\bibfnamefont {J.}~\bibnamefont {Dove}}, \bibinfo {author}
  {\bibfnamefont {S.}~\bibnamefont {Aaronson}}, \bibinfo {author}
  {\bibfnamefont {T.~C.}\ \bibnamefont {Ralph}}, and\ \bibinfo {author}
  {\bibfnamefont {A.~G.}\ \bibnamefont {White}}} (\bibinfo {year} {2013}),\
  \bibfield  {title} {\enquote {\bibinfo {title} {Photonic {{boson sampling}}
  in a {{tunable circuit}}},}\ }\href {https://doi.org/10.1126/science.1231440}
  {\bibfield  {journal} {\bibinfo  {journal} {Science}\ }\textbf {\bibinfo
  {volume} {339}},\ \bibinfo {pages} {794--798}}\BibitemShut {NoStop}%
\bibitem [{\citenamefont {Brouwer}\ and\ \citenamefont
  {Beenakker}(1996)}]{brouwer_diagrammatic_1996}%
  \BibitemOpen
  \bibfield  {author} {\bibinfo {author} {\bibnamefont {Brouwer}, \bibfnamefont
  {P~W}}, and\ \bibinfo {author} {\bibfnamefont {C.~W.~J.}\ \bibnamefont
  {Beenakker}}} (\bibinfo {year} {1996}),\ \bibfield  {title} {\enquote
  {\bibinfo {title} {Diagrammatic method of integration over the unitary group,
  with applications to quantum transport in mesoscopic systems},}\ }\href
  {https://doi.org/10.1063/1.531667} {\bibfield  {journal} {\bibinfo  {journal}
  {J. Math. Phys.}\ }\textbf {\bibinfo {volume} {37}},\ \bibinfo {pages}
  {4904--4934}}\BibitemShut {NoStop}%
\bibitem [{\citenamefont {Bulmer}\ \emph {et~al.}(2022)\citenamefont {Bulmer},
  \citenamefont {Bell}, \citenamefont {Chadwick}, \citenamefont {Jones},
  \citenamefont {Moise}, \citenamefont {Rigazzi}, \citenamefont {Thorbecke},
  \citenamefont {Haus}, \citenamefont {Van~Vaerenbergh}, \citenamefont {Patel},
  \citenamefont {Walmsley},\ and\ \citenamefont
  {Laing}}]{bulmer_boundary_2022}%
  \BibitemOpen
  \bibfield  {author} {\bibinfo {author} {\bibnamefont {Bulmer}, \bibfnamefont
  {J~F~F}}, \bibinfo {author} {\bibfnamefont {B.~A.}\ \bibnamefont {Bell}},
  \bibinfo {author} {\bibfnamefont {R.~S.}\ \bibnamefont {Chadwick}}, \bibinfo
  {author} {\bibfnamefont {A.~E.}\ \bibnamefont {Jones}}, \bibinfo {author}
  {\bibfnamefont {D.}~\bibnamefont {Moise}}, \bibinfo {author} {\bibfnamefont
  {A.}~\bibnamefont {Rigazzi}}, \bibinfo {author} {\bibfnamefont
  {J.}~\bibnamefont {Thorbecke}}, \bibinfo {author} {\bibfnamefont {U.-U.}\
  \bibnamefont {Haus}}, \bibinfo {author} {\bibfnamefont {T.}~\bibnamefont
  {Van~Vaerenbergh}}, \bibinfo {author} {\bibfnamefont {R.~B.}\ \bibnamefont
  {Patel}}, \bibinfo {author} {\bibfnamefont {I.~A.}\ \bibnamefont {Walmsley}},
  and\ \bibinfo {author} {\bibfnamefont {A.}~\bibnamefont {Laing}}} (\bibinfo
  {year} {2022}),\ \bibfield  {title} {\enquote {\bibinfo {title} {The boundary
  for quantum advantage in {{Gaussian}} boson sampling},}\ }\href
  {https://doi.org/10.1126/sciadv.abl9236} {\bibfield  {journal} {\bibinfo
  {journal} {Science Adv.}\ }\textbf {\bibinfo {volume} {8}},\ \bibinfo {pages}
  {eabl9236}}\BibitemShut {NoStop}%
\bibitem [{\citenamefont {Bultink}\ \emph {et~al.}(2018)\citenamefont
  {Bultink}, \citenamefont {Tarasinski}, \citenamefont {Haandb{\ae}k},
  \citenamefont {Poletto}, \citenamefont {Haider}, \citenamefont {Michalak},
  \citenamefont {Bruno},\ and\ \citenamefont {DiCarlo}}]{bultink_general_2018}%
  \BibitemOpen
  \bibfield  {author} {\bibinfo {author} {\bibnamefont {Bultink}, \bibfnamefont
  {C~C}}, \bibinfo {author} {\bibfnamefont {B.}~\bibnamefont {Tarasinski}},
  \bibinfo {author} {\bibfnamefont {N.}~\bibnamefont {Haandb{\ae}k}}, \bibinfo
  {author} {\bibfnamefont {S.}~\bibnamefont {Poletto}}, \bibinfo {author}
  {\bibfnamefont {N.}~\bibnamefont {Haider}}, \bibinfo {author} {\bibfnamefont
  {D.~J.}\ \bibnamefont {Michalak}}, \bibinfo {author} {\bibfnamefont
  {A.}~\bibnamefont {Bruno}}, and\ \bibinfo {author} {\bibfnamefont
  {L.}~\bibnamefont {DiCarlo}}} (\bibinfo {year} {2018}),\ \bibfield  {title}
  {\enquote {\bibinfo {title} {General method for extracting the quantum
  efficiency of dispersive qubit readout in circuit {{QED}}},}\ }\href
  {https://doi.org/10.1063/1.5015954} {\bibfield  {journal} {\bibinfo
  {journal} {Appl. Phys. Lett.}\ }\textbf {\bibinfo {volume} {112}},\ \bibinfo
  {pages} {092601}}\BibitemShut {NoStop}%
\bibitem [{\citenamefont {Cai}\ \emph {et~al.}(1999)\citenamefont {Cai},
  \citenamefont {Pavan},\ and\ \citenamefont {Sivakumar}}]{cai_hardness_1999}%
  \BibitemOpen
  \bibfield  {author} {\bibinfo {author} {\bibnamefont {Cai}, \bibfnamefont
  {J-Y}}, \bibinfo {author} {\bibfnamefont {A.}~\bibnamefont {Pavan}}, and\
  \bibinfo {author} {\bibfnamefont {D.}~\bibnamefont {Sivakumar}}} (\bibinfo
  {year} {1999}),\ \bibfield  {title} {\enquote {\bibinfo {title} {On the
  {{hardness}} of {{permanent}}},}\ }in\ \href
  {https://doi.org/10.1007/3-540-49116-3_8} {\emph {\bibinfo {booktitle}
  {{{STACS}} 99}}},\ \bibinfo {series and number} {Lecture {{Notes}} in
  {{Computer Science}}},\ \bibinfo {editor} {edited by\ \bibinfo {editor}
  {\bibfnamefont {C.}~\bibnamefont {Meinel}}\ and\ \bibinfo {editor}
  {\bibfnamefont {S.}~\bibnamefont {Tison}}}\ (\bibinfo  {publisher}
  {{Springer}},\ \bibinfo {address} {{Berlin, Heidelberg}})\ pp.\ \bibinfo
  {pages} {90--99}\BibitemShut {NoStop}%
\bibitem [{\citenamefont {Canonne}\ and\ \citenamefont
  {Wimmer}(2020)}]{canonne_testing_2020}%
  \BibitemOpen
  \bibfield  {author} {\bibinfo {author} {\bibnamefont {Canonne}, \bibfnamefont
  {C~L}}, and\ \bibinfo {author} {\bibfnamefont {K.}~\bibnamefont {Wimmer}}}
  (\bibinfo {year} {2020}),\ \bibfield  {title} {\enquote {\bibinfo {title}
  {Testing data binnings},}\ }\href@noop {} {\ }\Eprint
  {https://arxiv.org/abs/2004.12893} {arXiv:2004.12893} \BibitemShut {NoStop}%
\bibitem [{\citenamefont {Carolan}\ \emph {et~al.}(2014)\citenamefont
  {Carolan}, \citenamefont {Meinecke}, \citenamefont {Shadbolt}, \citenamefont
  {Russell}, \citenamefont {Ismail}, \citenamefont {W{\"o}rhoff}, \citenamefont
  {Rudolph}, \citenamefont {Thompson}, \citenamefont {O'Brien}, \citenamefont
  {Matthews},\ and\ \citenamefont {Laing}}]{carolan_experimental_2014}%
  \BibitemOpen
  \bibfield  {author} {\bibinfo {author} {\bibnamefont {Carolan}, \bibfnamefont
  {J}}, \bibinfo {author} {\bibfnamefont {J.~D.~A.}\ \bibnamefont {Meinecke}},
  \bibinfo {author} {\bibfnamefont {P.}~\bibnamefont {Shadbolt}}, \bibinfo
  {author} {\bibfnamefont {N.~J.}\ \bibnamefont {Russell}}, \bibinfo {author}
  {\bibfnamefont {N.}~\bibnamefont {Ismail}}, \bibinfo {author} {\bibfnamefont
  {K.}~\bibnamefont {W{\"o}rhoff}}, \bibinfo {author} {\bibfnamefont
  {T.}~\bibnamefont {Rudolph}}, \bibinfo {author} {\bibfnamefont {M.~G.}\
  \bibnamefont {Thompson}}, \bibinfo {author} {\bibfnamefont {J.~L.}\
  \bibnamefont {O'Brien}}, \bibinfo {author} {\bibfnamefont {J.~C.~F.}\
  \bibnamefont {Matthews}}, and\ \bibinfo {author} {\bibfnamefont
  {A.}~\bibnamefont {Laing}}} (\bibinfo {year} {2014}),\ \bibfield  {title}
  {\enquote {\bibinfo {title} {On the experimental verification of quantum
  complexity in linear optics},}\ }\href
  {https://doi.org/10.1038/nphoton.2014.152} {\bibfield  {journal} {\bibinfo
  {journal} {Nature Phot.}\ }\textbf {\bibinfo {volume} {8}},\ \bibinfo {pages}
  {621--626}}\BibitemShut {NoStop}%
\bibitem [{\citenamefont {Caves}\ \emph {et~al.}(2002)\citenamefont {Caves},
  \citenamefont {Fuchs},\ and\ \citenamefont {Schack}}]{caves_unknown_2002}%
  \BibitemOpen
  \bibfield  {author} {\bibinfo {author} {\bibnamefont {Caves}, \bibfnamefont
  {C~M}}, \bibinfo {author} {\bibfnamefont {C.~A.}\ \bibnamefont {Fuchs}}, and\
  \bibinfo {author} {\bibfnamefont {R.}~\bibnamefont {Schack}}} (\bibinfo
  {year} {2002}),\ \bibfield  {title} {\enquote {\bibinfo {title} {{Unknown
  quantum states: The quantum de Finetti representation}},}\ }\href
  {https://doi.org/10.1063/1.1494475} {\bibfield  {journal} {\bibinfo
  {journal} {J. Math. Phys.}\ }\textbf {\bibinfo {volume} {43}},\ \bibinfo
  {pages} {4537--4559}},\ \Eprint {https://arxiv.org/abs/quant-ph/0104088}
  {arxiv:quant-ph/0104088} \BibitemShut {NoStop}%
\bibitem [{\citenamefont {Cerfontaine}\ \emph {et~al.}(2020)\citenamefont
  {Cerfontaine}, \citenamefont {Otten},\ and\ \citenamefont
  {Bluhm}}]{cerfontaine_self-consistent_2019}%
  \BibitemOpen
  \bibfield  {author} {\bibinfo {author} {\bibnamefont {Cerfontaine},
  \bibfnamefont {P}}, \bibinfo {author} {\bibfnamefont {R.}~\bibnamefont
  {Otten}}, and\ \bibinfo {author} {\bibfnamefont {H.}~\bibnamefont {Bluhm}}}
  (\bibinfo {year} {2020}),\ \bibfield  {title} {\enquote {\bibinfo {title}
  {Self-consistent calibration of quantum-gate sets},}\ }\href
  {https://doi.org/10.1103/PhysRevApplied.13.044071} {\bibfield  {journal}
  {\bibinfo  {journal} {Phys. Rev. Appl.}\ }\textbf {\bibinfo {volume} {13}},\
  \bibinfo {pages} {044071}}\BibitemShut {NoStop}%
\bibitem [{\citenamefont {Chabaud}\ \emph {et~al.}(2020)\citenamefont
  {Chabaud}, \citenamefont {Douce}, \citenamefont {Grosshans}, \citenamefont
  {Kashefi},\ and\ \citenamefont {Markham}}]{chabaud_building_2020}%
  \BibitemOpen
  \bibfield  {author} {\bibinfo {author} {\bibnamefont {Chabaud}, \bibfnamefont
  {U}}, \bibinfo {author} {\bibfnamefont {T.}~\bibnamefont {Douce}}, \bibinfo
  {author} {\bibfnamefont {F.}~\bibnamefont {Grosshans}}, \bibinfo {author}
  {\bibfnamefont {E.}~\bibnamefont {Kashefi}}, and\ \bibinfo {author}
  {\bibfnamefont {D.}~\bibnamefont {Markham}}} (\bibinfo {year} {2020}),\
  \bibfield  {title} {\enquote {\bibinfo {title} {Building trust for continuous
  variable quantum states},}\ }\href
  {https://doi.org/10.4230/LIPIcs.TQC.2020.3} {\ 10.4230/LIPIcs.TQC.2020.3},\
  \Eprint {https://arxiv.org/abs/1905.12700} {arXiv:1905.12700 [quant-ph]}
  \BibitemShut {NoStop}%
\bibitem [{\citenamefont {Chabaud}\ \emph {et~al.}(2017)\citenamefont
  {Chabaud}, \citenamefont {Douce}, \citenamefont {Markham}, \citenamefont
  {{van Loock}}, \citenamefont {Kashefi},\ and\ \citenamefont
  {Ferrini}}]{chabaud_continuous-variable_2017}%
  \BibitemOpen
  \bibfield  {author} {\bibinfo {author} {\bibnamefont {Chabaud}, \bibfnamefont
  {U}}, \bibinfo {author} {\bibfnamefont {T.}~\bibnamefont {Douce}}, \bibinfo
  {author} {\bibfnamefont {D.}~\bibnamefont {Markham}}, \bibinfo {author}
  {\bibfnamefont {P.}~\bibnamefont {{van Loock}}}, \bibinfo {author}
  {\bibfnamefont {E.}~\bibnamefont {Kashefi}}, and\ \bibinfo {author}
  {\bibfnamefont {G.}~\bibnamefont {Ferrini}}} (\bibinfo {year} {2017}),\
  \bibfield  {title} {\enquote {\bibinfo {title} {Continuous-variable sampling
  from photon-added or photon-subtracted squeezed states},}\ }\href
  {https://doi.org/10.1103/PhysRevA.96.062307} {\bibfield  {journal} {\bibinfo
  {journal} {Phys. Rev. A}\ }\textbf {\bibinfo {volume} {96}}~(\bibinfo
  {number} {6}),\ \bibinfo {pages} {062307}},\ \Eprint
  {https://arxiv.org/abs/1707.09245} {arXiv:1707.09245} \BibitemShut {NoStop}%
\bibitem [{\citenamefont {Chabaud}\ \emph
  {et~al.}(2021{\natexlab{a}})\citenamefont {Chabaud}, \citenamefont {Ferrini},
  \citenamefont {Grosshans},\ and\ \citenamefont
  {Markham}}]{chabaud_classical_2021}%
  \BibitemOpen
  \bibfield  {author} {\bibinfo {author} {\bibnamefont {Chabaud}, \bibfnamefont
  {U}}, \bibinfo {author} {\bibfnamefont {G.}~\bibnamefont {Ferrini}}, \bibinfo
  {author} {\bibfnamefont {F.}~\bibnamefont {Grosshans}}, and\ \bibinfo
  {author} {\bibfnamefont {D.}~\bibnamefont {Markham}}} (\bibinfo {year}
  {2021}{\natexlab{a}}),\ \bibfield  {title} {\enquote {\bibinfo {title}
  {Classical simulation of {{Gaussian}} quantum circuits with non-{{Gaussian}}
  input states},}\ }\href {https://doi.org/10.1103/PhysRevResearch.3.033018}
  {\bibfield  {journal} {\bibinfo  {journal} {Phys. Rev. Research}\ }\textbf
  {\bibinfo {volume} {3}}~(\bibinfo {number} {3}),\ \bibinfo {pages}
  {033018}}\BibitemShut {NoStop}%
\bibitem [{\citenamefont {Chabaud}\ \emph
  {et~al.}(2021{\natexlab{b}})\citenamefont {Chabaud}, \citenamefont
  {Grosshans}, \citenamefont {Kashefi},\ and\ \citenamefont
  {Markham}}]{chabaud_efficient_2021}%
  \BibitemOpen
  \bibfield  {author} {\bibinfo {author} {\bibnamefont {Chabaud}, \bibfnamefont
  {U}}, \bibinfo {author} {\bibfnamefont {F.}~\bibnamefont {Grosshans}},
  \bibinfo {author} {\bibfnamefont {E.}~\bibnamefont {Kashefi}}, and\ \bibinfo
  {author} {\bibfnamefont {D.}~\bibnamefont {Markham}}} (\bibinfo {year}
  {2021}{\natexlab{b}}),\ \bibfield  {title} {\enquote {\bibinfo {title}
  {Efficient verification of {{boson sampling}}},}\ }\href
  {https://doi.org/10.22331/q-2021-11-15-578} {\bibfield  {journal} {\bibinfo
  {journal} {Quantum}\ }\textbf {\bibinfo {volume} {5}},\ \bibinfo {pages}
  {578}}\BibitemShut {NoStop}%
\bibitem [{\citenamefont {Chabaud}\ \emph
  {et~al.}(2021{\natexlab{c}})\citenamefont {Chabaud}, \citenamefont
  {Markham},\ and\ \citenamefont {Sohbi}}]{chabaud_quantum_2021}%
  \BibitemOpen
  \bibfield  {author} {\bibinfo {author} {\bibnamefont {Chabaud}, \bibfnamefont
  {U}}, \bibinfo {author} {\bibfnamefont {D.}~\bibnamefont {Markham}}, and\
  \bibinfo {author} {\bibfnamefont {A.}~\bibnamefont {Sohbi}}} (\bibinfo {year}
  {2021}{\natexlab{c}}),\ \bibfield  {title} {\enquote {\bibinfo {title}
  {Quantum machine learning with adaptive linear optics},}\ }\href
  {https://doi.org/10.22331/q-2021-07-05-496} {\bibfield  {journal} {\bibinfo
  {journal} {Quantum}\ }\textbf {\bibinfo {volume} {5}},\ \bibinfo {pages}
  {496}}\BibitemShut {NoStop}%
\bibitem [{\citenamefont {Chabaud}\ and\ \citenamefont
  {Walschaers}(2022)}]{chabaud_resources_2022}%
  \BibitemOpen
  \bibfield  {author} {\bibinfo {author} {\bibnamefont {Chabaud}, \bibfnamefont
  {U}}, and\ \bibinfo {author} {\bibfnamefont {M.}~\bibnamefont {Walschaers}}}
  (\bibinfo {year} {2022}),\ \bibfield  {title} {\enquote {\bibinfo {title}
  {Resources for bosonic quantum computational advantage},}\ }\href@noop {} {\
  }\Eprint {https://arxiv.org/abs/2207.11781} {arXiv:2207.11781 [quant-ph]}
  \BibitemShut {NoStop}%
\bibitem [{\citenamefont {Chakhmakhchyan}\ and\ \citenamefont
  {Cerf}(2017)}]{chakhmakhchyan_boson_2017}%
  \BibitemOpen
  \bibfield  {author} {\bibinfo {author} {\bibnamefont {Chakhmakhchyan},
  \bibfnamefont {L}}, and\ \bibinfo {author} {\bibfnamefont {N.~J.}\
  \bibnamefont {Cerf}}} (\bibinfo {year} {2017}),\ \bibfield  {title} {\enquote
  {\bibinfo {title} {Boson sampling with {{Gaussian}} measurements},}\ }\href
  {https://doi.org/10.1103/PhysRevA.96.032326} {\bibfield  {journal} {\bibinfo
  {journal} {Phys. Rev. A}\ }\textbf {\bibinfo {volume} {96}},\ \bibinfo
  {pages} {032326}}\BibitemShut {NoStop}%
\bibitem [{\citenamefont {Chen}\ \emph
  {et~al.}(2018{\natexlab{a}})\citenamefont {Chen}, \citenamefont {Zhang},
  \citenamefont {Huang}, \citenamefont {Newman},\ and\ \citenamefont
  {Shi}}]{chen_classical_2018-1}%
  \BibitemOpen
  \bibfield  {author} {\bibinfo {author} {\bibnamefont {Chen}, \bibfnamefont
  {J}}, \bibinfo {author} {\bibfnamefont {F.}~\bibnamefont {Zhang}}, \bibinfo
  {author} {\bibfnamefont {C.}~\bibnamefont {Huang}}, \bibinfo {author}
  {\bibfnamefont {M.}~\bibnamefont {Newman}}, and\ \bibinfo {author}
  {\bibfnamefont {Y.}~\bibnamefont {Shi}}} (\bibinfo {year}
  {2018}{\natexlab{a}}),\ \bibfield  {title} {\enquote {\bibinfo {title}
  {Classical {{simulation}} of {{intermediate-size quantum circuits}}},}\
  }\href@noop {} {\ }\Eprint {https://arxiv.org/abs/1805.01450}
  {arXiv:1805.01450} \BibitemShut {NoStop}%
\bibitem [{\citenamefont {Chen}\ \emph {et~al.}(2020)\citenamefont {Chen},
  \citenamefont {Li}, \citenamefont {Gan}, \citenamefont {Zhu}, \citenamefont
  {Yang}, \citenamefont {Lu},\ and\ \citenamefont
  {Pan}}]{chen_quantum-teleportation-inspired_2020}%
  \BibitemOpen
  \bibfield  {author} {\bibinfo {author} {\bibnamefont {Chen}, \bibfnamefont
  {M-C}}, \bibinfo {author} {\bibfnamefont {R.}~\bibnamefont {Li}}, \bibinfo
  {author} {\bibfnamefont {L.}~\bibnamefont {Gan}}, \bibinfo {author}
  {\bibfnamefont {X.}~\bibnamefont {Zhu}}, \bibinfo {author} {\bibfnamefont
  {G.}~\bibnamefont {Yang}}, \bibinfo {author} {\bibfnamefont {C.-Y.}\
  \bibnamefont {Lu}}, and\ \bibinfo {author} {\bibfnamefont {J.-W.}\
  \bibnamefont {Pan}}} (\bibinfo {year} {2020}),\ \bibfield  {title} {\enquote
  {\bibinfo {title} {Quantum-{{teleportation-inspired algorithm}} for
  {{sampling large random quantum circuits}}},}\ }\href
  {https://doi.org/10.1103/PhysRevLett.124.080502} {\bibfield  {journal}
  {\bibinfo  {journal} {Phys. Rev. Lett.}\ }\textbf {\bibinfo {volume} {124}},\
  \bibinfo {pages} {080502}}\BibitemShut {NoStop}%
\bibitem [{\citenamefont {Chen}\ \emph
  {et~al.}(2018{\natexlab{b}})\citenamefont {Chen}, \citenamefont {Zhou},
  \citenamefont {Xue}, \citenamefont {Yang}, \citenamefont {Guo},\ and\
  \citenamefont {Guo}}]{chen_64-qubit_2018}%
  \BibitemOpen
  \bibfield  {author} {\bibinfo {author} {\bibnamefont {Chen}, \bibfnamefont
  {Z-Y}}, \bibinfo {author} {\bibfnamefont {Q.}~\bibnamefont {Zhou}}, \bibinfo
  {author} {\bibfnamefont {C.}~\bibnamefont {Xue}}, \bibinfo {author}
  {\bibfnamefont {X.}~\bibnamefont {Yang}}, \bibinfo {author} {\bibfnamefont
  {G.-C.}\ \bibnamefont {Guo}}, and\ \bibinfo {author} {\bibfnamefont {G.-P.}\
  \bibnamefont {Guo}}} (\bibinfo {year} {2018}{\natexlab{b}}),\ \bibfield
  {title} {\enquote {\bibinfo {title} {64-qubit quantum circuit simulation},}\
  }\href {https://doi.org/10.1016/j.scib.2018.06.007} {\bibfield  {journal}
  {\bibinfo  {journal} {Science Bulletin}\ }\textbf {\bibinfo {volume} {63}},\
  \bibinfo {pages} {964--971}}\BibitemShut {NoStop}%
\bibitem [{\citenamefont {Childs}\ \emph {et~al.}(2021)\citenamefont {Childs},
  \citenamefont {Su}, \citenamefont {Tran}, \citenamefont {Wiebe},\ and\
  \citenamefont {Zhu}}]{childs_theory_2021}%
  \BibitemOpen
  \bibfield  {author} {\bibinfo {author} {\bibnamefont {Childs}, \bibfnamefont
  {A~M}}, \bibinfo {author} {\bibfnamefont {Y.}~\bibnamefont {Su}}, \bibinfo
  {author} {\bibfnamefont {M.~C.}\ \bibnamefont {Tran}}, \bibinfo {author}
  {\bibfnamefont {N.}~\bibnamefont {Wiebe}}, and\ \bibinfo {author}
  {\bibfnamefont {S.}~\bibnamefont {Zhu}}} (\bibinfo {year} {2021}),\ \bibfield
   {title} {\enquote {\bibinfo {title} {Theory of {{Trotter error}} with
  {{commutator scaling}}},}\ }\href
  {https://doi.org/10.1103/PhysRevX.11.011020} {\bibfield  {journal} {\bibinfo
  {journal} {Phys. Rev. X}\ }\textbf {\bibinfo {volume} {11}},\ \bibinfo
  {pages} {011020}}\BibitemShut {NoStop}%
\bibitem [{\citenamefont {Childs}\ and\ \citenamefont
  {Wiebe}(2012)}]{ChildsWiebe}%
  \BibitemOpen
  \bibfield  {author} {\bibinfo {author} {\bibnamefont {Childs}, \bibfnamefont
  {A~M}}, and\ \bibinfo {author} {\bibfnamefont {N.}~\bibnamefont {Wiebe}}}
  (\bibinfo {year} {2012}),\ \bibfield  {title} {\enquote {\bibinfo {title}
  {Hamiltonian simulation using linear combinations of unitary operations},}\
  }\href {https://doi.org/10.48550/arXiv.1202.5822} {\bibfield  {journal}
  {\bibinfo  {journal} {Quant. Inf. Comp.}\ }\textbf {\bibinfo {volume} {12}},\
  \bibinfo {pages} {901--924}}\BibitemShut {NoStop}%
\bibitem [{\citenamefont {Chin}\ and\ \citenamefont
  {Huh}(2018)}]{chin_generalized_2018-1}%
  \BibitemOpen
  \bibfield  {author} {\bibinfo {author} {\bibnamefont {Chin}, \bibfnamefont
  {S}}, and\ \bibinfo {author} {\bibfnamefont {J.}~\bibnamefont {Huh}}}
  (\bibinfo {year} {2018}),\ \bibfield  {title} {\enquote {\bibinfo {title}
  {Generalized concurrence in boson sampling},}\ }\href
  {https://doi.org/10.1038/s41598-018-24302-5} {\bibfield  {journal} {\bibinfo
  {journal} {Sci. Rep.}\ }\textbf {\bibinfo {volume} {8}},\ \bibinfo {pages}
  {6101}}\BibitemShut {NoStop}%
\bibitem [{\citenamefont {Choi}\ \emph {et~al.}(2021)\citenamefont {Choi},
  \citenamefont {Shaw}, \citenamefont {Madjarov}, \citenamefont {Xie},
  \citenamefont {Covey}, \citenamefont {Cotler}, \citenamefont {Mark},
  \citenamefont {Huang}, \citenamefont {Kale}, \citenamefont {Pichler},
  \citenamefont {Brand{\~a}o}, \citenamefont {Choi},\ and\ \citenamefont
  {Endres}}]{choi_emergent_2021}%
  \BibitemOpen
  \bibfield  {author} {\bibinfo {author} {\bibnamefont {Choi}, \bibfnamefont
  {J}}, \bibinfo {author} {\bibfnamefont {A.~L.}\ \bibnamefont {Shaw}},
  \bibinfo {author} {\bibfnamefont {I.~S.}\ \bibnamefont {Madjarov}}, \bibinfo
  {author} {\bibfnamefont {X.}~\bibnamefont {Xie}}, \bibinfo {author}
  {\bibfnamefont {J.~P.}\ \bibnamefont {Covey}}, \bibinfo {author}
  {\bibfnamefont {J.~S.}\ \bibnamefont {Cotler}}, \bibinfo {author}
  {\bibfnamefont {D.~K.}\ \bibnamefont {Mark}}, \bibinfo {author}
  {\bibfnamefont {H.-Y.}\ \bibnamefont {Huang}}, \bibinfo {author}
  {\bibfnamefont {A.}~\bibnamefont {Kale}}, \bibinfo {author} {\bibfnamefont
  {H.}~\bibnamefont {Pichler}}, \bibinfo {author} {\bibfnamefont {F.~G. S.~L.}\
  \bibnamefont {Brand{\~a}o}}, \bibinfo {author} {\bibfnamefont
  {S.}~\bibnamefont {Choi}}, and\ \bibinfo {author} {\bibfnamefont
  {M.}~\bibnamefont {Endres}}} (\bibinfo {year} {2021}),\ \bibfield  {title}
  {\enquote {\bibinfo {title} {Emergent {{randomness}} and {{benchmarking}}
  from {{many-body quantum chaos}}},}\ }\href@noop {} {\ }\Eprint
  {https://arxiv.org/abs/2103.03535} {arXiv:2103.03535} \BibitemShut {NoStop}%
\bibitem [{\citenamefont {Choi}\ \emph {et~al.}(2016)\citenamefont {Choi},
  \citenamefont {Hild}, \citenamefont {Zeiher}, \citenamefont {Schau{\ss}},
  \citenamefont {{Rubio-Abadal}}, \citenamefont {Yefsah}, \citenamefont
  {Khemani}, \citenamefont {Huse}, \citenamefont {Bloch},\ and\ \citenamefont
  {Gross}}]{choi_exploring_2016}%
  \BibitemOpen
  \bibfield  {author} {\bibinfo {author} {\bibnamefont {Choi}, \bibfnamefont
  {J-y}}, \bibinfo {author} {\bibfnamefont {S.}~\bibnamefont {Hild}}, \bibinfo
  {author} {\bibfnamefont {J.s}\ \bibnamefont {Zeiher}}, \bibinfo {author}
  {\bibfnamefont {P.}~\bibnamefont {Schau{\ss}}}, \bibinfo {author}
  {\bibfnamefont {A.}~\bibnamefont {{Rubio-Abadal}}}, \bibinfo {author}
  {\bibfnamefont {T.}~\bibnamefont {Yefsah}}, \bibinfo {author} {\bibfnamefont
  {V.}~\bibnamefont {Khemani}}, \bibinfo {author} {\bibfnamefont {D.~A.}\
  \bibnamefont {Huse}}, \bibinfo {author} {\bibfnamefont {I.}~\bibnamefont
  {Bloch}}, and\ \bibinfo {author} {\bibfnamefont {C.}~\bibnamefont {Gross}}}
  (\bibinfo {year} {2016}),\ \bibfield  {title} {\enquote {\bibinfo {title}
  {Exploring the many-body localization transition in two dimensions},}\ }\href
  {https://doi.org/10.1126/science.aaf8834} {\bibfield  {journal} {\bibinfo
  {journal} {Science}\ }\textbf {\bibinfo {volume} {352}},\ \bibinfo {pages}
  {1547--1552}},\ \Eprint {https://arxiv.org/abs/1604.04178} {arXiv:1604.04178}
  \BibitemShut {NoStop}%
\bibitem [{\citenamefont {Chung}\ \emph {et~al.}(2020)\citenamefont {Chung},
  \citenamefont {Lee}, \citenamefont {Lin},\ and\ \citenamefont
  {Wu}}]{chung_constant-round_2020}%
  \BibitemOpen
  \bibfield  {author} {\bibinfo {author} {\bibnamefont {Chung}, \bibfnamefont
  {K-M}}, \bibinfo {author} {\bibfnamefont {Y.}~\bibnamefont {Lee}}, \bibinfo
  {author} {\bibfnamefont {H.-H.}\ \bibnamefont {Lin}}, and\ \bibinfo {author}
  {\bibfnamefont {X.}~\bibnamefont {Wu}}} (\bibinfo {year} {2020}),\ \bibfield
  {title} {\enquote {\bibinfo {title} {Constant-round {{blind classical
  verification}} of {{quantum sampling}}},}\ }\href@noop {} {\ }\Eprint
  {https://arxiv.org/abs/2012.04848} {arXiv:2012.04848} \BibitemShut {NoStop}%
\bibitem [{\citenamefont {Clarke}\ and\ \citenamefont
  {Wilhelm}(2008)}]{clarke_superconducting_2008}%
  \BibitemOpen
  \bibfield  {author} {\bibinfo {author} {\bibnamefont {Clarke}, \bibfnamefont
  {J}}, and\ \bibinfo {author} {\bibfnamefont {F.~K.}\ \bibnamefont {Wilhelm}}}
  (\bibinfo {year} {2008}),\ \bibfield  {title} {\enquote {\bibinfo {title}
  {Superconducting quantum bits},}\ }\href
  {https://doi.org/10.1038/nature07128} {\bibfield  {journal} {\bibinfo
  {journal} {Nature}\ }\textbf {\bibinfo {volume} {453}},\ \bibinfo {pages}
  {1031--1042}}\BibitemShut {NoStop}%
\bibitem [{\citenamefont {Clements}\ \emph {et~al.}(2018)\citenamefont
  {Clements}, \citenamefont {Renema}, \citenamefont {Eckstein}, \citenamefont
  {Valido}, \citenamefont {Lita}, \citenamefont {Gerrits}, \citenamefont {Nam},
  \citenamefont {Kolthammer}, \citenamefont {Huh},\ and\ \citenamefont
  {Walmsley}}]{clements_approximating_2018}%
  \BibitemOpen
  \bibfield  {author} {\bibinfo {author} {\bibnamefont {Clements},
  \bibfnamefont {W~R}}, \bibinfo {author} {\bibfnamefont {J.~J.}\ \bibnamefont
  {Renema}}, \bibinfo {author} {\bibfnamefont {A.}~\bibnamefont {Eckstein}},
  \bibinfo {author} {\bibfnamefont {A.~A.}\ \bibnamefont {Valido}}, \bibinfo
  {author} {\bibfnamefont {A.}~\bibnamefont {Lita}}, \bibinfo {author}
  {\bibfnamefont {T.}~\bibnamefont {Gerrits}}, \bibinfo {author} {\bibfnamefont
  {S.~W.}\ \bibnamefont {Nam}}, \bibinfo {author} {\bibfnamefont {W.~S.}\
  \bibnamefont {Kolthammer}}, \bibinfo {author} {\bibfnamefont
  {J.}~\bibnamefont {Huh}}, and\ \bibinfo {author} {\bibfnamefont {I.~A.}\
  \bibnamefont {Walmsley}}} (\bibinfo {year} {2018}),\ \bibfield  {title}
  {\enquote {\bibinfo {title} {Approximating vibronic spectroscopy with
  imperfect quantum optics},}\ }\href
  {https://doi.org/10.1088/1361-6455/aaf031} {\bibfield  {journal} {\bibinfo
  {journal} {J. Phys. B}\ }\textbf {\bibinfo {volume} {51}},\ \bibinfo {pages}
  {245503}}\BibitemShut {NoStop}%
\bibitem [{\citenamefont {Clifford}\ and\ \citenamefont
  {Clifford}(2018)}]{clifford_classical_2018}%
  \BibitemOpen
  \bibfield  {author} {\bibinfo {author} {\bibnamefont {Clifford},
  \bibfnamefont {P}}, and\ \bibinfo {author} {\bibfnamefont {R.}~\bibnamefont
  {Clifford}}} (\bibinfo {year} {2018}),\ \bibfield  {title} {\enquote
  {\bibinfo {title} {The {{classical complexity}} of {{Boson sampling}}},}\
  }in\ \href@noop {} {\emph {\bibinfo {booktitle} {Proc. {{29th Ann. ACM-SIAM
  Symp.}} {{Disc. Alg.}}}}},\ \bibinfo {series and number} {{{SODA}} '18}\
  (\bibinfo  {publisher} {{Society for Industrial and Applied Mathematics}},\
  \bibinfo {address} {{Philadelphia, PA, USA}})\ pp.\ \bibinfo {pages}
  {146--155}\BibitemShut {NoStop}%
\bibitem [{\citenamefont {Clifford}\ and\ \citenamefont
  {Clifford}(2020)}]{clifford_faster_2020}%
  \BibitemOpen
  \bibfield  {author} {\bibinfo {author} {\bibnamefont {Clifford},
  \bibfnamefont {P}}, and\ \bibinfo {author} {\bibfnamefont {R.}~\bibnamefont
  {Clifford}}} (\bibinfo {year} {2020}),\ \bibfield  {title} {\enquote
  {\bibinfo {title} {Faster classical {{boson sampling}}},}\ }\href@noop {} {\
  }\Eprint {https://arxiv.org/abs/2005.04214} {arXiv:2005.04214} \BibitemShut
  {NoStop}%
\bibitem [{\citenamefont {Cramer}\ \emph {et~al.}(2010)\citenamefont {Cramer},
  \citenamefont {Plenio}, \citenamefont {Flammia}, \citenamefont {Somma},
  \citenamefont {Gross}, \citenamefont {Bartlett}, \citenamefont
  {{Landon-Cardinal}}, \citenamefont {Poulin},\ and\ \citenamefont
  {Liu}}]{cramer_efficient_2010}%
  \BibitemOpen
  \bibfield  {author} {\bibinfo {author} {\bibnamefont {Cramer}, \bibfnamefont
  {M}}, \bibinfo {author} {\bibfnamefont {M.~B.}\ \bibnamefont {Plenio}},
  \bibinfo {author} {\bibfnamefont {S.~T.}\ \bibnamefont {Flammia}}, \bibinfo
  {author} {\bibfnamefont {R.}~\bibnamefont {Somma}}, \bibinfo {author}
  {\bibfnamefont {D.}~\bibnamefont {Gross}}, \bibinfo {author} {\bibfnamefont
  {S.~D.}\ \bibnamefont {Bartlett}}, \bibinfo {author} {\bibfnamefont
  {O.}~\bibnamefont {{Landon-Cardinal}}}, \bibinfo {author} {\bibfnamefont
  {D.}~\bibnamefont {Poulin}}, and\ \bibinfo {author} {\bibfnamefont {Y.-K.}\
  \bibnamefont {Liu}}} (\bibinfo {year} {2010}),\ \bibfield  {title} {\enquote
  {\bibinfo {title} {Efficient quantum state tomography},}\ }\href
  {https://doi.org/10.1038/ncomms1147} {\bibfield  {journal} {\bibinfo
  {journal} {Nature Comm.}\ }\textbf {\bibinfo {volume} {1}},\ \bibinfo {pages}
  {149}}\BibitemShut {NoStop}%
\bibitem [{\citenamefont {Crespi}\ \emph {et~al.}(2013)\citenamefont {Crespi},
  \citenamefont {Osellame}, \citenamefont {Ramponi}, \citenamefont {Brod},
  \citenamefont {Galv{\~a}o}, \citenamefont {Spagnolo}, \citenamefont
  {Vitelli}, \citenamefont {Maiorino}, \citenamefont {Mataloni},\ and\
  \citenamefont {Sciarrino}}]{crespi_integrated_2013}%
  \BibitemOpen
  \bibfield  {author} {\bibinfo {author} {\bibnamefont {Crespi}, \bibfnamefont
  {A}}, \bibinfo {author} {\bibfnamefont {R.}~\bibnamefont {Osellame}},
  \bibinfo {author} {\bibfnamefont {R.}~\bibnamefont {Ramponi}}, \bibinfo
  {author} {\bibfnamefont {D.~J.}\ \bibnamefont {Brod}}, \bibinfo {author}
  {\bibfnamefont {E.~F.}\ \bibnamefont {Galv{\~a}o}}, \bibinfo {author}
  {\bibfnamefont {N.}~\bibnamefont {Spagnolo}}, \bibinfo {author}
  {\bibfnamefont {C.}~\bibnamefont {Vitelli}}, \bibinfo {author} {\bibfnamefont
  {E.}~\bibnamefont {Maiorino}}, \bibinfo {author} {\bibfnamefont
  {P.}~\bibnamefont {Mataloni}}, and\ \bibinfo {author} {\bibfnamefont
  {F.}~\bibnamefont {Sciarrino}}} (\bibinfo {year} {2013}),\ \bibfield  {title}
  {\enquote {\bibinfo {title} {Integrated multimode interferometers with
  arbitrary designs for photonic boson sampling},}\ }\href
  {https://doi.org/10.1038/nphoton.2013.112} {\bibfield  {journal} {\bibinfo
  {journal} {Nature Phot.}\ }\textbf {\bibinfo {volume} {7}},\ \bibinfo {pages}
  {545--549}}\BibitemShut {NoStop}%
\bibitem [{\citenamefont {De~las Cuevas}\ \emph {et~al.}(2011)\citenamefont
  {De~las Cuevas}, \citenamefont {D{\"u}r}, \citenamefont {Van~den Nest},\ and\
  \citenamefont {Martin-Delgado}}]{de_las_cuevas_quantum_2011}%
  \BibitemOpen
  \bibfield  {author} {\bibinfo {author} {\bibnamefont {De~las Cuevas},
  \bibfnamefont {G}}, \bibinfo {author} {\bibfnamefont {W.}~\bibnamefont
  {D{\"u}r}}, \bibinfo {author} {\bibfnamefont {M.}~\bibnamefont {Van~den
  Nest}}, and\ \bibinfo {author} {\bibfnamefont {M.~A.}\ \bibnamefont
  {Martin-Delgado}}} (\bibinfo {year} {2011}),\ \bibfield  {title} {\enquote
  {\bibinfo {title} {Quantum algorithms for classical lattice models},}\ }\href
  {https://doi.org/10.1088/1367-2630/13/9/093021} {\bibfield  {journal}
  {\bibinfo  {journal} {New J. Phys.}\ }\textbf {\bibinfo {volume} {13}},\
  \bibinfo {pages} {093021}}\BibitemShut {NoStop}%
\bibitem [{\citenamefont {Dalzell}\ \emph {et~al.}(2020)\citenamefont
  {Dalzell}, \citenamefont {Harrow}, \citenamefont {Koh},\ and\ \citenamefont
  {Placa}}]{dalzell_how_2018}%
  \BibitemOpen
  \bibfield  {author} {\bibinfo {author} {\bibnamefont {Dalzell}, \bibfnamefont
  {A~M}}, \bibinfo {author} {\bibfnamefont {A.~W.}\ \bibnamefont {Harrow}},
  \bibinfo {author} {\bibfnamefont {D.~E.}\ \bibnamefont {Koh}}, and\ \bibinfo
  {author} {\bibfnamefont {R.~L.~La}\ \bibnamefont {Placa}}} (\bibinfo {year}
  {2020}),\ \bibfield  {title} {\enquote {\bibinfo {title} {How many qubits are
  needed for quantum computational supremacy?}}\ }\href
  {https://doi.org/10.22331/q-2020-05-11-264} {\bibfield  {journal} {\bibinfo
  {journal} {Quantum}\ }\textbf {\bibinfo {volume} {4}},\ \bibinfo {pages}
  {264}}\BibitemShut {NoStop}%
\bibitem [{\citenamefont {Dalzell}\ \emph {et~al.}(2021)\citenamefont
  {Dalzell}, \citenamefont {{Hunter-Jones}},\ and\ \citenamefont
  {Brand{\~a}o}}]{dalzell_random_2021}%
  \BibitemOpen
  \bibfield  {author} {\bibinfo {author} {\bibnamefont {Dalzell}, \bibfnamefont
  {A~M}}, \bibinfo {author} {\bibfnamefont {N.}~\bibnamefont {{Hunter-Jones}}},
  and\ \bibinfo {author} {\bibfnamefont {F.~G. S.~L.}\ \bibnamefont
  {Brand{\~a}o}}} (\bibinfo {year} {2021}),\ \bibfield  {title} {\enquote
  {\bibinfo {title} {Random quantum circuits transform local noise into global
  white noise},}\ }\href@noop {} {\ }\Eprint {https://arxiv.org/abs/2111.14907}
  {arXiv:2111.14907} \BibitemShut {NoStop}%
\bibitem [{\citenamefont {Dalzell}\ \emph {et~al.}(2022)\citenamefont
  {Dalzell}, \citenamefont {{Hunter-Jones}},\ and\ \citenamefont
  {Brand{\~a}o}}]{dalzell_random_2020}%
  \BibitemOpen
  \bibfield  {author} {\bibinfo {author} {\bibnamefont {Dalzell}, \bibfnamefont
  {A~M}}, \bibinfo {author} {\bibfnamefont {N.}~\bibnamefont {{Hunter-Jones}}},
  and\ \bibinfo {author} {\bibfnamefont {F.~G. S.~L.}\ \bibnamefont
  {Brand{\~a}o}}} (\bibinfo {year} {2022}),\ \bibfield  {title} {\enquote
  {\bibinfo {title} {Random {{quantum circuits anticoncentrate}} in {{log
  depth}}},}\ }\href {https://doi.org/10.1103/PRXQuantum.3.010333} {\bibfield
  {journal} {\bibinfo  {journal} {PRX Quantum}\ }\textbf {\bibinfo {volume}
  {3}},\ \bibinfo {pages} {010333}},\ \Eprint
  {https://arxiv.org/abs/2011.12277} {arXiv:2011.12277} \BibitemShut {NoStop}%
\bibitem [{\citenamefont {Dawson}\ \emph {et~al.}(2005)\citenamefont {Dawson},
  \citenamefont {Haselgrove}, \citenamefont {Hines}, \citenamefont {Mortimer},
  \citenamefont {Nielsen},\ and\ \citenamefont
  {Osborne}}]{dawson_quantum_2005}%
  \BibitemOpen
  \bibfield  {author} {\bibinfo {author} {\bibnamefont {Dawson}, \bibfnamefont
  {C~M}}, \bibinfo {author} {\bibfnamefont {H.~L.}\ \bibnamefont {Haselgrove}},
  \bibinfo {author} {\bibfnamefont {A.~P.}\ \bibnamefont {Hines}}, \bibinfo
  {author} {\bibfnamefont {D.}~\bibnamefont {Mortimer}}, \bibinfo {author}
  {\bibfnamefont {M.~A.}\ \bibnamefont {Nielsen}}, and\ \bibinfo {author}
  {\bibfnamefont {T.~J.}\ \bibnamefont {Osborne}}} (\bibinfo {year} {2005}),\
  \bibfield  {title} {\enquote {\bibinfo {title} {Quantum computing and
  polynomial equations over the finite field $\mathbb{Z}_2$},}\ }\href
  {https://doi.org/10.26421/QIC5.2-2} {\bibfield  {journal} {\bibinfo
  {journal} {Quant. Inf. Comp.}\ }\textbf {\bibinfo {volume} {5}},\
  10.26421/QIC5.2-2}\BibitemShut {NoStop}%
\bibitem [{\citenamefont {De~Raedt}\ \emph {et~al.}(2019)\citenamefont
  {De~Raedt}, \citenamefont {Jin}, \citenamefont {Willsch}, \citenamefont
  {Willsch}, \citenamefont {Yoshioka}, \citenamefont {Ito}, \citenamefont
  {Yuan},\ and\ \citenamefont {Michielsen}}]{de_raedt_massively_2019}%
  \BibitemOpen
  \bibfield  {author} {\bibinfo {author} {\bibnamefont {De~Raedt},
  \bibfnamefont {H}}, \bibinfo {author} {\bibfnamefont {F.}~\bibnamefont
  {Jin}}, \bibinfo {author} {\bibfnamefont {D.}~\bibnamefont {Willsch}},
  \bibinfo {author} {\bibfnamefont {M.}~\bibnamefont {Willsch}}, \bibinfo
  {author} {\bibfnamefont {N.}~\bibnamefont {Yoshioka}}, \bibinfo {author}
  {\bibfnamefont {N.}~\bibnamefont {Ito}}, \bibinfo {author} {\bibfnamefont
  {S.}~\bibnamefont {Yuan}}, and\ \bibinfo {author} {\bibfnamefont
  {K.}~\bibnamefont {Michielsen}}} (\bibinfo {year} {2019}),\ \bibfield
  {title} {\enquote {\bibinfo {title} {Massively parallel quantum computer
  simulator, eleven years later},}\ }\href
  {https://doi.org/10.1016/j.cpc.2018.11.005} {\bibfield  {journal} {\bibinfo
  {journal} {Comp. Phys. Comm.}\ }\textbf {\bibinfo {volume} {237}},\ \bibinfo
  {pages} {47--61}}\BibitemShut {NoStop}%
\bibitem [{\citenamefont {De~Raedt}\ \emph {et~al.}(2007)\citenamefont
  {De~Raedt}, \citenamefont {Michielsen}, \citenamefont {De~Raedt},
  \citenamefont {Trieu}, \citenamefont {Arnold}, \citenamefont {Richter},
  \citenamefont {Lippert}, \citenamefont {Watanabe},\ and\ \citenamefont
  {Ito}}]{de_raedt_massively_2007}%
  \BibitemOpen
  \bibfield  {author} {\bibinfo {author} {\bibnamefont {De~Raedt},
  \bibfnamefont {K}}, \bibinfo {author} {\bibfnamefont {K.}~\bibnamefont
  {Michielsen}}, \bibinfo {author} {\bibfnamefont {H.}~\bibnamefont
  {De~Raedt}}, \bibinfo {author} {\bibfnamefont {B.}~\bibnamefont {Trieu}},
  \bibinfo {author} {\bibfnamefont {G.}~\bibnamefont {Arnold}}, \bibinfo
  {author} {\bibfnamefont {M.}~\bibnamefont {Richter}}, \bibinfo {author}
  {\bibfnamefont {Th.}\ \bibnamefont {Lippert}}, \bibinfo {author}
  {\bibfnamefont {H.}~\bibnamefont {Watanabe}}, and\ \bibinfo {author}
  {\bibfnamefont {N.}~\bibnamefont {Ito}}} (\bibinfo {year} {2007}),\ \bibfield
   {title} {\enquote {\bibinfo {title} {Massively parallel quantum computer
  simulator},}\ }\href {https://doi.org/10.1016/j.cpc.2006.08.007} {\bibfield
  {journal} {\bibinfo  {journal} {Comp. Phys. Comm.}\ }\textbf {\bibinfo
  {volume} {176}},\ \bibinfo {pages} {121--136}}\BibitemShut {NoStop}%
\bibitem [{\citenamefont {Debnath}\ \emph {et~al.}(2016)\citenamefont
  {Debnath}, \citenamefont {Linke}, \citenamefont {Figgatt}, \citenamefont
  {Landsman}, \citenamefont {Wright},\ and\ \citenamefont
  {Monroe}}]{debnath_demonstration_2016}%
  \BibitemOpen
  \bibfield  {author} {\bibinfo {author} {\bibnamefont {Debnath}, \bibfnamefont
  {S}}, \bibinfo {author} {\bibfnamefont {N.~M.}\ \bibnamefont {Linke}},
  \bibinfo {author} {\bibfnamefont {C.}~\bibnamefont {Figgatt}}, \bibinfo
  {author} {\bibfnamefont {K.~A.}\ \bibnamefont {Landsman}}, \bibinfo {author}
  {\bibfnamefont {K.}~\bibnamefont {Wright}}, and\ \bibinfo {author}
  {\bibfnamefont {C.}~\bibnamefont {Monroe}}} (\bibinfo {year} {2016}),\
  \bibfield  {title} {\enquote {\bibinfo {title} {Demonstration of a small
  programmable quantum computer with atomic qubits},}\ }\href
  {https://doi.org/10.1038/nature18648} {\bibfield  {journal} {\bibinfo
  {journal} {Nature}\ }\textbf {\bibinfo {volume} {536}},\ \bibinfo {pages}
  {63--66}}\BibitemShut {NoStop}%
\bibitem [{\citenamefont {Deshpande}\ \emph {et~al.}(2018)\citenamefont
  {Deshpande}, \citenamefont {Fefferman}, \citenamefont {Tran}, \citenamefont
  {{Foss-Feig}},\ and\ \citenamefont {Gorshkov}}]{deshpande_dynamical_2018}%
  \BibitemOpen
  \bibfield  {author} {\bibinfo {author} {\bibnamefont {Deshpande},
  \bibfnamefont {A}}, \bibinfo {author} {\bibfnamefont {B.}~\bibnamefont
  {Fefferman}}, \bibinfo {author} {\bibfnamefont {M.~C.}\ \bibnamefont {Tran}},
  \bibinfo {author} {\bibfnamefont {M.}~\bibnamefont {{Foss-Feig}}}, and\
  \bibinfo {author} {\bibfnamefont {A.~V.}\ \bibnamefont {Gorshkov}}} (\bibinfo
  {year} {2018}),\ \bibfield  {title} {\enquote {\bibinfo {title} {Dynamical
  {{phase transitions}} in {{sampling complexity}}},}\ }\href
  {https://doi.org/10.1103/PhysRevLett.121.030501} {\bibfield  {journal}
  {\bibinfo  {journal} {Phys. Rev. Lett.}\ }\textbf {\bibinfo {volume} {121}},\
  \bibinfo {pages} {030501}}\BibitemShut {NoStop}%
\bibitem [{\citenamefont {Deshpande}\ \emph {et~al.}(2022)\citenamefont
  {Deshpande}, \citenamefont {Mehta}, \citenamefont {Vincent}, \citenamefont
  {Quesada}, \citenamefont {Hinsche}, \citenamefont {Ioannou}, \citenamefont
  {Madsen}, \citenamefont {Lavoie}, \citenamefont {Qi}, \citenamefont {Eisert},
  \citenamefont {Hangleiter}, \citenamefont {Fefferman},\ and\ \citenamefont
  {Dhand}}]{deshpande_quantum_2022}%
  \BibitemOpen
  \bibfield  {author} {\bibinfo {author} {\bibnamefont {Deshpande},
  \bibfnamefont {A}}, \bibinfo {author} {\bibfnamefont {A.}~\bibnamefont
  {Mehta}}, \bibinfo {author} {\bibfnamefont {T.}~\bibnamefont {Vincent}},
  \bibinfo {author} {\bibfnamefont {N.}~\bibnamefont {Quesada}}, \bibinfo
  {author} {\bibfnamefont {M.}~\bibnamefont {Hinsche}}, \bibinfo {author}
  {\bibfnamefont {M.}~\bibnamefont {Ioannou}}, \bibinfo {author} {\bibfnamefont
  {L.}~\bibnamefont {Madsen}}, \bibinfo {author} {\bibfnamefont
  {J.}~\bibnamefont {Lavoie}}, \bibinfo {author} {\bibfnamefont
  {H.}~\bibnamefont {Qi}}, \bibinfo {author} {\bibfnamefont {J.}~\bibnamefont
  {Eisert}}, \bibinfo {author} {\bibfnamefont {D.}~\bibnamefont {Hangleiter}},
  \bibinfo {author} {\bibfnamefont {B.}~\bibnamefont {Fefferman}}, and\
  \bibinfo {author} {\bibfnamefont {I.}~\bibnamefont {Dhand}}} (\bibinfo {year}
  {2022}),\ \bibfield  {title} {\enquote {\bibinfo {title} {Quantum
  computational advantage via high-dimensional {{Gaussian}} boson sampling},}\
  }\href {https://doi.org/10.1126/sciadv.abi7894} {\bibfield  {journal}
  {\bibinfo  {journal} {Science Adv.}\ }\textbf {\bibinfo {volume} {8}},\
  \bibinfo {pages} {eabi7894}}\BibitemShut {NoStop}%
\bibitem [{\citenamefont {Deshpande}\ \emph {et~al.}(2021)\citenamefont
  {Deshpande}, \citenamefont {Niroula}, \citenamefont {Shtanko}, \citenamefont
  {Gorshkov}, \citenamefont {Fefferman},\ and\ \citenamefont
  {Gullans}}]{deshpande_tight_2022}%
  \BibitemOpen
  \bibfield  {author} {\bibinfo {author} {\bibnamefont {Deshpande},
  \bibfnamefont {A}}, \bibinfo {author} {\bibfnamefont {P.}~\bibnamefont
  {Niroula}}, \bibinfo {author} {\bibfnamefont {O.}~\bibnamefont {Shtanko}},
  \bibinfo {author} {\bibfnamefont {A.~V.}\ \bibnamefont {Gorshkov}}, \bibinfo
  {author} {\bibfnamefont {B.}~\bibnamefont {Fefferman}}, and\ \bibinfo
  {author} {\bibfnamefont {M.~J.}\ \bibnamefont {Gullans}}} (\bibinfo {year}
  {2021}),\ \bibfield  {title} {\enquote {\bibinfo {title} {Tight bounds on the
  convergence of noisy random circuits to the uniform distribution},}\
  }\href@noop {} {\ }\Eprint {https://arxiv.org/abs/2112.00716}
  {arXiv:2112.00716} \BibitemShut {NoStop}%
\bibitem [{\citenamefont {Deutsch}(1985)}]{deutsch_quantum_1985}%
  \BibitemOpen
  \bibfield  {author} {\bibinfo {author} {\bibnamefont {Deutsch}, \bibfnamefont
  {D}}} (\bibinfo {year} {1985}),\ \bibfield  {title} {\enquote {\bibinfo
  {title} {Quantum theory, the {{Church}}\textendash{{Turing}} principle and
  the universal quantum computer},}\ }\href
  {https://doi.org/10.1098/rspa.1985.0070} {\bibfield  {journal} {\bibinfo
  {journal} {Proc. R. Soc. Lond. A}\ }\textbf {\bibinfo {volume} {400}},\
  \bibinfo {pages} {97--117}}\BibitemShut {NoStop}%
\bibitem [{\citenamefont {Drummond}\ \emph {et~al.}(2022)\citenamefont
  {Drummond}, \citenamefont {Opanchuk}, \citenamefont {Dellios},\ and\
  \citenamefont {Reid}}]{drummond_simulating_2022}%
  \BibitemOpen
  \bibfield  {author} {\bibinfo {author} {\bibnamefont {Drummond},
  \bibfnamefont {P~D}}, \bibinfo {author} {\bibfnamefont {B.}~\bibnamefont
  {Opanchuk}}, \bibinfo {author} {\bibfnamefont {A.}~\bibnamefont {Dellios}},
  and\ \bibinfo {author} {\bibfnamefont {M.~D.}\ \bibnamefont {Reid}}}
  (\bibinfo {year} {2022}),\ \bibfield  {title} {\enquote {\bibinfo {title}
  {Simulating complex networks in phase space: {{Gaussian}} boson sampling},}\
  }\href {https://doi.org/10.1103/PhysRevA.105.012427} {\bibfield  {journal}
  {\bibinfo  {journal} {Phys. Rev. A}\ }\textbf {\bibinfo {volume} {105}},\
  \bibinfo {pages} {012427}}\BibitemShut {NoStop}%
\bibitem [{\citenamefont {Ebadi}\ \emph {et~al.}(2021)\citenamefont {Ebadi},
  \citenamefont {Wang}, \citenamefont {Levine}, \citenamefont {Keesling},
  \citenamefont {Semeghini}, \citenamefont {Omran}, \citenamefont {Bluvstein},
  \citenamefont {Samajdar}, \citenamefont {Pichler}, \citenamefont {Ho},
  \citenamefont {Choi}, \citenamefont {Sachdev}, \citenamefont {Greiner},
  \citenamefont {Vuleti{\'c}},\ and\ \citenamefont
  {Lukin}}]{ebadi_quantum_2021}%
  \BibitemOpen
  \bibfield  {author} {\bibinfo {author} {\bibnamefont {Ebadi}, \bibfnamefont
  {S}}, \bibinfo {author} {\bibfnamefont {T.~T.}\ \bibnamefont {Wang}},
  \bibinfo {author} {\bibfnamefont {H.}~\bibnamefont {Levine}}, \bibinfo
  {author} {\bibfnamefont {A.}~\bibnamefont {Keesling}}, \bibinfo {author}
  {\bibfnamefont {G.}~\bibnamefont {Semeghini}}, \bibinfo {author}
  {\bibfnamefont {A.}~\bibnamefont {Omran}}, \bibinfo {author} {\bibfnamefont
  {D.}~\bibnamefont {Bluvstein}}, \bibinfo {author} {\bibfnamefont
  {R.}~\bibnamefont {Samajdar}}, \bibinfo {author} {\bibfnamefont
  {H.}~\bibnamefont {Pichler}}, \bibinfo {author} {\bibfnamefont {W.~W.}\
  \bibnamefont {Ho}}, \bibinfo {author} {\bibfnamefont {S.}~\bibnamefont
  {Choi}}, \bibinfo {author} {\bibfnamefont {S.}~\bibnamefont {Sachdev}},
  \bibinfo {author} {\bibfnamefont {M.}~\bibnamefont {Greiner}}, \bibinfo
  {author} {\bibfnamefont {V.}~\bibnamefont {Vuleti{\'c}}}, and\ \bibinfo
  {author} {\bibfnamefont {M.~D.}\ \bibnamefont {Lukin}}} (\bibinfo {year}
  {2021}),\ \bibfield  {title} {\enquote {\bibinfo {title} {Quantum phases of
  matter on a 256-atom programmable quantum simulator},}\ }\href
  {https://doi.org/10.1038/s41586-021-03582-4} {\bibfield  {journal} {\bibinfo
  {journal} {Nature}\ }\textbf {\bibinfo {volume} {595}},\ \bibinfo {pages}
  {227--232}}\BibitemShut {NoStop}%
\bibitem [{\citenamefont {Egan}\ \emph {et~al.}(2021)\citenamefont {Egan},
  \citenamefont {Debroy}, \citenamefont {Noel}, \citenamefont {Risinger},
  \citenamefont {Zhu}, \citenamefont {Biswas}, \citenamefont {Newman},
  \citenamefont {Li}, \citenamefont {Brown}, \citenamefont {Cetina},\ and\
  \citenamefont {Monroe}}]{egan_fault-tolerant_2021}%
  \BibitemOpen
  \bibfield  {author} {\bibinfo {author} {\bibnamefont {Egan}, \bibfnamefont
  {L}}, \bibinfo {author} {\bibfnamefont {D.~M.}\ \bibnamefont {Debroy}},
  \bibinfo {author} {\bibfnamefont {C.}~\bibnamefont {Noel}}, \bibinfo {author}
  {\bibfnamefont {A.}~\bibnamefont {Risinger}}, \bibinfo {author}
  {\bibfnamefont {D.}~\bibnamefont {Zhu}}, \bibinfo {author} {\bibfnamefont
  {D.}~\bibnamefont {Biswas}}, \bibinfo {author} {\bibfnamefont
  {M.}~\bibnamefont {Newman}}, \bibinfo {author} {\bibfnamefont
  {M.}~\bibnamefont {Li}}, \bibinfo {author} {\bibfnamefont {K.~R.}\
  \bibnamefont {Brown}}, \bibinfo {author} {\bibfnamefont {M.}~\bibnamefont
  {Cetina}}, and\ \bibinfo {author} {\bibfnamefont {C.}~\bibnamefont {Monroe}}}
  (\bibinfo {year} {2021}),\ \bibfield  {title} {\enquote {\bibinfo {title}
  {Fault-tolerant control of an eerror-corrected qubit},}\ }\href
  {https://doi.org/10.1038/s41586-021-03928-y} {\bibfield  {journal} {\bibinfo
  {journal} {Nature}\ }\textbf {\bibinfo {volume} {598}},\ \bibinfo {pages}
  {281--286}}\BibitemShut {NoStop}%
\bibitem [{\citenamefont {Ehrenberg}\ \emph {et~al.}(2022)\citenamefont
  {Ehrenberg}, \citenamefont {Deshpande}, \citenamefont {Baldwin},
  \citenamefont {Abanin},\ and\ \citenamefont
  {Gorshkov}}]{ehrenberg_simulation_2022}%
  \BibitemOpen
  \bibfield  {author} {\bibinfo {author} {\bibnamefont {Ehrenberg},
  \bibfnamefont {A}}, \bibinfo {author} {\bibfnamefont {A.}~\bibnamefont
  {Deshpande}}, \bibinfo {author} {\bibfnamefont {C.~L.}\ \bibnamefont
  {Baldwin}}, \bibinfo {author} {\bibfnamefont {D.~A.}\ \bibnamefont {Abanin}},
  and\ \bibinfo {author} {\bibfnamefont {A.~V.}\ \bibnamefont {Gorshkov}}}
  (\bibinfo {year} {2022}),\ \bibfield  {title} {\enquote {\bibinfo {title}
  {Simulation {{complexity}} of {{many-body localized systems}}},}\ }\href@noop
  {} {\ }\Eprint {https://arxiv.org/abs/2205.12967} {arXiv:2205.12967}
  \BibitemShut {NoStop}%
\bibitem [{\citenamefont {Einstein}\ \emph {et~al.}(1935)\citenamefont
  {Einstein}, \citenamefont {Podolsky},\ and\ \citenamefont
  {Rosen}}]{einstein_can_1935}%
  \BibitemOpen
  \bibfield  {author} {\bibinfo {author} {\bibnamefont {Einstein},
  \bibfnamefont {A}}, \bibinfo {author} {\bibfnamefont {B.}~\bibnamefont
  {Podolsky}}, and\ \bibinfo {author} {\bibfnamefont {N.}~\bibnamefont
  {Rosen}}} (\bibinfo {year} {1935}),\ \bibfield  {title} {\enquote {\bibinfo
  {title} {Can {{quantum-mechanical description}} of {{physical reality be
  considered complete}}?}}\ }\href {https://doi.org/10.1103/PhysRev.47.777}
  {\bibfield  {journal} {\bibinfo  {journal} {Phys. Rev.}\ }\textbf {\bibinfo
  {volume} {47}},\ \bibinfo {pages} {777--780}}\BibitemShut {NoStop}%
\bibitem [{\citenamefont {Eisert}\ \emph {et~al.}(2020)\citenamefont {Eisert},
  \citenamefont {Hangleiter}, \citenamefont {Walk}, \citenamefont {Roth},
  \citenamefont {Markham}, \citenamefont {Parekh}, \citenamefont {Chabaud},\
  and\ \citenamefont {Kashefi}}]{BenchmarkingReview}%
  \BibitemOpen
  \bibfield  {author} {\bibinfo {author} {\bibnamefont {Eisert}, \bibfnamefont
  {J}}, \bibinfo {author} {\bibfnamefont {D.}~\bibnamefont {Hangleiter}},
  \bibinfo {author} {\bibfnamefont {N.}~\bibnamefont {Walk}}, \bibinfo {author}
  {\bibfnamefont {I.}~\bibnamefont {Roth}}, \bibinfo {author} {\bibfnamefont
  {D.}~\bibnamefont {Markham}}, \bibinfo {author} {\bibfnamefont
  {R.}~\bibnamefont {Parekh}}, \bibinfo {author} {\bibfnamefont
  {U.}~\bibnamefont {Chabaud}}, and\ \bibinfo {author} {\bibfnamefont
  {E.}~\bibnamefont {Kashefi}}} (\bibinfo {year} {2020}),\ \bibfield  {title}
  {\enquote {\bibinfo {title} {Quantum certification and benchmarking},}\
  }\href {https://doi.org/10.1038/s42254-020-0186-4} {\bibfield  {journal}
  {\bibinfo  {journal} {Nature Rev. Phys.}\ }\textbf {\bibinfo {volume} {2}},\
  \bibinfo {pages} {382--390}}\BibitemShut {NoStop}%
\bibitem [{\citenamefont {Eldar}\ and\ \citenamefont
  {Mehraban}(2018)}]{eldar_approximating_2018}%
  \BibitemOpen
  \bibfield  {author} {\bibinfo {author} {\bibnamefont {Eldar}, \bibfnamefont
  {L}}, and\ \bibinfo {author} {\bibfnamefont {S.}~\bibnamefont {Mehraban}}}
  (\bibinfo {year} {2018}),\ \bibfield  {title} {\enquote {\bibinfo {title}
  {Approximating the {{permanent}} of a {{random matrix}} with {{vanishing
  mean}}},}\ }in\ \href {https://doi.org/10.1109/FOCS.2018.00012} {\emph
  {\bibinfo {booktitle} {2018 {{IEEE}} 59th {{Ann. Symp.}} {{Found.}} {{Comp.
  Sc.}} ({{FOCS}})}}},\ pp.\ \bibinfo {pages} {23--34}\BibitemShut {NoStop}%
\bibitem [{\citenamefont {Erhard}\ \emph {et~al.}(2019)\citenamefont {Erhard},
  \citenamefont {Wallman}, \citenamefont {Postler}, \citenamefont {Meth},
  \citenamefont {Stricker}, \citenamefont {Martinez}, \citenamefont
  {Schindler}, \citenamefont {Monz}, \citenamefont {Emerson},\ and\
  \citenamefont {Blatt}}]{erhard_characterizing_2019}%
  \BibitemOpen
  \bibfield  {author} {\bibinfo {author} {\bibnamefont {Erhard}, \bibfnamefont
  {A}}, \bibinfo {author} {\bibfnamefont {J.~J.}\ \bibnamefont {Wallman}},
  \bibinfo {author} {\bibfnamefont {L.}~\bibnamefont {Postler}}, \bibinfo
  {author} {\bibfnamefont {M.}~\bibnamefont {Meth}}, \bibinfo {author}
  {\bibfnamefont {R.}~\bibnamefont {Stricker}}, \bibinfo {author}
  {\bibfnamefont {E.~A.}\ \bibnamefont {Martinez}}, \bibinfo {author}
  {\bibfnamefont {P.}~\bibnamefont {Schindler}}, \bibinfo {author}
  {\bibfnamefont {T.}~\bibnamefont {Monz}}, \bibinfo {author} {\bibfnamefont
  {J.}~\bibnamefont {Emerson}}, and\ \bibinfo {author} {\bibfnamefont
  {R.}~\bibnamefont {Blatt}}} (\bibinfo {year} {2019}),\ \bibfield  {title}
  {\enquote {\bibinfo {title} {Characterizing large-scale quantum computers via
  cycle benchmarking},}\ }\href {https://doi.org/10.1038/s41467-019-13068-7}
  {\bibfield  {journal} {\bibinfo  {journal} {Nat. Comm.}\ }\textbf {\bibinfo
  {volume} {10}},\ \bibinfo {pages} {5347}}\BibitemShut {NoStop}%
\bibitem [{\citenamefont {Farhi}\ \emph {et~al.}(2014)\citenamefont {Farhi},
  \citenamefont {Goldstone},\ and\ \citenamefont {Gutmann}}]{QAOA}%
  \BibitemOpen
  \bibfield  {author} {\bibinfo {author} {\bibnamefont {Farhi}, \bibfnamefont
  {E}}, \bibinfo {author} {\bibfnamefont {J.}~\bibnamefont {Goldstone}}, and\
  \bibinfo {author} {\bibfnamefont {S.}~\bibnamefont {Gutmann}}} (\bibinfo
  {year} {2014}),\ \bibfield  {title} {\enquote {\bibinfo {title} {A quantum
  approximate optimization algorithm},}\ }\href@noop {} {\ }\Eprint
  {https://arxiv.org/abs/1411.4028} {arXiv:1411.4028} \BibitemShut {NoStop}%
\bibitem [{\citenamefont {Farhi}\ \emph {et~al.}(2000)\citenamefont {Farhi},
  \citenamefont {Goldstone}, \citenamefont {Gutmann},\ and\ \citenamefont
  {Sipser}}]{farhi_quantum_2000}%
  \BibitemOpen
  \bibfield  {author} {\bibinfo {author} {\bibnamefont {Farhi}, \bibfnamefont
  {E}}, \bibinfo {author} {\bibfnamefont {J.}~\bibnamefont {Goldstone}},
  \bibinfo {author} {\bibfnamefont {S.}~\bibnamefont {Gutmann}}, and\ \bibinfo
  {author} {\bibfnamefont {M.}~\bibnamefont {Sipser}}} (\bibinfo {year}
  {2000}),\ \bibfield  {title} {\enquote {\bibinfo {title} {Quantum
  {{computation}} by {{adiabatic evolution}}},}\ }\href@noop {} {\ }\Eprint
  {https://arxiv.org/abs/quant-ph/0001106} {arXiv:quant-ph/0001106}
  \BibitemShut {NoStop}%
\bibitem [{\citenamefont {Fefferman}\ and\ \citenamefont
  {Umans}(2015)}]{fefferman_power_2015}%
  \BibitemOpen
  \bibfield  {author} {\bibinfo {author} {\bibnamefont {Fefferman},
  \bibfnamefont {B}}, and\ \bibinfo {author} {\bibfnamefont {C.}~\bibnamefont
  {Umans}}} (\bibinfo {year} {2015}),\ \bibfield  {title} {\enquote {\bibinfo
  {title} {The {{power}} of {{quantum Fourier sampling}}},}\ }\href@noop {} {\
  }\Eprint {https://arxiv.org/abs/1507.05592} {arXiv:1507.05592} \BibitemShut
  {NoStop}%
\bibitem [{\citenamefont {Fenner}\ \emph {et~al.}(1998)\citenamefont {Fenner},
  \citenamefont {Green}, \citenamefont {Homer},\ and\ \citenamefont
  {Pruim}}]{fenner_determining_1998}%
  \BibitemOpen
  \bibfield  {author} {\bibinfo {author} {\bibnamefont {Fenner}, \bibfnamefont
  {S}}, \bibinfo {author} {\bibfnamefont {F.}~\bibnamefont {Green}}, \bibinfo
  {author} {\bibfnamefont {S.}~\bibnamefont {Homer}}, and\ \bibinfo {author}
  {\bibfnamefont {R.}~\bibnamefont {Pruim}}} (\bibinfo {year} {1998}),\
  \bibfield  {title} {\enquote {\bibinfo {title} {Determining {{acceptance
  possibility}} for a {{quantum computation}} is {{hard}} for the {{polynomial
  hierarchy}}},}\ }\href@noop {} {\ }\Eprint
  {https://arxiv.org/abs/quant-ph/9812056} {arXiv:quant-ph/9812056}
  \BibitemShut {NoStop}%
\bibitem [{\citenamefont {Fenner}\ \emph {et~al.}(1994)\citenamefont {Fenner},
  \citenamefont {Fortnow},\ and\ \citenamefont
  {Kurtz}}]{fenner_gap-definable_1994}%
  \BibitemOpen
  \bibfield  {author} {\bibinfo {author} {\bibnamefont {Fenner}, \bibfnamefont
  {S~A}}, \bibinfo {author} {\bibfnamefont {L.~J.}\ \bibnamefont {Fortnow}},
  and\ \bibinfo {author} {\bibfnamefont {S.~A.}\ \bibnamefont {Kurtz}}}
  (\bibinfo {year} {1994}),\ \bibfield  {title} {\enquote {\bibinfo {title}
  {Gap-definable counting classes},}\ }\href
  {https://doi.org/10.1016/S0022-0000(05)80024-8} {\bibfield  {journal}
  {\bibinfo  {journal} {J. Comp. Sys. Sc.}\ }\textbf {\bibinfo {volume} {48}},\
  \bibinfo {pages} {116--148}}\BibitemShut {NoStop}%
\bibitem [{\citenamefont {Ferris}\ and\ \citenamefont
  {Vidal}(2012)}]{PhysRevB.85.165146}%
  \BibitemOpen
  \bibfield  {author} {\bibinfo {author} {\bibnamefont {Ferris}, \bibfnamefont
  {A~J}}, and\ \bibinfo {author} {\bibfnamefont {G.}~\bibnamefont {Vidal}}}
  (\bibinfo {year} {2012}),\ \bibfield  {title} {\enquote {\bibinfo {title}
  {Perfect sampling with unitary tensor networks},}\ }\href
  {https://doi.org/10.1103/PhysRevB.85.165146} {\bibfield  {journal} {\bibinfo
  {journal} {Phys. Rev. B}\ }\textbf {\bibinfo {volume} {85}},\ \bibinfo
  {pages} {165146}}\BibitemShut {NoStop}%
\bibitem [{\citenamefont {Feynman}(1982)}]{feynman_simulating_1982}%
  \BibitemOpen
  \bibfield  {author} {\bibinfo {author} {\bibnamefont {Feynman}, \bibfnamefont
  {R~P}}} (\bibinfo {year} {1982}),\ \bibfield  {title} {\enquote {\bibinfo
  {title} {Simulating physics with computers},}\ }\href
  {https://doi.org/10.1007/BF02650179} {\bibfield  {journal} {\bibinfo
  {journal} {Int. J. Theor. Phys.}\ }\textbf {\bibinfo {volume} {21}},\
  \bibinfo {pages} {467--488}}\BibitemShut {NoStop}%
\bibitem [{\citenamefont {Feynman}(1985)}]{feynman_quantum_1985}%
  \BibitemOpen
  \bibfield  {author} {\bibinfo {author} {\bibnamefont {Feynman}, \bibfnamefont
  {R~P}}} (\bibinfo {year} {1985}),\ \bibfield  {title} {\enquote {\bibinfo
  {title} {Quantum {{mechanical computers}}},}\ }\href
  {https://doi.org/10.1364/ON.11.2.000011} {\bibfield  {journal} {\bibinfo
  {journal} {Opt. News}\ }\textbf {\bibinfo {volume} {11}},\ \bibinfo {pages}
  {11--20}}\BibitemShut {NoStop}%
\bibitem [{\citenamefont {Finetti}(1937)}]{finetti_prevision_1937}%
  \BibitemOpen
  \bibfield  {author} {\bibinfo {author} {\bibnamefont {Finetti}, \bibfnamefont
  {B~De}}} (\bibinfo {year} {1937}),\ \bibfield  {title} {\enquote {\bibinfo
  {title} {La pr{\'e}vision: ses lois logiques, ses sources subjectives},}\
  }\href {https://doi.org/10.1007/s00023-012-0204-x} {\bibfield  {journal}
  {\bibinfo  {journal} {Ann. Inst. H. Poincar{\'e}}\ }\textbf {\bibinfo
  {volume} {7}},\ \bibinfo {pages} {1--68}}\BibitemShut {NoStop}%
\bibitem [{\citenamefont {Fitzsimons}\ \emph {et~al.}(2018)\citenamefont
  {Fitzsimons}, \citenamefont {Hajdusek},\ and\ \citenamefont
  {Morimae}}]{fitzsimons_post_2018}%
  \BibitemOpen
  \bibfield  {author} {\bibinfo {author} {\bibnamefont {Fitzsimons},
  \bibfnamefont {J~F}}, \bibinfo {author} {\bibfnamefont {M.}~\bibnamefont
  {Hajdusek}}, and\ \bibinfo {author} {\bibfnamefont {T.}~\bibnamefont
  {Morimae}}} (\bibinfo {year} {2018}),\ \bibfield  {title} {\enquote {\bibinfo
  {title} {Post hoc verification of quantum computation},}\ }\href
  {https://doi.org/t10.1103/PhysRevLett.120.040501} {\bibfield  {journal}
  {\bibinfo  {journal} {Phys. Rev. Lett.}\ }\textbf {\bibinfo {volume} {120}},\
  \bibinfo {pages} {040501}}\BibitemShut {NoStop}%
\bibitem [{\citenamefont {Fitzsimons}\ and\ \citenamefont
  {Kashefi}(2017)}]{fitzsimons_unconditionally_2017}%
  \BibitemOpen
  \bibfield  {author} {\bibinfo {author} {\bibnamefont {Fitzsimons},
  \bibfnamefont {J~F}}, and\ \bibinfo {author} {\bibfnamefont {E.}~\bibnamefont
  {Kashefi}}} (\bibinfo {year} {2017}),\ \bibfield  {title} {\enquote {\bibinfo
  {title} {Unconditionally verifiable blind quantum computation},}\ }\href
  {https://doi.org/10.1103/PhysRevA.96.012303} {\bibfield  {journal} {\bibinfo
  {journal} {Phys. Rev. A}\ }\textbf {\bibinfo {volume} {96}},\ \bibinfo
  {pages} {012303}}\BibitemShut {NoStop}%
\bibitem [{\citenamefont {Flammia}\ and\ \citenamefont
  {Liu}(2011)}]{flammia_direct_2011}%
  \BibitemOpen
  \bibfield  {author} {\bibinfo {author} {\bibnamefont {Flammia}, \bibfnamefont
  {S~T}}, and\ \bibinfo {author} {\bibfnamefont {Y.-K.}\ \bibnamefont {Liu}}}
  (\bibinfo {year} {2011}),\ \bibfield  {title} {\enquote {\bibinfo {title}
  {{Direct fidelity estimation from few Pauli measurements}},}\ }\href
  {https://doi.org/10.1103/PhysRevLett.106.230501} {\bibfield  {journal}
  {\bibinfo  {journal} {Phys. Rev. Lett.}\ }\textbf {\bibinfo {volume} {106}},\
  10.1103/PhysRevLett.106.230501}\BibitemShut {NoStop}%
\bibitem [{\citenamefont {Foxen}\ \emph {et~al.}(2020)\citenamefont {Foxen},
  \citenamefont {Neill}, \citenamefont {Dunsworth}, \citenamefont {Roushan},
  \citenamefont {Chiaro}, \citenamefont {Megrant}, \citenamefont {Kelly},
  \citenamefont {Chen}, \citenamefont {Satzinger}, \citenamefont {Barends},
  \citenamefont {Arute}, \citenamefont {Arya}, \citenamefont {Babbush},
  \citenamefont {Bacon}, \citenamefont {Bardin}, \citenamefont {Boixo},
  \citenamefont {Buell}, \citenamefont {Burkett}, \citenamefont {Chen},
  \citenamefont {Collins}, \citenamefont {Farhi}, \citenamefont {Fowler},
  \citenamefont {Gidney}, \citenamefont {Giustina}, \citenamefont {Graff},
  \citenamefont {Harrigan}, \citenamefont {Huang}, \citenamefont {Isakov},
  \citenamefont {Jeffrey}, \citenamefont {Jiang}, \citenamefont {Kafri},
  \citenamefont {Kechedzhi}, \citenamefont {Klimov}, \citenamefont {Korotkov},
  \citenamefont {Kostritsa}, \citenamefont {Landhuis}, \citenamefont {Lucero},
  \citenamefont {McClean}, \citenamefont {McEwen}, \citenamefont {Mi},
  \citenamefont {Mohseni}, \citenamefont {Mutus}, \citenamefont {Naaman},
  \citenamefont {Neeley}, \citenamefont {Niu}, \citenamefont {Petukhov},
  \citenamefont {Quintana}, \citenamefont {Rubin}, \citenamefont {Sank},
  \citenamefont {Smelyanskiy}, \citenamefont {Vainsencher}, \citenamefont
  {White}, \citenamefont {Yao}, \citenamefont {Yeh}, \citenamefont {Zalcman},
  \citenamefont {Neven},\ and\ \citenamefont
  {Martinis}}]{foxen_demonstrating_2020}%
  \BibitemOpen
  \bibfield  {author} {\bibinfo {author} {\bibnamefont {Foxen}, \bibfnamefont
  {B}}, \bibinfo {author} {\bibfnamefont {C.}~\bibnamefont {Neill}}, \bibinfo
  {author} {\bibfnamefont {A.}~\bibnamefont {Dunsworth}}, \bibinfo {author}
  {\bibfnamefont {P.}~\bibnamefont {Roushan}}, \bibinfo {author} {\bibfnamefont
  {B.}~\bibnamefont {Chiaro}}, \bibinfo {author} {\bibfnamefont
  {A.}~\bibnamefont {Megrant}}, \bibinfo {author} {\bibfnamefont
  {J.}~\bibnamefont {Kelly}}, \bibinfo {author} {\bibfnamefont {Zijun}\
  \bibnamefont {Chen}}, \bibinfo {author} {\bibfnamefont {K.}~\bibnamefont
  {Satzinger}}, \bibinfo {author} {\bibfnamefont {R.}~\bibnamefont {Barends}},
  \bibinfo {author} {\bibfnamefont {F.}~\bibnamefont {Arute}}, \bibinfo
  {author} {\bibfnamefont {K.}~\bibnamefont {Arya}}, \bibinfo {author}
  {\bibfnamefont {R.}~\bibnamefont {Babbush}}, \bibinfo {author} {\bibfnamefont
  {D.}~\bibnamefont {Bacon}}, \bibinfo {author} {\bibfnamefont {J.~C.}\
  \bibnamefont {Bardin}}, \bibinfo {author} {\bibfnamefont {S.}~\bibnamefont
  {Boixo}}, \bibinfo {author} {\bibfnamefont {D.}~\bibnamefont {Buell}},
  \bibinfo {author} {\bibfnamefont {B.}~\bibnamefont {Burkett}}, \bibinfo
  {author} {\bibfnamefont {Yu}~\bibnamefont {Chen}}, \bibinfo {author}
  {\bibfnamefont {R.}~\bibnamefont {Collins}}, \bibinfo {author} {\bibfnamefont
  {E.}~\bibnamefont {Farhi}}, \bibinfo {author} {\bibfnamefont
  {A.}~\bibnamefont {Fowler}}, \bibinfo {author} {\bibfnamefont
  {C.}~\bibnamefont {Gidney}}, \bibinfo {author} {\bibfnamefont
  {M.}~\bibnamefont {Giustina}}, \bibinfo {author} {\bibfnamefont
  {R.}~\bibnamefont {Graff}}, \bibinfo {author} {\bibfnamefont
  {M.}~\bibnamefont {Harrigan}}, \bibinfo {author} {\bibfnamefont
  {T.}~\bibnamefont {Huang}}, \bibinfo {author} {\bibfnamefont {S.~V.}\
  \bibnamefont {Isakov}}, \bibinfo {author} {\bibfnamefont {E.}~\bibnamefont
  {Jeffrey}}, \bibinfo {author} {\bibfnamefont {Z.}~\bibnamefont {Jiang}},
  \bibinfo {author} {\bibfnamefont {D.}~\bibnamefont {Kafri}}, \bibinfo
  {author} {\bibfnamefont {K.}~\bibnamefont {Kechedzhi}}, \bibinfo {author}
  {\bibfnamefont {P.}~\bibnamefont {Klimov}}, \bibinfo {author} {\bibfnamefont
  {A.}~\bibnamefont {Korotkov}}, \bibinfo {author} {\bibfnamefont
  {F.}~\bibnamefont {Kostritsa}}, \bibinfo {author} {\bibfnamefont
  {D.}~\bibnamefont {Landhuis}}, \bibinfo {author} {\bibfnamefont
  {E.}~\bibnamefont {Lucero}}, \bibinfo {author} {\bibfnamefont
  {J.}~\bibnamefont {McClean}}, \bibinfo {author} {\bibfnamefont
  {M.}~\bibnamefont {McEwen}}, \bibinfo {author} {\bibfnamefont
  {X.}~\bibnamefont {Mi}}, \bibinfo {author} {\bibfnamefont {M.}~\bibnamefont
  {Mohseni}}, \bibinfo {author} {\bibfnamefont {J.~Y.}\ \bibnamefont {Mutus}},
  \bibinfo {author} {\bibfnamefont {O.}~\bibnamefont {Naaman}}, \bibinfo
  {author} {\bibfnamefont {M.}~\bibnamefont {Neeley}}, \bibinfo {author}
  {\bibfnamefont {M.}~\bibnamefont {Niu}}, \bibinfo {author} {\bibfnamefont
  {A.}~\bibnamefont {Petukhov}}, \bibinfo {author} {\bibfnamefont
  {C.}~\bibnamefont {Quintana}}, \bibinfo {author} {\bibfnamefont
  {N.}~\bibnamefont {Rubin}}, \bibinfo {author} {\bibfnamefont
  {D.}~\bibnamefont {Sank}}, \bibinfo {author} {\bibfnamefont {V.}~\bibnamefont
  {Smelyanskiy}}, \bibinfo {author} {\bibfnamefont {A.}~\bibnamefont
  {Vainsencher}}, \bibinfo {author} {\bibfnamefont {T.~C.}\ \bibnamefont
  {White}}, \bibinfo {author} {\bibfnamefont {Z.}~\bibnamefont {Yao}}, \bibinfo
  {author} {\bibfnamefont {P.}~\bibnamefont {Yeh}}, \bibinfo {author}
  {\bibfnamefont {A.}~\bibnamefont {Zalcman}}, \bibinfo {author} {\bibfnamefont
  {H.}~\bibnamefont {Neven}}, and\ \bibinfo {author} {\bibfnamefont {J.~M.}\
  \bibnamefont {Martinis}}} (\bibinfo {year} {2020}),\ \bibfield  {title}
  {\enquote {\bibinfo {title} {Demonstrating a {{continuous set}} of
  {{two-qubit gates}} for {{near-term quantum algorithms}}},}\ }\href
  {https://doi.org/10.1103/PhysRevLett.125.120504} {\bibfield  {journal}
  {\bibinfo  {journal} {Phys. Rev. Lett.}\ }\textbf {\bibinfo {volume} {125}},\
  \bibinfo {pages} {120504}}\BibitemShut {NoStop}%
\bibitem [{\citenamefont {Fredkin}\ and\ \citenamefont
  {Toffoli}(1982)}]{fredkin_conservative_1982}%
  \BibitemOpen
  \bibfield  {author} {\bibinfo {author} {\bibnamefont {Fredkin}, \bibfnamefont
  {E}}, and\ \bibinfo {author} {\bibfnamefont {T.}~\bibnamefont {Toffoli}}}
  (\bibinfo {year} {1982}),\ \bibfield  {title} {\enquote {\bibinfo {title}
  {Conservative logic},}\ }\href {https://doi.org/10.1007/BF01857727}
  {\bibfield  {journal} {\bibinfo  {journal} {Int. J. Theor. Phys.}\ }\textbf
  {\bibinfo {volume} {21}},\ \bibinfo {pages} {219--253}}\BibitemShut {NoStop}%
\bibitem [{\citenamefont {Freedman}\ and\ \citenamefont
  {Clauser}(1972)}]{freedman_experimental_1972}%
  \BibitemOpen
  \bibfield  {author} {\bibinfo {author} {\bibnamefont {Freedman},
  \bibfnamefont {S~J}}, and\ \bibinfo {author} {\bibfnamefont {J.~F.}\
  \bibnamefont {Clauser}}} (\bibinfo {year} {1972}),\ \bibfield  {title}
  {\enquote {\bibinfo {title} {Experimental {{test}} of {{local hidden-variable
  theories}}},}\ }\href {https://doi.org/10.1103/PhysRevLett.28.938} {\bibfield
   {journal} {\bibinfo  {journal} {Phys. Rev. Lett.}\ }\textbf {\bibinfo
  {volume} {28}},\ \bibinfo {pages} {938--941}}\BibitemShut {NoStop}%
\bibitem [{\citenamefont {Friis}\ \emph {et~al.}(2018)\citenamefont {Friis},
  \citenamefont {Marty}, \citenamefont {Maier}, \citenamefont {Hempel},
  \citenamefont {Holz{\"a}pfel}, \citenamefont {Jurcevic}, \citenamefont
  {Plenio}, \citenamefont {Huber}, \citenamefont {Roos}, \citenamefont
  {Blatt},\ and\ \citenamefont {Lanyon}}]{friis_observation_2018}%
  \BibitemOpen
  \bibfield  {author} {\bibinfo {author} {\bibnamefont {Friis}, \bibfnamefont
  {N}}, \bibinfo {author} {\bibfnamefont {O.}~\bibnamefont {Marty}}, \bibinfo
  {author} {\bibfnamefont {C.}~\bibnamefont {Maier}}, \bibinfo {author}
  {\bibfnamefont {C.}~\bibnamefont {Hempel}}, \bibinfo {author} {\bibfnamefont
  {M.}~\bibnamefont {Holz{\"a}pfel}}, \bibinfo {author} {\bibfnamefont
  {P.}~\bibnamefont {Jurcevic}}, \bibinfo {author} {\bibfnamefont {M.~B.}\
  \bibnamefont {Plenio}}, \bibinfo {author} {\bibfnamefont {M.}~\bibnamefont
  {Huber}}, \bibinfo {author} {\bibfnamefont {C.}~\bibnamefont {Roos}},
  \bibinfo {author} {\bibfnamefont {R.}~\bibnamefont {Blatt}}, and\ \bibinfo
  {author} {\bibfnamefont {B.}~\bibnamefont {Lanyon}}} (\bibinfo {year}
  {2018}),\ \bibfield  {title} {\enquote {\bibinfo {title} {Observation of
  {{entangled states}} of a {{fully controlled}} 20-{{qubit system}}},}\ }\href
  {https://doi.org/10.1103/PhysRevX.8.021012} {\bibfield  {journal} {\bibinfo
  {journal} {Phys. Rev. X}\ }\textbf {\bibinfo {volume} {8}},\ \bibinfo {pages}
  {021012}}\BibitemShut {NoStop}%
\bibitem [{\citenamefont {Fujii}(2016)}]{fujii_noise_2016}%
  \BibitemOpen
  \bibfield  {author} {\bibinfo {author} {\bibnamefont {Fujii}, \bibfnamefont
  {K}}} (\bibinfo {year} {2016}),\ \bibfield  {title} {\enquote {\bibinfo
  {title} {Noise {{threshold}} of {{quantum supremacy}}},}\ }\href@noop {} {\
  }\Eprint {https://arxiv.org/abs/1610.03632} {arXiv:1610.03632} \BibitemShut
  {NoStop}%
\bibitem [{\citenamefont {Fujii}\ \emph {et~al.}(2018)\citenamefont {Fujii},
  \citenamefont {Kobayashi}, \citenamefont {Morimae}, \citenamefont
  {Nishimura}, \citenamefont {Tamate},\ and\ \citenamefont
  {Tani}}]{fujii_impossibility_2018}%
  \BibitemOpen
  \bibfield  {author} {\bibinfo {author} {\bibnamefont {Fujii}, \bibfnamefont
  {K}}, \bibinfo {author} {\bibfnamefont {H.}~\bibnamefont {Kobayashi}},
  \bibinfo {author} {\bibfnamefont {T.}~\bibnamefont {Morimae}}, \bibinfo
  {author} {\bibfnamefont {H.}~\bibnamefont {Nishimura}}, \bibinfo {author}
  {\bibfnamefont {S.}~\bibnamefont {Tamate}}, and\ \bibinfo {author}
  {\bibfnamefont {S.}~\bibnamefont {Tani}}} (\bibinfo {year} {2018}),\
  \bibfield  {title} {\enquote {\bibinfo {title} {Impossibility of classically
  simulating one-clean-qubit model with multiplicative error},}\ }\href
  {https://doi.org/10.1103/PhysRevLett.120.200502} {\bibfield  {journal}
  {\bibinfo  {journal} {Phys. Rev. Lett.}\ }\textbf {\bibinfo {volume} {120}},\
  \bibinfo {pages} {200502}}\BibitemShut {NoStop}%
\bibitem [{\citenamefont {Fujii}\ and\ \citenamefont
  {Morimae}(2017)}]{fujii_commuting_2017}%
  \BibitemOpen
  \bibfield  {author} {\bibinfo {author} {\bibnamefont {Fujii}, \bibfnamefont
  {K}}, and\ \bibinfo {author} {\bibfnamefont {T.}~\bibnamefont {Morimae}}}
  (\bibinfo {year} {2017}),\ \bibfield  {title} {\enquote {\bibinfo {title}
  {Commuting quantum circuits and complexity of {{Ising}} partition
  functions},}\ }\href {https://doi.org/10.1088/1367-2630/aa5fdb} {\bibfield
  {journal} {\bibinfo  {journal} {New J. Phys.}\ }\textbf {\bibinfo {volume}
  {19}},\ \bibinfo {pages} {033003}}\BibitemShut {NoStop}%
\bibitem [{\citenamefont {Gambetta}\ \emph {et~al.}(2006)\citenamefont
  {Gambetta}, \citenamefont {Blais}, \citenamefont {Schuster}, \citenamefont
  {Wallraff}, \citenamefont {Frunzio}, \citenamefont {Majer}, \citenamefont
  {Devoret}, \citenamefont {Girvin},\ and\ \citenamefont
  {Schoelkopf}}]{gambetta_qubit-photon_2006}%
  \BibitemOpen
  \bibfield  {author} {\bibinfo {author} {\bibnamefont {Gambetta},
  \bibfnamefont {J}}, \bibinfo {author} {\bibfnamefont {A.}~\bibnamefont
  {Blais}}, \bibinfo {author} {\bibfnamefont {D.~I.}\ \bibnamefont {Schuster}},
  \bibinfo {author} {\bibfnamefont {A.}~\bibnamefont {Wallraff}}, \bibinfo
  {author} {\bibfnamefont {L.}~\bibnamefont {Frunzio}}, \bibinfo {author}
  {\bibfnamefont {J.}~\bibnamefont {Majer}}, \bibinfo {author} {\bibfnamefont
  {M.~H.}\ \bibnamefont {Devoret}}, \bibinfo {author} {\bibfnamefont {S.~M.}\
  \bibnamefont {Girvin}}, and\ \bibinfo {author} {\bibfnamefont {R.~J.}\
  \bibnamefont {Schoelkopf}}} (\bibinfo {year} {2006}),\ \bibfield  {title}
  {\enquote {\bibinfo {title} {Qubit-photon interactions in a cavity:
  {{Measurement-induced}} dephasing and number splitting},}\ }\href
  {https://doi.org/10.1103/PhysRevA.74.042318} {\bibfield  {journal} {\bibinfo
  {journal} {Phys. Rev. A}\ }\textbf {\bibinfo {volume} {74}},\ \bibinfo
  {pages} {042318}}\BibitemShut {NoStop}%
\bibitem [{\citenamefont {Gao}\ and\ \citenamefont
  {Duan}(2018)}]{gao_efficient_2018}%
  \BibitemOpen
  \bibfield  {author} {\bibinfo {author} {\bibnamefont {Gao}, \bibfnamefont
  {X}}, and\ \bibinfo {author} {\bibfnamefont {L.}~\bibnamefont {Duan}}}
  (\bibinfo {year} {2018}),\ \bibfield  {title} {\enquote {\bibinfo {title}
  {Efficient classical simulation of noisy quantum computation},}\ }\href@noop
  {} {\ }\Eprint {https://arxiv.org/abs/1810.03176} {arxiv:1810.03176}
  \BibitemShut {NoStop}%
\bibitem [{\citenamefont {Gao}\ \emph {et~al.}(2021)\citenamefont {Gao},
  \citenamefont {Kalinowski}, \citenamefont {Chou}, \citenamefont {Lukin},
  \citenamefont {Barak},\ and\ \citenamefont {Choi}}]{gao_limitations_2021}%
  \BibitemOpen
  \bibfield  {author} {\bibinfo {author} {\bibnamefont {Gao}, \bibfnamefont
  {X}}, \bibinfo {author} {\bibfnamefont {M.}~\bibnamefont {Kalinowski}},
  \bibinfo {author} {\bibfnamefont {C.-N.}\ \bibnamefont {Chou}}, \bibinfo
  {author} {\bibfnamefont {M.~D.}\ \bibnamefont {Lukin}}, \bibinfo {author}
  {\bibfnamefont {B.}~\bibnamefont {Barak}}, and\ \bibinfo {author}
  {\bibfnamefont {S.}~\bibnamefont {Choi}}} (\bibinfo {year} {2021}),\
  \bibfield  {title} {\enquote {\bibinfo {title} {Limitations of {{linear
  cross-entropy}} as a {{measure}} for {{quantum advantage}}},}\ }\href@noop {}
  {\ }\Eprint {https://arxiv.org/abs/2112.01657} {arXiv:2112.01657}
  \BibitemShut {NoStop}%
\bibitem [{\citenamefont {Gao}\ \emph {et~al.}(2017)\citenamefont {Gao},
  \citenamefont {Wang},\ and\ \citenamefont {Duan}}]{gao_quantum_2017}%
  \BibitemOpen
  \bibfield  {author} {\bibinfo {author} {\bibnamefont {Gao}, \bibfnamefont
  {X}}, \bibinfo {author} {\bibfnamefont {S.-T.}\ \bibnamefont {Wang}}, and\
  \bibinfo {author} {\bibfnamefont {L.-M.}\ \bibnamefont {Duan}}} (\bibinfo
  {year} {2017}),\ \bibfield  {title} {\enquote {\bibinfo {title} {Quantum
  {{supremacy}} for {{simulating}} a {{translation}}-{{invariant Ising spin
  model}}},}\ }\href {https://doi.org/10.1103/PhysRevLett.118.040502}
  {\bibfield  {journal} {\bibinfo  {journal} {Phys. Rev. Lett.}\ }\textbf
  {\bibinfo {volume} {118}},\ \bibinfo {pages} {040502}}\BibitemShut {NoStop}%
\bibitem [{\citenamefont {{Garc{\'i}a-Patr{\'o}n}}\ \emph
  {et~al.}(2019)\citenamefont {{Garc{\'i}a-Patr{\'o}n}}, \citenamefont
  {Renema},\ and\ \citenamefont
  {Shchesnovich}}]{garcia-patron_simulating_2019}%
  \BibitemOpen
  \bibfield  {author} {\bibinfo {author} {\bibnamefont
  {{Garc{\'i}a-Patr{\'o}n}}, \bibfnamefont {R}}, \bibinfo {author}
  {\bibfnamefont {J.~J.}\ \bibnamefont {Renema}}, and\ \bibinfo {author}
  {\bibfnamefont {V.~S.}\ \bibnamefont {Shchesnovich}}} (\bibinfo {year}
  {2019}),\ \bibfield  {title} {\enquote {\bibinfo {title} {Simulating boson
  sampling in lossy architectures},}\ }\href
  {https://doi.org/10.22331/q-2019-08-05-169} {\bibfield  {journal} {\bibinfo
  {journal} {Quantum}\ }\textbf {\bibinfo {volume} {3}},\ \bibinfo {pages}
  {169}}\BibitemShut {NoStop}%
\bibitem [{\citenamefont {Gemmell}\ \emph {et~al.}(1991)\citenamefont
  {Gemmell}, \citenamefont {Lipton}, \citenamefont {Rubinfeld}, \citenamefont
  {Sudan},\ and\ \citenamefont
  {Wigderson}}]{gemmell_self-testingcorrecting_1991}%
  \BibitemOpen
  \bibfield  {author} {\bibinfo {author} {\bibnamefont {Gemmell}, \bibfnamefont
  {P}}, \bibinfo {author} {\bibfnamefont {R.}~\bibnamefont {Lipton}}, \bibinfo
  {author} {\bibfnamefont {R.}~\bibnamefont {Rubinfeld}}, \bibinfo {author}
  {\bibfnamefont {M.}~\bibnamefont {Sudan}}, and\ \bibinfo {author}
  {\bibfnamefont {A.}~\bibnamefont {Wigderson}}} (\bibinfo {year} {1991}),\
  \bibfield  {title} {\enquote {\bibinfo {title} {Self-testing/correcting for
  polynomials and for approximate functions},}\ }in\ \href
  {https://doi.org/10.1145/103418.103429} {\emph {\bibinfo {booktitle} {Proc.
  T23rd Ann. {{ACM}} Symp. {{Th.}} {{Comp.}}}}},\ \bibinfo {series and number}
  {{{STOC}} '91}\ (\bibinfo  {publisher} {{Association for Computing
  Machinery}},\ \bibinfo {address} {{New York, NY, USA}})\ pp.\ \bibinfo
  {pages} {33--42}\BibitemShut {NoStop}%
\bibitem [{\citenamefont {Gemmell}\ and\ \citenamefont
  {Sudan}(1992)}]{gemmell_highly_1992}%
  \BibitemOpen
  \bibfield  {author} {\bibinfo {author} {\bibnamefont {Gemmell}, \bibfnamefont
  {P}}, and\ \bibinfo {author} {\bibfnamefont {M.}~\bibnamefont {Sudan}}}
  (\bibinfo {year} {1992}),\ \bibfield  {title} {\enquote {\bibinfo {title}
  {Highly resilient correctors for polynomials},}\ }\href
  {https://doi.org/10.1016/0020-0190(92)90195-2} {\bibfield  {journal}
  {\bibinfo  {journal} {Inf. Proc. Lett.}\ }\textbf {\bibinfo {volume} {43}},\
  \bibinfo {pages} {169--174}}\BibitemShut {NoStop}%
\bibitem [{\citenamefont {Gheorghiu}\ and\ \citenamefont
  {Mosca}(2019)}]{gheorghiu_benchmarking_2019}%
  \BibitemOpen
  \bibfield  {author} {\bibinfo {author} {\bibnamefont {Gheorghiu},
  \bibfnamefont {V}}, and\ \bibinfo {author} {\bibfnamefont {M.}~\bibnamefont
  {Mosca}}} (\bibinfo {year} {2019}),\ \bibfield  {title} {\enquote {\bibinfo
  {title} {Benchmarking the quantum cryptanalysis of symmetric, public-key and
  hash-based cryptographic schemes},}\ }\href@noop {} {\ }\Eprint
  {https://arxiv.org/abs/1902.02332} {arXiv:1902.02332} \BibitemShut {NoStop}%
\bibitem [{\citenamefont {Gidney}\ and\ \citenamefont
  {Ekera}(2019)}]{gidney_how_2019}%
  \BibitemOpen
  \bibfield  {author} {\bibinfo {author} {\bibnamefont {Gidney}, \bibfnamefont
  {C}}, and\ \bibinfo {author} {\bibfnamefont {M.}~\bibnamefont {Ekera}}}
  (\bibinfo {year} {2019}),\ \bibfield  {title} {\enquote {\bibinfo {title}
  {How to factor 2048 bit {{RSA}} integers in 8 hours using 20 million noisy
  qubits},}\ }\href@noop {} {\ }\Eprint {https://arxiv.org/abs/1905.09749}
  {arXiv:1905.09749} \BibitemShut {NoStop}%
\bibitem [{\citenamefont {Gily{\'e}n}\ \emph {et~al.}(2019)\citenamefont
  {Gily{\'e}n}, \citenamefont {Su}, \citenamefont {Low},\ and\ \citenamefont
  {Wiebe}}]{gilyen_quantum_2019}%
  \BibitemOpen
  \bibfield  {author} {\bibinfo {author} {\bibnamefont {Gily{\'e}n},
  \bibfnamefont {A}}, \bibinfo {author} {\bibfnamefont {Y.}~\bibnamefont {Su}},
  \bibinfo {author} {\bibfnamefont {G.~H.}\ \bibnamefont {Low}}, and\ \bibinfo
  {author} {\bibfnamefont {N.}~\bibnamefont {Wiebe}}} (\bibinfo {year}
  {2019}),\ \bibfield  {title} {\enquote {\bibinfo {title} {Quantum singular
  value transformation and beyond: Exponential improvements for quantum matrix
  arithmetics},}\ }\href {https://doi.org/10.1145/3313276.3316366} {\bibfield
  {journal} {\bibinfo  {journal} {Proceedings of the 51st Annual ACM SIGACT
  Symposium on Theory of Computing}\ ,\ \bibinfo {pages} {193--204}}}\Eprint
  {https://arxiv.org/abs/1806.01838} {arXiv:1806.01838} \BibitemShut {NoStop}%
\bibitem [{\citenamefont {Giordani}\ \emph {et~al.}(2018)\citenamefont
  {Giordani}, \citenamefont {Flamini}, \citenamefont {Pompili}, \citenamefont
  {Viggianiello}, \citenamefont {Spagnolo}, \citenamefont {Crespi},
  \citenamefont {Osellame}, \citenamefont {Wiebe}, \citenamefont {Walschaers},
  \citenamefont {Buchleitner},\ and\ \citenamefont
  {Sciarrino}}]{giordani_experimental_2018}%
  \BibitemOpen
  \bibfield  {author} {\bibinfo {author} {\bibnamefont {Giordani},
  \bibfnamefont {T}}, \bibinfo {author} {\bibfnamefont {F.}~\bibnamefont
  {Flamini}}, \bibinfo {author} {\bibfnamefont {M.}~\bibnamefont {Pompili}},
  \bibinfo {author} {\bibfnamefont {N.}~\bibnamefont {Viggianiello}}, \bibinfo
  {author} {\bibfnamefont {N.}~\bibnamefont {Spagnolo}}, \bibinfo {author}
  {\bibfnamefont {A.}~\bibnamefont {Crespi}}, \bibinfo {author} {\bibfnamefont
  {R.}~\bibnamefont {Osellame}}, \bibinfo {author} {\bibfnamefont
  {N.}~\bibnamefont {Wiebe}}, \bibinfo {author} {\bibfnamefont
  {M.}~\bibnamefont {Walschaers}}, \bibinfo {author} {\bibfnamefont
  {A.}~\bibnamefont {Buchleitner}}, and\ \bibinfo {author} {\bibfnamefont
  {F.}~\bibnamefont {Sciarrino}}} (\bibinfo {year} {2018}),\ \bibfield  {title}
  {\enquote {\bibinfo {title} {Experimental statistical signature of many-body
  quantum interference},}\ }\href {https://doi.org/10.1038/s41566-018-0097-4}
  {\bibfield  {journal} {\bibinfo  {journal} {Nature Photon}\ }\textbf
  {\bibinfo {volume} {12}}~(\bibinfo {number} {3}),\ \bibinfo {pages}
  {173--178}},\ \Eprint {https://arxiv.org/abs/2103.16418} {arXiv:2103.16418
  [quant-ph]} \BibitemShut {NoStop}%
\bibitem [{\citenamefont {Giustina}\ \emph {et~al.}(2015)\citenamefont
  {Giustina}, \citenamefont {Versteegh}, \citenamefont {Wengerowsky},
  \citenamefont {Handsteiner}, \citenamefont {Hochrainer}, \citenamefont
  {Phelan}, \citenamefont {Steinlechner}, \citenamefont {Kofler}, \citenamefont
  {Larsson}, \citenamefont {Abell\'an}, \citenamefont {Amaya}, \citenamefont
  {Pruneri}, \citenamefont {Mitchell}, \citenamefont {Beyer}, \citenamefont
  {Gerrits}, \citenamefont {Lita}, \citenamefont {Shalm}, \citenamefont {Nam},
  \citenamefont {Scheidl}, \citenamefont {Ursin}, \citenamefont {Wittmann},\
  and\ \citenamefont {Zeilinger}}]{giustina_significant-loophole-free_2015}%
  \BibitemOpen
  \bibfield  {author} {\bibinfo {author} {\bibnamefont {Giustina},
  \bibfnamefont {M}}, \bibinfo {author} {\bibfnamefont {M.~A.~M.}\ \bibnamefont
  {Versteegh}}, \bibinfo {author} {\bibfnamefont {S.}~\bibnamefont
  {Wengerowsky}}, \bibinfo {author} {\bibfnamefont {J.}~\bibnamefont
  {Handsteiner}}, \bibinfo {author} {\bibfnamefont {A.}~\bibnamefont
  {Hochrainer}}, \bibinfo {author} {\bibfnamefont {K.}~\bibnamefont {Phelan}},
  \bibinfo {author} {\bibfnamefont {F.}~\bibnamefont {Steinlechner}}, \bibinfo
  {author} {\bibfnamefont {J.}~\bibnamefont {Kofler}}, \bibinfo {author}
  {\bibfnamefont {J.-A.}\ \bibnamefont {Larsson}}, \bibinfo {author}
  {\bibfnamefont {C.}~\bibnamefont {Abell\'an}}, \bibinfo {author}
  {\bibfnamefont {W.}~\bibnamefont {Amaya}}, \bibinfo {author} {\bibfnamefont
  {V.}~\bibnamefont {Pruneri}}, \bibinfo {author} {\bibfnamefont {M.~W.}\
  \bibnamefont {Mitchell}}, \bibinfo {author} {\bibfnamefont {J.}~\bibnamefont
  {Beyer}}, \bibinfo {author} {\bibfnamefont {T.}~\bibnamefont {Gerrits}},
  \bibinfo {author} {\bibfnamefont {A.~E.}\ \bibnamefont {Lita}}, \bibinfo
  {author} {\bibfnamefont {L.~K.}\ \bibnamefont {Shalm}}, \bibinfo {author}
  {\bibfnamefont {S.~W.}\ \bibnamefont {Nam}}, \bibinfo {author} {\bibfnamefont
  {T.}~\bibnamefont {Scheidl}}, \bibinfo {author} {\bibfnamefont
  {R.}~\bibnamefont {Ursin}}, \bibinfo {author} {\bibfnamefont
  {B.}~\bibnamefont {Wittmann}}, and\ \bibinfo {author} {\bibfnamefont
  {A.}~\bibnamefont {Zeilinger}}} (\bibinfo {year} {2015}),\ \bibfield  {title}
  {\enquote {\bibinfo {title} {{Significant-loophole-free test of Bell's
  theorem with entangled photons}},}\ }\href
  {https://doi.org/10.1103/PhysRevLett.115.250401} {\bibfield  {journal}
  {\bibinfo  {journal} {Phys. Rev. Lett.}\ }\textbf {\bibinfo {volume} {115}},\
  \bibinfo {pages} {250401}}\BibitemShut {NoStop}%
\bibitem [{\citenamefont {Glasser}\ \emph {et~al.}(2019)\citenamefont
  {Glasser}, \citenamefont {Sweke}, \citenamefont {Pancotti}, \citenamefont
  {Eisert},\ and\ \citenamefont {Cirac}}]{Expressive}%
  \BibitemOpen
  \bibfield  {author} {\bibinfo {author} {\bibnamefont {Glasser}, \bibfnamefont
  {I}}, \bibinfo {author} {\bibfnamefont {R.}~\bibnamefont {Sweke}}, \bibinfo
  {author} {\bibfnamefont {N.}~\bibnamefont {Pancotti}}, \bibinfo {author}
  {\bibfnamefont {J.}~\bibnamefont {Eisert}}, and\ \bibinfo {author}
  {\bibfnamefont {J.~I.}\ \bibnamefont {Cirac}}} (\bibinfo {year} {2019}),\
  \bibfield  {title} {\enquote {\bibinfo {title} {{Expressive power of
  tensor-network factorizations for probabilistic modelling, with applications
  from hidden Markov models to quantum machine learning}},}\ }\href@noop {} {\
  }\Eprint {https://arxiv.org/abs/1907.03741} {arXiv:1907.03741} \BibitemShut
  {NoStop}%
\bibitem [{\citenamefont {Gluza}\ \emph {et~al.}(2018)\citenamefont {Gluza},
  \citenamefont {Kliesch}, \citenamefont {Eisert},\ and\ \citenamefont
  {Aolita}}]{gluza_fidelity_2018}%
  \BibitemOpen
  \bibfield  {author} {\bibinfo {author} {\bibnamefont {Gluza}, \bibfnamefont
  {M}}, \bibinfo {author} {\bibfnamefont {M.}~\bibnamefont {Kliesch}}, \bibinfo
  {author} {\bibfnamefont {J.}~\bibnamefont {Eisert}}, and\ \bibinfo {author}
  {\bibfnamefont {L.}~\bibnamefont {Aolita}}} (\bibinfo {year} {2018}),\
  \bibfield  {title} {\enquote {\bibinfo {title} {Fidelity witnesses for
  fermionic quantum simulations},}\ }\href
  {https://doi.org/10.1103/PhysRevLett.120.190501} {\bibfield  {journal}
  {\bibinfo  {journal} {Phys. Rev. Lett.}\ }\textbf {\bibinfo {volume} {120}},\
  \bibinfo {pages} {190501}}\BibitemShut {NoStop}%
\bibitem [{\citenamefont {Glynn}(2010)}]{glynn_permanent_2010}%
  \BibitemOpen
  \bibfield  {author} {\bibinfo {author} {\bibnamefont {Glynn}, \bibfnamefont
  {D~G}}} (\bibinfo {year} {2010}),\ \bibfield  {title} {\enquote {\bibinfo
  {title} {The permanent of a square matrix},}\ }\href
  {https://doi.org/10.1016/j.ejc.2010.01.010} {\bibfield  {journal} {\bibinfo
  {journal} {Eur. J. Comb.}\ }\textbf {\bibinfo {volume} {31}},\ \bibinfo
  {pages} {1887--1891}}\BibitemShut {NoStop}%
\bibitem [{\citenamefont {Gogolin}\ \emph {et~al.}(2013)\citenamefont
  {Gogolin}, \citenamefont {Kliesch}, \citenamefont {Aolita},\ and\
  \citenamefont {Eisert}}]{gogolin_bosonsampling_2013}%
  \BibitemOpen
  \bibfield  {author} {\bibinfo {author} {\bibnamefont {Gogolin}, \bibfnamefont
  {C}}, \bibinfo {author} {\bibfnamefont {M.}~\bibnamefont {Kliesch}}, \bibinfo
  {author} {\bibfnamefont {L.}~\bibnamefont {Aolita}}, and\ \bibinfo {author}
  {\bibfnamefont {J.}~\bibnamefont {Eisert}}} (\bibinfo {year} {2013}),\
  \bibfield  {title} {\enquote {\bibinfo {title} {Boson-sampling in the light
  of sample complexity},}\ }\href@noop {} {\ }\Eprint
  {https://arxiv.org/abs/1306.3995} {arXiv:1306.3995} \BibitemShut {NoStop}%
\bibitem [{\citenamefont {Goldberg}\ and\ \citenamefont
  {Guo}(2014)}]{goldberg_complexity_2014}%
  \BibitemOpen
  \bibfield  {author} {\bibinfo {author} {\bibnamefont {Goldberg},
  \bibfnamefont {L~A}}, and\ \bibinfo {author} {\bibfnamefont {H.}~\bibnamefont
  {Guo}}} (\bibinfo {year} {2014}),\ \bibfield  {title} {\enquote {\bibinfo
  {title} {The complexity of approximating complex-valued {{Ising}} and
  {{Tutte}} partition functions},}\ }\href@noop {} {\ }\Eprint
  {https://arxiv.org/abs/1409.5627} {arXiv:1409.5627} \BibitemShut {NoStop}%
\bibitem [{\citenamefont {Goldreich}(2017)}]{goldreich_introduction_2017}%
  \BibitemOpen
  \bibfield  {author} {\bibinfo {author} {\bibnamefont {Goldreich},
  \bibfnamefont {Oded}}} (\bibinfo {year} {2017}),\ \href
  {https://doi.org/10.1017/9781108135252} {\emph {\bibinfo {title}
  {Introduction to {{Property Testing}}}}}\ (\bibinfo  {publisher} {{Cambridge
  University Press}},\ \bibinfo {address} {{Cambridge}})\BibitemShut {NoStop}%
\bibitem [{\citenamefont {Gottesman}(1997)}]{gottesman_stabilizer_1997}%
  \BibitemOpen
  \bibfield  {author} {\bibinfo {author} {\bibnamefont {Gottesman},
  \bibfnamefont {D}}} (\bibinfo {year} {1997}),\ \emph {\bibinfo {title}
  {Stabilizer {{codes}} and {{quantum error correction}}}},\ \href@noop {}
  {Ph.D. thesis}\ (\bibinfo  {school} {California Institute of Technology},
  \bibinfo {address} {{Pasadena, CA}}),\ \Eprint
  {https://arxiv.org/abs/quant-ph/9705052} {arXiv:quant-ph/9705052}
  \BibitemShut {NoStop}%
\bibitem [{\citenamefont {Gray}\ and\ \citenamefont
  {Kourtis}(2021)}]{gray_hyper-optimized_2021}%
  \BibitemOpen
  \bibfield  {author} {\bibinfo {author} {\bibnamefont {Gray}, \bibfnamefont
  {J}}, and\ \bibinfo {author} {\bibfnamefont {S.}~\bibnamefont {Kourtis}}}
  (\bibinfo {year} {2021}),\ \bibfield  {title} {\enquote {\bibinfo {title}
  {Hyper-optimized tensor network contraction},}\ }\href
  {https://doi.org/10.22331/q-2021-03-15-410} {\bibfield  {journal} {\bibinfo
  {journal} {Quantum}\ }\textbf {\bibinfo {volume} {5}},\ \bibinfo {pages}
  {410}}\BibitemShut {NoStop}%
\bibitem [{\citenamefont {{Greenbaum}}(2015)}]{Gre15}%
  \BibitemOpen
  \bibfield  {author} {\bibinfo {author} {\bibnamefont {{Greenbaum}},
  \bibfnamefont {D}}} (\bibinfo {year} {2015}),\ \bibfield  {title} {\enquote
  {\bibinfo {title} {{Introduction to quantum gate set tomography}},}\
  }\href@noop {} {\ }\Eprint {https://arxiv.org/abs/1509.02921}
  {arXiv:1509.02921} \BibitemShut {NoStop}%
\bibitem [{\citenamefont {Grier}\ \emph {et~al.}(2022)\citenamefont {Grier},
  \citenamefont {Brod}, \citenamefont {Arrazola}, \citenamefont {Alonso},\ and\
  \citenamefont {Quesada}}]{grier_complexity_2022}%
  \BibitemOpen
  \bibfield  {author} {\bibinfo {author} {\bibnamefont {Grier}, \bibfnamefont
  {D}}, \bibinfo {author} {\bibfnamefont {D.~J.}\ \bibnamefont {Brod}},
  \bibinfo {author} {\bibfnamefont {J.~M.}\ \bibnamefont {Arrazola}}, \bibinfo
  {author} {\bibfnamefont {M.~B.~A.}\ \bibnamefont {Alonso}}, and\ \bibinfo
  {author} {\bibfnamefont {N.}~\bibnamefont {Quesada}}} (\bibinfo {year}
  {2022}),\ \bibfield  {title} {\enquote {\bibinfo {title} {The {{complexity}}
  of {{bipartite Gaussian boson sampling}}},}\ }\href@noop {} {\ }\Eprint
  {https://arxiv.org/abs/2110.06964} {arXiv:2110.06964} \BibitemShut {NoStop}%
\bibitem [{\citenamefont {Grier}\ and\ \citenamefont
  {Schaeffer}(2018)}]{grier_new_2018}%
  \BibitemOpen
  \bibfield  {author} {\bibinfo {author} {\bibnamefont {Grier}, \bibfnamefont
  {D}}, and\ \bibinfo {author} {\bibfnamefont {L.}~\bibnamefont {Schaeffer}}}
  (\bibinfo {year} {2018}),\ \bibfield  {title} {\enquote {\bibinfo {title}
  {New {{hardness results}} for the {{permanent using linear optics}}},}\
  }\href@noop {} {\ }\Eprint {https://arxiv.org/abs/1610.04670}
  {arXiv:1610.04670} \BibitemShut {NoStop}%
\bibitem [{\citenamefont {Grover}(1996)}]{grover_fast_1996}%
  \BibitemOpen
  \bibfield  {author} {\bibinfo {author} {\bibnamefont {Grover}, \bibfnamefont
  {L~K}}} (\bibinfo {year} {1996}),\ \bibfield  {title} {\enquote {\bibinfo
  {title} {A fast quantum mechanical algorithm for database search},}\ }in\
  \href {https://doi.org/10.1145/237814.237866} {\emph {\bibinfo {booktitle}
  {Proce. 28th Ann. {{ACM}} Symp. {{Th.}} Comp. - {{STOC}} '96}}}\ (\bibinfo
  {publisher} {{ACM Press}},\ \bibinfo {address} {{Philadelphia, Pennsylvania,
  United States}})\ pp.\ \bibinfo {pages} {212--219}\BibitemShut {NoStop}%
\bibitem [{\citenamefont {Guanzon}\ \emph {et~al.}(2021)\citenamefont
  {Guanzon}, \citenamefont {Lund},\ and\ \citenamefont
  {Ralph}}]{PhysRevA.104.032607}%
  \BibitemOpen
  \bibfield  {author} {\bibinfo {author} {\bibnamefont {Guanzon}, \bibfnamefont
  {J~J}}, \bibinfo {author} {\bibfnamefont {A.~P.}\ \bibnamefont {Lund}}, and\
  \bibinfo {author} {\bibfnamefont {T.~C.}\ \bibnamefont {Ralph}}} (\bibinfo
  {year} {2021}),\ \bibfield  {title} {\enquote {\bibinfo {title} {Multimode
  metrology via scattershot sampling},}\ }\href
  {https://doi.org/10.1103/PhysRevA.104.032607} {\bibfield  {journal} {\bibinfo
   {journal} {Phys. Rev. A}\ }\textbf {\bibinfo {volume} {104}},\ \bibinfo
  {pages} {032607}}\BibitemShut {NoStop}%
\bibitem [{\citenamefont {G{\"u}hne}\ and\ \citenamefont
  {T{\'o}th}(2009)}]{guehne_entanglement_2009}%
  \BibitemOpen
  \bibfield  {author} {\bibinfo {author} {\bibnamefont {G{\"u}hne},
  \bibfnamefont {O}}, and\ \bibinfo {author} {\bibfnamefont {G.}~\bibnamefont
  {T{\'o}th}}} (\bibinfo {year} {2009}),\ \bibfield  {title} {\enquote
  {\bibinfo {title} {Entanglement detection},}\ }\href
  {https://doi.org/10.1016/j.physrep.2009.02.004} {\bibfield  {journal}
  {\bibinfo  {journal} {Phys. Rep.}\ }\textbf {\bibinfo {volume} {474}},\
  \bibinfo {pages} {1}}\BibitemShut {NoStop}%
\bibitem [{\citenamefont {Guo}\ \emph {et~al.}(2019)\citenamefont {Guo},
  \citenamefont {Liu}, \citenamefont {Xiong}, \citenamefont {Xue},
  \citenamefont {Fu}, \citenamefont {Huang}, \citenamefont {Qiang},
  \citenamefont {Xu}, \citenamefont {Liu}, \citenamefont {Zheng}, \citenamefont
  {Huang}, \citenamefont {Deng}, \citenamefont {Poletti}, \citenamefont {Bao},\
  and\ \citenamefont {Wu}}]{guo_general-purpose_2019}%
  \BibitemOpen
  \bibfield  {author} {\bibinfo {author} {\bibnamefont {Guo}, \bibfnamefont
  {C}}, \bibinfo {author} {\bibfnamefont {Y.}~\bibnamefont {Liu}}, \bibinfo
  {author} {\bibfnamefont {M.}~\bibnamefont {Xiong}}, \bibinfo {author}
  {\bibfnamefont {S.}~\bibnamefont {Xue}}, \bibinfo {author} {\bibfnamefont
  {X.}~\bibnamefont {Fu}}, \bibinfo {author} {\bibfnamefont {A.}~\bibnamefont
  {Huang}}, \bibinfo {author} {\bibfnamefont {X.}~\bibnamefont {Qiang}},
  \bibinfo {author} {\bibfnamefont {P.}~\bibnamefont {Xu}}, \bibinfo {author}
  {\bibfnamefont {J.}~\bibnamefont {Liu}}, \bibinfo {author} {\bibfnamefont
  {S.}~\bibnamefont {Zheng}}, \bibinfo {author} {\bibfnamefont {H.-L.}\
  \bibnamefont {Huang}}, \bibinfo {author} {\bibfnamefont {M.}~\bibnamefont
  {Deng}}, \bibinfo {author} {\bibfnamefont {D.}~\bibnamefont {Poletti}},
  \bibinfo {author} {\bibfnamefont {W.-S.}\ \bibnamefont {Bao}}, and\ \bibinfo
  {author} {\bibfnamefont {J.}~\bibnamefont {Wu}}} (\bibinfo {year} {2019}),\
  \bibfield  {title} {\enquote {\bibinfo {title} {General-{{purpose quantum
  circuit simulator}} with {{projected entangled-pair states}} and the
  {{quantum supremacy frontier}}},}\ }\href
  {https://doi.org/10.1103/PhysRevLett.123.190501} {\bibfield  {journal}
  {\bibinfo  {journal} {Phys. Rev. Lett.}\ }\textbf {\bibinfo {volume} {123}},\
  \bibinfo {pages} {190501}}\BibitemShut {NoStop}%
\bibitem [{\citenamefont {Guo}\ \emph {et~al.}(2021)\citenamefont {Guo},
  \citenamefont {Zhao},\ and\ \citenamefont {Huang}}]{guo_verifying_2021}%
  \BibitemOpen
  \bibfield  {author} {\bibinfo {author} {\bibnamefont {Guo}, \bibfnamefont
  {C}}, \bibinfo {author} {\bibfnamefont {Y.}~\bibnamefont {Zhao}}, and\
  \bibinfo {author} {\bibfnamefont {H.-L.}\ \bibnamefont {Huang}}} (\bibinfo
  {year} {2021}),\ \bibfield  {title} {\enquote {\bibinfo {title} {Verifying
  {{random quantum circuits}} with {{arbitrary geometry using tensor network
  states algorithm}}},}\ }\href
  {https://doi.org/10.1103/PhysRevLett.126.070502} {\bibfield  {journal}
  {\bibinfo  {journal} {Phys. Rev. Lett.}\ }\textbf {\bibinfo {volume} {126}},\
  \bibinfo {pages} {070502}}\BibitemShut {NoStop}%
\bibitem [{\citenamefont {Gupt}\ \emph {et~al.}(2020)\citenamefont {Gupt},
  \citenamefont {Arrazola}, \citenamefont {Quesada},\ and\ \citenamefont
  {Bromley}}]{gupt_classical_2020-1}%
  \BibitemOpen
  \bibfield  {author} {\bibinfo {author} {\bibnamefont {Gupt}, \bibfnamefont
  {B}}, \bibinfo {author} {\bibfnamefont {J.~M.}\ \bibnamefont {Arrazola}},
  \bibinfo {author} {\bibfnamefont {N.}~\bibnamefont {Quesada}}, and\ \bibinfo
  {author} {\bibfnamefont {T.~R.}\ \bibnamefont {Bromley}}} (\bibinfo {year}
  {2020}),\ \bibfield  {title} {\enquote {\bibinfo {title} {Classical
  benchmarking of {{Gaussian boson sampling}} on the {{Titan}}
  supercomputer},}\ }\href {https://doi.org/10.1007/s11128-020-02713-6}
  {\bibfield  {journal} {\bibinfo  {journal} {Quant. Inf. Proc.}\ }\textbf
  {\bibinfo {volume} {19}},\ \bibinfo {pages} {249}}\BibitemShut {NoStop}%
\bibitem [{\citenamefont {Gupt}\ \emph {et~al.}(2019)\citenamefont {Gupt},
  \citenamefont {Izaac},\ and\ \citenamefont {Quesada}}]{gupt_walrus_2019}%
  \BibitemOpen
  \bibfield  {author} {\bibinfo {author} {\bibnamefont {Gupt}, \bibfnamefont
  {B}}, \bibinfo {author} {\bibfnamefont {J.}~\bibnamefont {Izaac}}, and\
  \bibinfo {author} {\bibfnamefont {N.}~\bibnamefont {Quesada}}} (\bibinfo
  {year} {2019}),\ \bibfield  {title} {\enquote {\bibinfo {title} {{The
  {{Walrus}}: A library for the calculation of Hafnians, {{Hermite}}
  polynomials and {{Gaussian}} boson sampling}},}\ }\href
  {https://doi.org/10.21105/joss.01705} {\bibfield  {journal} {\bibinfo
  {journal} {J. Open Source Soft.}\ }\textbf {\bibinfo {volume} {4}},\ \bibinfo
  {pages} {1705}}\BibitemShut {NoStop}%
\bibitem [{\citenamefont {Guruswami}(2006)}]{guruswami_list_2006}%
  \BibitemOpen
  \bibfield  {author} {\bibinfo {author} {\bibnamefont {Guruswami},
  \bibfnamefont {V}}} (\bibinfo {year} {2006}),\ \bibfield  {title} {\enquote
  {\bibinfo {title} {List {{decoding}} in {{average-case complexity}} and
  {{pseudorandomness}}},}\ }in\ \href
  {https://doi.org/10.1109/ITW.2006.1633776} {\emph {\bibinfo {booktitle} {2006
  {{IEEE Information Theory Workshop}} - {{ITW}} '06 {{Punta}} Del
  {{Este}}}}},\ pp.\ \bibinfo {pages} {32--36}\BibitemShut {NoStop}%
\bibitem [{\citenamefont {Gurvits}(2003)}]{gurvits_classical_2003}%
  \BibitemOpen
  \bibfield  {author} {\bibinfo {author} {\bibnamefont {Gurvits}, \bibfnamefont
  {L}}} (\bibinfo {year} {2003}),\ \bibfield  {title} {\enquote {\bibinfo
  {title} {Classical deterministic complexity of {{Edmonds}}' problem and
  {{quantum entanglement}}},}\ }\href@noop {} {\ }\Eprint
  {https://arxiv.org/abs/quant-ph/0303055} {arXiv:quant-ph/0303055}
  \BibitemShut {NoStop}%
\bibitem [{\citenamefont {Gurvits}\ and\ \citenamefont
  {Samorodnitsky}(2002)}]{gurvits_deterministic_2002}%
  \BibitemOpen
  \bibfield  {author} {\bibinfo {author} {\bibnamefont {Gurvits}, \bibfnamefont
  {L}}, and\ \bibinfo {author} {\bibfnamefont {A.}~\bibnamefont
  {Samorodnitsky}}} (\bibinfo {year} {2002}),\ \bibfield  {title} {\enquote
  {\bibinfo {title} {A {{deterministic algorithm}} for {{approximating}} the
  {{mixed discriminant}} and {{mixed volume}}, and a {{combinatorial
  corollary}}},}\ }\href {https://doi.org/10.1007/s00454-001-0083-2} {\bibfield
   {journal} {\bibinfo  {journal} {Disc. Comp. Geom.}\ }\textbf {\bibinfo
  {volume} {27}},\ \bibinfo {pages} {531--550}}\BibitemShut {NoStop}%
\bibitem [{\citenamefont {Haake}(2010)}]{haake_quantum_2010}%
  \BibitemOpen
  \bibfield  {author} {\bibinfo {author} {\bibnamefont {Haake}, \bibfnamefont
  {F}}} (\bibinfo {year} {2010}),\ \href@noop {} {\emph {\bibinfo {title}
  {Quantum {{signatures}} of {{chaos}}}}},\ \bibinfo {series} {Springer
  {{Series}} in {{Synergetics}}}, Vol.~\bibinfo {volume} {54}\ (\bibinfo
  {publisher} {{Springer Berlin Heidelberg}},\ \bibinfo {address} {{Berlin,
  Heidelberg}})\BibitemShut {NoStop}%
\bibitem [{\citenamefont {Haferkamp}(2022)}]{haferkamp_random_2022}%
  \BibitemOpen
  \bibfield  {author} {\bibinfo {author} {\bibnamefont {Haferkamp},
  \bibfnamefont {J}}} (\bibinfo {year} {2022}),\ \bibfield  {title} {\enquote
  {\bibinfo {title} {Random quantum circuits are approximate unitary
  $t$-designs in depth $o(nt^{5+o(1)})$},}\ }\href@noop {} {\ }\Eprint
  {https://arxiv.org/abs/2203.16571} {arXiv:2203.16571} \BibitemShut {NoStop}%
\bibitem [{\citenamefont {Haferkamp}\ \emph
  {et~al.}(2020{\natexlab{a}})\citenamefont {Haferkamp}, \citenamefont
  {Hangleiter}, \citenamefont {Bouland}, \citenamefont {Fefferman},
  \citenamefont {Eisert},\ and\ \citenamefont
  {{Bermejo-Vega}}}]{haferkamp_closing_2020}%
  \BibitemOpen
  \bibfield  {author} {\bibinfo {author} {\bibnamefont {Haferkamp},
  \bibfnamefont {J}}, \bibinfo {author} {\bibfnamefont {D.}~\bibnamefont
  {Hangleiter}}, \bibinfo {author} {\bibfnamefont {A.}~\bibnamefont {Bouland}},
  \bibinfo {author} {\bibfnamefont {B.}~\bibnamefont {Fefferman}}, \bibinfo
  {author} {\bibfnamefont {J.}~\bibnamefont {Eisert}}, and\ \bibinfo {author}
  {\bibfnamefont {J.}~\bibnamefont {{Bermejo-Vega}}}} (\bibinfo {year}
  {2020}{\natexlab{a}}),\ \bibfield  {title} {\enquote {\bibinfo {title}
  {Closing {{gaps}} of a {{quantum advantage}} with {{short-time Hamiltonian
  dynamics}}},}\ }\href {https://doi.org/10.1103/PhysRevLett.125.250501}
  {\bibfield  {journal} {\bibinfo  {journal} {Phys. Rev. Lett.}\ }\textbf
  {\bibinfo {volume} {125}},\ \bibinfo {pages} {250501}}\BibitemShut {NoStop}%
\bibitem [{\citenamefont {Haferkamp}\ \emph
  {et~al.}(2020{\natexlab{b}})\citenamefont {Haferkamp}, \citenamefont
  {Hangleiter}, \citenamefont {Eisert},\ and\ \citenamefont
  {Gluza}}]{haferkamp_contracting_2020}%
  \BibitemOpen
  \bibfield  {author} {\bibinfo {author} {\bibnamefont {Haferkamp},
  \bibfnamefont {J}}, \bibinfo {author} {\bibfnamefont {D.}~\bibnamefont
  {Hangleiter}}, \bibinfo {author} {\bibfnamefont {J.}~\bibnamefont {Eisert}},
  and\ \bibinfo {author} {\bibfnamefont {M.}~\bibnamefont {Gluza}}} (\bibinfo
  {year} {2020}{\natexlab{b}}),\ \bibfield  {title} {\enquote {\bibinfo {title}
  {Contracting projected entangled pair states is average-case hard},}\ }\href
  {https://doi.org/10.1103/PhysRevResearch.2.013010} {\bibfield  {journal}
  {\bibinfo  {journal} {Phys. Rev. Res.}\ }\textbf {\bibinfo {volume} {2}},\
  \bibinfo {pages} {013010}}\BibitemShut {NoStop}%
\bibitem [{\citenamefont {Hamilton}\ \emph {et~al.}(2017)\citenamefont
  {Hamilton}, \citenamefont {Kruse}, \citenamefont {Sansoni}, \citenamefont
  {Barkhofen}, \citenamefont {Silberhorn},\ and\ \citenamefont
  {Jex}}]{hamilton_gaussian_2017}%
  \BibitemOpen
  \bibfield  {author} {\bibinfo {author} {\bibnamefont {Hamilton},
  \bibfnamefont {C~S}}, \bibinfo {author} {\bibfnamefont {R.}~\bibnamefont
  {Kruse}}, \bibinfo {author} {\bibfnamefont {L.}~\bibnamefont {Sansoni}},
  \bibinfo {author} {\bibfnamefont {S.}~\bibnamefont {Barkhofen}}, \bibinfo
  {author} {\bibfnamefont {C.}~\bibnamefont {Silberhorn}}, and\ \bibinfo
  {author} {\bibfnamefont {I.}~\bibnamefont {Jex}}} (\bibinfo {year} {2017}),\
  \bibfield  {title} {\enquote {\bibinfo {title} {Gaussian boson sampling},}\
  }\href {https://doi.org/10.1103/PhysRevLett.119.170501} {\bibfield  {journal}
  {\bibinfo  {journal} {Phys. Rev. Lett.}\ }\textbf {\bibinfo {volume} {119}},\
  \bibinfo {pages} {170501}}\BibitemShut {NoStop}%
\bibitem [{\citenamefont {H{\"a}ner}\ \emph {et~al.}(2017)\citenamefont
  {H{\"a}ner}, \citenamefont {Roetteler},\ and\ \citenamefont
  {Svore}}]{haner_factoring_2017}%
  \BibitemOpen
  \bibfield  {author} {\bibinfo {author} {\bibnamefont {H{\"a}ner},
  \bibfnamefont {T}}, \bibinfo {author} {\bibfnamefont {M.}~\bibnamefont
  {Roetteler}}, and\ \bibinfo {author} {\bibfnamefont {K.~M.}\ \bibnamefont
  {Svore}}} (\bibinfo {year} {2017}),\ \bibfield  {title} {\enquote {\bibinfo
  {title} {Factoring using $2n + 2$ qubits with {{Toffoli}} based modular
  multiplication},}\ }\href {https://doi.org/10.48550/arXiv.1611.07995}
  {\bibfield  {journal} {\bibinfo  {journal} {Quant. Inf. Comp.}\ }\textbf
  {\bibinfo {volume} {17}},\ \bibinfo {pages} {673--684}}\BibitemShut {NoStop}%
\bibitem [{\citenamefont {H{\"a}ner}\ and\ \citenamefont
  {Steiger}(2017)}]{haner_05_2017}%
  \BibitemOpen
  \bibfield  {author} {\bibinfo {author} {\bibnamefont {H{\"a}ner},
  \bibfnamefont {T}}, and\ \bibinfo {author} {\bibfnamefont {D.~S.}\
  \bibnamefont {Steiger}}} (\bibinfo {year} {2017}),\ \bibfield  {title}
  {\enquote {\bibinfo {title} {{0.5 Petabyte simulation of a 45-qubit quantum
  circuit}},}\ }in\ \href {https://doi.org/10.1145/3126908.3126947} {\emph
  {\bibinfo {booktitle} {Proceedings of the {{International Conference}} for
  {{High Performance Computing}}, {{Networking}}, {{Storage}} and
  {{Analysis}}}}},\ \bibinfo {series and number} {{{SC}} '17}\ (\bibinfo
  {publisher} {{Association for Computing Machinery}},\ \bibinfo {address}
  {{New York, NY, USA}})\ pp.\ \bibinfo {pages} {1--10}\BibitemShut {NoStop}%
\bibitem [{\citenamefont {Hangleiter}(2021)}]{hangleiter_sampling_2021}%
  \BibitemOpen
  \bibfield  {author} {\bibinfo {author} {\bibnamefont {Hangleiter},
  \bibfnamefont {D}}} (\bibinfo {year} {2021}),\ \emph {\bibinfo {title}
  {Sampling and the complexity of nature}},\ \href@noop {} {Ph.D. thesis}\
  (\bibinfo  {school} {Freie Universit\"at}, \bibinfo {address} {{Berlin}}),\
  \Eprint {https://arxiv.org/abs/2012.07905} {arXiv:2012.07905} \BibitemShut
  {NoStop}%
\bibitem [{\citenamefont {Hangleiter}\ \emph {et~al.}(2018)\citenamefont
  {Hangleiter}, \citenamefont {Bermejo-Vega}, \citenamefont {Schwarz},\ and\
  \citenamefont {Eisert}}]{hangleiter_anticoncentration_2018}%
  \BibitemOpen
  \bibfield  {author} {\bibinfo {author} {\bibnamefont {Hangleiter},
  \bibfnamefont {D}}, \bibinfo {author} {\bibfnamefont {J.}~\bibnamefont
  {Bermejo-Vega}}, \bibinfo {author} {\bibfnamefont {M.}~\bibnamefont
  {Schwarz}}, and\ \bibinfo {author} {\bibfnamefont {J.}~\bibnamefont
  {Eisert}}} (\bibinfo {year} {2018}),\ \bibfield  {title} {\enquote {\bibinfo
  {title} {Anticoncentration theorems for schemes showing a quantum speedup},}\
  }\href {https://doi.org/10.22331/q-2018-05-22-65} {\bibfield  {journal}
  {\bibinfo  {journal} {Quantum}\ }\textbf {\bibinfo {volume} {2}},\ \bibinfo
  {pages} {65}}\BibitemShut {NoStop}%
\bibitem [{\citenamefont {Hangleiter}\ \emph {et~al.}(2019)\citenamefont
  {Hangleiter}, \citenamefont {Kliesch}, \citenamefont {Eisert},\ and\
  \citenamefont {Gogolin}}]{hangleiter_sample_2019}%
  \BibitemOpen
  \bibfield  {author} {\bibinfo {author} {\bibnamefont {Hangleiter},
  \bibfnamefont {D}}, \bibinfo {author} {\bibfnamefont {M.}~\bibnamefont
  {Kliesch}}, \bibinfo {author} {\bibfnamefont {J.}~\bibnamefont {Eisert}},
  and\ \bibinfo {author} {\bibfnamefont {C.}~\bibnamefont {Gogolin}}} (\bibinfo
  {year} {2019}),\ \bibfield  {title} {\enquote {\bibinfo {title} {Sample
  complexity of device-independently certified quantum supremacy},}\ }\href
  {https://doi.org/10.1103/PhysRevLett.122.210502} {\bibfield  {journal}
  {\bibinfo  {journal} {Phys. Rev. Lett.}\ }\textbf {\bibinfo {volume} {122}},\
  \bibinfo {pages} {210502}}\BibitemShut {NoStop}%
\bibitem [{\citenamefont {Hangleiter}\ \emph {et~al.}(2017)\citenamefont
  {Hangleiter}, \citenamefont {Kliesch}, \citenamefont {Schwarz},\ and\
  \citenamefont {Eisert}}]{hangleiter_direct_2017}%
  \BibitemOpen
  \bibfield  {author} {\bibinfo {author} {\bibnamefont {Hangleiter},
  \bibfnamefont {D}}, \bibinfo {author} {\bibfnamefont {M.}~\bibnamefont
  {Kliesch}}, \bibinfo {author} {\bibfnamefont {M.}~\bibnamefont {Schwarz}},
  and\ \bibinfo {author} {\bibfnamefont {J.}~\bibnamefont {Eisert}}} (\bibinfo
  {year} {2017}),\ \bibfield  {title} {\enquote {\bibinfo {title} {Direct
  certification of a class of quantum simulations},}\ }\href
  {https://doi.org/10.1088/2058-9565/2/1/015004} {\bibfield  {journal}
  {\bibinfo  {journal} {Quant. Sci. Tech.}\ }\textbf {\bibinfo {volume} {2}},\
  \bibinfo {pages} {015004}}\BibitemShut {NoStop}%
\bibitem [{\citenamefont {Harrow}\ \emph {et~al.}(2009)\citenamefont {Harrow},
  \citenamefont {Hassidim},\ and\ \citenamefont
  {Lloyd}}]{PhysRevLett.103.150502}%
  \BibitemOpen
  \bibfield  {author} {\bibinfo {author} {\bibnamefont {Harrow}, \bibfnamefont
  {A~W}}, \bibinfo {author} {\bibfnamefont {A.}~\bibnamefont {Hassidim}}, and\
  \bibinfo {author} {\bibfnamefont {S.}~\bibnamefont {Lloyd}}} (\bibinfo {year}
  {2009}),\ \bibfield  {title} {\enquote {\bibinfo {title} {Quantum algorithm
  for linear systems of equations},}\ }\href
  {https://doi.org/10.1103/PhysRevLett.103.150502} {\bibfield  {journal}
  {\bibinfo  {journal} {Phys. Rev. Lett.}\ }\textbf {\bibinfo {volume} {103}},\
  \bibinfo {pages} {150502}}\BibitemShut {NoStop}%
\bibitem [{\citenamefont {Harrow}\ and\ \citenamefont
  {Low}(2009)}]{harrow_random_2009}%
  \BibitemOpen
  \bibfield  {author} {\bibinfo {author} {\bibnamefont {Harrow}, \bibfnamefont
  {A~W}}, and\ \bibinfo {author} {\bibfnamefont {R.~A.}\ \bibnamefont {Low}}}
  (\bibinfo {year} {2009}),\ \bibfield  {title} {\enquote {\bibinfo {title}
  {Random {quantum} {circuits} are approximate 2-designs},}\ }\href
  {https://doi.org/10.1007/s00220-009-0873-6} {\bibfield  {journal} {\bibinfo
  {journal} {Commun. Math. Phys.}\ }\textbf {\bibinfo {volume} {291}},\
  \bibinfo {pages} {257--302}}\BibitemShut {NoStop}%
\bibitem [{\citenamefont {Harrow}\ and\ \citenamefont
  {Mehraban}(2018)}]{harrow_approximate_2018}%
  \BibitemOpen
  \bibfield  {author} {\bibinfo {author} {\bibnamefont {Harrow}, \bibfnamefont
  {A~W}}, and\ \bibinfo {author} {\bibfnamefont {S.}~\bibnamefont {Mehraban}}}
  (\bibinfo {year} {2018}),\ \bibfield  {title} {\enquote {\bibinfo {title}
  {Approximate unitary $t$-designs by short random quantum circuits using
  nearest-neighbor and long-range gates},}\ }\href@noop {} {\ }\Eprint
  {https://arxiv.org/abs/1809.06957} {arXiv:1809.06957} \BibitemShut {NoStop}%
\bibitem [{\citenamefont {Harrow}\ and\ \citenamefont
  {Montanaro}(2017)}]{HarrowSupremacy}%
  \BibitemOpen
  \bibfield  {author} {\bibinfo {author} {\bibnamefont {Harrow}, \bibfnamefont
  {A~W}}, and\ \bibinfo {author} {\bibfnamefont {A.}~\bibnamefont {Montanaro}}}
  (\bibinfo {year} {2017}),\ \bibfield  {title} {\enquote {\bibinfo {title}
  {Quantum computational supremacy},}\ }\href
  {https://doi.org/10.1038/nature23458} {\bibfield  {journal} {\bibinfo
  {journal} {Nature}\ }\textbf {\bibinfo {volume} {549}},\ \bibinfo {pages}
  {203--209}}\BibitemShut {NoStop}%
\bibitem [{\citenamefont {Havl{\'i}cek}\ \emph {et~al.}(2019)\citenamefont
  {Havl{\'i}cek}, \citenamefont {C{\'o}rcoles}, \citenamefont {Temme},
  \citenamefont {Harrow}, \citenamefont {Kandala}, \citenamefont {Chow},\ and\
  \citenamefont {Gambetta}}]{havlicek_supervised_2019}%
  \BibitemOpen
  \bibfield  {author} {\bibinfo {author} {\bibnamefont {Havl{\'i}cek},
  \bibfnamefont {Vojtech}}, \bibinfo {author} {\bibfnamefont {Antonio~D.}\
  \bibnamefont {C{\'o}rcoles}}, \bibinfo {author} {\bibfnamefont {Kristan}\
  \bibnamefont {Temme}}, \bibinfo {author} {\bibfnamefont {Aram~W.}\
  \bibnamefont {Harrow}}, \bibinfo {author} {\bibfnamefont {Abhinav}\
  \bibnamefont {Kandala}}, \bibinfo {author} {\bibfnamefont {Jerry~M.}\
  \bibnamefont {Chow}}, and\ \bibinfo {author} {\bibfnamefont {Jay~M.}\
  \bibnamefont {Gambetta}}} (\bibinfo {year} {2019}),\ \bibfield  {title}
  {\enquote {\bibinfo {title} {Supervised learning with quantum-enhanced
  feature spaces},}\ }\href {https://doi.org/10.1038/s41586-019-0980-2}
  {\bibfield  {journal} {\bibinfo  {journal} {Nature}\ }\textbf {\bibinfo
  {volume} {567}}~(\bibinfo {number} {7747}),\ \bibinfo {pages}
  {209--212}}\BibitemShut {NoStop}%
\bibitem [{\citenamefont {Hayashi}\ and\ \citenamefont
  {Takeuchi}(2019)}]{hayashi_verifying_2019}%
  \BibitemOpen
  \bibfield  {author} {\bibinfo {author} {\bibnamefont {Hayashi}, \bibfnamefont
  {M}}, and\ \bibinfo {author} {\bibfnamefont {Y.}~\bibnamefont {Takeuchi}}}
  (\bibinfo {year} {2019}),\ \bibfield  {title} {\enquote {\bibinfo {title}
  {Verifying commuting quantum computations via fidelity estimation of weighted
  graph states},}\ }\href {https://doi.org/10.1088/1367-2630/ab3d88} {\bibfield
   {journal} {\bibinfo  {journal} {New J. Phys.}\ }\textbf {\bibinfo {volume}
  {21}},\ \bibinfo {pages} {093060}}\BibitemShut {NoStop}%
\bibitem [{\citenamefont {Hebenstreit}\ \emph {et~al.}(2019)\citenamefont
  {Hebenstreit}, \citenamefont {Jozsa}, \citenamefont {Kraus}, \citenamefont
  {Strelchuk},\ and\ \citenamefont {Yoganathan}}]{hebenstreit_all_2019}%
  \BibitemOpen
  \bibfield  {author} {\bibinfo {author} {\bibnamefont {Hebenstreit},
  \bibfnamefont {M}}, \bibinfo {author} {\bibfnamefont {R.}~\bibnamefont
  {Jozsa}}, \bibinfo {author} {\bibfnamefont {B.}~\bibnamefont {Kraus}},
  \bibinfo {author} {\bibfnamefont {S.}~\bibnamefont {Strelchuk}}, and\
  \bibinfo {author} {\bibfnamefont {M.}~\bibnamefont {Yoganathan}}} (\bibinfo
  {year} {2019}),\ \bibfield  {title} {\enquote {\bibinfo {title} {All {{pure
  fermionic non-Gaussian states are magic states}} for {{matchgate
  computations}}},}\ }\href {https://doi.org/10.1103/PhysRevLett.123.080503}
  {\bibfield  {journal} {\bibinfo  {journal} {Phys. Rev. Lett.}\ }\textbf
  {\bibinfo {volume} {123}},\ \bibinfo {pages} {080503}}\BibitemShut {NoStop}%
\bibitem [{\citenamefont {Helsen}\ \emph {et~al.}(2021)\citenamefont {Helsen},
  \citenamefont {Ioannou}, \citenamefont {Roth}, \citenamefont {Kitzinger},
  \citenamefont {Onorati}, \citenamefont {Werner},\ and\ \citenamefont
  {Eisert}}]{EstimatingGateSetProperties}%
  \BibitemOpen
  \bibfield  {author} {\bibinfo {author} {\bibnamefont {Helsen}, \bibfnamefont
  {J}}, \bibinfo {author} {\bibfnamefont {M.}~\bibnamefont {Ioannou}}, \bibinfo
  {author} {\bibfnamefont {I.}~\bibnamefont {Roth}}, \bibinfo {author}
  {\bibfnamefont {J.}~\bibnamefont {Kitzinger}}, \bibinfo {author}
  {\bibfnamefont {E.}~\bibnamefont {Onorati}}, \bibinfo {author} {\bibfnamefont
  {A.~H.}\ \bibnamefont {Werner}}, and\ \bibinfo {author} {\bibfnamefont
  {J.}~\bibnamefont {Eisert}}} (\bibinfo {year} {2021}),\ \bibfield  {title}
  {\enquote {\bibinfo {title} {Estimating gate-set properties from random
  sequences},}\ }\href@noop {} {\ }\Eprint {https://arxiv.org/abs/2110.13178}
  {arXiv:2110.13178} \BibitemShut {NoStop}%
\bibitem [{\citenamefont {Helsen}\ \emph {et~al.}(2022)\citenamefont {Helsen},
  \citenamefont {Roth}, \citenamefont {Onorati}, \citenamefont {Werner},\ and\
  \citenamefont {Eisert}}]{helsen_general_2020}%
  \BibitemOpen
  \bibfield  {author} {\bibinfo {author} {\bibnamefont {Helsen}, \bibfnamefont
  {J}}, \bibinfo {author} {\bibfnamefont {I.}~\bibnamefont {Roth}}, \bibinfo
  {author} {\bibfnamefont {E.}~\bibnamefont {Onorati}}, \bibinfo {author}
  {\bibfnamefont {A.~H.}\ \bibnamefont {Werner}}, and\ \bibinfo {author}
  {\bibfnamefont {J.}~\bibnamefont {Eisert}}} (\bibinfo {year} {2022}),\
  \bibfield  {title} {\enquote {\bibinfo {title} {A general framework for
  randomized benchmarking},}\ }\href
  {https://doi.org/10.1103/PRXQuantum.3.020357} {\bibfield  {journal} {\bibinfo
   {journal} {PRX Quantum}\ }\textbf {\bibinfo {volume} {3}},\ \bibinfo {pages}
  {020357}}\BibitemShut {NoStop}%
\bibitem [{\citenamefont {Helsen}\ \emph {et~al.}(2019)\citenamefont {Helsen},
  \citenamefont {Xue}, \citenamefont {Vandersypen},\ and\ \citenamefont
  {Wehner}}]{helsen_new_2019}%
  \BibitemOpen
  \bibfield  {author} {\bibinfo {author} {\bibnamefont {Helsen}, \bibfnamefont
  {J}}, \bibinfo {author} {\bibfnamefont {X.}~\bibnamefont {Xue}}, \bibinfo
  {author} {\bibfnamefont {L.~M.~K.}\ \bibnamefont {Vandersypen}}, and\
  \bibinfo {author} {\bibfnamefont {S.}~\bibnamefont {Wehner}}} (\bibinfo
  {year} {2019}),\ \bibfield  {title} {\enquote {\bibinfo {title} {A new class
  of efficient randomized benchmarking protocols},}\ }\href
  {https://doi.org/10.1038/s41534-019-0182-7} {\bibfield  {journal} {\bibinfo
  {journal} {npj Quant. Inf.}\ }\textbf {\bibinfo {volume} {5}},\ \bibinfo
  {pages} {1--9}}\BibitemShut {NoStop}%
\bibitem [{\citenamefont {Hensen}\ \emph {et~al.}(2015)\citenamefont {Hensen},
  \citenamefont {Bernien}, \citenamefont {Dréau}, \citenamefont {Reiserer},
  \citenamefont {Kalb}, \citenamefont {Blok}, \citenamefont {Ruitenberg},
  \citenamefont {Vermeulen}, \citenamefont {Schouten}, \citenamefont
  {Abellán}, \citenamefont {Amaya}, \citenamefont {Pruneri}, \citenamefont
  {Mitchell}, \citenamefont {Markham}, \citenamefont {Twitchen}, \citenamefont
  {Elkouss}, \citenamefont {Wehner}, \citenamefont {Taminiau},\ and\
  \citenamefont {Hanson}}]{hensen_loophole-free_2015}%
  \BibitemOpen
  \bibfield  {author} {\bibinfo {author} {\bibnamefont {Hensen}, \bibfnamefont
  {B}}, \bibinfo {author} {\bibfnamefont {H.}~\bibnamefont {Bernien}}, \bibinfo
  {author} {\bibfnamefont {A.~E.}\ \bibnamefont {Dréau}}, \bibinfo {author}
  {\bibfnamefont {A.}~\bibnamefont {Reiserer}}, \bibinfo {author}
  {\bibfnamefont {N.}~\bibnamefont {Kalb}}, \bibinfo {author} {\bibfnamefont
  {M.~S.}\ \bibnamefont {Blok}}, \bibinfo {author} {\bibfnamefont
  {J.}~\bibnamefont {Ruitenberg}}, \bibinfo {author} {\bibfnamefont {R.~F.~L.}\
  \bibnamefont {Vermeulen}}, \bibinfo {author} {\bibfnamefont {R.~N.}\
  \bibnamefont {Schouten}}, \bibinfo {author} {\bibfnamefont {C.}~\bibnamefont
  {Abellán}}, \bibinfo {author} {\bibfnamefont {W.}~\bibnamefont {Amaya}},
  \bibinfo {author} {\bibfnamefont {V.}~\bibnamefont {Pruneri}}, \bibinfo
  {author} {\bibfnamefont {M.~W.}\ \bibnamefont {Mitchell}}, \bibinfo {author}
  {\bibfnamefont {M.}~\bibnamefont {Markham}}, \bibinfo {author} {\bibfnamefont
  {D.~J.}\ \bibnamefont {Twitchen}}, \bibinfo {author} {\bibfnamefont
  {D.}~\bibnamefont {Elkouss}}, \bibinfo {author} {\bibfnamefont
  {S.}~\bibnamefont {Wehner}}, \bibinfo {author} {\bibfnamefont {T.~H.}\
  \bibnamefont {Taminiau}}, and\ \bibinfo {author} {\bibfnamefont
  {R.}~\bibnamefont {Hanson}}} (\bibinfo {year} {2015}),\ \bibfield  {title}
  {\enquote {\bibinfo {title} {Loophole-free bell inequality violation using
  electron spins separated by 1.3 kilometres},}\ }\href
  {https://doi.org/10.1038/nature15759} {\bibfield  {journal} {\bibinfo
  {journal} {Nature}\ }\textbf {\bibinfo {volume} {526}},\ \bibinfo {pages}
  {682--686}}\BibitemShut {NoStop}%
\bibitem [{\citenamefont {Hirahara}\ and\ \citenamefont
  {Le~Gall}(2021)}]{hirahara_test_2021}%
  \BibitemOpen
  \bibfield  {author} {\bibinfo {author} {\bibnamefont {Hirahara},
  \bibfnamefont {S}}, and\ \bibinfo {author} {\bibfnamefont {F.}~\bibnamefont
  {Le~Gall}}} (\bibinfo {year} {2021}),\ \bibfield  {title} {\enquote {\bibinfo
  {title} {Test of {{Quantumness}} with {{Small-Depth Quantum Circuits}}},}\
  }\bibfield  {booktitle} {\emph {\bibinfo {booktitle} {46th {{International
  Symposium}} on {{Mathematical Foundations}} of {{Computer Science}} ({{MFCS}}
  2021)}},\ }\href {https://doi.org/10.4230/LIPIcs.MFCS.2021.59} {\ \bibinfo
  {series} {Leibniz {{International Proceedings}} in {{Informatics}}
  ({{LIPIcs}})},\ \textbf {\bibinfo {volume} {202}},\ \bibinfo {pages}
  {59:1--59:15}},\ \Eprint {https://arxiv.org/abs/2105.05500}
  {arXiv:2105.05500} \BibitemShut {NoStop}%
\bibitem [{\citenamefont {Hong}\ \emph {et~al.}(1987)\citenamefont {Hong},
  \citenamefont {Ou},\ and\ \citenamefont {Mandel}}]{hong_measurement_1987}%
  \BibitemOpen
  \bibfield  {author} {\bibinfo {author} {\bibnamefont {Hong}, \bibfnamefont
  {C~K}}, \bibinfo {author} {\bibfnamefont {Z.~Y.}\ \bibnamefont {Ou}}, and\
  \bibinfo {author} {\bibfnamefont {L.}~\bibnamefont {Mandel}}} (\bibinfo
  {year} {1987}),\ \bibfield  {title} {\enquote {\bibinfo {title} {Measurement
  of sub-picosecond time intervals between two photons by interference},}\
  }\href {https://doi.org/10.1103/PhysRevLett.59.2044} {\bibfield  {journal}
  {\bibinfo  {journal} {Phys. Rev. Lett.}\ }\textbf {\bibinfo {volume} {59}},\
  \bibinfo {pages} {2044--2046}}\BibitemShut {NoStop}%
\bibitem [{\citenamefont {Huang}\ \emph
  {et~al.}(2020{\natexlab{a}})\citenamefont {Huang}, \citenamefont {Newman},\
  and\ \citenamefont {Szegedy}}]{huang_explicit_2020}%
  \BibitemOpen
  \bibfield  {author} {\bibinfo {author} {\bibnamefont {Huang}, \bibfnamefont
  {C}}, \bibinfo {author} {\bibfnamefont {M.}~\bibnamefont {Newman}}, and\
  \bibinfo {author} {\bibfnamefont {M.}~\bibnamefont {Szegedy}}} (\bibinfo
  {year} {2020}{\natexlab{a}}),\ \bibfield  {title} {\enquote {\bibinfo {title}
  {Explicit {{lower bounds}} on {{strong quantum simulation}}},}\ }\href
  {https://doi.org/10.1109/TIT.2020.3004427} {\bibfield  {journal} {\bibinfo
  {journal} {IEEE Trans. Inf. Th.}\ }\textbf {\bibinfo {volume} {66}},\
  \bibinfo {pages} {5585--5600}}\BibitemShut {NoStop}%
\bibitem [{\citenamefont {Huang}\ \emph
  {et~al.}(2020{\natexlab{b}})\citenamefont {Huang}, \citenamefont {Zhang},
  \citenamefont {Newman}, \citenamefont {Cai}, \citenamefont {Gao},
  \citenamefont {Tian}, \citenamefont {Wu}, \citenamefont {Xu}, \citenamefont
  {Yu}, \citenamefont {Yuan}, \citenamefont {Szegedy}, \citenamefont {Shi},\
  and\ \citenamefont {Chen}}]{huang_classical_2020}%
  \BibitemOpen
  \bibfield  {author} {\bibinfo {author} {\bibnamefont {Huang}, \bibfnamefont
  {C}}, \bibinfo {author} {\bibfnamefont {F.}~\bibnamefont {Zhang}}, \bibinfo
  {author} {\bibfnamefont {M.}~\bibnamefont {Newman}}, \bibinfo {author}
  {\bibfnamefont {J.}~\bibnamefont {Cai}}, \bibinfo {author} {\bibfnamefont
  {X.}~\bibnamefont {Gao}}, \bibinfo {author} {\bibfnamefont {Z.}~\bibnamefont
  {Tian}}, \bibinfo {author} {\bibfnamefont {J.}~\bibnamefont {Wu}}, \bibinfo
  {author} {\bibfnamefont {H.}~\bibnamefont {Xu}}, \bibinfo {author}
  {\bibfnamefont {H.}~\bibnamefont {Yu}}, \bibinfo {author} {\bibfnamefont
  {B.}~\bibnamefont {Yuan}}, \bibinfo {author} {\bibfnamefont {M.}~\bibnamefont
  {Szegedy}}, \bibinfo {author} {\bibfnamefont {Y.}~\bibnamefont {Shi}}, and\
  \bibinfo {author} {\bibfnamefont {J.}~\bibnamefont {Chen}}} (\bibinfo {year}
  {2020}{\natexlab{b}}),\ \bibfield  {title} {\enquote {\bibinfo {title}
  {Classical {{simulation}} of {{quantum supremacy circuits}}},}\ }\href@noop
  {} {\ }\Eprint {https://arxiv.org/abs/2005.06787} {arXiv:2005.06787}
  \BibitemShut {NoStop}%
\bibitem [{\citenamefont {Huang}\ \emph
  {et~al.}(2021{\natexlab{a}})\citenamefont {Huang}, \citenamefont {Zhang},
  \citenamefont {Newman}, \citenamefont {Ni}, \citenamefont {Ding},
  \citenamefont {Cai}, \citenamefont {Gao}, \citenamefont {Wang}, \citenamefont
  {Wu}, \citenamefont {Zhang}, \citenamefont {Ku}, \citenamefont {Tian},
  \citenamefont {Wu}, \citenamefont {Xu}, \citenamefont {Yu}, \citenamefont
  {Yuan}, \citenamefont {Szegedy}, \citenamefont {Shi}, \citenamefont {Zhao},
  \citenamefont {Deng},\ and\ \citenamefont {Chen}}]{huang_efficient_2021-1}%
  \BibitemOpen
  \bibfield  {author} {\bibinfo {author} {\bibnamefont {Huang}, \bibfnamefont
  {C}}, \bibinfo {author} {\bibfnamefont {F.}~\bibnamefont {Zhang}}, \bibinfo
  {author} {\bibfnamefont {M.}~\bibnamefont {Newman}}, \bibinfo {author}
  {\bibfnamefont {X.}~\bibnamefont {Ni}}, \bibinfo {author} {\bibfnamefont
  {D.}~\bibnamefont {Ding}}, \bibinfo {author} {\bibfnamefont {J.}~\bibnamefont
  {Cai}}, \bibinfo {author} {\bibfnamefont {X.}~\bibnamefont {Gao}}, \bibinfo
  {author} {\bibfnamefont {T.}~\bibnamefont {Wang}}, \bibinfo {author}
  {\bibfnamefont {F.}~\bibnamefont {Wu}}, \bibinfo {author} {\bibfnamefont
  {G.}~\bibnamefont {Zhang}}, \bibinfo {author} {\bibfnamefont {H.-S.}\
  \bibnamefont {Ku}}, \bibinfo {author} {\bibfnamefont {Z.}~\bibnamefont
  {Tian}}, \bibinfo {author} {\bibfnamefont {J.}~\bibnamefont {Wu}}, \bibinfo
  {author} {\bibfnamefont {H.}~\bibnamefont {Xu}}, \bibinfo {author}
  {\bibfnamefont {H.}~\bibnamefont {Yu}}, \bibinfo {author} {\bibfnamefont
  {B.}~\bibnamefont {Yuan}}, \bibinfo {author} {\bibfnamefont {M.}~\bibnamefont
  {Szegedy}}, \bibinfo {author} {\bibfnamefont {Y.}~\bibnamefont {Shi}},
  \bibinfo {author} {\bibfnamefont {H.-H.}\ \bibnamefont {Zhao}}, \bibinfo
  {author} {\bibfnamefont {Ch.}\ \bibnamefont {Deng}}, and\ \bibinfo {author}
  {\bibfnamefont {J.}~\bibnamefont {Chen}}} (\bibinfo {year}
  {2021}{\natexlab{a}}),\ \bibfield  {title} {\enquote {\bibinfo {title}
  {Efficient parallelization of tensor network contraction for simulating
  quantum computation},}\ }\href {https://doi.org/10.1038/s43588-021-00119-7}
  {\bibfield  {journal} {\bibinfo  {journal} {Nature Comp. Sci.}\ }\textbf
  {\bibinfo {volume} {1}},\ \bibinfo {pages} {578--587}}\BibitemShut {NoStop}%
\bibitem [{\citenamefont {Huang}\ \emph
  {et~al.}(2020{\natexlab{c}})\citenamefont {Huang}, \citenamefont {Kueng},\
  and\ \citenamefont {Preskill}}]{huang_predicting_2020}%
  \BibitemOpen
  \bibfield  {author} {\bibinfo {author} {\bibnamefont {Huang}, \bibfnamefont
  {H-Y}}, \bibinfo {author} {\bibfnamefont {R.}~\bibnamefont {Kueng}}, and\
  \bibinfo {author} {\bibfnamefont {J.}~\bibnamefont {Preskill}}} (\bibinfo
  {year} {2020}{\natexlab{c}}),\ \bibfield  {title} {\enquote {\bibinfo {title}
  {Predicting many properties of a quantum system from very few
  measurements},}\ }\href {https://doi.org/10.1038/s41567-020-0932-7}
  {\bibfield  {journal} {\bibinfo  {journal} {Nature Phys.}\ }\textbf {\bibinfo
  {volume} {16}},\ \bibinfo {pages} {1050--1057}}\BibitemShut {NoStop}%
\bibitem [{\citenamefont {Huang}\ \emph
  {et~al.}(2021{\natexlab{b}})\citenamefont {Huang}, \citenamefont {Kok},\ and\
  \citenamefont {Lupo}}]{huang_protecting_2020}%
  \BibitemOpen
  \bibfield  {author} {\bibinfo {author} {\bibnamefont {Huang}, \bibfnamefont
  {Z}}, \bibinfo {author} {\bibfnamefont {P.}~\bibnamefont {Kok}}, and\
  \bibinfo {author} {\bibfnamefont {C.}~\bibnamefont {Lupo}}} (\bibinfo {year}
  {2021}{\natexlab{b}}),\ \bibfield  {title} {\enquote {\bibinfo {title}
  {Fault-tolerant quantum data locking},}\ }\href
  {https://doi.org/10.1103/PhysRevA.103.052611} {\bibfield  {journal} {\bibinfo
   {journal} {Phys. Rev. A}\ }\textbf {\bibinfo {volume} {103}},\ \bibinfo
  {pages} {052611}}\BibitemShut {NoStop}%
\bibitem [{\citenamefont {Huang}\ \emph
  {et~al.}(2021{\natexlab{c}})\citenamefont {Huang}, \citenamefont {Rohde},
  \citenamefont {Berry}, \citenamefont {Kok}, \citenamefont {Dowling},\ and\
  \citenamefont {Lupo}}]{huang_boson_2019}%
  \BibitemOpen
  \bibfield  {author} {\bibinfo {author} {\bibnamefont {Huang}, \bibfnamefont
  {Z}}, \bibinfo {author} {\bibfnamefont {P.~P.}\ \bibnamefont {Rohde}},
  \bibinfo {author} {\bibfnamefont {D.~W.}\ \bibnamefont {Berry}}, \bibinfo
  {author} {\bibfnamefont {P.}~\bibnamefont {Kok}}, \bibinfo {author}
  {\bibfnamefont {J.~P.}\ \bibnamefont {Dowling}}, and\ \bibinfo {author}
  {\bibfnamefont {C.}~\bibnamefont {Lupo}}} (\bibinfo {year}
  {2021}{\natexlab{c}}),\ \bibfield  {title} {\enquote {\bibinfo {title}
  {Photonic quantum data locking},}\ }\href
  {https://doi.org/10.22331/q-2021-04-28-447} {\bibfield  {journal} {\bibinfo
  {journal} {Quantum}\ }\textbf {\bibinfo {volume} {5}},\ \bibinfo {pages}
  {447}}\BibitemShut {NoStop}%
\bibitem [{\citenamefont {Hudson}\ and\ \citenamefont
  {Moody}(1976)}]{hudson_locally_1976}%
  \BibitemOpen
  \bibfield  {author} {\bibinfo {author} {\bibnamefont {Hudson}, \bibfnamefont
  {R~L}}, and\ \bibinfo {author} {\bibfnamefont {G.~R.}\ \bibnamefont {Moody}}}
  (\bibinfo {year} {1976}),\ \bibfield  {title} {\enquote {\bibinfo {title}
  {{Locally normal symmetric states and an analogue of de Finetti's
  theorem}},}\ }\href {https://doi.org/10.1007/BF00534784} {\bibfield
  {journal} {\bibinfo  {journal} {Z. Wahrsch. verw. Geb.}\ }\textbf {\bibinfo
  {volume} {33}},\ \bibinfo {pages} {343--351}}\BibitemShut {NoStop}%
\bibitem [{\citenamefont {Huh}(2022)}]{huh_fast_2022}%
  \BibitemOpen
  \bibfield  {author} {\bibinfo {author} {\bibnamefont {Huh}, \bibfnamefont
  {J}}} (\bibinfo {year} {2022}),\ \bibfield  {title} {\enquote {\bibinfo
  {title} {A fast quantum algorithm for computing matrix permanent},}\
  }\href@noop {} {\ }\Eprint {https://arxiv.org/abs/2205.01328v2}
  {arXiv:2205.01328v2} \BibitemShut {NoStop}%
\bibitem [{\citenamefont {Huh}\ \emph {et~al.}(2015)\citenamefont {Huh},
  \citenamefont {Guerreschi}, \citenamefont {Peropadre}, \citenamefont
  {McClean},\ and\ \citenamefont {{Aspuru-Guzik}}}]{huh_boson_2015}%
  \BibitemOpen
  \bibfield  {author} {\bibinfo {author} {\bibnamefont {Huh}, \bibfnamefont
  {J}}, \bibinfo {author} {\bibfnamefont {G.~G.}\ \bibnamefont {Guerreschi}},
  \bibinfo {author} {\bibfnamefont {B.}~\bibnamefont {Peropadre}}, \bibinfo
  {author} {\bibfnamefont {J.~R.}\ \bibnamefont {McClean}}, and\ \bibinfo
  {author} {\bibfnamefont {A.}~\bibnamefont {{Aspuru-Guzik}}}} (\bibinfo {year}
  {2015}),\ \bibfield  {title} {\enquote {\bibinfo {title} {Boson sampling for
  molecular vibronic spectra},}\ }\href
  {https://doi.org/10.1038/nphoton.2015.153} {\bibfield  {journal} {\bibinfo
  {journal} {Nature Phot.}\ }\textbf {\bibinfo {volume} {9}},\ \bibinfo {pages}
  {615--620}}\BibitemShut {NoStop}%
\bibitem [{\citenamefont {{Hunter-Jones}}(2019)}]{hunter-jones_unitary_2019}%
  \BibitemOpen
  \bibfield  {author} {\bibinfo {author} {\bibnamefont {{Hunter-Jones}},
  \bibfnamefont {N}}} (\bibinfo {year} {2019}),\ \bibfield  {title} {\enquote
  {\bibinfo {title} {Unitary designs from statistical mechanics in random
  quantum circuits},}\ }\href@noop {} {\ }\Eprint
  {https://arxiv.org/abs/1905.12053} {arXiv:1905.12053} \BibitemShut {NoStop}%
\bibitem [{\citenamefont {Jahangiri}\ \emph {et~al.}(2020)\citenamefont
  {Jahangiri}, \citenamefont {Arrazola}, \citenamefont {Quesada},\ and\
  \citenamefont {Killoran}}]{PhysRevE.101.022134}%
  \BibitemOpen
  \bibfield  {author} {\bibinfo {author} {\bibnamefont {Jahangiri},
  \bibfnamefont {S}}, \bibinfo {author} {\bibfnamefont {J.~M.}\ \bibnamefont
  {Arrazola}}, \bibinfo {author} {\bibfnamefont {N.}~\bibnamefont {Quesada}},
  and\ \bibinfo {author} {\bibfnamefont {N.}~\bibnamefont {Killoran}}}
  (\bibinfo {year} {2020}),\ \bibfield  {title} {\enquote {\bibinfo {title}
  {{Point processes with Gaussian boson sampling}},}\ }\href
  {https://doi.org/10.1103/PhysRevE.101.022134} {\bibfield  {journal} {\bibinfo
   {journal} {Phys. Rev. E}\ }\textbf {\bibinfo {volume} {101}},\ \bibinfo
  {pages} {022134}}\BibitemShut {NoStop}%
\bibitem [{\citenamefont {Jaksch}\ \emph {et~al.}(1998)\citenamefont {Jaksch},
  \citenamefont {Bruder}, \citenamefont {Cirac}, \citenamefont {Gardiner},\
  and\ \citenamefont {Zoller}}]{jaksch_cold_1998}%
  \BibitemOpen
  \bibfield  {author} {\bibinfo {author} {\bibnamefont {Jaksch}, \bibfnamefont
  {D}}, \bibinfo {author} {\bibfnamefont {C.}~\bibnamefont {Bruder}}, \bibinfo
  {author} {\bibfnamefont {J.~I.}\ \bibnamefont {Cirac}}, \bibinfo {author}
  {\bibfnamefont {C.~W.}\ \bibnamefont {Gardiner}}, and\ \bibinfo {author}
  {\bibfnamefont {P.}~\bibnamefont {Zoller}}} (\bibinfo {year} {1998}),\
  \bibfield  {title} {\enquote {\bibinfo {title} {Cold bosonic atoms in optical
  lattices},}\ }\href {https://doi.org/10.1103/PhysRevLett.81.3108} {\bibfield
  {journal} {\bibinfo  {journal} {Phys. Rev. Lett.}\ }\textbf {\bibinfo
  {volume} {81}},\ \bibinfo {pages} {3108}}\BibitemShut {NoStop}%
\bibitem [{\citenamefont {Jerrum}\ and\ \citenamefont
  {Sinclair}(1993)}]{jerrum_polynomial-time_1990}%
  \BibitemOpen
  \bibfield  {author} {\bibinfo {author} {\bibnamefont {Jerrum}, \bibfnamefont
  {M~R}}, and\ \bibinfo {author} {\bibfnamefont {A.}~\bibnamefont {Sinclair}}}
  (\bibinfo {year} {1993}),\ \bibfield  {title} {\enquote {\bibinfo {title}
  {{Polynomial-time approximation algorithms for the Ising model}},}\ }\href
  {https://doi.org/10.1137/0222066} {\bibfield  {journal} {\bibinfo  {journal}
  {SIAM J. Comp.}\ }\textbf {\bibinfo {volume} {22}},\ \bibinfo {pages}
  {1087--1116}}\BibitemShut {NoStop}%
\bibitem [{\citenamefont {Jerrum}\ \emph {et~al.}(2004)\citenamefont {Jerrum},
  \citenamefont {Sinclair},\ and\ \citenamefont
  {Vigoda}}]{jerrum_polynomial-time_2004}%
  \BibitemOpen
  \bibfield  {author} {\bibinfo {author} {\bibnamefont {Jerrum}, \bibfnamefont
  {M~R}}, \bibinfo {author} {\bibfnamefont {A.}~\bibnamefont {Sinclair}}, and\
  \bibinfo {author} {\bibfnamefont {E.}~\bibnamefont {Vigoda}}} (\bibinfo
  {year} {2004}),\ \bibfield  {title} {\enquote {\bibinfo {title} {A
  polynomial-time approximation algorithm for the permanent of a matrix with
  non-negative entries},}\ }\href {https://doi.org/10.1145/1008731.1008738}
  {\bibfield  {journal} {\bibinfo  {journal} {J. {ACM}}\ }\textbf {\bibinfo
  {volume} {51}},\ \bibinfo {pages} {671--697}}\BibitemShut {NoStop}%
\bibitem [{\citenamefont {Jerrum}\ \emph {et~al.}(1986)\citenamefont {Jerrum},
  \citenamefont {Valiant},\ and\ \citenamefont
  {Vazirani}}]{jerrum_random_1986}%
  \BibitemOpen
  \bibfield  {author} {\bibinfo {author} {\bibnamefont {Jerrum}, \bibfnamefont
  {M~R}}, \bibinfo {author} {\bibfnamefont {L.~G.}\ \bibnamefont {Valiant}},
  and\ \bibinfo {author} {\bibfnamefont {V.~V.}\ \bibnamefont {Vazirani}}}
  (\bibinfo {year} {1986}),\ \bibfield  {title} {\enquote {\bibinfo {title}
  {Random generation of combinatorial structures from a uniform
  distribution},}\ }\href {https://doi.org/10.1016/0304-3975(86)90174-X}
  {\bibfield  {journal} {\bibinfo  {journal} {Th. Comp. Sc.}\ }\textbf
  {\bibinfo {volume} {43}},\ \bibinfo {pages} {169--188}}\BibitemShut {NoStop}%
\bibitem [{\citenamefont {Jian}\ \emph {et~al.}(2020)\citenamefont {Jian},
  \citenamefont {You}, \citenamefont {Vasseur},\ and\ \citenamefont
  {Ludwig}}]{jian_measurement-induced_2019}%
  \BibitemOpen
  \bibfield  {author} {\bibinfo {author} {\bibnamefont {Jian}, \bibfnamefont
  {C-M}}, \bibinfo {author} {\bibfnamefont {Y.-Z.}\ \bibnamefont {You}},
  \bibinfo {author} {\bibfnamefont {R.}~\bibnamefont {Vasseur}}, and\ \bibinfo
  {author} {\bibfnamefont {A.~W.~W.}\ \bibnamefont {Ludwig}}} (\bibinfo {year}
  {2020}),\ \bibfield  {title} {\enquote {\bibinfo {title} {Measurement-induced
  criticality in random quantum circuits},}\ }\href
  {https://doi.org/10.1103/PhysRevB.101.104302} {\bibfield  {journal} {\bibinfo
   {journal} {Phys. Rev. B}\ }\textbf {\bibinfo {volume} {101}},\ \bibinfo
  {pages} {104302}}\BibitemShut {NoStop}%
\bibitem [{\citenamefont {Jiang}(2006)}]{jiang_how_2006}%
  \BibitemOpen
  \bibfield  {author} {\bibinfo {author} {\bibnamefont {Jiang}, \bibfnamefont
  {T}}} (\bibinfo {year} {2006}),\ \bibfield  {title} {\enquote {\bibinfo
  {title} {How many entries of a typical orthogonal matrix can be approximated
  by independent normals?}}\ }\href
  {https://doi.org/10.1214/009117906000000205} {\bibfield  {journal} {\bibinfo
  {journal} {Ann. Probab.}\ }\textbf {\bibinfo {volume} {34}},\ \bibinfo
  {pages} {1497--1529}}\BibitemShut {NoStop}%
\bibitem [{\citenamefont {Jiang}(2009)}]{jiang_entries_2009}%
  \BibitemOpen
  \bibfield  {author} {\bibinfo {author} {\bibnamefont {Jiang}, \bibfnamefont
  {T}}} (\bibinfo {year} {2009}),\ \bibfield  {title} {\enquote {\bibinfo
  {title} {The entries of circular orthogonal ensembles},}\ }\href
  {https://doi.org/10.1063/1.3152217} {\bibfield  {journal} {\bibinfo
  {journal} {J. Math. Phys.}\ }\textbf {\bibinfo {volume} {50}},\ \bibinfo
  {pages} {063302}}\BibitemShut {NoStop}%
\bibitem [{\citenamefont {Jnane}\ \emph {et~al.}(2021)\citenamefont {Jnane},
  \citenamefont {Sawaya}, \citenamefont {Peropadre}, \citenamefont
  {Aspuru-Guzik}, \citenamefont {Garcia-Patron},\ and\ \citenamefont
  {Huh}}]{AnalogSimulation}%
  \BibitemOpen
  \bibfield  {author} {\bibinfo {author} {\bibnamefont {Jnane}, \bibfnamefont
  {H}}, \bibinfo {author} {\bibfnamefont {N.~P.~D.}\ \bibnamefont {Sawaya}},
  \bibinfo {author} {\bibfnamefont {B.}~\bibnamefont {Peropadre}}, \bibinfo
  {author} {\bibfnamefont {A.}~\bibnamefont {Aspuru-Guzik}}, \bibinfo {author}
  {\bibfnamefont {R.}~\bibnamefont {Garcia-Patron}}, and\ \bibinfo {author}
  {\bibfnamefont {J.}~\bibnamefont {Huh}}} (\bibinfo {year} {2021}),\ \bibfield
   {title} {\enquote {\bibinfo {title} {Analog quantum simulation of non-condon
  effects in molecular spectroscopy},}\ }\href
  {https://doi.org/10.1021/acsphotonics.1c00059} {\bibfield  {journal}
  {\bibinfo  {journal} {ACS Phot.}\ }\textbf {\bibinfo {volume} {8}},\ \bibinfo
  {pages} {2007--2016}}\BibitemShut {NoStop}%
\bibitem [{\citenamefont {Jurcevic}\ \emph {et~al.}(2021)\citenamefont
  {Jurcevic}, \citenamefont {{Javadi-Abhari}}, \citenamefont {Bishop},
  \citenamefont {Lauer}, \citenamefont {Bogorin}, \citenamefont {Brink},
  \citenamefont {Capelluto}, \citenamefont {G{\"u}nl{\"u}k}, \citenamefont
  {Itoko}, \citenamefont {Kanazawa}, \citenamefont {Kandala}, \citenamefont
  {Keefe}, \citenamefont {Krsulich}, \citenamefont {Landers}, \citenamefont
  {Lewandowski}, \citenamefont {McClure}, \citenamefont {Nannicini},
  \citenamefont {Narasgond}, \citenamefont {Nayfeh}, \citenamefont {Pritchett},
  \citenamefont {Rothwell}, \citenamefont {Srinivasan}, \citenamefont
  {Sundaresan}, \citenamefont {Wang}, \citenamefont {Wei}, \citenamefont
  {Wood}, \citenamefont {Yau}, \citenamefont {Zhang}, \citenamefont {Dial},
  \citenamefont {Chow},\ and\ \citenamefont
  {Gambetta}}]{jurcevic_demonstration_2021}%
  \BibitemOpen
  \bibfield  {author} {\bibinfo {author} {\bibnamefont {Jurcevic},
  \bibfnamefont {P}}, \bibinfo {author} {\bibfnamefont {A.}~\bibnamefont
  {{Javadi-Abhari}}}, \bibinfo {author} {\bibfnamefont {L.~S.}\ \bibnamefont
  {Bishop}}, \bibinfo {author} {\bibfnamefont {I.}~\bibnamefont {Lauer}},
  \bibinfo {author} {\bibfnamefont {D.~F.}\ \bibnamefont {Bogorin}}, \bibinfo
  {author} {\bibfnamefont {M.}~\bibnamefont {Brink}}, \bibinfo {author}
  {\bibfnamefont {L.}~\bibnamefont {Capelluto}}, \bibinfo {author}
  {\bibfnamefont {O.}~\bibnamefont {G{\"u}nl{\"u}k}}, \bibinfo {author}
  {\bibfnamefont {T.}~\bibnamefont {Itoko}}, \bibinfo {author} {\bibfnamefont
  {N.}~\bibnamefont {Kanazawa}}, \bibinfo {author} {\bibfnamefont
  {A.}~\bibnamefont {Kandala}}, \bibinfo {author} {\bibfnamefont {G.~A.}\
  \bibnamefont {Keefe}}, \bibinfo {author} {\bibfnamefont {K.}~\bibnamefont
  {Krsulich}}, \bibinfo {author} {\bibfnamefont {W.}~\bibnamefont {Landers}},
  \bibinfo {author} {\bibfnamefont {E.~P.}\ \bibnamefont {Lewandowski}},
  \bibinfo {author} {\bibfnamefont {D.~T.}\ \bibnamefont {McClure}}, \bibinfo
  {author} {\bibfnamefont {G.}~\bibnamefont {Nannicini}}, \bibinfo {author}
  {\bibfnamefont {A.}~\bibnamefont {Narasgond}}, \bibinfo {author}
  {\bibfnamefont {H.~M.}\ \bibnamefont {Nayfeh}}, \bibinfo {author}
  {\bibfnamefont {E.}~\bibnamefont {Pritchett}}, \bibinfo {author}
  {\bibfnamefont {M.~B.}\ \bibnamefont {Rothwell}}, \bibinfo {author}
  {\bibfnamefont {S.}~\bibnamefont {Srinivasan}}, \bibinfo {author}
  {\bibfnamefont {N.}~\bibnamefont {Sundaresan}}, \bibinfo {author}
  {\bibfnamefont {C.}~\bibnamefont {Wang}}, \bibinfo {author} {\bibfnamefont
  {K.~X.}\ \bibnamefont {Wei}}, \bibinfo {author} {\bibfnamefont {C.~J.}\
  \bibnamefont {Wood}}, \bibinfo {author} {\bibfnamefont {J.-B.}\ \bibnamefont
  {Yau}}, \bibinfo {author} {\bibfnamefont {E.~J.}\ \bibnamefont {Zhang}},
  \bibinfo {author} {\bibfnamefont {O.~E.}\ \bibnamefont {Dial}}, \bibinfo
  {author} {\bibfnamefont {J.~M.}\ \bibnamefont {Chow}}, and\ \bibinfo {author}
  {\bibfnamefont {J.~M.}\ \bibnamefont {Gambetta}}} (\bibinfo {year} {2021}),\
  \bibfield  {title} {\enquote {\bibinfo {title} {Demonstration of quantum
  volume 64 on a superconducting quantum computing system},}\ }\href
  {https://doi.org/10.1088/2058-9565/abe519} {\bibfield  {journal} {\bibinfo
  {journal} {Quantum Sci. Technol.}\ }\textbf {\bibinfo {volume} {6}},\
  \bibinfo {pages} {025020}}\BibitemShut {NoStop}%
\bibitem [{\citenamefont
  {Kahanamoku-Meyer}(2019)}]{kahanamoku-meyer_forging_2019}%
  \BibitemOpen
  \bibfield  {author} {\bibinfo {author} {\bibnamefont {Kahanamoku-Meyer},
  \bibfnamefont {G~D}}} (\bibinfo {year} {2019}),\ \bibfield  {title} {\enquote
  {\bibinfo {title} {Forging quantum data: classically defeating an {IQP}-based
  quantum test},}\ }\href@noop {} {\ }\Eprint
  {https://arxiv.org/abs/1912.05547} {arxiv:1912.05547} \BibitemShut {NoStop}%
\bibitem [{\citenamefont {{Kahanamoku-Meyer}}\ \emph
  {et~al.}(2022)\citenamefont {{Kahanamoku-Meyer}}, \citenamefont {Choi},
  \citenamefont {Vazirani},\ and\ \citenamefont
  {Yao}}]{kahanamoku-meyer_classically-verifiable_2021}%
  \BibitemOpen
  \bibfield  {author} {\bibinfo {author} {\bibnamefont {{Kahanamoku-Meyer}},
  \bibfnamefont {G~D}}, \bibinfo {author} {\bibfnamefont {S.}~\bibnamefont
  {Choi}}, \bibinfo {author} {\bibfnamefont {U.~V.}\ \bibnamefont {Vazirani}},
  and\ \bibinfo {author} {\bibfnamefont {N.~Y.}\ \bibnamefont {Yao}}} (\bibinfo
  {year} {2022}),\ \bibfield  {title} {\enquote {\bibinfo {title} {Classically
  verifiable quantum advantage from a computational {{Bell}} test},}\ }\href
  {https://doi.org/10.1038/s41567-022-01643-7} {\bibfield  {journal} {\bibinfo
  {journal} {Nature Phys.}\ }\textbf {\bibinfo {volume} {18}},\ \bibinfo
  {pages} {918--924}}\BibitemShut {NoStop}%
\bibitem [{\citenamefont {Kalachev}\ \emph
  {et~al.}(2021{\natexlab{a}})\citenamefont {Kalachev}, \citenamefont
  {Panteleev},\ and\ \citenamefont {Yung}}]{kalachev_recursive_2021}%
  \BibitemOpen
  \bibfield  {author} {\bibinfo {author} {\bibnamefont {Kalachev},
  \bibfnamefont {G}}, \bibinfo {author} {\bibfnamefont {P.}~\bibnamefont
  {Panteleev}}, and\ \bibinfo {author} {\bibfnamefont {M.-H.}\ \bibnamefont
  {Yung}}} (\bibinfo {year} {2021}{\natexlab{a}}),\ \bibfield  {title}
  {\enquote {\bibinfo {title} {Recursive {{multi-tensor contraction}} for {{XEB
  verification}} of {{quantum circuits}}},}\ }\href@noop {} {\ }\Eprint
  {https://arxiv.org/abs/2108.05665} {arXiv:2108.05665} \BibitemShut {NoStop}%
\bibitem [{\citenamefont {Kalachev}\ \emph
  {et~al.}(2021{\natexlab{b}})\citenamefont {Kalachev}, \citenamefont
  {Panteleev}, \citenamefont {Zhou},\ and\ \citenamefont
  {Yung}}]{kalachev_classical_2021}%
  \BibitemOpen
  \bibfield  {author} {\bibinfo {author} {\bibnamefont {Kalachev},
  \bibfnamefont {G}}, \bibinfo {author} {\bibfnamefont {P.}~\bibnamefont
  {Panteleev}}, \bibinfo {author} {\bibfnamefont {P.-F.}\ \bibnamefont {Zhou}},
  and\ \bibinfo {author} {\bibfnamefont {M.-H.}\ \bibnamefont {Yung}}}
  (\bibinfo {year} {2021}{\natexlab{b}}),\ \href@noop {} {\enquote {\bibinfo
  {title} {Classical {{sampling}} of {{random quantum circuits}} with {{bounded
  fidelity}}},}\ }\Eprint {https://arxiv.org/abs/2112.15083} {arXiv:2112.15083}
  \BibitemShut {NoStop}%
\bibitem [{\citenamefont {Kalai}\ and\ \citenamefont
  {Kindler}(2014)}]{kalai_gaussian_2014}%
  \BibitemOpen
  \bibfield  {author} {\bibinfo {author} {\bibnamefont {Kalai}, \bibfnamefont
  {G}}, and\ \bibinfo {author} {\bibfnamefont {G.}~\bibnamefont {Kindler}}}
  (\bibinfo {year} {2014}),\ \bibfield  {title} {\enquote {\bibinfo {title}
  {Gaussian {{noise sensitivity}} and {{BosonSampling}}},}\ }\href@noop {} {\
  }\Eprint {https://arxiv.org/abs/1409.3093} {arXiv:1409.3093} \BibitemShut
  {NoStop}%
\bibitem [{\citenamefont {Kalev}\ \emph {et~al.}(2019)\citenamefont {Kalev},
  \citenamefont {Kyrillidis},\ and\ \citenamefont
  {Linke}}]{kalev_validating_2019}%
  \BibitemOpen
  \bibfield  {author} {\bibinfo {author} {\bibnamefont {Kalev}, \bibfnamefont
  {A}}, \bibinfo {author} {\bibfnamefont {A.}~\bibnamefont {Kyrillidis}}, and\
  \bibinfo {author} {\bibfnamefont {N.~M.}\ \bibnamefont {Linke}}} (\bibinfo
  {year} {2019}),\ \bibfield  {title} {\enquote {\bibinfo {title} {Validating
  and certifying stabilizer states},}\ }\href
  {https://doi.org/10.1103/PhysRevA.99.042337} {\bibfield  {journal} {\bibinfo
  {journal} {Phys. Rev. A}\ }\textbf {\bibinfo {volume} {99}},\ \bibinfo
  {pages} {042337}}\BibitemShut {NoStop}%
\bibitem [{\citenamefont {Kane}\ \emph {et~al.}(2017)\citenamefont {Kane},
  \citenamefont {Karmalkar},\ and\ \citenamefont {Price}}]{kane_robust_2017}%
  \BibitemOpen
  \bibfield  {author} {\bibinfo {author} {\bibnamefont {Kane}, \bibfnamefont
  {D}}, \bibinfo {author} {\bibfnamefont {S.}~\bibnamefont {Karmalkar}}, and\
  \bibinfo {author} {\bibfnamefont {E.}~\bibnamefont {Price}}} (\bibinfo {year}
  {2017}),\ \bibfield  {title} {\enquote {\bibinfo {title} {Robust {{Polynomial
  Regression}} up to the {{information theoretic limit}}},}\ }in\ \href
  {https://doi.org/10.1109/FOCS.2017.43} {\emph {\bibinfo {booktitle} {2017
  {{IEEE}} 58th {{Annual Symposium}} on {{Foundations}} of {{Computer Science}}
  ({{FOCS}})}}}\ (\bibinfo  {publisher} {{IEEE}},\ \bibinfo {address}
  {{Berkeley, CA}})\ pp.\ \bibinfo {pages} {391--402}\BibitemShut {NoStop}%
\bibitem [{\citenamefont {Kapourniotis}\ and\ \citenamefont
  {Datta}(2019)}]{kapourniotis_nonadaptive_2019}%
  \BibitemOpen
  \bibfield  {author} {\bibinfo {author} {\bibnamefont {Kapourniotis},
  \bibfnamefont {T}}, and\ \bibinfo {author} {\bibfnamefont {A.}~\bibnamefont
  {Datta}}} (\bibinfo {year} {2019}),\ \bibfield  {title} {\enquote {\bibinfo
  {title} {Nonadaptive fault-tolerant verification of quantum supremacy with
  noise},}\ }\href {https://doi.org/10.22331/q-2019-07-12-164} {\bibfield
  {journal} {\bibinfo  {journal} {Quantum}\ }\textbf {\bibinfo {volume} {3}},\
  \bibinfo {pages} {164}}\BibitemShut {NoStop}%
\bibitem [{\citenamefont {Kiesel}\ \emph {et~al.}(2005)\citenamefont {Kiesel},
  \citenamefont {Schmid}, \citenamefont {Weber}, \citenamefont {T{\'o}th},
  \citenamefont {G{\"u}hne}, \citenamefont {Ursin},\ and\ \citenamefont
  {Weinfurter}}]{kiesel_experimental_2005}%
  \BibitemOpen
  \bibfield  {author} {\bibinfo {author} {\bibnamefont {Kiesel}, \bibfnamefont
  {N}}, \bibinfo {author} {\bibfnamefont {C.}~\bibnamefont {Schmid}}, \bibinfo
  {author} {\bibfnamefont {U.}~\bibnamefont {Weber}}, \bibinfo {author}
  {\bibfnamefont {G.}~\bibnamefont {T{\'o}th}}, \bibinfo {author}
  {\bibfnamefont {O.}~\bibnamefont {G{\"u}hne}}, \bibinfo {author}
  {\bibfnamefont {R.}~\bibnamefont {Ursin}}, and\ \bibinfo {author}
  {\bibfnamefont {H.}~\bibnamefont {Weinfurter}}} (\bibinfo {year} {2005}),\
  \bibfield  {title} {\enquote {\bibinfo {title} {Experimental {analysis} of a
  {four}-{qubit} {photon} {cluster} {state}},}\ }\href
  {https://doi.org/10.1103/PhysRevLett.95.210502} {\bibfield  {journal}
  {\bibinfo  {journal} {Phys. Rev. Lett.}\ }\textbf {\bibinfo {volume} {95}},\
  \bibinfo {pages} {210502}}\BibitemShut {NoStop}%
\bibitem [{\citenamefont {Kliesch}\ and\ \citenamefont
  {Roth}(2021)}]{kliesch_theory_2021}%
  \BibitemOpen
  \bibfield  {author} {\bibinfo {author} {\bibnamefont {Kliesch}, \bibfnamefont
  {M}}, and\ \bibinfo {author} {\bibfnamefont {I.}~\bibnamefont {Roth}}}
  (\bibinfo {year} {2021}),\ \bibfield  {title} {\enquote {\bibinfo {title}
  {Theory of {{quantum system certification}}},}\ }\href
  {https://doi.org/10.1103/PRXQuantum.2.010201} {\bibfield  {journal} {\bibinfo
   {journal} {PRX Quantum}\ }\textbf {\bibinfo {volume} {2}},\ \bibinfo {pages}
  {010201}}\BibitemShut {NoStop}%
\bibitem [{\citenamefont {Kok}\ and\ \citenamefont
  {Lovett}(2010)}]{kok_introduction_2010}%
  \BibitemOpen
  \bibfield  {author} {\bibinfo {author} {\bibnamefont {Kok}, \bibfnamefont
  {P}}, and\ \bibinfo {author} {\bibfnamefont {B.~W.}\ \bibnamefont {Lovett}}}
  (\bibinfo {year} {2010}),\ \href@noop {} {\emph {\bibinfo {title}
  {Introduction to optical quantum information processing}}}\ (\bibinfo
  {publisher} {{Cambridge University Press}},\ \bibinfo {address} {{Cambridge;
  New York}})\BibitemShut {NoStop}%
\bibitem [{\citenamefont {Kok}\ \emph {et~al.}(2007)\citenamefont {Kok},
  \citenamefont {Munro}, \citenamefont {Nemoto}, \citenamefont {Ralph},
  \citenamefont {Dowling},\ and\ \citenamefont {Milburn}}]{kok_linear_2007}%
  \BibitemOpen
  \bibfield  {author} {\bibinfo {author} {\bibnamefont {Kok}, \bibfnamefont
  {P}}, \bibinfo {author} {\bibfnamefont {W.~J.}\ \bibnamefont {Munro}},
  \bibinfo {author} {\bibfnamefont {K.}~\bibnamefont {Nemoto}}, \bibinfo
  {author} {\bibfnamefont {T.~C.}\ \bibnamefont {Ralph}}, \bibinfo {author}
  {\bibfnamefont {J.~P.}\ \bibnamefont {Dowling}}, and\ \bibinfo {author}
  {\bibfnamefont {G.~J.}\ \bibnamefont {Milburn}}} (\bibinfo {year} {2007}),\
  \bibfield  {title} {\enquote {\bibinfo {title} {Linear optical quantum
  computing with photonic qubits},}\ }\href
  {https://doi.org/10.1103/RevModPhys.79.135} {\bibfield  {journal} {\bibinfo
  {journal} {Rev. Mod. Phys.}\ }\textbf {\bibinfo {volume} {79}},\ \bibinfo
  {pages} {135--174}}\BibitemShut {NoStop}%
\bibitem [{\citenamefont {Kondo}\ \emph {et~al.}(2022)\citenamefont {Kondo},
  \citenamefont {Mori},\ and\ \citenamefont
  {Movassagh}}]{kondo_fine-grained_2021}%
  \BibitemOpen
  \bibfield  {author} {\bibinfo {author} {\bibnamefont {Kondo}, \bibfnamefont
  {Y}}, \bibinfo {author} {\bibfnamefont {R.}~\bibnamefont {Mori}}, and\
  \bibinfo {author} {\bibfnamefont {R.}~\bibnamefont {Movassagh}}} (\bibinfo
  {year} {2022}),\ \bibfield  {title} {\enquote {\bibinfo {title} {Quantum
  supremacy and hardness of estimating output probabilities of quantum
  circuits},}\ }in\ \href {https://doi.org/10.1109/FOCS52979.2021.00126} {\emph
  {\bibinfo {booktitle} {2021 {{IEEE}} 62nd {{Annual Symposium}} on
  {{Foundations}} of {{Computer Science}} ({{FOCS}})}}}\ (\bibinfo  {publisher}
  {{IEEE}},\ \bibinfo {address} {{Berkeley, CA}})\ pp.\ \bibinfo {pages}
  {1296--1307}\BibitemShut {NoStop}%
\bibitem [{\citenamefont {K{\"o}nig}\ and\ \citenamefont
  {Renner}(2005)}]{konig_finetti_2005}%
  \BibitemOpen
  \bibfield  {author} {\bibinfo {author} {\bibnamefont {K{\"o}nig},
  \bibfnamefont {R}}, and\ \bibinfo {author} {\bibfnamefont {R.}~\bibnamefont
  {Renner}}} (\bibinfo {year} {2005}),\ \bibfield  {title} {\enquote {\bibinfo
  {title} {{A de Finetti representation for finite symmetric quantum
  states}},}\ }\href {https://doi.org/10.1063/1.2146188} {\bibfield  {journal}
  {\bibinfo  {journal} {J. Math. Phys.}\ }\textbf {\bibinfo {volume} {46}},\
  \bibinfo {pages} {122108}}\BibitemShut {NoStop}%
\bibitem [{\citenamefont {Krinner}\ \emph {et~al.}(2022)\citenamefont
  {Krinner}, \citenamefont {Lacroix}, \citenamefont {Remm}, \citenamefont
  {Di~Paolo}, \citenamefont {Genois}, \citenamefont {Leroux}, \citenamefont
  {Hellings}, \citenamefont {Lazar}, \citenamefont {Swiadek}, \citenamefont
  {Herrmann}, \citenamefont {Norris}, \citenamefont {Andersen}, \citenamefont
  {M{\"u}ller}, \citenamefont {Blais}, \citenamefont {Eichler},\ and\
  \citenamefont {Wallraff}}]{krinner_realizing_2022}%
  \BibitemOpen
  \bibfield  {author} {\bibinfo {author} {\bibnamefont {Krinner}, \bibfnamefont
  {S}}, \bibinfo {author} {\bibfnamefont {N.}~\bibnamefont {Lacroix}}, \bibinfo
  {author} {\bibfnamefont {A.}~\bibnamefont {Remm}}, \bibinfo {author}
  {\bibfnamefont {A.}~\bibnamefont {Di~Paolo}}, \bibinfo {author}
  {\bibfnamefont {E.}~\bibnamefont {Genois}}, \bibinfo {author} {\bibfnamefont
  {C.}~\bibnamefont {Leroux}}, \bibinfo {author} {\bibfnamefont
  {C.}~\bibnamefont {Hellings}}, \bibinfo {author} {\bibfnamefont
  {S.}~\bibnamefont {Lazar}}, \bibinfo {author} {\bibfnamefont
  {F.}~\bibnamefont {Swiadek}}, \bibinfo {author} {\bibfnamefont
  {J.}~\bibnamefont {Herrmann}}, \bibinfo {author} {\bibfnamefont {G.~J.}\
  \bibnamefont {Norris}}, \bibinfo {author} {\bibfnamefont {C.~K.}\
  \bibnamefont {Andersen}}, \bibinfo {author} {\bibfnamefont {M.}~\bibnamefont
  {M{\"u}ller}}, \bibinfo {author} {\bibfnamefont {A.}~\bibnamefont {Blais}},
  \bibinfo {author} {\bibfnamefont {C.}~\bibnamefont {Eichler}}, and\ \bibinfo
  {author} {\bibfnamefont {A.}~\bibnamefont {Wallraff}}} (\bibinfo {year}
  {2022}),\ \bibfield  {title} {\enquote {\bibinfo {title} {Realizing repeated
  quantum error correction in a distance-three surface code},}\ }\href
  {https://doi.org/10.1038/s41586-022-04566-8} {\bibfield  {journal} {\bibinfo
  {journal} {Nature}\ }\textbf {\bibinfo {volume} {605}},\ \bibinfo {pages}
  {669--674}}\BibitemShut {NoStop}%
\bibitem [{\citenamefont {Krovi}(2022)}]{krovi_average-case_2022}%
  \BibitemOpen
  \bibfield  {author} {\bibinfo {author} {\bibnamefont {Krovi}, \bibfnamefont
  {H}}} (\bibinfo {year} {2022}),\ \href@noop {} {\enquote {\bibinfo {title}
  {Average-case hardness of estimating probabilities of random quantum circuits
  with a linear scaling in the error exponent},}\ }\Eprint
  {https://arxiv.org/abs/2206.05642} {arXiv:2206.05642} \BibitemShut {NoStop}%
\bibitem [{\citenamefont {Kruse}\ \emph {et~al.}(2019)\citenamefont {Kruse},
  \citenamefont {Hamilton}, \citenamefont {Sansoni}, \citenamefont {Barkhofen},
  \citenamefont {Silberhorn},\ and\ \citenamefont {Jex}}]{kruse_detailed_2019}%
  \BibitemOpen
  \bibfield  {author} {\bibinfo {author} {\bibnamefont {Kruse}, \bibfnamefont
  {R}}, \bibinfo {author} {\bibfnamefont {C.~S.}\ \bibnamefont {Hamilton}},
  \bibinfo {author} {\bibfnamefont {L.}~\bibnamefont {Sansoni}}, \bibinfo
  {author} {\bibfnamefont {S.}~\bibnamefont {Barkhofen}}, \bibinfo {author}
  {\bibfnamefont {C.}~\bibnamefont {Silberhorn}}, and\ \bibinfo {author}
  {\bibfnamefont {I.}~\bibnamefont {Jex}}} (\bibinfo {year} {2019}),\ \bibfield
   {title} {\enquote {\bibinfo {title} {{A detailed study of Gaussian boson
  sampling}},}\ }\href {https://doi.org/10.1103/PhysRevA.100.032326} {\bibfield
   {journal} {\bibinfo  {journal} {Phys. Rev. A}\ }\textbf {\bibinfo {volume}
  {100}},\ \bibinfo {pages} {032326}}\BibitemShut {NoStop}%
\bibitem [{\citenamefont {Kuperberg}(2015)}]{kuperberg_how_2015}%
  \BibitemOpen
  \bibfield  {author} {\bibinfo {author} {\bibnamefont {Kuperberg},
  \bibfnamefont {G}}} (\bibinfo {year} {2015}),\ \bibfield  {title} {\enquote
  {\bibinfo {title} {{How hard is it to approximate the Jones polynomial?}}}\
  }\href {https://doi.org/10.4086/toc.2015.v011a006} {\bibfield  {journal}
  {\bibinfo  {journal} {Th. Comp.}\ }\textbf {\bibinfo {volume} {11}},\
  \bibinfo {pages} {183--219}}\BibitemShut {NoStop}%
\bibitem [{\citenamefont {Kushilevitz}\ and\ \citenamefont
  {Mansour}(1993)}]{kushilevitz_learning_1993}%
  \BibitemOpen
  \bibfield  {author} {\bibinfo {author} {\bibnamefont {Kushilevitz},
  \bibfnamefont {E}}, and\ \bibinfo {author} {\bibfnamefont {Y.}~\bibnamefont
  {Mansour}}} (\bibinfo {year} {1993}),\ \bibfield  {title} {\enquote {\bibinfo
  {title} {Learning {{decision trees using}} the {{Fourier spectrum}}},}\
  }\href {https://doi.org/10.1137/0222080} {\bibfield  {journal} {\bibinfo
  {journal} {SIAM J. Comput.}\ }\textbf {\bibinfo {volume} {22}},\ \bibinfo
  {pages} {1331--1348}}\BibitemShut {NoStop}%
\bibitem [{\citenamefont {Lautemann}(1983)}]{lautemann_bpp_1983}%
  \BibitemOpen
  \bibfield  {author} {\bibinfo {author} {\bibnamefont {Lautemann},
  \bibfnamefont {C}}} (\bibinfo {year} {1983}),\ \bibfield  {title} {\enquote
  {\bibinfo {title} {{{BPP}} and the polynomial hierarchy},}\ }\href
  {https://doi.org/10.1016/0020-0190(83)90044-3} {\bibfield  {journal}
  {\bibinfo  {journal} {Inf. Proc. Lett.}\ }\textbf {\bibinfo {volume}
  {17}}~(\bibinfo {number} {4}),\ \bibinfo {pages} {215--217}}\BibitemShut
  {NoStop}%
\bibitem [{\citenamefont {Ledoux}(2005)}]{ledoux_concentration_2005}%
  \BibitemOpen
  \bibfield  {author} {\bibinfo {author} {\bibnamefont {Ledoux}, \bibfnamefont
  {M}}} (\bibinfo {year} {2005}),\ \href {https://doi.org/10.1090/surv/089}
  {\emph {\bibinfo {title} {The {{concentration}} of {{measure
  phenomenon}}}}},\ \bibinfo {series} {Mathematical {{Surveys}} and
  {{Monographs}}}, Vol.~\bibinfo {volume} {89}\ (\bibinfo  {publisher}
  {{American Mathematical Society}},\ \bibinfo {address} {{Providence, Rhode
  Island}})\BibitemShut {NoStop}%
\bibitem [{\citenamefont {Lee}\ \emph {et~al.}(2010)\citenamefont {Lee},
  \citenamefont {Ruan}, \citenamefont {Jin},\ and\ \citenamefont
  {Aggarwal}}]{GraphData}%
  \BibitemOpen
  \bibfield  {author} {\bibinfo {author} {\bibnamefont {Lee}, \bibfnamefont
  {V~E}}, \bibinfo {author} {\bibfnamefont {N.}~\bibnamefont {Ruan}}, \bibinfo
  {author} {\bibfnamefont {R.}~\bibnamefont {Jin}}, and\ \bibinfo {author}
  {\bibfnamefont {C.}~\bibnamefont {Aggarwal}}} (\bibinfo {year} {2010}),\ in\
  \href@noop {} {\emph {\bibinfo {booktitle} {Managing and mining graph
  data}}}\ (\bibinfo  {publisher} {Springer},\ \bibinfo {address} {Berlin})\
  pp.\ \bibinfo {pages} {303--336}\BibitemShut {NoStop}%
\bibitem [{\citenamefont {Leone}\ \emph {et~al.}(2022)\citenamefont {Leone},
  \citenamefont {Oliviero},\ and\ \citenamefont {Hamma}}]{leone_magic_2022}%
  \BibitemOpen
  \bibfield  {author} {\bibinfo {author} {\bibnamefont {Leone}, \bibfnamefont
  {L}}, \bibinfo {author} {\bibfnamefont {S.~F.~E.}\ \bibnamefont {Oliviero}},
  and\ \bibinfo {author} {\bibfnamefont {A.}~\bibnamefont {Hamma}}} (\bibinfo
  {year} {2022}),\ \bibfield  {title} {\enquote {\bibinfo {title} {Magic
  hinders quantum certification},}\ }\href@noop {} {\ }\Eprint
  {https://arxiv.org/abs/2204.02995} {arXiv:2204.02995} \BibitemShut {NoStop}%
\bibitem [{\citenamefont {Leverrier}\ and\ \citenamefont
  {{Garc{\'i}a-Patr{\'o}n}}(2015)}]{leverrier_analysis_2015}%
  \BibitemOpen
  \bibfield  {author} {\bibinfo {author} {\bibnamefont {Leverrier},
  \bibfnamefont {A}}, and\ \bibinfo {author} {\bibfnamefont {R.}~\bibnamefont
  {{Garc{\'i}a-Patr{\'o}n}}}} (\bibinfo {year} {2015}),\ \bibfield  {title}
  {\enquote {\bibinfo {title} {Analysis of circuit imperfections in
  {{BosonSampling}}},}\ }\href {https://doi.org/10.26421/QIC15.5-6-8}
  {\bibfield  {journal} {\bibinfo  {journal} {Quant. Inf. Comp.}\ }\textbf
  {\bibinfo {volume} {15}},\ \bibinfo {pages} {0489--0512}}\BibitemShut
  {NoStop}%
\bibitem [{\citenamefont {Levin}(1986)}]{levin_average_1986}%
  \BibitemOpen
  \bibfield  {author} {\bibinfo {author} {\bibnamefont {Levin}, \bibfnamefont
  {L~A}}} (\bibinfo {year} {1986}),\ \bibfield  {title} {\enquote {\bibinfo
  {title} {Average case complete problems},}\ }\href
  {https://doi.org/10.1137/0215020} {\bibfield  {journal} {\bibinfo  {journal}
  {{SIAM} J. Comput.}\ }\textbf {\bibinfo {volume} {15}},\ \bibinfo {pages}
  {285--286}}\BibitemShut {NoStop}%
\bibitem [{\citenamefont {Li}\ and\ \citenamefont
  {Vit{\'a}nyi}(1992)}]{li_average_1992}%
  \BibitemOpen
  \bibfield  {author} {\bibinfo {author} {\bibnamefont {Li}, \bibfnamefont
  {M}}, and\ \bibinfo {author} {\bibfnamefont {P.~M.~B.}\ \bibnamefont
  {Vit{\'a}nyi}}} (\bibinfo {year} {1992}),\ \bibfield  {title} {\enquote
  {\bibinfo {title} {Average case complexity under the universal distribution
  equals worst-case complexity},}\ }\href
  {https://doi.org/10.1016/0020-0190(92)90138-L} {\bibfield  {journal}
  {\bibinfo  {journal} {Inf. Proc. Lett.}\ }\textbf {\bibinfo {volume} {42}},\
  \bibinfo {pages} {145--149}}\BibitemShut {NoStop}%
\bibitem [{\citenamefont {Li}\ \emph {et~al.}(2018)\citenamefont {Li},
  \citenamefont {Wu}, \citenamefont {Ying}, \citenamefont {Sun},\ and\
  \citenamefont {Yang}}]{li_quantum_2018}%
  \BibitemOpen
  \bibfield  {author} {\bibinfo {author} {\bibnamefont {Li}, \bibfnamefont
  {R}}, \bibinfo {author} {\bibfnamefont {B.}~\bibnamefont {Wu}}, \bibinfo
  {author} {\bibfnamefont {M.}~\bibnamefont {Ying}}, \bibinfo {author}
  {\bibfnamefont {X.}~\bibnamefont {Sun}}, and\ \bibinfo {author}
  {\bibfnamefont {G.}~\bibnamefont {Yang}}} (\bibinfo {year} {2018}),\
  \bibfield  {title} {\enquote {\bibinfo {title} {Quantum {{supremacy circuit
  simulation}} on {{Sunway TaihuLight}}},}\ }\href@noop {} {\ }\Eprint
  {https://arxiv.org/abs/1804.04797} {arXiv:1804.04797} \BibitemShut {NoStop}%
\bibitem [{\citenamefont {Linial}\ \emph {et~al.}(1998)\citenamefont {Linial},
  \citenamefont {Samorodnitsky},\ and\ \citenamefont
  {Wigderson}}]{linial_deterministic_1998}%
  \BibitemOpen
  \bibfield  {author} {\bibinfo {author} {\bibnamefont {Linial}, \bibfnamefont
  {N}}, \bibinfo {author} {\bibfnamefont {A.}~\bibnamefont {Samorodnitsky}},
  and\ \bibinfo {author} {\bibfnamefont {A.}~\bibnamefont {Wigderson}}}
  (\bibinfo {year} {1998}),\ \bibfield  {title} {\enquote {\bibinfo {title} {A
  deterministic strongly polynomial algorithm for matrix scaling and
  approximate permanents},}\ }in\ \href {https://doi.org/10.1145/276698.276880}
  {\emph {\bibinfo {booktitle} {Proc. 13th Ann. {{ACM}} Symp. {{Th.}}
  Comp.}}},\ \bibinfo {series and number} {{{STOC}} '98}\ (\bibinfo
  {publisher} {{Association for Computing Machinery}},\ \bibinfo {address}
  {{New York, NY, USA}})\ pp.\ \bibinfo {pages} {644--652}\BibitemShut
  {NoStop}%
\bibitem [{\citenamefont {Lipton}(1991)}]{lipton_permanent_1991}%
  \BibitemOpen
  \bibfield  {author} {\bibinfo {author} {\bibnamefont {Lipton}, \bibfnamefont
  {R}}} (\bibinfo {year} {1991}),\ \enquote {\bibinfo {title} {New directions
  in testing},}\ in\ \href {https://bookstore.ams.org/dimacs-2} {\emph
  {\bibinfo {booktitle} {Distributed Computing and Cryptography}}},\
  Vol.~\bibinfo {volume} {2}\ (\bibinfo  {publisher} {AMS})\ pp.\ \bibinfo
  {pages} {191--202}\BibitemShut {NoStop}%
\bibitem [{\citenamefont {Liu}\ \emph {et~al.}(2021{\natexlab{a}})\citenamefont
  {Liu}, \citenamefont {Kolden}, \citenamefont {Krovi}, \citenamefont
  {Loureiro}, \citenamefont {Trivisa},\ and\ \citenamefont
  {Childs}}]{liu_efficient_2021}%
  \BibitemOpen
  \bibfield  {author} {\bibinfo {author} {\bibnamefont {Liu}, \bibfnamefont
  {J-P}}, \bibinfo {author} {\bibfnamefont {H.~{\O}.}\ \bibnamefont {Kolden}},
  \bibinfo {author} {\bibfnamefont {H.~K.}\ \bibnamefont {Krovi}}, \bibinfo
  {author} {\bibfnamefont {N.~F.}\ \bibnamefont {Loureiro}}, \bibinfo {author}
  {\bibfnamefont {K.}~\bibnamefont {Trivisa}}, and\ \bibinfo {author}
  {\bibfnamefont {A.~M.}\ \bibnamefont {Childs}}} (\bibinfo {year}
  {2021}{\natexlab{a}}),\ \bibfield  {title} {\enquote {\bibinfo {title}
  {Efficient quantum algorithm for dissipative nonlinear differential
  equations},}\ }\href {https://doi.org/10.1073/pnas.2026805118} {\bibfield
  {journal} {\bibinfo  {journal} {PNAS}\ }\textbf {\bibinfo {volume} {118}},\
  \bibinfo {pages} {e2026805118}}\BibitemShut {NoStop}%
\bibitem [{\citenamefont {Liu}\ \emph {et~al.}(2021{\natexlab{b}})\citenamefont
  {Liu}, \citenamefont {Guo}, \citenamefont {Liu}, \citenamefont {Yang},
  \citenamefont {Song}, \citenamefont {Gao}, \citenamefont {Wang},
  \citenamefont {Wu}, \citenamefont {Peng}, \citenamefont {Zhao}, \citenamefont
  {Li}, \citenamefont {Huang}, \citenamefont {Fu},\ and\ \citenamefont
  {Chen}}]{liu_redefining_2021}%
  \BibitemOpen
  \bibfield  {author} {\bibinfo {author} {\bibnamefont {Liu}, \bibfnamefont
  {X}}, \bibinfo {author} {\bibfnamefont {C.}~\bibnamefont {Guo}}, \bibinfo
  {author} {\bibfnamefont {Y.}~\bibnamefont {Liu}}, \bibinfo {author}
  {\bibfnamefont {Y.}~\bibnamefont {Yang}}, \bibinfo {author} {\bibfnamefont
  {J.}~\bibnamefont {Song}}, \bibinfo {author} {\bibfnamefont {J.}~\bibnamefont
  {Gao}}, \bibinfo {author} {\bibfnamefont {Z.}~\bibnamefont {Wang}}, \bibinfo
  {author} {\bibfnamefont {W.}~\bibnamefont {Wu}}, \bibinfo {author}
  {\bibfnamefont {D.}~\bibnamefont {Peng}}, \bibinfo {author} {\bibfnamefont
  {P.}~\bibnamefont {Zhao}}, \bibinfo {author} {\bibfnamefont {F.}~\bibnamefont
  {Li}}, \bibinfo {author} {\bibfnamefont {H.-L.}\ \bibnamefont {Huang}},
  \bibinfo {author} {\bibfnamefont {H.}~\bibnamefont {Fu}}, and\ \bibinfo
  {author} {\bibfnamefont {D.}~\bibnamefont {Chen}}} (\bibinfo {year}
  {2021}{\natexlab{b}}),\ \bibfield  {title} {\enquote {\bibinfo {title}
  {Redefining the {{quantum supremacy baseline with}} a {{new generation Sunway
  supercomputer}}},}\ }\href@noop {} {\ }\Eprint
  {https://arxiv.org/abs/2111.01066} {arXiv:2111.01066} \BibitemShut {NoStop}%
\bibitem [{\citenamefont {Liu}\ \emph {et~al.}(2021{\natexlab{c}})\citenamefont
  {Liu}, \citenamefont {Otten}, \citenamefont {Bassirianjahromi}, \citenamefont
  {Jiang},\ and\ \citenamefont {Fefferman}}]{liu_benchmarking_2021}%
  \BibitemOpen
  \bibfield  {author} {\bibinfo {author} {\bibnamefont {Liu}, \bibfnamefont
  {Y}}, \bibinfo {author} {\bibfnamefont {M.}~\bibnamefont {Otten}}, \bibinfo
  {author} {\bibfnamefont {R.}~\bibnamefont {Bassirianjahromi}}, \bibinfo
  {author} {\bibfnamefont {L.}~\bibnamefont {Jiang}}, and\ \bibinfo {author}
  {\bibfnamefont {B.}~\bibnamefont {Fefferman}}} (\bibinfo {year}
  {2021}{\natexlab{c}}),\ \bibfield  {title} {\enquote {\bibinfo {title}
  {Benchmarking near-term quantum computers via random circuit sampling},}\
  }\href@noop {} {\ }\Eprint {https://arxiv.org/abs/2105.05232}
  {arXiv:2105.05232} \BibitemShut {NoStop}%
\bibitem [{\citenamefont {Liu}\ \emph {et~al.}(2021{\natexlab{d}})\citenamefont
  {Liu}, \citenamefont {Liu}, \citenamefont {Li}, \citenamefont {Fu},
  \citenamefont {Yang}, \citenamefont {Song}, \citenamefont {Zhao},
  \citenamefont {Wang}, \citenamefont {Peng}, \citenamefont {Chen},
  \citenamefont {Guo}, \citenamefont {Huang}, \citenamefont {Wu},\ and\
  \citenamefont {Chen}}]{liu_closing_2021-1}%
  \BibitemOpen
  \bibfield  {author} {\bibinfo {author} {\bibnamefont {Liu}, \bibfnamefont
  {Y~(A)}}, \bibinfo {author} {\bibfnamefont {X.~(L.)}\ \bibnamefont {Liu}},
  \bibinfo {author} {\bibfnamefont {F.~(N.)}\ \bibnamefont {Li}}, \bibinfo
  {author} {\bibfnamefont {H.}~\bibnamefont {Fu}}, \bibinfo {author}
  {\bibfnamefont {Y.}~\bibnamefont {Yang}}, \bibinfo {author} {\bibfnamefont
  {J.}~\bibnamefont {Song}}, \bibinfo {author} {\bibfnamefont {P.}~\bibnamefont
  {Zhao}}, \bibinfo {author} {\bibfnamefont {Z.}~\bibnamefont {Wang}}, \bibinfo
  {author} {\bibfnamefont {D.}~\bibnamefont {Peng}}, \bibinfo {author}
  {\bibfnamefont {H.}~\bibnamefont {Chen}}, \bibinfo {author} {\bibfnamefont
  {C.}~\bibnamefont {Guo}}, \bibinfo {author} {\bibfnamefont {H.}~\bibnamefont
  {Huang}}, \bibinfo {author} {\bibfnamefont {W.}~\bibnamefont {Wu}}, and\
  \bibinfo {author} {\bibfnamefont {D.}~\bibnamefont {Chen}}} (\bibinfo {year}
  {2021}{\natexlab{d}}),\ \bibfield  {title} {\enquote {\bibinfo {title}
  {Closing the "quantum supremacy" gap: Achieving real-time simulation of a
  random quantum circuit using a new {{Sunway}} supercomputer},}\ }in\ \href
  {https://doi.org/10.1145/3458817.3487399} {\emph {\bibinfo {booktitle}
  {Proceedings of the {{International Conference}} for {{High Performance
  Computing}}, {{Networking}}, {{Storage}} and {{Analysis}}}}},\ \bibinfo
  {series and number} {{{SC}} '21}\ (\bibinfo  {publisher} {{Association for
  Computing Machinery}},\ \bibinfo {address} {{New York, NY, USA}})\ pp.\
  \bibinfo {pages} {1--12}\BibitemShut {NoStop}%
\bibitem [{\citenamefont {Liu}\ and\ \citenamefont
  {Gheorghiu}(2022)}]{liu_depth-efficient_2021}%
  \BibitemOpen
  \bibfield  {author} {\bibinfo {author} {\bibnamefont {Liu}, \bibfnamefont
  {Z}}, and\ \bibinfo {author} {\bibfnamefont {A.}~\bibnamefont {Gheorghiu}}}
  (\bibinfo {year} {2022}),\ \bibfield  {title} {\enquote {\bibinfo {title}
  {Depth-efficient proofs of quantumness},}\ }\href
  {https://doi.org/10.22331/q-2022-09-19-807} {\bibfield  {journal} {\bibinfo
  {journal} {Quantum}\ }\textbf {\bibinfo {volume} {6}},\ \bibinfo {pages}
  {807}},\ \Eprint {https://arxiv.org/abs/2107.02163} {arXiv:2107.02163}
  \BibitemShut {NoStop}%
\bibitem [{\citenamefont {Lloyd}(1996)}]{lloyd_universal_1996}%
  \BibitemOpen
  \bibfield  {author} {\bibinfo {author} {\bibnamefont {Lloyd}, \bibfnamefont
  {S}}} (\bibinfo {year} {1996}),\ \bibfield  {title} {\enquote {\bibinfo
  {title} {Universal {{quantum simulators}}},}\ }\href
  {https://doi.org/10.1126/science.273.5278.1073} {\bibfield  {journal}
  {\bibinfo  {journal} {Science}\ }\textbf {\bibinfo {volume} {273}},\ \bibinfo
  {pages} {1073--1078}}\BibitemShut {NoStop}%
\bibitem [{\citenamefont {Lokshtanov}\ \emph {et~al.}(2017)\citenamefont
  {Lokshtanov}, \citenamefont {Paturi}, \citenamefont {Tamaki}, \citenamefont
  {Williams},\ and\ \citenamefont {Yu}}]{lokshtanov_beating_2017}%
  \BibitemOpen
  \bibfield  {author} {\bibinfo {author} {\bibnamefont {Lokshtanov},
  \bibfnamefont {D}}, \bibinfo {author} {\bibfnamefont {R.}~\bibnamefont
  {Paturi}}, \bibinfo {author} {\bibfnamefont {S.}~\bibnamefont {Tamaki}},
  \bibinfo {author} {\bibfnamefont {R.}~\bibnamefont {Williams}}, and\ \bibinfo
  {author} {\bibfnamefont {H.}~\bibnamefont {Yu}}} (\bibinfo {year} {2017}),\
  \bibfield  {title} {\enquote {\bibinfo {title} {Beating {{brute force}} for
  {{systems}} of {{polynomial equations}} over {{finite fields}}},}\ }in\ \href
  {https://doi.org/10.1137/1.9781611974782.143} {\emph {\bibinfo {booktitle}
  {Proc. 2017 {{Ann. ACM-SIAM Symp.}} {{Disc. Alg.}} ({{SODA}})}}},\ \bibinfo
  {series and number} {Proceedings}\ (\bibinfo  {publisher} {{Society for
  Industrial and Applied Mathematics}})\ pp.\ \bibinfo {pages}
  {2190--2202}\BibitemShut {NoStop}%
\bibitem [{\citenamefont {Loredo}\ \emph {et~al.}(2017)\citenamefont {Loredo},
  \citenamefont {Broome}, \citenamefont {Hilaire}, \citenamefont {Gazzano},
  \citenamefont {Sagnes}, \citenamefont {Lemaitre}, \citenamefont {Almeida},
  \citenamefont {Senellart},\ and\ \citenamefont {White}}]{loredo_boson_2017}%
  \BibitemOpen
  \bibfield  {author} {\bibinfo {author} {\bibnamefont {Loredo}, \bibfnamefont
  {J~C}}, \bibinfo {author} {\bibfnamefont {M.~A.}\ \bibnamefont {Broome}},
  \bibinfo {author} {\bibfnamefont {P.}~\bibnamefont {Hilaire}}, \bibinfo
  {author} {\bibfnamefont {O.}~\bibnamefont {Gazzano}}, \bibinfo {author}
  {\bibfnamefont {I.}~\bibnamefont {Sagnes}}, \bibinfo {author} {\bibfnamefont
  {A.}~\bibnamefont {Lemaitre}}, \bibinfo {author} {\bibfnamefont {M.~P.}\
  \bibnamefont {Almeida}}, \bibinfo {author} {\bibfnamefont {P.}~\bibnamefont
  {Senellart}}, and\ \bibinfo {author} {\bibfnamefont {A.~G.}\ \bibnamefont
  {White}}} (\bibinfo {year} {2017}),\ \bibfield  {title} {\enquote {\bibinfo
  {title} {Boson {{sampling}} with {{single-photon Fock states}} from a
  {{bright solid-state source}}},}\ }\href
  {https://doi.org/10.1103/PhysRevLett.118.130503} {\bibfield  {journal}
  {\bibinfo  {journal} {Phys. Rev. Lett.}\ }\textbf {\bibinfo {volume} {118}},\
  \bibinfo {pages} {130503}}\BibitemShut {NoStop}%
\bibitem [{\citenamefont {Low}\ and\ \citenamefont
  {Chuang}(2017)}]{low_optimal_2017}%
  \BibitemOpen
  \bibfield  {author} {\bibinfo {author} {\bibnamefont {Low}, \bibfnamefont
  {G~H}}, and\ \bibinfo {author} {\bibfnamefont {I.~L.}\ \bibnamefont
  {Chuang}}} (\bibinfo {year} {2017}),\ \bibfield  {title} {\enquote {\bibinfo
  {title} {Optimal {{Hamiltonian simulation}} by {{quantum signal
  processing}}},}\ }\href {https://doi.org/10.1103/PhysRevLett.118.010501}
  {\bibfield  {journal} {\bibinfo  {journal} {Phys. Rev. Lett.}\ }\textbf
  {\bibinfo {volume} {118}},\ \bibinfo {pages} {010501}}\BibitemShut {NoStop}%
\bibitem [{\citenamefont {Low}\ and\ \citenamefont
  {Chuang}(2019)}]{low_hamiltonian_2019}%
  \BibitemOpen
  \bibfield  {author} {\bibinfo {author} {\bibnamefont {Low}, \bibfnamefont
  {G~H}}, and\ \bibinfo {author} {\bibfnamefont {I.~L.}\ \bibnamefont
  {Chuang}}} (\bibinfo {year} {2019}),\ \bibfield  {title} {\enquote {\bibinfo
  {title} {Hamiltonian {{simulation}} by {{qubitization}}},}\ }\href
  {https://doi.org/10.22331/q-2019-07-12-163} {\bibfield  {journal} {\bibinfo
  {journal} {Quantum}\ }\textbf {\bibinfo {volume} {3}},\ \bibinfo {pages}
  {163}}\BibitemShut {NoStop}%
\bibitem [{\citenamefont {Lund}\ \emph
  {et~al.}(2017{\natexlab{a}})\citenamefont {Lund}, \citenamefont {Bremner},\
  and\ \citenamefont {Ralph}}]{lund_quantum_2017}%
  \BibitemOpen
  \bibfield  {author} {\bibinfo {author} {\bibnamefont {Lund}, \bibfnamefont
  {A~P}}, \bibinfo {author} {\bibfnamefont {M.~J.}\ \bibnamefont {Bremner}},
  and\ \bibinfo {author} {\bibfnamefont {T.~C.}\ \bibnamefont {Ralph}}}
  (\bibinfo {year} {2017}{\natexlab{a}}),\ \bibfield  {title} {\enquote
  {\bibinfo {title} {Quantum sampling problems, {BosonSampling} and quantum
  supremacy},}\ }\href {https://doi.org/10.1038/s41534-017-0018-2} {\bibfield
  {journal} {\bibinfo  {journal} {npj Quant.\ Inf.}\ }\textbf {\bibinfo
  {volume} {3}},\ \bibinfo {pages} {15}}\BibitemShut {NoStop}%
\bibitem [{\citenamefont {Lund}\ \emph {et~al.}(2014)\citenamefont {Lund},
  \citenamefont {Laing}, \citenamefont {Rahimi-Keshari}, \citenamefont
  {Rudolph}, \citenamefont {O'Brien},\ and\ \citenamefont
  {Ralph}}]{lund_boson_2014}%
  \BibitemOpen
  \bibfield  {author} {\bibinfo {author} {\bibnamefont {Lund}, \bibfnamefont
  {A~P}}, \bibinfo {author} {\bibfnamefont {A.}~\bibnamefont {Laing}}, \bibinfo
  {author} {\bibfnamefont {S.}~\bibnamefont {Rahimi-Keshari}}, \bibinfo
  {author} {\bibfnamefont {T.}~\bibnamefont {Rudolph}}, \bibinfo {author}
  {\bibfnamefont {J.~L.}\ \bibnamefont {O'Brien}}, and\ \bibinfo {author}
  {\bibfnamefont {T.~C.}\ \bibnamefont {Ralph}}} (\bibinfo {year} {2014}),\
  \bibfield  {title} {\enquote {\bibinfo {title} {{Boson sampling from a
  Gaussian state}},}\ }\href {https://doi.org/10.1103/PhysRevLett.113.100502}
  {\bibfield  {journal} {\bibinfo  {journal} {Phys. Rev. Lett.}\ }\textbf
  {\bibinfo {volume} {113}},\ \bibinfo {pages} {100502}}\BibitemShut {NoStop}%
\bibitem [{\citenamefont {Lund}\ \emph
  {et~al.}(2017{\natexlab{b}})\citenamefont {Lund}, \citenamefont
  {{Rahimi-Keshari}},\ and\ \citenamefont {Ralph}}]{lund_exact_2017}%
  \BibitemOpen
  \bibfield  {author} {\bibinfo {author} {\bibnamefont {Lund}, \bibfnamefont
  {A~P}}, \bibinfo {author} {\bibfnamefont {S.}~\bibnamefont
  {{Rahimi-Keshari}}}, and\ \bibinfo {author} {\bibfnamefont {T.~C.}\
  \bibnamefont {Ralph}}} (\bibinfo {year} {2017}{\natexlab{b}}),\ \bibfield
  {title} {\enquote {\bibinfo {title} {Exact {{boson sampling}} using
  {{Gaussian}} continuous variable measurements},}\ }\href
  {https://doi.org/10.1103/PhysRevA.96.022301} {\bibfield  {journal} {\bibinfo
  {journal} {Phys. Rev. A}\ }\textbf {\bibinfo {volume} {96}},\ \bibinfo
  {pages} {022301}}\BibitemShut {NoStop}%
\bibitem [{\citenamefont {Lundow}\ and\ \citenamefont
  {Markstr{\"o}m}(2022)}]{lundow_efficient_2022}%
  \BibitemOpen
  \bibfield  {author} {\bibinfo {author} {\bibnamefont {Lundow}, \bibfnamefont
  {P~H}}, and\ \bibinfo {author} {\bibfnamefont {K.}~\bibnamefont
  {Markstr{\"o}m}}} (\bibinfo {year} {2022}),\ \bibfield  {title} {\enquote
  {\bibinfo {title} {Efficient computation of permanents, with applications to
  {{Boson}} sampling and random matrices},}\ }\href
  {https://doi.org/10.1016/j.jcp.2022.110990} {\bibfield  {journal} {\bibinfo
  {journal} {J. Comp. Phys.}\ }\textbf {\bibinfo {volume} {455}},\ \bibinfo
  {pages} {110990}}\BibitemShut {NoStop}%
\bibitem [{\citenamefont {Madsen}\ \emph {et~al.}(2022)\citenamefont {Madsen},
  \citenamefont {Laudenbach}, \citenamefont {Askarani}, \citenamefont
  {Rortais}, \citenamefont {Vincent}, \citenamefont {Bulmer}, \citenamefont
  {Miatto}, \citenamefont {Neuhaus}, \citenamefont {Helt}, \citenamefont
  {Collins}, \citenamefont {Lita}, \citenamefont {Gerrits}, \citenamefont
  {Nam}, \citenamefont {Vaidya}, \citenamefont {Menotti}, \citenamefont
  {Dhand}, \citenamefont {Vernon}, \citenamefont {Quesada},\ and\ \citenamefont
  {Lavoie}}]{madsen_quantum_2022}%
  \BibitemOpen
  \bibfield  {author} {\bibinfo {author} {\bibnamefont {Madsen}, \bibfnamefont
  {L~S}}, \bibinfo {author} {\bibfnamefont {F.}~\bibnamefont {Laudenbach}},
  \bibinfo {author} {\bibfnamefont {M.~F.}\ \bibnamefont {Askarani}}, \bibinfo
  {author} {\bibfnamefont {F.}~\bibnamefont {Rortais}}, \bibinfo {author}
  {\bibfnamefont {T.}~\bibnamefont {Vincent}}, \bibinfo {author} {\bibfnamefont
  {J.~F.~F.}\ \bibnamefont {Bulmer}}, \bibinfo {author} {\bibfnamefont {F.~M.}\
  \bibnamefont {Miatto}}, \bibinfo {author} {\bibfnamefont {L.}~\bibnamefont
  {Neuhaus}}, \bibinfo {author} {\bibfnamefont {L.~G.}\ \bibnamefont {Helt}},
  \bibinfo {author} {\bibfnamefont {M.~J.}\ \bibnamefont {Collins}}, \bibinfo
  {author} {\bibfnamefont {A.~E.}\ \bibnamefont {Lita}}, \bibinfo {author}
  {\bibfnamefont {T.}~\bibnamefont {Gerrits}}, \bibinfo {author} {\bibfnamefont
  {S.~W.}\ \bibnamefont {Nam}}, \bibinfo {author} {\bibfnamefont {V.~D.}\
  \bibnamefont {Vaidya}}, \bibinfo {author} {\bibfnamefont {M.}~\bibnamefont
  {Menotti}}, \bibinfo {author} {\bibfnamefont {I.}~\bibnamefont {Dhand}},
  \bibinfo {author} {\bibfnamefont {Z.}~\bibnamefont {Vernon}}, \bibinfo
  {author} {\bibfnamefont {N.}~\bibnamefont {Quesada}}, and\ \bibinfo {author}
  {\bibfnamefont {J.}~\bibnamefont {Lavoie}}} (\bibinfo {year} {2022}),\
  \bibfield  {title} {\enquote {\bibinfo {title} {Quantum computational
  advantage with a programmable photonic processor},}\ }\href
  {https://doi.org/10.1038/s41586-022-04725-x} {\bibfield  {journal} {\bibinfo
  {journal} {Nature}\ }\textbf {\bibinfo {volume} {606}},\ \bibinfo {pages}
  {75--81}}\BibitemShut {NoStop}%
\bibitem [{\citenamefont {Mahadev}(2018)}]{mahadev_classical_2018}%
  \BibitemOpen
  \bibfield  {author} {\bibinfo {author} {\bibnamefont {Mahadev}, \bibfnamefont
  {U}}} (\bibinfo {year} {2018}),\ \bibfield  {title} {\enquote {\bibinfo
  {title} {Classical {verification} of {quantum} {computations}},}\ }\href@noop
  {} {\ }\Eprint {https://arxiv.org/abs/1804.01082} {arXiv:1804.01082}
  \BibitemShut {NoStop}%
\bibitem [{\citenamefont {Mandel}\ \emph {et~al.}(2003)\citenamefont {Mandel},
  \citenamefont {Greiner}, \citenamefont {Widera}, \citenamefont {Rom},
  \citenamefont {H{\"a}nsch},\ and\ \citenamefont {Bloch}}]{ColdCollisions}%
  \BibitemOpen
  \bibfield  {author} {\bibinfo {author} {\bibnamefont {Mandel}, \bibfnamefont
  {O}}, \bibinfo {author} {\bibfnamefont {M.}~\bibnamefont {Greiner}}, \bibinfo
  {author} {\bibfnamefont {A.}~\bibnamefont {Widera}}, \bibinfo {author}
  {\bibfnamefont {T.}~\bibnamefont {Rom}}, \bibinfo {author} {\bibfnamefont
  {T.~W.}\ \bibnamefont {H{\"a}nsch}}, and\ \bibinfo {author} {\bibfnamefont
  {I.}~\bibnamefont {Bloch}}} (\bibinfo {year} {2003}),\ \bibfield  {title}
  {\enquote {\bibinfo {title} {Controlled collisions for multi-particle
  entanglement of optically trapped atoms},}\ }\href
  {https://doi.org/10.1038/nature02008} {\bibfield  {journal} {\bibinfo
  {journal} {Nature}\ }\textbf {\bibinfo {volume} {425}},\ \bibinfo {pages}
  {937--940}}\BibitemShut {NoStop}%
\bibitem [{\citenamefont {Mann}\ and\ \citenamefont
  {Bremner}(2017)}]{mann_complexity_2017}%
  \BibitemOpen
  \bibfield  {author} {\bibinfo {author} {\bibnamefont {Mann}, \bibfnamefont
  {R~L}}, and\ \bibinfo {author} {\bibfnamefont {M.~J.}\ \bibnamefont
  {Bremner}}} (\bibinfo {year} {2017}),\ \bibfield  {title} {\enquote {\bibinfo
  {title} {On the {{complexity}} of {{random quantum computations}} and the
  {{Jones polynomial}}},}\ }\href@noop {} {\ }\Eprint
  {https://arxiv.org/abs/1711.00686} {arxiv:1711.00686} \BibitemShut {NoStop}%
\bibitem [{\citenamefont {Mantri}\ \emph {et~al.}(2017)\citenamefont {Mantri},
  \citenamefont {Demarie},\ and\ \citenamefont
  {Fitzsimons}}]{mantri_universality_2017}%
  \BibitemOpen
  \bibfield  {author} {\bibinfo {author} {\bibnamefont {Mantri}, \bibfnamefont
  {A}}, \bibinfo {author} {\bibfnamefont {T.~F.}\ \bibnamefont {Demarie}}, and\
  \bibinfo {author} {\bibfnamefont {J.~F.}\ \bibnamefont {Fitzsimons}}}
  (\bibinfo {year} {2017}),\ \bibfield  {title} {\enquote {\bibinfo {title}
  {{Universality of quantum computation with cluster states and $(X, Y)$-plane
  measurements}},}\ }\href {https://doi.org/10.1038/srep42861} {\bibfield
  {journal} {\bibinfo  {journal} {Sci. Rep.}\ }\textbf {\bibinfo {volume}
  {7}},\ \bibinfo {pages} {1--7}}\BibitemShut {NoStop}%
\bibitem [{\citenamefont {Markham}\ and\ \citenamefont
  {Krause}(2020)}]{markham_simple_2020}%
  \BibitemOpen
  \bibfield  {author} {\bibinfo {author} {\bibnamefont {Markham}, \bibfnamefont
  {D}}, and\ \bibinfo {author} {\bibfnamefont {A.}~\bibnamefont {Krause}}}
  (\bibinfo {year} {2020}),\ \bibfield  {title} {\enquote {\bibinfo {title} {A
  simple protocol for certifying graph states and applications in quantum
  networks},}\ }\href {https://doi.org/10.3390/cryptography4010003} {\bibfield
  {journal} {\bibinfo  {journal} {Cryptography}\ }\textbf {\bibinfo {volume}
  {4}},\ \bibinfo {pages} {3}}\BibitemShut {NoStop}%
\bibitem [{\citenamefont {Markov}\ \emph {et~al.}(2018)\citenamefont {Markov},
  \citenamefont {Fatima}, \citenamefont {Isakov},\ and\ \citenamefont
  {Boixo}}]{markov_quantum_2018}%
  \BibitemOpen
  \bibfield  {author} {\bibinfo {author} {\bibnamefont {Markov}, \bibfnamefont
  {I~L}}, \bibinfo {author} {\bibfnamefont {A.}~\bibnamefont {Fatima}},
  \bibinfo {author} {\bibfnamefont {S.~V.}\ \bibnamefont {Isakov}}, and\
  \bibinfo {author} {\bibfnamefont {S.}~\bibnamefont {Boixo}}} (\bibinfo {year}
  {2018}),\ \bibfield  {title} {\enquote {\bibinfo {title} {Quantum {{supremacy
  is both closer}} and {{farther}} than {{it appears}}},}\ }\href@noop {} {\
  }\Eprint {https://arxiv.org/abs/1807.10749} {1807.10749} \BibitemShut
  {NoStop}%
\bibitem [{\citenamefont {Markov}\ and\ \citenamefont
  {Shi}(2008)}]{markov_simulating_2008}%
  \BibitemOpen
  \bibfield  {author} {\bibinfo {author} {\bibnamefont {Markov}, \bibfnamefont
  {I~L}}, and\ \bibinfo {author} {\bibfnamefont {Y.}~\bibnamefont {Shi}}}
  (\bibinfo {year} {2008}),\ \bibfield  {title} {\enquote {\bibinfo {title}
  {Simulating {{quantum computation}} by {{contracting tensor networks}}},}\
  }\href {https://doi.org/10.1137/050644756} {\bibfield  {journal} {\bibinfo
  {journal} {SIAM J. Comput.}\ }\textbf {\bibinfo {volume} {38}},\ \bibinfo
  {pages} {963--981}}\BibitemShut {NoStop}%
\bibitem [{\citenamefont {{Mart{\'i}nez-Cifuentes}}\ \emph
  {et~al.}(2022)\citenamefont {{Mart{\'i}nez-Cifuentes}}, \citenamefont
  {{Fonseca-Romero}},\ and\ \citenamefont
  {Quesada}}]{martinez-cifuentes_classical_2022}%
  \BibitemOpen
  \bibfield  {author} {\bibinfo {author} {\bibnamefont
  {{Mart{\'i}nez-Cifuentes}}, \bibfnamefont {J}}, \bibinfo {author}
  {\bibfnamefont {K.~M.}\ \bibnamefont {{Fonseca-Romero}}}, and\ \bibinfo
  {author} {\bibfnamefont {N.}~\bibnamefont {Quesada}}} (\bibinfo {year}
  {2022}),\ \bibfield  {title} {\enquote {\bibinfo {title} {Classical models
  are a better explanation of the {{Jiuzhang Gaussian Boson Samplers}} than
  their targeted squeezed light models},}\ }\href@noop {} {\ }\Eprint
  {https://arxiv.org/abs/2207.10058} {arXiv:2207.10058} \BibitemShut {NoStop}%
\bibitem [{\citenamefont {Martyn}\ \emph {et~al.}(2021)\citenamefont {Martyn},
  \citenamefont {Rossi}, \citenamefont {Tan},\ and\ \citenamefont
  {Chuang}}]{martyn_grand_2021}%
  \BibitemOpen
  \bibfield  {author} {\bibinfo {author} {\bibnamefont {Martyn}, \bibfnamefont
  {J~M}}, \bibinfo {author} {\bibfnamefont {Z.~M.}\ \bibnamefont {Rossi}},
  \bibinfo {author} {\bibfnamefont {A.~K.}\ \bibnamefont {Tan}}, and\ \bibinfo
  {author} {\bibfnamefont {I.~L.}\ \bibnamefont {Chuang}}} (\bibinfo {year}
  {2021}),\ \bibfield  {title} {\enquote {\bibinfo {title} {Grand
  {{unification}} of {{quantum algorithms}}},}\ }\href
  {https://doi.org/10.1103/PRXQuantum.2.040203} {\bibfield  {journal} {\bibinfo
   {journal} {PRX Quantum}\ }\textbf {\bibinfo {volume} {2}},\ \bibinfo {pages}
  {040203}},\ \Eprint {https://arxiv.org/abs/2105.02859} {arXiv:2105.02859}
  \BibitemShut {NoStop}%
\bibitem [{\citenamefont {Maskara}\ \emph {et~al.}(2019)\citenamefont
  {Maskara}, \citenamefont {Deshpande}, \citenamefont {Tran}, \citenamefont
  {Ehrenberg}, \citenamefont {Fefferman},\ and\ \citenamefont
  {Gorshkov}}]{maskara_complexity_2019}%
  \BibitemOpen
  \bibfield  {author} {\bibinfo {author} {\bibnamefont {Maskara}, \bibfnamefont
  {N}}, \bibinfo {author} {\bibfnamefont {A.}~\bibnamefont {Deshpande}},
  \bibinfo {author} {\bibfnamefont {M.~C.}\ \bibnamefont {Tran}}, \bibinfo
  {author} {\bibfnamefont {A.}~\bibnamefont {Ehrenberg}}, \bibinfo {author}
  {\bibfnamefont {B.}~\bibnamefont {Fefferman}}, and\ \bibinfo {author}
  {\bibfnamefont {A.~V.}\ \bibnamefont {Gorshkov}}} (\bibinfo {year} {2019}),\
  \bibfield  {title} {\enquote {\bibinfo {title} {Complexity phase diagram for
  interacting and long-range bosonic {{Hamiltonians}}},}\ }\href@noop {} {\
  }\Eprint {https://arxiv.org/abs/1906.04178} {arXiv:1906.04178} \BibitemShut
  {NoStop}%
\bibitem [{\citenamefont {McCaskey}\ \emph {et~al.}(2018)\citenamefont
  {McCaskey}, \citenamefont {Dumitrescu}, \citenamefont {Chen}, \citenamefont
  {Lyakh},\ and\ \citenamefont {Humble}}]{mccaskey_validating_2018}%
  \BibitemOpen
  \bibfield  {author} {\bibinfo {author} {\bibnamefont {McCaskey},
  \bibfnamefont {A}}, \bibinfo {author} {\bibfnamefont {E.}~\bibnamefont
  {Dumitrescu}}, \bibinfo {author} {\bibfnamefont {M.}~\bibnamefont {Chen}},
  \bibinfo {author} {\bibfnamefont {D.}~\bibnamefont {Lyakh}}, and\ \bibinfo
  {author} {\bibfnamefont {T.}~\bibnamefont {Humble}}} (\bibinfo {year}
  {2018}),\ \bibfield  {title} {\enquote {\bibinfo {title} {Validating
  quantum-classical programming models with tensor network simulations},}\
  }\href {https://doi.org/10.1371/journal.pone.0206704} {\bibfield  {journal}
  {\bibinfo  {journal} {PLOS ONE}\ }\textbf {\bibinfo {volume} {13}},\ \bibinfo
  {pages} {e0206704}}\BibitemShut {NoStop}%
\bibitem [{\citenamefont {McClean}\ \emph {et~al.}(2016)\citenamefont
  {McClean}, \citenamefont {Romero}, \citenamefont {Babbush},\ and\
  \citenamefont {Aspuru-Guzik}}]{McClean_2016}%
  \BibitemOpen
  \bibfield  {author} {\bibinfo {author} {\bibnamefont {McClean}, \bibfnamefont
  {J~R}}, \bibinfo {author} {\bibfnamefont {J.}~\bibnamefont {Romero}},
  \bibinfo {author} {\bibfnamefont {R.}~\bibnamefont {Babbush}}, and\ \bibinfo
  {author} {\bibfnamefont {A.}~\bibnamefont {Aspuru-Guzik}}} (\bibinfo {year}
  {2016}),\ \bibfield  {title} {\enquote {\bibinfo {title} {The theory of
  variational hybrid quantum-classical algorithms},}\ }\href
  {https://doi.org/10.1088/1367-2630/18/2/023023} {\bibfield  {journal}
  {\bibinfo  {journal} {New J. Phys.}\ }\textbf {\bibinfo {volume} {18}},\
  \bibinfo {pages} {023023}}\BibitemShut {NoStop}%
\bibitem [{\citenamefont {{Merkel}}\ \emph {et~al.}(2013)\citenamefont
  {{Merkel}}, \citenamefont {{Gambetta}}, \citenamefont {{Smolin}},
  \citenamefont {{Poletto}}, \citenamefont {{C{\'o}rcoles}}, \citenamefont
  {{Johnson}}, \citenamefont {{Ryan}},\ and\ \citenamefont
  {{Steffen}}}]{MerGamSmo13}%
  \BibitemOpen
  \bibfield  {author} {\bibinfo {author} {\bibnamefont {{Merkel}},
  \bibfnamefont {S~T}}, \bibinfo {author} {\bibfnamefont {J.~M.}\ \bibnamefont
  {{Gambetta}}}, \bibinfo {author} {\bibfnamefont {J.~A.}\ \bibnamefont
  {{Smolin}}}, \bibinfo {author} {\bibfnamefont {S.}~\bibnamefont {{Poletto}}},
  \bibinfo {author} {\bibfnamefont {A.~D.}\ \bibnamefont {{C{\'o}rcoles}}},
  \bibinfo {author} {\bibfnamefont {B.~R.}\ \bibnamefont {{Johnson}}}, \bibinfo
  {author} {\bibfnamefont {C.~A.}\ \bibnamefont {{Ryan}}}, and\ \bibinfo
  {author} {\bibfnamefont {M.}~\bibnamefont {{Steffen}}}} (\bibinfo {year}
  {2013}),\ \bibfield  {title} {\enquote {\bibinfo {title} {Self-consistent
  quantum process tomography},}\ }\href
  {https://doi.org/10.1103/PhysRevA.87.062119} {\bibfield  {journal} {\bibinfo
  {journal} {Phys. Rev. A}\ }\textbf {\bibinfo {volume} {87}},\ \bibinfo
  {pages} {062119}}\BibitemShut {NoStop}%
\bibitem [{\citenamefont {Mezher}\ \emph {et~al.}(2020)\citenamefont {Mezher},
  \citenamefont {Ghalbouni}, \citenamefont {Dgheim},\ and\ \citenamefont
  {Markham}}]{mezher_fault-tolerant_2020}%
  \BibitemOpen
  \bibfield  {author} {\bibinfo {author} {\bibnamefont {Mezher}, \bibfnamefont
  {R}}, \bibinfo {author} {\bibfnamefont {J.}~\bibnamefont {Ghalbouni}},
  \bibinfo {author} {\bibfnamefont {J.}~\bibnamefont {Dgheim}}, and\ \bibinfo
  {author} {\bibfnamefont {D.}~\bibnamefont {Markham}}} (\bibinfo {year}
  {2020}),\ \bibfield  {title} {\enquote {\bibinfo {title} {Fault-tolerant
  quantum speedup from constant depth quantum circuits},}\ }\href
  {https://doi.org/10.1103/PhysRevResearch.2.033444} {\bibfield  {journal}
  {\bibinfo  {journal} {Phys. Rev. Res.}\ }\textbf {\bibinfo {volume} {2}},\
  \bibinfo {pages} {033444}}\BibitemShut {NoStop}%
\bibitem [{\citenamefont {Miller}\ \emph {et~al.}(2017)\citenamefont {Miller},
  \citenamefont {Sanders},\ and\ \citenamefont {Miyake}}]{miller_quantum_2017}%
  \BibitemOpen
  \bibfield  {author} {\bibinfo {author} {\bibnamefont {Miller}, \bibfnamefont
  {J}}, \bibinfo {author} {\bibfnamefont {S.}~\bibnamefont {Sanders}}, and\
  \bibinfo {author} {\bibfnamefont {A.}~\bibnamefont {Miyake}}} (\bibinfo
  {year} {2017}),\ \bibfield  {title} {\enquote {\bibinfo {title} {Quantum
  supremacy in constant-time measurement-based computation: A unified
  architecture for sampling and verification},}\ }\href
  {https://doi.org/10.1103/PhysRevA.96.062320} {\bibfield  {journal} {\bibinfo
  {journal} {Phys. Rev. A}\ }\textbf {\bibinfo {volume} {96}},\ \bibinfo
  {pages} {062320}}\BibitemShut {NoStop}%
\bibitem [{\citenamefont {Montanaro}(2016{\natexlab{a}})}]{QuantumAlgorithms}%
  \BibitemOpen
  \bibfield  {author} {\bibinfo {author} {\bibnamefont {Montanaro},
  \bibfnamefont {A}}} (\bibinfo {year} {2016}{\natexlab{a}}),\ \bibfield
  {title} {\enquote {\bibinfo {title} {Quantum algorithms: an overview},}\
  }\href {https://doi.org/10.48550/arXiv.1511.04206} {\bibfield  {journal}
  {\bibinfo  {journal} {npj Quant. Inf.}\ }\textbf {\bibinfo {volume} {2}},\
  \bibinfo {pages} {15023}}\BibitemShut {NoStop}%
\bibitem [{\citenamefont
  {Montanaro}(2016{\natexlab{b}})}]{montanaro_quantum_2017}%
  \BibitemOpen
  \bibfield  {author} {\bibinfo {author} {\bibnamefont {Montanaro},
  \bibfnamefont {A}}} (\bibinfo {year} {2016}{\natexlab{b}}),\ \bibfield
  {title} {\enquote {\bibinfo {title} {Quantum circuits and low-degree
  polynomials over f\_2},}\ }\href {https://doi.org/10.1088/1751-8121/aa565f}
  {\bibfield  {journal} {\bibinfo  {journal} {J. Phys. A}\ }\textbf {\bibinfo
  {volume} {50}},\ \bibinfo {pages} {084002}}\BibitemShut {NoStop}%
\bibitem [{\citenamefont {Morimae}(2017)}]{morimae_hardness_2017}%
  \BibitemOpen
  \bibfield  {author} {\bibinfo {author} {\bibnamefont {Morimae}, \bibfnamefont
  {T}}} (\bibinfo {year} {2017}),\ \bibfield  {title} {\enquote {\bibinfo
  {title} {Hardness of classically sampling the one-clean-qubit model with
  constant total variation distance error},}\ }\href
  {https://doi.org/10.1103/PhysRevA.96.040302} {\bibfield  {journal} {\bibinfo
  {journal} {Phys. Rev. A}\ }\textbf {\bibinfo {volume} {96}},\ \bibinfo
  {pages} {040302}},\ \Eprint {https://arxiv.org/abs/1704.03640}
  {arXiv:1704.03640} \BibitemShut {NoStop}%
\bibitem [{\citenamefont {Morimae}\ \emph {et~al.}(2014)\citenamefont
  {Morimae}, \citenamefont {Fujii},\ and\ \citenamefont
  {Fitzsimons}}]{morimae_hardness_2014}%
  \BibitemOpen
  \bibfield  {author} {\bibinfo {author} {\bibnamefont {Morimae}, \bibfnamefont
  {T}}, \bibinfo {author} {\bibfnamefont {K.}~\bibnamefont {Fujii}}, and\
  \bibinfo {author} {\bibfnamefont {J.~F.}\ \bibnamefont {Fitzsimons}}}
  (\bibinfo {year} {2014}),\ \bibfield  {title} {\enquote {\bibinfo {title}
  {Hardness of {{classically simulating}} the {{one}}-{{clean}}-{{qubit
  model}}},}\ }\href {https://doi.org/10.1103/PhysRevLett.112.130502}
  {\bibfield  {journal} {\bibinfo  {journal} {Phys. Rev. Lett.}\ }\textbf
  {\bibinfo {volume} {112}},\ \bibinfo {pages} {130502}}\BibitemShut {NoStop}%
\bibitem [{\citenamefont {Morimae}\ \emph {et~al.}(2019)\citenamefont
  {Morimae}, \citenamefont {Takeuchi},\ and\ \citenamefont
  {Hayashi}}]{morimae_verification_2017}%
  \BibitemOpen
  \bibfield  {author} {\bibinfo {author} {\bibnamefont {Morimae}, \bibfnamefont
  {T}}, \bibinfo {author} {\bibfnamefont {Y.}~\bibnamefont {Takeuchi}}, and\
  \bibinfo {author} {\bibfnamefont {M.}~\bibnamefont {Hayashi}}} (\bibinfo
  {year} {2019}),\ \bibfield  {title} {\enquote {\bibinfo {title} {Verification
  of hypergraph states},}\ }\href {https://doi.org/10.1103/PhysRevA.96.062321}
  {\bibfield  {journal} {\bibinfo  {journal} {Phys. Rev. A}\ }\textbf {\bibinfo
  {volume} {96}},\ \bibinfo {pages} {062321}}\BibitemShut {NoStop}%
\bibitem [{\citenamefont {Morimae}\ and\ \citenamefont
  {Tamaki}(2019)}]{morimae_fine-grained_2019}%
  \BibitemOpen
  \bibfield  {author} {\bibinfo {author} {\bibnamefont {Morimae}, \bibfnamefont
  {T}}, and\ \bibinfo {author} {\bibfnamefont {S.}~\bibnamefont {Tamaki}}}
  (\bibinfo {year} {2019}),\ \bibfield  {title} {\enquote {\bibinfo {title}
  {Fine-grained quantum computational supremacy},}\ }\href
  {https://doi.org/10.26421/QIC19.13-14-2} {\bibfield  {journal} {\bibinfo
  {journal} {Quant. Inf. Comp.}\ }\textbf {\bibinfo {volume} {19}},\ \bibinfo
  {pages} {1089--1115}}\BibitemShut {NoStop}%
\bibitem [{\citenamefont {Movassagh}(2018)}]{movassagh_efficient_2018}%
  \BibitemOpen
  \bibfield  {author} {\bibinfo {author} {\bibnamefont {Movassagh},
  \bibfnamefont {R}}} (\bibinfo {year} {2018}),\ \bibfield  {title} {\enquote
  {\bibinfo {title} {Efficient unitary paths and quantum computational
  supremacy: {{A}} proof of average-case hardness of {{random circuit
  sampling}}},}\ }\href@noop {} {\ }\Eprint {https://arxiv.org/abs/1810.04681}
  {arXiv:1810.04681} \BibitemShut {NoStop}%
\bibitem [{\citenamefont {Movassagh}(2020)}]{movassagh_quantum_2020}%
  \BibitemOpen
  \bibfield  {author} {\bibinfo {author} {\bibnamefont {Movassagh},
  \bibfnamefont {R}}} (\bibinfo {year} {2020}),\ \bibfield  {title} {\enquote
  {\bibinfo {title} {Quantum supremacy and random circuits},}\ }\href@noop {}
  {\ }\Eprint {https://arxiv.org/abs/1909.06210} {arXiv:1909.06210}
  \BibitemShut {NoStop}%
\bibitem [{\citenamefont {Moylett}\ \emph {et~al.}(2019)\citenamefont
  {Moylett}, \citenamefont {Garc{\'i}a-Patr{\'o}n}, \citenamefont {Renema},\
  and\ \citenamefont {Turner}}]{moylett_classically_2019}%
  \BibitemOpen
  \bibfield  {author} {\bibinfo {author} {\bibnamefont {Moylett}, \bibfnamefont
  {A~E}}, \bibinfo {author} {\bibfnamefont {R.}~\bibnamefont
  {Garc{\'i}a-Patr{\'o}n}}, \bibinfo {author} {\bibfnamefont {J.~J.}\
  \bibnamefont {Renema}}, and\ \bibinfo {author} {\bibfnamefont {P.~S.}\
  \bibnamefont {Turner}}} (\bibinfo {year} {2019}),\ \bibfield  {title}
  {\enquote {\bibinfo {title} {Classically simulating near-term
  partially-distinguishable and lossy boson sampling},}\ }\href
  {https://doi.org/10.1088/2058-9565/ab5555} {\bibfield  {journal} {\bibinfo
  {journal} {Quant. Sci. Tech.}\ }\textbf {\bibinfo {volume} {5}},\ \bibinfo
  {pages} {015001}}\BibitemShut {NoStop}%
\bibitem [{\citenamefont {Muller}(1954)}]{muller_application_1954}%
  \BibitemOpen
  \bibfield  {author} {\bibinfo {author} {\bibnamefont {Muller}, \bibfnamefont
  {D~E}}} (\bibinfo {year} {1954}),\ \bibfield  {title} {\enquote {\bibinfo
  {title} {{Application of Boolean algebra to switching circuit design and to
  error detection}},}\ }\href {https://doi.org/10.1109/IREPGELC.1954.6499441}
  {\bibfield  {journal} {\bibinfo  {journal} {Trans. I.R.E. Prof. Group Elec.
  Comp.}\ }\textbf {\bibinfo {volume} {{EC}-3}},\ \bibinfo {pages}
  {6--12}}\BibitemShut {NoStop}%
\bibitem [{\citenamefont {Muraleedharan}\ \emph {et~al.}(2019)\citenamefont
  {Muraleedharan}, \citenamefont {Miyake},\ and\ \citenamefont
  {Deutsch}}]{Deutsch}%
  \BibitemOpen
  \bibfield  {author} {\bibinfo {author} {\bibnamefont {Muraleedharan},
  \bibfnamefont {G}}, \bibinfo {author} {\bibfnamefont {A.}~\bibnamefont
  {Miyake}}, and\ \bibinfo {author} {\bibfnamefont {I.~H.}\ \bibnamefont
  {Deutsch}}} (\bibinfo {year} {2019}),\ \bibfield  {title} {\enquote {\bibinfo
  {title} {Quantum computational supremacy in the sampling of bosonic random
  walkers on a one-dimensional lattice},}\ }\href
  {https://doi.org/10.1088/1367-2630/ab0610} {\bibfield  {journal} {\bibinfo
  {journal} {New J. Phys.}\ }\textbf {\bibinfo {volume} {21}},\ \bibinfo
  {pages} {055003}}\BibitemShut {NoStop}%
\bibitem [{\citenamefont {Murphy}(2012)}]{Murphy}%
  \BibitemOpen
  \bibfield  {author} {\bibinfo {author} {\bibnamefont {Murphy}, \bibfnamefont
  {K~P}}} (\bibinfo {year} {2012}),\ \href@noop {} {\emph {\bibinfo {title}
  {Machine learning: a probabilistic perspective}}},\ Adaptive computation and
  machine learning series\ (\bibinfo  {publisher} {MIT Press},\ \bibinfo
  {address} {Cambridge, MA, USA})\BibitemShut {NoStop}%
\bibitem [{\citenamefont {Nachtergaele}(1996)}]{nachtergaele_spectral_1996}%
  \BibitemOpen
  \bibfield  {author} {\bibinfo {author} {\bibnamefont {Nachtergaele},
  \bibfnamefont {B}}} (\bibinfo {year} {1996}),\ \bibfield  {title} {\enquote
  {\bibinfo {title} {The spectral gap for some spin chains with discrete
  symmetry breaking},}\ }\href {https://doi.org/10.1007/BF02099509} {\bibfield
  {journal} {\bibinfo  {journal} {Commun.Math. Phys.}\ }\textbf {\bibinfo
  {volume} {175}},\ \bibinfo {pages} {565--606}}\BibitemShut {NoStop}%
\bibitem [{\citenamefont {Nagaj}(2012)}]{nagaj_universal_2012}%
  \BibitemOpen
  \bibfield  {author} {\bibinfo {author} {\bibnamefont {Nagaj}, \bibfnamefont
  {D}}} (\bibinfo {year} {2012}),\ \bibfield  {title} {\enquote {\bibinfo
  {title} {Universal two-body-{{Hamiltonian}} quantum computing},}\ }\href
  {https://doi.org/10.1103/PhysRevA.85.032330} {\bibfield  {journal} {\bibinfo
  {journal} {Phys. Rev. A}\ }\textbf {\bibinfo {volume} {85}},\ \bibinfo
  {pages} {032330}}\BibitemShut {NoStop}%
\bibitem [{\citenamefont {Nagaj}\ and\ \citenamefont
  {Wocjan}(2008)}]{nagaj_hamiltonian_2008}%
  \BibitemOpen
  \bibfield  {author} {\bibinfo {author} {\bibnamefont {Nagaj}, \bibfnamefont
  {D}}, and\ \bibinfo {author} {\bibfnamefont {P.}~\bibnamefont {Wocjan}}}
  (\bibinfo {year} {2008}),\ \bibfield  {title} {\enquote {\bibinfo {title}
  {Hamiltonian quantum cellular automata in one dimension},}\ }\href
  {https://doi.org/10.1103/PhysRevA.78.032311} {\bibfield  {journal} {\bibinfo
  {journal} {Phys. Rev. A}\ }\textbf {\bibinfo {volume} {78}},\ \bibinfo
  {pages} {032311}}\BibitemShut {NoStop}%
\bibitem [{\citenamefont {Nakata}\ \emph {et~al.}(2014)\citenamefont {Nakata},
  \citenamefont {Koashi},\ and\ \citenamefont
  {Murao}}]{nakata_generating_2014}%
  \BibitemOpen
  \bibfield  {author} {\bibinfo {author} {\bibnamefont {Nakata}, \bibfnamefont
  {Y}}, \bibinfo {author} {\bibfnamefont {M.}~\bibnamefont {Koashi}}, and\
  \bibinfo {author} {\bibfnamefont {M.}~\bibnamefont {Murao}}} (\bibinfo {year}
  {2014}),\ \bibfield  {title} {\enquote {\bibinfo {title} {Generating a state
  $t$-design by diagonal quantum circuits},}\ }\href
  {https://doi.org/10.1088/1367-2630/16/5/053043} {\bibfield  {journal}
  {\bibinfo  {journal} {New J. Phys.}\ }\textbf {\bibinfo {volume} {16}},\
  \bibinfo {pages} {053043}}\BibitemShut {NoStop}%
\bibitem [{\citenamefont {Napp}\ \emph {et~al.}(2022)\citenamefont {Napp},
  \citenamefont {La~Placa}, \citenamefont {Dalzell}, \citenamefont
  {Brand{\~a}o},\ and\ \citenamefont {Harrow}}]{napp_efficient_2022}%
  \BibitemOpen
  \bibfield  {author} {\bibinfo {author} {\bibnamefont {Napp}, \bibfnamefont
  {J~C}}, \bibinfo {author} {\bibfnamefont {R.~L.}\ \bibnamefont {La~Placa}},
  \bibinfo {author} {\bibfnamefont {A.~M.}\ \bibnamefont {Dalzell}}, \bibinfo
  {author} {\bibfnamefont {F.~G. S.~L.}\ \bibnamefont {Brand{\~a}o}}, and\
  \bibinfo {author} {\bibfnamefont {A.~W.}\ \bibnamefont {Harrow}}} (\bibinfo
  {year} {2022}),\ \bibfield  {title} {\enquote {\bibinfo {title} {Efficient
  {{classical simulation}} of {{random shallow 2D quantum circuits}}},}\ }\href
  {https://doi.org/10.1103/PhysRevX.12.021021} {\bibfield  {journal} {\bibinfo
  {journal} {Phys. Rev. X}\ }\textbf {\bibinfo {volume} {12}},\ \bibinfo
  {pages} {021021}}\BibitemShut {NoStop}%
\bibitem [{\citenamefont {Negrevergne}\ \emph {et~al.}(2020)\citenamefont
  {Negrevergne}, \citenamefont {Somma}, \citenamefont {Ortiz}, \citenamefont
  {Knill},\ and\ \citenamefont {Laflamme}}]{negrevergne_liquid-state_2005}%
  \BibitemOpen
  \bibfield  {author} {\bibinfo {author} {\bibnamefont {Negrevergne},
  \bibfnamefont {C}}, \bibinfo {author} {\bibfnamefont {R.}~\bibnamefont
  {Somma}}, \bibinfo {author} {\bibfnamefont {G.}~\bibnamefont {Ortiz}},
  \bibinfo {author} {\bibfnamefont {E.}~\bibnamefont {Knill}}, and\ \bibinfo
  {author} {\bibfnamefont {R.}~\bibnamefont {Laflamme}}} (\bibinfo {year}
  {2020}),\ \bibfield  {title} {\enquote {\bibinfo {title} {Liquid-state {NMR}
  simulations of quantum many-body problems},}\ }\href
  {https://doi.org/10.1103/PhysRevA.71.032344} {\bibfield  {journal} {\bibinfo
  {journal} {Phys. Rev. A}\ }\textbf {\bibinfo {volume} {71}},\ \bibinfo
  {pages} {032344}}\BibitemShut {NoStop}%
\bibitem [{\citenamefont {Neill}\ \emph {et~al.}(2018)\citenamefont {Neill},
  \citenamefont {Roushan}, \citenamefont {Kechedzhi}, \citenamefont {Boixo},
  \citenamefont {Isakov}, \citenamefont {Smelyanskiy}, \citenamefont {Megrant},
  \citenamefont {Chiaro}, \citenamefont {Dunsworth}, \citenamefont {Arya},
  \citenamefont {Barends}, \citenamefont {Burkett}, \citenamefont {Chen},
  \citenamefont {Chen}, \citenamefont {Fowler}, \citenamefont {Foxen},
  \citenamefont {Giustina}, \citenamefont {Graff}, \citenamefont {Jeffrey},
  \citenamefont {Huang}, \citenamefont {Kelly}, \citenamefont {Klimov},
  \citenamefont {Lucero}, \citenamefont {Mutus}, \citenamefont {Neeley},
  \citenamefont {Quintana}, \citenamefont {Sank}, \citenamefont {Vainsencher},
  \citenamefont {Wenner}, \citenamefont {White}, \citenamefont {Neven},\ and\
  \citenamefont {Martinis}}]{neill_blueprint_2018}%
  \BibitemOpen
  \bibfield  {author} {\bibinfo {author} {\bibnamefont {Neill}, \bibfnamefont
  {C}}, \bibinfo {author} {\bibfnamefont {P.}~\bibnamefont {Roushan}}, \bibinfo
  {author} {\bibfnamefont {K.}~\bibnamefont {Kechedzhi}}, \bibinfo {author}
  {\bibfnamefont {S.}~\bibnamefont {Boixo}}, \bibinfo {author} {\bibfnamefont
  {S.~V.}\ \bibnamefont {Isakov}}, \bibinfo {author} {\bibfnamefont
  {V.}~\bibnamefont {Smelyanskiy}}, \bibinfo {author} {\bibfnamefont
  {A.}~\bibnamefont {Megrant}}, \bibinfo {author} {\bibfnamefont
  {B.}~\bibnamefont {Chiaro}}, \bibinfo {author} {\bibfnamefont
  {A.}~\bibnamefont {Dunsworth}}, \bibinfo {author} {\bibfnamefont
  {K.}~\bibnamefont {Arya}}, \bibinfo {author} {\bibfnamefont {R.}~\bibnamefont
  {Barends}}, \bibinfo {author} {\bibfnamefont {B.}~\bibnamefont {Burkett}},
  \bibinfo {author} {\bibfnamefont {Y.}~\bibnamefont {Chen}}, \bibinfo {author}
  {\bibfnamefont {Z.}~\bibnamefont {Chen}}, \bibinfo {author} {\bibfnamefont
  {A.}~\bibnamefont {Fowler}}, \bibinfo {author} {\bibfnamefont
  {B.}~\bibnamefont {Foxen}}, \bibinfo {author} {\bibfnamefont
  {M.}~\bibnamefont {Giustina}}, \bibinfo {author} {\bibfnamefont
  {R.}~\bibnamefont {Graff}}, \bibinfo {author} {\bibfnamefont
  {E.}~\bibnamefont {Jeffrey}}, \bibinfo {author} {\bibfnamefont
  {T.}~\bibnamefont {Huang}}, \bibinfo {author} {\bibfnamefont
  {J.}~\bibnamefont {Kelly}}, \bibinfo {author} {\bibfnamefont
  {P.}~\bibnamefont {Klimov}}, \bibinfo {author} {\bibfnamefont
  {E.}~\bibnamefont {Lucero}}, \bibinfo {author} {\bibfnamefont
  {J.}~\bibnamefont {Mutus}}, \bibinfo {author} {\bibfnamefont
  {M.}~\bibnamefont {Neeley}}, \bibinfo {author} {\bibfnamefont
  {C.}~\bibnamefont {Quintana}}, \bibinfo {author} {\bibfnamefont
  {D.}~\bibnamefont {Sank}}, \bibinfo {author} {\bibfnamefont {A.}~\bibnamefont
  {Vainsencher}}, \bibinfo {author} {\bibfnamefont {J.}~\bibnamefont {Wenner}},
  \bibinfo {author} {\bibfnamefont {T.~C.}\ \bibnamefont {White}}, \bibinfo
  {author} {\bibfnamefont {H.}~\bibnamefont {Neven}}, and\ \bibinfo {author}
  {\bibfnamefont {J.~M.}\ \bibnamefont {Martinis}}} (\bibinfo {year} {2018}),\
  \bibfield  {title} {\enquote {\bibinfo {title} {A blueprint for demonstrating
  quantum supremacy with superconducting qubits},}\ }\href
  {https://doi.org/10.1126/science.aao4309} {\bibfield  {journal} {\bibinfo
  {journal} {Science}\ }\textbf {\bibinfo {volume} {360}},\ \bibinfo {pages}
  {195--199}}\BibitemShut {NoStop}%
\bibitem [{\citenamefont {Neville}\ \emph {et~al.}(2017)\citenamefont
  {Neville}, \citenamefont {Sparrow}, \citenamefont {Clifford}, \citenamefont
  {Johnston}, \citenamefont {Birchall}, \citenamefont {Montanaro},\ and\
  \citenamefont {Laing}}]{neville_classical_2017}%
  \BibitemOpen
  \bibfield  {author} {\bibinfo {author} {\bibnamefont {Neville}, \bibfnamefont
  {A}}, \bibinfo {author} {\bibfnamefont {C.}~\bibnamefont {Sparrow}}, \bibinfo
  {author} {\bibfnamefont {R.}~\bibnamefont {Clifford}}, \bibinfo {author}
  {\bibfnamefont {E.}~\bibnamefont {Johnston}}, \bibinfo {author}
  {\bibfnamefont {P.~M.}\ \bibnamefont {Birchall}}, \bibinfo {author}
  {\bibfnamefont {A.}~\bibnamefont {Montanaro}}, and\ \bibinfo {author}
  {\bibfnamefont {A.}~\bibnamefont {Laing}}} (\bibinfo {year} {2017}),\
  \bibfield  {title} {\enquote {\bibinfo {title} {Classical boson sampling
  algorithms with superior performance to near-term experiments},}\ }\href
  {https://doi.org/10.1038/nphys4270} {\bibfield  {journal} {\bibinfo
  {journal} {Nature Phys.}\ }\textbf {\bibinfo {volume} {13}},\ \bibinfo
  {pages} {1153--1157}}\BibitemShut {NoStop}%
\bibitem [{\citenamefont {Nezami}(2021)}]{nezami_permanent_2021}%
  \BibitemOpen
  \bibfield  {author} {\bibinfo {author} {\bibnamefont {Nezami}, \bibfnamefont
  {S}}} (\bibinfo {year} {2021}),\ \bibfield  {title} {\enquote {\bibinfo
  {title} {Permanent of random matrices from representation theory: Moments,
  numerics, concentration, and comments on hardness of boson-sampling},}\
  }\href@noop {} {\ }\Eprint {https://arxiv.org/abs/2104.06423}
  {arXiv:2104.06423} \BibitemShut {NoStop}%
\bibitem [{\citenamefont {Nielsen}\ and\ \citenamefont
  {Chuang}(2010)}]{nielsen_quantum_2010}%
  \BibitemOpen
  \bibfield  {author} {\bibinfo {author} {\bibnamefont {Nielsen}, \bibfnamefont
  {M~A}}, and\ \bibinfo {author} {\bibfnamefont {I.~L.}\ \bibnamefont
  {Chuang}}} (\bibinfo {year} {2010}),\ \href@noop {} {\emph {\bibinfo {title}
  {Quantum Computation and Quantum Information}}},\ \bibinfo {edition} {10th}\
  ed.\ (\bibinfo  {publisher} {{Cambridge University Press}},\ \bibinfo
  {address} {{Cambridge ; New York}})\BibitemShut {NoStop}%
\bibitem [{\citenamefont
  {Nikolopoulos}(2019)}]{nikolopoulos_cryptographic_2019}%
  \BibitemOpen
  \bibfield  {author} {\bibinfo {author} {\bibnamefont {Nikolopoulos},
  \bibfnamefont {G~M}}} (\bibinfo {year} {2019}),\ \bibfield  {title} {\enquote
  {\bibinfo {title} {Cryptographic one-way function based on boson sampling},}\
  }\href {https://doi.org/10.1007/s11128-019-2372-9} {\bibfield  {journal}
  {\bibinfo  {journal} {Quant. Inf. Proc.}\ }\textbf {\bibinfo {volume} {18}},\
  \bibinfo {pages} {259}}\BibitemShut {NoStop}%
\bibitem [{\citenamefont {Novo}\ \emph {et~al.}(2021)\citenamefont {Novo},
  \citenamefont {Bermejo-Vega},\ and\ \citenamefont
  {Garc{\'i}a-Patr{\'o}n}}]{novo_quantum_2019}%
  \BibitemOpen
  \bibfield  {author} {\bibinfo {author} {\bibnamefont {Novo}, \bibfnamefont
  {L}}, \bibinfo {author} {\bibfnamefont {J.}~\bibnamefont {Bermejo-Vega}},
  and\ \bibinfo {author} {\bibfnamefont {R.}~\bibnamefont
  {Garc{\'i}a-Patr{\'o}n}}} (\bibinfo {year} {2021}),\ \bibfield  {title}
  {\enquote {\bibinfo {title} {Quantum advantage from energy measurements of
  many-body quantum systems},}\ }\href
  {https://doi.org/10.22331/q-2021-06-02-465} {\bibfield  {journal} {\bibinfo
  {journal} {Quantum}\ }\textbf {\bibinfo {volume} {5}},\ \bibinfo {pages}
  {465}}\BibitemShut {NoStop}%
\bibitem [{\citenamefont {Ofek}\ \emph {et~al.}(2016)\citenamefont {Ofek},
  \citenamefont {Petrenko}, \citenamefont {Heeres}, \citenamefont {Reinhold},
  \citenamefont {Leghtas}, \citenamefont {Vlastakis}, \citenamefont {Liu},
  \citenamefont {Frunzio}, \citenamefont {Girvin}, \citenamefont {Jiang},
  \citenamefont {Mirrahimi}, \citenamefont {Devoret},\ and\ \citenamefont
  {Schoelkopf}}]{ofek_extending_2016}%
  \BibitemOpen
  \bibfield  {author} {\bibinfo {author} {\bibnamefont {Ofek}, \bibfnamefont
  {N}}, \bibinfo {author} {\bibfnamefont {A.}~\bibnamefont {Petrenko}},
  \bibinfo {author} {\bibfnamefont {R.}~\bibnamefont {Heeres}}, \bibinfo
  {author} {\bibfnamefont {P.}~\bibnamefont {Reinhold}}, \bibinfo {author}
  {\bibfnamefont {Z.}~\bibnamefont {Leghtas}}, \bibinfo {author} {\bibfnamefont
  {B.}~\bibnamefont {Vlastakis}}, \bibinfo {author} {\bibfnamefont
  {Y.}~\bibnamefont {Liu}}, \bibinfo {author} {\bibfnamefont {L.}~\bibnamefont
  {Frunzio}}, \bibinfo {author} {\bibfnamefont {S.~M.}\ \bibnamefont {Girvin}},
  \bibinfo {author} {\bibfnamefont {L.}~\bibnamefont {Jiang}}, \bibinfo
  {author} {\bibfnamefont {M.}~\bibnamefont {Mirrahimi}}, \bibinfo {author}
  {\bibfnamefont {M.~H.}\ \bibnamefont {Devoret}}, and\ \bibinfo {author}
  {\bibfnamefont {R.~J.}\ \bibnamefont {Schoelkopf}}} (\bibinfo {year}
  {2016}),\ \bibfield  {title} {\enquote {\bibinfo {title} {Extending the
  lifetime of a quantum bit with error correction in superconducting
  circuits},}\ }\href {https://doi.org/10.1038/nature18949} {\bibfield
  {journal} {\bibinfo  {journal} {Nature}\ }\textbf {\bibinfo {volume} {536}},\
  \bibinfo {pages} {441--445}}\BibitemShut {NoStop}%
\bibitem [{\citenamefont {O'Gorman}\ and\ \citenamefont
  {Campbell}(2017)}]{ogorman_quantum_2017}%
  \BibitemOpen
  \bibfield  {author} {\bibinfo {author} {\bibnamefont {O'Gorman},
  \bibfnamefont {J}}, and\ \bibinfo {author} {\bibfnamefont {E.~T.}\
  \bibnamefont {Campbell}}} (\bibinfo {year} {2017}),\ \bibfield  {title}
  {\enquote {\bibinfo {title} {Quantum computation with realistic magic-state
  factories},}\ }\href {https://doi.org/10.1103/PhysRevA.95.032338} {\bibfield
  {journal} {\bibinfo  {journal} {Phys. Rev. A}\ }\textbf {\bibinfo {volume}
  {95}},\ \bibinfo {pages} {032338}}\BibitemShut {NoStop}%
\bibitem [{\citenamefont {Oh}\ \emph {et~al.}(2023)\citenamefont {Oh},
  \citenamefont {Jiang},\ and\ \citenamefont {Fefferman}}]{oh_classical_2023}%
  \BibitemOpen
  \bibfield  {author} {\bibinfo {author} {\bibnamefont {Oh}, \bibfnamefont
  {C}}, \bibinfo {author} {\bibfnamefont {L.}~\bibnamefont {Jiang}}, and\
  \bibinfo {author} {\bibfnamefont {B.}~\bibnamefont {Fefferman}}} (\bibinfo
  {year} {2023}),\ \bibfield  {title} {\enquote {\bibinfo {title} {On classical
  simulation algorithms for noisy {{Boson Sampling}}},}\ }\href@noop {} {\
  }\Eprint {https://arxiv.org/abs/2301.11532} {arXiv:2301.11532} \BibitemShut
  {NoStop}%
\bibitem [{\citenamefont {Oh}\ \emph {et~al.}(2022{\natexlab{a}})\citenamefont
  {Oh}, \citenamefont {Lim}, \citenamefont {Fefferman},\ and\ \citenamefont
  {Jiang}}]{oh_classical_2022}%
  \BibitemOpen
  \bibfield  {author} {\bibinfo {author} {\bibnamefont {Oh}, \bibfnamefont
  {C}}, \bibinfo {author} {\bibfnamefont {Y.}~\bibnamefont {Lim}}, \bibinfo
  {author} {\bibfnamefont {B.}~\bibnamefont {Fefferman}}, and\ \bibinfo
  {author} {\bibfnamefont {L.}~\bibnamefont {Jiang}}} (\bibinfo {year}
  {2022}{\natexlab{a}}),\ \bibfield  {title} {\enquote {\bibinfo {title}
  {Classical simulation of boson sampling based on graph structure},}\ }\href
  {https://doi.org/10.1103/PhysRevLett.128.190501} {\bibfield  {journal}
  {\bibinfo  {journal} {Phys. Rev. Lett.}\ }\textbf {\bibinfo {volume} {128}},\
  \bibinfo {pages} {190501}}\BibitemShut {NoStop}%
\bibitem [{\citenamefont {Oh}\ \emph {et~al.}(2022{\natexlab{b}})\citenamefont
  {Oh}, \citenamefont {Lim}, \citenamefont {Wong}, \citenamefont {Fefferman},\
  and\ \citenamefont {Jiang}}]{oh_quantum-inspired_2022}%
  \BibitemOpen
  \bibfield  {author} {\bibinfo {author} {\bibnamefont {Oh}, \bibfnamefont
  {C}}, \bibinfo {author} {\bibfnamefont {Y.}~\bibnamefont {Lim}}, \bibinfo
  {author} {\bibfnamefont {Y.}~\bibnamefont {Wong}}, \bibinfo {author}
  {\bibfnamefont {B.}~\bibnamefont {Fefferman}}, and\ \bibinfo {author}
  {\bibfnamefont {L.}~\bibnamefont {Jiang}}} (\bibinfo {year}
  {2022}{\natexlab{b}}),\ \bibfield  {title} {\enquote {\bibinfo {title}
  {Quantum-inspired classical algorithm for molecular vibronic spectra},}\
  }\href@noop {} {\ }\Eprint {https://arxiv.org/abs/2202.01861}
  {arXiv:2202.01861} \BibitemShut {NoStop}%
\bibitem [{\citenamefont {Oh}\ \emph {et~al.}(2021)\citenamefont {Oh},
  \citenamefont {Noh}, \citenamefont {Fefferman},\ and\ \citenamefont
  {Jiang}}]{oh_classical_2021-2}%
  \BibitemOpen
  \bibfield  {author} {\bibinfo {author} {\bibnamefont {Oh}, \bibfnamefont
  {C}}, \bibinfo {author} {\bibfnamefont {K.}~\bibnamefont {Noh}}, \bibinfo
  {author} {\bibfnamefont {B.}~\bibnamefont {Fefferman}}, and\ \bibinfo
  {author} {\bibfnamefont {L.}~\bibnamefont {Jiang}}} (\bibinfo {year}
  {2021}),\ \bibfield  {title} {\enquote {\bibinfo {title} {Classical
  simulation of lossy boson sampling using matrix product operators},}\ }\href
  {https://doi.org/10.1103/PhysRevA.104.022407} {\bibfield  {journal} {\bibinfo
   {journal} {Phys. Rev. A}\ }\textbf {\bibinfo {volume} {104}},\ \bibinfo
  {pages} {022407}}\BibitemShut {NoStop}%
\bibitem [{\citenamefont {Oszmaniec}\ \emph {et~al.}(2016)\citenamefont
  {Oszmaniec}, \citenamefont {Augusiak}, \citenamefont {Gogolin}, \citenamefont
  {Kolodynski}, \citenamefont {Ac\'{\i}n},\ and\ \citenamefont
  {Lewenstein}}]{PhysRevX.6.041044}%
  \BibitemOpen
  \bibfield  {author} {\bibinfo {author} {\bibnamefont {Oszmaniec},
  \bibfnamefont {M}}, \bibinfo {author} {\bibfnamefont {R.}~\bibnamefont
  {Augusiak}}, \bibinfo {author} {\bibfnamefont {C.}~\bibnamefont {Gogolin}},
  \bibinfo {author} {\bibfnamefont {J.}~\bibnamefont {Kolodynski}}, \bibinfo
  {author} {\bibfnamefont {A.}~\bibnamefont {Ac\'{\i}n}}, and\ \bibinfo
  {author} {\bibfnamefont {M.}~\bibnamefont {Lewenstein}}} (\bibinfo {year}
  {2016}),\ \bibfield  {title} {\enquote {\bibinfo {title} {Random bosonic
  states for robust quantum metrology},}\ }\href
  {https://doi.org/10.1103/PhysRevX.6.041044} {\bibfield  {journal} {\bibinfo
  {journal} {Phys. Rev. X}\ }\textbf {\bibinfo {volume} {6}},\ \bibinfo {pages}
  {041044}}\BibitemShut {NoStop}%
\bibitem [{\citenamefont {Oszmaniec}\ and\ \citenamefont
  {Brod}(2018)}]{oszmaniec_classical_2018}%
  \BibitemOpen
  \bibfield  {author} {\bibinfo {author} {\bibnamefont {Oszmaniec},
  \bibfnamefont {M}}, and\ \bibinfo {author} {\bibfnamefont {D.~J.}\
  \bibnamefont {Brod}}} (\bibinfo {year} {2018}),\ \bibfield  {title} {\enquote
  {\bibinfo {title} {Classical simulation of photonic linear optics with lost
  particles},}\ }\href {https://doi.org/10.1088/1367-2630/aadfa8} {\bibfield
  {journal} {\bibinfo  {journal} {New J. Phys.}\ }\textbf {\bibinfo {volume}
  {20}},\ \bibinfo {pages} {092002}}\BibitemShut {NoStop}%
\bibitem [{\citenamefont {Oszmaniec}\ \emph {et~al.}(2022)\citenamefont
  {Oszmaniec}, \citenamefont {Dangniam}, \citenamefont {Morales},\ and\
  \citenamefont {Zimbor{\'a}s}}]{oszmaniec_fermion_2020}%
  \BibitemOpen
  \bibfield  {author} {\bibinfo {author} {\bibnamefont {Oszmaniec},
  \bibfnamefont {M}}, \bibinfo {author} {\bibfnamefont {N.}~\bibnamefont
  {Dangniam}}, \bibinfo {author} {\bibfnamefont {M.~E.~S.}\ \bibnamefont
  {Morales}}, and\ \bibinfo {author} {\bibfnamefont {Z.}~\bibnamefont
  {Zimbor{\'a}s}}} (\bibinfo {year} {2022}),\ \bibfield  {title} {\enquote
  {\bibinfo {title} {Fermion {{sampling}}: {{A robust quantum computational
  advantage scheme using fermionic linear optics}} and {{magic input
  states}}},}\ }\href {https://doi.org/10.1103/PRXQuantum.3.020328} {\bibfield
  {journal} {\bibinfo  {journal} {PRX Quantum}\ }\textbf {\bibinfo {volume}
  {3}},\ \bibinfo {pages} {020328}}\BibitemShut {NoStop}%
\bibitem [{\citenamefont {Paesani}\ \emph {et~al.}(2019)\citenamefont
  {Paesani}, \citenamefont {Ding}, \citenamefont {Santagati}, \citenamefont
  {Chakhmakhchyan}, \citenamefont {Vigliar}, \citenamefont {Rottwitt},
  \citenamefont {Oxenl{\o}we}, \citenamefont {Wang}, \citenamefont {Thompson},\
  and\ \citenamefont {Laing}}]{paesani_generation_2019}%
  \BibitemOpen
  \bibfield  {author} {\bibinfo {author} {\bibnamefont {Paesani}, \bibfnamefont
  {S}}, \bibinfo {author} {\bibfnamefont {Y.}~\bibnamefont {Ding}}, \bibinfo
  {author} {\bibfnamefont {R.}~\bibnamefont {Santagati}}, \bibinfo {author}
  {\bibfnamefont {L.}~\bibnamefont {Chakhmakhchyan}}, \bibinfo {author}
  {\bibfnamefont {C.}~\bibnamefont {Vigliar}}, \bibinfo {author} {\bibfnamefont
  {K.}~\bibnamefont {Rottwitt}}, \bibinfo {author} {\bibfnamefont {L.~K.}\
  \bibnamefont {Oxenl{\o}we}}, \bibinfo {author} {\bibfnamefont
  {J.}~\bibnamefont {Wang}}, \bibinfo {author} {\bibfnamefont {M.~G.}\
  \bibnamefont {Thompson}}, and\ \bibinfo {author} {\bibfnamefont
  {A.}~\bibnamefont {Laing}}} (\bibinfo {year} {2019}),\ \bibfield  {title}
  {\enquote {\bibinfo {title} {Generation and sampling of quantum states of
  light in a silicon chip},}\ }\href
  {https://doi.org/10.1038/s41567-019-0567-8} {\bibfield  {journal} {\bibinfo
  {journal} {Nat. Phys.}\ }\textbf {\bibinfo {volume} {15}},\ \bibinfo {pages}
  {925--929}}\BibitemShut {NoStop}%
\bibitem [{\citenamefont {Pallister}\ \emph {et~al.}(2018)\citenamefont
  {Pallister}, \citenamefont {Linden},\ and\ \citenamefont
  {Montanaro}}]{pallister_optimal_2018}%
  \BibitemOpen
  \bibfield  {author} {\bibinfo {author} {\bibnamefont {Pallister},
  \bibfnamefont {S}}, \bibinfo {author} {\bibfnamefont {N.}~\bibnamefont
  {Linden}}, and\ \bibinfo {author} {\bibfnamefont {A.}~\bibnamefont
  {Montanaro}}} (\bibinfo {year} {2018}),\ \bibfield  {title} {\enquote
  {\bibinfo {title} {Optimal {verification} of {entangled} {states} with
  {local} {measurements}},}\ }\href
  {https://doi.org/10.1103/PhysRevLett.120.170502} {\bibfield  {journal}
  {\bibinfo  {journal} {Phys.\ Rev.\ Lett.}\ }\textbf {\bibinfo {volume}
  {120}},\ \bibinfo {pages} {170502}}\BibitemShut {NoStop}%
\bibitem [{\citenamefont {Pan}\ \emph {et~al.}(2022)\citenamefont {Pan},
  \citenamefont {Chen},\ and\ \citenamefont {Zhang}}]{pan_solving_2022}%
  \BibitemOpen
  \bibfield  {author} {\bibinfo {author} {\bibnamefont {Pan}, \bibfnamefont
  {F}}, \bibinfo {author} {\bibfnamefont {K.}~\bibnamefont {Chen}}, and\
  \bibinfo {author} {\bibfnamefont {P.}~\bibnamefont {Zhang}}} (\bibinfo {year}
  {2022}),\ \bibfield  {title} {\enquote {\bibinfo {title} {Solving the
  sampling problem of the {{Sycamore}} quantum circuits},}\ }\href
  {https://doi.org/10.1103/PhysRevLett.129.090502} {\bibfield  {journal}
  {\bibinfo  {journal} {Phys. Rev. Lett.}\ }\textbf {\bibinfo {volume} {129}},\
  \bibinfo {pages} {090502}}\BibitemShut {NoStop}%
\bibitem [{\citenamefont {Pan}\ and\ \citenamefont
  {Zhang}(2022)}]{pan_simulation_2022}%
  \BibitemOpen
  \bibfield  {author} {\bibinfo {author} {\bibnamefont {Pan}, \bibfnamefont
  {F}}, and\ \bibinfo {author} {\bibfnamefont {P.}~\bibnamefont {Zhang}}}
  (\bibinfo {year} {2022}),\ \bibfield  {title} {\enquote {\bibinfo {title}
  {Simulation of {{quantum circuits using}} the {{big-batch tensor network
  method}}},}\ }\href {https://doi.org/10.1103/PhysRevLett.128.030501}
  {\bibfield  {journal} {\bibinfo  {journal} {Phys. Rev. Lett.}\ }\textbf
  {\bibinfo {volume} {128}},\ \bibinfo {pages} {030501}}\BibitemShut {NoStop}%
\bibitem [{\citenamefont {Pan}\ \emph {et~al.}(2020)\citenamefont {Pan},
  \citenamefont {Zhou}, \citenamefont {Li},\ and\ \citenamefont
  {Zhang}}]{pan_contracting_2020}%
  \BibitemOpen
  \bibfield  {author} {\bibinfo {author} {\bibnamefont {Pan}, \bibfnamefont
  {F}}, \bibinfo {author} {\bibfnamefont {P.}~\bibnamefont {Zhou}}, \bibinfo
  {author} {\bibfnamefont {S.}~\bibnamefont {Li}}, and\ \bibinfo {author}
  {\bibfnamefont {P.}~\bibnamefont {Zhang}}} (\bibinfo {year} {2020}),\
  \bibfield  {title} {\enquote {\bibinfo {title} {Contracting {{arbitrary
  tensor networks}}: {{General approximate algorithm}} and {{applications}} in
  {{graphical models}} and {{quantum circuit simulations}}},}\ }\href
  {https://doi.org/10.1103/PhysRevLett.125.060503} {\bibfield  {journal}
  {\bibinfo  {journal} {Phys. Rev. Lett.}\ }\textbf {\bibinfo {volume} {125}},\
  \bibinfo {pages} {060503}}\BibitemShut {NoStop}%
\bibitem [{\citenamefont {Paturi}(1992)}]{paturi_degree_1992}%
  \BibitemOpen
  \bibfield  {author} {\bibinfo {author} {\bibnamefont {Paturi}, \bibfnamefont
  {R}}} (\bibinfo {year} {1992}),\ \bibfield  {title} {\enquote {\bibinfo
  {title} {{On the degree of polynomials that approximate symmetric Boolean
  functions (preliminary version)}},}\ }in\ \href
  {https://doi.org/10.1145/129712.129758} {\emph {\bibinfo {booktitle} {Proc.
  24fourth Ann. {ACM} Symp. The. Comp.}}},\ \bibinfo {series and number}
  {{STOC} '92}\ (\bibinfo  {publisher} {Association for Computing Machinery})\
  pp.\ \bibinfo {pages} {468--474}\BibitemShut {NoStop}%
\bibitem [{\citenamefont {Pednault}\ \emph {et~al.}(2017)\citenamefont
  {Pednault}, \citenamefont {Gunnels}, \citenamefont {Nannicini}, \citenamefont
  {Horesh}, \citenamefont {Magerlein}, \citenamefont {Solomonik}, \citenamefont
  {Draeger}, \citenamefont {Holland},\ and\ \citenamefont
  {Wisnieff}}]{pednault_pareto-efficient_2017}%
  \BibitemOpen
  \bibfield  {author} {\bibinfo {author} {\bibnamefont {Pednault},
  \bibfnamefont {E}}, \bibinfo {author} {\bibfnamefont {J.~A.}\ \bibnamefont
  {Gunnels}}, \bibinfo {author} {\bibfnamefont {G.}~\bibnamefont {Nannicini}},
  \bibinfo {author} {\bibfnamefont {L.}~\bibnamefont {Horesh}}, \bibinfo
  {author} {\bibfnamefont {T.}~\bibnamefont {Magerlein}}, \bibinfo {author}
  {\bibfnamefont {E.}~\bibnamefont {Solomonik}}, \bibinfo {author}
  {\bibfnamefont {E.~W.}\ \bibnamefont {Draeger}}, \bibinfo {author}
  {\bibfnamefont {E.~T.}\ \bibnamefont {Holland}}, and\ \bibinfo {author}
  {\bibfnamefont {R.}~\bibnamefont {Wisnieff}}} (\bibinfo {year} {2017}),\
  \bibfield  {title} {\enquote {\bibinfo {title} {Pareto-{{efficient quantum
  circuit simulation using tensor contraction deferral}}},}\ }\href@noop {} {\
  }\Eprint {https://arxiv.org/abs/1710.05867} {arXiv:1710.05867} \BibitemShut
  {NoStop}%
\bibitem [{\citenamefont {Pednault}\ \emph {et~al.}(2019)\citenamefont
  {Pednault}, \citenamefont {Gunnels}, \citenamefont {Nannicini}, \citenamefont
  {Horesh},\ and\ \citenamefont {Wisnieff}}]{pednault_leveraging_2019}%
  \BibitemOpen
  \bibfield  {author} {\bibinfo {author} {\bibnamefont {Pednault},
  \bibfnamefont {E}}, \bibinfo {author} {\bibfnamefont {J.~A.}\ \bibnamefont
  {Gunnels}}, \bibinfo {author} {\bibfnamefont {G.}~\bibnamefont {Nannicini}},
  \bibinfo {author} {\bibfnamefont {L.}~\bibnamefont {Horesh}}, and\ \bibinfo
  {author} {\bibfnamefont {R.}~\bibnamefont {Wisnieff}}} (\bibinfo {year}
  {2019}),\ \bibfield  {title} {\enquote {\bibinfo {title} {Leveraging
  secondary storage to simulate deep 54-qubit sycamore circuits},}\ }\href@noop
  {} {\ }\Eprint {https://arxiv.org/abs/1910.09534} {arXiv:1910.09534}
  \BibitemShut {NoStop}%
\bibitem [{\citenamefont {Peruzzo}\ \emph {et~al.}(2014)\citenamefont
  {Peruzzo}, \citenamefont {McClean}, \citenamefont {Shadbolt}, \citenamefont
  {Yung}, \citenamefont {Zhou}, \citenamefont {Love}, \citenamefont
  {Aspuru-Guzik},\ and\ \citenamefont {O'Brien}}]{Peruzzo}%
  \BibitemOpen
  \bibfield  {author} {\bibinfo {author} {\bibnamefont {Peruzzo}, \bibfnamefont
  {A}}, \bibinfo {author} {\bibfnamefont {J.}~\bibnamefont {McClean}}, \bibinfo
  {author} {\bibfnamefont {P.}~\bibnamefont {Shadbolt}}, \bibinfo {author}
  {\bibfnamefont {M.-H.}\ \bibnamefont {Yung}}, \bibinfo {author}
  {\bibfnamefont {X.-Q.}\ \bibnamefont {Zhou}}, \bibinfo {author}
  {\bibfnamefont {P.~J.}\ \bibnamefont {Love}}, \bibinfo {author}
  {\bibfnamefont {A.}~\bibnamefont {Aspuru-Guzik}}, and\ \bibinfo {author}
  {\bibfnamefont {J.~L.}\ \bibnamefont {O'Brien}}} (\bibinfo {year} {2014}),\
  \bibfield  {title} {\enquote {\bibinfo {title} {A variational eigenvalue
  solver on a photonic quantum processor},}\ }\href
  {https://doi.org/10.1038/ncomms5213} {\bibfield  {journal} {\bibinfo
  {journal} {Nature Comm.}\ }\textbf {\bibinfo {volume} {5}},\ \bibinfo {pages}
  {4213}}\BibitemShut {NoStop}%
\bibitem [{\citenamefont {Phillips}\ \emph {et~al.}(2019)\citenamefont
  {Phillips}, \citenamefont {Walschaers}, \citenamefont {Renema}, \citenamefont
  {Walmsley}, \citenamefont {Treps},\ and\ \citenamefont
  {Sperling}}]{phillips_certification_2019}%
  \BibitemOpen
  \bibfield  {author} {\bibinfo {author} {\bibnamefont {Phillips},
  \bibfnamefont {D~S}}, \bibinfo {author} {\bibfnamefont {M.}~\bibnamefont
  {Walschaers}}, \bibinfo {author} {\bibfnamefont {J.~J.}\ \bibnamefont
  {Renema}}, \bibinfo {author} {\bibfnamefont {I.~A.}\ \bibnamefont
  {Walmsley}}, \bibinfo {author} {\bibfnamefont {N.}~\bibnamefont {Treps}},
  and\ \bibinfo {author} {\bibfnamefont {J.}~\bibnamefont {Sperling}}}
  (\bibinfo {year} {2019}),\ \bibfield  {title} {\enquote {\bibinfo {title}
  {{Benchmarking of Gaussian boson sampling using two-point correlators}},}\
  }\href {https://doi.org/10.1103/PhysRevA.99.023836} {\bibfield  {journal}
  {\bibinfo  {journal} {Phys. Rev. A}\ }\textbf {\bibinfo {volume} {99}},\
  \bibinfo {pages} {023836}}\BibitemShut {NoStop}%
\bibitem [{\citenamefont {Popova}\ and\ \citenamefont
  {Rubtsov}(2021)}]{popova_cracking_2021}%
  \BibitemOpen
  \bibfield  {author} {\bibinfo {author} {\bibnamefont {Popova}, \bibfnamefont
  {A~S}}, and\ \bibinfo {author} {\bibfnamefont {A.~N.}\ \bibnamefont
  {Rubtsov}}} (\bibinfo {year} {2021}),\ \bibfield  {title} {\enquote {\bibinfo
  {title} {Cracking the {{quantum advantage}} threshold for {{Gaussian boson
  sampling}}},}\ }\href@noop {} {\ }\Eprint {https://arxiv.org/abs/2106.01445}
  {arXiv:2106.01445} \BibitemShut {NoStop}%
\bibitem [{\citenamefont {Porter}\ and\ \citenamefont
  {Thomas}(1956)}]{porter_fluctuations_1956}%
  \BibitemOpen
  \bibfield  {author} {\bibinfo {author} {\bibnamefont {Porter}, \bibfnamefont
  {C~E}}, and\ \bibinfo {author} {\bibfnamefont {R.~G.}\ \bibnamefont
  {Thomas}}} (\bibinfo {year} {1956}),\ \bibfield  {title} {\enquote {\bibinfo
  {title} {Fluctuations of {{nuclear reaction widths}}},}\ }\href
  {https://doi.org/10.1103/PhysRev.104.483} {\bibfield  {journal} {\bibinfo
  {journal} {Phys. Rev.}\ }\textbf {\bibinfo {volume} {104}},\ \bibinfo {pages}
  {483--491}}\BibitemShut {NoStop}%
\bibitem [{\citenamefont {Preskill}(2012)}]{preskill2012quantum}%
  \BibitemOpen
  \bibfield  {author} {\bibinfo {author} {\bibnamefont {Preskill},
  \bibfnamefont {J}}} (\bibinfo {year} {2012}),\ \bibfield  {title} {\enquote
  {\bibinfo {title} {Quantum computing and the entanglement frontier},}\
  }\href@noop {} {\ }\Eprint {https://arxiv.org/abs/1203.5813}
  {arXiv:1203.5813} \BibitemShut {NoStop}%
\bibitem [{\citenamefont {Qassim}\ \emph {et~al.}(2021)\citenamefont {Qassim},
  \citenamefont {Pashayan},\ and\ \citenamefont
  {Gosset}}]{qassim_improved_2021}%
  \BibitemOpen
  \bibfield  {author} {\bibinfo {author} {\bibnamefont {Qassim}, \bibfnamefont
  {H}}, \bibinfo {author} {\bibfnamefont {H.}~\bibnamefont {Pashayan}}, and\
  \bibinfo {author} {\bibfnamefont {D.}~\bibnamefont {Gosset}}} (\bibinfo
  {year} {2021}),\ \bibfield  {title} {\enquote {\bibinfo {title} {Improved
  upper bounds on the stabilizer rank of magic states},}\ }\href
  {https://doi.org/10.22331/q-2021-12-20-606} {\bibfield  {journal} {\bibinfo
  {journal} {Quantum}\ }\textbf {\bibinfo {volume} {5}},\ \bibinfo {pages}
  {606}}\BibitemShut {NoStop}%
\bibitem [{\citenamefont {Qi}\ \emph {et~al.}(2020{\natexlab{a}})\citenamefont
  {Qi}, \citenamefont {Brod}, \citenamefont {Quesada},\ and\ \citenamefont
  {{Garc{\'i}a-Patr{\'o}n}}}]{qi_regimes_2020}%
  \BibitemOpen
  \bibfield  {author} {\bibinfo {author} {\bibnamefont {Qi}, \bibfnamefont
  {H}}, \bibinfo {author} {\bibfnamefont {D.~J.}\ \bibnamefont {Brod}},
  \bibinfo {author} {\bibfnamefont {N.}~\bibnamefont {Quesada}}, and\ \bibinfo
  {author} {\bibfnamefont {R.}~\bibnamefont {{Garc{\'i}a-Patr{\'o}n}}}}
  (\bibinfo {year} {2020}{\natexlab{a}}),\ \bibfield  {title} {\enquote
  {\bibinfo {title} {Regimes of {{classical simulability}} for {{noisy Gaussian
  boson sampling}}},}\ }\href {https://doi.org/10.1103/PhysRevLett.124.100502}
  {\bibfield  {journal} {\bibinfo  {journal} {Phys. Rev. Lett.}\ }\textbf
  {\bibinfo {volume} {124}},\ \bibinfo {pages} {100502}}\BibitemShut {NoStop}%
\bibitem [{\citenamefont {Qi}\ \emph {et~al.}(2020{\natexlab{b}})\citenamefont
  {Qi}, \citenamefont {Cifuentes}, \citenamefont {Br{\'a}dler}, \citenamefont
  {Israel}, \citenamefont {Kalajdzievski},\ and\ \citenamefont
  {Quesada}}]{qi_efficient_2020}%
  \BibitemOpen
  \bibfield  {author} {\bibinfo {author} {\bibnamefont {Qi}, \bibfnamefont
  {H}}, \bibinfo {author} {\bibfnamefont {D.}~\bibnamefont {Cifuentes}},
  \bibinfo {author} {\bibfnamefont {K.}~\bibnamefont {Br{\'a}dler}}, \bibinfo
  {author} {\bibfnamefont {R.}~\bibnamefont {Israel}}, \bibinfo {author}
  {\bibfnamefont {T.}~\bibnamefont {Kalajdzievski}}, and\ \bibinfo {author}
  {\bibfnamefont {N.}~\bibnamefont {Quesada}}} (\bibinfo {year}
  {2020}{\natexlab{b}}),\ \bibfield  {title} {\enquote {\bibinfo {title}
  {Efficient sampling from shallow {{Gaussian}} quantum-optical circuits with
  local interactions},}\ }\href@noop {} {\ }\Eprint
  {https://arxiv.org/abs/2009.11824} {arXiv:2009.11824} \BibitemShut {NoStop}%
\bibitem [{\citenamefont {Quesada}(2019)}]{quesada_franck-condon_2019}%
  \BibitemOpen
  \bibfield  {author} {\bibinfo {author} {\bibnamefont {Quesada}, \bibfnamefont
  {N}}} (\bibinfo {year} {2019}),\ \bibfield  {title} {\enquote {\bibinfo
  {title} {Franck-{{Condon}} factors by counting perfect matchings of graphs
  with loops},}\ }\href {https://doi.org/10.1063/1.5086387} {\bibfield
  {journal} {\bibinfo  {journal} {J. Chem. Phys.}\ }\textbf {\bibinfo {volume}
  {150}},\ \bibinfo {pages} {164113}}\BibitemShut {NoStop}%
\bibitem [{\citenamefont {Quesada}\ and\ \citenamefont
  {Arrazola}(2020)}]{quesada_exact_2020}%
  \BibitemOpen
  \bibfield  {author} {\bibinfo {author} {\bibnamefont {Quesada}, \bibfnamefont
  {N}}, and\ \bibinfo {author} {\bibfnamefont {J.~M.}\ \bibnamefont
  {Arrazola}}} (\bibinfo {year} {2020}),\ \bibfield  {title} {\enquote
  {\bibinfo {title} {Exact simulation of {{Gaussian}} boson sampling in
  polynomial space and exponential time},}\ }\href
  {https://doi.org/10.1103/PhysRevResearch.2.023005} {\bibfield  {journal}
  {\bibinfo  {journal} {Phys. Rev. Res.}\ }\textbf {\bibinfo {volume} {2}},\
  \bibinfo {pages} {023005}}\BibitemShut {NoStop}%
\bibitem [{\citenamefont {Quesada}\ \emph {et~al.}(2018)\citenamefont
  {Quesada}, \citenamefont {Arrazola},\ and\ \citenamefont
  {Killoran}}]{quesada_gaussian_2018}%
  \BibitemOpen
  \bibfield  {author} {\bibinfo {author} {\bibnamefont {Quesada}, \bibfnamefont
  {N}}, \bibinfo {author} {\bibfnamefont {J.~M.}\ \bibnamefont {Arrazola}},
  and\ \bibinfo {author} {\bibfnamefont {N.}~\bibnamefont {Killoran}}}
  (\bibinfo {year} {2018}),\ \bibfield  {title} {\enquote {\bibinfo {title}
  {{Gaussian Boson sampling using threshold detectors}},}\ }\href
  {https://doi.org/10.1103/PhysRevA.98.062322} {\bibfield  {journal} {\bibinfo
  {journal} {Phys. Rev. A}\ }\textbf {\bibinfo {volume} {98}},\ \bibinfo
  {pages} {062322}}\BibitemShut {NoStop}%
\bibitem [{\citenamefont {Quesada}\ \emph {et~al.}(2022)\citenamefont
  {Quesada}, \citenamefont {Chadwick}, \citenamefont {Bell}, \citenamefont
  {Arrazola}, \citenamefont {Vincent}, \citenamefont {Qi},\ and\ \citenamefont
  {Garc{\'i}a-Patr{\'o}n}}]{quesada_quadratic_2022}%
  \BibitemOpen
  \bibfield  {author} {\bibinfo {author} {\bibnamefont {Quesada}, \bibfnamefont
  {N}}, \bibinfo {author} {\bibfnamefont {R.~S.}\ \bibnamefont {Chadwick}},
  \bibinfo {author} {\bibfnamefont {B.~A.}\ \bibnamefont {Bell}}, \bibinfo
  {author} {\bibfnamefont {J.~M.}\ \bibnamefont {Arrazola}}, \bibinfo {author}
  {\bibfnamefont {T.}~\bibnamefont {Vincent}}, \bibinfo {author} {\bibfnamefont
  {H.}~\bibnamefont {Qi}}, and\ \bibinfo {author} {\bibfnamefont
  {R.}~\bibnamefont {Garc{\'i}a-Patr{\'o}n}}} (\bibinfo {year} {2022}),\
  \bibfield  {title} {\enquote {\bibinfo {title} {Quadratic {{speed-up}} for
  {{simulating Gaussian boson sampling}}},}\ }\href
  {https://doi.org/10.1103/PRXQuantum.3.010306} {\bibfield  {journal} {\bibinfo
   {journal} {PRX Quantum}\ }\textbf {\bibinfo {volume} {3}},\ \bibinfo {pages}
  {010306}}\BibitemShut {NoStop}%
\bibitem [{\citenamefont {Quesada}\ \emph {et~al.}(2019)\citenamefont
  {Quesada}, \citenamefont {Helt}, \citenamefont {Izaac}, \citenamefont
  {Arrazola}, \citenamefont {Shahrokhshahi}, \citenamefont {Myers},\ and\
  \citenamefont {Sabapathy}}]{quesada_simulating_2019}%
  \BibitemOpen
  \bibfield  {author} {\bibinfo {author} {\bibnamefont {Quesada}, \bibfnamefont
  {N}}, \bibinfo {author} {\bibfnamefont {L.~G.}\ \bibnamefont {Helt}},
  \bibinfo {author} {\bibfnamefont {J.}~\bibnamefont {Izaac}}, \bibinfo
  {author} {\bibfnamefont {J.~M.}\ \bibnamefont {Arrazola}}, \bibinfo {author}
  {\bibfnamefont {R.}~\bibnamefont {Shahrokhshahi}}, \bibinfo {author}
  {\bibfnamefont {C.~R.}\ \bibnamefont {Myers}}, and\ \bibinfo {author}
  {\bibfnamefont {K.~K.}\ \bibnamefont {Sabapathy}}} (\bibinfo {year} {2019}),\
  \bibfield  {title} {\enquote {\bibinfo {title} {Simulating realistic
  non-{{Gaussian}} state preparation},}\ }\href
  {https://doi.org/10.1103/PhysRevA.100.022341} {\bibfield  {journal} {\bibinfo
   {journal} {Phys. Rev. A}\ }\textbf {\bibinfo {volume} {100}},\ \bibinfo
  {pages} {022341}}\BibitemShut {NoStop}%
\bibitem [{\citenamefont {{Rahimi-Keshari}}\ \emph {et~al.}(2015)\citenamefont
  {{Rahimi-Keshari}}, \citenamefont {Lund},\ and\ \citenamefont
  {Ralph}}]{rahimi-keshari_what_2015}%
  \BibitemOpen
  \bibfield  {author} {\bibinfo {author} {\bibnamefont {{Rahimi-Keshari}},
  \bibfnamefont {S}}, \bibinfo {author} {\bibfnamefont {A.~P.}\ \bibnamefont
  {Lund}}, and\ \bibinfo {author} {\bibfnamefont {T.~C.}\ \bibnamefont
  {Ralph}}} (\bibinfo {year} {2015}),\ \bibfield  {title} {\enquote {\bibinfo
  {title} {What can quantum optics say about computational complexity
  theory?}}\ }\href {https://doi.org/10.1103/PhysRevLett.114.060501} {\bibfield
   {journal} {\bibinfo  {journal} {Phys. Rev. Lett.}\ }\textbf {\bibinfo
  {volume} {114}},\ \bibinfo {pages} {060501}}\BibitemShut {NoStop}%
\bibitem [{\citenamefont {{Rahimi-Keshari}}\ \emph {et~al.}(2016)\citenamefont
  {{Rahimi-Keshari}}, \citenamefont {Ralph},\ and\ \citenamefont
  {Caves}}]{rahimi-keshari_sufficient_2016}%
  \BibitemOpen
  \bibfield  {author} {\bibinfo {author} {\bibnamefont {{Rahimi-Keshari}},
  \bibfnamefont {S}}, \bibinfo {author} {\bibfnamefont {T.~C.}\ \bibnamefont
  {Ralph}}, and\ \bibinfo {author} {\bibfnamefont {C.~M.}\ \bibnamefont
  {Caves}}} (\bibinfo {year} {2016}),\ \bibfield  {title} {\enquote {\bibinfo
  {title} {Sufficient {{conditions}} for {{efficient classical simulation}} of
  {{quantum optics}}},}\ }\href {https://doi.org/10.1103/PhysRevX.6.021039}
  {\bibfield  {journal} {\bibinfo  {journal} {Phys. Rev. X}\ }\textbf {\bibinfo
  {volume} {6}},\ \bibinfo {pages} {021039}}\BibitemShut {NoStop}%
\bibitem [{\citenamefont {Rakhmanov}(2007)}]{rakhmanov_bounds_2007}%
  \BibitemOpen
  \bibfield  {author} {\bibinfo {author} {\bibnamefont {Rakhmanov},
  \bibfnamefont {E~A}}} (\bibinfo {year} {2007}),\ \bibfield  {title} {\enquote
  {\bibinfo {title} {Bounds for polynomials with a unit discrete norm},}\
  }\href {https://www.jstor.org/stable/20160024} {\bibfield  {journal}
  {\bibinfo  {journal} {Ann. Math.}\ }\textbf {\bibinfo {volume} {165}},\
  \bibinfo {pages} {55--88}}\BibitemShut {NoStop}%
\bibitem [{\citenamefont {Raussendorf}\ \emph {et~al.}(2005)\citenamefont
  {Raussendorf}, \citenamefont {Bravyi},\ and\ \citenamefont
  {Harrington}}]{PhysRevA.71.062313}%
  \BibitemOpen
  \bibfield  {author} {\bibinfo {author} {\bibnamefont {Raussendorf},
  \bibfnamefont {R}}, \bibinfo {author} {\bibfnamefont {S.}~\bibnamefont
  {Bravyi}}, and\ \bibinfo {author} {\bibfnamefont {J.}~\bibnamefont
  {Harrington}}} (\bibinfo {year} {2005}),\ \bibfield  {title} {\enquote
  {\bibinfo {title} {Long-range quantum entanglement in noisy cluster
  states},}\ }\href {https://doi.org/10.1103/PhysRevA.71.062313} {\bibfield
  {journal} {\bibinfo  {journal} {Phys. Rev. A}\ }\textbf {\bibinfo {volume}
  {71}},\ \bibinfo {pages} {062313}}\BibitemShut {NoStop}%
\bibitem [{\citenamefont {Raussendorf}\ and\ \citenamefont
  {Briegel}(2001)}]{raussendorf_one-way_2001}%
  \BibitemOpen
  \bibfield  {author} {\bibinfo {author} {\bibnamefont {Raussendorf},
  \bibfnamefont {R}}, and\ \bibinfo {author} {\bibfnamefont {H.~J.}\
  \bibnamefont {Briegel}}} (\bibinfo {year} {2001}),\ \bibfield  {title}
  {\enquote {\bibinfo {title} {A one-way quantum computer},}\ }\href
  {https://doi.org/10.1103/PhysRevLett.86.5188} {\bibfield  {journal} {\bibinfo
   {journal} {Phys. Rev. Lett.}\ }\textbf {\bibinfo {volume} {86}},\ \bibinfo
  {pages} {5188--5191}}\BibitemShut {NoStop}%
\bibitem [{\citenamefont {Raussendorf}\ \emph {et~al.}(2003)\citenamefont
  {Raussendorf}, \citenamefont {Browne},\ and\ \citenamefont
  {Briegel}}]{raussendorf_measurement-based_2003}%
  \BibitemOpen
  \bibfield  {author} {\bibinfo {author} {\bibnamefont {Raussendorf},
  \bibfnamefont {R}}, \bibinfo {author} {\bibfnamefont {D.~E.}\ \bibnamefont
  {Browne}}, and\ \bibinfo {author} {\bibfnamefont {H.~J.}\ \bibnamefont
  {Briegel}}} (\bibinfo {year} {2003}),\ \bibfield  {title} {\enquote {\bibinfo
  {title} {Measurement-based quantum computation on cluster states},}\ }\href
  {https://doi.org/10.1103/PhysRevA.68.022312} {\bibfield  {journal} {\bibinfo
  {journal} {Phys. Rev. A}\ }\textbf {\bibinfo {volume} {68}},\ \bibinfo
  {pages} {022312}}\BibitemShut {NoStop}%
\bibitem [{\citenamefont {Raussendorf}\ \emph {et~al.}(2006)\citenamefont
  {Raussendorf}, \citenamefont {Harrington},\ and\ \citenamefont
  {Goyal}}]{raussendorf_fault-tolerant_2006}%
  \BibitemOpen
  \bibfield  {author} {\bibinfo {author} {\bibnamefont {Raussendorf},
  \bibfnamefont {R}}, \bibinfo {author} {\bibfnamefont {J.}~\bibnamefont
  {Harrington}}, and\ \bibinfo {author} {\bibfnamefont {K.}~\bibnamefont
  {Goyal}}} (\bibinfo {year} {2006}),\ \bibfield  {title} {\enquote {\bibinfo
  {title} {A fault-tolerant one-way quantum computer},}\ }\href
  {https://doi.org/10.1016/j.aop.2006.01.012} {\bibfield  {journal} {\bibinfo
  {journal} {Ann. Phys.}\ }\textbf {\bibinfo {volume} {321}},\ \bibinfo {pages}
  {2242--2270}}\BibitemShut {NoStop}%
\bibitem [{\citenamefont {Raz}\ and\ \citenamefont
  {Tal}(2019)}]{raz_oracle_2019}%
  \BibitemOpen
  \bibfield  {author} {\bibinfo {author} {\bibnamefont {Raz}, \bibfnamefont
  {R}}, and\ \bibinfo {author} {\bibfnamefont {A.}~\bibnamefont {Tal}}}
  (\bibinfo {year} {2019}),\ \bibfield  {title} {\enquote {\bibinfo {title}
  {Oracle separation of {{BQP}} and {{PH}}},}\ }in\ \href
  {https://doi.org/10.1145/3313276.3316315} {\emph {\bibinfo {booktitle} {Proc.
  51st {{Ann. ACM SIGACT Symp.}} {{Th.}} {{Comp.}} - {{STOC}} 2019}}}\
  (\bibinfo  {publisher} {{ACM Press}},\ \bibinfo {address} {{Phoenix, AZ,
  USA}})\ pp.\ \bibinfo {pages} {13--23}\BibitemShut {NoStop}%
\bibitem [{\citenamefont {Reagor}\ \emph {et~al.}(2018)\citenamefont {Reagor},
  \citenamefont {Osborn}, \citenamefont {Tezak}, \citenamefont {Staley},
  \citenamefont {Prawiroatmodjo}, \citenamefont {Scheer}, \citenamefont
  {Alidoust}, \citenamefont {Sete}, \citenamefont {Didier}, \citenamefont
  {da~Silva}, \citenamefont {Acala}, \citenamefont {Angeles}, \citenamefont
  {Bestwick}, \citenamefont {Block}, \citenamefont {Bloom}, \citenamefont
  {Bradley}, \citenamefont {Bui}, \citenamefont {Caldwell}, \citenamefont
  {Capelluto}, \citenamefont {Chilcott}, \citenamefont {Cordova}, \citenamefont
  {Crossman}, \citenamefont {Curtis}, \citenamefont {Deshpande}, \citenamefont
  {Bouayadi}, \citenamefont {Girshovich}, \citenamefont {Hong}, \citenamefont
  {Hudson}, \citenamefont {Karalekas}, \citenamefont {Kuang}, \citenamefont
  {Lenihan}, \citenamefont {Manenti}, \citenamefont {Manning}, \citenamefont
  {Marshall}, \citenamefont {Mohan}, \citenamefont {O'Brien}, \citenamefont
  {Otterbach}, \citenamefont {Papageorge}, \citenamefont {Paquette},
  \citenamefont {Pelstring}, \citenamefont {Polloreno}, \citenamefont {Rawat},
  \citenamefont {Ryan}, \citenamefont {Renzas}, \citenamefont {Rubin},
  \citenamefont {Russel}, \citenamefont {Rust}, \citenamefont {Scarabelli},
  \citenamefont {Selvanayagam}, \citenamefont {Sinclair}, \citenamefont
  {Smith}, \citenamefont {Suska}, \citenamefont {To}, \citenamefont
  {Vahidpour}, \citenamefont {Vodrahalli}, \citenamefont {Whyland},
  \citenamefont {Yadav}, \citenamefont {Zeng},\ and\ \citenamefont
  {Rigetti}}]{Rigetti2018}%
  \BibitemOpen
  \bibfield  {author} {\bibinfo {author} {\bibnamefont {Reagor}, \bibfnamefont
  {M}}, \bibinfo {author} {\bibfnamefont {C.~B.}\ \bibnamefont {Osborn}},
  \bibinfo {author} {\bibfnamefont {N.}~\bibnamefont {Tezak}}, \bibinfo
  {author} {\bibfnamefont {A.}~\bibnamefont {Staley}}, \bibinfo {author}
  {\bibfnamefont {G.}~\bibnamefont {Prawiroatmodjo}}, \bibinfo {author}
  {\bibfnamefont {M.}~\bibnamefont {Scheer}}, \bibinfo {author} {\bibfnamefont
  {N.}~\bibnamefont {Alidoust}}, \bibinfo {author} {\bibfnamefont {E.~A}\
  \bibnamefont {Sete}}, \bibinfo {author} {\bibfnamefont {N.}~\bibnamefont
  {Didier}}, \bibinfo {author} {\bibfnamefont {M.~P.}\ \bibnamefont
  {da~Silva}}, \bibinfo {author} {\bibfnamefont {E.}~\bibnamefont {Acala}},
  \bibinfo {author} {\bibfnamefont {J.}~\bibnamefont {Angeles}}, \bibinfo
  {author} {\bibfnamefont {A.}~\bibnamefont {Bestwick}}, \bibinfo {author}
  {\bibfnamefont {M.}~\bibnamefont {Block}}, \bibinfo {author} {\bibfnamefont
  {B.}~\bibnamefont {Bloom}}, \bibinfo {author} {\bibfnamefont
  {A.}~\bibnamefont {Bradley}}, \bibinfo {author} {\bibfnamefont
  {C.}~\bibnamefont {Bui}}, \bibinfo {author} {\bibfnamefont {S.}~\bibnamefont
  {Caldwell}}, \bibinfo {author} {\bibfnamefont {L.}~\bibnamefont {Capelluto}},
  \bibinfo {author} {\bibfnamefont {R.}~\bibnamefont {Chilcott}}, \bibinfo
  {author} {\bibfnamefont {J.}~\bibnamefont {Cordova}}, \bibinfo {author}
  {\bibfnamefont {G.}~\bibnamefont {Crossman}}, \bibinfo {author}
  {\bibfnamefont {M.}~\bibnamefont {Curtis}}, \bibinfo {author} {\bibfnamefont
  {S.}~\bibnamefont {Deshpande}}, \bibinfo {author} {\bibfnamefont {T.~El}\
  \bibnamefont {Bouayadi}}, \bibinfo {author} {\bibfnamefont {D.}~\bibnamefont
  {Girshovich}}, \bibinfo {author} {\bibfnamefont {S.}~\bibnamefont {Hong}},
  \bibinfo {author} {\bibfnamefont {A.}~\bibnamefont {Hudson}}, \bibinfo
  {author} {\bibfnamefont {P.}~\bibnamefont {Karalekas}}, \bibinfo {author}
  {\bibfnamefont {K.}~\bibnamefont {Kuang}}, \bibinfo {author} {\bibfnamefont
  {M.}~\bibnamefont {Lenihan}}, \bibinfo {author} {\bibfnamefont
  {R.}~\bibnamefont {Manenti}}, \bibinfo {author} {\bibfnamefont
  {T.}~\bibnamefont {Manning}}, \bibinfo {author} {\bibfnamefont
  {J.}~\bibnamefont {Marshall}}, \bibinfo {author} {\bibfnamefont
  {Y.}~\bibnamefont {Mohan}}, \bibinfo {author} {\bibfnamefont
  {W.}~\bibnamefont {O'Brien}}, \bibinfo {author} {\bibfnamefont
  {J.}~\bibnamefont {Otterbach}}, \bibinfo {author} {\bibfnamefont
  {A.}~\bibnamefont {Papageorge}}, \bibinfo {author} {\bibfnamefont {J.-P.}\
  \bibnamefont {Paquette}}, \bibinfo {author} {\bibfnamefont {M.}~\bibnamefont
  {Pelstring}}, \bibinfo {author} {\bibfnamefont {A.}~\bibnamefont
  {Polloreno}}, \bibinfo {author} {\bibfnamefont {V.}~\bibnamefont {Rawat}},
  \bibinfo {author} {\bibfnamefont {C.~A.}\ \bibnamefont {Ryan}}, \bibinfo
  {author} {\bibfnamefont {R.}~\bibnamefont {Renzas}}, \bibinfo {author}
  {\bibfnamefont {N.}~\bibnamefont {Rubin}}, \bibinfo {author} {\bibfnamefont
  {D.}~\bibnamefont {Russel}}, \bibinfo {author} {\bibfnamefont
  {M.}~\bibnamefont {Rust}}, \bibinfo {author} {\bibfnamefont {D.}~\bibnamefont
  {Scarabelli}}, \bibinfo {author} {\bibfnamefont {M.}~\bibnamefont
  {Selvanayagam}}, \bibinfo {author} {\bibfnamefont {R.}~\bibnamefont
  {Sinclair}}, \bibinfo {author} {\bibfnamefont {R.}~\bibnamefont {Smith}},
  \bibinfo {author} {\bibfnamefont {M.}~\bibnamefont {Suska}}, \bibinfo
  {author} {\bibfnamefont {T.-W.}\ \bibnamefont {To}}, \bibinfo {author}
  {\bibfnamefont {M.}~\bibnamefont {Vahidpour}}, \bibinfo {author}
  {\bibfnamefont {N.}~\bibnamefont {Vodrahalli}}, \bibinfo {author}
  {\bibfnamefont {T.}~\bibnamefont {Whyland}}, \bibinfo {author} {\bibfnamefont
  {K.}~\bibnamefont {Yadav}}, \bibinfo {author} {\bibfnamefont
  {W.}~\bibnamefont {Zeng}}, and\ \bibinfo {author} {\bibfnamefont {C.~T}\
  \bibnamefont {Rigetti}}} (\bibinfo {year} {2018}),\ \bibfield  {title}
  {\enquote {\bibinfo {title} {Demonstration of universal parametric entangling
  gates on a multi-qubit lattice},}\ }\href
  {https://doi.org/10.1126/sciadv.aao3603} {\bibfield  {journal} {\bibinfo
  {journal} {Science Adv.}\ }\textbf {\bibinfo {volume} {4}},\ \bibinfo {pages}
  {eaao3603}}\BibitemShut {NoStop}%
\bibitem [{\citenamefont {Reed}(1954)}]{reed_class_1954}%
  \BibitemOpen
  \bibfield  {author} {\bibinfo {author} {\bibnamefont {Reed}, \bibfnamefont
  {I~S}}} (\bibinfo {year} {1954}),\ \bibfield  {title} {\enquote {\bibinfo
  {title} {A class of multiple-error-correcting codes and the decoding
  scheme},}\ }\href {https://doi.org/10.1109/TIT.1954.1057465} {\bibfield
  {journal} {\bibinfo  {journal} {Trans. {IRE} Prof. Gr. Inf. Th.}\ }\textbf
  {\bibinfo {volume} {4}},\ \bibinfo {pages} {38--49}}\BibitemShut {NoStop}%
\bibitem [{\citenamefont {Reed}\ and\ \citenamefont
  {Solomon}(1960)}]{reed_polynomial_1960}%
  \BibitemOpen
  \bibfield  {author} {\bibinfo {author} {\bibnamefont {Reed}, \bibfnamefont
  {I~S}}, and\ \bibinfo {author} {\bibfnamefont {G.}~\bibnamefont {Solomon}}}
  (\bibinfo {year} {1960}),\ \bibfield  {title} {\enquote {\bibinfo {title}
  {Polynomial codes over certain finite fields},}\ }\href
  {https://doi.org/10.1137/0108018} {\bibfield  {journal} {\bibinfo  {journal}
  {J. Soc. Ind. Appl. Math.}\ }\textbf {\bibinfo {volume} {8}},\ \bibinfo
  {pages} {300--304}}\BibitemShut {NoStop}%
\bibitem [{\citenamefont
  {Renema}(2020{\natexlab{a}})}]{renema_marginal_2020-1}%
  \BibitemOpen
  \bibfield  {author} {\bibinfo {author} {\bibnamefont {Renema}, \bibfnamefont
  {J~J}}} (\bibinfo {year} {2020}{\natexlab{a}}),\ \bibfield  {title} {\enquote
  {\bibinfo {title} {Marginal probabilities in boson samplers with arbitrary
  input states},}\ }\href@noop {} {\ }\Eprint
  {https://arxiv.org/abs/2012.14917} {arXiv:2012.14917} \BibitemShut {NoStop}%
\bibitem [{\citenamefont
  {Renema}(2020{\natexlab{b}})}]{renema_simulability_2020}%
  \BibitemOpen
  \bibfield  {author} {\bibinfo {author} {\bibnamefont {Renema}, \bibfnamefont
  {J~J}}} (\bibinfo {year} {2020}{\natexlab{b}}),\ \bibfield  {title} {\enquote
  {\bibinfo {title} {Simulability of partially distinguishable superposition
  and {{Gaussian}} boson sampling},}\ }\href
  {https://doi.org/10.1103/PhysRevA.101.063840} {\bibfield  {journal} {\bibinfo
   {journal} {Phys. Rev. A}\ }\textbf {\bibinfo {volume} {101}},\ \bibinfo
  {pages} {063840}}\BibitemShut {NoStop}%
\bibitem [{\citenamefont {Renema}\ \emph {et~al.}(2018)\citenamefont {Renema},
  \citenamefont {Menssen}, \citenamefont {Clements}, \citenamefont {Triginer},
  \citenamefont {Kolthammer},\ and\ \citenamefont
  {Walmsley}}]{renema_efficient_2018}%
  \BibitemOpen
  \bibfield  {author} {\bibinfo {author} {\bibnamefont {Renema}, \bibfnamefont
  {J~J}}, \bibinfo {author} {\bibfnamefont {A.}~\bibnamefont {Menssen}},
  \bibinfo {author} {\bibfnamefont {W.~R.}\ \bibnamefont {Clements}}, \bibinfo
  {author} {\bibfnamefont {G.}~\bibnamefont {Triginer}}, \bibinfo {author}
  {\bibfnamefont {W.~S.}\ \bibnamefont {Kolthammer}}, and\ \bibinfo {author}
  {\bibfnamefont {I.~A.}\ \bibnamefont {Walmsley}}} (\bibinfo {year} {2018}),\
  \bibfield  {title} {\enquote {\bibinfo {title} {Efficient classical algorithm
  for boson sampling with partially distinguishable photons},}\ }\href
  {https://doi.org/10.1103/PhysRevLett.120.220502} {\bibfield  {journal}
  {\bibinfo  {journal} {Phys. Rev. Lett.}\ }\textbf {\bibinfo {volume} {120}},\
  \bibinfo {pages} {220502}}\BibitemShut {NoStop}%
\bibitem [{\citenamefont {Renema}\ \emph {et~al.}(2019)\citenamefont {Renema},
  \citenamefont {Shchesnovich},\ and\ \citenamefont
  {Garc{\'i}a-Patr{\'o}n}}]{renema_classical_2019}%
  \BibitemOpen
  \bibfield  {author} {\bibinfo {author} {\bibnamefont {Renema}, \bibfnamefont
  {J~J}}, \bibinfo {author} {\bibfnamefont {V.~S.}\ \bibnamefont
  {Shchesnovich}}, and\ \bibinfo {author} {\bibfnamefont {R.}~\bibnamefont
  {Garc{\'i}a-Patr{\'o}n}}} (\bibinfo {year} {2019}),\ \bibfield  {title}
  {\enquote {\bibinfo {title} {Classical simulability of noisy boson
  sampling},}\ }\href@noop {} {\ }\Eprint {https://arxiv.org/abs/1809.01953v2}
  {arxiv:1809.01953v2} \BibitemShut {NoStop}%
\bibitem [{\citenamefont {Ringbauer}\ \emph {et~al.}(2022)\citenamefont
  {Ringbauer}, \citenamefont {Hinsche}, \citenamefont {Feldker}, \citenamefont
  {Faehrmann}, \citenamefont {Bermejo-Vega}, \citenamefont {Edmunds},
  \citenamefont {Stricker}, \citenamefont {Marciniak}, \citenamefont {Meth},
  \citenamefont {Pogorelov}, \citenamefont {Postler}, \citenamefont {Blatt},
  \citenamefont {Schindler}, \citenamefont {Eisert}, \citenamefont {Monz},\
  and\ \citenamefont {Hangleiter}}]{IonSampling}%
  \BibitemOpen
  \bibfield  {author} {\bibinfo {author} {\bibnamefont {Ringbauer},
  \bibfnamefont {M}}, \bibinfo {author} {\bibfnamefont {M.}~\bibnamefont
  {Hinsche}}, \bibinfo {author} {\bibfnamefont {T.}~\bibnamefont {Feldker}},
  \bibinfo {author} {\bibfnamefont {P.~K.}\ \bibnamefont {Faehrmann}}, \bibinfo
  {author} {\bibfnamefont {J.}~\bibnamefont {Bermejo-Vega}}, \bibinfo {author}
  {\bibfnamefont {C.}~\bibnamefont {Edmunds}}, \bibinfo {author} {\bibfnamefont
  {R.}~\bibnamefont {Stricker}}, \bibinfo {author} {\bibfnamefont {C.~D.}\
  \bibnamefont {Marciniak}}, \bibinfo {author} {\bibfnamefont {M.}~\bibnamefont
  {Meth}}, \bibinfo {author} {\bibfnamefont {I.}~\bibnamefont {Pogorelov}},
  \bibinfo {author} {\bibfnamefont {L.}~\bibnamefont {Postler}}, \bibinfo
  {author} {\bibfnamefont {R.}~\bibnamefont {Blatt}}, \bibinfo {author}
  {\bibfnamefont {P.}~\bibnamefont {Schindler}}, \bibinfo {author}
  {\bibfnamefont {J.}~\bibnamefont {Eisert}}, \bibinfo {author} {\bibfnamefont
  {T.}~\bibnamefont {Monz}}, and\ \bibinfo {author} {\bibfnamefont
  {D.}~\bibnamefont {Hangleiter}}} (\bibinfo {year} {2022}),\ \bibfield
  {title} {\enquote {\bibinfo {title} {Verifiable measurement-based quantum
  random sampling with trapped ions},}\ }\href@noop {} {\ }\bibinfo {note}
  {(forthcoming)}\BibitemShut {NoStop}%
\bibitem [{\citenamefont {Rinott}\ \emph {et~al.}(2022)\citenamefont {Rinott},
  \citenamefont {Shoham},\ and\ \citenamefont
  {Kalai}}]{rinott_statistical_2022}%
  \BibitemOpen
  \bibfield  {author} {\bibinfo {author} {\bibnamefont {Rinott}, \bibfnamefont
  {Y}}, \bibinfo {author} {\bibfnamefont {T.}~\bibnamefont {Shoham}}, and\
  \bibinfo {author} {\bibfnamefont {G.}~\bibnamefont {Kalai}}} (\bibinfo {year}
  {2022}),\ \bibfield  {title} {\enquote {\bibinfo {title} {Statistical
  {{aspects}} of the {{quantum supremacy demonstration}}},}\ }\href
  {https://doi.org/10.1214/21-STS836} {\bibfield  {journal} {\bibinfo
  {journal} {Stat. Sc.}\ }\textbf {\bibinfo {volume} {37}}~(\bibinfo {number}
  {3}),\ \bibinfo {pages} {322--347}},\ \Eprint
  {https://arxiv.org/abs/2008.05177} {arXiv:2008.05177} \BibitemShut {NoStop}%
\bibitem [{\citenamefont {Roga}\ and\ \citenamefont
  {Takeoka}(2020)}]{RogaSimulation}%
  \BibitemOpen
  \bibfield  {author} {\bibinfo {author} {\bibnamefont {Roga}, \bibfnamefont
  {W}}, and\ \bibinfo {author} {\bibfnamefont {M.}~\bibnamefont {Takeoka}}}
  (\bibinfo {year} {2020}),\ \bibfield  {title} {\enquote {\bibinfo {title}
  {Classical simulation of boson sampling with sparse output},}\ }\href
  {https://doi.org/10.1038/s41598-020-71892-0} {\bibfield  {journal} {\bibinfo
  {journal} {Sci. Rep.}\ }\textbf {\bibinfo {volume} {10}},\ \bibinfo {pages}
  {14739}}\BibitemShut {NoStop}%
\bibitem [{\citenamefont {{Ryan-Anderson}}\ \emph {et~al.}(2022)\citenamefont
  {{Ryan-Anderson}}, \citenamefont {Brown}, \citenamefont {Allman},
  \citenamefont {Arkin}, \citenamefont {{Asa-Attuah}}, \citenamefont {Baldwin},
  \citenamefont {Berg}, \citenamefont {Bohnet}, \citenamefont {Braxton},
  \citenamefont {Burdick}, \citenamefont {Campora}, \citenamefont
  {Chernoguzov}, \citenamefont {Esposito}, \citenamefont {Evans}, \citenamefont
  {Francois}, \citenamefont {Gaebler}, \citenamefont {Gatterman}, \citenamefont
  {Gerber}, \citenamefont {Gilmore}, \citenamefont {Gresh}, \citenamefont
  {Hall}, \citenamefont {Hankin}, \citenamefont {Hostetter}, \citenamefont
  {Lucchetti}, \citenamefont {Mayer}, \citenamefont {Myers}, \citenamefont
  {Neyenhuis}, \citenamefont {Santiago}, \citenamefont {Sedlacek},
  \citenamefont {Skripka}, \citenamefont {Slattery}, \citenamefont {Stutz},
  \citenamefont {Tait}, \citenamefont {Tobey}, \citenamefont {Vittorini},
  \citenamefont {Walker},\ and\ \citenamefont
  {Hayes}}]{ryan-anderson_implementing_2022}%
  \BibitemOpen
  \bibfield  {author} {\bibinfo {author} {\bibnamefont {{Ryan-Anderson}},
  \bibfnamefont {C}}, \bibinfo {author} {\bibfnamefont {N.~C.}\ \bibnamefont
  {Brown}}, \bibinfo {author} {\bibfnamefont {M.~S.}\ \bibnamefont {Allman}},
  \bibinfo {author} {\bibfnamefont {B.}~\bibnamefont {Arkin}}, \bibinfo
  {author} {\bibfnamefont {G.}~\bibnamefont {{Asa-Attuah}}}, \bibinfo {author}
  {\bibfnamefont {C.}~\bibnamefont {Baldwin}}, \bibinfo {author} {\bibfnamefont
  {J.}~\bibnamefont {Berg}}, \bibinfo {author} {\bibfnamefont {J.~G.}\
  \bibnamefont {Bohnet}}, \bibinfo {author} {\bibfnamefont {S.}~\bibnamefont
  {Braxton}}, \bibinfo {author} {\bibfnamefont {N.}~\bibnamefont {Burdick}},
  \bibinfo {author} {\bibfnamefont {J.~P.}\ \bibnamefont {Campora}}, \bibinfo
  {author} {\bibfnamefont {A.}~\bibnamefont {Chernoguzov}}, \bibinfo {author}
  {\bibfnamefont {J.}~\bibnamefont {Esposito}}, \bibinfo {author}
  {\bibfnamefont {B.}~\bibnamefont {Evans}}, \bibinfo {author} {\bibfnamefont
  {D.}~\bibnamefont {Francois}}, \bibinfo {author} {\bibfnamefont {J.~P.}\
  \bibnamefont {Gaebler}}, \bibinfo {author} {\bibfnamefont {T.~M.}\
  \bibnamefont {Gatterman}}, \bibinfo {author} {\bibfnamefont {J.}~\bibnamefont
  {Gerber}}, \bibinfo {author} {\bibfnamefont {K.}~\bibnamefont {Gilmore}},
  \bibinfo {author} {\bibfnamefont {D.}~\bibnamefont {Gresh}}, \bibinfo
  {author} {\bibfnamefont {A.}~\bibnamefont {Hall}}, \bibinfo {author}
  {\bibfnamefont {A.}~\bibnamefont {Hankin}}, \bibinfo {author} {\bibfnamefont
  {J.}~\bibnamefont {Hostetter}}, \bibinfo {author} {\bibfnamefont
  {D.}~\bibnamefont {Lucchetti}}, \bibinfo {author} {\bibfnamefont
  {K.}~\bibnamefont {Mayer}}, \bibinfo {author} {\bibfnamefont
  {J.}~\bibnamefont {Myers}}, \bibinfo {author} {\bibfnamefont
  {B.}~\bibnamefont {Neyenhuis}}, \bibinfo {author} {\bibfnamefont
  {J.}~\bibnamefont {Santiago}}, \bibinfo {author} {\bibfnamefont
  {J.}~\bibnamefont {Sedlacek}}, \bibinfo {author} {\bibfnamefont
  {T.}~\bibnamefont {Skripka}}, \bibinfo {author} {\bibfnamefont
  {A.}~\bibnamefont {Slattery}}, \bibinfo {author} {\bibfnamefont {R.~P.}\
  \bibnamefont {Stutz}}, \bibinfo {author} {\bibfnamefont {J.}~\bibnamefont
  {Tait}}, \bibinfo {author} {\bibfnamefont {R.}~\bibnamefont {Tobey}},
  \bibinfo {author} {\bibfnamefont {G.}~\bibnamefont {Vittorini}}, \bibinfo
  {author} {\bibfnamefont {J.}~\bibnamefont {Walker}}, and\ \bibinfo {author}
  {\bibfnamefont {D.}~\bibnamefont {Hayes}}} (\bibinfo {year} {2022}),\
  \bibfield  {title} {\enquote {\bibinfo {title} {Implementing {{fault-tolerant
  entangling gates}} on the {{five-qubit code}} and the {{color code}}},}\
  }\href@noop {} {\ }\Eprint {https://arxiv.org/abs/2208.01863}
  {arXiv:2208.01863} \BibitemShut {NoStop}%
\bibitem [{\citenamefont {Ryser}(1963)}]{ryser_combinatorial_1963-1}%
  \BibitemOpen
  \bibfield  {author} {\bibinfo {author} {\bibnamefont {Ryser}, \bibfnamefont
  {H~J}}} (\bibinfo {year} {1963}),\ \href
  {https://doi.org/10.5948/UPO9781614440147} {\emph {\bibinfo {title}
  {Combinatorial {{mathematics}}}}}\ (\bibinfo  {publisher} {{American
  Mathematical Soc.}})\BibitemShut {NoStop}%
\bibitem [{\citenamefont {Satzinger}\ \emph {et~al.}(2021)\citenamefont
  {Satzinger}, \citenamefont {Liu}, \citenamefont {Smith}, \citenamefont
  {Knapp}, \citenamefont {Newman}, \citenamefont {Jones}, \citenamefont {Chen},
  \citenamefont {Quintana}, \citenamefont {Mi}, \citenamefont {Dunsworth},
  \citenamefont {Gidney}, \citenamefont {Aleiner}, \citenamefont {Arute},
  \citenamefont {Arya}, \citenamefont {Atalaya}, \citenamefont {Babbush},
  \citenamefont {Bardin}, \citenamefont {Barends}, \citenamefont {Basso},
  \citenamefont {Bengtsson}, \citenamefont {Bilmes}, \citenamefont {Broughton},
  \citenamefont {Buckley}, \citenamefont {Buell}, \citenamefont {Burkett},
  \citenamefont {Bushnell}, \citenamefont {Chiaro}, \citenamefont {Collins},
  \citenamefont {Courtney}, \citenamefont {Demura}, \citenamefont {Derk},
  \citenamefont {Eppens}, \citenamefont {Erickson}, \citenamefont {Faoro},
  \citenamefont {Farhi}, \citenamefont {Fowler}, \citenamefont {Foxen},
  \citenamefont {Giustina}, \citenamefont {Greene}, \citenamefont {Gross},
  \citenamefont {Harrigan}, \citenamefont {Harrington}, \citenamefont {Hilton},
  \citenamefont {Hong}, \citenamefont {Huang}, \citenamefont {Huggins},
  \citenamefont {Ioffe}, \citenamefont {Isakov}, \citenamefont {Jeffrey},
  \citenamefont {Jiang}, \citenamefont {Kafri}, \citenamefont {Kechedzhi},
  \citenamefont {Khattar}, \citenamefont {Kim}, \citenamefont {Klimov},
  \citenamefont {Korotkov}, \citenamefont {Kostritsa}, \citenamefont
  {Landhuis}, \citenamefont {Laptev}, \citenamefont {Locharla}, \citenamefont
  {Lucero}, \citenamefont {Martin}, \citenamefont {McClean}, \citenamefont
  {McEwen}, \citenamefont {Miao}, \citenamefont {Mohseni}, \citenamefont
  {Montazeri}, \citenamefont {Mruczkiewicz}, \citenamefont {Mutus},
  \citenamefont {Naaman}, \citenamefont {Neeley}, \citenamefont {Neill},
  \citenamefont {Niu}, \citenamefont {O'Brien}, \citenamefont {Opremcak},
  \citenamefont {Pat{\'o}}, \citenamefont {Petukhov}, \citenamefont {Rubin},
  \citenamefont {Sank}, \citenamefont {Shvarts}, \citenamefont {Strain},
  \citenamefont {Szalay}, \citenamefont {Villalonga}, \citenamefont {White},
  \citenamefont {Yao}, \citenamefont {Yeh}, \citenamefont {Yoo}, \citenamefont
  {Zalcman}, \citenamefont {Neven}, \citenamefont {Boixo}, \citenamefont
  {Megrant}, \citenamefont {Chen}, \citenamefont {Kelly}, \citenamefont
  {Smelyanskiy}, \citenamefont {Kitaev}, \citenamefont {Knap}, \citenamefont
  {Pollmann},\ and\ \citenamefont {Roushan}}]{satzinger_realizing_2021}%
  \BibitemOpen
  \bibfield  {author} {\bibinfo {author} {\bibnamefont {Satzinger},
  \bibfnamefont {K~J}}, \bibinfo {author} {\bibfnamefont {Y.-J}\ \bibnamefont
  {Liu}}, \bibinfo {author} {\bibfnamefont {A.}~\bibnamefont {Smith}}, \bibinfo
  {author} {\bibfnamefont {C.}~\bibnamefont {Knapp}}, \bibinfo {author}
  {\bibfnamefont {M.}~\bibnamefont {Newman}}, \bibinfo {author} {\bibfnamefont
  {C.}~\bibnamefont {Jones}}, \bibinfo {author} {\bibfnamefont
  {Z.}~\bibnamefont {Chen}}, \bibinfo {author} {\bibfnamefont {C.}~\bibnamefont
  {Quintana}}, \bibinfo {author} {\bibfnamefont {X.}~\bibnamefont {Mi}},
  \bibinfo {author} {\bibfnamefont {A.}~\bibnamefont {Dunsworth}}, \bibinfo
  {author} {\bibfnamefont {C.}~\bibnamefont {Gidney}}, \bibinfo {author}
  {\bibfnamefont {I.}~\bibnamefont {Aleiner}}, \bibinfo {author} {\bibfnamefont
  {F.}~\bibnamefont {Arute}}, \bibinfo {author} {\bibfnamefont
  {K.}~\bibnamefont {Arya}}, \bibinfo {author} {\bibfnamefont {J.}~\bibnamefont
  {Atalaya}}, \bibinfo {author} {\bibfnamefont {R.}~\bibnamefont {Babbush}},
  \bibinfo {author} {\bibfnamefont {J.~C.}\ \bibnamefont {Bardin}}, \bibinfo
  {author} {\bibfnamefont {R.}~\bibnamefont {Barends}}, \bibinfo {author}
  {\bibfnamefont {J.}~\bibnamefont {Basso}}, \bibinfo {author} {\bibfnamefont
  {A.}~\bibnamefont {Bengtsson}}, \bibinfo {author} {\bibfnamefont
  {A.}~\bibnamefont {Bilmes}}, \bibinfo {author} {\bibfnamefont
  {M.}~\bibnamefont {Broughton}}, \bibinfo {author} {\bibfnamefont {B.~B.}\
  \bibnamefont {Buckley}}, \bibinfo {author} {\bibfnamefont {D.~A.}\
  \bibnamefont {Buell}}, \bibinfo {author} {\bibfnamefont {B.}~\bibnamefont
  {Burkett}}, \bibinfo {author} {\bibfnamefont {N.}~\bibnamefont {Bushnell}},
  \bibinfo {author} {\bibfnamefont {B.}~\bibnamefont {Chiaro}}, \bibinfo
  {author} {\bibfnamefont {R.}~\bibnamefont {Collins}}, \bibinfo {author}
  {\bibfnamefont {W.}~\bibnamefont {Courtney}}, \bibinfo {author}
  {\bibfnamefont {S.}~\bibnamefont {Demura}}, \bibinfo {author} {\bibfnamefont
  {A.~R.}\ \bibnamefont {Derk}}, \bibinfo {author} {\bibfnamefont
  {D.}~\bibnamefont {Eppens}}, \bibinfo {author} {\bibfnamefont
  {C.}~\bibnamefont {Erickson}}, \bibinfo {author} {\bibfnamefont
  {L.}~\bibnamefont {Faoro}}, \bibinfo {author} {\bibfnamefont
  {E.}~\bibnamefont {Farhi}}, \bibinfo {author} {\bibfnamefont {A.~G.}\
  \bibnamefont {Fowler}}, \bibinfo {author} {\bibfnamefont {B.}~\bibnamefont
  {Foxen}}, \bibinfo {author} {\bibfnamefont {M.}~\bibnamefont {Giustina}},
  \bibinfo {author} {\bibfnamefont {A.}~\bibnamefont {Greene}}, \bibinfo
  {author} {\bibfnamefont {J.~A.}\ \bibnamefont {Gross}}, \bibinfo {author}
  {\bibfnamefont {M.~P.}\ \bibnamefont {Harrigan}}, \bibinfo {author}
  {\bibfnamefont {S.~D.}\ \bibnamefont {Harrington}}, \bibinfo {author}
  {\bibfnamefont {J.}~\bibnamefont {Hilton}}, \bibinfo {author} {\bibfnamefont
  {S.}~\bibnamefont {Hong}}, \bibinfo {author} {\bibfnamefont {T.}~\bibnamefont
  {Huang}}, \bibinfo {author} {\bibfnamefont {W.~J.}\ \bibnamefont {Huggins}},
  \bibinfo {author} {\bibfnamefont {L.~B.}\ \bibnamefont {Ioffe}}, \bibinfo
  {author} {\bibfnamefont {S.~V.}\ \bibnamefont {Isakov}}, \bibinfo {author}
  {\bibfnamefont {E.}~\bibnamefont {Jeffrey}}, \bibinfo {author} {\bibfnamefont
  {Z.}~\bibnamefont {Jiang}}, \bibinfo {author} {\bibfnamefont
  {D.}~\bibnamefont {Kafri}}, \bibinfo {author} {\bibfnamefont
  {K.}~\bibnamefont {Kechedzhi}}, \bibinfo {author} {\bibfnamefont
  {T.}~\bibnamefont {Khattar}}, \bibinfo {author} {\bibfnamefont
  {S.}~\bibnamefont {Kim}}, \bibinfo {author} {\bibfnamefont {P.~V.}\
  \bibnamefont {Klimov}}, \bibinfo {author} {\bibfnamefont {A.~N.}\
  \bibnamefont {Korotkov}}, \bibinfo {author} {\bibfnamefont {F.}~\bibnamefont
  {Kostritsa}}, \bibinfo {author} {\bibfnamefont {D.}~\bibnamefont {Landhuis}},
  \bibinfo {author} {\bibfnamefont {P.}~\bibnamefont {Laptev}}, \bibinfo
  {author} {\bibfnamefont {A.}~\bibnamefont {Locharla}}, \bibinfo {author}
  {\bibfnamefont {E.}~\bibnamefont {Lucero}}, \bibinfo {author} {\bibfnamefont
  {O.}~\bibnamefont {Martin}}, \bibinfo {author} {\bibfnamefont {J.~R.}\
  \bibnamefont {McClean}}, \bibinfo {author} {\bibfnamefont {M.}~\bibnamefont
  {McEwen}}, \bibinfo {author} {\bibfnamefont {K.~C.}\ \bibnamefont {Miao}},
  \bibinfo {author} {\bibfnamefont {M.}~\bibnamefont {Mohseni}}, \bibinfo
  {author} {\bibfnamefont {S.}~\bibnamefont {Montazeri}}, \bibinfo {author}
  {\bibfnamefont {W.}~\bibnamefont {Mruczkiewicz}}, \bibinfo {author}
  {\bibfnamefont {J.}~\bibnamefont {Mutus}}, \bibinfo {author} {\bibfnamefont
  {O.}~\bibnamefont {Naaman}}, \bibinfo {author} {\bibfnamefont
  {M.}~\bibnamefont {Neeley}}, \bibinfo {author} {\bibfnamefont
  {C.}~\bibnamefont {Neill}}, \bibinfo {author} {\bibfnamefont {M.~Y.}\
  \bibnamefont {Niu}}, \bibinfo {author} {\bibfnamefont {T.~E.}\ \bibnamefont
  {O'Brien}}, \bibinfo {author} {\bibfnamefont {A.}~\bibnamefont {Opremcak}},
  \bibinfo {author} {\bibfnamefont {B.}~\bibnamefont {Pat{\'o}}}, \bibinfo
  {author} {\bibfnamefont {A.}~\bibnamefont {Petukhov}}, \bibinfo {author}
  {\bibfnamefont {N.~C.}\ \bibnamefont {Rubin}}, \bibinfo {author}
  {\bibfnamefont {D.}~\bibnamefont {Sank}}, \bibinfo {author} {\bibfnamefont
  {V.}~\bibnamefont {Shvarts}}, \bibinfo {author} {\bibfnamefont
  {D.}~\bibnamefont {Strain}}, \bibinfo {author} {\bibfnamefont
  {M.}~\bibnamefont {Szalay}}, \bibinfo {author} {\bibfnamefont
  {B.}~\bibnamefont {Villalonga}}, \bibinfo {author} {\bibfnamefont {T.~C.}\
  \bibnamefont {White}}, \bibinfo {author} {\bibfnamefont {Z.}~\bibnamefont
  {Yao}}, \bibinfo {author} {\bibfnamefont {P.}~\bibnamefont {Yeh}}, \bibinfo
  {author} {\bibfnamefont {J.}~\bibnamefont {Yoo}}, \bibinfo {author}
  {\bibfnamefont {A.}~\bibnamefont {Zalcman}}, \bibinfo {author} {\bibfnamefont
  {H.}~\bibnamefont {Neven}}, \bibinfo {author} {\bibfnamefont
  {S.}~\bibnamefont {Boixo}}, \bibinfo {author} {\bibfnamefont
  {A.}~\bibnamefont {Megrant}}, \bibinfo {author} {\bibfnamefont
  {Y.}~\bibnamefont {Chen}}, \bibinfo {author} {\bibfnamefont {J.}~\bibnamefont
  {Kelly}}, \bibinfo {author} {\bibfnamefont {V.}~\bibnamefont {Smelyanskiy}},
  \bibinfo {author} {\bibfnamefont {A.}~\bibnamefont {Kitaev}}, \bibinfo
  {author} {\bibfnamefont {M.}~\bibnamefont {Knap}}, \bibinfo {author}
  {\bibfnamefont {F.}~\bibnamefont {Pollmann}}, and\ \bibinfo {author}
  {\bibfnamefont {P.}~\bibnamefont {Roushan}}} (\bibinfo {year} {2021}),\
  \bibfield  {title} {\enquote {\bibinfo {title} {Realizing topologically
  ordered states on a quantum processor},}\ }\href
  {https://doi.org/10.1126/science.abi8378} {\bibfield  {journal} {\bibinfo
  {journal} {Science}\ }\textbf {\bibinfo {volume} {374}},\ \bibinfo {pages}
  {1237--1241}}\BibitemShut {NoStop}%
\bibitem [{\citenamefont {Scheel}(2008)}]{scheel_permanents_2004}%
  \BibitemOpen
  \bibfield  {author} {\bibinfo {author} {\bibnamefont {Scheel}, \bibfnamefont
  {S}}} (\bibinfo {year} {2008}),\ \bibfield  {title} {\enquote {\bibinfo
  {title} {Permanents in linear optical networks},}\ }\href
  {https://doi.org/10.2478/v10155-010-0092-x} {\bibfield  {journal} {\bibinfo
  {journal} {Acta Phys. Slov.}\ }\textbf {\bibinfo {volume} {58}},\ \bibinfo
  {pages} {675}}\BibitemShut {NoStop}%
\bibitem [{\citenamefont
  {Schollw{\"o}ck}(2005)}]{schollwock_density-matrix_2005}%
  \BibitemOpen
  \bibfield  {author} {\bibinfo {author} {\bibnamefont {Schollw{\"o}ck},
  \bibfnamefont {U}}} (\bibinfo {year} {2005}),\ \bibfield  {title} {\enquote
  {\bibinfo {title} {The density-matrix renormalization group},}\ }\href
  {https://doi.org/10.1103/RevModPhys.77.259} {\bibfield  {journal} {\bibinfo
  {journal} {Rev. Mod. Phys.}\ }\textbf {\bibinfo {volume} {77}},\ \bibinfo
  {pages} {259--315}}\BibitemShut {NoStop}%
\bibitem [{\citenamefont {Schuch}\ \emph {et~al.}(2007)\citenamefont {Schuch},
  \citenamefont {Wolf}, \citenamefont {Verstraete},\ and\ \citenamefont
  {Cirac}}]{schuch_computational_2007}%
  \BibitemOpen
  \bibfield  {author} {\bibinfo {author} {\bibnamefont {Schuch}, \bibfnamefont
  {N}}, \bibinfo {author} {\bibfnamefont {M.~M.}\ \bibnamefont {Wolf}},
  \bibinfo {author} {\bibfnamefont {F.}~\bibnamefont {Verstraete}}, and\
  \bibinfo {author} {\bibfnamefont {J.~I.}\ \bibnamefont {Cirac}}} (\bibinfo
  {year} {2007}),\ \bibfield  {title} {\enquote {\bibinfo {title}
  {Computational complexity of projected entangled pair states},}\ }\href
  {https://doi.org/10.1103/PhysRevLett.98.140506} {\bibfield  {journal}
  {\bibinfo  {journal} {Phys. Rev. Lett.}\ }\textbf {\bibinfo {volume} {98}},\
  \bibinfo {pages} {140506}}\BibitemShut {NoStop}%
\bibitem [{\citenamefont {Schuld}\ \emph {et~al.}(2020)\citenamefont {Schuld},
  \citenamefont {Br{\'a}dler}, \citenamefont {Israel}, \citenamefont {Su},\
  and\ \citenamefont {Gupt}}]{schuld_quantum_2020}%
  \BibitemOpen
  \bibfield  {author} {\bibinfo {author} {\bibnamefont {Schuld}, \bibfnamefont
  {M}}, \bibinfo {author} {\bibfnamefont {K.}~\bibnamefont {Br{\'a}dler}},
  \bibinfo {author} {\bibfnamefont {R.}~\bibnamefont {Israel}}, \bibinfo
  {author} {\bibfnamefont {D.}~\bibnamefont {Su}}, and\ \bibinfo {author}
  {\bibfnamefont {B.}~\bibnamefont {Gupt}}} (\bibinfo {year} {2020}),\
  \bibfield  {title} {\enquote {\bibinfo {title} {{Measuring the similarity of
  graphs with a Gaussian boson sampler}},}\ }\href
  {https://doi.org/10.1103/PhysRevA.101.032314} {\bibfield  {journal} {\bibinfo
   {journal} {Phys. Rev. A}\ }\textbf {\bibinfo {volume} {101}},\ \bibinfo
  {pages} {032314}}\BibitemShut {NoStop}%
\bibitem [{\citenamefont {Schutski}\ \emph {et~al.}(2020)\citenamefont
  {Schutski}, \citenamefont {Khakhulin}, \citenamefont {Oseledets},\ and\
  \citenamefont {Kolmakov}}]{schutski_simple_2020}%
  \BibitemOpen
  \bibfield  {author} {\bibinfo {author} {\bibnamefont {Schutski},
  \bibfnamefont {R}}, \bibinfo {author} {\bibfnamefont {T.}~\bibnamefont
  {Khakhulin}}, \bibinfo {author} {\bibfnamefont {I.}~\bibnamefont
  {Oseledets}}, and\ \bibinfo {author} {\bibfnamefont {D.}~\bibnamefont
  {Kolmakov}}} (\bibinfo {year} {2020}),\ \bibfield  {title} {\enquote
  {\bibinfo {title} {Simple heuristics for efficient parallel tensor
  contraction and quantum circuit simulation},}\ }\href
  {https://doi.org/10.1103/PhysRevA.102.062614} {\bibfield  {journal} {\bibinfo
   {journal} {Phys. Rev. A}\ }\textbf {\bibinfo {volume} {102}},\ \bibinfo
  {pages} {062614}}\BibitemShut {NoStop}%
\bibitem [{\citenamefont {Schwarz}\ and\ \citenamefont {{van den
  Nest}}(2013)}]{schwarz_simulating_2013}%
  \BibitemOpen
  \bibfield  {author} {\bibinfo {author} {\bibnamefont {Schwarz}, \bibfnamefont
  {M}}, and\ \bibinfo {author} {\bibfnamefont {M.}~\bibnamefont {{van den
  Nest}}}} (\bibinfo {year} {2013}),\ \bibfield  {title} {\enquote {\bibinfo
  {title} {Simulating {{quantum circuits}} with {{sparse output
  distributions}}},}\ }\href@noop {} {\ }\Eprint
  {https://arxiv.org/abs/1310.6749} {arXiv:1310.6749} \BibitemShut {NoStop}%
\bibitem [{\citenamefont {Sekatski}\ \emph {et~al.}(2018)\citenamefont
  {Sekatski}, \citenamefont {Bancal}, \citenamefont {Wagner},\ and\
  \citenamefont {Sangouard}}]{Sekatski:2018eo}%
  \BibitemOpen
  \bibfield  {author} {\bibinfo {author} {\bibnamefont {Sekatski},
  \bibfnamefont {P}}, \bibinfo {author} {\bibfnamefont {J.-D.}\ \bibnamefont
  {Bancal}}, \bibinfo {author} {\bibfnamefont {S.}~\bibnamefont {Wagner}}, and\
  \bibinfo {author} {\bibfnamefont {N.}~\bibnamefont {Sangouard}}} (\bibinfo
  {year} {2018}),\ \bibfield  {title} {\enquote {\bibinfo {title} {{Certifying
  the building blocks of quantum computers from Bell{\textquoteright}s
  theorem}},}\ }\href {https://doi.org/10.1103/PhysRevLett.121.180505}
  {\bibfield  {journal} {\bibinfo  {journal} {Phys. Rev. Lett.}\ }\textbf
  {\bibinfo {volume} {121}},\ \bibinfo {pages} {180505}}\BibitemShut {NoStop}%
\bibitem [{\citenamefont {{Sempere-Llagostera}}\ \emph
  {et~al.}(2022)\citenamefont {{Sempere-Llagostera}}, \citenamefont {Patel},
  \citenamefont {Walmsley},\ and\ \citenamefont
  {Kolthammer}}]{sempere-llagostera_experimentally_2022}%
  \BibitemOpen
  \bibfield  {author} {\bibinfo {author} {\bibnamefont {{Sempere-Llagostera}},
  \bibfnamefont {S}}, \bibinfo {author} {\bibfnamefont {R.~B.}\ \bibnamefont
  {Patel}}, \bibinfo {author} {\bibfnamefont {I.~A.}\ \bibnamefont {Walmsley}},
  and\ \bibinfo {author} {\bibfnamefont {W.~S.}\ \bibnamefont {Kolthammer}}}
  (\bibinfo {year} {2022}),\ \bibfield  {title} {\enquote {\bibinfo {title}
  {Experimentally finding dense subgraphs using a time-bin encoded {{Gaussian}}
  boson sampling device},}\ }\href@noop {} {\ }\Eprint
  {https://arxiv.org/abs/2204.05254} {arXiv:2204.05254} \BibitemShut {NoStop}%
\bibitem [{\citenamefont {Shalm}\ \emph {et~al.}(2015)\citenamefont {Shalm},
  \citenamefont {{Meyer-Scott}}, \citenamefont {Christensen}, \citenamefont
  {Bierhorst}, \citenamefont {Wayne}, \citenamefont {Stevens}, \citenamefont
  {Gerrits}, \citenamefont {Glancy}, \citenamefont {Hamel}, \citenamefont
  {Allman}, \citenamefont {Coakley}, \citenamefont {Dyer}, \citenamefont
  {Hodge}, \citenamefont {Lita}, \citenamefont {Verma}, \citenamefont
  {Lambrocco}, \citenamefont {Tortorici}, \citenamefont {Migdall},
  \citenamefont {Zhang}, \citenamefont {Kumor}, \citenamefont {Farr},
  \citenamefont {Marsili}, \citenamefont {Shaw}, \citenamefont {Stern},
  \citenamefont {Abell{\'a}n}, \citenamefont {Amaya}, \citenamefont {Pruneri},
  \citenamefont {Jennewein}, \citenamefont {Mitchell}, \citenamefont {Kwiat},
  \citenamefont {Bienfang}, \citenamefont {Mirin}, \citenamefont {Knill},\ and\
  \citenamefont {Nam}}]{shalm_strong_2015}%
  \BibitemOpen
  \bibfield  {author} {\bibinfo {author} {\bibnamefont {Shalm}, \bibfnamefont
  {L~K}}, \bibinfo {author} {\bibfnamefont {E.}~\bibnamefont {{Meyer-Scott}}},
  \bibinfo {author} {\bibfnamefont {B.~G.}\ \bibnamefont {Christensen}},
  \bibinfo {author} {\bibfnamefont {P.}~\bibnamefont {Bierhorst}}, \bibinfo
  {author} {\bibfnamefont {M.~A.}\ \bibnamefont {Wayne}}, \bibinfo {author}
  {\bibfnamefont {M.~J.}\ \bibnamefont {Stevens}}, \bibinfo {author}
  {\bibfnamefont {T.}~\bibnamefont {Gerrits}}, \bibinfo {author} {\bibfnamefont
  {S.}~\bibnamefont {Glancy}}, \bibinfo {author} {\bibfnamefont {D.~R.}\
  \bibnamefont {Hamel}}, \bibinfo {author} {\bibfnamefont {M.~S.}\ \bibnamefont
  {Allman}}, \bibinfo {author} {\bibfnamefont {K.~J.}\ \bibnamefont {Coakley}},
  \bibinfo {author} {\bibfnamefont {S.~D.}\ \bibnamefont {Dyer}}, \bibinfo
  {author} {\bibfnamefont {C.}~\bibnamefont {Hodge}}, \bibinfo {author}
  {\bibfnamefont {A.~E.}\ \bibnamefont {Lita}}, \bibinfo {author}
  {\bibfnamefont {V.~B.}\ \bibnamefont {Verma}}, \bibinfo {author}
  {\bibfnamefont {C.}~\bibnamefont {Lambrocco}}, \bibinfo {author}
  {\bibfnamefont {E.}~\bibnamefont {Tortorici}}, \bibinfo {author}
  {\bibfnamefont {A.~L.}\ \bibnamefont {Migdall}}, \bibinfo {author}
  {\bibfnamefont {Y.}~\bibnamefont {Zhang}}, \bibinfo {author} {\bibfnamefont
  {D.~R.}\ \bibnamefont {Kumor}}, \bibinfo {author} {\bibfnamefont {W.~H.}\
  \bibnamefont {Farr}}, \bibinfo {author} {\bibfnamefont {F.}~\bibnamefont
  {Marsili}}, \bibinfo {author} {\bibfnamefont {M.~D.}\ \bibnamefont {Shaw}},
  \bibinfo {author} {\bibfnamefont {J.~A.}\ \bibnamefont {Stern}}, \bibinfo
  {author} {\bibfnamefont {C.}~\bibnamefont {Abell{\'a}n}}, \bibinfo {author}
  {\bibfnamefont {W.}~\bibnamefont {Amaya}}, \bibinfo {author} {\bibfnamefont
  {V.}~\bibnamefont {Pruneri}}, \bibinfo {author} {\bibfnamefont
  {T.}~\bibnamefont {Jennewein}}, \bibinfo {author} {\bibfnamefont {M.~W.}\
  \bibnamefont {Mitchell}}, \bibinfo {author} {\bibfnamefont {P.~G.}\
  \bibnamefont {Kwiat}}, \bibinfo {author} {\bibfnamefont {J.~C.}\ \bibnamefont
  {Bienfang}}, \bibinfo {author} {\bibfnamefont {R.~P.}\ \bibnamefont {Mirin}},
  \bibinfo {author} {\bibfnamefont {E.}~\bibnamefont {Knill}}, and\ \bibinfo
  {author} {\bibfnamefont {S.~W.}\ \bibnamefont {Nam}}} (\bibinfo {year}
  {2015}),\ \bibfield  {title} {\enquote {\bibinfo {title} {Strong
  {{loophole-free test}} of {{local realism}}},}\ }\href
  {https://doi.org/10.1103/PhysRevLett.115.250402} {\bibfield  {journal}
  {\bibinfo  {journal} {Phys. Rev. Lett.}\ }\textbf {\bibinfo {volume} {115}},\
  \bibinfo {pages} {250402}}\BibitemShut {NoStop}%
\bibitem [{\citenamefont {Shchesnovich}(2022)}]{shchesnovich_boson_2022}%
  \BibitemOpen
  \bibfield  {author} {\bibinfo {author} {\bibnamefont {Shchesnovich},
  \bibfnamefont {V}}} (\bibinfo {year} {2022}),\ \bibfield  {title} {\enquote
  {\bibinfo {title} {Boson sampling cannot be faithfully simulated by only the
  lower-order multi-boson interferences},}\ }\href@noop {} {\ }\Eprint
  {https://arxiv.org/abs/2204.07792} {arXiv:2204.07792} \BibitemShut {NoStop}%
\bibitem [{\citenamefont {Shchesnovich}(2013)}]{shchesnovich_asymptotic_2013}%
  \BibitemOpen
  \bibfield  {author} {\bibinfo {author} {\bibnamefont {Shchesnovich},
  \bibfnamefont {V~S}}} (\bibinfo {year} {2013}),\ \bibfield  {title} {\enquote
  {\bibinfo {title} {Asymptotic evaluation of bosonic probability amplitudes in
  linear unitary networks in the case of large number of bosons},}\ }\href
  {https://doi.org/10.1142/S0219749913500457} {\bibfield  {journal} {\bibinfo
  {journal} {Int. J. Quant. Inf.}\ }\textbf {\bibinfo {volume} {11}},\ \bibinfo
  {pages} {1350045}}\BibitemShut {NoStop}%
\bibitem [{\citenamefont {Shchesnovich}(2014)}]{shchesnovich_sufficient_2014}%
  \BibitemOpen
  \bibfield  {author} {\bibinfo {author} {\bibnamefont {Shchesnovich},
  \bibfnamefont {V~S}}} (\bibinfo {year} {2014}),\ \bibfield  {title} {\enquote
  {\bibinfo {title} {Sufficient condition for the mode mismatch of single
  photons for scalability of the boson-sampling computer},}\ }\href
  {https://doi.org/10.1103/PhysRevA.89.022333} {\bibfield  {journal} {\bibinfo
  {journal} {Phys. Rev. A}\ }\textbf {\bibinfo {volume} {89}},\ \bibinfo
  {pages} {022333}}\BibitemShut {NoStop}%
\bibitem [{\citenamefont {Shchesnovich}(2019)}]{shchesnovich_noise_2019}%
  \BibitemOpen
  \bibfield  {author} {\bibinfo {author} {\bibnamefont {Shchesnovich},
  \bibfnamefont {V~S}}} (\bibinfo {year} {2019}),\ \bibfield  {title} {\enquote
  {\bibinfo {title} {Noise in boson sampling and the threshold of efficient
  classical simulatability},}\ }\href
  {https://doi.org/10.1103/PhysRevA.100.012340} {\bibfield  {journal} {\bibinfo
   {journal} {Phys. Rev. A}\ }\textbf {\bibinfo {volume} {100}},\ \bibinfo
  {pages} {012340}}\BibitemShut {NoStop}%
\bibitem [{\citenamefont
  {Shchesnovich}(2021)}]{shchesnovich_distinguishing_2021}%
  \BibitemOpen
  \bibfield  {author} {\bibinfo {author} {\bibnamefont {Shchesnovich},
  \bibfnamefont {V~S}}} (\bibinfo {year} {2021}),\ \bibfield  {title} {\enquote
  {\bibinfo {title} {Distinguishing noisy boson sampling from classical
  simulations},}\ }\href {https://doi.org/10.22331/q-2021-03-29-423} {\bibfield
   {journal} {\bibinfo  {journal} {Quantum}\ }\textbf {\bibinfo {volume} {5}},\
  \bibinfo {pages} {423}}\BibitemShut {NoStop}%
\bibitem [{\citenamefont {Shepherd}\ and\ \citenamefont
  {Bremner}(2009)}]{shepherd_temporally_2009}%
  \BibitemOpen
  \bibfield  {author} {\bibinfo {author} {\bibnamefont {Shepherd},
  \bibfnamefont {D}}, and\ \bibinfo {author} {\bibfnamefont {M.~J.}\
  \bibnamefont {Bremner}}} (\bibinfo {year} {2009}),\ \bibfield  {title}
  {\enquote {\bibinfo {title} {Temporally unstructured quantum computation},}\
  }\href {https://doi.org/10.1098/rspa.2008.0443} {\bibfield  {journal}
  {\bibinfo  {journal} {Proc. Roy. Soc. A}\ }\textbf {\bibinfo {volume}
  {465}},\ \bibinfo {pages} {1413--1439}},\ \Eprint
  {https://arxiv.org/abs/0809.0847} {arXiv:0809.0847} \BibitemShut {NoStop}%
\bibitem [{\citenamefont {Shi}(2002)}]{shi_both_2002}%
  \BibitemOpen
  \bibfield  {author} {\bibinfo {author} {\bibnamefont {Shi}, \bibfnamefont
  {Y}}} (\bibinfo {year} {2002}),\ \bibfield  {title} {\enquote {\bibinfo
  {title} {{Both Toffoli and controlled-{NOT} need little help to do universal
  quantum computation}},}\ }\href@noop {} {\ }\Eprint
  {https://arxiv.org/abs/quant-ph/0205115} {arXiv:quant-ph/0205115}
  \BibitemShut {NoStop}%
\bibitem [{\citenamefont {Shor}(1994)}]{shor_algorithms_1994}%
  \BibitemOpen
  \bibfield  {author} {\bibinfo {author} {\bibnamefont {Shor}, \bibfnamefont
  {P~W}}} (\bibinfo {year} {1994}),\ \bibfield  {title} {\enquote {\bibinfo
  {title} {Algorithms for quantum computation: Discrete logarithms and
  factoring},}\ }in\ \href {https://doi.org/10.1109/SFCS.1994.365700} {\emph
  {\bibinfo {booktitle} {Proce. 35th {{Ann. Symp.}} {{Found.}} {{Comp.
  Sc.}}}}},\ pp.\ \bibinfo {pages} {124--134}\BibitemShut {NoStop}%
\bibitem [{\citenamefont {Shor}(1996)}]{shor_fault-tolerant_1996}%
  \BibitemOpen
  \bibfield  {author} {\bibinfo {author} {\bibnamefont {Shor}, \bibfnamefont
  {P~W}}} (\bibinfo {year} {1996}),\ \bibfield  {title} {\enquote {\bibinfo
  {title} {Fault-tolerant quantum computation},}\ }in\ \href
  {https://doi.org/10.1109/SFCS.1996.548464} {\emph {\bibinfo {booktitle}
  {Proc. 37th {{Conf.}} {{Found.}} {{Comp. Sc.}}}}},\ pp.\ \bibinfo {pages}
  {56--65}\BibitemShut {NoStop}%
\bibitem [{\citenamefont {Shor}(1997)}]{shor_polynomial-time_1997}%
  \BibitemOpen
  \bibfield  {author} {\bibinfo {author} {\bibnamefont {Shor}, \bibfnamefont
  {P~W}}} (\bibinfo {year} {1997}),\ \bibfield  {title} {\enquote {\bibinfo
  {title} {Polynomial-{{time algorithms}} for {{prime factorization}} and
  {{discrete logarithms}} on a {{quantum computer}}},}\ }\href
  {https://doi.org/10.1137/S0097539795293172} {\bibfield  {journal} {\bibinfo
  {journal} {SIAM J. Comput.}\ }\textbf {\bibinfo {volume} {26}},\ \bibinfo
  {pages} {1484--1509}}\BibitemShut {NoStop}%
\bibitem [{\citenamefont {Simon}(1994)}]{simon_power_1994}%
  \BibitemOpen
  \bibfield  {author} {\bibinfo {author} {\bibnamefont {Simon}, \bibfnamefont
  {D~R}}} (\bibinfo {year} {1994}),\ \bibfield  {title} {\enquote {\bibinfo
  {title} {On the power of quantum computation},}\ }in\ \href
  {https://doi.org/10.1109/SFCS.1994.365701} {\emph {\bibinfo {booktitle}
  {Proceedings of the 35th {{Annual Symposium}} on {{Foundations}} of
  {{Computer Science}}}}},\ \bibinfo {series and number} {{{SFCS}} '94}\
  (\bibinfo  {publisher} {{IEEE Computer Society}},\ \bibinfo {address}
  {{USA}})\ pp.\ \bibinfo {pages} {116--123}\BibitemShut {NoStop}%
\bibitem [{\citenamefont {Simon}(1997)}]{simon_power_1997}%
  \BibitemOpen
  \bibfield  {author} {\bibinfo {author} {\bibnamefont {Simon}, \bibfnamefont
  {D~R}}} (\bibinfo {year} {1997}),\ \bibfield  {title} {\enquote {\bibinfo
  {title} {On the {{power}} of {{quantum computation}}},}\ }\href
  {https://doi.org/10.1137/S0097539796298637} {\bibfield  {journal} {\bibinfo
  {journal} {SIAM J. Comput.}\ }\textbf {\bibinfo {volume} {26}},\ \bibinfo
  {pages} {1474--1483}}\BibitemShut {NoStop}%
\bibitem [{\citenamefont {Smelyanskiy}\ \emph {et~al.}(2016)\citenamefont
  {Smelyanskiy}, \citenamefont {Sawaya},\ and\ \citenamefont
  {{Aspuru-Guzik}}}]{smelyanskiy_qhipster_2016}%
  \BibitemOpen
  \bibfield  {author} {\bibinfo {author} {\bibnamefont {Smelyanskiy},
  \bibfnamefont {M}}, \bibinfo {author} {\bibfnamefont {N.~P.~D.}\ \bibnamefont
  {Sawaya}}, and\ \bibinfo {author} {\bibfnamefont {A.}~\bibnamefont
  {{Aspuru-Guzik}}}} (\bibinfo {year} {2016}),\ \bibfield  {title} {\enquote
  {\bibinfo {title} {{{qHiPSTER}}: {{The quantum high performance software
  testing environment}}},}\ }\href@noop {} {\ }\Eprint
  {https://arxiv.org/abs/1601.07195} {arXiv:1601.07195} \BibitemShut {NoStop}%
\bibitem [{\citenamefont {Somhorst}\ \emph {et~al.}(2022)\citenamefont
  {Somhorst}, \citenamefont {van~der Meer}, \citenamefont {Anguita},
  \citenamefont {Schadow}, \citenamefont {Snijders}, \citenamefont {de~Goede},
  \citenamefont {Kassenberg}, \citenamefont {Venderbosch}, \citenamefont
  {Taballione}, \citenamefont {Epping}, \citenamefont {van~den Vlekkert},
  \citenamefont {Bulmer}, \citenamefont {Lugani}, \citenamefont {Walmsley},
  \citenamefont {Pinkse}, \citenamefont {Eisert}, \citenamefont {Walk},\ and\
  \citenamefont {Renema}}]{QuantumPhotoThermodynamics}%
  \BibitemOpen
  \bibfield  {author} {\bibinfo {author} {\bibnamefont {Somhorst},
  \bibfnamefont {F~H~B}}, \bibinfo {author} {\bibfnamefont {R.}~\bibnamefont
  {van~der Meer}}, \bibinfo {author} {\bibfnamefont {M.~Correa}\ \bibnamefont
  {Anguita}}, \bibinfo {author} {\bibfnamefont {R.}~\bibnamefont {Schadow}},
  \bibinfo {author} {\bibfnamefont {H.~J.}\ \bibnamefont {Snijders}}, \bibinfo
  {author} {\bibfnamefont {M.}~\bibnamefont {de~Goede}}, \bibinfo {author}
  {\bibfnamefont {B.}~\bibnamefont {Kassenberg}}, \bibinfo {author}
  {\bibfnamefont {P.}~\bibnamefont {Venderbosch}}, \bibinfo {author}
  {\bibfnamefont {C.}~\bibnamefont {Taballione}}, \bibinfo {author}
  {\bibfnamefont {J.~P.}\ \bibnamefont {Epping}}, \bibinfo {author}
  {\bibfnamefont {H.~H.}\ \bibnamefont {van~den Vlekkert}}, \bibinfo {author}
  {\bibfnamefont {J.~F.~F.}\ \bibnamefont {Bulmer}}, \bibinfo {author}
  {\bibfnamefont {J.}~\bibnamefont {Lugani}}, \bibinfo {author} {\bibfnamefont
  {I.~A.}\ \bibnamefont {Walmsley}}, \bibinfo {author} {\bibfnamefont
  {P.~W.~H.}\ \bibnamefont {Pinkse}}, \bibinfo {author} {\bibfnamefont
  {J.}~\bibnamefont {Eisert}}, \bibinfo {author} {\bibfnamefont
  {N.}~\bibnamefont {Walk}}, and\ \bibinfo {author} {\bibfnamefont {J.~J.}\
  \bibnamefont {Renema}}} (\bibinfo {year} {2022}),\ \bibfield  {title}
  {\enquote {\bibinfo {title} {Quantum photo-thermodynamics on a programmable
  photonic quantum processor},}\ }\href@noop {} {\ }\Eprint
  {https://arxiv.org/abs/2201.00049} {arXiv:2201.00049} \BibitemShut {NoStop}%
\bibitem [{\citenamefont {Spagnolo}\ \emph {et~al.}(2014)\citenamefont
  {Spagnolo}, \citenamefont {Vitelli}, \citenamefont {Bentivegna},
  \citenamefont {Brod}, \citenamefont {Crespi}, \citenamefont {Flamini},
  \citenamefont {Giacomini}, \citenamefont {Milani}, \citenamefont {Ramponi},
  \citenamefont {Mataloni}, \citenamefont {Osellame}, \citenamefont {Galvao},\
  and\ \citenamefont {Sciarrino}}]{spagnolo_efficient_2014}%
  \BibitemOpen
  \bibfield  {author} {\bibinfo {author} {\bibnamefont {Spagnolo},
  \bibfnamefont {N}}, \bibinfo {author} {\bibfnamefont {C.}~\bibnamefont
  {Vitelli}}, \bibinfo {author} {\bibfnamefont {M.}~\bibnamefont {Bentivegna}},
  \bibinfo {author} {\bibfnamefont {D.~J.}\ \bibnamefont {Brod}}, \bibinfo
  {author} {\bibfnamefont {A.}~\bibnamefont {Crespi}}, \bibinfo {author}
  {\bibfnamefont {F.}~\bibnamefont {Flamini}}, \bibinfo {author} {\bibfnamefont
  {S.}~\bibnamefont {Giacomini}}, \bibinfo {author} {\bibfnamefont
  {G.}~\bibnamefont {Milani}}, \bibinfo {author} {\bibfnamefont
  {R.}~\bibnamefont {Ramponi}}, \bibinfo {author} {\bibfnamefont
  {P.}~\bibnamefont {Mataloni}}, \bibinfo {author} {\bibfnamefont
  {R.}~\bibnamefont {Osellame}}, \bibinfo {author} {\bibfnamefont {E.~F.}\
  \bibnamefont {Galvao}}, and\ \bibinfo {author} {\bibfnamefont
  {F.}~\bibnamefont {Sciarrino}}} (\bibinfo {year} {2014}),\ \bibfield  {title}
  {\enquote {\bibinfo {title} {Efficient experimental validation of photonic
  boson sampling against the uniform distribution},}\ }\href
  {https://doi.org/10.1038/nphoton.2014.135} {\bibfield  {journal} {\bibinfo
  {journal} {Nature Phot.}\ }\textbf {\bibinfo {volume} {8}},\ \bibinfo {pages}
  {615--620}}\BibitemShut {NoStop}%
\bibitem [{\citenamefont {Spring}\ \emph {et~al.}(2013)\citenamefont {Spring},
  \citenamefont {Metcalf}, \citenamefont {Humphreys}, \citenamefont
  {Kolthammer}, \citenamefont {Jin}, \citenamefont {Barbieri}, \citenamefont
  {Datta}, \citenamefont {{Thomas-Peter}}, \citenamefont {Langford},
  \citenamefont {Kundys}, \citenamefont {Gates}, \citenamefont {Smith},
  \citenamefont {Smith},\ and\ \citenamefont {Walmsley}}]{spring_boson_2013}%
  \BibitemOpen
  \bibfield  {author} {\bibinfo {author} {\bibnamefont {Spring}, \bibfnamefont
  {J~B}}, \bibinfo {author} {\bibfnamefont {B.~J.}\ \bibnamefont {Metcalf}},
  \bibinfo {author} {\bibfnamefont {P.~C.}\ \bibnamefont {Humphreys}}, \bibinfo
  {author} {\bibfnamefont {W.~S.}\ \bibnamefont {Kolthammer}}, \bibinfo
  {author} {\bibfnamefont {Xian-Min}\ \bibnamefont {Jin}}, \bibinfo {author}
  {\bibfnamefont {M.}~\bibnamefont {Barbieri}}, \bibinfo {author}
  {\bibfnamefont {A.}~\bibnamefont {Datta}}, \bibinfo {author} {\bibfnamefont
  {N.}~\bibnamefont {{Thomas-Peter}}}, \bibinfo {author} {\bibfnamefont
  {N.~K.}\ \bibnamefont {Langford}}, \bibinfo {author} {\bibfnamefont
  {D.}~\bibnamefont {Kundys}}, \bibinfo {author} {\bibfnamefont {J.~C.}\
  \bibnamefont {Gates}}, \bibinfo {author} {\bibfnamefont {B.~J.}\ \bibnamefont
  {Smith}}, \bibinfo {author} {\bibfnamefont {P.~G.~R.}\ \bibnamefont {Smith}},
  and\ \bibinfo {author} {\bibfnamefont {I.~A.}\ \bibnamefont {Walmsley}}}
  (\bibinfo {year} {2013}),\ \bibfield  {title} {\enquote {\bibinfo {title}
  {Boson {{sampling}} on a {{photonic chip}}},}\ }\href
  {https://doi.org/10.1126/science.1231692} {\bibfield  {journal} {\bibinfo
  {journal} {Science}\ }\textbf {\bibinfo {volume} {339}},\ \bibinfo {pages}
  {798--801}}\BibitemShut {NoStop}%
\bibitem [{\citenamefont {{Stilck Fran{\c c}a}}\ and\ \citenamefont
  {{Garc{\'i}a-Patr{\'o}n}}(2021)}]{stilck_franca_limitations_2021}%
  \BibitemOpen
  \bibfield  {author} {\bibinfo {author} {\bibnamefont {{Stilck Fran{\c c}a}},
  \bibfnamefont {D}}, and\ \bibinfo {author} {\bibfnamefont {R.}~\bibnamefont
  {{Garc{\'i}a-Patr{\'o}n}}}} (\bibinfo {year} {2021}),\ \bibfield  {title}
  {\enquote {\bibinfo {title} {Limitations of optimization algorithms on noisy
  quantum devices},}\ }\href {https://doi.org/10.1038/s41567-021-01356-3}
  {\bibfield  {journal} {\bibinfo  {journal} {Nat. Phys.}\ }\textbf {\bibinfo
  {volume} {17}},\ \bibinfo {pages} {1221--1227}}\BibitemShut {NoStop}%
\bibitem [{\citenamefont {Stilck~Fran{\c c}a}\ and\ \citenamefont
  {{Garcia-Patron}}(2022)}]{franca_game_2020}%
  \BibitemOpen
  \bibfield  {author} {\bibinfo {author} {\bibnamefont {Stilck~Fran{\c c}a},
  \bibfnamefont {D}}, and\ \bibinfo {author} {\bibfnamefont {R.}~\bibnamefont
  {{Garcia-Patron}}}} (\bibinfo {year} {2022}),\ \bibfield  {title} {\enquote
  {\bibinfo {title} {A game of quantum advantage: Linking verification and
  simulation},}\ }\href {https://doi.org/10.22331/q-2022-06-30-753} {\bibfield
  {journal} {\bibinfo  {journal} {Quantum}\ }\textbf {\bibinfo {volume} {6}},\
  \bibinfo {pages} {753}}\BibitemShut {NoStop}%
\bibitem [{\citenamefont {Stockmeyer}(1983)}]{stockmeyer_complexity_1983}%
  \BibitemOpen
  \bibfield  {author} {\bibinfo {author} {\bibnamefont {Stockmeyer},
  \bibfnamefont {L}}} (\bibinfo {year} {1983}),\ \bibfield  {title} {\enquote
  {\bibinfo {title} {The complexity of approximate counting},}\ }\bibfield
  {booktitle} {\emph {\bibinfo {booktitle} {Proc. 15th Ann. ACM Symp. Th.
  Comp.}},\ }\href {https://doi.org/10.1145/800061.808740} {\ \bibinfo {series}
  {STOC '83},\ \bibinfo {pages} {118--126}}\BibitemShut {NoStop}%
\bibitem [{\citenamefont {Sudan}(1997)}]{sudan_decoding_1997}%
  \BibitemOpen
  \bibfield  {author} {\bibinfo {author} {\bibnamefont {Sudan}, \bibfnamefont
  {M}}} (\bibinfo {year} {1997}),\ \bibfield  {title} {\enquote {\bibinfo
  {title} {{Decoding of Reed Solomon codes beyond the error-correction
  bound}},}\ }\href {https://doi.org/10.1006/jcom.1997.0439} {\bibfield
  {journal} {\bibinfo  {journal} {J. Complex.}\ }\textbf {\bibinfo {volume}
  {13}},\ \bibinfo {pages} {180--193}}\BibitemShut {NoStop}%
\bibitem [{\citenamefont {Takeuchi}\ \emph {et~al.}(2019)\citenamefont
  {Takeuchi}, \citenamefont {Mantri}, \citenamefont {Morimae}, \citenamefont
  {Mizutani},\ and\ \citenamefont
  {Fitzsimons}}]{takeuchi_resource-efficient_2019}%
  \BibitemOpen
  \bibfield  {author} {\bibinfo {author} {\bibnamefont {Takeuchi},
  \bibfnamefont {Y}}, \bibinfo {author} {\bibfnamefont {A.}~\bibnamefont
  {Mantri}}, \bibinfo {author} {\bibfnamefont {T.}~\bibnamefont {Morimae}},
  \bibinfo {author} {\bibfnamefont {A.}~\bibnamefont {Mizutani}}, and\ \bibinfo
  {author} {\bibfnamefont {J.~F.}\ \bibnamefont {Fitzsimons}}} (\bibinfo {year}
  {2019}),\ \bibfield  {title} {\enquote {\bibinfo {title} {Resource-efficient
  verification of quantum computing using serfling's bound},}\ }\href
  {https://doi.org/10.1038/s41534-019-0142-2} {\bibfield  {journal} {\bibinfo
  {journal} {npj Quant. Inf.}\ }\textbf {\bibinfo {volume} {5}},\ \bibinfo
  {pages} {1--8}}\BibitemShut {NoStop}%
\bibitem [{\citenamefont {Takeuchi}\ and\ \citenamefont
  {Morimae}(2018)}]{takeuchi_verification_2018}%
  \BibitemOpen
  \bibfield  {author} {\bibinfo {author} {\bibnamefont {Takeuchi},
  \bibfnamefont {Y}}, and\ \bibinfo {author} {\bibfnamefont {T.}~\bibnamefont
  {Morimae}}} (\bibinfo {year} {2018}),\ \bibfield  {title} {\enquote {\bibinfo
  {title} {Verification of {many}-{qubit} {states}},}\ }\href
  {https://doi.org/10.1103/PhysRevX.8.021060} {\bibfield  {journal} {\bibinfo
  {journal} {Phys. Rev. X}\ }\textbf {\bibinfo {volume} {8}},\ \bibinfo {pages}
  {021060}}\BibitemShut {NoStop}%
\bibitem [{\citenamefont {Tao}\ and\ \citenamefont
  {Vu}(2008)}]{tao_permanent_2008}%
  \BibitemOpen
  \bibfield  {author} {\bibinfo {author} {\bibnamefont {Tao}, \bibfnamefont
  {T}}, and\ \bibinfo {author} {\bibfnamefont {V.}~\bibnamefont {Vu}}}
  (\bibinfo {year} {2008}),\ \bibfield  {title} {\enquote {\bibinfo {title} {On
  the permanent of random {{Bernoulli}} matrices},}\ }\href@noop {} {\ }\Eprint
  {https://arxiv.org/abs/0804.2362} {arXiv:0804.2362} \BibitemShut {NoStop}%
\bibitem [{\citenamefont {Terhal}\ and\ \citenamefont
  {DiVincenzo}(2004)}]{terhal_adaptive_2004}%
  \BibitemOpen
  \bibfield  {author} {\bibinfo {author} {\bibnamefont {Terhal}, \bibfnamefont
  {B~M}}, and\ \bibinfo {author} {\bibfnamefont {D.~P.}\ \bibnamefont
  {DiVincenzo}}} (\bibinfo {year} {2004}),\ \bibfield  {title} {\enquote
  {\bibinfo {title} {Adaptive {quantum} {computation}, {constant} {depth}
  {quantum} {circuits} and {Arthur}-{Merlin} {games}},}\ }\href
  {https://doi.org/10.26421/QIC4.2-5} {\bibfield  {journal} {\bibinfo
  {journal} {Quant. Inf. Comp.}\ }\textbf {\bibinfo {volume} {4}},\ \bibinfo
  {pages} {134--145}}\BibitemShut {NoStop}%
\bibitem [{\citenamefont {Thekkadath}\ \emph {et~al.}(2022)\citenamefont
  {Thekkadath}, \citenamefont {{Sempere-Llagostera}}, \citenamefont {Bell},
  \citenamefont {Patel}, \citenamefont {Kim},\ and\ \citenamefont
  {Walmsley}}]{thekkadath_experimental_2022}%
  \BibitemOpen
  \bibfield  {author} {\bibinfo {author} {\bibnamefont {Thekkadath},
  \bibfnamefont {G~S}}, \bibinfo {author} {\bibfnamefont {S.}~\bibnamefont
  {{Sempere-Llagostera}}}, \bibinfo {author} {\bibfnamefont {B.A.}\
  \bibnamefont {Bell}}, \bibinfo {author} {\bibfnamefont {R.B.}\ \bibnamefont
  {Patel}}, \bibinfo {author} {\bibfnamefont {M.S.}\ \bibnamefont {Kim}}, and\
  \bibinfo {author} {\bibfnamefont {I.A.}\ \bibnamefont {Walmsley}}} (\bibinfo
  {year} {2022}),\ \bibfield  {title} {\enquote {\bibinfo {title} {Experimental
  {{demonstration}} of {{Gaussian boson sampling}} with {{displacement}}},}\
  }\href {https://doi.org/10.1103/PRXQuantum.3.020336} {\bibfield  {journal}
  {\bibinfo  {journal} {PRX Quantum}\ }\textbf {\bibinfo {volume} {3}},\
  \bibinfo {pages} {020336}}\BibitemShut {NoStop}%
\bibitem [{\citenamefont {Tichy}(2014)}]{tichy_interference_2014}%
  \BibitemOpen
  \bibfield  {author} {\bibinfo {author} {\bibnamefont {Tichy}, \bibfnamefont
  {M~C}}} (\bibinfo {year} {2014}),\ \bibfield  {title} {\enquote {\bibinfo
  {title} {Interference of identical particles from entanglement to
  boson-sampling},}\ }\href {https://doi.org/10.1088/0953-4075/47/10/103001}
  {\bibfield  {journal} {\bibinfo  {journal} {J. Phys. B}\ }\textbf {\bibinfo
  {volume} {47}},\ \bibinfo {pages} {103001}}\BibitemShut {NoStop}%
\bibitem [{\citenamefont {Tichy}(2015)}]{tichy_sampling_2015}%
  \BibitemOpen
  \bibfield  {author} {\bibinfo {author} {\bibnamefont {Tichy}, \bibfnamefont
  {M~C}}} (\bibinfo {year} {2015}),\ \bibfield  {title} {\enquote {\bibinfo
  {title} {Sampling of partially distinguishable bosons and the relation to the
  multidimensional permanent},}\ }\href
  {https://doi.org/10.1103/PhysRevA.91.022316} {\bibfield  {journal} {\bibinfo
  {journal} {Phys. Rev. A}\ }\textbf {\bibinfo {volume} {91}},\ \bibinfo
  {pages} {022316}}\BibitemShut {NoStop}%
\bibitem [{\citenamefont {Tichy}\ \emph {et~al.}(2014)\citenamefont {Tichy},
  \citenamefont {Mayer}, \citenamefont {Buchleitner},\ and\ \citenamefont
  {M\o{}lmer}}]{TicMayBuc14}%
  \BibitemOpen
  \bibfield  {author} {\bibinfo {author} {\bibnamefont {Tichy}, \bibfnamefont
  {M~C}}, \bibinfo {author} {\bibfnamefont {K.}~\bibnamefont {Mayer}}, \bibinfo
  {author} {\bibfnamefont {A.}~\bibnamefont {Buchleitner}}, and\ \bibinfo
  {author} {\bibfnamefont {K.}~\bibnamefont {M\o{}lmer}}} (\bibinfo {year}
  {2014}),\ \bibfield  {title} {\enquote {\bibinfo {title} {Stringent and
  efficient assessment of boson-sampling devices},}\ }\href
  {https://doi.org/10.1103/PhysRevLett.113.020502} {\bibfield  {journal}
  {\bibinfo  {journal} {Phys. Rev. Lett.}\ }\textbf {\bibinfo {volume} {113}},\
  \bibinfo {pages} {020502}}\BibitemShut {NoStop}%
\bibitem [{\citenamefont {Tillmann}\ \emph {et~al.}(2013)\citenamefont
  {Tillmann}, \citenamefont {Daki{\'c}}, \citenamefont {Heilmann},
  \citenamefont {Nolte}, \citenamefont {Szameit},\ and\ \citenamefont
  {Walther}}]{tillmann_experimental_2013}%
  \BibitemOpen
  \bibfield  {author} {\bibinfo {author} {\bibnamefont {Tillmann},
  \bibfnamefont {M}}, \bibinfo {author} {\bibfnamefont {B.}~\bibnamefont
  {Daki{\'c}}}, \bibinfo {author} {\bibfnamefont {R.}~\bibnamefont {Heilmann}},
  \bibinfo {author} {\bibfnamefont {S.}~\bibnamefont {Nolte}}, \bibinfo
  {author} {\bibfnamefont {A.}~\bibnamefont {Szameit}}, and\ \bibinfo {author}
  {\bibfnamefont {P.}~\bibnamefont {Walther}}} (\bibinfo {year} {2013}),\
  \bibfield  {title} {\enquote {\bibinfo {title} {Experimental boson
  sampling},}\ }\href {https://doi.org/10.1038/nphoton.2013.102} {\bibfield
  {journal} {\bibinfo  {journal} {Nature Phot.}\ }\textbf {\bibinfo {volume}
  {7}},\ \bibinfo {pages} {540--544}}\BibitemShut {NoStop}%
\bibitem [{\citenamefont {Toda}(1991)}]{toda_pp_1991}%
  \BibitemOpen
  \bibfield  {author} {\bibinfo {author} {\bibnamefont {Toda}, \bibfnamefont
  {S}}} (\bibinfo {year} {1991}),\ \bibfield  {title} {\enquote {\bibinfo
  {title} {{PP} is as hard as the polynomial-time hierarchy},}\ }\href
  {https://doi.org/10.1137/0220053} {\bibfield  {journal} {\bibinfo  {journal}
  {{SIAM} J. Comput.}\ }\textbf {\bibinfo {volume} {20}},\ \bibinfo {pages}
  {865--877}}\BibitemShut {NoStop}%
\bibitem [{\citenamefont {Toda}\ and\ \citenamefont
  {Ogiwara}(1992)}]{toda_counting_1992}%
  \BibitemOpen
  \bibfield  {author} {\bibinfo {author} {\bibnamefont {Toda}, \bibfnamefont
  {S}}, and\ \bibinfo {author} {\bibfnamefont {M.}~\bibnamefont {Ogiwara}}}
  (\bibinfo {year} {1992}),\ \bibfield  {title} {\enquote {\bibinfo {title}
  {Counting {{classes}} are at {{least}} as {{hard}} as the {{polynomial-time
  hierarchy}}},}\ }\href {https://doi.org/10.1137/0221023} {\bibfield
  {journal} {\bibinfo  {journal} {SIAM J. Comput.}\ }\textbf {\bibinfo {volume}
  {21}},\ \bibinfo {pages} {316--328}}\BibitemShut {NoStop}%
\bibitem [{\citenamefont {Toffoli}(1980)}]{toffoli_reversible_1980}%
  \BibitemOpen
  \bibfield  {author} {\bibinfo {author} {\bibnamefont {Toffoli}, \bibfnamefont
  {T}}} (\bibinfo {year} {1980}),\ \bibfield  {title} {\enquote {\bibinfo
  {title} {Reversible computing},}\ }in\ \href
  {https://doi.org/10.1007/3-540-10003-2_104} {\emph {\bibinfo {booktitle}
  {International Colloquium on Automata, Languages and Programming ({ICALP}
  80)}}},\ \bibinfo {series} {Lecture Notes in Computer Science}, Vol.~\bibinfo
  {volume} {85},\ \bibinfo {editor} {edited by\ \bibinfo {editor}
  {\bibfnamefont {Jaco}\ \bibnamefont {de~Bakker}}\ and\ \bibinfo {editor}
  {\bibfnamefont {Jan}\ \bibnamefont {van Leeuwen}}}\ (\bibinfo  {publisher}
  {Springer})\ pp.\ \bibinfo {pages} {632--644}\BibitemShut {NoStop}%
\bibitem [{\citenamefont {T{\'o}th}\ and\ \citenamefont
  {G{\"u}hne}(2005)}]{toth_entanglement_2005}%
  \BibitemOpen
  \bibfield  {author} {\bibinfo {author} {\bibnamefont {T{\'o}th},
  \bibfnamefont {G}}, and\ \bibinfo {author} {\bibfnamefont {O.}~\bibnamefont
  {G{\"u}hne}}} (\bibinfo {year} {2005}),\ \bibfield  {title} {\enquote
  {\bibinfo {title} {Entanglement detection in the stabilizer formalism},}\
  }\href {https://doi.org/10.1103/PhysRevA.72.022340} {\bibfield  {journal}
  {\bibinfo  {journal} {Phys. Rev. A}\ }\textbf {\bibinfo {volume} {72}},\
  \bibinfo {pages} {022340}}\BibitemShut {NoStop}%
\bibitem [{\citenamefont {Trevisan}(2008)}]{trevisan_lecture_2008}%
  \BibitemOpen
  \bibfield  {author} {\bibinfo {author} {\bibnamefont {Trevisan},
  \bibfnamefont {L}}} (\bibinfo {year} {2008}),\ \bibfield  {title} {\enquote
  {\bibinfo {title} {Lecture 6: {{Approximate counting}}},}\ }in\ \href
  {https://lucatrevisan.github.io/cs278-08/lecture06.pdf} {\emph {\bibinfo
  {booktitle} {Lecture {{Notes}} on {{Computational Complexity}}}}},\ \bibinfo
  {note} {accessed: 4/2/2022}\BibitemShut {NoStop}%
\bibitem [{\citenamefont {Trotzky}\ \emph {et~al.}(2012)\citenamefont
  {Trotzky}, \citenamefont {Chen}, \citenamefont {Flesch}, \citenamefont
  {McCulloch}, \citenamefont {Schollw\"ock}, \citenamefont {Eisert},\ and\
  \citenamefont {Bloch}}]{Trotzky}%
  \BibitemOpen
  \bibfield  {author} {\bibinfo {author} {\bibnamefont {Trotzky}, \bibfnamefont
  {S}}, \bibinfo {author} {\bibfnamefont {Y.-A.}\ \bibnamefont {Chen}},
  \bibinfo {author} {\bibfnamefont {A.}~\bibnamefont {Flesch}}, \bibinfo
  {author} {\bibfnamefont {I.~P.}\ \bibnamefont {McCulloch}}, \bibinfo {author}
  {\bibfnamefont {U.}~\bibnamefont {Schollw\"ock}}, \bibinfo {author}
  {\bibfnamefont {J.}~\bibnamefont {Eisert}}, and\ \bibinfo {author}
  {\bibfnamefont {I.}~\bibnamefont {Bloch}}} (\bibinfo {year} {2012}),\
  \bibfield  {title} {\enquote {\bibinfo {title} {Probing the relaxation
  towards equilibrium in an isolated strongly correlated one-dimensional {B}ose
  gas},}\ }\href {https://doi.org/doi:10.1038/nphys2232} {\bibfield  {journal}
  {\bibinfo  {journal} {Nature Phys.}\ }\textbf {\bibinfo {volume} {8}},\
  \bibinfo {pages} {325--330}}\BibitemShut {NoStop}%
\bibitem [{\citenamefont {Trotzky}\ \emph {et~al.}(2010)\citenamefont
  {Trotzky}, \citenamefont {Pollet}, \citenamefont {Gerbier}, \citenamefont
  {Schnorrberger}, \citenamefont {Bloch}, \citenamefont {Prokof'ev},
  \citenamefont {Svistunov},\ and\ \citenamefont
  {Troyer}}]{MonteCarloValidator}%
  \BibitemOpen
  \bibfield  {author} {\bibinfo {author} {\bibnamefont {Trotzky}, \bibfnamefont
  {S}}, \bibinfo {author} {\bibfnamefont {L.}~\bibnamefont {Pollet}}, \bibinfo
  {author} {\bibfnamefont {F.}~\bibnamefont {Gerbier}}, \bibinfo {author}
  {\bibfnamefont {U.}~\bibnamefont {Schnorrberger}}, \bibinfo {author}
  {\bibfnamefont {I.}~\bibnamefont {Bloch}}, \bibinfo {author} {\bibfnamefont
  {N.V.}\ \bibnamefont {Prokof'ev}}, \bibinfo {author} {\bibfnamefont
  {B.}~\bibnamefont {Svistunov}}, and\ \bibinfo {author} {\bibfnamefont
  {M.}~\bibnamefont {Troyer}}} (\bibinfo {year} {2010}),\ \bibfield  {title}
  {\enquote {\bibinfo {title} {{Suppression of the critical temperature for
  superfluidity near the Mott transition: validating a quantum simulator}},}\
  }\href {https://doi.org/10.1038/nphys1799} {\bibfield  {journal} {\bibinfo
  {journal} {Nature Phys.}\ }\textbf {\bibinfo {volume} {6}},\ \bibinfo {pages}
  {998}}\BibitemShut {NoStop}%
\bibitem [{\citenamefont {Valiant}\ and\ \citenamefont
  {Valiant}(2017)}]{valiant_automatic_2017}%
  \BibitemOpen
  \bibfield  {author} {\bibinfo {author} {\bibnamefont {Valiant}, \bibfnamefont
  {G}}, and\ \bibinfo {author} {\bibfnamefont {P.}~\bibnamefont {Valiant}}}
  (\bibinfo {year} {2017}),\ \bibfield  {title} {\enquote {\bibinfo {title} {An
  {automatic} {inequality} {prover} and {instance} {optimal} {identity}
  {testing}},}\ }\href {https://doi.org/10.1137/151002526} {\bibfield
  {journal} {\bibinfo  {journal} {SIAM J. Comput.}\ }\textbf {\bibinfo {volume}
  {46}},\ \bibinfo {pages} {429--455}},\ \bibinfo {note} {eCCC,
  TR13-111}\BibitemShut {NoStop}%
\bibitem [{\citenamefont {Valiant}(1979)}]{valiant_complexity_1979}%
  \BibitemOpen
  \bibfield  {author} {\bibinfo {author} {\bibnamefont {Valiant}, \bibfnamefont
  {L~G}}} (\bibinfo {year} {1979}),\ \bibfield  {title} {\enquote {\bibinfo
  {title} {The complexity of computing the permanent},}\ }\href
  {https://doi.org/10.1016/0304-3975(79)90044-6} {\bibfield  {journal}
  {\bibinfo  {journal} {Th. Comp. Sc.}\ }\textbf {\bibinfo {volume} {8}},\
  \bibinfo {pages} {189--201}}\BibitemShut {NoStop}%
\bibitem [{\citenamefont {Valido}\ and\ \citenamefont
  {Garc\'{\i}a-Ripoll}(2021)}]{PhysRevA.103.032613}%
  \BibitemOpen
  \bibfield  {author} {\bibinfo {author} {\bibnamefont {Valido}, \bibfnamefont
  {A~A}}, and\ \bibinfo {author} {\bibfnamefont {J.~J.}\ \bibnamefont
  {Garc\'{\i}a-Ripoll}}} (\bibinfo {year} {2021}),\ \bibfield  {title}
  {\enquote {\bibinfo {title} {Gaussian phase sensitivity of
  boson-sampling-inspired strategies},}\ }\href
  {https://doi.org/10.1103/PhysRevA.103.032613} {\bibfield  {journal} {\bibinfo
   {journal} {Phys. Rev. A}\ }\textbf {\bibinfo {volume} {103}},\ \bibinfo
  {pages} {032613}}\BibitemShut {NoStop}%
\bibitem [{\citenamefont {Vandersypen}\ \emph {et~al.}(2000)\citenamefont
  {Vandersypen}, \citenamefont {Steffen}, \citenamefont {Breyta}, \citenamefont
  {Yannoni}, \citenamefont {Cleve},\ and\ \citenamefont
  {Chuang}}]{vandersypen_experimental_2000}%
  \BibitemOpen
  \bibfield  {author} {\bibinfo {author} {\bibnamefont {Vandersypen},
  \bibfnamefont {L~M~K}}, \bibinfo {author} {\bibfnamefont {M.}~\bibnamefont
  {Steffen}}, \bibinfo {author} {\bibfnamefont {G.}~\bibnamefont {Breyta}},
  \bibinfo {author} {\bibfnamefont {C.~S.}\ \bibnamefont {Yannoni}}, \bibinfo
  {author} {\bibfnamefont {R.}~\bibnamefont {Cleve}}, and\ \bibinfo {author}
  {\bibfnamefont {I.~L.}\ \bibnamefont {Chuang}}} (\bibinfo {year} {2000}),\
  \bibfield  {title} {\enquote {\bibinfo {title} {{Experimental realization of
  an order-finding algorithm with an NMR quantum computer}},}\ }\href
  {https://doi.org/10.1103/PhysRevLett.85.5452} {\bibfield  {journal} {\bibinfo
   {journal} {Phys. Rev. Lett.}\ }\textbf {\bibinfo {volume} {85}},\ \bibinfo
  {pages} {5452--5455}}\BibitemShut {NoStop}%
\bibitem [{\citenamefont {Vandersypen}\ \emph {et~al.}(2001)\citenamefont
  {Vandersypen}, \citenamefont {Steffen}, \citenamefont {Breyta}, \citenamefont
  {Yannoni}, \citenamefont {Sherwood},\ and\ \citenamefont
  {Chuang}}]{vandersypen_experimental_2001}%
  \BibitemOpen
  \bibfield  {author} {\bibinfo {author} {\bibnamefont {Vandersypen},
  \bibfnamefont {L~M~K}}, \bibinfo {author} {\bibfnamefont {M.}~\bibnamefont
  {Steffen}}, \bibinfo {author} {\bibfnamefont {G.}~\bibnamefont {Breyta}},
  \bibinfo {author} {\bibfnamefont {C.~S.}\ \bibnamefont {Yannoni}}, \bibinfo
  {author} {\bibfnamefont {M.~H.}\ \bibnamefont {Sherwood}}, and\ \bibinfo
  {author} {\bibfnamefont {I.~L.}\ \bibnamefont {Chuang}}} (\bibinfo {year}
  {2001}),\ \bibfield  {title} {\enquote {\bibinfo {title} {Experimental
  realization of {{Shor}}'s quantum factoring algorithm using nuclear magnetic
  resonance},}\ }\href {https://doi.org/10.1038/414883a} {\bibfield  {journal}
  {\bibinfo  {journal} {Nature}\ }\textbf {\bibinfo {volume} {414}},\ \bibinfo
  {pages} {883--887}}\BibitemShut {NoStop}%
\bibitem [{\citenamefont {Vergis}\ \emph {et~al.}(1986)\citenamefont {Vergis},
  \citenamefont {Steiglitz},\ and\ \citenamefont
  {Dickinson}}]{vergis_complexity_1986}%
  \BibitemOpen
  \bibfield  {author} {\bibinfo {author} {\bibnamefont {Vergis}, \bibfnamefont
  {A}}, \bibinfo {author} {\bibfnamefont {K.}~\bibnamefont {Steiglitz}}, and\
  \bibinfo {author} {\bibfnamefont {B.}~\bibnamefont {Dickinson}}} (\bibinfo
  {year} {1986}),\ \bibfield  {title} {\enquote {\bibinfo {title} {The
  complexity of analog computation},}\ }\href
  {https://doi.org/10.1016/0378-4754(86)90105-9} {\bibfield  {journal}
  {\bibinfo  {journal} {Math. Comp. Sim.}\ }\textbf {\bibinfo {volume} {28}},\
  \bibinfo {pages} {91--113}}\BibitemShut {NoStop}%
\bibitem [{\citenamefont {Verstraete}\ \emph {et~al.}(2008)\citenamefont
  {Verstraete}, \citenamefont {Cirac},\ and\ \citenamefont
  {Murg}}]{VerstraeteBig}%
  \BibitemOpen
  \bibfield  {author} {\bibinfo {author} {\bibnamefont {Verstraete},
  \bibfnamefont {F}}, \bibinfo {author} {\bibfnamefont {J.~I.}\ \bibnamefont
  {Cirac}}, and\ \bibinfo {author} {\bibfnamefont {V.}~\bibnamefont {Murg}}}
  (\bibinfo {year} {2008}),\ \bibfield  {title} {\enquote {\bibinfo {title}
  {Matrix product states, projected entangled pair states, and variational
  renormalization group methods for quantum spin systems},}\ }\href
  {https://doi.org/10.1080/14789940801912366} {\bibfield  {journal} {\bibinfo
  {journal} {Adv. Phys.}\ }\textbf {\bibinfo {volume} {57}},\ \bibinfo {pages}
  {143}}\BibitemShut {NoStop}%
\bibitem [{\citenamefont {Villalonga}\ \emph {et~al.}(2019)\citenamefont
  {Villalonga}, \citenamefont {Boixo}, \citenamefont {Nelson}, \citenamefont
  {Henze}, \citenamefont {Rieffel}, \citenamefont {Biswas},\ and\ \citenamefont
  {Mandr{\`a}}}]{villalonga_flexible_2019}%
  \BibitemOpen
  \bibfield  {author} {\bibinfo {author} {\bibnamefont {Villalonga},
  \bibfnamefont {B}}, \bibinfo {author} {\bibfnamefont {S.}~\bibnamefont
  {Boixo}}, \bibinfo {author} {\bibfnamefont {B.}~\bibnamefont {Nelson}},
  \bibinfo {author} {\bibfnamefont {C.}~\bibnamefont {Henze}}, \bibinfo
  {author} {\bibfnamefont {E.}~\bibnamefont {Rieffel}}, \bibinfo {author}
  {\bibfnamefont {R.}~\bibnamefont {Biswas}}, and\ \bibinfo {author}
  {\bibfnamefont {S.}~\bibnamefont {Mandr{\`a}}}} (\bibinfo {year} {2019}),\
  \bibfield  {title} {\enquote {\bibinfo {title} {A flexible high-performance
  simulator for verifying and benchmarking quantum circuits implemented on real
  hardware},}\ }\href {https://doi.org/10.1038/s41534-019-0196-1} {\bibfield
  {journal} {\bibinfo  {journal} {npj Quant. Inf.}\ }\textbf {\bibinfo {volume}
  {5}},\ \bibinfo {pages} {1--16}}\BibitemShut {NoStop}%
\bibitem [{\citenamefont {Villalonga}\ \emph {et~al.}(2020)\citenamefont
  {Villalonga}, \citenamefont {Lyakh}, \citenamefont {Boixo}, \citenamefont
  {Neven}, \citenamefont {Humble}, \citenamefont {Biswas}, \citenamefont
  {Rieffel}, \citenamefont {Ho},\ and\ \citenamefont
  {Mandr{\`a}}}]{villalonga_establishing_2020}%
  \BibitemOpen
  \bibfield  {author} {\bibinfo {author} {\bibnamefont {Villalonga},
  \bibfnamefont {B}}, \bibinfo {author} {\bibfnamefont {D.}~\bibnamefont
  {Lyakh}}, \bibinfo {author} {\bibfnamefont {S.}~\bibnamefont {Boixo}},
  \bibinfo {author} {\bibfnamefont {H.}~\bibnamefont {Neven}}, \bibinfo
  {author} {\bibfnamefont {T.~S.}\ \bibnamefont {Humble}}, \bibinfo {author}
  {\bibfnamefont {R.}~\bibnamefont {Biswas}}, \bibinfo {author} {\bibfnamefont
  {E.~G.}\ \bibnamefont {Rieffel}}, \bibinfo {author} {\bibfnamefont
  {A.}~\bibnamefont {Ho}}, and\ \bibinfo {author} {\bibfnamefont
  {S.}~\bibnamefont {Mandr{\`a}}}} (\bibinfo {year} {2020}),\ \bibfield
  {title} {\enquote {\bibinfo {title} {Establishing the quantum supremacy
  frontier with a 281 {{Pflop}}/s simulation},}\ }\href
  {https://doi.org/10.1088/2058-9565/ab7eeb} {\bibfield  {journal} {\bibinfo
  {journal} {Quant. Sci. Tech.}\ }\textbf {\bibinfo {volume} {5}},\ \bibinfo
  {pages} {034003}}\BibitemShut {NoStop}%
\bibitem [{\citenamefont {Villalonga}\ \emph {et~al.}(2021)\citenamefont
  {Villalonga}, \citenamefont {Niu}, \citenamefont {Li}, \citenamefont {Neven},
  \citenamefont {Platt}, \citenamefont {Smelyanskiy},\ and\ \citenamefont
  {Boixo}}]{villalonga_efficient_2021}%
  \BibitemOpen
  \bibfield  {author} {\bibinfo {author} {\bibnamefont {Villalonga},
  \bibfnamefont {B}}, \bibinfo {author} {\bibfnamefont {M.~Y.}\ \bibnamefont
  {Niu}}, \bibinfo {author} {\bibfnamefont {L.}~\bibnamefont {Li}}, \bibinfo
  {author} {\bibfnamefont {H.}~\bibnamefont {Neven}}, \bibinfo {author}
  {\bibfnamefont {J.~C.}\ \bibnamefont {Platt}}, \bibinfo {author}
  {\bibfnamefont {V.~N.}\ \bibnamefont {Smelyanskiy}}, and\ \bibinfo {author}
  {\bibfnamefont {S.}~\bibnamefont {Boixo}}} (\bibinfo {year} {2021}),\
  \bibfield  {title} {\enquote {\bibinfo {title} {Efficient approximation of
  experimental {{Gaussian}} boson sampling},}\ }\href@noop {} {\ }\Eprint
  {https://arxiv.org/abs/2109.11525} {arXiv:2109.11525} \BibitemShut {NoStop}%
\bibitem [{\citenamefont {Vollbrecht}\ and\ \citenamefont
  {Cirac}(2008)}]{Vollbrecht}%
  \BibitemOpen
  \bibfield  {author} {\bibinfo {author} {\bibnamefont {Vollbrecht},
  \bibfnamefont {K~G~H}}, and\ \bibinfo {author} {\bibfnamefont {J.~I.}\
  \bibnamefont {Cirac}}} (\bibinfo {year} {2008}),\ \bibfield  {title}
  {\enquote {\bibinfo {title} {Quantum simulators, continuous-time automata,
  and translationally invariant systems},}\ }\href
  {https://doi.org/10.1103/PhysRevLett.100.010501} {\bibfield  {journal}
  {\bibinfo  {journal} {Phys. Rev. Lett.}\ }\textbf {\bibinfo {volume} {100}},\
  \bibinfo {pages} {010501}}\BibitemShut {NoStop}%
\bibitem [{\citenamefont {Walschaers}\ \emph {et~al.}(2016)\citenamefont
  {Walschaers}, \citenamefont {Kuipers}, \citenamefont {Urbina}, \citenamefont
  {Mayer}, \citenamefont {Tichy}, \citenamefont {{K. Richter}},\ and\
  \citenamefont {Buchleitner}}]{walschaers_statistical_2016}%
  \BibitemOpen
  \bibfield  {author} {\bibinfo {author} {\bibnamefont {Walschaers},
  \bibfnamefont {M}}, \bibinfo {author} {\bibfnamefont {J.}~\bibnamefont
  {Kuipers}}, \bibinfo {author} {\bibfnamefont {J.-D.}\ \bibnamefont {Urbina}},
  \bibinfo {author} {\bibfnamefont {K.}~\bibnamefont {Mayer}}, \bibinfo
  {author} {\bibfnamefont {M.~C.}\ \bibnamefont {Tichy}}, \bibinfo {author}
  {\bibnamefont {{K. Richter}}}, and\ \bibinfo {author} {\bibfnamefont
  {A.}~\bibnamefont {Buchleitner}}} (\bibinfo {year} {2016}),\ \bibfield
  {title} {\enquote {\bibinfo {title} {Statistical benchmark for
  {{BosonSampling}}},}\ }\href {https://doi.org/10.1088/1367-2630/18/3/032001}
  {\bibfield  {journal} {\bibinfo  {journal} {New J. Phys.}\ }\textbf {\bibinfo
  {volume} {18}}~(\bibinfo {number} {3}),\ \bibinfo {pages}
  {032001}}\BibitemShut {NoStop}%
\bibitem [{\citenamefont {Wang}\ \emph {et~al.}(2020)\citenamefont {Wang},
  \citenamefont {Curtis}, \citenamefont {Lester}, \citenamefont {Zhang},
  \citenamefont {Gao}, \citenamefont {Freeze}, \citenamefont {Batista},
  \citenamefont {Vaccaro}, \citenamefont {Chuang}, \citenamefont {Frunzio},
  \citenamefont {Jiang}, \citenamefont {Girvin},\ and\ \citenamefont
  {Schoelkopf}}]{wang_efficient_2020}%
  \BibitemOpen
  \bibfield  {author} {\bibinfo {author} {\bibnamefont {Wang}, \bibfnamefont
  {C~S}}, \bibinfo {author} {\bibfnamefont {J.~C.}\ \bibnamefont {Curtis}},
  \bibinfo {author} {\bibfnamefont {B.~J.}\ \bibnamefont {Lester}}, \bibinfo
  {author} {\bibfnamefont {Y.}~\bibnamefont {Zhang}}, \bibinfo {author}
  {\bibfnamefont {Y.~Y.}\ \bibnamefont {Gao}}, \bibinfo {author} {\bibfnamefont
  {J.}~\bibnamefont {Freeze}}, \bibinfo {author} {\bibfnamefont {V.~S.}\
  \bibnamefont {Batista}}, \bibinfo {author} {\bibfnamefont {P.~H.}\
  \bibnamefont {Vaccaro}}, \bibinfo {author} {\bibfnamefont {I.~L.}\
  \bibnamefont {Chuang}}, \bibinfo {author} {\bibfnamefont {L.}~\bibnamefont
  {Frunzio}}, \bibinfo {author} {\bibfnamefont {L.}~\bibnamefont {Jiang}},
  \bibinfo {author} {\bibfnamefont {S.~M.}\ \bibnamefont {Girvin}}, and\
  \bibinfo {author} {\bibfnamefont {R.~J.}\ \bibnamefont {Schoelkopf}}}
  (\bibinfo {year} {2020}),\ \bibfield  {title} {\enquote {\bibinfo {title}
  {Efficient {{multiphoton sampling}} of {{molecular vibronic spectra}} on a
  {{superconducting bosonic processor}}},}\ }\href
  {https://doi.org/10.1103/PhysRevX.10.021060} {\bibfield  {journal} {\bibinfo
  {journal} {Phys. Rev. X}\ }\textbf {\bibinfo {volume} {10}},\ \bibinfo
  {pages} {021060}}\BibitemShut {NoStop}%
\bibitem [{\citenamefont {Wang}\ \emph {et~al.}(2017)\citenamefont {Wang},
  \citenamefont {He}, \citenamefont {Li}, \citenamefont {Su}, \citenamefont
  {Li}, \citenamefont {Huang}, \citenamefont {Ding}, \citenamefont {Chen},
  \citenamefont {Liu}, \citenamefont {Qin}, \citenamefont {Li}, \citenamefont
  {He}, \citenamefont {Schneider}, \citenamefont {Kamp}, \citenamefont {Peng},
  \citenamefont {H{\"o}fling}, \citenamefont {Lu},\ and\ \citenamefont
  {Pan}}]{wang_high-efficiency_2017}%
  \BibitemOpen
  \bibfield  {author} {\bibinfo {author} {\bibnamefont {Wang}, \bibfnamefont
  {H}}, \bibinfo {author} {\bibfnamefont {Y.}~\bibnamefont {He}}, \bibinfo
  {author} {\bibfnamefont {Y.-H.}\ \bibnamefont {Li}}, \bibinfo {author}
  {\bibfnamefont {Z.-E.}\ \bibnamefont {Su}}, \bibinfo {author} {\bibfnamefont
  {B.}~\bibnamefont {Li}}, \bibinfo {author} {\bibfnamefont {H.-L.}\
  \bibnamefont {Huang}}, \bibinfo {author} {\bibfnamefont {X.}~\bibnamefont
  {Ding}}, \bibinfo {author} {\bibfnamefont {M.-C.}\ \bibnamefont {Chen}},
  \bibinfo {author} {\bibfnamefont {C.}~\bibnamefont {Liu}}, \bibinfo {author}
  {\bibfnamefont {J.}~\bibnamefont {Qin}}, \bibinfo {author} {\bibfnamefont
  {J.-P.}\ \bibnamefont {Li}}, \bibinfo {author} {\bibfnamefont {Y.-M.}\
  \bibnamefont {He}}, \bibinfo {author} {\bibfnamefont {C.}~\bibnamefont
  {Schneider}}, \bibinfo {author} {\bibfnamefont {M.}~\bibnamefont {Kamp}},
  \bibinfo {author} {\bibfnamefont {C.-Z.}\ \bibnamefont {Peng}}, \bibinfo
  {author} {\bibfnamefont {S.}~\bibnamefont {H{\"o}fling}}, \bibinfo {author}
  {\bibfnamefont {C.-Y.}\ \bibnamefont {Lu}}, and\ \bibinfo {author}
  {\bibfnamefont {J.-W.}\ \bibnamefont {Pan}}} (\bibinfo {year} {2017}),\
  \bibfield  {title} {\enquote {\bibinfo {title} {High-efficiency multiphoton
  boson sampling},}\ }\href {https://doi.org/10.1038/nphoton.2017.63}
  {\bibfield  {journal} {\bibinfo  {journal} {Nature Phot.}\ }\textbf {\bibinfo
  {volume} {11}},\ \bibinfo {pages} {361--365}}\BibitemShut {NoStop}%
\bibitem [{\citenamefont {Wang}\ \emph {et~al.}(2019)\citenamefont {Wang},
  \citenamefont {Qin}, \citenamefont {Ding}, \citenamefont {Chen},
  \citenamefont {Chen}, \citenamefont {You}, \citenamefont {He}, \citenamefont
  {Jiang}, \citenamefont {You}, \citenamefont {Wang}, \citenamefont
  {Schneider}, \citenamefont {Renema}, \citenamefont {H\"ofling}, \citenamefont
  {Lu},\ and\ \citenamefont {Pan}}]{PanSampling}%
  \BibitemOpen
  \bibfield  {author} {\bibinfo {author} {\bibnamefont {Wang}, \bibfnamefont
  {H}}, \bibinfo {author} {\bibfnamefont {J.}~\bibnamefont {Qin}}, \bibinfo
  {author} {\bibfnamefont {X.}~\bibnamefont {Ding}}, \bibinfo {author}
  {\bibfnamefont {M.-C.}\ \bibnamefont {Chen}}, \bibinfo {author}
  {\bibfnamefont {S.}~\bibnamefont {Chen}}, \bibinfo {author} {\bibfnamefont
  {X.}~\bibnamefont {You}}, \bibinfo {author} {\bibfnamefont {Y.-M.}\
  \bibnamefont {He}}, \bibinfo {author} {\bibfnamefont {X.}~\bibnamefont
  {Jiang}}, \bibinfo {author} {\bibfnamefont {L.}~\bibnamefont {You}}, \bibinfo
  {author} {\bibfnamefont {Z.}~\bibnamefont {Wang}}, \bibinfo {author}
  {\bibfnamefont {C.}~\bibnamefont {Schneider}}, \bibinfo {author}
  {\bibfnamefont {J.~J.}\ \bibnamefont {Renema}}, \bibinfo {author}
  {\bibfnamefont {S.}~\bibnamefont {H\"ofling}}, \bibinfo {author}
  {\bibfnamefont {C.-Y.}\ \bibnamefont {Lu}}, and\ \bibinfo {author}
  {\bibfnamefont {J.-W.}\ \bibnamefont {Pan}}} (\bibinfo {year} {2019}),\
  \bibfield  {title} {\enquote {\bibinfo {title} {{Boson sampling with 20 input
  photons and a 60-mode interferometer in a $1{0}^{14}$-dimensional Hilbert
  space}},}\ }\href {https://doi.org/10.1103/PhysRevLett.123.250503} {\bibfield
   {journal} {\bibinfo  {journal} {Phys. Rev. Lett.}\ }\textbf {\bibinfo
  {volume} {123}},\ \bibinfo {pages} {250503}}\BibitemShut {NoStop}%
\bibitem [{\citenamefont {Wang}\ and\ \citenamefont
  {Duan}(2016)}]{wang_certification_2016}%
  \BibitemOpen
  \bibfield  {author} {\bibinfo {author} {\bibnamefont {Wang}, \bibfnamefont
  {S-T}}, and\ \bibinfo {author} {\bibfnamefont {L.-M.}\ \bibnamefont {Duan}}}
  (\bibinfo {year} {2016}),\ \bibfield  {title} {\enquote {\bibinfo {title}
  {Certification of {boson} {sampling} {devices} with {coarse}-{grained}
  {measurements}},}\ }\href@noop {} {\ }\Eprint
  {https://arxiv.org/abs/1601.02627} {arXiv:1601.02627} \BibitemShut {NoStop}%
\bibitem [{\citenamefont {Watrous}(2018)}]{watrous_theory_2018}%
  \BibitemOpen
  \bibfield  {author} {\bibinfo {author} {\bibnamefont {Watrous}, \bibfnamefont
  {J}}} (\bibinfo {year} {2018}),\ \href
  {https://doi.org/10.1017/9781316848142} {\emph {\bibinfo {title} {The
  {{theory}} of {{quantum information}}}}},\ \bibinfo {edition} {1st}\ ed.\
  (\bibinfo  {publisher} {{Cambridge University Press}})\BibitemShut {NoStop}%
\bibitem [{\citenamefont {Welch}\ and\ \citenamefont
  {Berlekamp}(1986)}]{welch_error_1986}%
  \BibitemOpen
  \bibfield  {author} {\bibinfo {author} {\bibnamefont {Welch}, \bibfnamefont
  {L~R}}, and\ \bibinfo {author} {\bibfnamefont {E.~R.}\ \bibnamefont
  {Berlekamp}}} (\bibinfo {year} {1986}),\ \href
  {https://patents.google.com/patent/US4633470A/en} {\enquote {\bibinfo {title}
  {Error correction for algebraic block codes},}\ }\bibinfo {note}
  {Patent}\BibitemShut {NoStop}%
\bibitem [{\citenamefont {Williams}(2015)}]{williams_hardness_2015}%
  \BibitemOpen
  \bibfield  {author} {\bibinfo {author} {\bibnamefont {Williams},
  \bibfnamefont {V~V}}} (\bibinfo {year} {2015}),\ \bibfield  {title} {\enquote
  {\bibinfo {title} {Hardness of {{easy problems}}: {{Basing hardness}} on
  {{popular conjectures}} such as the {{strong exponential time hypothesis}}
  ({{invited talk}})},}\ }in\ \href
  {https://doi.org/10.4230/LIPIcs.IPEC.2015.17} {\emph {\bibinfo {booktitle}
  {10th {{International Symposium}} on {{Parameterized}} and {{Exact
  Computation}} ({{IPEC}} 2015)}}},\ \bibinfo {series} {Leibniz {{International
  Proceedings}} in {{Informatics}} ({{LIPIcs}})}, Vol.~\bibinfo {volume} {43},\
  \bibinfo {editor} {edited by\ \bibinfo {editor} {\bibfnamefont
  {T.}~\bibnamefont {Husfeldt}}\ and\ \bibinfo {editor} {\bibfnamefont
  {I.}~\bibnamefont {Kanj}}}\ (\bibinfo  {publisher} {{Schloss
  Dagstuhl\textendash Leibniz-Zentrum fuer Informatik}},\ \bibinfo {address}
  {{Dagstuhl, Germany}})\ pp.\ \bibinfo {pages} {17--29}\BibitemShut {NoStop}%
\bibitem [{\citenamefont {Wu}\ \emph {et~al.}(2018)\citenamefont {Wu},
  \citenamefont {Liu}, \citenamefont {Zhang}, \citenamefont {Jin},
  \citenamefont {Wang}, \citenamefont {Wang},\ and\ \citenamefont
  {Yang}}]{wu_benchmark_2018}%
  \BibitemOpen
  \bibfield  {author} {\bibinfo {author} {\bibnamefont {Wu}, \bibfnamefont
  {J}}, \bibinfo {author} {\bibfnamefont {Y.}~\bibnamefont {Liu}}, \bibinfo
  {author} {\bibfnamefont {B.}~\bibnamefont {Zhang}}, \bibinfo {author}
  {\bibfnamefont {X.}~\bibnamefont {Jin}}, \bibinfo {author} {\bibfnamefont
  {Y.}~\bibnamefont {Wang}}, \bibinfo {author} {\bibfnamefont {H.}~\bibnamefont
  {Wang}}, and\ \bibinfo {author} {\bibfnamefont {X.}~\bibnamefont {Yang}}}
  (\bibinfo {year} {2018}),\ \bibfield  {title} {\enquote {\bibinfo {title} {A
  benchmark test of boson sampling on {{Tianhe-2}} supercomputer},}\ }\href
  {https://doi.org/10.1093/nsr/nwy079} {\bibfield  {journal} {\bibinfo
  {journal} {Natl. Sc. Rev.}\ }\textbf {\bibinfo {volume} {5}},\ \bibinfo
  {pages} {715--720}}\BibitemShut {NoStop}%
\bibitem [{\citenamefont {Wu}\ \emph {et~al.}(2021)\citenamefont {Wu},
  \citenamefont {Bao}, \citenamefont {Cao}, \citenamefont {Chen}, \citenamefont
  {Chen}, \citenamefont {Chen}, \citenamefont {Chung}, \citenamefont {Deng},
  \citenamefont {Du}, \citenamefont {Fan}, \citenamefont {Gong}, \citenamefont
  {Guo}, \citenamefont {Guo}, \citenamefont {Guo}, \citenamefont {Han},
  \citenamefont {Hong}, \citenamefont {Huang}, \citenamefont {Huo},
  \citenamefont {Li}, \citenamefont {Li}, \citenamefont {Li}, \citenamefont
  {Li}, \citenamefont {Liang}, \citenamefont {Lin}, \citenamefont {Lin},
  \citenamefont {Qian}, \citenamefont {Qiao}, \citenamefont {Rong},
  \citenamefont {Su}, \citenamefont {Sun}, \citenamefont {Wang}, \citenamefont
  {Wang}, \citenamefont {Wu}, \citenamefont {Xu}, \citenamefont {Yan},
  \citenamefont {Yang}, \citenamefont {Yang}, \citenamefont {Ye}, \citenamefont
  {Yin}, \citenamefont {Ying}, \citenamefont {Yu}, \citenamefont {Zha},
  \citenamefont {Zhang}, \citenamefont {Zhang}, \citenamefont {Zhang},
  \citenamefont {Zhang}, \citenamefont {Zhao}, \citenamefont {Zhao},
  \citenamefont {Zhou}, \citenamefont {Zhu}, \citenamefont {Lu}, \citenamefont
  {Peng}, \citenamefont {Zhu},\ and\ \citenamefont {Pan}}]{wu_strong_2021}%
  \BibitemOpen
  \bibfield  {author} {\bibinfo {author} {\bibnamefont {Wu}, \bibfnamefont
  {Y}}, \bibinfo {author} {\bibfnamefont {W.-S.}\ \bibnamefont {Bao}}, \bibinfo
  {author} {\bibfnamefont {S.}~\bibnamefont {Cao}}, \bibinfo {author}
  {\bibfnamefont {F.}~\bibnamefont {Chen}}, \bibinfo {author} {\bibfnamefont
  {M.-C.}\ \bibnamefont {Chen}}, \bibinfo {author} {\bibfnamefont
  {X.}~\bibnamefont {Chen}}, \bibinfo {author} {\bibfnamefont {T.-H.}\
  \bibnamefont {Chung}}, \bibinfo {author} {\bibfnamefont {H.}~\bibnamefont
  {Deng}}, \bibinfo {author} {\bibfnamefont {Y.}~\bibnamefont {Du}}, \bibinfo
  {author} {\bibfnamefont {D.}~\bibnamefont {Fan}}, \bibinfo {author}
  {\bibfnamefont {M.}~\bibnamefont {Gong}}, \bibinfo {author} {\bibfnamefont
  {C.}~\bibnamefont {Guo}}, \bibinfo {author} {\bibfnamefont {C.}~\bibnamefont
  {Guo}}, \bibinfo {author} {\bibfnamefont {S.}~\bibnamefont {Guo}}, \bibinfo
  {author} {\bibfnamefont {L.}~\bibnamefont {Han}}, \bibinfo {author}
  {\bibfnamefont {L.}~\bibnamefont {Hong}}, \bibinfo {author} {\bibfnamefont
  {H.-L.}\ \bibnamefont {Huang}}, \bibinfo {author} {\bibfnamefont {Y.-H.}\
  \bibnamefont {Huo}}, \bibinfo {author} {\bibfnamefont {L.}~\bibnamefont
  {Li}}, \bibinfo {author} {\bibfnamefont {N.}~\bibnamefont {Li}}, \bibinfo
  {author} {\bibfnamefont {S.}~\bibnamefont {Li}}, \bibinfo {author}
  {\bibfnamefont {Y.}~\bibnamefont {Li}}, \bibinfo {author} {\bibfnamefont
  {F.}~\bibnamefont {Liang}}, \bibinfo {author} {\bibfnamefont
  {C.}~\bibnamefont {Lin}}, \bibinfo {author} {\bibfnamefont {J.}~\bibnamefont
  {Lin}}, \bibinfo {author} {\bibfnamefont {H.}~\bibnamefont {Qian}}, \bibinfo
  {author} {\bibfnamefont {D.}~\bibnamefont {Qiao}}, \bibinfo {author}
  {\bibfnamefont {H.}~\bibnamefont {Rong}}, \bibinfo {author} {\bibfnamefont
  {H.}~\bibnamefont {Su}}, \bibinfo {author} {\bibfnamefont {L.}~\bibnamefont
  {Sun}}, \bibinfo {author} {\bibfnamefont {L.}~\bibnamefont {Wang}}, \bibinfo
  {author} {\bibfnamefont {S.}~\bibnamefont {Wang}}, \bibinfo {author}
  {\bibfnamefont {D.}~\bibnamefont {Wu}}, \bibinfo {author} {\bibfnamefont
  {Y.}~\bibnamefont {Xu}}, \bibinfo {author} {\bibfnamefont {K.}~\bibnamefont
  {Yan}}, \bibinfo {author} {\bibfnamefont {W.}~\bibnamefont {Yang}}, \bibinfo
  {author} {\bibfnamefont {Y.}~\bibnamefont {Yang}}, \bibinfo {author}
  {\bibfnamefont {Y.}~\bibnamefont {Ye}}, \bibinfo {author} {\bibfnamefont
  {J.}~\bibnamefont {Yin}}, \bibinfo {author} {\bibfnamefont {C.}~\bibnamefont
  {Ying}}, \bibinfo {author} {\bibfnamefont {J.}~\bibnamefont {Yu}}, \bibinfo
  {author} {\bibfnamefont {C.}~\bibnamefont {Zha}}, \bibinfo {author}
  {\bibfnamefont {C.}~\bibnamefont {Zhang}}, \bibinfo {author} {\bibfnamefont
  {H.}~\bibnamefont {Zhang}}, \bibinfo {author} {\bibfnamefont
  {K.}~\bibnamefont {Zhang}}, \bibinfo {author} {\bibfnamefont
  {Y.}~\bibnamefont {Zhang}}, \bibinfo {author} {\bibfnamefont
  {H.}~\bibnamefont {Zhao}}, \bibinfo {author} {\bibfnamefont {Y.}~\bibnamefont
  {Zhao}}, \bibinfo {author} {\bibfnamefont {L.}~\bibnamefont {Zhou}}, \bibinfo
  {author} {\bibfnamefont {Q.}~\bibnamefont {Zhu}}, \bibinfo {author}
  {\bibfnamefont {C.-Y.}\ \bibnamefont {Lu}}, \bibinfo {author} {\bibfnamefont
  {C.-Z.}\ \bibnamefont {Peng}}, \bibinfo {author} {\bibfnamefont
  {X.}~\bibnamefont {Zhu}}, and\ \bibinfo {author} {\bibfnamefont {J.-W.}\
  \bibnamefont {Pan}}} (\bibinfo {year} {2021}),\ \bibfield  {title} {\enquote
  {\bibinfo {title} {Strong quantum computational advantage using a
  superconducting quantum processor},}\ }\href
  {https://doi.org/10.1103/PhysRevLett.127.180501} {\bibfield  {journal}
  {\bibinfo  {journal} {Phys. Rev. Lett.}\ }\textbf {\bibinfo {volume} {127}},\
  \bibinfo {pages} {180501}}\BibitemShut {NoStop}%
\bibitem [{\citenamefont {Yamakawa}\ and\ \citenamefont
  {Zhandry}(2022)}]{yamakawa_verifiable_2022}%
  \BibitemOpen
  \bibfield  {author} {\bibinfo {author} {\bibnamefont {Yamakawa},
  \bibfnamefont {T}}, and\ \bibinfo {author} {\bibfnamefont {M.}~\bibnamefont
  {Zhandry}}} (\bibinfo {year} {2022}),\ \bibfield  {title} {\enquote {\bibinfo
  {title} {Verifiable {{quantum advantage}} without {{structure}}},}\
  }\href@noop {} {\ }\Eprint {https://arxiv.org/abs/2204.02063}
  {arXiv:2204.02063} \BibitemShut {NoStop}%
\bibitem [{\citenamefont {Yoganathan}\ \emph {et~al.}(2019)\citenamefont
  {Yoganathan}, \citenamefont {Jozsa},\ and\ \citenamefont
  {Strelchuk}}]{yoganathan_quantum_2018}%
  \BibitemOpen
  \bibfield  {author} {\bibinfo {author} {\bibnamefont {Yoganathan},
  \bibfnamefont {M}}, \bibinfo {author} {\bibfnamefont {R.}~\bibnamefont
  {Jozsa}}, and\ \bibinfo {author} {\bibfnamefont {S.}~\bibnamefont
  {Strelchuk}}} (\bibinfo {year} {2019}),\ \bibfield  {title} {\enquote
  {\bibinfo {title} {{Quantum advantage of unitary {Clifford} circuits with
  magic state inputs}},}\ }\href {https://doi.org/10.1098/rspa.2018.0427}
  {\bibfield  {journal} {\bibinfo  {journal} {Proc. R. Soc. A}\ }\textbf
  {\bibinfo {volume} {475}},\ \bibinfo {pages} {20180427}}\BibitemShut
  {NoStop}%
\bibitem [{\citenamefont {Yung}\ and\ \citenamefont
  {Gao}(2017)}]{yung_can_2017}%
  \BibitemOpen
  \bibfield  {author} {\bibinfo {author} {\bibnamefont {Yung}, \bibfnamefont
  {M-H}}, and\ \bibinfo {author} {\bibfnamefont {X.}~\bibnamefont {Gao}}}
  (\bibinfo {year} {2017}),\ \bibfield  {title} {\enquote {\bibinfo {title}
  {Can chaotic quantum circuits maintain quantum supremacy under noise?}}\
  }\href {http://arxiv.org/abs/1706.08913} {\ }\Eprint
  {https://arxiv.org/abs/1706.08913} {arxiv:1706.08913} \BibitemShut {NoStop}%
\bibitem [{\citenamefont {Yung}\ \emph {et~al.}(2016)\citenamefont {Yung},
  \citenamefont {Gao},\ and\ \citenamefont {Huh}}]{yung_universal_2016}%
  \BibitemOpen
  \bibfield  {author} {\bibinfo {author} {\bibnamefont {Yung}, \bibfnamefont
  {M-H}}, \bibinfo {author} {\bibfnamefont {X.}~\bibnamefont {Gao}}, and\
  \bibinfo {author} {\bibfnamefont {J.}~\bibnamefont {Huh}}} (\bibinfo {year}
  {2016}),\ \bibfield  {title} {\enquote {\bibinfo {title} {Universal {{bound}}
  on {{sampling bosons}} in {{linear optics}}},}\ }\href@noop {} {\ }\Eprint
  {https://arxiv.org/abs/1608.00383} {arXiv:1608.00383} \BibitemShut {NoStop}%
\bibitem [{\citenamefont {Zhong}\ \emph {et~al.}(2021)\citenamefont {Zhong},
  \citenamefont {Deng}, \citenamefont {Qin}, \citenamefont {Wang},
  \citenamefont {Chen}, \citenamefont {Peng}, \citenamefont {Luo},
  \citenamefont {Wu}, \citenamefont {Gong}, \citenamefont {Su}, \citenamefont
  {Hu}, \citenamefont {Hu}, \citenamefont {Yang}, \citenamefont {Zhang},
  \citenamefont {Li}, \citenamefont {Li}, \citenamefont {Jiang}, \citenamefont
  {Gan}, \citenamefont {Yang}, \citenamefont {You}, \citenamefont {Wang},
  \citenamefont {Li}, \citenamefont {Liu}, \citenamefont {Renema},
  \citenamefont {Lu},\ and\ \citenamefont
  {Pan}}]{zhong_phase-programmable_2021}%
  \BibitemOpen
  \bibfield  {author} {\bibinfo {author} {\bibnamefont {Zhong}, \bibfnamefont
  {H-S}}, \bibinfo {author} {\bibfnamefont {Y.-H.}\ \bibnamefont {Deng}},
  \bibinfo {author} {\bibfnamefont {J.}~\bibnamefont {Qin}}, \bibinfo {author}
  {\bibfnamefont {H.}~\bibnamefont {Wang}}, \bibinfo {author} {\bibfnamefont
  {M.-C.}\ \bibnamefont {Chen}}, \bibinfo {author} {\bibfnamefont {L.-C.}\
  \bibnamefont {Peng}}, \bibinfo {author} {\bibfnamefont {Y.-H.}\ \bibnamefont
  {Luo}}, \bibinfo {author} {\bibfnamefont {D.}~\bibnamefont {Wu}}, \bibinfo
  {author} {\bibfnamefont {S.-Q.}\ \bibnamefont {Gong}}, \bibinfo {author}
  {\bibfnamefont {H.}~\bibnamefont {Su}}, \bibinfo {author} {\bibfnamefont
  {Y.}~\bibnamefont {Hu}}, \bibinfo {author} {\bibfnamefont {P.}~\bibnamefont
  {Hu}}, \bibinfo {author} {\bibfnamefont {X.-Y.}\ \bibnamefont {Yang}},
  \bibinfo {author} {\bibfnamefont {W.-J.}\ \bibnamefont {Zhang}}, \bibinfo
  {author} {\bibfnamefont {H.}~\bibnamefont {Li}}, \bibinfo {author}
  {\bibfnamefont {Y.}~\bibnamefont {Li}}, \bibinfo {author} {\bibfnamefont
  {X.}~\bibnamefont {Jiang}}, \bibinfo {author} {\bibfnamefont
  {L.}~\bibnamefont {Gan}}, \bibinfo {author} {\bibfnamefont {G.}~\bibnamefont
  {Yang}}, \bibinfo {author} {\bibfnamefont {L.}~\bibnamefont {You}}, \bibinfo
  {author} {\bibfnamefont {Z.}~\bibnamefont {Wang}}, \bibinfo {author}
  {\bibfnamefont {L.}~\bibnamefont {Li}}, \bibinfo {author} {\bibfnamefont
  {N.-L.}\ \bibnamefont {Liu}}, \bibinfo {author} {\bibfnamefont {J.~J.}\
  \bibnamefont {Renema}}, \bibinfo {author} {\bibfnamefont {C.-Y.}\
  \bibnamefont {Lu}}, and\ \bibinfo {author} {\bibfnamefont {J.-W.}\
  \bibnamefont {Pan}}} (\bibinfo {year} {2021}),\ \bibfield  {title} {\enquote
  {\bibinfo {title} {Phase-{{programmable Gaussian boson sampling using
  stimulated squeezed light}}},}\ }\href
  {https://doi.org/10.1103/PhysRevLett.127.180502} {\bibfield  {journal}
  {\bibinfo  {journal} {Phys. Rev. Lett.}\ }\textbf {\bibinfo {volume} {127}},\
  \bibinfo {pages} {180502}}\BibitemShut {NoStop}%
\bibitem [{\citenamefont {Zhong}\ \emph {et~al.}(2018)\citenamefont {Zhong},
  \citenamefont {Li}, \citenamefont {Li}, \citenamefont {Peng}, \citenamefont
  {Su}, \citenamefont {Hu}, \citenamefont {He}, \citenamefont {Ding},
  \citenamefont {Zhang}, \citenamefont {Li}, \citenamefont {Zhang},
  \citenamefont {Wang}, \citenamefont {You}, \citenamefont {Wang},
  \citenamefont {Jiang}, \citenamefont {Li}, \citenamefont {Chen},
  \citenamefont {Liu}, \citenamefont {Lu},\ and\ \citenamefont
  {Pan}}]{zhong_12-photon_2018}%
  \BibitemOpen
  \bibfield  {author} {\bibinfo {author} {\bibnamefont {Zhong}, \bibfnamefont
  {H-S}}, \bibinfo {author} {\bibfnamefont {Y.}~\bibnamefont {Li}}, \bibinfo
  {author} {\bibfnamefont {W.}~\bibnamefont {Li}}, \bibinfo {author}
  {\bibfnamefont {L.-C.}\ \bibnamefont {Peng}}, \bibinfo {author}
  {\bibfnamefont {Z.-E.}\ \bibnamefont {Su}}, \bibinfo {author} {\bibfnamefont
  {Y.}~\bibnamefont {Hu}}, \bibinfo {author} {\bibfnamefont {Y.-M.}\
  \bibnamefont {He}}, \bibinfo {author} {\bibfnamefont {X.}~\bibnamefont
  {Ding}}, \bibinfo {author} {\bibfnamefont {W.}~\bibnamefont {Zhang}},
  \bibinfo {author} {\bibfnamefont {H.}~\bibnamefont {Li}}, \bibinfo {author}
  {\bibfnamefont {L.}~\bibnamefont {Zhang}}, \bibinfo {author} {\bibfnamefont
  {Z.}~\bibnamefont {Wang}}, \bibinfo {author} {\bibfnamefont {L.}~\bibnamefont
  {You}}, \bibinfo {author} {\bibfnamefont {X.-L.}\ \bibnamefont {Wang}},
  \bibinfo {author} {\bibfnamefont {X.}~\bibnamefont {Jiang}}, \bibinfo
  {author} {\bibfnamefont {L.}~\bibnamefont {Li}}, \bibinfo {author}
  {\bibfnamefont {Y.-A.}\ \bibnamefont {Chen}}, \bibinfo {author}
  {\bibfnamefont {N.-L.}\ \bibnamefont {Liu}}, \bibinfo {author} {\bibfnamefont
  {C.-Y.}\ \bibnamefont {Lu}}, and\ \bibinfo {author} {\bibfnamefont {J.-W.}\
  \bibnamefont {Pan}}} (\bibinfo {year} {2018}),\ \bibfield  {title} {\enquote
  {\bibinfo {title} {12-{{photon entanglement}} and {{scalable scattershot
  boson sampling}} with {{optimal entangled-photon pairs}} from {{parametric
  down-conversion}}},}\ }\href {https://doi.org/10.1103/PhysRevLett.121.250505}
  {\bibfield  {journal} {\bibinfo  {journal} {Phys. Rev. Lett.}\ }\textbf
  {\bibinfo {volume} {121}},\ \bibinfo {pages} {250505}}\BibitemShut {NoStop}%
\bibitem [{\citenamefont {Zhong}\ \emph {et~al.}(2020)\citenamefont {Zhong},
  \citenamefont {Wang}, \citenamefont {Deng}, \citenamefont {Chen},
  \citenamefont {Peng}, \citenamefont {Luo}, \citenamefont {Qin}, \citenamefont
  {Wu}, \citenamefont {Ding}, \citenamefont {Hu}, \citenamefont {Hu},
  \citenamefont {Yang}, \citenamefont {Zhang}, \citenamefont {Li},
  \citenamefont {Li}, \citenamefont {Jiang}, \citenamefont {Gan}, \citenamefont
  {Yang}, \citenamefont {You}, \citenamefont {Wang}, \citenamefont {Li},
  \citenamefont {Liu}, \citenamefont {Lu},\ and\ \citenamefont
  {Pan}}]{zhong_quantum_2020}%
  \BibitemOpen
  \bibfield  {author} {\bibinfo {author} {\bibnamefont {Zhong}, \bibfnamefont
  {H-S}}, \bibinfo {author} {\bibfnamefont {H.}~\bibnamefont {Wang}}, \bibinfo
  {author} {\bibfnamefont {Y.-H.}\ \bibnamefont {Deng}}, \bibinfo {author}
  {\bibfnamefont {M.-C.}\ \bibnamefont {Chen}}, \bibinfo {author}
  {\bibfnamefont {L.-C.}\ \bibnamefont {Peng}}, \bibinfo {author}
  {\bibfnamefont {Y.-H.}\ \bibnamefont {Luo}}, \bibinfo {author} {\bibfnamefont
  {J.}~\bibnamefont {Qin}}, \bibinfo {author} {\bibfnamefont {D.}~\bibnamefont
  {Wu}}, \bibinfo {author} {\bibfnamefont {X.}~\bibnamefont {Ding}}, \bibinfo
  {author} {\bibfnamefont {Y.}~\bibnamefont {Hu}}, \bibinfo {author}
  {\bibfnamefont {P.}~\bibnamefont {Hu}}, \bibinfo {author} {\bibfnamefont
  {X.-Y.}\ \bibnamefont {Yang}}, \bibinfo {author} {\bibfnamefont {W.-J.}\
  \bibnamefont {Zhang}}, \bibinfo {author} {\bibfnamefont {H.}~\bibnamefont
  {Li}}, \bibinfo {author} {\bibfnamefont {Y.}~\bibnamefont {Li}}, \bibinfo
  {author} {\bibfnamefont {X.}~\bibnamefont {Jiang}}, \bibinfo {author}
  {\bibfnamefont {L.}~\bibnamefont {Gan}}, \bibinfo {author} {\bibfnamefont
  {G.}~\bibnamefont {Yang}}, \bibinfo {author} {\bibfnamefont {L.}~\bibnamefont
  {You}}, \bibinfo {author} {\bibfnamefont {Z.}~\bibnamefont {Wang}}, \bibinfo
  {author} {\bibfnamefont {L.}~\bibnamefont {Li}}, \bibinfo {author}
  {\bibfnamefont {N.-L.}\ \bibnamefont {Liu}}, \bibinfo {author} {\bibfnamefont
  {C.-Y.}\ \bibnamefont {Lu}}, and\ \bibinfo {author} {\bibfnamefont {J.-W.}\
  \bibnamefont {Pan}}} (\bibinfo {year} {2020}),\ \bibfield  {title} {\enquote
  {\bibinfo {title} {Quantum computational advantage using photons},}\ }\href
  {https://doi.org/10.1126/science.abe8770} {\bibfield  {journal} {\bibinfo
  {journal} {Science}\ }\textbf {\bibinfo {volume} {370}},\ \bibinfo {pages}
  {1460--1463}}\BibitemShut {NoStop}%
\bibitem [{\citenamefont {Zhou}\ \emph
  {et~al.}(2020{\natexlab{a}})\citenamefont {Zhou}, \citenamefont {Wang},
  \citenamefont {Choi}, \citenamefont {Pichler},\ and\ \citenamefont
  {Lukin}}]{PhysRevX.10.021067}%
  \BibitemOpen
  \bibfield  {author} {\bibinfo {author} {\bibnamefont {Zhou}, \bibfnamefont
  {L}}, \bibinfo {author} {\bibfnamefont {S.-T.}\ \bibnamefont {Wang}},
  \bibinfo {author} {\bibfnamefont {S.}~\bibnamefont {Choi}}, \bibinfo {author}
  {\bibfnamefont {H.}~\bibnamefont {Pichler}}, and\ \bibinfo {author}
  {\bibfnamefont {M.~D.}\ \bibnamefont {Lukin}}} (\bibinfo {year}
  {2020}{\natexlab{a}}),\ \bibfield  {title} {\enquote {\bibinfo {title}
  {Quantum approximate optimization algorithm: Performance, mechanism, and
  implementation on near-term devices},}\ }\href
  {https://doi.org/10.1103/PhysRevX.10.021067} {\bibfield  {journal} {\bibinfo
  {journal} {Phys. Rev. X}\ }\textbf {\bibinfo {volume} {10}},\ \bibinfo
  {pages} {021067}}\BibitemShut {NoStop}%
\bibitem [{\citenamefont {Zhou}\ and\ \citenamefont
  {Nahum}(2019)}]{zhou_emergent_2019}%
  \BibitemOpen
  \bibfield  {author} {\bibinfo {author} {\bibnamefont {Zhou}, \bibfnamefont
  {T}}, and\ \bibinfo {author} {\bibfnamefont {A.}~\bibnamefont {Nahum}}}
  (\bibinfo {year} {2019}),\ \bibfield  {title} {\enquote {\bibinfo {title}
  {Emergent statistical mechanics of entanglement in random unitary
  circuits},}\ }\href {https://doi.org/10.1103/PhysRevB.99.174205} {\bibfield
  {journal} {\bibinfo  {journal} {Phys. Rev. B}\ }\textbf {\bibinfo {volume}
  {99}},\ \bibinfo {pages} {174205}}\BibitemShut {NoStop}%
\bibitem [{\citenamefont {Zhou}\ \emph
  {et~al.}(2020{\natexlab{b}})\citenamefont {Zhou}, \citenamefont
  {Stoudenmire},\ and\ \citenamefont {Waintal}}]{zhou_what_2020-1}%
  \BibitemOpen
  \bibfield  {author} {\bibinfo {author} {\bibnamefont {Zhou}, \bibfnamefont
  {Y}}, \bibinfo {author} {\bibfnamefont {E.~M.}\ \bibnamefont {Stoudenmire}},
  and\ \bibinfo {author} {\bibfnamefont {X.}~\bibnamefont {Waintal}}} (\bibinfo
  {year} {2020}{\natexlab{b}}),\ \bibfield  {title} {\enquote {\bibinfo {title}
  {What {{limits}} the {{simulation}} of {{quantum computers}}?}}\ }\href
  {https://doi.org/10.1103/PhysRevX.10.041038} {\bibfield  {journal} {\bibinfo
  {journal} {Phys. Rev. X}\ }\textbf {\bibinfo {volume} {10}},\ \bibinfo
  {pages} {041038}}\BibitemShut {NoStop}%
\bibitem [{\citenamefont {Zhu}\ \emph {et~al.}(2021)\citenamefont {Zhu},
  \citenamefont {{Kahanamoku-Meyer}}, \citenamefont {Lewis}, \citenamefont
  {Noel}, \citenamefont {Katz}, \citenamefont {Harraz}, \citenamefont {Wang},
  \citenamefont {Risinger}, \citenamefont {Feng}, \citenamefont {Biswas},
  \citenamefont {Egan}, \citenamefont {Gheorghiu}, \citenamefont {Nam},
  \citenamefont {Vidick}, \citenamefont {Vazirani}, \citenamefont {Yao},
  \citenamefont {Cetina},\ and\ \citenamefont {Monroe}}]{zhu_interactive_2021}%
  \BibitemOpen
  \bibfield  {author} {\bibinfo {author} {\bibnamefont {Zhu}, \bibfnamefont
  {D}}, \bibinfo {author} {\bibfnamefont {G.~D.}\ \bibnamefont
  {{Kahanamoku-Meyer}}}, \bibinfo {author} {\bibfnamefont {L.}~\bibnamefont
  {Lewis}}, \bibinfo {author} {\bibfnamefont {C.}~\bibnamefont {Noel}},
  \bibinfo {author} {\bibfnamefont {O.}~\bibnamefont {Katz}}, \bibinfo {author}
  {\bibfnamefont {B.}~\bibnamefont {Harraz}}, \bibinfo {author} {\bibfnamefont
  {Q.}~\bibnamefont {Wang}}, \bibinfo {author} {\bibfnamefont {A.}~\bibnamefont
  {Risinger}}, \bibinfo {author} {\bibfnamefont {L.}~\bibnamefont {Feng}},
  \bibinfo {author} {\bibfnamefont {D.}~\bibnamefont {Biswas}}, \bibinfo
  {author} {\bibfnamefont {L.}~\bibnamefont {Egan}}, \bibinfo {author}
  {\bibfnamefont {A.}~\bibnamefont {Gheorghiu}}, \bibinfo {author}
  {\bibfnamefont {Y.}~\bibnamefont {Nam}}, \bibinfo {author} {\bibfnamefont
  {T.}~\bibnamefont {Vidick}}, \bibinfo {author} {\bibfnamefont
  {U.}~\bibnamefont {Vazirani}}, \bibinfo {author} {\bibfnamefont {N.~Y.}\
  \bibnamefont {Yao}}, \bibinfo {author} {\bibfnamefont {M.}~\bibnamefont
  {Cetina}}, and\ \bibinfo {author} {\bibfnamefont {C.}~\bibnamefont {Monroe}}}
  (\bibinfo {year} {2021}),\ \bibfield  {title} {\enquote {\bibinfo {title}
  {Interactive {{protocols}} for {{classically-verifiable quantum
  advantage}}},}\ }\href@noop {} {\ }\Eprint {https://arxiv.org/abs/2112.05156}
  {arXiv:2112.05156} \BibitemShut {NoStop}%
\bibitem [{\citenamefont {Zhu}\ and\ \citenamefont
  {Hayashi}(2019)}]{zhu_efficient_2019}%
  \BibitemOpen
  \bibfield  {author} {\bibinfo {author} {\bibnamefont {Zhu}, \bibfnamefont
  {H}}, and\ \bibinfo {author} {\bibfnamefont {M.}~\bibnamefont {Hayashi}}}
  (\bibinfo {year} {2019}),\ \bibfield  {title} {\enquote {\bibinfo {title}
  {Efficient verification of hypergraph states},}\ }\href
  {https://doi.org/10.1103/PhysRevApplied.12.054047} {\bibfield  {journal}
  {\bibinfo  {journal} {Phys. Rev. Appl.}\ }\textbf {\bibinfo {volume} {12}},\
  \bibinfo {pages} {054047}}\BibitemShut {NoStop}%
\bibitem [{\citenamefont {Zhu}\ \emph {et~al.}(2022)\citenamefont {Zhu},
  \citenamefont {Cao}, \citenamefont {Chen}, \citenamefont {Chen},
  \citenamefont {Chen}, \citenamefont {Chung}, \citenamefont {Deng},
  \citenamefont {Du}, \citenamefont {Fan}, \citenamefont {Gong}, \citenamefont
  {Guo}, \citenamefont {Guo}, \citenamefont {Guo}, \citenamefont {Han},
  \citenamefont {Hong}, \citenamefont {Huang}, \citenamefont {Huo},
  \citenamefont {Li}, \citenamefont {Li}, \citenamefont {Li}, \citenamefont
  {Li}, \citenamefont {Liang}, \citenamefont {Lin}, \citenamefont {Lin},
  \citenamefont {Qian}, \citenamefont {Qiao}, \citenamefont {Rong},
  \citenamefont {Su}, \citenamefont {Sun}, \citenamefont {Wang}, \citenamefont
  {Wang}, \citenamefont {Wu}, \citenamefont {Wu}, \citenamefont {Xu},
  \citenamefont {Yan}, \citenamefont {Yang}, \citenamefont {Yang},
  \citenamefont {Ye}, \citenamefont {Yin}, \citenamefont {Ying}, \citenamefont
  {Yu}, \citenamefont {Zha}, \citenamefont {Zhang}, \citenamefont {Zhang},
  \citenamefont {Zhang}, \citenamefont {Zhang}, \citenamefont {Zhao},
  \citenamefont {Zhao}, \citenamefont {Zhou}, \citenamefont {Lu}, \citenamefont
  {Peng}, \citenamefont {Zhu},\ and\ \citenamefont {Pan}}]{zhu_quantum_2022}%
  \BibitemOpen
  \bibfield  {author} {\bibinfo {author} {\bibnamefont {Zhu}, \bibfnamefont
  {Q}}, \bibinfo {author} {\bibfnamefont {S.}~\bibnamefont {Cao}}, \bibinfo
  {author} {\bibfnamefont {F.}~\bibnamefont {Chen}}, \bibinfo {author}
  {\bibfnamefont {M.-C.}\ \bibnamefont {Chen}}, \bibinfo {author}
  {\bibfnamefont {X.}~\bibnamefont {Chen}}, \bibinfo {author} {\bibfnamefont
  {T.-H.}\ \bibnamefont {Chung}}, \bibinfo {author} {\bibfnamefont
  {H.}~\bibnamefont {Deng}}, \bibinfo {author} {\bibfnamefont {Y.}~\bibnamefont
  {Du}}, \bibinfo {author} {\bibfnamefont {D.}~\bibnamefont {Fan}}, \bibinfo
  {author} {\bibfnamefont {M.}~\bibnamefont {Gong}}, \bibinfo {author}
  {\bibfnamefont {C.}~\bibnamefont {Guo}}, \bibinfo {author} {\bibfnamefont
  {C.}~\bibnamefont {Guo}}, \bibinfo {author} {\bibfnamefont {S.}~\bibnamefont
  {Guo}}, \bibinfo {author} {\bibfnamefont {L.}~\bibnamefont {Han}}, \bibinfo
  {author} {\bibfnamefont {L.}~\bibnamefont {Hong}}, \bibinfo {author}
  {\bibfnamefont {H.-L.}\ \bibnamefont {Huang}}, \bibinfo {author}
  {\bibfnamefont {Y.-H.}\ \bibnamefont {Huo}}, \bibinfo {author} {\bibfnamefont
  {L.}~\bibnamefont {Li}}, \bibinfo {author} {\bibfnamefont {N.}~\bibnamefont
  {Li}}, \bibinfo {author} {\bibfnamefont {S.}~\bibnamefont {Li}}, \bibinfo
  {author} {\bibfnamefont {Y.}~\bibnamefont {Li}}, \bibinfo {author}
  {\bibfnamefont {F.}~\bibnamefont {Liang}}, \bibinfo {author} {\bibfnamefont
  {C.}~\bibnamefont {Lin}}, \bibinfo {author} {\bibfnamefont {J.}~\bibnamefont
  {Lin}}, \bibinfo {author} {\bibfnamefont {H.}~\bibnamefont {Qian}}, \bibinfo
  {author} {\bibfnamefont {D.}~\bibnamefont {Qiao}}, \bibinfo {author}
  {\bibfnamefont {H.}~\bibnamefont {Rong}}, \bibinfo {author} {\bibfnamefont
  {H.}~\bibnamefont {Su}}, \bibinfo {author} {\bibfnamefont {L.}~\bibnamefont
  {Sun}}, \bibinfo {author} {\bibfnamefont {L.}~\bibnamefont {Wang}}, \bibinfo
  {author} {\bibfnamefont {S.}~\bibnamefont {Wang}}, \bibinfo {author}
  {\bibfnamefont {D.}~\bibnamefont {Wu}}, \bibinfo {author} {\bibfnamefont
  {Y.}~\bibnamefont {Wu}}, \bibinfo {author} {\bibfnamefont {Y.}~\bibnamefont
  {Xu}}, \bibinfo {author} {\bibfnamefont {K.}~\bibnamefont {Yan}}, \bibinfo
  {author} {\bibfnamefont {W.}~\bibnamefont {Yang}}, \bibinfo {author}
  {\bibfnamefont {Y.}~\bibnamefont {Yang}}, \bibinfo {author} {\bibfnamefont
  {Y.}~\bibnamefont {Ye}}, \bibinfo {author} {\bibfnamefont {J.}~\bibnamefont
  {Yin}}, \bibinfo {author} {\bibfnamefont {C.}~\bibnamefont {Ying}}, \bibinfo
  {author} {\bibfnamefont {J.}~\bibnamefont {Yu}}, \bibinfo {author}
  {\bibfnamefont {C.}~\bibnamefont {Zha}}, \bibinfo {author} {\bibfnamefont
  {C.}~\bibnamefont {Zhang}}, \bibinfo {author} {\bibfnamefont
  {H.}~\bibnamefont {Zhang}}, \bibinfo {author} {\bibfnamefont
  {K.}~\bibnamefont {Zhang}}, \bibinfo {author} {\bibfnamefont
  {Y.}~\bibnamefont {Zhang}}, \bibinfo {author} {\bibfnamefont
  {H.}~\bibnamefont {Zhao}}, \bibinfo {author} {\bibfnamefont {Y.}~\bibnamefont
  {Zhao}}, \bibinfo {author} {\bibfnamefont {L.}~\bibnamefont {Zhou}}, \bibinfo
  {author} {\bibfnamefont {C.-Y.}\ \bibnamefont {Lu}}, \bibinfo {author}
  {\bibfnamefont {C.-Z.}\ \bibnamefont {Peng}}, \bibinfo {author}
  {\bibfnamefont {X.}~\bibnamefont {Zhu}}, and\ \bibinfo {author}
  {\bibfnamefont {J.-W.}\ \bibnamefont {Pan}}} (\bibinfo {year} {2022}),\
  \bibfield  {title} {\enquote {\bibinfo {title} {Quantum computational
  advantage via 60-qubit 24-cycle random circuit sampling},}\ }\href
  {https://doi.org/10.1016/j.scib.2021.10.017} {\bibfield  {journal} {\bibinfo
  {journal} {Science Bulletin}\ }\textbf {\bibinfo {volume} {67}},\ \bibinfo
  {pages} {240--245}}\BibitemShut {NoStop}%
\bibitem [{\citenamefont {Zlokapa}\ \emph {et~al.}(2020)\citenamefont
  {Zlokapa}, \citenamefont {Boixo},\ and\ \citenamefont
  {Lidar}}]{zlokapa_boundaries_2020}%
  \BibitemOpen
  \bibfield  {author} {\bibinfo {author} {\bibnamefont {Zlokapa}, \bibfnamefont
  {A}}, \bibinfo {author} {\bibfnamefont {S.}~\bibnamefont {Boixo}}, and\
  \bibinfo {author} {\bibfnamefont {D.}~\bibnamefont {Lidar}}} (\bibinfo {year}
  {2020}),\ \bibfield  {title} {\enquote {\bibinfo {title} {Boundaries of
  quantum supremacy via random circuit sampling},}\ }\href@noop {} {\ }\Eprint
  {https://arxiv.org/abs/2005.02464} {arXiv:2005.02464} \BibitemShut {NoStop}%
\bibitem [{\citenamefont {Zwanenburg}\ \emph {et~al.}(2013)\citenamefont
  {Zwanenburg}, \citenamefont {Dzurak}, \citenamefont {Morello}, \citenamefont
  {Simmons}, \citenamefont {Hollenberg}, \citenamefont {Klimeck}, \citenamefont
  {Rogge}, \citenamefont {Coppersmith},\ and\ \citenamefont
  {Eriksson}}]{zwanenburg_silicon_2013}%
  \BibitemOpen
  \bibfield  {author} {\bibinfo {author} {\bibnamefont {Zwanenburg},
  \bibfnamefont {F~A}}, \bibinfo {author} {\bibfnamefont {A.~S.}\ \bibnamefont
  {Dzurak}}, \bibinfo {author} {\bibfnamefont {A.}~\bibnamefont {Morello}},
  \bibinfo {author} {\bibfnamefont {M.~Y.}\ \bibnamefont {Simmons}}, \bibinfo
  {author} {\bibfnamefont {L.~C.~L.}\ \bibnamefont {Hollenberg}}, \bibinfo
  {author} {\bibfnamefont {G.}~\bibnamefont {Klimeck}}, \bibinfo {author}
  {\bibfnamefont {S.}~\bibnamefont {Rogge}}, \bibinfo {author} {\bibfnamefont
  {S.~N.}\ \bibnamefont {Coppersmith}}, and\ \bibinfo {author} {\bibfnamefont
  {M.~A.}\ \bibnamefont {Eriksson}}} (\bibinfo {year} {2013}),\ \bibfield
  {title} {\enquote {\bibinfo {title} {Silicon quantum electronics},}\ }\href
  {https://doi.org/10.1103/RevModPhys.85.961} {\bibfield  {journal} {\bibinfo
  {journal} {Rev. Mod. Phys.}\ }\textbf {\bibinfo {volume} {85}},\ \bibinfo
  {pages} {961--1019}}\BibitemShut {NoStop}%
\end{thebibliography}%

\end{document}